\documentclass[a4paper, twoside, 11pt]{iiscthes}

\pdfoutput=1

\usepackage{amsmath}
\usepackage{amsfonts}
\usepackage{amssymb}
\usepackage{graphicx}
\usepackage{subfigure}
\usepackage{amsthm}
\usepackage{appendix}
\usepackage[Sonny]{fncychap}
\usepackage{fancyhdr}
\usepackage{fixltx2e}
\usepackage{longtable}
\usepackage[usenames,dvipsnames,svgnames,table]{xcolor}
\usepackage{hyperref}
\usepackage[small]{caption}

\theoremstyle{definition}

\newtheorem{theorem}{Theorem}[section]
\newtheorem{lemma}[theorem]{Lemma}
\newtheorem{proposition}[theorem]{Proposition}
\newtheorem{corollary}[theorem]{Corollary}

\newtheorem{remark}[theorem]{Remark}

\newcommand{\ceiling}[1]{\left\lceil{#1}\right\rceil}
\newcommand{\floor}[1]{\left\lfloor{#1}\right\rfloor}
\newcommand{\nfrac}[2]{\left( \frac{#1}{#2} \right)}
\newcommand{\brac}[1]{\left\{ #1 \right\}}
\newcommand{\bras}[1]{\left[ #1 \right]}
\newcommand{\brap}[1]{\left( #1 \right)}
\newcommand{\fpow}[3]{\left(\frac{#1}{#2}\right)^{#3}}
\newcommand{\dep}{\frac{\delta\epsilon_{a}}{\delta\epsilon_{a} + \epsilon_{V}}}
\newcommand{\depn}{\frac{\delta\epsilon_{a}}{\delta\epsilon_{a} + d}}
\newcommand{\Tau}{\mathcal{T}}
\newcommand{\Exp}{\mathbb{E}}
\newcommand{\Expp}{\mathbb{E}_{\pi}}
\newcommand{\cM}{|\mathcal{M}|}
\newcommand{\mQ}{\mathcal{Q}}

\newcommand{\pig}{\pi_{\gamma}}
\newcommand{\sZ}{\mathbb{Z}_{+}}
\newcommand{\sR}{\mathbb{R}_{+}}
\newcommand{\ExpH}{\mathbb{E}_{\pi_{H}}}
\newcommand{\ExpT}{\mathbb{E}_{\mathcal{T} \vert Q}}
\newcommand{\ExpS}{\mathbb{E}_{\mathcal{S} \vert Q}}
\newcommand{\ExpST}{\mathbb{E}_{\mathcal{S, T} \vert Q}}
\newcommand{\Expsqh}{\mathbb{E}_{S \vert Q, H}}
\newcommand{\Deq}{\stackrel{\Delta} = }
\newcommand{\Qg}{\overline{Q}(\gamma)}
\newcommand{\Pg}{\overline{P}(\gamma)}
\newcommand{\Cg}{\overline{C}(\gamma)}
\newcommand{\Ug}{\overline{U}(\gamma)}
\newcommand{\Ag}{\overline{A}(\gamma)}
\newcommand{\Sg}{\overline{S}(\gamma)}
\newcommand{\Qgk}{\overline{Q}(\gamma_{k})}
\newcommand{\Pgk}{\overline{P}(\gamma_{k})}
\newcommand{\Ugk}{\overline{U}(\gamma_{k})}
\newcommand{\Agk}{\overline{A}(\gamma_{k})}
\newcommand{\Sgk}{\overline{S}(\gamma_{k})}
\newcommand{\Cgk}{\overline{C}(\gamma_{k})}
\newcommand{\sq}{\overline{s}(q)}
\newcommand{\sQ}{\overline{s}(Q)}
\newcommand{\bS}[1]{\boldsymbol{#1}}
\newcommand{\hpq}{\mathcal{Q}_{h}}

\newcommand{\FMC}{FINITE-$\mu$CHOICE}
\newcommand{\INTMC}{INTERVAL-$\mu$CHOICE}
\newcommand{\INTLMC}{INTERVAL-$\lambda\mu$CHOICE}
\newcommand{\INTLC}{INTERVAL-$\lambda$CHOICE}

\DeclareMathOperator*{\argmin}{arg\,min}
\DeclareMathOperator*{\argmax}{arg\,max}
\DeclareMathOperator*{\mini}{minimize}
\DeclareMathOperator*{\maxi}{maximize}

\long\def\symbolfootnote[#1]#2{\begingroup\def\thefootnote{\fnsymbol{footnote}}\footnote[#1]{#2}\endgroup} 

\newcommand{\blankpage}{\newpage
\thispagestyle{empty}
\mbox{}
\newpage
}

\begin{document}

\begin{frontmatter}

\title{On the tradeoff of average delay, average service cost, and average utility for single server queues with monotone policies}
\author{Vineeth Bala Sukumaran}
\submitdate{June 2013}
\dept{Department of Electrical Communication Engineering}
\enggfaculty
\iisclogotrue
\tablespagetrue
\maketitle
\blankpage

\begin{dedication}
\begin{center}
  \emph{\large{To my family.}}
\end{center}
\end{dedication}

\blankpage
\makecontents
\setlength{\parindent}{0in}
\addtolength{\parskip}{2mm}

\blankpage
\acknowledgements

I am very much grateful to my wonderful family, without which this thesis would not have been possible.

I am indebted to \mbox{Prof. Utpal Mukherji} for his generous advice, guidance, and careful review of my work.
I have benefitted a lot, both professionally and personally, by being his student.

I am very grateful to \mbox{Prof. Anurag Kumar} and \mbox{Prof. Rajesh Sundaresan} for their help and insightful comments regarding the work done during my PhD as well as for some of the nicest courses I have attended during my stay at IISc.
I also thank \mbox{Prof. Vinod Sharma} and \mbox{Prof. Vijay Kumar} for their insightful comments and suggestions regarding my PhD work.

Akhil, Arjun, Birenjith, Divya, Deepak (L), Deepak (S), Jithin, Naveen, Nidhin, Sojan, Sreeram, Venu, and Vinodh are wonderful friends and have made my stay here at IISc memorable.
I am grateful to Birenjith, Naveen, Venu, and Vinodh for their generous advice, great technical and even greater non-technical discussions, and being the sink nodes for my non-stop griping.
My lab and department colleagues Arpan, Ashok, Avijeeth, Avijit, Avishek, Bharath, Chandramani, Deekshith, Jobin, Karthik, Krishna Chaitanya, Prakash, Prasad, Lalitha, Manoj, Prem, Rahul, Sayee, Srinidhi, Srinivasan, and Venkatesh have been very generous with their support along the way.
I also thank Ashwin, Jaideep Sir, Manoj, Manu, Parameshwaran, Ravi, Rahul, Simil, Sunil, Sunilkumar Sir, Sandeep, and Vaisakh for the unforgettable times in A mess and C mess.

I am grateful to Govind and Nikhil for their support for all these years.
I sincerely thank Hari, Sreeram, Sunil, Shri Indran Gurukkal, \mbox{Shri Kumar}, \mbox{Shri Manjunath}, Rahul, Guruprasad, Senthil, Srinidhi, Sukesh, Midhun, Mahesh, Sreevalsa, Ravi, \mbox{Mohanan Sir}, and Naveen for encouraging me do things I would have never thought of doing before.
I am grateful for the great food and hospitality of Divya, Birenjith, \mbox{Lekha chechi}, and \mbox{Ajayan Sir}.
I also thank the members of SIMA for their companionship.

I owe a good deal to the staff of A and C messes, Prakurthi, Nesara, J.B., F.C., and IISc gymkhana for keeping me well fed and healthy throughout these years.
I am very thankful for the help given by the staff of ERNET office, Network office, and DRDO-IISc office, especially Boregowda, Mrs. Chandrika, Mahesh, Priyanka, and Savitha.
I am also grateful to the office staff of the Department of Electrical Communication Engineering, especially Mr. Srinivasa Murthy and Mr. Nagaraj, for their help and support.
I also acknowledge the generous financial support from Indian Institute of Science, Defence Research and Development Organization (GoI), and the Ministry of Human Resource Development (GoI).

\begin{abstract}
  In this thesis, we study the optimal tradeoff of average delay, average service cost, and average utility for single server queueing models, with and without admission control.
The continuous time and discrete time queueing models that we consider are motivated by cross-layer models for noisy point-to-point links, with random packet arrivals.
We study the above tradeoff problem for a class of admissible policies, which are monotone and stationary.

The solutions that we obtain for the above tradeoff problem are asymptotic in nature.
For example, suppose we are interested in minimizing the average delay of packets, subject to a constraint on the average service cost of serving the packets.
It is intuitive that to keep the queue \emph{stable} the time average service rate of packets has to equal the time average arrival rate of packets.
This in turn implies that queue stability requires a positive minimum average service cost expenditure.
We obtain asymptotic bounds on the minimum average delay in the asymptotic regime $\Re$ where the average service cost constraint is a small positive $V$ more than the above minimum average service cost required for queue stability.
We note that such asymptotic bounds can be used to obtain a first order characterization of the tradeoff curve, and are useful in identifying \emph{good families} of scheduling policies, such as \emph{buffer partitioning} policies.

In this thesis, we obtain asymptotic lower bounds on the minimum average delay in the regime $\Re$, for the cases for which lower bounds were previously not known, for admissible policies.
The asymptotic characterization of the minimum average delay for admissible policies, for both continuous time and discrete time models, is obtained via new geometric bounds on the stationary probability distribution of the queue length, in the regime $\Re$. 
The restriction to admissible policies, also enables us to obtain an intuitive explanation for the behaviour of the asymptotic lower bounds, using the above geometric bounds on the stationary probability distribution of the queue length.
We observe that the \emph{shape} of the stationary probability distribution, in the regime $\Re$, determines the form of the asymptotic behaviour.

It is common practice to approximate a queueing model, where the queue length evolution is on the non-negative integers, with a queueing model where the queue length evolution is on the non-negative real numbers and the service cost function being strictly convex, for analytical tractability.
We compare the asymptotic bounds which are obtained for the approximate real valued queue evolution model with that of the original integer valued queue evolution model.
We observe that for some cases the average delay does not grow to infinity, in the regime $\Re$, although the real valued approximate queueing model, with a strictly convex cost function, suggests that the average delay should grow without bound in the regime $\Re$.
In other cases where the average delay does grow to infinity in the regime $\Re$, our results illustrate that the approximate model strictly underestimates the behaviour of the tradeoff for the original model unless the service cost function is modelled as the piecewise linear lower convex envelope of the service cost function for the original integer valued queueing model.

The geometric bounds on the stationary probability distribution of the queue length also lead to asymptotic bounds on any optimal admissible policy, in the regime $\Re$.
The asymptotic order bounds are independent of the exact service cost function, and are not available in previous work.
For buffer partitioning policies, the bounds also show how buffer partitions have to scale with $V$.

We then apply the above asymptotic lower bounds to the motivating applications, discussed above.
We develop geometric bounds on the stationary probability distribution for admissible policies to analyse the tradeoff problem in other scenarios, such as: (i)queueing models for N-user single hop communication networks, (ii)queueing models with non-convex service cost functions, and (iii)queueing models with general holding costs.

\end{abstract}

\noindent

\abbreviations
\begin{description}
  \item[IID ] independent and identically distributed
  \item[DMC ] discrete memoryless channel
  \item[TPM ] transition probability matrix
  \item[MC or DTMC ] discrete time Markov chain
  \item[CTMC ] continuous time Markov chain
  \item[EMC ] embedded Markov chain
  \item[SMP ] semi Markov process
  \item[MDP ] Markov decision process
  \item[SMDP ] semi Markov decision process
  \item[CMDP ] constrained Markov decision process
  % \item[OE ] optimality equation
  % \item[DCOE ] discounted cost optimality equation
  \item[ACOE ] average cost optimality equation
  \item[EXH ] exhaustive service policy
  % \item[APP ] approximately optimal policy
  % \item[PGF ] probability generating function
  % \item[HSDPA ] high speed downlink packet access
  % \item[CDMA ] code division multiple access
\end{description}

\conventions
\begin{itemize}
\item An increasing (decreasing) function is a non-decreasing (non-increasing) function. The property of strictly increasing or decreasing is explicitly stated with the qualifier strict.
\item Random variables are denoted by capital letters.
\item Realizations of random variables are denoted by the corresponding small letters or by specifying a sample point/path $\omega$. For example, $X$ and $x$ or $X(\omega)$ respectively.
\item A random or deterministic vector is typeset in \textbf{bold} face.
\item Derivatives will be explicitly shown, e.g. $\frac{d}{dx}$, rather than by using primes.
\item All logarithms are natural logarithms, unless specified.
\end{itemize}

\notations
\begin{longtable}{ll}
$\mathbb{R}_{+}$ & the set of all non-negative real numbers \\
$\mathbb{Z}_{+}$ &  the set of all non-negative integers \\
$\mathbb{I}_{A}$ & indicator function for the event $A$ \\
$\mathbb{E}$ &  expectation with respect to a distribution which is clear from the context \\
$\mathbb{E}_{p}$ & expectation with respect to the distribution p \\
$Pr\{E\}$ &  probability of event $E$ \\
$t$ & continuous time variable \\
$T$ & upper limit for $t, \in \mathbb{R}_+ $ \\
$n$ & discrete time index \\
$N$ & upper limit for $n, \in \mathbb{Z}_+$ \\
$m$ & index variable for special embedded epochs (such as decision instants) \\
$M$ & upper limit for $m, \in \mathbb{Z}_+$ \\
$i$ & index variable for customers \\
$I$ & upper limit for $i, \in \mathbb{Z}_+$ \\
$P_{A}(a)$ & distribution of the random variable A \\
$P_{A|B}(a|b)$ & the distribution of A being a conditioned on B being b \\
$\mathcal{O}$ & $f(x)$ is $\mathcal{O}(g(x))$ if there exists a $c > 0$ such that $\lim_{x \rightarrow 0} \frac{f(x)}{g(x)} \leq c$; $f(x), g(x) \geq 0$ \\
$o$ & $f(x)$ is $o(g(x))$ if for every $c > 0$, $\lim_{x \rightarrow 0} \frac{f(x)}{g(x)} \leq c$; $f(x), g(x) \geq 0$\\
$\Omega$ & $f(x)$ is $\Omega(g(x))$ if there exists a $c > 0$ such that $\lim_{x \rightarrow 0} \frac{f(x)}{g(x)} \geq c$; $f(x), g(x) \geq 0$ \\
$\omega$ & $f(x)$ is $\omega(g(x))$ if for every $c > 0$, $\lim_{x \rightarrow 0} \frac{f(x)}{g(x)} \geq c$ or $g(x) = o(f(x))$; $f(x), g(x) \geq 0$ \\
$\Theta$ & $f(x)$ is $\Theta(g(x))$ if $f(x) = \mathcal{O}(g(x))$ and $f(x) = \Omega(g(x))$; $f(x), g(x) \geq 0$ \\
\end{longtable}

\end{frontmatter}

\setlength{\parindent}{0in}
\addtolength{\parskip}{2mm}

\addtocontents{toc}{\protect\setcounter{tocdepth}{2}}

\blankpage
\setcounter{page}{1}
\chapter{\textbf{Introduction}}
In this thesis, we study the optimal tradeoff of average delay, average service cost, and average utility for single server queueing models, with and without admission control.
The continuous time and discrete time queueing models that we consider are motivated by cross-layer models for noisy point-to-point links.
The features that we model are: (i)random packet arrivals, (ii)control of service and/or arrival rates, (iii)packet service costs and/or utility, and (iv)fading at slow and fast time scales. 
Our objective is to characterize the minimum average delay of the packets, under an upper bound constraint on the average service cost and/or a lower bound constraint on the average throughput for systems with admission control.
We are also motivated by the problem of characterizing the minimum average delay of randomly arriving message symbols which are transmitted over a noisy point-to-point link with no admission control, under an upper bound constraint on the average error rate of the message symbols.
Such tradeoff problems arise in the study of cross layer scheduling algorithms for wireless communication networks \cite{neely} or in the study of processor speed scaling \cite{chen}.

In this thesis, we consider the performance of scheduling algorithms which optimally trade off average delay with other performance measures, as in \cite{berry}, \cite{neely}, and \cite{munish}.
Related problems include the design of cross layer scheduling algorithms for: (a) \emph{stabilizing} a communication network, as in \cite{ephremides}, \cite{ephremides_tassiulas}, \cite{neely_thesis}, \cite{neely_tassiulas}, \cite{chaporkar}, \cite{utpalunequal}, \cite{sayee_thesis}, \cite{sayeepap1}, \cite{sayeepap2}, and \cite{sayeeallerton}, or (b) minimizing a delay measure, as in \cite{yehdelay}, \cite{musy}, and \cite{javidi_ehsan}.

We study the above tradeoff problem for a class of monotone policies, which we call admissible policies.
Monotone policies are stationary policies, i.e., the service rate (and the number of packets admitted, for queues with admission control) at a time is a function\footnote{this function could be randomized.} only of the current state of the queue\footnote{e.g., the current queue length or the current queue length and an auxiliary state variable such as the fade state.} rather than a function of the whole history of evolution as well as the current state of the queue.
For monotone policies, the expected service rate for a queue length is non-decreasing as a function of the queue length.
Intuitively, if the average delay is to be minimized subject to a constraint on the average service cost, for deterministic stationary policies, as the queue length increases, the service rate should also increase.
We note that intuitively although the service rate should increase, the amount by which the service rate increases depends on the corresponding increase in the service cost.
So in practice, monotone policies are usually used.
Furthermore, in certain cases it can be shown that the \emph{optimal} policy for the above tradeoff problem is in fact monotone\footnote{The monotonicity property is obtained using a Markov decision theoretic formulation of the tradeoff problem as in \cite{weber}, \cite{george}, or \cite{berry}}.
Motivated by this reason, as well as the above intuition, we consider the tradeoff problem for the class of admissible policies only.
The class of admissible policies is a subset of the class of monotone policies, possessing some additional properties, which makes their analysis more amenable.

We consider several variations of the above tradeoff problem.
The solutions to the tradeoff problems that we address in this thesis are of the following form.
Consider an example of an infinite buffer queueing model with service rate control, but no admission control.
We are interested in minimizing the average delay subject to a constraint on the average service cost.
It is intuitive that to keep the queue \emph{stable} the time average service rate of packets has to equal the time average arrival rate of packets.
This in turn implies that queue stability requires a positive minimum average service cost expenditure.
We obtain asymptotic bounds on the minimum average delay in the asymptotic regime $\Re$ where the average service cost constraint is a small positive $V$ more than the above minimum average service cost required for queue stability.
We note that such asymptotic bounds can be used to obtain a first order characterization of the tradeoff curve.
Furthermore, such bounds can also be used to identify good families of scheduling policies, as in \cite{berry}.

We note that asymptotic upper bounds (see \cite{neely}) as well as asymptotic lower bounds (see \cite{berry} and \cite{neely_mac}) on the minimum average delay in the regime $\Re$ are available for a variety of queueing models.
However, asymptotic lower bounds are not known in many cases.
We obtain asymptotic lower bounds for the minimum average delay in the regime $\Re$, for these cases, for admissible policies.
Additionally, we also obtain \emph{asymptotic bounds} on the \emph{structure} of admissible policies which achieve the above asymptotic lower bounds.
The method by which we derive these bounds, which is different from previous approaches, also leads to \emph{geometric} bounds on the stationary probability of the queue length, in the regime $\Re$.
These bounds have the added advantage of directly providing intuition for the behaviour of asymptotic lower bounds in the regime $\Re$.

Using the above asymptotic lower bounds on the minimum average delay and the already available asymptotic upper bounds we obtain a complete asymptotic characterization of the tradeoff between average delay and average service cost for several single server queueing models, in the regime $\Re$, for admissible policies.
We start with an informal introduction to the queueing models and the tradeoff problems considered in this thesis.

\section{Introduction to the queueing models and the tradeoff problem}
We note that the primary features of the point-to-point links are: (i)random packet arrivals, (ii)control of service and/or arrival rates, (iii)packet service costs and/or utility, and (iv)fading at slow and fast time scales. 
These features are captured by two single server controlled queueing models in this thesis.
Initial insights into the asymptotic behaviour of the tradeoff are obtained by studying a continuous time queueing model with exponential interarrival times and exponential service requirements, in Chapters 2 and 3.
These insights are used in characterizing the asymptotic behaviour of the tradeoff for discrete time models in the following chapters.

For the purposes of this introductory discussion, a queue is defined to be stable (or more precisely, mean rate stable as in \cite{neely}) under a \emph{policy} if the time average service rate equals the time average arrival rate, although in later chapters we use stronger notions of stability.

In the asymptotic regime $\Re$, asymptotic upper and lower bounds on the average delay can be obtained from asymptotic upper and lower bounds on the average queue length by applying Little's law with lower and upper bounds on the average throughput respectively.
Therefore, we focus on the average queue length instead of average delay throughout this thesis.

\subsection{A continuous time state dependent M/M/1 model}
\begin{figure}[h]
  \centering
  \includegraphics[width=120mm,height=60mm]{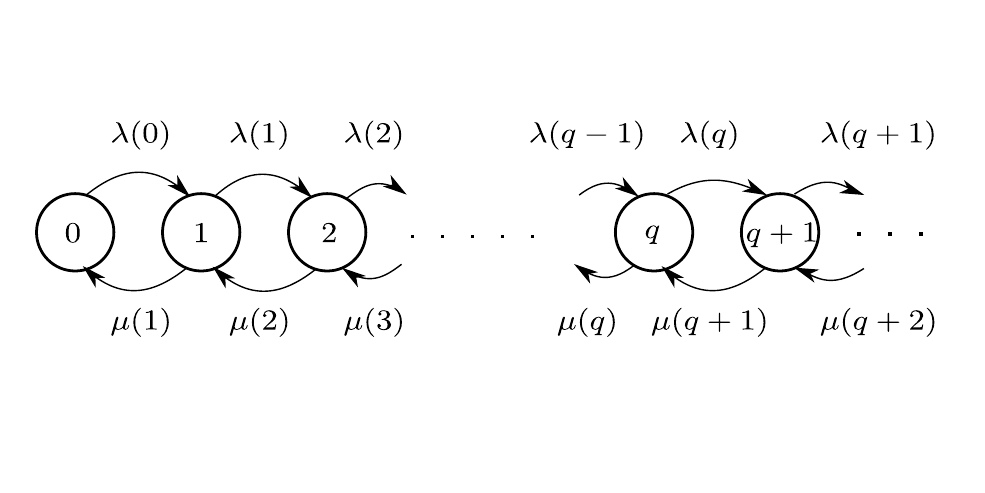}
  \vspace{-0.5in}
  \caption{Transition diagram for the continuous birth-death queueing model}
  \label{introduction:fig:mm1}
  \vspace{0.2in}
\end{figure}

The state dependent M/M/1 queueing model is a birth death process with state being the queue length, and with state dependent birth (or arrival) rates and death (or service) rates, as shown in the transition diagram in Figure \ref{introduction:fig:mm1}.
We note that the control policy $\gamma$ is the choice of the arrival rates $(\lambda(q), q \in \sZ)$ and the service rates $(\mu(q), q \in \sZ \setminus \brac{0})$, as a function of the queue length $q$, from sets $\mathcal{X}_{\lambda}$ and $\mathcal{X}_{\mu}$ respectively.

The average queue length $\Qg$ for a particular policy $\gamma$ is the time average of the expectation of the queue length $Q(t)$, where $Q(t)$ is the state of the birth death process at time $t$ under $\gamma$.
We assume that utility is accrued at the rate of $u(\lambda(Q(t)))$ and service cost is incurred at the rate of $c(\mu(Q(t)))$, where $u(\lambda)$ is a non-decreasing concave function of $\lambda$ and $c(\mu)$ is a non-decreasing convex function of $\mu$.
For the policy $\gamma$, the average utility rate $\Ug$ and average service cost rate $\Cg$ are defined as the time averages of the expectation of the utility rate $u(\lambda(Q(t)))$ and the expectation of the service cost rate $c(\mu(Q(t)))$ respectively.
The general tradeoff problem that we consider is the minimization of $\Qg$, subject to a lower bound constraint $u_{c}$ on $\Ug$, and an upper bound constraint $c_{c}$ on $\Cg$, over all policies $\gamma$, i.e.,
\begin{eqnarray}
  \mini_{\gamma} & & \Qg, \nonumber \\
  \text{ such that } & & \Ug \geq u_{c} \text{ and } \Cg \leq c_{c}.
  \label{introduction:eq:mm1_genTradeoff}
\end{eqnarray}
In the following, problems in the above constrained optimization form are called constrained tradeoff problems.
We note that one way to analyse the above constrained tradeoff problem is to consider its unconstrained Lagrange dual, the dual function of which is as follows:
\begin{eqnarray} 
  \mini_{\gamma} & & \Qg + \beta_{1} (\Cg - c_{c}) - \beta_{2} (\Ug - u_{c}),
  \label{introduction:eq:mm1_genTradeoff_dual}
\end{eqnarray}
where $\beta_{1}$ and $\beta_{2}$ are non-negative Lagrange multipliers.
In the following, such unconstrained dual problems are called unconstrained tradeoff problems.

We primarily consider the case where $\lambda(q)$ is fixed to be a $\lambda \in \mathcal{X}_{\lambda}$ such that $u(\lambda) \geq u_{c}$, so that the tradeoff problem reduces to
\begin{eqnarray}
  \mini_{\gamma} & & \Qg, \nonumber \\
  \text{ such that } & & \Cg \leq c_{c}.
  \label{introduction:eq:mm1_specialTradeoff}
\end{eqnarray}
We note that as in the case of \eqref{introduction:eq:mm1_genTradeoff}, we have the following unconstrained dual function:
\begin{eqnarray}
  \mini_{\gamma} & & \Qg + \beta_{1} (\Cg - c_{c}),
  \label{introduction:eq:mm1_specialTradeoff_dual}
\end{eqnarray}
where $\beta_{1} \geq 0$.

The optimal values of \eqref{introduction:eq:mm1_genTradeoff} and \eqref{introduction:eq:mm1_specialTradeoff}, as a function of their respective constraints, are referred to as the tradeoff curve in the following discussion.
We note that in the context of communication networks, the function $u(.)$ is usually assumed to be linear, so that $\Ug$ is the average throughput.
But for other applications, such as those in \cite{atamm1}, $u(.)$ could be a strictly concave function.

The optimization problem \eqref{introduction:eq:mm1_genTradeoff_dual} and its variants (such as \eqref{introduction:eq:mm1_specialTradeoff_dual}) have been formulated as Markov decision problems (MDP) and analyzed by many authors, e.g. \cite{atamm1}, \cite{weber}, and \cite{george}.
They show that there exists a monotone optimal policy $\gamma^*(\beta_{1}, \beta_{2})$ for \eqref{introduction:eq:mm1_genTradeoff_dual} ($\gamma^*(\beta_{1}, \beta_{2})$ is such that $\lambda(q)$ is a non-increasing function and $\mu(q)$ is a non-decreasing function of $q$).
Also, from \cite{ma}, we have that if $c_{c} = \overline{C}(\gamma^*(\beta_{1}, \beta_{2}))$ and $u_{c} = \overline{U}(\gamma^*(\beta_{1}, \beta_{2}))$, then $\gamma^*(\beta_{1}, \beta_{2})$ is also optimal for \eqref{introduction:eq:mm1_genTradeoff}.
Therefore, for at least such values of $c_{c}$ and $u_{c}$ there exist monotone optimal policies for \eqref{introduction:eq:mm1_genTradeoff}.
This motivates us to consider \eqref{introduction:eq:mm1_genTradeoff} for a class of admissible policies, which are monotone.

\textbf{The asymptotic regime $\Re$ :}
We consider the tradeoff problems \eqref{introduction:eq:mm1_genTradeoff} and \eqref{introduction:eq:mm1_specialTradeoff} in the asymptotic regime $\Re$ where the average service cost constraint $c_{c}$ is arbitrarily close to the minimum average service cost required for stability.
It turns out that the minimum average service cost required for stability is $c(u^{-1}(u_{c}))$ (where the inverse $u^{-1}(.)$ of $u(.)$ is assumed to exist) for \eqref{introduction:eq:mm1_genTradeoff} and $c(\lambda)$ for \eqref{introduction:eq:mm1_specialTradeoff}.
Therefore, the asymptotic regimes $\Re$ for problems \eqref{introduction:eq:mm1_genTradeoff} and \eqref{introduction:eq:mm1_specialTradeoff} are defined as the regime in which $c_{c} \downarrow c(u^{-1}(u_{c}))$ and $c_{c} \downarrow c(\lambda)$ respectively.

\subsection{A discrete time queueing model}
\begin{figure}[h]
  \centering
  \includegraphics[width=120mm,height=60mm]{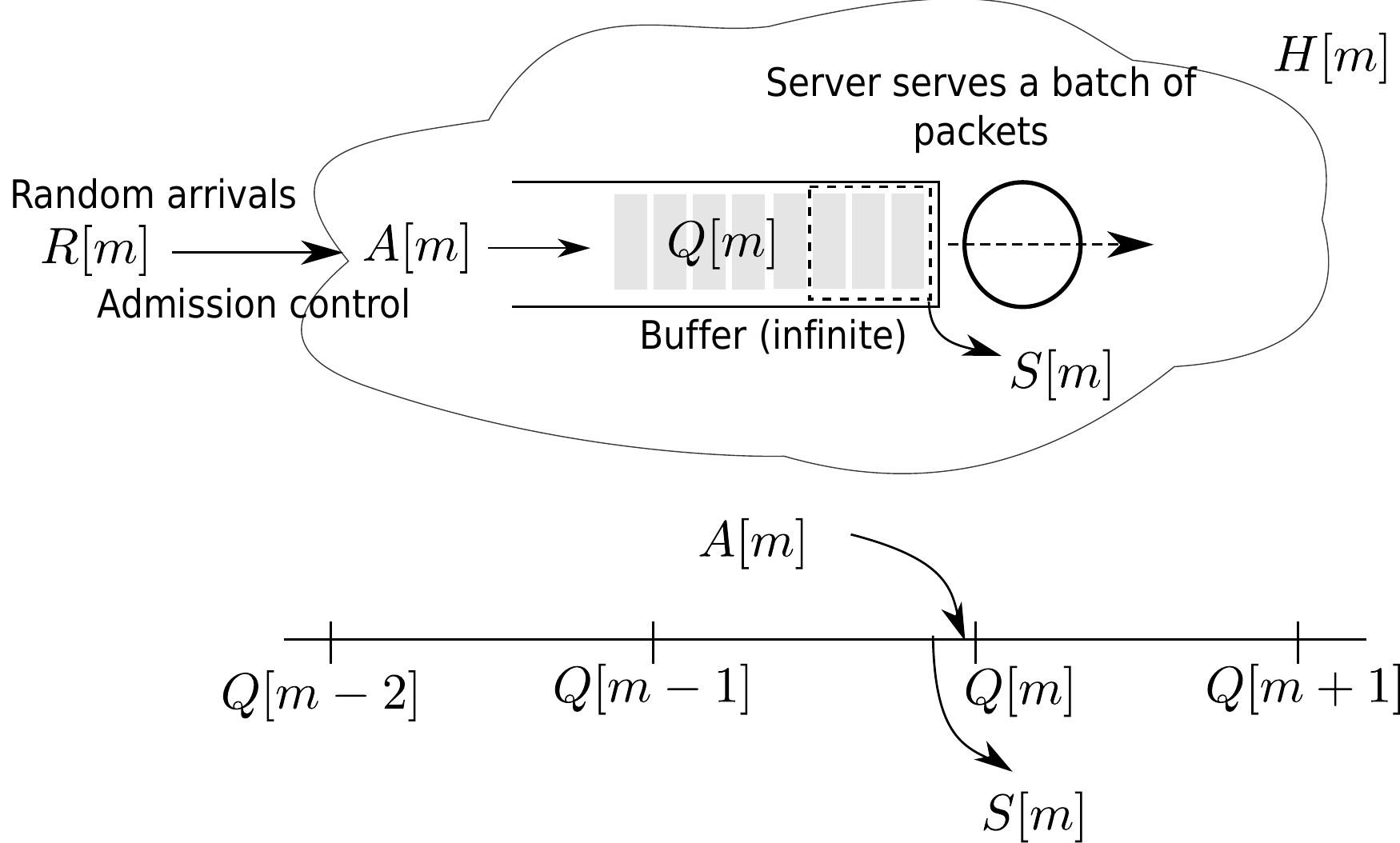}
  \caption{The discrete time single server queueing model with a single queue}
  \label{introduction:fig:discretetimemodel}
\end{figure}

In Chapters 4, 5, and 6, we consider discrete time single server queueing models with random batch arrivals and batch service.
We now introduce a general form of this model, shown in Figure \ref{introduction:fig:discretetimemodel}.
In each slot $m \geq 1$, we assume that a random number $R[m]$ of packets arrives into the system, with an arrival rate of $\lambda$ per slot.
In the $m^{th}$ slot, $A[m] \leq R[m]$ arrivals are admitted into the queue, which is assumed to have infinite buffer space.
We assume that there is a random process $(H[m])$ which models the environment in which the queue is evolving, e.g., this could be the fade state for a point-to-point wireless link, which stays constant for the duration of a slot.
We assume that $H[m]$ is known at the start of every slot $m$.
The queue length, at the slot boundaries, evolves according to the evolution equation:
\begin{eqnarray*}
  Q[m] = Q[m - 1] - S[m] + A[m], m \geq 1,
\end{eqnarray*}
where $Q[0] = q_{0}$, $S[m] \leq \min(S_{max}, Q[m - 1])$, and $A[m] \leq R[m]$.
We note that $S[m]$ and $A[m]$ are the control variables.
The service batch size $S[m]$ in slot $m$ is assumed to be chosen as a randomized function of the history
\begin{eqnarray*}
  \sigma[m] = (q_{0}, H[1], S[1], R[1], Q[1], H[2], S[2], R[2], Q[2], \dots, Q[m - 2], H[m - 1]),
\end{eqnarray*}
and the current queue length $Q[m - 1]$ and fade state $H[m]$.
while the arrival batch size $A[m]$ in slot $m$ is assumed to be chosen as a randomized function of the history $\sigma[m]$, $Q[m - 1]$, $H[m]$, and the current number of arrivals $R[m]$.
The choice of the sequence $((S[m], A[m]), m \geq 1)$ constitutes the policy $\gamma$ for this discrete time model.

The average queue length $\Qg$ for a policy $\gamma$ is defined to be the time average of the expectation of the queue length $Q[m]$.
We assume that service cost is incurred at the rate of $P(H[m], S[m])$ in slot $m$, where $P(h, s)$ is a service cost function which is assumed to be non-decreasing and convex in $s$ for every $h$.
We note that the cost function $P(h, s)$ models the cost incurred in transmission of packets, e.g., $P(h, s)$ could be the expected number of packets that are received in error when a batch of $s$ packets are jointly encoded and transmitted when the environment state is $h$, or $P(h, s)$ could be the power expended in transmission of $s$ packets when the fade state is $h$.
The average service cost $\Pg$ for $\gamma$ is defined as the time average of the expectation of the service cost $P(H[m], S[m])$.

We define the average throughput $\overline{A}(\gamma)$ for $\gamma$ as the time average of the expectation of the admitted arrival batch size $A[m]$.
The performance measure that we are interested in is the utility of the average throughput, $u(\overline{A}(\gamma))$, achieved by $\gamma$, where $u(.)$ is a non-decreasing and concave utility function.
The general tradeoff problem that we consider is the minimization of $\Qg$ subject to a lower bound constraint $u_{c}$ on $u(\overline{A}(\gamma))$ and an upper bound constraint $P_{c}$ on $\Pg$ over all policies $\gamma$, i.e.,
\begin{eqnarray}
  \mini_{\gamma} & & \Qg, \nonumber \\
  \text{ such that } & & u(\Ag) \geq u_{c} \text{ and } \Pg \leq P_{c}.
  \label{introduction:eq:dt_genTradeoff}
\end{eqnarray}
As for the state dependent M/M/1 model, we have the following unconstrained dual function:
\begin{eqnarray}
  \mini_{\gamma} & & \Qg + \beta_{1}\brap{\Pg - P_{c}} - \beta_{2}\brap{u(\Ag) - u_{c}},
  \label{introduction:eq:dt_genTradeoff_dual}
\end{eqnarray}
where $\beta_{1}$ and $\beta_{2}$ are non-negative.

As for the state dependent M/M/1 model, we primarily consider the case where $A[m] = R[m]$, i.e., with no admission control.
Then the tradeoff problem is:
\begin{eqnarray}
  \mini_{\gamma} & & \Qg, \nonumber \\
  \text{ such that } & & \Pg \leq P_{c},
  \label{introduction:eq:dt_specialTradeoff}
\end{eqnarray}
where we have assumed that $u_{c}$ is such that $\lambda \geq u^{-1}(u_{c})$ (where the inverse $u^{-1}(.)$ of $u(.)$ is assumed to exist), so that for stable policies the utility constraint is satisfied.
Similar to \eqref{introduction:eq:dt_genTradeoff_dual}, we have the unconstrained dual function:
\begin{eqnarray}
  \mini_{\gamma} & & \Qg + \beta_{1}\brap{\Pg - P_{c}},
  \label{introduction:eq:dt_specialTradeoff_dual}
\end{eqnarray}
where $\beta_{1} \geq 0$.
The optimal values of \eqref{introduction:eq:dt_genTradeoff} and \eqref{introduction:eq:dt_specialTradeoff} as a function of their respective constraints are referred to as the tradeoff curve in the following discussion.

As for the state dependent M/M/1 model, for the discrete time model, again using a MDP formulation, it is possible to show that (e.g. \cite{berry}, \cite{munish}, and \cite{agarwal}) there exists a stationary monotone optimal policy for \eqref{introduction:eq:dt_genTradeoff_dual} for each pair of $\beta_{1}$ and $\beta_{2}$.
Again, if $u_{c}$ and $P_{c}$ are respectively equal to the utility and average service cost for the above optimal policy (for some $\beta_{1}$ and $\beta_{2}$), then this monotone policy is also optimal (see \cite{ma}) for \eqref{introduction:eq:dt_genTradeoff}.
This motivates us to consider \eqref{introduction:eq:dt_genTradeoff} for a class of admissible policies, which are monotone.

\textbf{The asymptotic regime $\Re$ :}
We consider the tradeoff problems \eqref{introduction:eq:dt_genTradeoff} and \eqref{introduction:eq:dt_specialTradeoff} in the asymptotic regime $\Re$ where the average service cost constraint $P_{c}$ is arbitrarily close to the minimum average service cost required for stability.
It turns out that this minimum average service cost required for stability is a function of $u_{c}$ and $\lambda$ for \eqref{introduction:eq:dt_genTradeoff} and \eqref{introduction:eq:dt_specialTradeoff} respectively.
Because of the similarities in its properties with those of $c(.)$ for the state dependent M/M/1 model, the minimum average service cost required for stability is denoted as $c(.)$ for \eqref{introduction:eq:dt_genTradeoff} and \eqref{introduction:eq:dt_specialTradeoff} also.
We note that if there is only one environment state, say $h_{0}$, then $c(s) = P(h_{0}, s)$.
The asymptotic regime $\Re$ for problems \eqref{introduction:eq:dt_genTradeoff} and \eqref{introduction:eq:dt_specialTradeoff} is defined as the regime in which $P_{c} \downarrow c(u^{-1}(u_{c}))$ and $P_{c} \downarrow c(\lambda)$ respectively.

\begin{figure}[h]
  \centering
  \includegraphics[width=120mm,height=60mm]{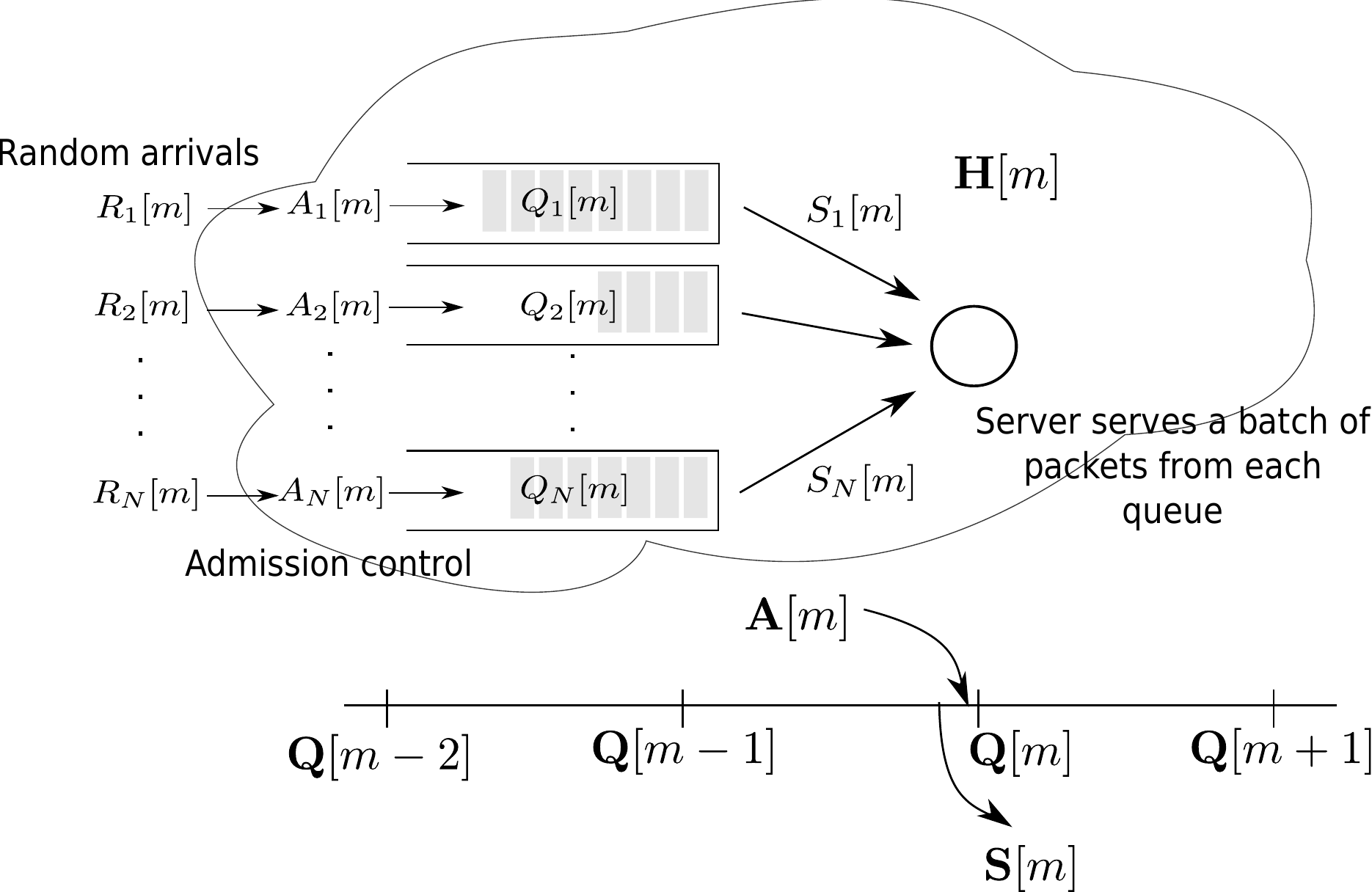}
  \caption{The discrete time single server queueing model with $N$ queues }
  \label{introduction:fig:discretetimemodel_mac}
\end{figure}

\textbf{A multiqueue model :}
We also consider a multiqueue single server queueing model, with $N$ queues being served by a single server as shown in Figure \ref{introduction:fig:discretetimemodel_mac}.
The model is a straightforward generalization of the single queue model discussed above.
We assume that there is an environment variable $H_{n}[m]$ associated with the $n^{th}$ queue.
The vector of $N$ environment variables is denoted as $\bS{H}[m] = (H_{1}[m], \dots, H_{N}[m])$.
The vector of $N$ queue lengths at the start of slot $m$ is denoted as $\bS{Q}[m - 1] = (Q_{1}[m - 1], \dots, Q_{N}[m - 1])$.
We note that in this case the batch service vector, $\bS{S}[m]$, is a vector function of the history $\sigma[m]$ for the $N$ queues, the current queue length vector $\bS{Q}[m - 1]$, and the current environment state $\bS{H}[m]$ as in the single queue case.
The batch of arrivals which are admitted, $\bS{A}[m]$, is a vector function of the history $\sigma[m]$ for the $N$ queues, $\bS{Q}[m - 1]$, $\bS{H}[m]$, and the vector of current arrivals $\bS{R}[m]$.

The service cost is a scalar function $P(\bS{h}, \bS{s})$ of the service vector $\bS{s}$ and the environment vector $\bS{h}$.
For a particular policy $\gamma$, we are interested in the total average queue length $\overline{Q}(\gamma)$, which is the sum of the average queue lengths for the $N$ queues.
The average service cost is the time average of the expectation of $P(\bS{H}[m], \bS{S}[m])$.
We also assume that there are individual lower bound constraints on the $u(\overline{A}_{n}(\gamma))$ for each queue.
Other definitions are straightforward extensions of the definitions for the single queue model.
The asymptotic regime $\Re$ for the multiqueue model is similar to that for the single queue case.
We consider the problem of minimizing the average queue length in the asymptotic regime $\Re$, where the scalar cost constraint approaches the minimum average power required for mean rate stability.

\section{Literature survey}
Tradeoff problems for both continuous time and discrete time single server queueing models has been addressed by many researchers.
To review what is known for such tradeoff problems, let us consider the problem \eqref{introduction:eq:mm1_specialTradeoff}.
One of the first questions that can be asked is that of feasibility; for what values of $c_{c}$ are there feasible solutions to \eqref{introduction:eq:mm1_specialTradeoff}?
Such questions are commonly answered using Lyapunov drift arguments as in, \cite{neely} and \cite{neely_thesis}.
The next question that can be asked is that of the existence of an optimal policy for \eqref{introduction:eq:mm1_specialTradeoff}.
Such questions are commonly answered by posing the problem \eqref{introduction:eq:mm1_specialTradeoff} as a constrained Markov decision problem (CMDP) as in \cite{sennott} or \cite{altman}.
Under certain technical assumptions, the existence of an optimal policy which is also stationary can be shown using results as in \cite{altman}, \cite{hernandez}, or \cite{hernandez_2}.
In certain cases, it may be possible to show that there is a stationary deterministic optimal policy for \eqref{introduction:eq:mm1_specialTradeoff}\footnote{Stationary deterministic optimal policies are such that at a time, the service rate (and the arrival rate for models with admission control) is chosen as a deterministic function of the system state.}.
In such cases, it is also possible to convert the CMDP \eqref{introduction:eq:mm1_specialTradeoff} into an unconstrained Markov decision problem (MDP) using suitable Lagrange multipliers as in \eqref{introduction:eq:mm1_specialTradeoff_dual}.
For the state dependent M/M/1 model, the approach is then to find an optimal policy for the above MDP.
We note that the optimal policy for \eqref{introduction:eq:mm1_specialTradeoff} specifies the optimal service rate $\mu^*(q)$ and arrival rate $\lambda^*(q)$ as a function of queue length, to operate the system in order to minimize the time average of the single stage cost rate.
Characterization of the optimal policy using the MDP approach yields structural properties, which are useful in reducing the search space for the optimal policy, for example see \cite{weber}, \cite{george}, \cite{ata_pcstatic}, and \cite{atamm1}.
Surveys of the above approach can be found in \cite{koole} and \cite{sennott_text}.

We first review the results which are available for the continuous time model.
In most cases, a monotonicity property of the optimal policy for \eqref{introduction:eq:mm1_genTradeoff_dual} is obtained, i.e., $\mu^*(q)$ is a non-decreasing function of $q$ and $\lambda^*(q)$ is a non-increasing function of $q$.
Stidham and Weber \cite{weber} show that $\mu^*(q)$ is non-decreasing and $\lambda^*(q)$ is non-increasing, for a state dependent M/G/1 model, where the objective is to minimize the expected total cost from any initial queue length by serving customers until the queue length is zero, which is related to the average cost criterion.
George and Harrison \cite{george} show that $\mu^*(q)$ is non-decreasing in $q$, for a state dependent M/M/1 model with a Poisson arrival process of rate $1$, for an MDP of the form in \eqref{introduction:eq:mm1_specialTradeoff_dual}.
Similar results have also been obtained by Ata in \cite{ata_pcstatic}, and Ata and Shneorson in \cite{atamm1}.
This monotonicity property of $\mu^*(q)$ and $\lambda^*(q)$ is used to motivate the definition of \emph{admissible}\footnote{Admissible policies are monotone policies. Since admissible policies are stationary, the queue evolution process under admissible policies is a Markov process. Admissible policies are monotone policies which are such that the Markov queue evolution process possesses \emph{nice} properties such as aperiodicity, irreducibility, positive recurrence, and finite mean queue length.} policies in Chapters 2 and 3, which are policies with non-decreasing service rates $\mu(q)$ and non-increasing arrival rates $\lambda(q)$ as a function of $q$.
We then obtain an asymptotic characterization of the tradeoff problems \eqref{introduction:eq:mm1_genTradeoff} and \eqref{introduction:eq:mm1_specialTradeoff} for the class of admissible policies.
We note that whenever the solution of \eqref{introduction:eq:mm1_genTradeoff} coincides with that of its dual (obtained by optimizing \eqref{introduction:eq:mm1_genTradeoff_dual} over $\beta_{1}$ and $\beta_{2} \geq 0$), then the optimal policy for \eqref{introduction:eq:mm1_genTradeoff} is admissible.

For the discrete time model, from a CMDP formulation for \eqref{introduction:eq:dt_specialTradeoff}, it can be shown that there exists an optimal policy which chooses the service batch size $S[m]$ as a function $S^*(Q[m - 1], H[m])$ of the current queue length and environment state, if the arrival process $(R[m])$ and the environment process $(H[m])$ are IID.
The optimal rate or batch size $S^*(q, h)$ which has to be served as a function of the current queue length $q$ and environment state $h$, so as to minimize the average queue length for a given constraint on the average service cost, can be characterized.
Berry and Gallager \cite{berry}, Collins and Cruz \cite{collins}, Agarwal et al. \cite{agarwal}, and Goyal et.al. \cite{munish} consider a fading point to point link with no admission control, where they study the tradeoff problem \eqref{introduction:eq:dt_specialTradeoff}.
They use a Lagrangian relaxation of the CMDP as in \eqref{introduction:eq:dt_specialTradeoff_dual} to obtain that in many cases, $S^*(q, h)$ is a deterministic function $s^*(q, h)$ which is monotonically non-decreasing in $q$ for every $h$.
The monotonicity property of $s^*(q, h)$ is used in motivating the definition of admissible policies in Chapters 4, 5, and 6.
We obtain an asymptotic characterization of the tradeoff problems \eqref{introduction:eq:dt_genTradeoff} and \eqref{introduction:eq:dt_specialTradeoff} for the class of admissible policies.
We note that whenever the solution of \eqref{introduction:eq:dt_genTradeoff} coincides with that of its dual (obtained by optimizing \eqref{introduction:eq:dt_genTradeoff_dual} over $\beta_{1}$ and $\beta_{2} \geq 0$), then the optimal policy for \eqref{introduction:eq:dt_genTradeoff} is admissible. 
A similar observation holds for the solution of the tradeoff problem \eqref{introduction:eq:dt_specialTradeoff}.

Going beyond the above monotonicity property of $s^*(q, h)$, Berry and Gallager \cite{berry} also characterized the asymptotic order behaviour of the tradeoff curve in the regime of large average delay.
It was shown that when the average delay is allowed to be large, the average transmitter power can be made arbitrarily close to $c(\lambda)$.
It was also shown that if the power $P(h, s)$ expended in transmission of $s$ packets is a strictly convex function of $s$ for fixed fade state $h$, then the average queue length is $\Omega\nfrac{1}{\sqrt{V}}$ if the average transmitter power constraint is $V$ more than $c(\lambda)$ as $V \downarrow 0$ (this is known as the Berry-Gallager lower bound).
This asymptotic characterization of the average transmitter power was motivated by the asymptotic characterization of the average distortion of an information source obtained by Tse \cite{tse_thesis}.

We note that the order behaviour provides a first order characterization of the tradeoff curve.
Furthermore, the order characterization provides a criterion to identify a \emph{good} family\footnote{A family of transmission policies is a set of policies with common structure, e.g., a set of parametrized policies that do not serve below a threshold (parameter) while serving a particular batch size (another parameter) above the threshold.} of transmission policies.
The authors in \cite{berry} suggest that the family of \emph{buffer partitioning} policies achieves the $\frac{1}{D^{2}}$ order bound but were unable to prove this.
Buffer partitioning policies partition the buffer into two regions, and use an average batch service rate less than $\lambda$ in the lower region and an average batch service rate greater than $\lambda$ in the higher region to drive the average queue length towards the partitioning value.

Neely \cite{neely_mac} extended the Berry-Gallager lower bound to single hop networks and presented a backpressure based (TOCA) algorithm which achieves the lower bound to within a logarithmic factor, i.e., the algorithm achieves an average delay of $\mathcal{O}\brap{\frac{1}{\sqrt{V}} \log\nfrac{1}{V}}$ when the average transmitter power is at most $V$ more than $c(\lambda)$.
Neely \cite{neely_mac} also observed that if the transmitter power $P(h, s)$ is a piecewise linear function of $s$ for every $h$, then the above rate of increase of the average delay can be improved.
We note that if $P(h, s)$ is piecewise linear in $s, \forall h$, then the function $c(\lambda)$ is a piecewise linear function of $\lambda$.
If $(\lambda,c(\lambda))$ lies on a linear portion of the function $c(.)$, it was observed \cite{neely_mac} that there is a family of policies for which the average delay is $\mathcal{O}\brap{\log\nfrac{1}{V}}$ if the average transmitter power is $V$ more than $c(\lambda)$.
It was also observed \cite{neely_mac} that for all values of $\lambda$ and $P(h, s)$ convex in $s, \forall h$, there is a family of policies for which the average delay is $\mathcal{O}\nfrac{1}{V}$ if the average transmitter power is $V$ more than $c(\lambda)$.

In \cite{neely_utility}, Neely considered the problem \eqref{introduction:eq:dt_genTradeoff}, with $u(x) = x$ and $P(h, s)$ a strictly convex function in $s$ for every $h$.
He observed that if the transmitter is allowed to drop a non-zero fraction of the customers arriving into the queue, such that $\Ag \geq \rho\lambda, 0 < \rho < 1$, then the average queue length grows only as $O\left(\log\left(\frac{1}{V}\right)\right)$ rather than $\mathcal{O}\brap{\frac{1}{\sqrt{V}} \log\nfrac{1}{V}}$ if the average transmitter power is $V$ more than $c(\lambda)$.
Extensions to more general networks and other formulations can be found in \cite{neely}.

We note that an order optimality result was obtained by Ramaiyan \cite{venkithesis} for a particular birth death queueing model, for the average queue length at the relay node for a two-way relay link using network coding, in the regime $\Re$, for monotone policies.
Ramaiyan et al. \cite{venkialtman} also obtained the optimal tradeoff of average queueing delay and average transit delay for a two-hop vehicular relay network.
Asymptotic upper bounds on the minimum average delay for general wireless networks using network coding, under an average power constraint, was obtained in \cite{chaporkar_alex}.

An asymptotic upper bound for the tradeoff curve corresponding to \eqref{introduction:eq:dt_specialTradeoff} has been obtained for the case where $(A[m])$ and $(H[m])$ are ergodic Markov processes in \cite[Section 4.9]{neely} and \cite{huang_neely}.
In \cite[Theorem 4.12]{neely} and \cite{huang_neely}, it has been shown that if $(A[m])$ and $(H[m])$ are ergodic Markov processes, then for a sequence of Quadratic Lyapunov Algorithm (QLA) policies, parametrized by a sequence $V \downarrow 0$, the average queue length is $\mathcal{O}\nfrac{1}{V}$ for an average power $V$ more than $c(\lambda)$.
Order optimality has also been explored for finite buffer systems.
In \cite[Chapter 6]{berry_thesis} it is shown that for a finite buffer discrete time queueing model, as the buffer size $B$ goes to infinity, for any sequence of policies such that the buffer overflow probability is $o\brap{\frac{1}{B^{2}}}$, the average service cost is at least $\Omega\brap{\frac{1}{B^{2}}}$ more than $c(\lambda)$.

We note that, for the tradeoff problems \eqref{introduction:eq:dt_genTradeoff} and \eqref{introduction:eq:dt_specialTradeoff}, although asymptotic upper bounds on the minimum average queue length are known, {asymptotic lower bounds are not available in many cases}.
We note that such asymptotic lower bounds are significant, since they may help in determining the best possible tradeoff.
Let $V$ be the difference between the average service cost constraint $P_{c}$ and the minimum average service cost for stability, $c(\lambda)$, in the asymptotic regime $\Re$, where $V \downarrow 0$.
The known asymptotic lower bounds on the minimum average queue length, along with the details of the models analysed, and the asymptotic upper bounds on the minimum average queue length are summarized in Table \ref{introduction:table:existinglb}.

\begin{table}[h]
  \centering
  \begin{tabular}{|l|l|c|c|}
    \hline
     & 
    \textbf{Model details} & 
    \begin{minipage}{0.25\textwidth}
      \textbf{Asymptotic upper bound (Regime $\Re$)}
    \end{minipage} &
    \begin{minipage}{0.25\textwidth}
      \textbf{Asymptotic lower bound (Regime $\Re$)}
    \end{minipage} \\
    \hline
    1 &
    \begin{minipage}{0.4\textwidth}
      \vspace{0.2em}
      Berry-Gallager power delay tradeoff \cite{berry}; $P(h, s)$ strictly convex in $s, \forall h$ 
      \vspace{0.2em}
    \end{minipage} & $\mathcal{O}\brap{\frac{1}{\sqrt{V}}\log\nfrac{1}{V}}$ & $\Omega\nfrac{1}{\sqrt{V}}$ \\
    \hline
    2 &
    \begin{minipage}{0.4\textwidth}
      \vspace{0.2em}
      Multiuser Berry-Gallager power delay tradeoff \cite{neely_mac}; $P(\bS{h}, \bS{s})$ strictly convex in $\bS{s}, \forall \bS{h}$
      \vspace{0.2em}
    \end{minipage} & $\mathcal{O}\brap{\frac{1}{\sqrt{V}}\log\nfrac{1}{V}}$ & $\Omega\nfrac{1}{\sqrt{V}}$ \\
    \hline
    3 &
    \begin{minipage}{0.4\textwidth}
      \vspace{0.2em}
      Multiuser Berry-Gallager power delay tradeoff \cite{neely_mac}; piecewise linear $c(.)$, $\lambda$ is such that $c(\lambda)$ is on a piecewise linear portion of $c(.)$
      \vspace{0.2em}
    \end{minipage} & $\mathcal{O}\brap{\log\nfrac{1}{V}}$ &
    \begin{minipage}{0.25\textwidth}
      \vspace{0.2em}
      $\Omega\brap{\log\nfrac{1}{V}}$ shown for a specific example, not known in general
      \vspace{0.2em}
    \end{minipage} \\
    \hline
    4 &
    \begin{minipage}{0.4\textwidth}
      \vspace{0.2em}
      Multiuser Berry-Gallager power delay tradeoff \cite{neely_mac}; piecewise linear $c(.)$, $\lambda$ is  any abscissa at which the slope of $c(.)$ changes
      \vspace{0.2em}
    \end{minipage} & $\mathcal{O}\nfrac{1}{V}$ &
    \begin{minipage}{0.2\textwidth}
      \begin{center}
        Not known
      \end{center}
    \end{minipage} \\
    \hline
    5 &
    \begin{minipage}{0.4\textwidth}
      \vspace{0.2em}
      Power delay tradeoff with lower bound constraint on average throughput \cite{neely_utility}
      \vspace{0.2em}
    \end{minipage} & $\mathcal{O}\brap{\log\nfrac{1}{V}}$ &
    \begin{minipage}{0.25\textwidth}
      $\Omega\brap{\log\nfrac{1}{V}}$ but with single fade state
    \end{minipage} \\
    \hline
    6 &
    Utility delay tradeoff \cite{neely_superfast} & $\mathcal{O}\brap{\log\nfrac{1}{V}}$ & $\Omega\brap{\log\nfrac{1}{V}}$ \\
    \hline
    7 & 
    \begin{minipage}{0.4\textwidth}
      \vspace{0.2em}
      Power delay tradeoff with Markov arrival and fading process \cite{huang_neely}
      \vspace{0.2em}
    \end{minipage} & $\mathcal{O}\nfrac{1}{V}$ & 
    \begin{minipage}{0.2\textwidth}
      \begin{center}
        Not known
      \end{center}
    \end{minipage} \\    
    \hline
  \end{tabular}
  \caption{Some of the available asymptotic bounds on the minimum average queue length; except for case 7 all other models assume that the arrival process and the fade process are IID, and except for cases 5 and 6 all models have $A[m] = R[m]$. Also, all lower bounds are derived under the assumption that the queue length can take real values.}
  \label{introduction:table:existinglb}
\end{table}

We note that several asymptotic lower bounds in Table \ref{introduction:table:existinglb} have been derived under the assumption that the queue length and service batch size take values in $\sR$ and the service cost function is strictly convex.
In certain cases, these real valued queueing models are used as approximate models for queueing models where the queue length and the service batch size take values in $\sZ$.
We note that there are also scenarios, where modelling the queue length evolution to be on $\sR$ is natural, such as when the queue is assumed to buffer a certain amount of \emph{error exponent} as in \cite{berry}.

An approximate solution to the tradeoff problem has been obtained by Ata et al. \cite{ata} by approximating $(Q[m])$ by a diffusion process, which enables them to find the optimal policy for the control of the approximating diffusion process in closed form.
The complete characterization of the optimal admission control policy for a continuous time queueing model was obtained in \cite{perkins}.

Bettesh and Shamai \cite{bettesh} obtain approximations for $s^*(q, h)$ for every $h$, in the regime of large $q$, by solving the average cost optimality equation associated with the MDP \eqref{introduction:eq:dt_specialTradeoff_dual}.
For the MDP \eqref{introduction:eq:dt_specialTradeoff_dual}, Chen et al. \cite{chen} obtain approximations for $s^*(q)$ from a fluid approximation.
However, we note that the bounds on $s^*(q)$ depend on the form of the service cost functions.

Motivated by the above survey of known results, we ask and try to answer the following questions in this thesis:
\begin{enumerate}
\item We note that, in Table \ref{introduction:table:existinglb}, there are several cases in which asymptotic lower bounds are not known. What are these asymptotic lower bounds?
\item From Table \ref{introduction:table:existinglb}, we observe that asymptotic lower bounds have the form of $\log\nfrac{1}{V}$ or $\frac{1}{\sqrt{V}}$. What is the intuition behind such a behaviour?
\item As stated before, for certain cases, asymptotic lower bounds have been derived for an approximate queueing model, where the queue evolution is real valued. Are the asymptotic lower or upper bounds different for the original integer valued queue evolution model?
\item Bounds on $s^*(q, h_{0})$ have been obtained, which are dependent on the service cost function. However, the asymptotic order bounds on the minimum average queue length are dependent only on certain properties of the service cost function, rather than its exact form. Can we obtain asymptotic order bounds on the policy which are independent of the exact form of the service cost function?
\end{enumerate}

In the next section, we briefly survey how these issues have been addressed in this thesis.

\section{Overview of the thesis and contributions}
This thesis consists of two parts.
In the first part, comprising Chapters 2, 3, and 4, we consider the tradeoff problems \eqref{introduction:eq:mm1_genTradeoff}, \eqref{introduction:eq:mm1_specialTradeoff}, \eqref{introduction:eq:dt_genTradeoff}, and \eqref{introduction:eq:dt_specialTradeoff} in their respective asymptotic regimes $\Re$, for the class of admissible policies.
The second part, comprising Chapters 5 and 6, primarily illustrates the application of the results obtained in the first part to the motivating resource tradeoff problems arising in point-to-point communication links.

In this introductory discussion, the results are stated informally.
We note that the results hold under further technical assumptions, which are stated in the respective chapters.

We study the continuous time state dependent M/M/1 queueing model in Chapters 2 and 3.
The main results which are obtained in Chapters 2 and 3 are summarized in Table \ref{introduction:table:inthisthesis_ct}.
For the first two cases, let $V$ be the difference between $c_{c}$ and $c(\lambda)$ in the asymptotic regime $\Re$, where $V \downarrow 0$.
For the third case $V$ is the difference between $u(\mu)$ and $u_{c}$, while for the fourth case $V$ is the difference between $c_{c}$ and $c(u^{-1}(u_{c}))$.
We note that the first two cases are instances of \eqref{introduction:eq:mm1_specialTradeoff} while the fourth case is an instance of \eqref{introduction:eq:mm1_genTradeoff}.
The third case is similar to \eqref{introduction:eq:mm1_specialTradeoff}, except that the roles of $\lambda(q)$ and $\mu(q)$ are interchanged, i.e., there is admission control with a fixed service rate.
To the best of our knowledge, such asymptotic results for the tradeoff for the continuous time state dependent M/M/1 model, which may be of independent interest, are new.
Thus, we obtain answers for the first question that we posed, for the continuous time queueing models.
\begin{table}
  \centering
  \begin{tabular}{|l|l|c|}
    \hline
    \textbf{Control} & \textbf{Service cost and utility functions} & 
    \begin{minipage}{0.3\textwidth}
      \textbf{Results (in the regime $\Re$, for admissible policies)}
    \end{minipage} \\
    \hline
    \begin{minipage}{0.3\textwidth}
      $\mathcal{X}_{\mu}$ is discrete, $\mathcal{X}_{\lambda} = \brac{\lambda}$
    \end{minipage} & 
    \begin{minipage}{0.3\textwidth}
      $c(.)$ is piecewise linear 
    \end{minipage} & 
    \begin{minipage}{0.3\textwidth}
      \vspace{0.1em}
      Depending on $\lambda$, minimum average queue length either increases to a finite value, is $\Theta\brap{\log\nfrac{1}{V}}$, or $\Theta\nfrac{1}{V}$
    \end{minipage} \\
    \hline
    \begin{minipage}{0.3\textwidth}
      $\mathcal{X}_{\mu}$ is a finite interval, $\mathcal{X}_{\lambda} = \brac{\lambda}$
    \end{minipage} & 
    \begin{minipage}{0.3\textwidth}
      $c(.)$ is strictly convex
    \end{minipage} &
    \begin{minipage}{0.3\textwidth}
      \vspace{0.1em}
      Minimum average queue length is $\Omega\nfrac{1}{\sqrt{V}}$ 
    \end{minipage} \\
    &
    \begin{minipage}{0.3\textwidth}
      $c(.)$ is piecewise linear
    \end{minipage} &
    \begin{minipage}{0.3\textwidth}
      \vspace{0.2em}
      Depending on $\lambda$, minimum average queue length either increases to a finite value, is $\Theta\brap{\log\nfrac{1}{V}}$, or $\Theta\nfrac{1}{V}$
    \end{minipage} \\
    \hline
    \begin{minipage}{0.3\textwidth}
      $\mathcal{X}_{\mu} = \brac{\mu}$, $\mathcal{X}_{\lambda}$ is a finite interval
    \end{minipage} & 
    \begin{minipage}{0.3\textwidth}
      $u(.)$ is strictly concave
    \end{minipage} &
    \begin{minipage}{0.3\textwidth}
      \vspace{0.1em}
      Minimum average queue length is $\Omega\nfrac{1}{\sqrt{V}}$ 
    \end{minipage} \\
    &
    \begin{minipage}{0.3\textwidth}
      $u(.)$ is piecewise linear
    \end{minipage} &
    \begin{minipage}{0.3\textwidth}
      \vspace{0.2em}
      Depending on $\lambda$, minimum average queue length is $\Omega\brap{\log\nfrac{1}{V}}$, or $\Omega\nfrac{1}{V}$
    \end{minipage} \\
    \hline
    \begin{minipage}{0.3\textwidth}
      $\mathcal{X}_{\mu}$ is a finite interval, $\mathcal{X}_{\lambda}$ is a finite interval
    \end{minipage} &
    \begin{minipage}{0.3\textwidth}
      $c(.)$ is strictly convex, $u(.)$ is strictly concave
    \end{minipage}
    &
    \begin{minipage}{0.3\textwidth}
      Minimum average queue length is $\Theta\brap{\log\nfrac{1}{V}}$
    \end{minipage} \\
    \hline
  \end{tabular}
  \caption{List of asymptotic results derived for continuous time queueing models in Chapters 2 and 3.}
  \label{introduction:table:inthisthesis_ct}
\end{table}

The insights obtained from these two chapters are then used in deriving asymptotic lower bounds for a discrete time model in Chapter 4.
The correspondence between discrete time models and the continuous time models can be achieved by the choice of $\mathcal{X}_{\mu}$, $\mathcal{X}_{\lambda}$, and the form of the functions $c(.)$ and $u(.)$.
The motivation behind the choice of $\mathcal{X}_{\mu}$, $\mathcal{X}_{\lambda}$, and the form of the functions $c(.)$ and $u(.)$ is explained in Chapter 3.

From the analysis in Chapters 2 and 3, we obtain the following intuition for the behaviour of the asymptotic lower bounds, which partly answers the second question that we posed.
We note that for the state dependent M/M/1 model, for admissible policies, it can be shown that the stationary probability distribution of the queue length exists.
The intuition for the behaviour of the asymptotic lower bounds is based on the \emph{shape} of this stationary probability distribution in the asymptotic regime $\Re$.
We discuss the behaviour of the stationary probability distribution only for the cases where the minimum average queue length increases to infinity in the regime $\Re$.

\begin{figure}
  \centering
  \includegraphics[width=120mm,height=60mm]{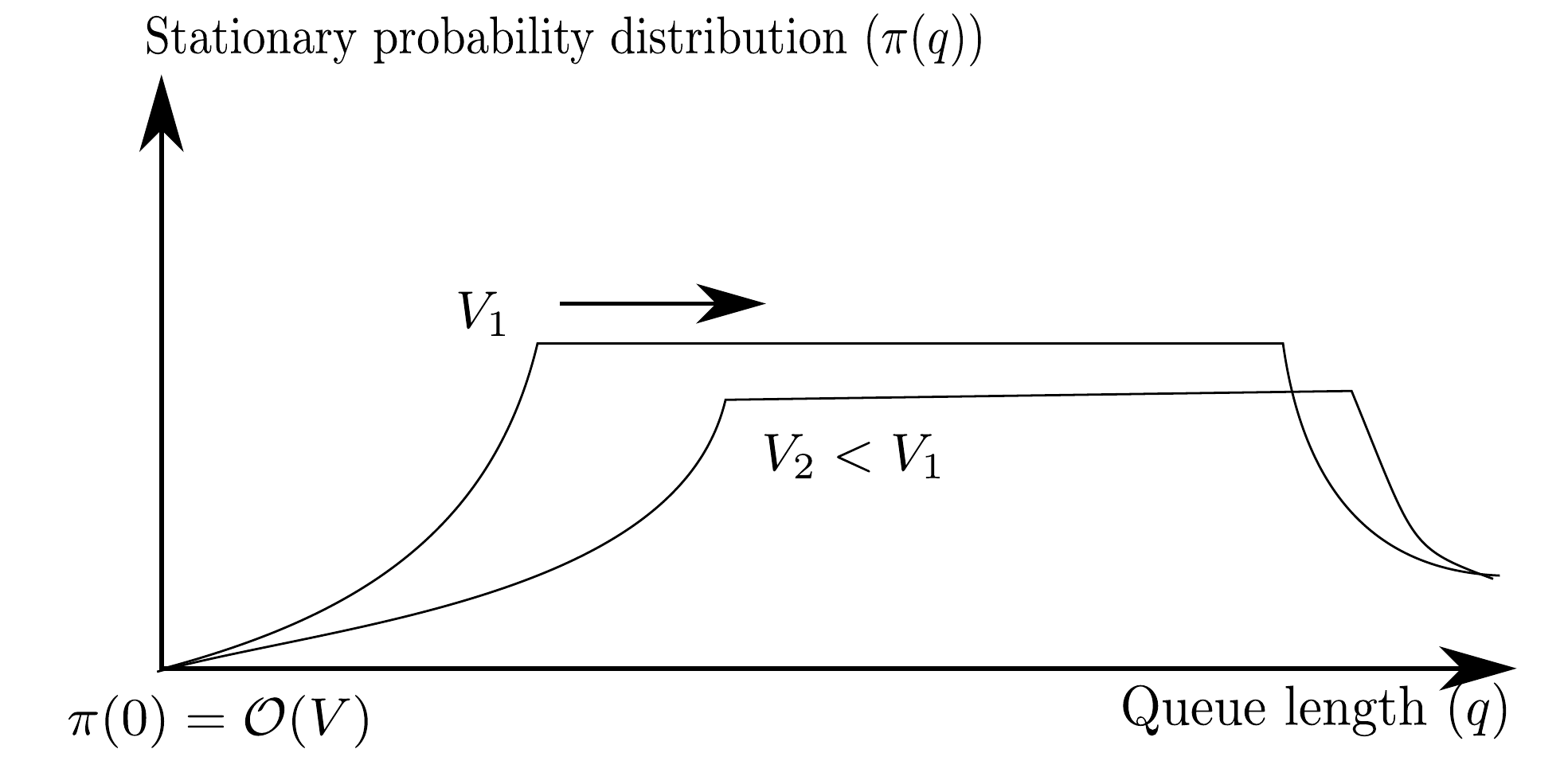}
  \caption{Intuition for the behaviour of the stationary probability distribution in the regime $\Re$}
  \label{introduction:figure:intuition_1}
\end{figure}

The behaviour of the stationary probability distribution of the queue length is determined by the behaviour of the stationary probability distribution of the service rates, which is in turn decided by the nature of the function $c(\mu)$ at $\mu = \lambda$ or $u^{-1}(u_{c})$ as the case may be.
We consider non-idling admissible policies for the purpose of discussion.
We find that in the regime $\Re$, the stationary probability of using a service rate of zero, goes to zero as $\mathcal{O}(V)$.
Then, intuitively, since the probability of this queue being empty goes to zero, the stationary probability distribution \emph{shifts to the right} as shown in Figure \ref{introduction:figure:intuition_1}.
Therefore, the average queue length has to increase.
We note that this intuition has been used in the design of tradeoff optimal policies in \cite{neely_mac}.

We note that for monotone admissible policies, the stationary probability distribution has the following \emph{shape} (as in Figure \ref{introduction:figure:intuition_1}). The stationary probability distribution is monotonically increasing, then may or may not be constant for a set of queue lengths, and then is monotonically decreasing.

In the asymptotic regime $\Re$, the probability of using certain service rates decreases to zero as $\mathcal{O}(V)$, while the probability of using certain service rates increases.
Consider the set of queue lengths, $\mathcal{Q}_{h}$, which are such that the stationary probability of using the service rates $\mu(q), q \in \mathcal{Q}_{h}$ does not decrease to zero.
The different behaviours for the minimum average queue length, depends on (i) the shape of the stationary probability distribution for the set $\mathcal{Q}_{h}$, and (ii) the stationary probability of the smallest queue length in $\mathcal{Q}_{h}$.

\begin{figure}
  \centering
  \includegraphics[width=120mm,height=60mm]{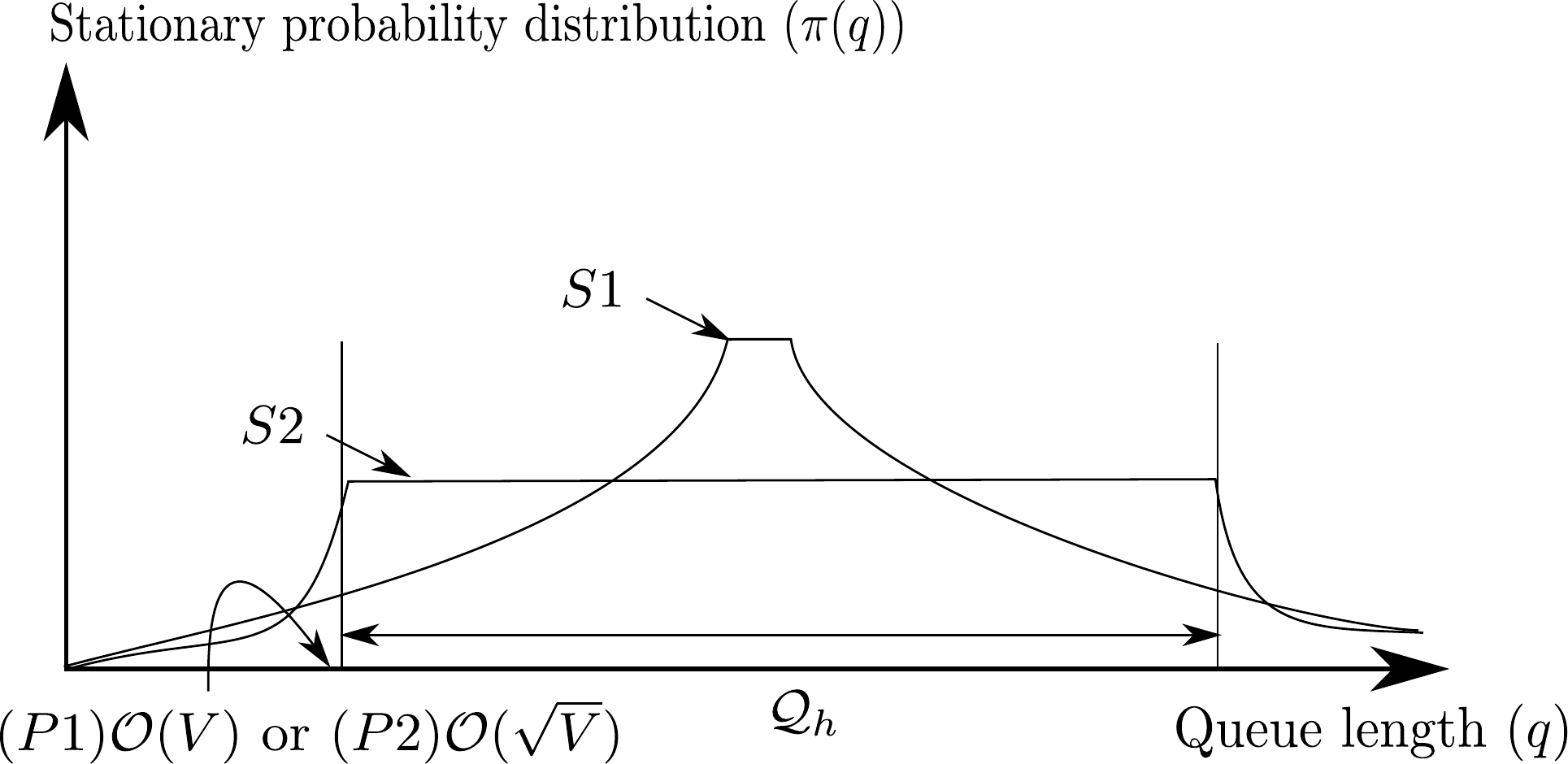}
  \caption{Possibilities for the behaviour of the stationary probability distribution in the regime $\Re$}
  \label{introduction:figure:intuition_2}
\end{figure}

In the asymptotic regime $\Re$, the shape of the stationary probability distribution for the set of queue lengths $\mathcal{Q}_{h}$, can be either (S1) monotonically increasing, constant, and then monotonically decreasing, or (S2) constant.
We shall see that this is decided by the extent of freedom that we have in the choice of $\lambda(q)$ and $\mu(q)$.
In the asymptotic regime $\Re$, the stationary probability of the smallest queue length in $\mathcal{Q}_{h}$ is either (P1) $\mathcal{O}(V)$ or (P2) $\mathcal{O}(\sqrt{V})$.
We shall see that this is decided by the form of the function $c(.)$ at $\mu = \lambda$ or $u^{-1}(u_{c})$ as the case may be.
The various possibilities are illustrated in Figure \ref{introduction:figure:intuition_2}.

Then, the $\log\nfrac{1}{V}$ behaviour for the minimum average queue length arises because the stationary probability distribution of the queue length is (S1) and the stationary probability of the smallest queue length in $\mathcal{Q}_{h}$ is (P1).
The asymptotic $\frac{1}{V}$ behaviour arises with (S2) and (P1).
The asymptotic $\frac{1}{\sqrt{V}}$ behaviour arises with (S2) and (P2).
Although we discuss this in more detail in Chapters 2 and 3, here we provide an example to illustrate the derivation of the $\log\nfrac{1}{V}$ behaviour.
For a particular policy, we obtain a geometric upper bound of the form $\pi(0)\rho^{q}$ ($\rho > 1$) on the stationary probability $\pi(q)$ of the queue length.
In the asymptotic regime $\Re$ we also show that $\pi(0) = \mathcal{O}(V)$.
Then applying Markov inequality we obtain that the average queue length is a constant times $\log\nfrac{1}{V}$.

We note that these results are obtained using geometric upper bounds on the stationary probability of the queue length.
Under the assumptions that we have made regarding the form of $c(.)$ and $u(.)$, we therefore observe that the possible forms for asymptotic lower bounds are $\log\nfrac{1}{V}$, $\frac{1}{V}$, or $\frac{1}{\sqrt{V}}$.
By providing this intuition, we have successfully answered the second question, for admissible policies, for the continuous time model.

Using bounds on the stationary probability distribution of queue length, we also obtain asymptotic bounds on the structure of \emph{order-optimal} admissible policies\footnote{these are admissible policies which achieve the asymptotic lower bounds in Table \ref{introduction:table:inthisthesis_ct}.} in the regime $\Re$.
Since admissible policies are monotone, we note that corresponding to a set of service rates of $S$, there is a contiguous set or interval, $\mathcal{Q}_{S}$, of queue lengths such that $\mu(q) \in S, \forall q \in \mathcal{Q}_{S}$.
In both Chapter 2 and Chapter 3, we obtain asymptotic bounds on the cardinality of $\mathcal{Q}_{S}$.
These bounds are independent of the exact form of the functions $c(.)$ and $u(.)$.
Thus, we obtain some answers for the fourth question that we have posed, for continuous time queueing models.

In Chapter 2, we also apply the analysis to a flow-level resource allocation model for a wireless downlink and obtain asymptotic bounds on the tradeoff of average power and average number of flows.

In Chapter 4, we consider the problem \eqref{introduction:eq:dt_specialTradeoff}, with a single environment state ($h_{0}$), in the asymptotic regime $\Re$.
We note that in this case the service cost $c(s) = P(h_{0}, s)$ is a function of the batch size $s$ only.
For admissible policies, it turns out that the stationary distribution of the queue length exists.
Using the insights about the shape of the stationary distribution of the queue length in the regime $\Re$ from Chapters 2 and 3, we obtain two upper bounds on the stationary probability distribution of the queue length, one of which is an extension of the bound on stationary probability distribution obtained in Bertsimas et al. \cite{gamarnik_1}  to the case where the service rate is dependent on the queue length.
The bounds can be used to obtain the same intuition, as explained earlier for the continuous time model, for the discrete time model.
The bounds are used to obtain the results that are summarized in Table \ref{introduction:table:inthisthesis_ch4}.
Thus, we answer the first and second questions that we have posed, for admissible policies.
The bounds on the stationary probability distribution are also used to obtain asymptotic bounds on the cardinality of $\mathcal{Q}_{S}$, leading to some answers for the fourth question that we have posed.

We note that our asymptotic results apply to the tradeoff problems in Table \ref{introduction:table:existinglb} under the assumption that the optimal policy lies in the class of admissible policies, which is true for many cases \footnote{For example, whenever the solution to \eqref{introduction:eq:dt_specialTradeoff} coincides with that of its dual (obtained from \eqref{introduction:eq:dt_specialTradeoff_dual}).}.

We also show that approximating the original integer valued queueing model by a real valued model with a strictly convex function, leads to the average queue length and average service cost being underestimated, in certain cases.
We also analyse a case, which have not been hitherto identified, where the average queue length increases to only a finite value in the asymptotic regime $\Re$, for the original integer valued queueing model.
Thus, we obtain some answers to the third question that we have posed.
We also show that a more appropriate real valued approximation is one in which the service cost function is approximated as the piecewise linear lower envelope of the service cost function for the original integer valued queueing model.

\begin{table}
  \centering
  \begin{tabular}{|l|l|l|}
    \hline
    \textbf{Model details} & \textbf{Service cost function} & 
    \begin{minipage}{0.3\textwidth}
      \textbf{Results (in the regime $\Re$, for admissible policies)}
    \end{minipage} \\
    \hline
    \begin{minipage}{0.3\textwidth}
      $Q[m]$ evolves on $\sZ$
    \end{minipage} & 
    \begin{minipage}{0.3\textwidth}
      $P(h_{0}, s)$ is piecewise linear in $s$
    \end{minipage} & 
    \begin{minipage}{0.3\textwidth}
      \vspace{0.2em}
      Depending on $\lambda$, minimum average queue length either increases to a finite value,
      is $\Theta\brap{\log\nfrac{1}{V}}$, or $\Theta\nfrac{1}{V}$
      \vspace{0.2em}
    \end{minipage} \\
    \hline
    \begin{minipage}{0.3\textwidth}
      $Q[m]$ evolves on $\sZ$; $A[m]$ is ergodic
    \end{minipage} & 
    \begin{minipage}{0.3\textwidth}
      $P(h_{0}, s)$ is piecewise linear in $s$
    \end{minipage} & 
    \begin{minipage}{0.3\textwidth}
      \vspace{0.2em}
      Depending on $\lambda$, minimum average queue length either increases to a finite value or is $\Omega\brap{\log\nfrac{1}{V}}$
      \vspace{0.2em}
    \end{minipage} \\
    \hline
    \begin{minipage}{0.3\textwidth}
      \vspace{0.2em}
      $Q[m]$ evolves on $\sR$
      \vspace{0.2em}
    \end{minipage} & 
    \begin{minipage}{0.3\textwidth}
      $P(h_{0}, s)$ is strictly convex in $s$
    \end{minipage} & 
    \begin{minipage}{0.3\textwidth}
      \vspace{0.1em}
      Minimum average queue length is $\Omega\nfrac{1}{\sqrt{V}}$ (previously known \cite{berry} but re-derived here using our method).
    \end{minipage} \\
    \hline
  \end{tabular}
  \caption{List of asymptotic results derived for tradeoff problem \eqref{introduction:eq:dt_specialTradeoff} in Chapter 4 with single environment state $h_{0}$.}
  \label{introduction:table:inthisthesis_ch4}
\end{table}

In Chapter 5, we consider problems \eqref{introduction:eq:dt_specialTradeoff} and \eqref{introduction:eq:dt_genTradeoff}, in the context of a point-to-point link with fast fading.
In this chapter, the environment variable $h$ models the fade state and $P(h, s)$ is the power expended in transmitting $s$ packets in fade state $h$.
Let $V$ be the difference between the power constraint $P_{c}$ and $c(\lambda)$ (or $P_{c}$ and $c(u^{-1}(u_{c}))$) in the asymptotic regime $\Re$, where $V \downarrow 0$.
Using the results in Chapter 4, we obtain an asymptotic characterization of the tradeoff.
The main results obtained in this chapter are summarized in Table \ref{introduction:table:inthisthesis_dt}.
We also comment on the extension of these asymptotic results to: (a) a $N$ user single hop network model (as in Figure \ref{introduction:fig:discretetimemodel_mac}) with $Q_{n}[m]$ assumed to evolve on $\sZ$, (b) a model with admission control and ergodic arrival and fading processes, and (c) a model with no service cost, but for which we are interested in the tradeoff of utility and delay as in \cite{neely_superfast}.
We note that except for the case where the minimum average queue length increases to only a finite value (which has been hitherto not identified in literature) we are able to obtain asymptotic lower bounds for all of the models in Table \ref{introduction:table:existinglb} for admissible policies.

\begin{table}[!h]
  \centering
  \begin{tabular}{|l|l|l|}
    \hline
    \textbf{Model details} & \textbf{Service cost function} & 
    \begin{minipage}{0.3\textwidth}
      \textbf{Results (in the regime $\Re$, for admissible policies)}
    \end{minipage} \\
    \hline
    \begin{minipage}{0.3\textwidth}
      \vspace{0.2em}
      $Q[m]$ evolves on $\sZ$, $A[m] = R[m]$, as in \cite{neely_mac}
      \vspace{0.2em}
    \end{minipage} & 
    \begin{minipage}{0.3\textwidth}
      $P(h,s)$ is piecewise linear in $s, \forall h$
    \end{minipage} &
    \begin{minipage}{0.3\textwidth}
      Depending on the arrival rate $\lambda$, minimum average queue length either increases to only a finite value, or is $\Theta\brap{\log\nfrac{1}{V}}$ or is $\Theta\nfrac{1}{{V}}$
    \end{minipage} \\
    \hline
    \begin{minipage}{0.3\textwidth}
      \vspace{0.2em}
      $Q[m]$ evolves on $\sR$, $A[m] = R[m]$, same as the Berry-Gallager tradeoff problem \cite{berry}
      \vspace{0.2em}
    \end{minipage} & 
    \begin{minipage}{0.3\textwidth}
      $P(h,s)$ is strictly convex in $s, \forall h$
    \end{minipage} &
    \begin{minipage}{0.3\textwidth}
      Minimum average queue length is $\Omega\nfrac{1}{\sqrt{V}}$ (previously known \cite{berry} but re-derived here using our method).
    \end{minipage} \\
    \hline
    \begin{minipage}{0.3\textwidth}
      \vspace{0.2em}
      $Q[m]$ evolves on $\sR$, with admission control, same as the model in \cite{neely_utility}
      \vspace{0.2em}
    \end{minipage} & 
    \begin{minipage}{0.3\textwidth}
      $P(h,s)$ is strictly convex in $s, \forall h$
    \end{minipage} &
    \begin{minipage}{0.3\textwidth}
      Minimum average queue length is $\Theta\brap{\log\nfrac{1}{{V}}}$
    \end{minipage} \\
    \hline
    \begin{minipage}{0.3\textwidth}
      \vspace{0.2em}
      $Q[m]$ evolves on $\sR$, $A[m] = R[m]$, same as the Berry-Gallager tradeoff problem \cite{berry}, but with ergodic arrival and fading process
      \vspace{0.2em}
    \end{minipage} & 
    \begin{minipage}{0.3\textwidth}
      $P(h,s)$ is strictly convex in $s, \forall h$
    \end{minipage} &
    \begin{minipage}{0.3\textwidth}
      Depending on the value of $\lambda$, minimum average queue length either increases to only a finite value or is $\Omega\brap{\log\nfrac{1}{{V}}}$
    \end{minipage} \\
    \hline
    \begin{minipage}{0.3\textwidth}
      \vspace{0.2em}
      $Q_{n}[m]$ evolves on $\sR$ for every user $n$, no admission control, same as the multiuser Berry-Gallager tradeoff problem \cite{neely_mac}
      \vspace{0.2em}
    \end{minipage} & 
    \begin{minipage}{0.3\textwidth}
      $P(\bS{h},\bS{s})$ is strictly convex in the vector $\bS{s}, \forall \bS{h}$
    \end{minipage} &
    \begin{minipage}{0.3\textwidth}
      Minimum average total queue length is $\Omega\nfrac{1}{\sqrt{V}}$, individual average queue length is also $\Omega\nfrac{1}{\sqrt{V}}$  (previously known \cite{neely_mac} but re-derived here using our method).
    \end{minipage} \\
    \hline
    \begin{minipage}{0.3\textwidth}
      \vspace{0.2em}
      $Q_{n}[m]$ evolves on $\sR$ for every user $n$, admission control
      \vspace{0.2em}
    \end{minipage} & 
    \begin{minipage}{0.3\textwidth}
      $P(\bS{h},\bS{s})$ is strictly convex in the vector $\bS{s}, \forall \bS{h}$
    \end{minipage} &
    \begin{minipage}{0.3\textwidth}
      Minimum average queue length is $\Theta\brap{\log\nfrac{1}{{V}}}$
    \end{minipage} \\
    \hline
  \end{tabular}
  \caption{The main asymptotic results derived for discrete time queueing models with fading; unless stated otherwise the models assume that $(A[m])$ and $(H[m])$ are IID}
  \label{introduction:table:inthisthesis_dt}
\end{table}

We consider the tradeoff of average delay with average error rate for a point-to-point link in Chapter 6.
The transmitter is assumed to use fixed or variable-length block coding.
In this chapter, we interpret a packet as an information message symbol, which could be a bit.
For fixed length block coding, we assume that $s$ message symbols are encoded into a codeword of length $N_{c}$ channel uses.
We assume that $N_{c}$ channel uses correspond to one slot.
In the context of our discrete time model, the environment state is fixed ($h_{0}$) and $c(s) = P(h_{0}, s)$ is the expected number of message symbols which are decoded in error.
The function $c(s)$ is approximated as $s$ times the average block error probability when $s$ message symbols are transmitted using a random block code of length $N_{c}$, where the average block error probability is further approximated by using Gallager's random coding upper bound \cite[Chapter 5]{gallager_text}.
Asymptotic bounds to the optimal average delay for a given average error rate constraint are obtained as in Chapter 4, although in this case $c(s)$ is a non-convex function of $s$.
The asymptotic lower bounds obtained in Chapter 4 can be applied to non-convex $c(s)$ through the use of the lower convex envelope of $c(s)$.

In Chapter 6, we also consider a single server queueing model, where the codeword length $N_{c}$ is a parameter for the policy, i.e., different policies can choose different codeword lengths, but every transmission uses codewords with the same length.
For such models, it is intuitive that by using arbitrarily large block lengths the average message symbol error rate can be made arbitrarily close to zero.
We show that the exponential decay rate of the average error rate with average queueing delay is at most $\frac{2}{3} E_{R}(\lambda)$ where $E_{R}(\lambda)$ is the Gallager random coding exponent and $\lambda$ is the arrival rate of packets per slot.
Furthermore for fixed length block codes, for $\lambda$ sufficiently close to the capacity of the point-to-point link, a class of fixed rate service policies is shown to achieve the decay rate $\frac{2}{3}E_{R}(\lambda)$.
We then consider a single server queueing model where the service time can also be varied, to model scenarios where variable length coding is used.
For variable length block codes which constrain the average message symbol error rate by a constant bound on the block error probability, the class of exhaustive service policies, which transmit all the message symbols in the queue at a transmission instant, is shown to achieve the decay rate $\frac{2}{3}E_{R}(\lambda)$ for any $\lambda$.

We summarize the thesis in Chapter 7 and discuss some problems with scope for future work that are motivated by the analysis carried out in the thesis.
The notation that is common to all the chapters in this thesis is summarized on page xii.
The notation that is used in each chapter is summarized in each chapter.

\blankpage
\chapter[On the tradeoff of average queue length, average service cost, and average utility for the state dependent M/M/1 queue: Part I]{\textbf{On the tradeoff of average queue length, average service cost, and \\average utility for the state dependent M/M/1 queue: Part I}}

\section{Introduction}
\label{chap4:sec:introduction}

We consider the tradeoff between average queue length, average service cost, and average utility for the continuous time single server queueing model in this chapter and the next.
The mathematical model considered captures the problem of how a constrained/scarce resource should be dynamically allocated to randomly arriving demands, which may be subjected to admission control, in order that the system is operated \emph{optimally}.
Herein, this dynamic allocation problem is modelled using the simple state dependent M/M/1 model discussed in Chapter 1.
Our primary motivation for modelling and studying this tradeoff problem as such, is the variety of tradeoff problems that arise in resource allocation problems in wireless networks.

The state dependent M/M/1 model that we consider in this chapter is a birth death process with the state corresponding to the queue length, as reviewed in Chapter 1.
In this chapter, we consider the problem of making the optimal choice of the arrival rate and the service rate at each queue length, such that the time average queue length is minimized subject to constraints on both the time average service cost and the time average utility, associated with the service of customers.
We recall that such problems have been analysed in \cite{weber}, \cite{george}, \cite{ata_pcstatic}, and \cite{atamm1}, the results of which have been discussed in Chapter 1.
But unlike the approach in these papers, in this chapter we obtain an asymptotic characterization of the tradeoff in the regime $\Re$.
The asymptotic characterization of the tradeoff is discussed in this chapter and the next.
The notation that we use in these two chapters are summarized in Table \ref{chapter2:notationtable}.
We first summarize the methodology that is used for obtaining the asymptotic bounds.

\begin{table}
\centering
\begin{tabular}{|l|l|}
\hline
Symbol & Description \\
\hline
$t$ & time index \\
$Q(t)$ & queue length at time $t$ \\
$\mu(q)$ & service rate used at queue length $q$ \\
$\lambda(q)$ & arrival rate used at queue length $q$ \\
$r_{q_{1}, q_{2}}$ & transition rate from state $q_{1}$ to $q_{2}$ for a CTMC \\
$u(.)$ & utility rate function \\
$c(.)$ & service cost rate function \\
$\mathcal{X}_{\mu}$ & set of all possible service rates \\
$\mathcal{X}_{\lambda}$ & set of all possible arrival rates \\
$[r_{a,min}, r_{a,max}]$ & range of values for $\lambda(q)$ \\
$[0, r_{max}]$ & range of values for $\mu(q)$ \\
$\gamma$ & a policy \\
$\Gamma$ & set of all policies \\
$\Qg$ & average queue length \\
$\Cg$ & average service cost rate \\
$\Ug$ & average utility rate \\
$\Gamma_{a}$ & set of admissible policies \\
$\pi_{\gamma}$ & stationary probability distribution for policy $\gamma$ \\
$\Gamma_{a, M}$ & set of all mixtures of admissible policies \\
$\gamma_{M}$ & a mixture policy \\
$u_{c}$ & constraint on average utility rate \\
$c_{c}$ & constraint on average service cost rate \\
$\Re$ & asymptotic regime in which $c_{c} \rightarrow u^{-1}(u_{c})$ \\
$Q^*_{M}(c_{c}, u_{c})$ & minimum average queue length over $\Gamma_{a, M}$ under constraints $c_{c}$ and $u_{c}$ \\
$Q^*(c_{c}, u_{c})$ & minimum average queue length over $\Gamma_{a}$ under constraints $c_{c}$ and $u_{c}$ \\
$\beta_{c}, \beta_{u}$ & non-negative Lagrange multipliers corresponding to service cost and utility constraints \\
$\mathcal{O}^{u}$ & set of constraint value pairs $(c_{c}, u_{c})$ for which admissible policies are optimal \\
$\gamma^*(c_{c}, u_{c})$ & optimal policy for constraints $c_{c}$ and $u_{c}$ \\
$Q^*(c_{c})$ & minimum average queue length for the set $\Gamma_{a}$ service cost constraint $c_{c}$ \\
$\pi_{\mu}(k)$ & stationary probability of service rate $\mu_{k}$ \\
$\mu_{l}$ & largest service rate $\leq \lambda$ at which the slope of $c(.)$ changes \\
$\mu_{u}$ & smallest service rate $\geq \lambda$ at which the slope of $c(.)$ changes \\
$Q^*_{M}(c_{c})$ & minimum average queue length for the set $\Gamma_{a, M}$ under service cost constraint $c_{c}$ \\
$P_{l}\brac{.}$ & lower bound on probability of an event \\
$P_{u}\brac{.}$ & upper bound on probability of an event \\
\hline
\end{tabular}
\caption{Notation used in Chapters 2 and 3.}
\label{chapter2:notationtable}
\end{table}

\subsection{Methodology}
\label{chap2:methodology}

We note that if $\pi(q)$ is the stationary distribution of the queue length for a policy $\gamma$, then $\Qg = \Exp_{\pi} Q$.
Suppose $Pr_{u}\brac{Q = q}$ is any upper bound on $\pi(q)$.
If $\overline{q}$ is the largest $q$ such that $Pr_{u}\brac{Q < q} \leq \frac{1}{2}$, then $\Exp_{\pi} Q \geq \frac{\overline{q}}{2}$.
We obtain $Pr_{u}\brac{Q = q}$ for different cases to obtain $\overline{q}$.
For example, if $c(\mu)$ is piecewise linear then it can be shown that $Pr_{u}\brac{Q < q} = \pi(0) \rho^{q}$, where $\rho > 1$.
We note that then $\overline{q}$ is a function of $\pi(0)$.
However, for the tradeoff problem that we consider, it can be shown that $\pi(0)$ is proportional to $c_{c} - c(\lambda)$, i.e. the difference between the average service cost constraint and the infimum of the average service cost.
Then, $\frac{\overline{q}}{2} \geq \text{a constant} \times {\log\nfrac{1}{c_{c} - c(\lambda)}}$ and therefore so is $\Qg$.
This leads to the $\Omega\brap{\log\nfrac{1}{c_{c} - c(\lambda)}}$ asymptotic lower bound for a particular case.

For deriving asymptotic upper bounds, we consider a sequence of admissible policies.
For a particular policy in the sequence, the derivation of asymptotic upper bounds on average service cost rate and average utility rate uses upper bounds on stationary probability distribution of the queue length, whereas in all except one case asymptotic upper bounds on the average queue length are derived using the Lyapunov comparison theorem \cite[Theorem A.4.3]{meynctcn}.
For several cases, we identify sequences of order-optimal admissible policies $\gamma_{k}$, for which the asymptotic growth rates of $\Qgk$ matches with the corresponding asymptotic lower bounds.
Now we will discuss the model that we consider in this chapter.

\subsection{System model}
\label{chap4:sec:statedepmm1}

\begin{figure}
  \centering
  \includegraphics[width=120mm,height=60mm]{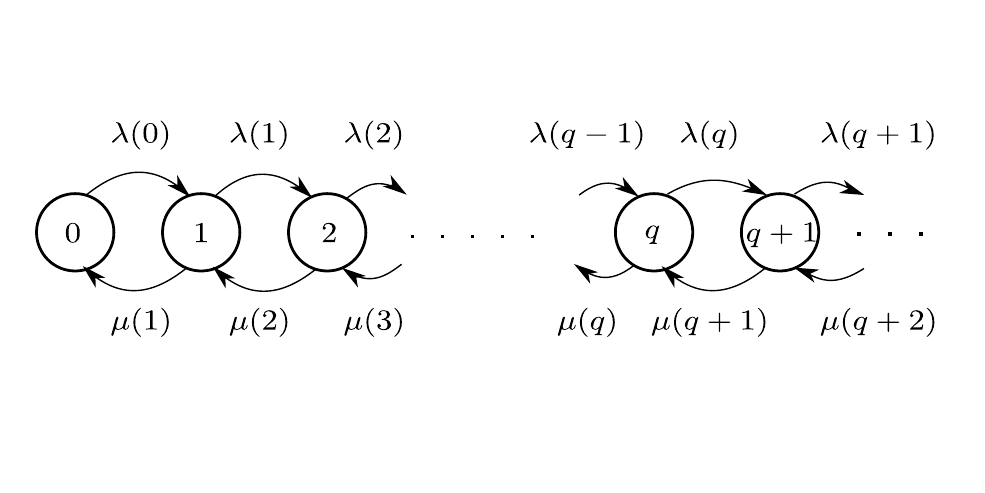}
  \vspace{-0.5in}
  \caption{The birth death process under a particular policy $\gamma$}
  \label{chap4:fig:mm1model}
\end{figure}

The queue evolves in continuous time, which is denoted by $t \in \mathbb{R}_{+}$.
The number of customers in the queue at time $t$ (including the one in service, if any)  is denoted by $Q(t) \in \mathbb{Z}_{+}$.
The state dependent M/M/1 model for the process $Q(t)$ is a birth death process with birth rate $r_{q,q + 1} = \lambda(q)$, death rate $r_{q,q - 1} = \mu(q)$ for $q \in \{1,\cdots\}$, and birth rate when there are zero customers in the queue, $r_{0,1} = \lambda(0)$.
The state transition diagram of the birth-death process for a policy $\gamma$ is shown in Figure \ref{chap4:fig:mm1model}.
A policy $\gamma$ is the sequence $(\mu(0) = 0, \lambda(0), \mu(1), \lambda(1) \cdots)$
\footnote{We note that we are restricting to policies which are stationary. Such a restriction is reasonable for the class of tradeoff problems that we are interested in.}.
The set of all policies is denoted as $\Gamma$.

We associate an utility rate function $u(.)$ with the arrival of customers and a cost rate function $c(.)$ with their service.
The utility rate function models the benefit in serving customers, while the cost rate function models the cost incurred in serving customers.
We assume that utility is accrued at the rate of $u(\lambda(Q(t)))$ at time $t$ and cost is incurred at the rate of $c(\mu(Q(t)))$ at time $t$.
The functions $u(.)$ and $c(.)$ are assumed to satisfy the following properties :
\begin{description}
\item[U1 :]{The function $u(\lambda) : \mathcal{X}_{\lambda} \rightarrow \mathbb{R}_{+}$ is strictly increasing and concave in $\lambda$, with $u(0) = 0$ and $\mathcal{X}_{\lambda}$ the set of arrival rates.}
\item[C1 :]{The function $c(\mu) : \mathcal{X}_{\mu} \rightarrow \mathbb{R}_{+}$ is strictly increasing and convex in $\mu$, with $c(0) = 0$ and $\mathcal{X}_{\mu}$ the set of service rates.}
\end{description}
The set $\mathcal{X}_{\lambda}$ is assumed to be either a) a finite set of discrete points $(\lambda_{0},\lambda_{1},\dots,\lambda_{K})$ or b) an interval $[r_{a,min},r_{a,max}]$ of the real line.
Similarly the set $\mathcal{X}_{\mu}$ is assumed to be either a) a finite set of discrete points $(\mu_{0},\mu_{1},\dots,\mu_{K})$ or b) an interval $[r_{min},r_{max}]$ of the real line.
If the set $\mathcal{X}_{\lambda}$ is a set of discrete points, then we extend the definition of $u(\lambda)$ to $[r_{a,min} = \lambda_{0}, r_{a,max} = \lambda_{K}]$ by linear interpolation.
A similar extension is done for the function $c(\mu)$.
Note that the linear interpolation preserves the concavity of $u(.)$ and the convexity of $c(.)$.
Let $u^{-1} : \mathbb{R}_{+} \rightarrow [r_{a,min}, r_{a,max}]$ and $c^{-1} : \mathbb{R}_{+} \rightarrow [r_{min}, r_{max}]$ be the inverse functions of $u(.)$ and $c(.)$ respectively.

The average service cost for the policy $\gamma$, $\overline{C}(\gamma)$ is defined as
\begin{equation}
  \overline{C}(\gamma) = \limsup_{T \rightarrow \infty}  \frac{1}{T} \mathbb{E} \bras{\int_{0}^{T} c(\mu(Q(t)))dt \middle \vert Q(0) = q_{0}}.
  \label{chap4:eq:sysmodel:sercostdef}
\end{equation}
The average utility for the policy $\gamma$, $\overline{U}(\gamma)$ is defined as
\begin{equation}
  \overline{U}(\gamma) = \limsup_{T \rightarrow \infty} \frac{1}{T} \mathbb{E} \bras{\int_{0}^{T} u(\lambda(Q(t)))dt \middle \vert Q(0) = q_{0}}.
  \label{chap4:eq:sysmodel:utilitydef}
\end{equation}
The average queue length for the policy $\gamma$, $\overline{Q}(\gamma)$ is defined as
\begin{equation}
  \overline{Q}(\gamma) = \limsup_{T \rightarrow \infty} \frac{1}{T} \mathbb{E} \bras{\int_{0}^{T} Q(t)dt \middle \vert Q(0) = q_{0}}.
  \label{chap4:eq:sysmodel:qlength}
\end{equation}
In this chapter, we restrict attention to policies for which the above three performance measures are independent of the initial state $q_{0}$, hence in the above definitions the dependence of these quantities on $q_{0}$ is not made explicit.
We note that the above definition of average utility (as in \cite{atamm1}) is much more general and encompasses scenarios where the utility of average throughput is of interest (e.g. as in \cite{neely_utility}).

We note that the state dependent M/M/1 model can be directly applied to study resource allocation in modern high rate data networks.
We consider such a motivational example in the next section.
Thus, the tradeoff problem for the state dependent M/M/1 model can be studied in its own right.
Furthermore, in Chapters 3 and 4, we shall see that the ideas developed for this simple state dependent M/M/1 model can be used in the study of discrete time queueing models, which in some cases are more representative of the resource allocation problems in wireless networks.

\subsection{A motivational example}
\label{chap4:sec:motivatingeg}
We discuss a motivating example in this section, which is based on the problem considered by Borst \cite{borst}.
We consider the downlink of a base station, operating in slotted time, with each slot of duration $1.67ms$.
Flows, each a file of size $480Kb$, arrive at the downlink scheduler queue for transmission to different users.
We assume that at most one flow arrives in a slot, and the flow arrival process is an IID Bernoulli process.
Each flow is destined to an user, which is one of two types, T1 or T2 with uniform probability.
The base station uses a round robin scheduler, which transmits bits from each flow, in order of their arrival instants.
We assume that, if the transmitter is transmitting at a power level of $10W$, then the transmission rate is $50 Kb/s$ for T1 flows, while for T2 flows, it is $150 Kb/s$.
Let $r_{1} = 50 Kb/s$ and $r_{2} = 150 Kb/s$.
The transmitter may dynamically vary its power to change a multiplier $m$ of the transmission rate whenever a flow arrives or a flow leaves the system.
The transmitter may choose $m \in \mathcal{M} = \{0, 0.25, 0.5, 0.75, 1, 1.25, 1.5, 1.75, 2\}$.
The multiplier $m$ is used to model the constraint that each flow may choose a rate corresponding to the choice of a codebook from a finite set of codebooks.
Then the transmission rate is $m r_{1}$ and $m r_{2}$ for T1 and T2 flows.
As in Borst \cite{borst}, we assume that the transmission rate as a function of the received SNR is $800\log_{10} (1 + SNR) Kb/s$.
We note that this formula for the transmission rate models a case where the fading gain is fixed or a case where transmission is done at a fixed rate only if the fading gain is above a certain threshold.
Then the transmitter power is $P^{1}(m) = 64.6 \brap{10^{\nfrac{50m}{800}} - 1} W$ and $P^{2}(m) = 18.5 \brap{10^{\nfrac{150m}{800}} - 1} W$, when transmitting to receivers with T1 and T2 flows respectively.
We note that as a function of the number of T1 and T2 flows, and therefore the total number of flows in the system, the actual transmitter power used varies within a round-robin scheduling cycle.
We are interested in dynamically controlling $m$ as a function of the current total number of flows to minimize the average number of flows in the system (or for a fixed arrival rate, the average flow transfer latency) subject to a constraint on the average transmitter power.

As in Borst \cite{borst} we note that, as the minimum flow transfer time ($\frac{480 Kb}{300 Kb/s}$) for any flow is of the order of seconds while the slot duration is of the order of milliseconds, we can model the system as a M/G/1-PS (processor sharing) queue, but with control on the total rate of service, through the choice of $m$ as a function $m(q)$ of the current number of flows $q$ in the system.
The Bernoulli arrival process is approximated as a Poisson process of rate $\lambda$.
Let us consider the normalized service requirement for T1 and T2 flows.
If a T1-flow is the only flow present in the system, then it requires a time of $\frac{480}{50} s$ if $m = 1$.
Similarly a T2-flow requires a time of $\frac{480}{150} s$ if $m = 1$.
So the normalized service requirement of the flows arriving are distributed as $\frac{480}{50}$ with probability $0.5$ and $\frac{480}{150}$ with probability $0.5$.
The time sharing amongst users manifests itself as processor sharing in the continuous time model.
At a time $t$, if there are $q$ flows in the M/G/1-PS model, the remaining service requirement of each flow is reduced at the rate $\frac{m(q)}{q}$.
From Bonald \cite[Theorem 2]{bonald}, we note that for any policy $\gamma$, the stationary probability of the M/G/1-PS queue with control on the service rate, is independent of the service requirement distribution.
Therefore, following Borst \cite{borst}, we consider a M/M/1-PS queue with control on the service rate, where the service requirement distribution is exponential, but with the same mean as the service requirement distribution in the M/G/1-PS queue, i.e, a mean service requirement of $\frac{1}{2}\brap{\frac{480}{50} + \frac{480}{150}} = 6.4 s$.
At a time $t$, if there are $q$ flows in this model, the remaining service requirement of each flow is reduced at the rate $\frac{m(q)}{q}$.
Therefore the rate at which a flow leaves the system is $q \frac{m(q)}{q} \frac{1}{6.4}$.
We then note that the M/M/1-PS queue with state dependent total service rate, determined by the policy $\gamma$, is the same as the model considered in Section \ref{chap4:sec:statedepmm1}.
The number of flows in the system is modelled by the state of the birth-death process.
The birth rate is $\lambda$, while the death rate which is dependent on $q$ can take values in $\mathcal{S} = \frac{1}{6.4} \times \brac{0, 0.25, 0.50, 0.75, 1 , 1.25, 1.50, 1.75, 2}$.

We note that the average transmitter power used when there are $q$ flows in the system depends on the proportion of T1 flows and T2 flows.
If there are $q_{1}$ T1 flows when there are $q$ flows in the system, then the average transmitter power is $\frac{q_{1} P^{1}(m) + (q - q_{1}) P^{2}(m)}{q}$.
To obtain \emph{good} policies which tradeoff average power with average number of flows for the discrete time system, we analyse the M/M/1-PS model, under the assumption that the transmitter power is $P(m) = P^{1}(m)$, irrespective of the flow which is being transmitted. 
The model obtained under this assumption is denoted as M/M/1-PS (I).
We also repeat the analysis under the assumption that $P(m) = P^{2}(m)$, in which case the model is denoted as M/M/1-PS (II).
The results obtained from the analysis of both M/M/1-PS (I) and M/M/1-PS (II) models are found to be useful in designing policies for the discrete time system.

\subsection{Overview}
We formulate the tradeoff problem in Section \ref{chap4:sec:probform} for a restricted class of admissible policies.
A non-idling property of any optimal admissible policy is also shown in the same section.
We consider three cases of the tradeoff problem: FINITE-$\mu$CHOICE, INTERVAL-$\mu$CHOICE, and INTERVAL-$\lambda\mu$CHOICE, which correspond to different choices of the sets $\mathcal{X}_{\lambda}$ and $\mathcal{X}_{\mu}$.
The analysis of FINITE-$\mu$CHOICE in the asymptotic regime $\Re$ is carried out in Section \ref{chap4:sec:problem1} while INTERVAL-$\mu$CHOICE, and INTERVAL-$\lambda\mu$CHOICE are analysed in Chapter 3.
For FINITE-$\mu$CHOICE, where $\mathcal{X}_{\lambda} = \brac{\lambda}$, we identify three cases based on the value of $\lambda$ and the set of available service rates $\brac{\mu_{0}, \dots, \mu_{K}}$, for which the asymptotic behaviour of the tradeoff curve in the asymptotic regime $\Re$ is different.
Asymptotic lower bounds and upper bounds to the tradeoff problem for these three cases are then obtained in Section \ref{chap4:sec:p1_asymp_lb} and Section \ref{chap4:sec:asympbehaviour} respectively.
An asymptotic characterization of optimal policies is presented in Section \ref{chap4:sec:optimalpolicy_aschar}.
We then numerically illustrate the asymptotic behaviour of the solution to FINITE-$\mu$CHOICE for several examples in Section \ref{chap4:sec:numexp1}.
Asymptotic bounds to the tradeoff curve for the example in Section \ref{chap4:sec:motivatingeg} are then presented in Section \ref{sec:application} using the results derived in this chapter.

\section{Problem formulation}
\label{chap4:sec:probform}
In this chapter and the next, we consider the tradeoff problems \eqref{introduction:eq:mm1_genTradeoff} and \eqref{introduction:eq:mm1_specialTradeoff} for a restricted class of admissible policies $\Gamma_{a}$.
The set of admissible policies is defined as follows.

\textbf{Stability :} 
A policy $\gamma$ is defined to be stable if the birth death process $Q(t)$ under policy $\gamma$ is irreducible and positive recurrent with stationary distribution $\pi_{\gamma}$.

\textbf{Admissibility :}
A policy $\gamma$ is admissible, if 
\begin{description}
\item[G1 :] it is stable,
\item[G2 :] the sequence $(\mu(0), \mu(1), \mu(2), \cdots)$ is non-decreasing, and, 
\item[G3 :] the sequence $(\lambda(0), \lambda(1), \lambda(2), \cdots)$ is non-increasing.
\end{description}
Then we define the set of admissible policies as
\begin{equation*}
  \Gamma_{a} \Deq \{ \gamma : \gamma \in {\Gamma}, \gamma \text{ is admissible}\}.
\end{equation*}
\begin{remark}
  We note that restricting attention to $\Gamma_{a}$ is reasonable, as the optimal policy which minimizes the average queue length subject to constraints on the average service cost and average utility possesses the properties G1, G2, and G3 in many cases (see Chapter 1)
\footnote{We note that there exists an admissible policy which achieves the minimum for the constrained optimization problems \eqref{introduction:eq:mm1_genTradeoff} or \eqref{introduction:eq:mm1_specialTradeoff} for certain values of the constraints $c_{c}$ and $u_{c}$. 
For example, for \eqref{introduction:eq:mm1_genTradeoff}, these values of $c_{c}$ and $u_{c}$ are such that there exists Lagrange multipliers $\beta_{1}$ and $\beta_{2}$ for which average service cost and average utility of any admissible optimal policy for the dual problem \eqref{introduction:eq:mm1_genTradeoff_dual} are equal to $c_{c}$ and $u_{c}$.}.
\end{remark}
We note that for any admissible policy, we have that 
\begin{eqnarray*}
  \overline{C}(\gamma) & = & \mathbb{E}_{\pi_{\gamma}} c(\mu(Q)), \\
  \overline{U}(\gamma) & = & \mathbb{E}_{\pi_{\gamma}} u(\lambda(Q)), \text{ and } \\
  \overline{Q}(\gamma) & = & \mathbb{E}_{\pi_{\gamma}} Q,
\end{eqnarray*}
where $Q \sim \pi_{\gamma}$ and the performance measures are independent of the initial state $q_{0}$.
We note that for a policy $\gamma$ if G2 and G3 holds and if at any finite $q'$, $\mu(q') - \lambda(q') = \epsilon > 0$, then $\gamma$ is stable and therefore admissible.

We note that a larger set of policies can be obtained by mixing the \emph{pure} policies in $\Gamma_{a}$.
We note that a mixture policy corresponds to time sharing of pure policies, with the time period, in which a particular pure policy is used, tending to infinity.
The set of policies which are obtained by a finite mixture of the policies in $\Gamma_{a}$ is denoted as $\Gamma_{a,M}$.
We note that associated with a $\gamma_{M} \in \Gamma_{a,M}$ we have a set $\Gamma(\gamma_{M}) \subseteq \Gamma_{a}$, which is the set of policies which are mixed according to a probability mass function $p_{\gamma}$, for $\gamma \in \Gamma(\gamma_{M})$.
For a $\gamma_{M} \in \Gamma_{a,M}$, $\overline{Q}(\gamma_{M}) = \sum_{\gamma} p_{\gamma} \Qg$.
The average service cost rate and average utility rate are defined similarly for $\gamma \in \Gamma_{a,M}$.

\subsection{Problem}
\label{chap4:sec:mm1problemstatement}
Our objective is to solve the following optimization problem, TRADEOFF-M:
\begin{eqnarray}
  \text{ minimize }_{\gamma \in \Gamma_{a,M}} & \overline{Q}(\gamma) \nonumber \\
  \text{ such that } & \overline{C}(\gamma) \leq c_{c}, \nonumber \\
  \text{ and } & \overline{U}(\gamma) \geq u_{c},
  \label{chap4:eq:sysmodel:probstat}
\end{eqnarray}
where $c_{c}$ and $u_{c}$ are constraints on the average service cost and average utility respectively.
The optimal value of the above problem is denoted by $Q^*_{M}(c_{c},u_{c})$.
We note that above constrained minimization of the average queue length corresponds to the constrained minimization of average delay if the average arrival rate is fixed.

In the following lemma, we show that TRADEOFF-M can be solved, only for certain values of $c_{c}$ and $u_{c}$.
\begin{lemma}
  If TRADEOFF-M has any feasible solutions, then $u^{-1}(u_{c}) \leq c^{-1}(c_{c})$.
  \label{chap4:lemma:feasible_solns}
\end{lemma}
\begin{proof}
  Assume that there is an policy $\gamma_{M} \in \Gamma_{a,M}$ which is feasible for TRADEOFF.
  Then from Jensen's inequality we have that $c(\Exp_{p_{\gamma}} \bras{\Exp_{\pi_{\gamma}} \bras{\mu(Q)}}) \leq \Exp_{p_{\gamma}} \bras{\Exp_{\pi_{\gamma}} \bras{c(\mu(Q))}}$ and $\Exp_{p_{\gamma}} \bras{\Exp_{\pi_{\gamma}} \bras{u(\lambda(Q))}} \leq u(\Exp_{p_{\gamma}}\bras{ \Exp_{\pi_{\gamma}}\bras{ \lambda(Q))}}$.
  For brevity let us denote $\Exp_{p_{\gamma}} \bras{\Exp_{\pi_{\gamma}}\bras{.}}$ by just $\Exp\bras{.}$ in this proof.
  Therefore $\Exp \mu(Q) \leq c^{-1}(\Exp c(\mu(Q)))$ and $u^{-1}(\Exp u(\lambda(Q))) \leq \Exp \lambda(Q)$.
  As $\Exp_{\pi_{\gamma}} Q < \infty$, $\Exp_{\pi_{\gamma}} \mu(Q) = \Exp_{\pi_{gamma}} \lambda(Q)$, $\forall \gamma \in \Gamma(\gamma_{M})$.
  Therefore for $\gamma$, $u^{-1}(\Exp u(\lambda(Q))) \leq c^{-1}(\Exp c(\mu(Q)))$.
  From the non-decreasing properties of $c(.)$ and $u(.)$ we have that $c^{-1}(.)$ and $u^{-1}(.)$ are also non-decreasing.
  Hence if there is any one feasible policy $\gamma$, $u^{-1}(u_{c}) \leq c^{-1}(c_{c})$.
\end{proof}

If $c^{-1}(c_{c}) > u^{-1}(u_{c})$, then we show that there exists a feasible policy $\gamma \in \Gamma_{a}$ on a case by case basis in the following discussion.
We note that if $c^{-1}(c_{c}) > u^{-1}(u_{c})$, then it is not guaranteed that an optimal policy $\gamma^*(c_{c}, u_{c}) \in \Gamma_{a,M}$ exists for the above problem.
However, in the following discussion we identify a set of $(c_{c}, u_{c})$ for which the existence of an optimal policy in $\Gamma_{a,M}$ is guaranteed.

\begin{remark}
  Let $\beta_{c}$ and $\beta_{u} \in \sR$.
  Consider an unconstrained MDP denoted as $MDP(\beta_{c}, \beta_{u})$ as in \cite{atamm1} which is obtained by uniformization at rate $r_{u}$ with single stage cost $\frac{q + \beta_{c} c(\mu) - \beta_{u} u(\lambda)}{r_{u}}$.
  Then from \cite{atamm1} we know that an optimal policy $\gamma^*(\beta_{c}, \beta_{u}) \in \Gamma_{a}$ exists for $MDP(\beta_{c}, \beta_{u})$.
  Let $\Gamma^*(\beta_{c}, \beta_{u})$ be the set of all optimal $\Gamma_{a}$ policies for $MDP(\beta_{c}, \beta_{u})$.
  Also let $\Gamma^*_{M}(\beta_{c}, \beta_{u})$ be the set of all mixed policies obtained by a finite mixture of $\gamma \in \Gamma^*(\beta_{c}, \beta_{u})$.
  Let $\mathcal{O}^{u} = \brac{(\Cg, \Ug), \gamma \in \Gamma^*_{M}(\beta_{c}, \beta_{u}), \forall \beta_{c}, \beta_{u} \geq 0}$.
  Then from \cite{ma}, if $(c_{c}, u_{c}) \in \mathcal{O}^{u}$, then there exists an optimal policy in $\Gamma_{a, M}$ for TRADEOFF-M.
\end{remark}

\begin{remark}
  Suppose $(c_{c}, u_{c}) \not \in \mathcal{O}^{u}$, but $c^{-1}(c_{c}) > u^{-1}(u_{c})$.
  We note that then for all $\gamma_{M}$ which are feasible for TRADEOFF-M, $\overline{Q}(\gamma_{M}) > 0$.
  Hence, for every $\epsilon > 0$, there exists some feasible $\gamma_{M}$ such that $\overline{Q}(\gamma_{M}) < Q^*_{M}(c_{c}, u_{c}) + \epsilon$. 
  We call such policies $\epsilon$-optimal for $c_{c}$.
\end{remark}

We now show that any optimal policy for TRADEOFF-M is \emph{non-idling}, if it exists.
\begin{lemma}
  Any optimal mixed policy $\gamma^*(c_{c}, u_{c})$ for TRADEOFF-M, has $\mu(q) > 0$ for every $q \geq 1$, for every $\gamma \in \Gamma(\gamma^*(c_{c}, u_{c}))$.
  \label{chap4:prop:optpolicy_nonidling}
\end{lemma}
\begin{proof}
  Let $\gamma \in \Gamma(\gamma^*(c_{c}, u_{c}))$ be an admissible policy with service rate and arrival rate given by $\mu(q)$ and $\lambda(q)$ respectively for $q \geq 0$.
  Let $q_{0} = \max\{ q : \mu(q) = 0\}$.
  Assume that for $\gamma$, $q_{0} > 0$.
  As $\gamma$ is admissible, $\mu(q) = 0$, for all $q \leq q_{0}$.
  Then the states $\brac{0,\dots,q_{0} - 1}$ are transient under policy $\gamma$.
  Let $\gamma'$ be another policy such that at a queue length $q$ the service rate and arrival rate are $\mu'(q)$ and $\lambda'(q)$ respectively.
  For $\forall q$, let $\mu'(q) = \mu(q + q_{0})$ and $\lambda'(q) = \lambda(q + q_{0})$.
  We note that the birth-death process under $\gamma'$ is obtained by a relabelling of the states under the policy $\gamma$.
  And $\gamma'$ is admissible as $\gamma$ is admissible.
  It is clear that $\overline{U}(\gamma') = \overline{U}(\gamma)$ and $\overline{C}(\gamma') = \overline{C}(\gamma)$, but $\overline{Q}(\gamma') = \overline{Q}(\gamma) - q_{0}$.
  Thus any $\gamma$ such that $q_{0} > 0$ cannot be an element of $\Gamma(\gamma^*(c_{c}, u_{c}))$.
\end{proof}
Thus in the following we need only consider non-idling admissible policies.

In the following discussion we consider the problem TRADEOFF,
\begin{eqnarray}
  \text{ minimize }_{\gamma \in \Gamma_{a}} & \overline{Q}(\gamma) \nonumber \\
  \text{ such that } & \overline{C}(\gamma) \leq c_{c}, \nonumber \\
  \text{ and } & \overline{U}(\gamma) \geq u_{c},
\end{eqnarray}
where we minimize over the set $\Gamma_{a}$ only.
The asymptotic bounds on the optimal value for TRADEOFF-M can be obtained easily from the analysis of TRADEOFF.
We now consider three special cases of TRADEOFF, which are either representative of the problems that arise in the context of communication networks or are useful in understanding the tradeoff for discrete time queues.

\paragraph{FINITE-$\mu$CHOICE:}
\begin{enumerate}
\item We restrict to policies $\gamma$ such that $\lambda(q) = \lambda, \forall q \in \mathbb{Z}_{+}$.
\item For any such policy $\gamma$, $\overline{U}(\gamma) = u(\lambda)$. We choose $\lambda$ such that $u(\lambda) \geq u_{c}$.
\item We also restrict to policies $\gamma$ such that $\mu(q) \in \mathcal{X}_{\mu} = \{\mu_{0} = 0, \mu_{1}, \mu_{2}, \cdots, \mu_{K}\}$, where $\mu_{i} < \mu_{i + 1}$, $\mu_{K} = r_{max} < \infty$. Thus the available service rates take values from a finite discrete set. We assume that $\lambda < r_{max}$.
\item The optimal value of the tradeoff problem is denoted by $Q^*(c_{c})$.
\end{enumerate}

\paragraph{INTERVAL-$\mu$CHOICE:}
\begin{enumerate}
\item We restrict to policies $\gamma$ such that $\lambda(q) = \lambda, \forall q \in \mathbb{Z}_{+}$.
\item For any such policy $\gamma, \overline{U}(\gamma) = u(\lambda)$. We choose $\lambda$ such that $u(\lambda) \geq u_{c}$.
\item We restrict to policies $\gamma$ such that $\mu(q) \in [0,r_{max}]$. Thus the available service rates take values in a finite interval. We assume that $\lambda < r_{max}$.
\item The optimal value of the tradeoff problem is denoted by $Q^*(c_{c})$.
\end{enumerate}

\paragraph{INTERVAL-$\lambda\mu$CHOICE:}
\begin{enumerate}
\item We restrict to policies $\gamma$ such that $\lambda(q) \in [r_{a,min},r_{a,max}]$, where $r_{a,min} \geq 0, r_{a,max} < \infty$.
\item We restrict to policies $\gamma$ such that $\mu(q) \in [0,r_{max}]$. We assume that $r_{a,min} < r_{max}$.
\end{enumerate}

We note that for FINITE-$\mu$CHOICE and INTERVAL-$\mu$CHOICE, the constraint on the average utility in TRADEOFF is satisfied by the choice of $\lambda$, and therefore this constraint is not explicitly mentioned (as in \eqref{introduction:eq:mm1_specialTradeoff}).
In the following, we obtain an asymptotic characterization of $Q^*(c_{c})$ for FINITE-$\mu$CHOICE, while INTERVAL-$\mu$CHOICE and INTERVAL-$\lambda\mu$CHOICE are analysed in Chapter 3.

\section{Analysis of FINITE-$\mu$CHOICE}
\label{chap4:sec:problem1}
In the following, we state the motivation for considering FINITE-$\mu$CHOICE.
\begin{remark}
  FINITE-$\mu$CHOICE is motivated by the tradeoff problem \eqref{introduction:eq:dt_specialTradeoff} for the following discrete time queueing model.
  Customers arrive in a batch of random size, in every slot, into an infinite length queue.
  All the customers which arrive in a slot are admitted into the queue.
  The number of customers, which are served in each slot, or the service batch size, is chosen as a deterministic function, of the current queue length.
  This feature of the discrete time queue is modelled by the choice of the service rate, $\mu(q)$, as a function of $q$ in FINITE-$\mu$CHOICE.
  We assume that for the discrete time model, the queue evolves on the set of non-negative integers.
  Hence, the service batch size also takes values in the set of non-negative integers.
  The essential feature here is that the set of batch sizes is discrete and therefore we assume that $\mu(q)$ takes values in a finite discrete set.
  As there is no admission control in the discrete time model, we assume that the arrival rate is a fixed $\lambda$ for every $q$ for FINITE-$\mu$CHOICE.
  In each slot, assume that there is a service cost incurred in serving the customers.
  This is modelled by the service cost rate function $c(.)$ in FINITE-$\mu$CHOICE.
  By analysing FINITE-$\mu$CHOICE we illustrate the basic techniques which are used in the asymptotic analysis of TRADEOFF, which turn out to be useful in the analysis of problem \eqref{introduction:eq:dt_specialTradeoff}.
\end{remark}

We now present an asymptotic analysis of FINITE-$\mu$CHOICE in the regime $\Re$.
We note that part of this analysis was presented in \cite{vineeth_ncc13_ct}.
For brevity, we use $\pi$ rather than $\pi_{\gamma}$ to denote the stationary distribution corresponding to an admissible policy $\gamma$ in places where there is no source for confusion.
The stationary probability of queue length being $q$ is denoted by $\pi(q)$.
The stationary probability of using a rate $\mu_{k}$ is denoted by $\pi_{\mu}(k)$. 
We note that $\pi_{\mu}(k) = \sum_{\{q : \mu(q) = \mu_{k}\}} \pi(q)$, and  $\overline{C}(\gamma) = \sum_{k = 0}^{K} \pi_{\mu}(k) c(\mu_{k})$.
Since $c(\mu)$ is convex in $\mu$, by Jensen's inequality, we have that the average service cost $\overline{C}(\gamma) \geq c(\lambda)$, for any admissible policy $\gamma$.
We note that $c(\lambda)$ is the minimum average service cost which has to be expended for the average service rate to be equal to the average arrival rate, as noted in Chapter 1.

We first obtain an asymptotic lower bound to $Q^*(c_{c})$ in the regime $\Re$ as $c_{c} \downarrow c(\lambda)$, by finding a lower bound on $\overline{Q}(\gamma_{k})$, as a function of $\overline{C}(\gamma_{k}) - c(\lambda)$, for any sequence of feasible admissible policies $\gamma_{k}$ with $\overline{C}(\gamma_{k}) \downarrow c(\lambda)$.
Subsequently, we show that there exists a sequence of admissible policies $\gamma_{k}$ for which $\overline{C}(\gamma_{k})$ approaches $c(\lambda)$ arbitrarily closely, so that $c(\lambda) = \inf_{\gamma \in \Gamma_{a}} \overline{C}(\gamma)$.

\begin{figure}[h]
  \centering
  \includegraphics[width=170mm,height=35mm]{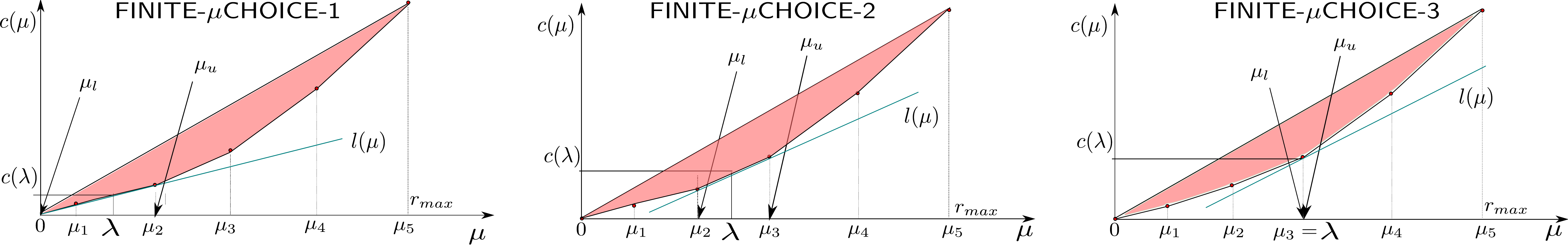}
  \caption{Illustration of the relationship between $\lambda$, $\mu_{l}$, and $\mu_{u}$ along with the minimum average service cost $c(\lambda)$ and the line $l(\mu)$ for the three cases that arises for FINITE-$\mu$CHOICE problem}
  \label{chap4:fig:lambdacases}
\end{figure}

The asymptotic behaviour of $Q^*(c_{c})$ for FINITE-$\mu$CHOICE depends on the behaviour of $c(\mu)$ in the neighbourhood of $\mu = \lambda$.
We now define quantities $\mu_{u}, k_{u}, \mu_{l},$ and $k_{l}$, which are related to this behaviour.
\begin{eqnarray}
  \mu_{u} = & 
  \begin{cases}
    \min\left\{\mu_{k} : k \leq K - 1, \mu_{k} \geq \lambda, \frac{c(\mu_{k + 1}) - c(\mu_{k})}{\mu_{k + 1} - \mu_{k}} > \frac{c(\mu_{k}) - c(\mu_{k - 1})}{\mu_{k} - \mu_{k - 1}} \right\} & \text{ if this set is non-empty}, \nonumber \\
    \mu_{K} & \text{ otherwise}.
  \end{cases} \\
  \mu_{l} = & 
  \begin{cases}
    \max\left\{\mu_{k} : k \geq 1, \mu_{k} \leq \lambda, \frac{c(\mu_{k + 1}) - c(\mu_{k})}{\mu_{k + 1} - \mu_{k}} > \frac{c(\mu_{k}) - c(\mu_{k - 1})}{\mu_{k} - \mu_{k - 1}}\right\} & \text{ if this set is non-empty}, \nonumber \\
    0 & \text{ otherwise}.
  \end{cases}
\end{eqnarray}
Let $\mu_{k_{u}} = \mu_{u}$ and $\mu_{k_{l}} = \mu_{l}$. 
In words, $\mu_{u}$ is the service rate $\mu_{k_{u}}$ greater than or equal to $\lambda$ at which the slope of $c(.)$, considered for service rates in $\mathcal{X}_{\mu}$, changes. 
A similar interpretation can be given for $\mu_{l}$.
Depending upon the value of $\mu_{l}$, $\mu_{u}$, and their relationship with $\lambda$, there are three different cases that need to be considered (also see Figure \ref{chap4:fig:lambdacases}):
\begin{description}
\item[FINITE-$\mu$CHOICE-1 : ]{$\mu_{l} = 0, \lambda < \mu_{u}$,}
\item[FINITE-$\mu$CHOICE-2 : ]{$\mu_{l} \geq \mu_{1}, \mu_{l} < \lambda < \mu_{u}$, and }
\item[FINITE-$\mu$CHOICE-3 : ]{$\mu_{l} \geq \mu_{1}, \mu_{l} = \lambda = \mu_{u}$.}
\end{description}
We now state the motivation for this classification.
We note that as the constraint $c_{c}$ approaches $c(\lambda)$, if there exists a feasible policy $\gamma$ with $\overline{C}(\gamma) \leq c_{c}$, then for that policy the stationary probability of certain service rates should go to zero.
The classification is based on the set of service rates, whose stationary probability goes to zero.
For example, for FINITE-$\mu$CHOICE-3, as $(\lambda, c(\lambda))$ is a \emph{corner} point, as $c_{c} \downarrow c(\lambda)$, the stationary probability that any service rate other than $\lambda$ is used approaches zero.
For FINITE-$\mu$CHOICE-1 and FINITE-$\mu$CHOICE-2, as $c_{c} \downarrow c(\lambda)$, the stationary probability that any service rate which is less than $\mu_{l}$ or greater than $\mu_{u}$ is used, approaches zero.
We note that $\mu_{u} > \lambda$ in the case of FINITE-$\mu$CHOICE-1 and FINITE-$\mu$CHOICE-2, and as $c_{c} \downarrow c(\lambda)$ the service rate $\mu_{u} > \lambda$ could be used, unlike in the case of FINITE-$\mu$CHOICE-3 where only $\mu = \lambda$ can be used.
Furthermore we note that for FINITE-$\mu$CHOICE-1, $\mu_{l} = 0$ and for both FINITE-$\mu$CHOICE-2 and \FMC-3, $\mu_{l} > 0$.
Then as $c_{c} \downarrow c(\lambda)$, for \FMC-1, for a non-idling $\gamma \in \Gamma_{a}$, the queue becomes empty infinitely often, while this cannot happen for both \FMC-2 and \FMC-3.

\subsection{Asymptotic lower bounds}
\label{chap4:sec:p1_asymp_lb}

For an admissible policy $\gamma$, for obtaining an asymptotic lower bound on $\overline{Q}(\gamma)$ we: a) obtain an upper bound on $\pi_{\mu}(k)$ for certain values of $k$ in terms of $\overline{C}(\gamma)$ and $c(\lambda)$, b) relate the stationary probability $\pi_{\mu}(k)$ to the stationary probability of the queue $\pi(q)$, and c) obtain a lower bound on $\overline{Q}(\gamma)$ in terms of $\pi(q)$.
For the cases FINITE-$\mu$CHOICE-1 and FINITE-$\mu$CHOICE-2, define the line $l(\mu)$ as the line through the points $(\mu_{l}, c(\mu_{l}))$ and $(\mu_{u}, c(\mu_{u}))$.
For the case FINITE-$\mu$CHOICE-3, let $l(\mu)$ be any line through $(\lambda, c(\lambda))$ with slope greater than $\frac{c(\lambda) - c(\mu_{k_{l} - 1})}{\lambda - \mu_{k_{l} - 1}}$ and less than $\frac{c(\mu_{k_{l} + 1}) - c(\lambda)}{\mu_{k_{l} + 1} - \lambda}$.
The line $l(\mu)$ is illustrated for the three cases in Figure \ref{chap4:fig:lambdacases}.
We note that $c(\lambda) = l(\lambda)$ by construction.
Furthermore $\mathbb{E}_{\pi_{\mu}}l(\mu(Q)) = l(\lambda)$ as $l$ is linear.

We now present an upper bound on the stationary probability of certain service rates as the average service cost approaches $c(\lambda)$.
\begin{lemma}
  Let $\mathcal{R}_{k} = \{k : \mu_{k} < \mu_{l} \text{ or } \mu_{k} > \mu_{u}\}$.
  For an admissible policy $\gamma$, for all $k \in \mathcal{R}_{k}$, 
  \begin{equation*}
    \pi_{\mu}(k) \leq \frac{\overline{C}(\gamma) - c(\lambda)}{c(\mu_{k}) - l(\mu_{k})}.
  \end{equation*}
  \label{chap4:prop:pimu_ub}
\end{lemma}
\begin{proof}
  We have that $\overline{C}(\gamma) - c(\lambda) = \sum_{k = 0}^{K} \pi_{\mu}(k)[c(\mu_{k}) - l(\mu_{k})]$.
  For $k \in \mathcal{R}_{k}^{c}$, $c(\mu_{k}) - l(\mu_{k}) = 0$.
  Since $c(\mu_{k}) > l(\mu_{k})$, for all $k \in \mathcal{R}_{k}$, we have that $\pi_{\mu}(k) \leq \frac{\Cg - c(\lambda)}{c(\mu_{k}) - l(\mu_{k})}$.
\end{proof}
A non-idling admissible policy $\gamma$ is specified by the sequence $(q_0 = 0, q_{1}, q_{2}, \cdots, q_{K - 1}, q_{K} = \infty), q_{k} \leq q_{k + 1}$, which is such that 
\begin{eqnarray*}
  \mu(q_{0}) & = & \mu(0) = 0 \nonumber \\
  \mu(q) & = & \mu_{k}, \text{ if } q \in \{q_{k - 1} + 1, \cdots, q_{k}\}, \text{ for } k \in \{1,\cdots,K\}.
  \label{chap4:eq:P1policyspec}
\end{eqnarray*}
For any $k$, if $q_{k - 1} = q_{k}$, then the rate $\mu_{k}$ is not used by the policy $\gamma$.
From the definition of $\pi_{\mu}(k)$ we have that
\begin{eqnarray}
  \pi_{\mu}(0) & = & \pi(0), \nonumber \\
  \pi_{\mu}(k) & = & \sum_{q = q_{k - 1} + 1}^{q_{k}} \pi(q), \forall k \geq 1.
  \label{chap4:eq:piz_pimu_relation}
\end{eqnarray}
In the following we obtain lower bounds on the average queue length $\overline{Q}(\gamma)$ as a function of the upper bounds in Lemma \ref{chap4:prop:pimu_ub} on stationary probabilities of service rates, when $c_{c} - c(\lambda) = V \downarrow 0$ for the three cases FINITE-$\mu$CHOICE-1, FINITE-$\mu$CHOICE-2, and FINITE-$\mu$CHOICE-3 respectively.

We first consider the case FINITE-$\mu$CHOICE-1, where $k_{l} = 0$ and $\mu_{l} = 0$.
\begin{remark}
  If $\mu_{u} = \mu_{K}$, then we note that $c(\mu)$ is a linear function of $\mu$.
  Then the admissible policy $\gamma$, with $\mu(0) = 0$, and $\mu(q) = \mu_{K}$ for all $q \geq 1$, has $\overline{C}(\gamma) = \Exp c(\mu(Q)) = c(\Exp \mu(Q)) = c(\lambda)$.
  Furthermore, $\gamma$ has the minimum average queue length $\frac{\lambda}{\mu_{K} - \lambda}$.
  Hence in this case there is no tradeoff between the average queue length and average service cost.
  So in the following, we assume $c(\mu)$ is such that $\mu_{u} < \mu_{K}$.
\end{remark}

We note that an admissible policy $\gamma$, which uses only the service rates $\brac{\mu_{k}, k \in \{0,\cdots, k_{u}\}}$ has $\overline{C}(\gamma) = \sum_{k = 0}^{k_{u}} \pi_{\mu}(k) c(\mu_{k}) = \sum_{k = 0}^{k_{u}} \pi_{\mu}(k) l(\mu_{k}) = c(\lambda)$.
Furthermore, $\overline{C}(\gamma) = c(\lambda)$ is achieved only by admissible policies which uses only the service rates $\brac{\mu_{k}, k \in \{0,\cdots, k_{u}\}}$.
Hence, as $\lambda < \mu_{u}$, the policy $\gamma_{u}$, that uses $\mu(0) = 0$ and $\mu(q) = \mu_{u}$ for $q \geq 1$ has a service cost of $c(\lambda)$.
We note that $\gamma_{u}$ has the minimum average queue length $\frac{\lambda}{\mu_{u} - \lambda}$, among all policies $\gamma$ for which $\overline{C}(\gamma) = c(\lambda)$.
Thus the minimum average queue length, among policies $\gamma$ for which $\overline{C}(\gamma) \leq c_{c}$, where $c_{c}  > c(\lambda)$ is at most $\frac{\lambda}{\mu_{u} - \lambda}$.
We note that if $\overline{C}(\gamma) > c(\lambda)$, then service rates $\mu_{k}$ with $k > k_{u}$ could be used, which could yield an average queue length less than $\frac{\lambda}{\mu_{u} - \lambda}$.
In the following lower bound on $\overline{Q}(\gamma)$, we observe that if $c_{c} > c(\lambda)$, then the average queue length can be less than $\frac{\lambda}{\mu_{u} - \lambda}$, but has the limit $\frac{\lambda}{\mu_{u} - \lambda}$, as $\overline{C}(\gamma) \rightarrow c(\lambda)$.
\begin{lemma}
  For any sequence of non-idling admissible policies $\gamma_{k}$ with $\overline{C}(\gamma_{k}) - c(\lambda) = V_{k} \downarrow 0$, we have that
  \begin{equation*}
    \frac{\lambda}{\mu_{u} - \lambda} - \overline{Q}(\gamma_{k}) = \mathcal{O}\brap{V_{k}\log\left(\frac{1}{V_{k}}\right)}.
  \end{equation*}
  \label{chap4:prop:p11lb}
\end{lemma}
In the proof, for any sequence of non-idling admissible policies $\gamma_{k}$, we first show that as $V_{k} = \overline{C}(\gamma_{k}) - c(\lambda) \downarrow 0$, the largest queue length $q_{k_{u}}$ at which service rate $\mu_{u}$ is used, increases to infinity as $\log\nfrac{1}{V_{k}}$.
This asymptotic lower bound is obtained by showing that a lower bound $q_{k_{u},l}$ to $q_{k_{u}}$ increases as $\log\nfrac{1}{V_{k}}$.
For a $V_{k}$, the lower bound $q_{k_{u},l}$ is used to define the policy $\gamma'_{k}$, which has $\mu(0) = 0, \mu(q) = \mu_{u}$ for $1 \leq q \leq q_{k_u,l}$, and $\mu(q) = \mu_{K}$ for $q > q_{k_{u},l}$.
We note that for any policy $\gamma_{k}$ with $\overline{C}(\gamma) - c(\lambda) = V_{k}$, $\overline{Q}(\gamma_{k}) \geq \overline{Q}(\gamma'_{k})$.
The sequence $\overline{Q}(\gamma'_{k})$, obtained as $V_{k} \downarrow 0$, is shown to have the asymptotic behaviour in the above lemma.
\begin{proof}
  Let us consider a particular policy $\gamma$ in the sequence $\gamma_{k}$ with $\overline{C}(\gamma) - c(\lambda) = V$.
  As $Q(t)$ is a birth-death process we have that
  \begin{equation*}
    \pi(q - 1) \leq \pi(q)\frac{\mu_{u}}{\lambda}, \forall q \in \brac{1,\cdots, q_{k_{u}}}.
  \end{equation*}
  By induction, we obtain that
  \begin{equation*}
    \pi(q_{k_{u}} - m) \leq \pi(q_{k_{u}})\left(\frac{\mu_{u}}{\lambda}\right)^{m} \leq \pi(q_{k_{u}} + 1)\frac{\mu_{K}}{\lambda}\left(\frac{\mu_{u}}{\lambda}\right)^{m}, \forall m \in \{0,\cdots,q_{k_{u}}\}.
  \end{equation*}
  Now we note that
  \begin{eqnarray*}
    \sum_{q = 0}^{q_{k_{u}}} \pi(q) & = & \sum_{k = 0}^{k_{u}} \pi_{\mu}(k) = 1 - \sum_{k = k_{u} + 1}^{K} \pi_{\mu}(k) \\
    \text{but } \sum_{q = 0}^{q_{k_{u}}} \pi(q) & = & \sum_{m = 0}^{q_{k_{u}}} \pi(q_{k_{u}} - m) \leq \pi(q_{k_{u}} + 1)\frac{\mu_{K}}{\lambda} \sum_{m = 0}^{q_{k_{u}}} \left( \frac{\mu_{k_{u}}}{\lambda}\right)^{m}, \\
    \text{and from Lemma \ref{chap4:prop:pimu_ub}, } 1 - \sum_{k = k_{u} + 1}^{K} \pi_{\mu}(k) & \geq & 1 - \sum_{k = k_{u} + 1}^{K} \frac{V}{c(\mu_{k}) - l(\mu_{k})}.
  \end{eqnarray*}
  Hence
  \begin{eqnarray*}
    1 - \sum_{k = k_{u} + 1}^{K} \frac{V}{c(\mu_{k}) - l(\mu_{k})} \leq \pi(q_{k_{u}} + 1)\frac{\mu_{K}}{\lambda} \sum_{m = 0}^{q_{k_{u}}} \left(\frac{\mu_{k_{u}}}{\lambda}\right)^{m}, \\
    \text{ or } 1 - \sum_{k = k_{u} + 1}^{K} \frac{V}{c(\mu_{k}) - l(\mu_{k})} \leq \pi(q_{k_{u}} + 1)\mu_{K}\frac{\left(\frac{\mu_{k_{u}}}{\lambda}\right)^{q_{k_{u}} + 1} - 1}{\mu_{k_{u}} - \lambda}, \\
    \text{ or } \frac{\mu_{u} - \lambda}{\mu_{K}} \frac{1}{\pi(q_{k_{u}} + 1)} \left[1 - \sum_{k = k_{u} + 1}^{K} \frac{V}{c(\mu_{k}) - l(\mu_{k})} \right] + 1 \leq \left(\frac{\mu_{u}}{\lambda}\right)^{q_{k_{u}} + 1}.
  \end{eqnarray*}
  Therefore
  \begin{eqnarray*}
    q_{k_{u}} \geq \log_{\left(\frac{\mu_{u}}{\lambda}\right)} \left[ \frac{\mu_{u} - \lambda}{\mu_{K}} \frac{1}{\pi(q_{k_{u}} + 1)} \left[1 - \sum_{k = k_{u} + 1}^{K} \frac{V}{c(\mu_{k}) - l(\mu_{k})} \right] + 1 \right] - 1.
  \end{eqnarray*}
  But we note that $\pi(q_{k_{u}} + 1) \leq \sum_{k = k_{u} + 1}^{K} \pi_{\mu}(k) \leq \sum_{k = k_{u} + 1}^{K} \frac{V}{c(\mu_{k}) - l(\mu_{k})} \stackrel{\Delta} = \frac{V}{c_{1}}$.
  So that
  \begin{eqnarray}
    q_{k_{u}} \geq \log_{\left(\frac{\mu_{u}}{\lambda}\right)} \left[ \frac{\mu_{u} - \lambda}{\mu_{K}} \brap{\frac{c_{1}}{V}  - 1} + 1 \right] - 1.
    \label{chap4:eq:zkulb}
  \end{eqnarray}
  Therefore, for any non-idling admissible policy $\gamma$ we have that $q_{k_{u}} \geq q_{k_{u},l}$, where
  \begin{eqnarray*}
    q_{k_{u},l} \stackrel{\Delta} = \left \lceil \log_{\left(\frac{\mu_{u}}{\lambda}\right)} \left[ \frac{\mu_{u} - \lambda}{\mu_{K}} \frac{c_{1}}{V} \left[1 - \frac{V}{c_{1}} \right] + 1 \right] - 1 \right \rceil.
  \end{eqnarray*}
  Now we note that for the policy $\gamma$ under consideration, $\mu(q) \leq \mu_{u}$ for $q \in \{0,\cdots, q_{k_{u}}\}$,  and $\mu(q) \leq \mu_{K}$ for $q \in \{q_{k_{u}} + 1, \cdots\}$.
  Let policy $\gamma'$ be defined as follows :
   \begin{eqnarray*}
    \mu'(q) = &
    \begin{cases}
      0 \text{ if } q = 0, \\
      \mu_{u} \text{ if } q \in \{1, \cdots, q_{k_{u},l}\}, \\
      \mu_{K} \text{ otherwise}.
    \end{cases}
  \end{eqnarray*}
  We note that $\overline{Q}(\gamma') \leq \overline{Q}(\gamma)$.
  Let the stationary distribution of CTMC under $\gamma'$ be denoted as $\pi'(q)$.
  Let $a \stackrel{\Delta} = \frac{\lambda}{\mu_{u}}$ and $b \stackrel{\Delta} = \frac{\lambda}{\mu_{K}}$.
  Then we have that
    \begin{eqnarray*}
    \pi'(q) = &
    \begin{cases}
      \pi'(0)a^{q} \text{ if } q \in \{0,\cdots, q_{k_{u},l}\} \\
      \pi'(0)a^{q_{k_{u},l}} b^{q - q_{k_{u},l}} \text{ otherwise.}
    \end{cases}
  \end{eqnarray*}
  We note that $\gamma'$ is also admissible.
  As $\sum_{q = 0}^{\infty} {\pi'}(q) = 1$, we have that
  \begin{eqnarray}
    {\pi'}(0)\left[ 1 + \sum_{q = 1}^{q_{k_{u},l}} a^{q} + \sum_{q = q_{k_{u},l} + 1}^{\infty} a^{q_{k_{u},l}} b^{q - q_{k_{u},l}} \right] & = & 1, \nonumber \\
    {\pi'}(0)\left[1 + \frac{a}{1 - a}\left(1 - a^{q_{k_{u},l}}\right) + a^{q_{k_{u},l}}\frac{b}{1 - b} \right] & = & 1.
    \label{chap4:eq:tildepi0}
  \end{eqnarray}
  We note that ${\pi'}(0)$ can be obtained from \eqref{chap4:eq:tildepi0} in terms of $q_{k_{u},l}$.
  We have
  \begin{eqnarray*}
    \overline{Q}(\gamma') =  \sum_{q = 0}^{\infty} {\pi'}(q) q = {\pi'}(0)\left[ \sum_{q = 1}^{q_{k_{u},l}} q a^{q} + \sum_{q = q_{k_{u},l} + 1}^{\infty} q a^{q_{k_{u},l}} b^{q - q_{k_{u},l}} \right].
  \end{eqnarray*}
  Simplification leads to
  \small
  \begin{eqnarray}
    \overline{Q}(\gamma') & = & {\pi'}(0)\left[ \frac{a}{(1 - a)^{2}}\left( (1 - a)(1 - (q_{k_{u},l} + 1)a^{q_{k_{u},l}}) + a(1 - a^{q_{k_{u},l}})\right) + a^{q_{k_{u},l}}\left(q_{k_{u},l} \frac{b}{1 - b} + \frac{b}{(1 - b)^{2}} \right) \right], \nonumber \\
    & = & \frac{ \left[ \frac{a}{(1 - a)^{2}}\left( (1-a)(1 - (q_{k_{u},l} + 1)a^{q_{k_{u},l}}) + a(1 - a^{q_{k_{u},l}})\right) + a^{q_{k_{u},l}}\left(q_{k_{u},l} \frac{b}{1 - b} + \frac{b}{(1 - b)^{2}} \right) \right]}{ \left[1 + \frac{a}{1 - a}\left(1 - a^{q_{k_{u},l}}\right) + a^{q_{k_{u},l}}\frac{b}{1 - b} \right]}
    \label{chap4:eq:ezdomexpression}
  \end{eqnarray}
  \normalsize
  where ${\pi'}(0)$ was obtained from \eqref{chap4:eq:tildepi0}.
  At this point, we note that as $V \rightarrow 0$, $q_{k_{u},l} \rightarrow \infty$ and this lower bound $\overline{Q}(\gamma')$ to $\overline{Q}(\gamma)$ approaches $\frac{a}{1 - a}$, which is what we expect.
  However, in order to ascertain the behaviour of average queue length as $V$ approaches zero, we need to lower bound the right-hand side of \eqref{chap4:eq:ezdomexpression} for $V > 0$.

  The denominator of \eqref{chap4:eq:ezdomexpression} can be bounded above as follows :
  \begin{eqnarray*}
    & & 1 + \frac{a}{1 - a} - \frac{a}{1 - a} a^{q_{k_{u},l}} + \frac{b}{1 - b} a^{q_{k_{u},l}} \nonumber \\
    & = & 1 + \frac{a}{1 - a} + a^{q_{k_{u},l}}\left[ \frac{1}{\frac{1}{b} - 1} - \frac{1}{\frac{1}{a} - 1} \right].
  \end{eqnarray*}
  As $b < a$, $\frac{1}{\frac{1}{b} - 1} < \frac{1}{\frac{1}{a} - 1}$, so that the denominator of \eqref{chap4:eq:ezdomexpression} $\leq \frac{1}{1 - a}$.
  After substituting this upper bound for the denominator in \eqref{chap4:eq:ezdomexpression}, we have
  \small
  \begin{eqnarray}
    \overline{Q}(\gamma') & \geq & (1 - a)\left[ \frac{a}{(1 - a)^{2}} \left\{1 - (1 - a)(q_{k_{u},l} + 1) a^{q_{k_{u},l}} - a.a^{q_{k_{u},l}}\right\} + a^{q_{k_{u},l}}\left\{ q_{k_{u},l} \frac{b}{1 - b} + \frac{b}{(1 - b)^{2}} \right\} \right], \nonumber \\
    & = & \frac{a}{1 - a} + (1 - a)\left[ a^{q_{k_{u},l}}\left\{ q_{k_{u},l} \frac{b}{1 - b} + \frac{b}{(1 - b)^{2}}\right\} - \frac{a}{(1 - a)^{2}} \left\{ (1 - a)(q_{k_{u},l} + 1)a^{q_{k_{u},l}} + a.a^{q_{k_{u},l}} \right\}\right], \nonumber \\
    & \geq & \frac{a}{1 - a} - \frac{a}{1 - a} a^{q_{k_{u},l}}\left[1 + (1 - a)q_{k_{u},l}\right].
    \label{chap4:eq:p13ezlb}
  \end{eqnarray}	
  \normalsize
  From \eqref{chap4:eq:zkulb}, with $a < 1$, we have that
  \begin{eqnarray*}
    a^{q_{k_{u},l}} & \leq & \frac{1}{a} a^{\log_{\frac{1}{a}} \left[ \frac{\mu_{u} - \lambda}{\mu_{K}} \brap{\frac{c_{1}}{V} - 1} + 1 \right]}, \\
    & = & \frac{1}{a} \left[ a^{\log_{{a}} \left[ \frac{\mu_{u} - \lambda}{\mu_{K}} \brap{\frac{c_1}{V} - 1} + 1 \right]}\right]^{\frac{1}{\log_{a} (1/a)}}, \\
    & = & \frac{1}{a} \left[ \frac{\frac{\mu_{u} - \lambda}{\mu_{K}} \brap{c_{1} - V} + V}{V} \right]^{-1}, \\
    & = & \frac{1}{a}\frac{1}{\frac{\mu_{u} - \lambda}{\mu_{K}}\left(\frac{c_{1}}{V} - 1 \right) + 1}.
  \end{eqnarray*}
  From the definition of $q_{k_{u},l}$ we have that
  \begin{eqnarray*}
    q_{k_u,l} \leq \log_{\left(\frac{\mu_{u}}{\lambda}\right)} \left[ \frac{\mu_{u} - \lambda}{\mu_{K}} \brap{\frac{c_{1}}{V} - 1} + 1 \right]
  \end{eqnarray*}
  Therefore, $a^{q_{k_{u},l}}(1 + (1 - a)q_{k_{u},l}) \leq$
  \small
  \begin{eqnarray*}
    \frac{1}{a}\frac{1}{\frac{\mu_{u} - \lambda}{\mu_{K}}\left(\frac{c_{1}}{V} - 1 \right) + 1}\left( 1 + (1 - a) \left\{ \log_{\left(\frac{\mu_{u}}{\lambda}\right)} \left[ \frac{\mu_{u} - \lambda}{\mu_{K}} \brap{\frac{c_{1}}{V}  - 1} + 1 \right]
      \right\}\right)
  \end{eqnarray*}
  \normalsize
  Substituting this in \eqref{chap4:eq:p13ezlb} we have that $\frac{a}{1 - a} - \overline{Q}(\gamma') \leq$ 
  \small
  \begin{eqnarray}
    \frac{a}{1-a}\left\{\frac{V}{a}\frac{1}{\frac{\mu_{u} - \lambda}{\mu_{K}}c_{1}\left(1 - \frac{V}{c_1} \right) + V}\left( 1 + (1 - a) \left\{ \log_{\left(\frac{\mu_{u}}{\lambda}\right)} \left[ \frac{\mu_{u} - \lambda}{\mu_{K}} \brap{\frac{c_{1}}{V} - 1} + 1 \right]\right\}\right)\right\}.
    \label{chap4:eq:p11finalbound}
  \end{eqnarray}
  \normalsize
  Thus, for a sequence $\gamma_k$ such that $V_{k} \downarrow 0$, we have that $\frac{a}{1 - a} - \overline{Q}(\gamma_{k}) = \mathcal{O}\brap{V_{k}\log\left(\frac{1}{V_{k}}\right)}$.
\end{proof}

\begin{corollary}
  For any sequence of non-idling admissible policies $\gamma_{M,k} \in \Gamma_{a,M}$ with $\overline{C}(\gamma_{M,k}) - c(\lambda) = V_{k} \downarrow 0$, we have that
  \begin{equation*}
    \frac{\lambda}{\mu_{u} - \lambda} - \overline{Q}(\gamma_{M,k}) = \mathcal{O}\brap{V_{k}\log\left(\frac{1}{V_{k}}\right)}.
  \end{equation*}
  \label{chap4:prop:p11lb_mixed}
\end{corollary}
\begin{proof}
  For a $k$, if $\gamma_{M,k}$ is such that $\overline{C}(\gamma_{M,k}) - c(\lambda) = V_{k}$, then we have that for every $\gamma_{k} \in \Gamma(\gamma_{M,k})$, $\overline{C}(\gamma_{k}) - c(\lambda) \leq \frac{V_{k}}{p_{\gamma_{k}}} = U_{\gamma_{k}}$.
  Then \[\frac{\lambda}{\mu_{u} - \lambda} - \overline{Q}(\gamma_{k}) = \mathcal{O}\brap{U_{\gamma_{k}}\log\left(\frac{1}{U_{\gamma_{k}}}\right)}.\]
  We note that $\overline{Q}(\gamma_{k,M}) = \Exp_{p_{\gamma_{k}}} \overline{Q}(\gamma_{k})$.
  Then applying $\Exp_{p_{\gamma_{k}}}$ to LHS and RHS of the above equation we obtain that
  \[\frac{\lambda}{\mu_{u} - \lambda} - \overline{Q}(\gamma_{k,M}) = \sum {p_{\gamma_{k}}} \bras{\mathcal{O}\brap{U_{\gamma_{k}}\log\left(\frac{1}{U_{\gamma_{k}}}\right)}}.\]
  We note that by definition, there exists some constant $c > 0$ such that the RHS is
  \[\leq \sum {p_{\gamma_{k}}} \bras{c\bras{U_{\gamma_{k}}\log\left(\frac{1}{U_{\gamma_{k}}}\right)}}.\]
  Since the function $u\log\nfrac{1}{u}$ is concave, we have that
  \[\sum {p_{\gamma_{k}}} \bras{c\bras{U_{\gamma_{k}}\log\left(\frac{1}{U_{\gamma_{k}}}\right)}} \leq c \sum {p_{\gamma_{k}}}\bras{U_{\gamma_{k}}}\log\left(\frac{1}{\sum {p_{\gamma_{k}}}\bras{U_{\gamma_{k}}}}\right) .\]
  Since $\sum {p_{\gamma_{k}}}\bras{U_{\gamma_{k}}} = V_{k}$ we have that 
  \begin{equation*}
    \frac{\lambda}{\mu_{u} - \lambda} - \overline{Q}(\gamma_{M,k}) = \mathcal{O}\brap{V_{k}\log\left(\frac{1}{V_{k}}\right)}.
  \end{equation*}
\end{proof}

We now present asymptotic lower bounds for FINITE-$\mu$CHOICE-2 and FINITE-$\mu$CHOICE-3.
For an admissible policy $\gamma$, to relate the stationary probability distribution $\pi(q)$ to the average queue length $\overline{Q}(\gamma)$, as noted in Section \ref{chap2:methodology} we make use of the fact that if $\overline{q}$ is such that $\sum_{q = 0}^{\overline{q}} \pi(q) \leq \frac{1}{2}$, then $\overline{Q}(\gamma) \geq \frac{\overline{q}}{2}$.
The choice of $\frac{1}{2}$ here is arbitrary.
The best lower bound on $\overline{Q}(\gamma)$ is given by the largest $\overline{q}$ such that $\sum_{q = 0}^{\overline{q}} \pi(q) \leq \frac{1}{2}$.

\begin{lemma}
  For any sequence of non-idling admissible policies $\gamma_{k}$ with $\overline{C}(\gamma_{k}) - c(\lambda) = V_{k} \downarrow 0$, we have that 
  \begin{equation}
    \overline{Q}(\gamma_{k}) = 
    \begin{cases}
      \Omega\left(\log\left(\frac{1}{V_{k}}\right)\right) \text{ for FINITE-$\mu$CHOICE-2,} \\
      \Omega\left(\frac{1}{V_{k}}\right) \text{ for FINITE-$\mu$CHOICE-3.}
    \end{cases}
  \end{equation}
  \label{chap4:prop:p1213lb}
\end{lemma}
\begin{proof}
  We note that $k_{l} \geq 1$ for the cases FINITE-$\mu$CHOICE-2 and FINITE-$\mu$CHOICE-3.
  Consider a particular policy $\gamma$ in the given sequence $\gamma_{k}$, with $\overline{C}(\gamma) - c(\lambda) = V$.
  Then we note that $\sum_{k = 0}^{k_{l} - 1} \pi_{\mu}(k) \leq \sum_{k = 0}^{k_{l} - 1} \frac{V}{c(\mu_{k}) - l(\mu_{k})} \stackrel{\Delta} = \frac{V}{c'_{1}}$, from the upper bound in Lemma \ref{chap4:prop:pimu_ub}.
  Therefore, $\sum_{q = 0}^{q_{k_{l} - 1}} \pi(q) = \sum_{k = 0}^{k_l - 1} \pi_{\mu}(k) \leq \frac{V}{c_{1}'}$.
  Also, for every $q \in \brac{0, \cdots, q_{k_{l} - 1}}$, $\pi(q) \leq \frac{V}{c'_{1}}$.
  Now we intend to find the largest $\overline{q}$ such that $\sum_{q = 0}^{\overline{q}} \pi(q) \leq \frac{1}{2}$.
  But as $\sum_{q = 0}^{q_{k_{l} - 1}} \pi(q) \leq \frac{V}{c'_{1}}$ and $\pi(q_{k_{l} - 1} + 1) \leq \pi(q_{k_{l} - 1}) \frac{\lambda}{\mu_{l}} \leq \frac{\lambda V}{\mu_{l} c'_{1}}$, the largest such $\overline{q}$ satisfies 
  \begin{equation*}
    \sum_{q = q_{k_{l} - 1} + 1}^{\overline{q}} \pi(q) \leq \frac{1}{2} - \frac{V}{c_{1}}
  \end{equation*}
  In the following, we use an upper bound on $\pi(q)$ which leads to a lower bound $\overline{q}_{1}$ on $\overline{q}$.
  Since $\pi(q)\lambda = \pi(q + 1)\mu(q)$, and for $q > q_{k_{l}}$, $\frac{\lambda}{\mu(q)} < \frac{\lambda}{\mu_{l}}$ (from the admissibility of $\gamma$), we have that 
  \begin{eqnarray*}
    \sum_{q = q_{k_{l} - 1} + 1}^{\overline{q}} \pi(q) < \pi(q_{k_{l} - 1}) \sum_{m = 1}^{\overline{q} - q_{k_{l}-1}} \left(\frac{\lambda}{\mu_{l}}\right)^{m}.
  \end{eqnarray*}
  If $\overline{q}_{1}$ is the largest integer such that
  \begin{equation}
    \pi(q_{k_{l} - 1}) \sum_{m = 1}^{\overline{q}_{1} - q_{k_{l}-1}} \left(\frac{\lambda}{\mu_{l}}\right)^{m} \leq \frac{1}{2} - \frac{V}{c'_{1}},
    \label{chap4:eq:pizub}
  \end{equation}
  then $\sum_{q = 0}^{\overline{q}_{1}} \pi(q) \leq \frac{1}{2}$ and $\overline{q} \geq \overline{q}_{1}$.

  In the case FINITE-$\mu$CHOICE-2, $\lambda > \mu_{l}$, so that summing the geometric series in \eqref{chap4:eq:pizub}, we have that
  \begin{eqnarray}
    \fpow{\lambda}{\mu_{l}}{\overline{q}_{1} - q_{k_{l} - 1}} - 1 \leq \frac{\lambda - \mu_{l}}{\lambda \pi(q_{k_{l} - 1})} \left(\frac{1}{2} - \frac{V}{c'_1}\right) \text{ or }
    \label{chap4:eq:p12basic} \\
    \overline{q}_{1} \leq q_{k_{l} - 1} + \bras{\log_{\left( \frac{\lambda}{\mu_{l}} \right)}\left(1 + \frac{\lambda - \mu_{l}}{\lambda \pi(q_{k_{l} - 1})} \left(\frac{1}{2} - \frac{V}{c'_{1}}\right)\right)}, \nonumber
  \end{eqnarray}
  and, in fact, 
  \begin{eqnarray*}
    \overline{q}_{1} = \left\lfloor q_{k_{l} - 1} + \log_{\left( \frac{\lambda}{\mu_{l}} \right)}\left(1 + \frac{\lambda - \mu_{l}}{\lambda \pi(q_{k_{l} - 1})} \left(\frac{1}{2} - \frac{V}{c'_{1}}\right)\right)\right\rfloor.
  \end{eqnarray*}
  Since $q_{k_{l} - 1} \geq 0$, we have that
  \begin{eqnarray*}
    \overline{q}_{1} \geq \log_{\left( \frac{\lambda}{\mu_{l}} \right)}\left(1 + \frac{\lambda - \mu_{l}}{\lambda \pi(q_{k_{l} - 1})} \left(\frac{1}{2} - \frac{V}{c'_{1}}\right)\right) - 1
  \end{eqnarray*}
  Now since $\pi(q_{k_{l} - 1}) \leq \pi_{\mu}(k_{l}) \leq \frac{V}{c({\mu_{k_{l} - 1}}) - l(\mu_{k_{l} - 1})}$ we have that
  \begin{equation*}
    \overline{q}_{1} \geq \log_{\left( \frac{\lambda}{\mu_{l}} \right)} \left(1 + \frac{(\lambda - \mu_{l})(c({\mu_{k_{l} - 1}}) - l(\mu_{k_{l} - 1}))}{\lambda V} \left(\frac{1}{2} - \frac{V}{c'_{1}}\right)\right) - 1
  \end{equation*}
  Then we have that
  \begin{equation}
    \overline{Q}(\gamma) \geq \frac{\overline{q}}{2} \geq \frac{\overline{q}_{1}}{2} \geq \frac{1}{2} \log_{\left( \frac{\lambda}{\mu_{l}} \right)} \left(1 + \frac{(\lambda - \mu_{l})(c({\mu_{k_{l} - 1}}) - l(\mu_{k_{l} - 1}))}{\lambda V} \left(\frac{1}{2} - \frac{V}{c'_{1}}\right)\right) - \frac{1}{2}
    \label{chap4:eq:p12finalbound}
  \end{equation}
  Thus for FINITE-$\mu$CHOICE-2, we have that $\overline{Q}(\gamma_{k}) = \Omega\left(\log\left(\frac{1}{V_{k}}\right)\right)$.

  For FINITE-$\mu$CHOICE-3, $\lambda = \mu_{l}$, so that from \eqref{chap4:eq:pizub}, instead of \eqref{chap4:eq:p12basic} we have that
  \begin{eqnarray*}
    \overline{q}_{1} - q_{k_{l} - 1} \leq \frac{1}{\pi(q_{k_{l} - 1})} \left(\frac{1}{2} - \frac{V}{c'_{1}}\right),
  \end{eqnarray*}
  and, in fact,
  \begin{eqnarray*}
    \overline{q}_{1} = \left\lfloor q_{k_{l} - 1} + \frac{1}{\pi(q_{k_{l} - 1})} \left(\frac{1}{2} - \frac{V}{c'_{1}}\right) \right\rfloor.
  \end{eqnarray*}
  Proceeding as for FINITE-$\mu$CHOICE-2, we have that
  \begin{eqnarray*}
    \overline{q}_{1} \geq \frac{c({\mu_{k_{l} - 1}}) - l(\mu_{k_{l} - 1})}{ V} \left(\frac{1}{2} - \frac{V}{c'_{1}}\right).
  \end{eqnarray*}
  Therefore we have that
  \begin{eqnarray}
    \overline{Q}(\gamma) \geq \frac{\overline{q}}{2} \geq \frac{\overline{q}_{1}}{2} \geq \frac{1}{2}\bras{1 + \frac{c({\mu_{k_{l} - 1}}) - l(\mu_{k_{l} - 1})}{ V} \left(\frac{1}{2} - \frac{V}{c'_{1}}\right)}.
    \label{chap4:eq:p13finalbound}
  \end{eqnarray}
  Hence for FINITE-$\mu$CHOICE-3, we conclude that $\overline{Q}(\gamma_{k}) = \Omega\left(\frac{1}{V_{k}}\right)$.
\end{proof}

\begin{corollary}
  For any sequence of non-idling admissible policies $\gamma_{k,M} \in \Gamma_{a,M}$ with $\overline{C}(\gamma_{k}) - c(\lambda) = V_{k} \downarrow 0$, we have that 
  \begin{equation}
    \overline{Q}(\gamma_{k,M}) = 
    \begin{cases}
      \Omega\left(\log\left(\frac{1}{V_{k}}\right)\right) \text{ for FINITE-$\mu$CHOICE-2,} \\
      \Omega\left(\frac{1}{V_{k}}\right) \text{ for FINITE-$\mu$CHOICE-3.}
    \end{cases}
  \end{equation}
  \label{chap4:prop:p1213lb_mixed}
\end{corollary}
The proof of this corollary is very similar to that of Corollary \ref{chap4:prop:p11lb_mixed}, except that the convexity property of the functions $\log\nfrac{1}{u}$ and $\frac{1}{u}$ are used instead of the concavity of the function $u\log\nfrac{1}{u}$.

\subsection{Asymptotic characterization of $Q^*_M(c_{c})$}
\label{chap4:sec:asympbehaviour}
In this section we obtain asymptotic upper bounds for TRADEOFF-M.
The sequence of policies that is constructed for FINITE-$\mu$CHOICE-1 is motivated by the policy $\gamma'$ that was used in the proof of the lower bound.
We shall see that a sequence of policies with $q_{k_{u}}$ scaling as $\log\nfrac{1}{V}$ gives the correct asymptotic upper bound.
\begin{lemma}
  For FINITE-$\mu$CHOICE-1, there exists a sequence of non-idling admissible policies $\gamma_{k}$ with a sequence $V_{k} \downarrow 0$ such that $\frac{\lambda}{\mu_{u} - \lambda} - \overline{Q}(\gamma_{k}) = \Theta\brap{V_{k}\log\nfrac{1}{V_{k}}}$ and $\overline{C}(\gamma_{k}) - c(\lambda) = V_{k}$.
  \label{chap4:prop:p11ub}
\end{lemma}
\begin{proof}
  We first consider a policy $\gamma$ in the sequence of policies $\gamma_{k}$.
  The policy $\gamma$ is defined as follows:
  \begin{eqnarray*}
    \mu(0) & = & 0, \\
    \mu(q) & = & \mu_{u}, \text{ for } q \in \{1,\dots, q_{k_{u}}\}, \\
    \mu(q) & = & \mu_{K}, \text{ for } q \in \{q_{k_{u}} + 1, \dots\}.
  \end{eqnarray*}
  The sequence of policies $\gamma_{k}, k \geq 1$ is obtained by choosing $q_{k_{u}} = k$.
  Consider the policy $\gamma$.
  As $\sum_{q = 0}^{\infty} \pi(q) = 1$ we have that
  \begin{eqnarray*}
    \pi(0)\bras{1 + \frac{\lambda}{\mu_{u} - \lambda}\brap{1 - \fpow{\lambda}{\mu_{u}}{q_{k_{u}}}} + \fpow{\lambda}{\mu_{u}}{q_{k_{u}}}\frac{\lambda}{\mu_{K} - \lambda}} = 1
  \end{eqnarray*}
  Therefore, we obtain that 
  \begin{eqnarray}
    \pi(0) & = & \frac{1}{1 + \frac{\lambda}{\mu_{u} - \lambda} + \fpow{\lambda}{\mu_{u}}{q_{k_u}}\bras{\frac{\lambda}{\mu_{K} - \lambda} - \frac{\lambda}{\mu_{u} - \lambda}}}, \text{ and } \nonumber \\
    \pi_{\mu}(K) = \sum_{q = q_{k_u} + 1}^{\infty} \pi(q) & = & \frac{\fpow{\lambda}{\mu_{u}}{q_{k_u}} \frac{\lambda}{\mu_{K} - \lambda}}{1 + \frac{\lambda}{\mu_{u} - \lambda} + \fpow{\lambda}{\mu_{u}}{q_{k_u}}\bras{\frac{\lambda}{\mu_{K} - \lambda} - \frac{\lambda}{\mu_{u} - \lambda}}}.
    \label{chap4:eq:p11ub1}
  \end{eqnarray}
  From the definition of $\overline{C}(\gamma)$ we have that 
  \begin{eqnarray}
    \overline{C}(\gamma) & = & \pi_{\mu}(k_{u}) c(\mu_{u}) + \pi_{\mu}(K)c(\mu_{K}), \nonumber \\
    & = & \pi_{\mu}(k_{u})\brap{c(\lambda) + m_{1}(\mu_{u} - \lambda)} + \pi_{\mu}(K)\brap{c(\lambda) + m_{2}(\mu_{K} - \lambda)}, \nonumber\\
    & = & c(\lambda)(1 - \pi(0)) + m_{1}(\mu_{u} - \lambda)\pi_{\mu}(k_{u}) + m_{2}(\mu_K - \lambda)\pi_{\mu}(K), \nonumber\\
    & = & c(\lambda) - \pi(0) c(\lambda) + m_{1}(\mu_{u} - \lambda)(1 - \pi_{\mu}(K) - \pi(0)) + m_{2}(\mu_K - \lambda)\pi_{\mu}(K). \nonumber\\
    \overline{C}(\gamma) - c(\lambda) & = & m_{1}(\mu_{u} - \lambda) + \pi_{\mu}(K)\bras{m_{2}(\mu_{K} - \lambda) - m_{1}(\mu_{u} - \lambda)} - \pi(0)\bras{c(\lambda) + m_{1}(\mu_{u} - \lambda)}, \nonumber \\
    \label{chap4:eq:p11ub2}
  \end{eqnarray}
  where $m_{1} = \frac{c(\mu_{u}) - c(\lambda)}{\mu_{u} - \lambda}$ and $m_{2} = \frac{c(\mu_{K}) - c(\lambda)}{\mu_{K} - \lambda}$.  
  We denote $\overline{C}(\gamma) -  c(\lambda)$ by $V$.
  We note that $c(\lambda) + m_{1}(\mu_{u} - \lambda) = c(\mu_{u})$ and $m_{2}(\mu_{K} - \lambda) - m_{1}(\mu_{u} - \lambda) = c(\mu_{K}) - c(\mu_{u})$.
  Then \eqref{chap4:eq:p11ub2} can be written as 
  \begin{eqnarray*}
    V & = & m_{1}(\mu_{u} - \lambda) + \pi_{\mu}(K)\brap{c(\mu_{K}) - c(\mu_{u})} - \pi(0)c(\mu_{u}), \\
    V - m_{1}(\mu_{u} - \lambda) & = & \frac{\fpow{\lambda}{\mu_{u}}{q_{k_{u}}} \frac{\lambda}{\mu_{K} - \lambda}\brap{c(\mu_{K}) - c(\mu_{u})} - c(\mu_{u})}{1 + \frac{\lambda}{\mu_{u} - \lambda} + \fpow{\lambda}{\mu_{u}}{q_{k_{u}}}\bras{\frac{\lambda}{\mu_{K} - \lambda} - \frac{\lambda}{\mu_{u} - \lambda}}}.
  \end{eqnarray*}
  Simplifying, we obtain, with $c(\lambda) = m_{1} \lambda$,
  \begin{eqnarray*}
    \frac{\bras{V - m_{1}(\mu_{u} - \lambda)}\bras{1 + \frac{\lambda}{\mu_{u} - \lambda}} + c(\mu_{u})}{\frac{\lambda}{\mu_{K} - \lambda} \brap{c(\mu_{K}) - c(\mu_{u})} - \bras{\frac{\lambda}{\mu_{K} - \lambda} - \frac{\lambda}{\mu_{u} - \lambda}}(V - m_{1}(\mu_{u} - \lambda))} = \fpow{\lambda}{\mu_{u}}{q_{k_{u}}}, \\
    \frac{\overline{C}(\gamma) \frac{\mu_{u}}{\mu_{u} - \lambda} - \frac{\lambda}{\mu_{u} - \lambda}\brap{c(\lambda) + m_{1}(\mu_{u} - \lambda)}}{\lambda(m_{2} - m_{1}) - \brap{\frac{\lambda}{\mu_{K} - \lambda} - \frac{\lambda}{\mu_{u} - \lambda}}(\overline{C}(\gamma) - c(\lambda))} = \fpow{\lambda}{\mu_{u}}{q_{k_{u}}},
  \end{eqnarray*}
  so that,
  \begin{eqnarray*}
    q_{k_{u}} = \log_{\nfrac{\mu_{u}}{\lambda}}\bras{\frac{\lambda(m_{2} - m_{1}) - \brap{\frac{\lambda}{\mu_{K} - \lambda} - \frac{\lambda}{\mu_{u} - \lambda}}V}{V \frac{\mu_{u}}{\mu_{u} - \lambda}}}.
  \end{eqnarray*}
  We note that the average queue length for the policy $\gamma$ is of the form given in \eqref{chap4:eq:ezdomexpression}, wherein $a = \frac{\lambda}{\mu_{u}}$ and $b = \frac{\lambda}{\mu_{K}}$.
  Simplifying this expression, we obtain that
  \begin{eqnarray}
    \overline{Q}(\gamma) = \frac{\frac{a}{(1 - a)^{2}} + q_{k_{u}} a^{q_{k_{u}}}\bras{\frac{b}{1 - b} - \frac{a}{1 - a}} + a^{q_{k_{u}}}\bras{\frac{b}{(1 - b)^{2}} - \frac{a}{(1 - a)^{2}}}}{\frac{1}{1 - a} + a^{q_{k_{u}}}\bras{\frac{b}{1 - b} - \frac{a}{1 - a}}}.
    \label{chap4:eq:p11finalub}
  \end{eqnarray}
  We are interested in only order approximations of $\overline{Q}(\gamma)$; so we proceed by considering large $k$, so that $V \downarrow 0$ and $a^{q_{k_{u}}} \downarrow 0$.
  We obtain that (only the dominant terms as $V \downarrow 0$)
  \begin{eqnarray*}
    \overline{Q}(\gamma) \approx \brap{\frac{a}{1 - a} + q_{k_{u}} a^{q_{k_{u}}}(1 - a)\bras{\frac{b}{1 - b} - \frac{a}{1 - a}} + a^{q_{k_{u}}}(1 - a)\bras{\frac{b}{(1 - b)^{2}} - \frac{a}{(1 - a)^{2}}}} \nonumber \\
    \times \brap{1 + a^{q_{k_{u}}}(1 - a)\bras{\frac{a}{1 - a} - \frac{b}{1 - b}}}.
  \end{eqnarray*}
  Expanding, we obtain
  \begin{eqnarray}
    \overline{Q}(\gamma) \approx \frac{a}{1 - a} + \frac{a}{1 - a}a^{q_{k_u}}(1 - a)\bras{\frac{a}{1 - a} - \frac{b}{1 - b}} + q_{k_{u}}a^{q_{k_u}}(1 - a)\bras{\frac{b}{1 - b} - \frac{a}{1 - a}} \label{chap4:eq:p11ub3} \\
    + a^{q_{k_u}}(1 - a)\bras{\frac{b}{(1 - b)^{2}} - \frac{a}{(1 - a)^{2}}} \label{chap4:eq:p11ub4} \\
    - q_{k_u}a^{2q_{k_u}}(1 - a)^{2}\bras{\frac{a}{1 - a} - \frac{b}{1 - b}}^{2} + a^{2q_{k_u}}(1 - a)^{2}\bras{\frac{a}{1 - a} - \frac{b}{1 - b}}\bras{\frac{b}{(1 - b)^{2}} - \frac{a}{(1 - a)^{2}}} \label{chap4:eq:p11ub41}.
  \end{eqnarray}
  We note that the second term in \eqref{chap4:eq:p11ub3} is positive and $\Theta(V)$, while the third term is negative and $\Theta\brap{V\log\nfrac{1}{V}}$.
  The term in \eqref{chap4:eq:p11ub4} is negative and $\Theta(V)$.
  The first term in \eqref{chap4:eq:p11ub41} is negative and $\Theta\brap{V^{2}\log\nfrac{1}{V}}$, while the second term is negative and $\Theta(V^{2})$. 
  Thus, the dominating term in $\frac{a}{1 - a} - \overline{Q}(\gamma)$ is positive and $\Theta\brap{V\log\nfrac{1}{V}}$ as $V \downarrow 0$.
  Hence we have that $\frac{a}{1 - a} - \overline{Q}(\gamma) = \Theta\brap{V\log\nfrac{1}{V}}$.
  
  We note that every $\gamma_{k}$ constructed, by choosing $q_{k_{u}} = k$, is non-idling and admissible.
  So, there exists a sequence of non-idling admissible policies $\gamma_{k}$ such that $\frac{\lambda}{\mu_{u} - \lambda} - \overline{Q}(\gamma_{k}) = \Theta\brap{V_{k}\log\nfrac{1}{V_{k}}}$ and $\overline{C}(\gamma) - c(\lambda) = V_{k}$.
\end{proof}

Using the asymptotic lower bound on $\overline{Q}(\gamma_{k})$ from Corollary \ref{chap4:prop:p11lb_mixed}, and the above asymptotic upper bound, we have the following result.

\begin{proposition}
  For FINITE-$\mu$CHOICE-1, we have that the optimal value of the tradeoff problem $Q^*_{M}(c_{c,k})$ is $\frac{\lambda}{\mu_{u} - \lambda} - \Theta\left((c_{c,k} - c(\lambda))\log\nfrac{1}{c_{c,k} - c(\lambda)}\right)$, for a sequence $c_{c,k} = \overline{C}(\gamma_{k})$ for the sequence of policies $\gamma_{k}$ in Lemma \ref{chap4:prop:p11ub}.
  \label{chap4:prop:p11}
\end{proposition}
\begin{proof}
  Consider the sequence $c_{c,k} = V_{k} + c(\lambda)$.
  Let us choose a sequence $\epsilon_{V} \Deq V_{k}\log\nfrac{1}{V_{k}}$ which decreases to zero as $V_{k} \downarrow 0$.
  Let $\gamma_{k} \in \Gamma_{a,M}$ be a sequence of $\epsilon_{V}$-optimal policies for FINITE-$\mu$CHOICE such that $\overline{Q}(\gamma_{k}) \leq Q^*_{M}(c_{c,k}) + \epsilon_{V}$.
  Then, applying Corollary \ref{chap4:prop:p11lb_mixed} we have that $\frac{\lambda}{\mu_{u} - \lambda} - \overline{Q}(\gamma_{k}) = \mathcal{O}\left((c_{c,k} - c(\lambda))\log\nfrac{1}{c_{c,k} - c(\lambda)}\right)$.
  Then, as $V_{k} \downarrow 0$, there exists some constant $c_{1}$ such that 
  \[\overline{Q}(\gamma_{k}) \geq \frac{\lambda}{\mu_{u} - \lambda} - c_{1}\left((c_{c,k} - c(\lambda))\log\nfrac{1}{c_{c,k} - c(\lambda)}\right).\]
  Since $\overline{Q}(\gamma_{k}) \leq Q^*_{M}(c_{c,k}) + \epsilon_{V}$, we have that
  \[Q^*_{M}(c_{c,k}) + \epsilon_{V} \geq \frac{\lambda}{\mu_{u} - \lambda} - c_{1}\left((c_{c,k} - c(\lambda))\log\nfrac{1}{c_{c,k} - c(\lambda)}\right).\]
  Then we have that 
  \[Q^*_{M}(c_{c,k}) \geq \frac{\lambda}{\mu_{u} - \lambda} - (c_{1} + 1)\left((c_{c,k} - c(\lambda))\log\nfrac{1}{c_{c,k} - c(\lambda)}\right).\]
  We have that $Q^*_{M}(c_{c,k}) \leq \overline{Q}(\gamma_{k})$, where $\gamma_{k}$ is the sequence of policies constructed in Lemma \ref{chap4:prop:p11ub}.
  Therefore $Q^*_{M}(c_{c,k}) = \overline{Q}(\gamma^*_{k}) = \frac{\lambda}{\mu_{u} - \lambda} - \Theta\left((c_{c,k} - c(\lambda))\log\nfrac{1}{c_{c,k} - c(\lambda)}\right)$
\end{proof}

\begin{remark}
  We note that the asymptotic characterization of $Q^*(c_{c})$ has been obtained only for a particular sequence $V_{k} = \overline{C}(\gamma_{k}) - c(\lambda)$, where $\gamma_{k}$ is as in Lemma \ref{chap4:prop:p11ub}.
  The set of average service cost values that can be achieved depends upon the set of service rates, $\brac{0,\mu_{1}, \dots, \mu_{K}}$, available for control.
  For example, if the set of service rates available for control is $\brac{0, \mu_{u}, \mu_{K}}$, then the average service cost always corresponds to the set of values $V_{k}$ in Proposition \ref{chap4:prop:p11}.
  In fact, the asymptotic $\Theta$ characterization of $Q^*_{M}(c_{c})$ can be obtained for any sequence of $c_{c,k}$ such that there exists a sequence of non-idling admissible $\gamma_{k}$ such that $\Cgk - c(\lambda) = \Theta(c_{c,k} - c(\lambda))$.
\end{remark}

\begin{remark}
  In this thesis, any sequence of admissible policies $\gamma_{k}$, which achieve the asymptotic lower bound is called an \emph{order-optimal} sequence of policies.
  For example, the sequence of policies $\gamma_{k}$ in Lemma \ref{chap4:prop:p11ub} is order-optimal.
\end{remark}

For obtaining bounds on the average queue length for FINITE-$\mu$CHOICE-2 and FINITE-$\mu$CHOICE-3, we use a result, presented in Appendix \ref{chap4:app:prop:dotzupperbound}, that uses a quadratic Lyapunov function to obtain bounds on the average queue length.

\begin{lemma}
  For FINITE-$\mu$CHOICE-2, there exists a sequence of non-idling admissible policies $\gamma_{k}$, with a sequence $V_{k} \downarrow 0$ such that $\overline{Q}(\gamma_{k}) = \mathcal{O}\left(\log\left(\frac{1}{V_{k}}\right)\right)$ and $\overline{C}(\gamma_{k}) - c(\lambda) =  V_{k}$.
  \label{chap4:prop:p12ub}
\end{lemma}
\begin{proof}
  Consider a policy $\gamma$ defined as follows :
  \begin{eqnarray*}
    \mu(0) & = & 0, \\
    \mu(q) & = & \mu_{l}, \text{ for } q \in \{1,\dots,q_{k_l}\}, \\
    \mu(q) & = & \mu_{u}, \text{ for } q \in \{q_{k_l} + 1, \dots\}.
  \end{eqnarray*}
  where $q_{k_l} \stackrel{\Delta} = \left\lceil \log_{\left(\frac{\lambda}{\mu_{l}}\right)} \left(1 + \frac{\lambda - \mu_{l}}{\lambda} \frac{1}{U} \right) \right \rceil$, with $U > 0$.
  The sequence of policies $\gamma_{k}$ is obtained by choosing $U$ from a sequence $U_{k}$ that decreases to zero.
  
  Now we note that for $q \in \{1,\dots, q_{k_l}\}$,
  \begin{eqnarray*}
    \pi(q) = \pi(0) \left(\frac{\lambda}{\mu_{l}}\right)^{q},
  \end{eqnarray*}
  and for $q \in \{q_{k_l} + 1, \dots\}$
  \begin{eqnarray*}
    \pi(q) = \pi(0) \left(\frac{\lambda}{\mu_{l}}\right)^{q_{k_l}} \left(\frac{\lambda}{\mu_{u}}\right)^{q - q_{k_l}}.
  \end{eqnarray*}
  As $\sum_{q = 0}^{\infty} \pi(q) = 1$, we have that
  \begin{eqnarray}
    \pi(0)\left[ 1 + \sum_{q = 1}^{q_{k_l}} \left( \frac{\lambda}{\mu_{l}} \right)^{q} + \left(\frac{\lambda}{\mu_{l}}\right)^{q_{k_l}} \sum_{q = 1}^{\infty} \left( \frac{\lambda}{\mu_{u}} \right)^{q} \right] & = & 1, \nonumber \\
    \pi(0)\left[ 1 + \frac{\lambda}{\lambda - \mu_{l}}\left[ \left(\frac{\lambda}{\mu_{l}}\right)^{q_{k_l}} - 1 \right] + \fpow{\lambda}{\mu_{l}}{q_{k_l}}\frac{\lambda}{\mu_{u} - \lambda} \right] & = & 1.
    \label{chap4:eq:p12ub0}
  \end{eqnarray}
  We note that $q_{k_l} \geq \log_{\nfrac{\lambda}{\mu_{l}}} \left(1 + \frac{\lambda - \mu_{l}}{\lambda}\frac{{1}}{U} \right)$.
  Since $\frac{\lambda}{\mu_{l}} > 1$ we have that
  \begin{eqnarray*}
    \fpow{\lambda}{\mu_{l}}{q_{k_l}} \geq 1 + \frac{\lambda - \mu_{l}}{\lambda} \frac{{1}}{U}.
  \end{eqnarray*}
  Substituting the above lower bound, in \eqref{chap4:eq:p12ub0}, we have that
  \begin{eqnarray}
    \pi(0) \leq \frac{U}{U\nfrac{\mu_{u}}{\mu_{u} - \lambda} + {1} \nfrac{\mu_{u} - \mu_{l}}{\mu_{u} - \lambda}}, \nonumber \\
    \text{or, } \pi(0) \leq \frac{U}{\nfrac{\mu_{u} - \mu_{l}}{\mu_{u} - \lambda}}.
    \label{chap4:eq:p12pi0bound}
  \end{eqnarray}
  
  We note that for $\gamma, \overline{C}(\gamma) = \pi(0).0 + \pi_{\mu}(k_{l}) c(\mu_{l}) + \pi_{\mu}(k_{u}) c(\mu_{u})$.
  Also $c(\mu_{l}) = c(\lambda) + (\mu_{l} - \lambda)m$ and $c(\mu_{u}) = c(\lambda) + (\mu_{u} - \lambda)m$, where $m = \frac{c(\mu_{u}) - c(\mu_{l})}{\mu_{u} - \mu_{l}}$.
  Then we have that
  \begin{eqnarray*}
    \overline{C}(\gamma) & = & \pi_{\mu}(k_{l}) (c(\lambda) + (\mu_{l} - \lambda)m) + \pi_{\mu}(k_{u}) (c(\lambda) + (\mu_{u} - \lambda)m), \\
    & \leq & c(\lambda) + m (\pi_{\mu}(k_{l}) (\mu_{l} - \lambda) + \pi_{\mu}(k_{u})(\mu_{u} - \lambda)).
  \end{eqnarray*}
  We note that $\gamma$ is admissible, therefore, we have that $\pi_{\mu}(k_{l}) \mu_{l} + \pi_{\mu}(k_{u}) \mu_{u} = \lambda$.
  Hence,
  \begin{eqnarray*}
    \pi_{\mu}(k_{l}) (\mu_{l} - \lambda) + \pi_{\mu}(k_{u})(\mu_{u} - \lambda) = \pi(0) \lambda
  \end{eqnarray*}
  Then,
  \begin{eqnarray}
    \overline{C}(\gamma) & \leq & c(\lambda) + m \pi(0)\lambda
    \label{chap4:eq:p12finalubc}
  \end{eqnarray}
  From \eqref{chap4:eq:p12pi0bound}, we have that 
  \begin{eqnarray*}
    \overline{C}(\gamma) - c(\lambda) = \mathcal{O}(U).
  \end{eqnarray*}
  Let $V \stackrel{\Delta} = \overline{C}(\gamma) - c(\lambda)$, then $V = \mathcal{O}(U)$.

  From Proposition \ref{chap4:app:prop:dotzupperbound}, with $q_{\epsilon} = q_{k_l} + 1$ and $\epsilon = \mu_{u} - \lambda$, we obtain that
  \begin{eqnarray}
    \overline{Q}(\gamma) \leq \frac{(q_{k_l} + 1)\mu_{u}}{\mu_{u} - \lambda} + \frac{\lambda + r_{max}}{2(\mu_{u} - \lambda)}.
    \label{chap4:eq:p12finalubq}
  \end{eqnarray}
  Now, for the sequence of policies $\gamma_{k}$ with $U_k \downarrow 0$, $q_{k_l} = \mathcal{O}\left(\log\left(\frac{1}{U_k}\right)\right)$.
  Hence $\overline{Q}(\gamma_k) = \mathcal{O}\left(\log\left(\frac{1}{U_k}\right)\right)$ and, since $V_{k} = \mathcal{O}(U_{k})$, $\overline{Q}(\gamma_k) = \mathcal{O}\left(\log\left(\frac{1}{V_k}\right)\right)$.
  So there exists a sequence of policies $\gamma_{k}$ such that $\overline{Q}(\gamma_k) = \mathcal{O}\brap{\log\nfrac{1}{V_k}}$ and $\overline{C}(\gamma_{k}) - c(\lambda) = V_k$.
\end{proof}

Using the asymptotic lower bound on $\overline{Q}(\gamma)$ from Corollary \ref{chap4:prop:p1213lb_mixed}, and the asymptotic upper bound above, and proceeding as for Proposition \ref{chap4:prop:p11} (except that $\epsilon_{V}$ is a constant $\epsilon > 0$) we obtain the following result.

\begin{proposition}
  For FINITE-$\mu$CHOICE-2, we have that the optimal value of the tradeoff problem $Q^*_{M}(c_{c,k})$ is $\Theta\left(\log\nfrac{1}{c_{c,k} - c(\lambda)}\right)$, for a sequence $c_{c,k} = V_{k} + c(\lambda)$, where $V_{k} = \overline{C}(\gamma_{k}) - c(\lambda)$ for the sequence of policies $\gamma_{k}$ in Lemma \ref{chap4:prop:p12ub}.
  \label{chap4:prop:p12}
\end{proposition}

The following asymptotic upper bound for FINITE-$\mu$CHOICE-3, is obtained using a procedure similar to that for FINITE-$\mu$CHOICE-2 in Lemma \ref{chap4:prop:p12ub}.
\begin{lemma}
  For FINITE-$\mu$CHOICE-3, there exists a sequence of non-idling admissible policies $\gamma_{k}$ with a sequence $V_{k} \downarrow 0$ such that $\overline{Q}(\gamma_{k}) = \mathcal{O}\left(\frac{1}{V_k}\right)$ with $\overline{C}(\gamma_{k}) - c(\lambda) = V_{k}$.
  \label{chap4:prop:p13ub}
\end{lemma}

\begin{proof}
  Consider a policy $\gamma$ defined as follows :
  \begin{eqnarray*}
    \mu(0) & = & 0, \\
    \mu(q) & = & \lambda, \text{ for } q \in \{1,\dots,q_{\lambda}\}, \\
    \mu(q) & = & \mu', \text{ for } q \in \{q_{\lambda} + 1, \dots\},
  \end{eqnarray*}
  where $\mu' = \min\brac{\mu_k : \mu_{k} > \lambda}$ and  $q_{\lambda}$ is chosen as $\left\lceil \frac{1}{U} \right \rceil$, with $U > 0$.
  The sequence of policies $\gamma_{k}$ is obtained by choosing $U$ from a sequence $U_{k} \downarrow 0$.
  
  We note that for $q \in \{1,\dots,q_{\lambda}\}$, as $\mu(q) = \lambda$, we have that $\pi(q) = \pi(0)$.
  And for $q \in \{q_{\lambda} + 1, \dots\}$, we have that $\pi(q) = \pi(0)\fpow{\lambda}{\mu'}{q - q_{\lambda}}$.
  As $\sum_{q = 0}^{\infty} \pi(q) = 1$, we have that
  \begin{eqnarray}
    \pi(0) \brap{1 + q_{\lambda} + \frac{\lambda}{\mu' - \lambda}} = 1, \nonumber \\
    \pi(0) \leq \frac{U(\mu' - \lambda)}{U\mu' + \mu' - \lambda} \leq U.
    \label{chap4:eq:p13pi0bound}
  \end{eqnarray}
  We note that $\overline{C}(\gamma) = \pi(0).0 + \pi_{\mu}(\lambda)c(\lambda) + \pi_{\mu}(\mu')c(\mu')$.
  We have that $c(\mu') = c(\lambda) + m (\mu' - \lambda)$ where $m = \frac{c(\mu') - c(\lambda)}{\mu' - \lambda}$.
  We note that $\gamma$ is admissible.
  Therefore, $\pi(0).0 + \pi_{\mu}(\lambda)\lambda + \pi_{\mu}(\mu')\mu' = \lambda$, or $\pi_{\mu}(\mu')(\mu' - \lambda) = \pi(0)\lambda$.
  Thus
  \begin{eqnarray}
    \overline{C}(\gamma) & = & \pi_{\mu}(\lambda) c(\lambda) + \pi_{\mu}(\mu')(c(\lambda) + m (\mu' - \lambda)), \nonumber \\
    & = & c(\lambda) + m \lambda \pi(0) - c(\lambda) \pi(0), \nonumber \\
    & \leq & c(\lambda) + (m\lambda - c(\lambda)) U.
    \label{chap4:eq:p13finalubc}
  \end{eqnarray}
  Using \eqref{chap4:eq:p13pi0bound} we obtain that, for the policy $\gamma$, $\overline{C}(\gamma) - c(\lambda) = \mathcal{O}\brap{U}$.
  Let $V \stackrel{\Delta} = \overline{C}(\gamma) - c(\lambda)$.
  Then $V = \mathcal{O}(U)$.
  
  To use Proposition \ref{chap4:app:prop:dotzupperbound}, we set $q_{\epsilon} = q_{\lambda} + 1$ and $\epsilon = \mu' - \lambda$.
  We obtain that 
  \begin{eqnarray}
    \overline{Q}(\gamma) \leq \frac{(q_{\lambda} + 1)\mu'}{\mu' - \lambda} + \frac{\lambda + r_{max}}{2(\mu' - \lambda)}.
    \label{chap4:eq:p13finalubq}
  \end{eqnarray}

  For the policy $\gamma_{k}$, as $q_{\lambda} = \mathcal{O}\nfrac{1}{U_{k}}$, we have that $\overline{Q}(\gamma_{k}) = \mathcal{O}\nfrac{1}{U_{k}}$.
  Since $V_{k} = \mathcal{O}(U_{k})$, we have that $\overline{Q}(\gamma_{k}) = \mathcal{O}\nfrac{1}{V_{k}}$.
  Hence there exists a sequence of non-idling admissible policies $\gamma_{k}$, with $\overline{Q}(\gamma_{k}) = \mathcal{O}\nfrac{1}{V_{k}}$ and $\overline{C}(\gamma_{k}) - c(\lambda) = V_{k}$.
\end{proof}

Using the asymptotic lower bound on $\overline{Q}(\gamma_{k})$ from Corollary \ref{chap4:prop:p1213lb_mixed}, and the asymptotic upper bound above, and proceeding as in Proposition \ref{chap4:prop:p11} (except that $\epsilon_{V}$ is a constant $\epsilon > 0$), we obtain the following result.

\begin{proposition}
  For FINITE-$\mu$CHOICE-3, we have that the optimal value of the tradeoff problem $Q^*_{M}(c_{c,k})$ is $\Theta\nfrac{1}{c_{c,k} - c(\lambda)}$, for a sequence $c_{c,k} = V_{k} + c(\lambda)$, where $V_{k} = \overline{C}(\gamma_{k}) - c(\lambda)$ for the sequence of policies $\gamma_{k}$ in Lemma \ref{chap4:prop:p13ub}.
  \label{chap4:prop:p13}
\end{proposition}

\subsection{Asymptotic characterization of order optimal admissible policies}
\label{chap4:sec:optimalpolicy_aschar}
The above approach that characterizes the asymptotic behaviour of the average queue length, via an upper bound on the stationary probability distribution of the queue length for admissible policies, is used throughout this thesis.
In this section, we discuss an advantage that this approach has over methods proposed in \cite{berry} or \cite{neely_utility}.
We illustrate how the above approach can be used to obtain an \emph{asymptotic characterization} of any sequence of order-optimal admissible policies $\gamma_{k}$ for which $\Cgk \downarrow c(\lambda)$.
Such an asymptotic characterization may lead to a reduction in the search space for the $\epsilon$-optimal admissible policy for \eqref{chap4:eq:sysmodel:probstat} as $c_{c} \approx c(\lambda)$.

We note that an admissible policy is equivalently described by the sequence $(q_{0} = 0, q_{1}, \dots, q_{K- 1}, q_{K} = \infty)$.
In this section, we discuss how asymptotic bounds on $q_{k}$ can be obtained for an admissible policy with $\Cg \approx c(\lambda)$.
We consider only \FMC-2 and \FMC-3 in this section, since these are the cases for which the design of policies is more critical (since $Q^*(c_{c})$ increases to infinity as $c_{c} \downarrow c(\lambda)$).

We note that for special classes of admissible policies, such as buffer-partitioning policies proposed in \cite{berry}, intuition about $q_{k}$ can be obtained from the asymptotic characterization of $\Qg$ for a $\gamma$ such that $\Cg \approx c(\lambda)$.
We note that buffer partitioning policies use only a specific set of rates, e.g., for \FMC-2, only the rates $\brac{0, \mu_{l}, \mu_{u}}$ are used.
Since $\mu_{l} < \lambda < \mu_{u}$, intuitively, we expect that the partition $q_{k_{l}} \approx \Qg$, which implies that $q_{k_{l}} = \Omega\brap{\log\nfrac{1}{c_{c} - c(\lambda)}}$ for any feasible policy as $c_{c} \downarrow c(\lambda)$.
We note that the above asymptotic behaviour for the partition can be surmised from the asymptotic behaviour of $\Qg$, which may be obtained via methods as in \cite{neely_mac} or \cite{neely_utility}.
However, in the following, we discuss how bounds on the stationary probability distribution of the queue length are useful in obtaining a much more refined asymptotic characterization of the policy.

\subsubsection{Two inequalities:}
\label{chapter2:policymethod}
In this section, we discuss two inequalities, which form the basis for the asymptotic characterization of any sequence of order-optimal policies.
Let $q^{1} \leq q^{2}$.
Let $P_{l}\brac{q^{1} \leq Q \leq q^{2}}$ and $P_{u}\brac{q^{1} \leq Q \leq q^{2}}$ be any lower bound and upper bound to $Pr\brac{q^{1} \leq Q \leq q^{2}}$, i.e., $P_{l}\brac{q^{1} \leq Q \leq q^{2}} \leq Pr\brac{q^{1} \leq Q \leq q^{2}} \leq P_{u}\brac{q^{1} \leq Q \leq q^{2}}$.
Also let $\pi_{l}(q)$ and $\pi_{u}(q)$ be any lower bound and upper bound to $\pi(q)$, i.e., $\pi_{l}(q) \leq \pi(q) \leq \pi_{u}(q)$.
The asymptotic characterization of any policy is obtained using the following two observations.
If $q_{l}$ is the largest integer such that
\begin{eqnarray}
  \sum_{q = q^{1}}^{q^{1} + q_{l}} \pi_{u}(q) \leq P_{l}\brac{q^{1} \leq Q \leq q^{2}},
  \label{chap4:eq:ascharpolicy_obs1}
\end{eqnarray}
then $q_{l} \leq q^{2} - q^{1}$.
If $q_{u}$ is the smallest integer such that
\begin{eqnarray}
  \sum_{q = q^{1}}^{q^{1} + q_{u}} \pi_{l}(q) \geq P_{u}\brac{q^{1} \leq Q \leq q^{2}},
  \label{chap4:eq:ascharpolicy_obs2}
\end{eqnarray}
then $q_{l} \geq q^{2} - q^{1}$.
We note that bounds on $q_{k} - q_{k - 1} + 1$, which is the set of queue lengths for which the service rate $\mu_{k}$ is used, can be obtained by using $q^{1} = q_{k - 1} + 1$ and $q^{2} = q_{k}$.

\subsubsection{The bounds - $P_{l}\brac{.}$, $P_{u}\brac{.}$, $\pi_{l}(.)$, and $\pi_{u}(.)$ :}
In the following we discuss how $P_{l}\brac{.}$, $P_{u}\brac{.}$, $\pi_{l}(.)$, and $\pi_{u}(.)$ can be obtained.
We note that we consider only cases where $q^{1} = q_{k'} + 1$ and $q^{2} = q_{k}$, where $k' < k$.
Then $Pr\brac{q^{1} \leq Q \leq q^{2}} = Pr\brac{\mu_{k' + 1} \leq \mu(Q) \leq \mu_{k}} = \sum_{n = k' + 1}^{k} \pi_{\mu}(n)$.

Consider any admissible policy $\gamma$.
Then for any $k$, we can obtain a lower bound on $\pi_{\mu}(k)$ as the optimal value of the linear program:
\begin{eqnarray}
  \min & & \pi_{\mu}(k)  \label{chap4:eq:ascharpolicy_lpbound} \\
  \text{such that } & & \sum_{k = 0}^{K} \pi_{\mu}(k) = 1, \label{chap4:eq:ascharpolicy_lpbound_c1}\\
  & & \sum_{k = 0}^{K} \pi_{\mu}(k) \mu_{k} = \lambda, \label{chap4:eq:ascharpolicy_lpbound_c2}\\
  & & \sum_{k = 0}^{K} \pi_{\mu}(k)\brap{c(\mu_{k}) - l(\mu_{k})} = \Cg - c(\lambda). \label{chap4:eq:ascharpolicy_lpbound_c3}
\end{eqnarray}
An upper bound on $\pi_{\mu}(k)$ can be obtained by maximising $\pi_{\mu}(k)$ in the above linear program.
However, we note that finding the above bounds analytically is difficult.
Hence, in the following we find other bounds on $\pi_{\mu}(k)$ or $\sum_{n = k' + 1}^{k} \pi_{\mu}(n)$, which can be expressed analytically.
We note that these analytical bounds are obtained from the constraints in the above linear program.

We note that for $k < k_{l}$ and $k > k_{u}$, from Lemma \ref{chap4:prop:pimu_ub} we have that $\pi_{\mu}(k) \leq \frac{\Cg - c(\lambda)}{c(\mu_{k}) - l(\mu_{k})}$.
Proceeding as in the proof of Lemma \ref{chap4:prop:pimu_ub} we can also show that if $k < k_{l}$ or $k' \geq k_{u}$, then
\begin{equation}
  \sum_{n = k' + 1}^{k} \pi_{\mu}(k) \leq \frac{\Cg - c(\lambda)}{\min_{n \in \brac{k' + 1, \dots, k}} \brap{c(\mu_{n}) - l(\mu_{n})}}.
  \label{chap4:eq:ascharpolicy_sbound1}
\end{equation}

If $k_{l} = k_{u}$, as in \FMC-3, then from \eqref{chap4:eq:ascharpolicy_lpbound_c1} and \eqref{chap4:eq:ascharpolicy_lpbound_c3}, we have that
\begin{equation}
  \pi_{\mu}(k_{l}) = 1 - \sum_{k \neq k_{l}} \pi_{\mu}(k) \geq 1 - \frac{\Cg - c(\lambda)}{\min_{k \neq k_{l}} \brap{c(\mu_{k}) - l(\mu_{k})}}.
\end{equation}
If $k_{l} < k_{u}$, as in \FMC-2, then again from \eqref{chap4:eq:ascharpolicy_lpbound_c1} and \eqref{chap4:eq:ascharpolicy_lpbound_c3}, we have that
\begin{equation}
  \sum_{k = k_{l}}^{k_{u}} \pi_{\mu}(k) \geq 1 - \frac{\Cg - c(\lambda)}{\min_{k < k_{l}, k > k_{u}} \brap{c(\mu_{k}) - l(\mu_{k})}}.
\end{equation}
\newcommand{\tk}{\tilde{k}}
\newcommand{\pim}{\pi_{\mu}}

Suppose we are interested in $\sum_{k = k_{l}}^{\tk} \pim(k)$.
Then we have that
\begin{equation}
  \sum_{k = k_{l}}^{\tk} \pi_{\mu}(k) + \sum_{k = \tk + 1}^{k_{u}} \pim(k) \geq 1 - \frac{\Cg - c(\lambda)}{\min_{k < k_{l}, k > k_{u}} \brap{c(\mu_{k}) - l(\mu_{k})}}.
\end{equation}
If $\tk = k_{u}$, then we note that the RHS is a lower bound.
If $\tk < k_{u}$, then we proceed as follows
\begin{equation}
  \sum_{k = k_{l}}^{\tk} \pi_{\mu}(k) \geq 1 - \sum_{k = \tk + 1}^{k_{u}} \pim(k) - \frac{\Cg - c(\lambda)}{\min_{k < k_{l}, k > k_{u}} \brap{c(\mu_{k}) - l(\mu_{k})}}.
\end{equation}
From \eqref{chap4:eq:ascharpolicy_lpbound_c2}, we have that
\begin{eqnarray*}
  \sum_{k = k_{l}}^{\tk} \pi_{\mu}(k)\mu_{k} + \sum_{k = \tk + 1}^{k_{u}} \pim(k)\mu_{k} & \leq & \lambda, \\
  \mu_{l} \sum_{k = k_{l}}^{\tk} \pi_{\mu}(k) + \mu_{\tk} \sum_{k = \tk + 1}^{k_{u}} \pim(k) & \leq & \lambda,
\end{eqnarray*}
Or,
\begin{eqnarray*}
  \sum_{k = \tk + 1}^{k_{u}} \pim(k) & \leq & \frac{\lambda - \mu_{l} \sum_{k = k_{l}}^{\tk} \pi_{\mu}(k)}{\mu_{\tk}}.
\end{eqnarray*}
Therefore, we have that
\begin{eqnarray}
  \sum_{k = k_{l}}^{\tk} \pi_{\mu}(k) & \geq & 1 - \frac{\lambda - \mu_{l} \sum_{k = k_{l}}^{\tk} \pi_{\mu}(k)}{\mu_{\tk}} - \frac{\Cg - c(\lambda)}{\min_{k < k_{l}, k > k_{u}} \brap{c(\mu_{k}) - l(\mu_{k})}}, \nonumber \\
  \sum_{k = k_{l}}^{\tk} \pi_{\mu}(k) & \geq & \frac{\mu_{\tk} - \lambda}{\mu_{\tk} - \mu_{l}} - {\mu_{\tk}}\frac{\Cg - c(\lambda)}{\min_{k < k_{l}, k > k_{u}} \brap{c(\mu_{k}) - l(\mu_{k})}}.
  \label{chap4:eq:ascharpolicy_lbmu_g_lambda}
\end{eqnarray}
We note that this lower bound is non-negative only if $\mu_{\tk} > \lambda$.
We note that
\begin{eqnarray}
  \sum_{k < k_{l}, k > k_{u}} \pim(k)\brap{c(\mu_{k}) - l(\mu_{k})} = \Cg - c(\lambda), \nonumber \\
  \sum_{k < k_{l}, k > k_{u}} \pim(k) \geq \frac{\Cg - c(\lambda)}{\max_{k < k_{l}, k > k_{u}} \brap{c(\mu_{k}) - l(\mu_{k})}}.
  \label{chap4:eq:ascharpolicy_outbound2}
\end{eqnarray}
We note that for \FMC-2, the best lower bound that can be obtained for $\sum_{k < k_{l}} \pim(k)$ and $\sum_{k > k_{u}} \pim(k)$ separately is zero.
For example, if we try to obtain a lower bound on $\sum_{k < k_{l}} \pim(k)$, since the constraints \eqref{chap4:eq:ascharpolicy_lpbound_c1}, \eqref{chap4:eq:ascharpolicy_lpbound_c2}, and \eqref{chap4:eq:ascharpolicy_lpbound_c3} can be met by assigning positive probability to $\mu_{k} \geq k_{l}$ only, we obtain the trivial lower bound that $\sum_{k < k_{l}} \pim(k) \geq 0$.
However, for \FMC-3 we have that
\begin{eqnarray}
  \sum_{k < k_{l}} \pim(k)\brap{\mu_{k} - \lambda} + \sum_{k > k_{u}} \pim(k)\brap{\mu_{k} - \lambda} & = & 0, \text{ from \eqref{chap4:eq:ascharpolicy_lpbound_c2}}, \nonumber \\
  \sum_{k > k_{u}} \pim(k)\brap{\mu_{k} - \lambda} & = & \sum_{k < k_{l}} \pim(k)\brap{\lambda - \mu_{k}}.
  \label{chap4:eq:ascharpolicy_outbound1}
\end{eqnarray}
To obtain a lower bound on $\sum_{k > k_{u}} \pim(k)$ we proceed as follows.
From \eqref{chap4:eq:ascharpolicy_outbound1}
\begin{eqnarray}
  \brap{\mu_{K} - \lambda}\sum_{k > k_{u}} \pim(k) & \geq & \brap{\lambda - \mu_{k_{l} - 1}}\sum_{k < k_{l}} \pim(k), \nonumber \\
  \frac{\mu_{K} - \lambda}{\lambda - \mu_{k_{l} - 1}} \sum_{k > k_{u}} \pim(k) + \sum_{k > k_{u}} \pim(k) & \geq & \frac{\Cg - c(\lambda)}{\max_{k < k_{l}, k > k_{u}} \brap{c(\mu_{k}) - l(\mu_{k})}}, \text{ from \eqref{chap4:eq:ascharpolicy_outbound2}}, \nonumber 
\end{eqnarray}
Or, we have that
\begin{eqnarray}
  \sum_{k > k_{u}} \pim(k) & \geq & \frac{\lambda - \mu_{k_{l} - 1}}{\mu_{K} - \mu_{k_{l} - 1}}\brap{\frac{\Cg - c(\lambda)}{\max_{k < k_{l}, k > k_{u}} \brap{c(\mu_{k}) - l(\mu_{k})}}}.
  \label{chap4:eq:ascharpolicy_outbound3}
\end{eqnarray}
To obtain a lower bound on $\sum_{k < k_{l}} \pim(k)$ we proceed as follows from \eqref{chap4:eq:ascharpolicy_outbound1} 
\begin{eqnarray}
  \brap{\mu_{k_{u} + 1} - \lambda}\sum_{k > k_{u}} \pim(k) & \leq & \lambda\sum_{k < k_{l}} \pim(k), \nonumber \\
  \frac{\lambda}{\mu_{k_{u} + 1} - \lambda} \sum_{k < k_{l}} \pim(k) + \sum_{k < k_{l}} \pim(k) & \geq & \frac{\Cg - c(\lambda)}{\max_{k < k_{l}, k > k_{u}} \brap{c(\mu_{k}) - l(\mu_{k})}}, \text{ from \eqref{chap4:eq:ascharpolicy_outbound2}}, \nonumber
\end{eqnarray}
Or, we have that
\begin{eqnarray}
  \sum_{k < k_{l}} \pim(k) & \geq & \frac{\mu_{k_{u} + 1} - \lambda}{\mu_{k_{u} + 1}}\brap{\frac{\Cg - c(\lambda)}{\max_{k < k_{l}, k > k_{u}} \brap{c(\mu_{k}) - l(\mu_{k})}}}.
  \label{chap4:eq:ascharpolicy_outbound4}
\end{eqnarray}
We now consider a method to obtain an upper bound on $\sum_{k_{l} \leq k \leq \tk} \pim(k)$.
We note that from \eqref{chap4:eq:ascharpolicy_lpbound_c1} we have
\begin{eqnarray}
  \sum_{k < k_{l}, k > k_{u}} \pim(k) + \sum_{k_{l} \leq k \leq \tk} \pim(k) + \sum_{\tk < k \leq k_{u}} \pim(k) & = & 1, \nonumber \\
  \sum_{k_{l} \leq k \leq \tk} \pim(k) + \sum_{\tk < k \leq k_{u}} \pim(k) & \leq & 1 - \frac{\Cg - c(\lambda)}{\max_{k < k_{l}, k > k_{u}} \brap{c(\mu_{k}) - l(\mu_{k})}}, \text{ from \eqref{chap4:eq:ascharpolicy_outbound2}}. \nonumber \\
  \label{chap4:eq:ascharpolicy_outbound5}
\end{eqnarray}
We note that RHS is the upper bound if $k_{l} = k_{u}$ or $\tk = k_{u}$.
Suppose $\tk < k_{u}$, then we proceed as follows.
From \eqref{chap4:eq:ascharpolicy_lpbound_c2}, we have that
\begin{eqnarray*}
  \sum_{k < k_{l}} \pim(k)\mu_{k} + \sum_{k_{l} \leq k \leq \tk} \pim(k)\mu_{k} + \sum_{\tk < k \leq k_{u}} \pim(k)\mu_{k} + \sum_{k_{u} < k} \pim(k)\mu_{k} & = & \lambda, \\
  \mu_{k_{l} - 1}\sum_{k < k_{l}} \pim(k) + \mu_{\tk} \sum_{k_{l} \leq k \leq \tk} \pim(k) + \mu_{u} \sum_{\tk < k \leq k_{u}} \pim(k) + \mu_{K} \sum_{k_{u} < k} \pim(k) & \geq & \lambda, \\
\end{eqnarray*}
Or we have that
\begin{eqnarray*}
  \sum_{\tk < k \leq k_{u}} \pim(k) & \geq & \frac{\lambda}{\mu_{u}} - \frac{\mu_{\tk}}{\mu_{u}} \sum_{k_{l} \leq k \leq \tk} \pim(k) - \frac{\mu_{k_{l} - 1}}{\mu_{u}} \sum_{k < k_{l}} \pim(k) - \frac{\mu_{K}}{\mu_{u}} \sum_{k_{u} < k} \pim(k), \\
  \sum_{\tk < k \leq k_{u}} \pim(k) & \geq & \frac{\lambda}{\mu_{u}} - \frac{\mu_{\tk}}{\mu_{u}} \sum_{k_{l} \leq k \leq \tk} \pim(k) - \frac{\mu_{k_{l} - 1}}{\mu_{u}} \frac{\Cg - c(\lambda)}{\min_{n < k_{l}} \brap{c(\mu_{n}) - l(\mu_{n})}} - \frac{\mu_{K}}{\mu_{u}} \frac{\Cg - c(\lambda)}{\min_{n > k_{u}} \brap{c(\mu_{n}) - l(\mu_{n})}}.
\end{eqnarray*}
From \eqref{chap4:eq:ascharpolicy_outbound5} we have that
\begin{eqnarray}
  \sum_{k_{l} \leq k \leq \tk} \pim(k) & \leq & 1 - \sum_{\tk < k \leq k_{u}} \pim(k) - \frac{\Cg - c(\lambda)}{\max_{k < k_{l}, k > k_{u}} \brap{c(\mu_{k}) - l(\mu_{k})}}, \nonumber \\
  & \leq & 1 - \frac{\lambda}{\mu_{u}} + \frac{\mu_{\tk}}{\mu_{u}} \sum_{k_{l} \leq k \leq \tk} \pim(k) + \frac{\mu_{k_{l} - 1}}{\mu_{u}} \frac{\Cg - c(\lambda)}{\min_{n < k_{l}} \brap{c(\mu_{n}) - l(\mu_{n})}} + \nonumber \\
  & & \frac{\mu_{K}}{\mu_{u}} \frac{\Cg - c(\lambda)}{\min_{n > k_{u}} \brap{c(\mu_{n}) - l(\mu_{n})}}- \frac{\Cg - c(\lambda)}{\max_{k < k_{l}, k > k_{u}} \brap{c(\mu_{k}) - l(\mu_{k})}}, \nonumber
\end{eqnarray}
Or, we have that
\begin{eqnarray}
  \brap{1 - \frac{\mu_{\tk}}{\mu_{u}}}\sum_{k_{l} \leq k \leq \tk} \pim(k) & \leq & 1 - \frac{\lambda}{\mu_{u}} + \frac{\mu_{k_{l} - 1}}{\mu_{u}} \frac{\Cg - c(\lambda)}{\min_{n < k_{l}} \brap{c(\mu_{n}) - l(\mu_{n})}} + \nonumber \\
  & & \frac{\mu_{K}}{\mu_{u}} \frac{\Cg - c(\lambda)}{\min_{n > k_{u}} \brap{c(\mu_{n}) - l(\mu_{n})}}- \frac{\Cg - c(\lambda)}{\max_{k < k_{l}, k > k_{u}} \brap{c(\mu_{k}) - l(\mu_{k})}}, \nonumber \\
  \sum_{k_{l} \leq k \leq \tk} \pim(k) & \leq & \frac{\mu_{u} - \lambda}{\mu_{u} - \mu_{\tk}} + {\mu_{k_{l} - 1}} \frac{\Cg - c(\lambda)}{\min_{n < k_{l}} \brap{c(\mu_{n}) - l(\mu_{n})}} +  \label{chap4:eq:ascharpolicy_outbound6}\\
  & & {\mu_{K}} \frac{\Cg - c(\lambda)}{\min_{n > k_{u}} \brap{c(\mu_{n}) - l(\mu_{n})}} - \mu_{u} \frac{\Cg - c(\lambda)}{\max_{k < k_{l}, k > k_{u}} \brap{c(\mu_{k}) - l(\mu_{k})}}. \nonumber 
\end{eqnarray}
We note that the above bound is less than one, in the limit as $\Cg \downarrow c(\lambda)$ only if $\mu_{\tk} < \lambda$.

We note that the above bounds can be used to obtain upper and lower bounds on $\sum_{k = 0}^{\tk} \pim(k)$ in many cases.
We then obtain the lower and upper bounds $P_{l}\brac{.}$ and $P_{u}\brac{.}$ using $\sum_{k = 0}^{\tk} \pim(k)$.
Now we discuss how $\pi_{l}(q)$ and $\pi_{u}(q)$ can be obtained.

We note that if $q^{1} = q_{k'} + 1$ and $q^{2} = q_{k}$ for $k' < k$, then for any $q$ such that $q^{1} \leq q \leq q^{2}$, we have that
\begin{eqnarray*}
  \pi(q) & \leq & \pi(q_{k'}) \fpow{\lambda}{\mu_{k' + 1}}{q - q_{k'}}, \text{ or,}\\
  \pi(q) & \leq & \pi(q_{k}) \fpow{\mu_{k}}{\lambda}{q_{k} - q}.
\end{eqnarray*}
We can then bound $\pi(q_{k'})$ by $\pim(k')$ or $\pi(q_{k})$ by $\pim(k)$, which leads to an upper bound $\pi_{u}(q)$ for $q^{1} \leq q \leq q^{2}$.
We note that a similar upper bound on $\pi(q)$ has been used in the asymptotic analysis of $Q^*(c_{c})$ above.

We are only able to obtain asymptotic lower bounds on $\pi(0)$, for any sequence of non-idling order optimal admissible policies $\gamma_{k}$.
The asymptotic lower bounds are obtained using the same method as in the proof of \cite[Theorem 2]{neely_utility}.
We consider the DTMC $Q_{d}[m]$ which is obtained by uniformizing $Q(t)$ at rate $r_{u}$, as in Appendix \ref{chap4:app:uniformization}.
Then, we note that the stationary distribution of $Q_{d}[m]$ is the same as that of $Q(t)$ under the policy $\gamma$.
We proceed as in \cite{neely_utility} by assuming that the process is stationary at $m = 0$.
Then, from Markov inequality we have that the probability that $Q_{d}[0] \leq \ceiling{2\Qg}$ and there are no arrivals (or up-transitions for $Q_{d}[m]$) in $k$ successive slots is at least $\frac{1}{2} \brap{1 - \frac{\lambda}{r_{u}}}^{k}$.
Suppose $k = \ceiling{2\Qg} + 1$.
Then we have that the probability that there is no service (or down-transition for $Q_{d}[m]$) in $k$ slots is at least $\frac{1}{2} \brap{1 - \frac{\lambda}{r_{u}}}^{k}$.
Therefore, we have that $\pi(0) \geq \frac{\brap{1 - \frac{\lambda}{r_{u}}}^{k}}{2k}$.
With our choice of $k$, we have that
\begin{equation}
  \pi(0) \geq \frac{\brap{1 - \frac{\lambda}{r_{u}}}^{\brap{\ceiling{2\Qg} + 1}}}{2\brap{\ceiling{2\Qg} + 1}}.
\end{equation}
We note that for the policy under consideration, for \FMC-2 $\Qg = \Theta\brap{\log\nfrac{1}{V}}$ and for \FMC-3 $\Qg = \Theta\nfrac{1}{V}$, as $V \downarrow 0$.
Therefore for small enough $V$, we have that
\begin{eqnarray}
  \pi(0) & \geq & \frac{\brap{1 - \frac{\lambda}{r_{u}}}^{\brap{2\Qg + 2}}}{4\Qg}, \nonumber \\
  \pi(0) & \geq & \pi_{l}(0) = 
  \begin{cases}
    \Omega\nfrac{V}{\log\nfrac{1}{V}}, \text{ for \FMC-2}, \\
    \Omega\brap{V\brap{1 - \frac{\lambda}{r_{u}}}^{\frac{1}{V}}}, \text{ for \FMC-3}.
  \end{cases}
  \label{chap4:eq:ascharpolicy_pi0lb}
\end{eqnarray}
We note that an asymptotic lower bound $\pi_{l}(q)$ can then be obtained since $\pi(q) \geq \pi_{l}(q) = \pi_{l}(0)\fpow{\lambda}{\mu_{K}}{q}$, but this bound is very weak in most cases.

\subsubsection{Asymptotic characterization:}
In this section, we obtain asymptotic bounds on $q_{k}$.
We note that since we are not able to obtain analytical forms for $P_{l}\brac{.}$, $P_{u}\brac{.}$, $\pi_{l}(.)$, and $\pi_{u}(.)$ in all cases, we are not able to obtain asymptotic bounds on $q_{k}, \forall k \in \brac{1, \dots, K}$.
The asymptotic bounds on $q_{k}$ are obtained using the methodology described in Section \ref{chapter2:policymethod}.
\begin{proposition}
  For \FMC-2, for any sequence of non-idling order-optimal admissible policies $\gamma_{k}$, with $\Cgk - c(\lambda) = V_{k}$, we have that
  \begin{eqnarray*}
    q_{\tk} & = & 
    \begin{cases}
      \mathcal{O}\brap{\log\brap{\log\nfrac{1}{V_{k}}}}, \text{ if } 1 \leq \tilde{k} \leq k_{l} - 1, \\
      \mathcal{O}\brap{\log\nfrac{1}{V_{k}}}, \text{ if } k_{l} \leq \tilde{k} \text{ and } \mu_{\tk} \leq \lambda, \\
      \Omega\brap{\log\nfrac{1}{V_{k}}}, \text{ if } \tilde{k} \text{ is such that } \mu_{\tk} > \lambda.
    \end{cases}
  \end{eqnarray*}
  \label{chap4:lemma:asympcharpolicy_fmc-2}
\end{proposition}
\begin{proof}
  Consider any policy $\gamma$ in the sequence with $V_{k} = V$.
  Consider any $\tilde{k}$ such that $1 \leq \tilde{k} \leq k_{l} - 1$.
  From \eqref{chap4:eq:ascharpolicy_sbound1}, we have that
  \begin{equation*}
    \sum_{k = 0}^{\tk} \pi_{\mu}(k) \leq \frac{V}{\min_{n \in \brac{0, \dots, \tk}} \brap{c(\mu_{n}) - l(\mu_{n})}} \Deq \frac{V}{c_{1}}.
  \end{equation*}
  From \eqref{chap4:eq:ascharpolicy_obs2}, we have that if $q_{\tk,u}$ is the smallest integer such that
  \begin{equation*}
    \sum_{q = 0}^{q_{\tk,u}} \pi_{l}(q) \geq \frac{V}{c_{1}},
  \end{equation*}
  then $q_{\tk} \leq q_{\tk,u}$.
  Substituting $\pi_{l}(q) = \pi_{l}(0)\fpow{\lambda}{\mu_{\tk}}{q}$, we have that
  \begin{eqnarray*}
    \pi_{l}(0)\sum_{q = 0}^{q_{k,u}} \fpow{\lambda}{\mu_{\tk}}{q} & \geq & \frac{V}{c_{1}}, \\
    \pi_{l}(0)\brap{\fpow{\lambda}{\mu_{\tk}}{q_{\tk,u} + 1} - 1} & \geq & \frac{\lambda - \mu_{\tk}}{\mu_{\tk}} \frac{V}{c_{1}}. \\
  \end{eqnarray*}  
  Or we have that $q_{\tk,u}$ is the smallest integer such that
  \begin{eqnarray*}
    q_{\tk,u} \geq \log_{\nfrac{\lambda}{\mu_{\tk}}} \bras{1 + \frac{\lambda - \mu_{\tk}}{\mu_{\tk}} \frac{V}{c_{1} \pi_{l}(0)}}.
  \end{eqnarray*}
  Since $\pi_{l}(0) = \Omega\nfrac{V}{\log\nfrac{1}{V}}$ from \eqref{chap4:eq:ascharpolicy_pi0lb}, we have that $q_{\tk} \leq q_{\tk,u} = \mathcal{O}\brap{\log\brap{\log\nfrac{1}{V}}}$.
  We do not have any asymptotic lower bounds for $q_{\tk}, 1 \leq \tk \leq k_{l} - 1$.

  Let us now consider $\tk$ such that $k_{l} \leq \tk$ and $\mu_{\tk} < \lambda$.
  Then from \eqref{chap4:eq:ascharpolicy_sbound1} and \eqref{chap4:eq:ascharpolicy_outbound6}, we have that
  \begin{eqnarray*}
    & & \Pr\brac{0 \leq Q \leq q_{\tk}} = \sum_{k = 0}^{\tk} \pi_{\mu}(k) = \sum_{k = 0}^{k_{l} - 1} \pim(k) + \sum_{k = k_{l}}^{\tk} \pim(k), \\
    & & \leq \frac{V}{\min_{n \in \brac{0, \dots, k_{l} - 1}} \brap{c(\mu_{n}) - l(\mu_{n})}} + \frac{\mu_{u} - \lambda}{\mu_{u} - \mu_{\tk}} + {\mu_{k_{l} - 1}} \frac{V}{\min_{n < k_{l}} \brap{c(\mu_{n}) - l(\mu_{n})}} +  \\
    & & {\mu_{K}} \frac{V}{\min_{n > k_{u}} \brap{c(\mu_{n}) - l(\mu_{n})}} - \mu_{u} \frac{V}{\max_{k < k_{l}, k > k_{u}} \brap{c(\mu_{k}) - l(\mu_{k})}} = P_{u}\brac{0 \leq Q \leq q_{\tk}}.
  \end{eqnarray*}
  We also have that $\forall q \leq q_{\tk}$, $\pi_{l}(q) = \pi_{l}(0) \fpow{\lambda}{\mu_{\tk}}{q}$.
  Then, from \eqref{chap4:eq:ascharpolicy_obs2}, if $q_{\tk,u}$ is the smallest integer such that
  \begin{equation*}
    \sum_{q = 0}^{q_{\tk,u}} \pi_{l}(q) \geq \frac{\mu_{u} - \lambda}{\mu_{u} - \mu_{\tk}} + \mathcal{O}(V),
  \end{equation*}
  then $q_{\tk} \leq q_{\tk,u}$.
  Then, we obtain that $q_{\tk,u} = \mathcal{O}\brap{\log\nfrac{1}{V}}$.
  Therefore, for $\tk$ such that $k_{l} \leq \tk < \lambda$, $q_{\tk} = \mathcal{O}\brap{\log\nfrac{1}{V}}$.

  Now consider $\tk$ such that $\tk \leq k_{u}$ and $\lambda < \mu_{\tk}$.
  From \eqref{chap4:eq:ascharpolicy_lbmu_g_lambda}, we have that
  \begin{equation*}
    \sum_{k = k_{l}}^{\tk} \pi_{\mu}(k) \geq \frac{\mu_{\tk} - \lambda}{\mu_{\tk} - \mu_{l}} - \frac{\mu_{\tk}}{c_{1}} V,
  \end{equation*}
  where $c_{1} = \min_{k < k_{l}, k > k_{u}} \brap{c(\mu_{k}) - l(\mu_{k})}$.
  Then 
  \begin{equation*}
    \sum_{k = 0}^{\tk} \pi_{\mu}(k) \geq P_{l}\brac{0 \leq Q \leq q_{\tk}} \Deq \frac{\mu_{\tk} - \lambda}{\mu_{\tk} - \mu_{l}} - \frac{\mu_{\tk}}{c_{1}} V,
  \end{equation*}
  For $0 \leq q \leq q_{\tk}$, we have that $\pi(q) \leq \pi_{u}(q) = \pi_{u}(0)\fpow{\lambda}{\mu_{1}}{q}$.
  From \eqref{chap4:eq:ascharpolicy_obs1} if $q_{\tk,l}$ is the largest integer such that
  \begin{equation*}
    \sum_{q = 0}^{q_{\tk,l}} \pi_{u}(q) \leq P_{l}\brac{0 \leq Q \leq q_{\tk}},
  \end{equation*}
  then $q_{\tk,l} \leq q_{\tk}$.
  Substituting for $\pi_{u}(q)$ and $P_{l}\brac{0 \leq Q \leq q_{\tk}}$, we have that
  \begin{equation*}
    \sum_{q = 0}^{q_{\tk,l}} \pi_{u}(0)\fpow{\lambda}{\mu_{1}}{q} \leq \frac{\mu_{\tk} - \lambda}{\mu_{\tk} - \mu_{l}} - \frac{\mu_{\tk}}{c_{1}} V.
  \end{equation*}
  Since $\pi_{u}(0) = \mathcal{O}(V)$, we have that $q_{\tk,l} = \Omega\brap{\log\nfrac{1}{V}}$.
\end{proof}

\begin{proposition}
  For \FMC-3, for any sequence of non-idling order-optimal admissible policies $\gamma_{k}$, with $\Cgk - c(\lambda) = V_{k}$, we have that
  \begin{eqnarray*}
    q_{\tk} & = & 
    \begin{cases}
      \mathcal{O}\nfrac{1}{V_{k}}, \text{ if } \mu_{\tk} < \lambda, \\
      {\Omega}\brap{1}, \text{ if } \mu_{\tk} = \mu_{k_{l} - 1}, \\
      \mathcal{O}\brap{\frac{1}{V\brap{1 - \frac{\lambda}{r_{u}}}^{\frac{1}{V}}}}, \text{ if } \mu_{\tk} = \lambda, \\
      {\Omega}\nfrac{1}{V_{k}}, \text{ if } \mu_{\tk} \geq \lambda.
    \end{cases}
  \end{eqnarray*}
  \label{chap4:lemma:asympcharpolicy_fmc-3}
\end{proposition}
\begin{proof}
  The methods used in this proof are similar to that used for the proof of Proposition \ref{chap4:lemma:asympcharpolicy_fmc-2}.
  We first consider $\tk$ such that $\mu_{\tk} < \lambda$.
  Since 
  \begin{equation*}
    \sum_{k = 0}^{\tk} \pi_{\mu}(k) \leq \frac{V}{\min_{n \in \brac{0, \dots, \tk}} \brap{c(\mu_{n}) - l(\mu_{n})}} \Deq \frac{V}{c_{1}},
  \end{equation*}
  from \eqref{chap4:eq:ascharpolicy_sbound1} and $\pi_{l}(0) = \Omega\brap{V\brap{1 - \frac{\lambda}{r_{u}}}^{\frac{1}{V}}}$ from \eqref{chap4:eq:ascharpolicy_pi0lb}, we have that $q_{\tk,u} = \mathcal{O}\brap{\frac{1}{V}}$.
  From \eqref{chap4:eq:ascharpolicy_outbound4}, for $\tk = k_{l} - 1$, we have that
  \begin{eqnarray*}
    \sum_{0 \leq k \leq k_{l} - 1} \pi_{\mu}(k) \geq \frac{\mu_{k_{u} + 1} - \lambda}{\mu_{k_{u} + 1}}\brap{\frac{V}{c_{2}}},
  \end{eqnarray*}
  where $c_{2} = \max_{k < k_{l}, k > k_{u}} \brap{c(\mu_{k}) - l(\mu_{k})}$.
  Then using $\pi_{u}(q) = \pi_{u}(0) \fpow{\lambda}{\mu_{1}}{q}$, we have that $q_{\tk,l} = \Omega(1)$, so that $q_{\tk} = \Omega(1)$.

  Consider $\tk$ such that $\mu_{\tk} = \lambda$.
  We note that $\pim(\lambda) \leq 1$, then $\pi_{l}(q) = \pi_{l}(0), \forall q \leq q_{\tk}$.
  Then we obtain that $q_{\tk,u} = \mathcal{O}\brap{\frac{1}{V\brap{1 - \frac{\lambda}{r_{u}}}^{\frac{1}{V}}}}$.

  Now consider any $\tk$ such that $\mu_{\tk} \geq \lambda$.
  We have that $\sum_{k = 0}^{\tk} \pim(k) \geq \pim(\lambda) \geq 1 - \frac{V}{c_{1}}$, where $c_{1} = \min_{k < k_{l}, k > k_{l}} \brac{c(\mu_{k}) - l(\mu_{k})}$.
  Since $\pi_{u}(q) = \pi_{u}(q_{k_{l} - 1} + 1)$, for all $q > q_{k_{l} - 1}$, we have that $q_{k_{l},l} = \Omega\nfrac{1}{V}$.
\end{proof}

\subsection{Numerical examples}
\label{chap4:sec:numexp1}
In this section, we consider some numerical examples for FINITE-$\mu$CHOICE.
In the examples, we compare the bounds on $Q^*(c_{c})$ which were obtained above, with the optimal tradeoff curve for FINITE-$\mu$CHOICE, which is obtained by the numerical solution of an MDP, obtained by uniformization as in \cite{atamm1}.
We now state, the chosen parameters for each numerical example and the quantities plotted in the corresponding figures.
Each numerical example is identified by ``E-abc'', where a,b, and c are numbers.
For each numerical example, we choose the set of service rates and the service cost function $c(\mu)$.
Then we consider a set of arrival rates, for each of which the bounds and the optimal tradeoff are plotted.
We note that for each value of $\lambda$, we obtain a corresponding minimum average service cost $c(\lambda)$.
All the numerical examples that we consider in this section, along with their parameters, are given in Tables \ref{chap4:table:ex-set11}, \ref{chap4:table:ex-set12}, \ref{chap4:table:ex-set21}, and \ref{chap4:table:ex-set22}, along with references to their corresponding plots.
\begin{table}
  \centering
  \begin{tabular}{|c|c|c|c|c|}
    \hline
    Ex. Identifier & $\lambda$ & $c(\lambda)$ & Type & Tradeoff plot \\
    \hline
    E-111 & 0.25 & 0.125 & FINITE-$\mu$CHOICE-1 & \ref{chap4:fig:111} \\
    E-112 & 0.50 & 0.250 & FINITE-$\mu$CHOICE-3 & \ref{chap4:fig:112} \\
    E-113 & 0.75 & 0.625 & FINITE-$\mu$CHOICE-2 & \ref{chap4:fig:113} \\
    \hline
  \end{tabular}
  \caption{Numerical examples with $c(\mu) = \mu^{2}$ for $\mu \in \brac{0, 0.5, 1}$.}
  \label{chap4:table:ex-set11}
\end{table}

\begin{table}
  \centering
  \begin{tabular}{|c|c|c|c|c|}
    \hline
    Ex. Identifier & $\lambda$ & $c(\lambda)$ & Type & Tradeoff plot \\
    \hline
    E-121 & 0.25 & 0.015625 & FINITE-$\mu$CHOICE-1 & \ref{chap4:fig:121} \\
    E-122 & 0.50 & 0.031250 & FINITE-$\mu$CHOICE-3 & \ref{chap4:fig:122} \\
    E-123 & 0.75 & 0.515600 & FINITE-$\mu$CHOICE-2 & \ref{chap4:fig:123} \\
    \hline
  \end{tabular}
  \caption{Numerical examples with $c(\mu) = \mu^{5}$ for $\mu \in \brac{0, 0.5, 1}$.}
  \label{chap4:table:ex-set12}
\end{table}

\begin{table}
  \centering
  \begin{tabular}{|c|c|c|c|c|}
    \hline
    Ex. Identifier & $\lambda$ & $c(\lambda)$ & Type & Tradeoff plot \\
    \hline
    E-211 & 0.10 & 0.02 & FINITE-$\mu$CHOICE-1 & \ref{chap4:fig:211} \\
    E-212 & 0.20 & 0.04 & FINITE-$\mu$CHOICE-3 & \ref{chap4:fig:212} \\
    E-213 & 0.25 & 0.07 & FINITE-$\mu$CHOICE-2 & \ref{chap4:fig:213} \\
    E-214 & 0.70 & 0.50 & FINITE-$\mu$CHOICE-2 & \ref{chap4:fig:214} \\
%   E-215 & 0.80 & 0.64 & FINITE-$\mu$CHOICE-3 & \ref{chap4:fig:215} \\
    \hline
  \end{tabular}
  \caption{Numerical examples with $c(\mu) = \mu^{2}$ for $\mu \in \brac{0, 0.2, 0.4, 0.5, 0.6, 0.8, 1}$.}
  \label{chap4:table:ex-set21}
\end{table}

\begin{table}
  \centering
  \begin{tabular}{|c|c|c|c|c|}
    \hline
    Ex. Identifier & $\lambda$ & $c(\lambda)$ & Type & Tradeoff plot \\
    \hline
    E-221 & 0.10 & 0.00016 & FINITE-$\mu$CHOICE-1 & \ref{chap4:fig:221} \\
    E-222 & 0.20 & 0.00032 & FINITE-$\mu$CHOICE-3 & \ref{chap4:fig:222} \\
    E-223 & 0.25 & 0.00280 & FINITE-$\mu$CHOICE-2 & \ref{chap4:fig:223} \\
    E-224 & 0.70 & 0.20272 & FINITE-$\mu$CHOICE-2 & \ref{chap4:fig:224} \\
%   E-225 & 0.80 & 0.32768 & FINITE-$\mu$CHOICE-3 & \ref{chap4:fig:225} \\
    \hline
  \end{tabular}
  \caption{Numerical examples with $c(\mu) = \mu^{5}$ for $\mu \in \brac{0, 0.2, 0.4, 0.5, 0.6, 0.8, 1}$.}
  \label{chap4:table:ex-set22}
\end{table}

The \emph{Optimal} set of points in each plot is obtained by considering a MDP with the single stage cost given by $q + \beta c(\mu)$, and solving for the optimal infinite horizon average cost policy.
The state of the MDP corresponds to the queue length $q$, and the set of actions taken at each state $q$ is the set of service rates.
Here $\beta$ is a positive Lagrange multiplier.
The MDP is obtained by uniformization at rate $r_{u} = \lambda + r_{max}$.
The transitions in the uniformized MDP are as follows : a) for $q = 0$, the Markov chain moves to $q = 1$ with probability $\frac{\lambda}{r_{u}}$, and stays in $q = 0$ with probability $1 - \frac{\lambda}{r_{u}}$, and b) for $q > 0$, the Markov chain moves to $q + 1$ with probability $\frac{\lambda}{r_{u}}$, to $q - 1$ with probability $\frac{\mu(q)}{r_{u}}$, and stays in $q$ with probability $1 - \frac{\lambda + \mu(q)}{r_{u}}$.
We note that the state space of the MDP is truncated at a maximum queue length, which is such that the optimal value does not change appreciably with further increase in this maximum queue length.
The Optimal points are obtained by varying $\beta$. The $x$-coordinate of a point corresponding to a value of $\beta$ is the difference between the average service cost, for the $\beta$-optimal policy for the MDP, and $c(\lambda)$ while the $y$-coordinate is the average queue length for the $\beta$-optimal policy.
The \emph{Lower bound (Analytical)} curve in each plot is: a) \eqref{chap4:eq:p11finalbound} for FINITE-$\mu$CHOICE-1, b) \eqref{chap4:eq:p12finalbound} for FINITE-$\mu$CHOICE-2, and c) \eqref{chap4:eq:p13finalbound} for FINITE-$\mu$CHOICE-3.
The \emph{Upper bound (Analytical)} curve in each plot is obtained as follows: a) for FINITE-$\mu$CHOICE-1, we choose the sequence $q_{k_{u}}$ to be a sequence of increasing positive integers, and for each $q_{k_{u}}$ obtain the bound on the average queue length from \eqref{chap4:eq:p11finalub}, and the average service cost from \eqref{chap4:eq:p11ub2}, b) for FINITE-$\mu$CHOICE-2, we choose the sequence $U_{k}$ to be a decreasing sequence, and for each $U_{k}$ obtain the bound on the average queue length from \eqref{chap4:eq:p12finalubq} and the average service cost from \eqref{chap4:eq:p12finalubc}, c) for FINITE-$\mu$CHOICE-3, we choose the sequence $U_{k}$ to be a decreasing sequence, and for each $U_{k}$ obtain the bound on the average queue length and the average service cost from \eqref{chap4:eq:p13finalubq} and \eqref{chap4:eq:p13finalubc} respectively.
We note that the bounds on the average queue length which are obtained using the Lyapunov drift method in Proposition \ref{chap4:app:prop:dotzupperbound} are usually weak (although they give the correct order behaviour). Therefore, for the sequence of policies which we have considered for \emph{Upper bound (Analytical)}, we also evaluate the actual average service cost and average queue length, by obtaining the stationary probability of the queue length for the system with truncated state space.
This curve is denoted as \emph{Upper bound} in the plots.
We note that discrete points are obtained while varying $\beta$, time sharing of the policies corresponding to these points leads to the continuous curves shown in the figures.
\begin{figure}
	\centering
	\includegraphics[width=140mm,height=70mm]{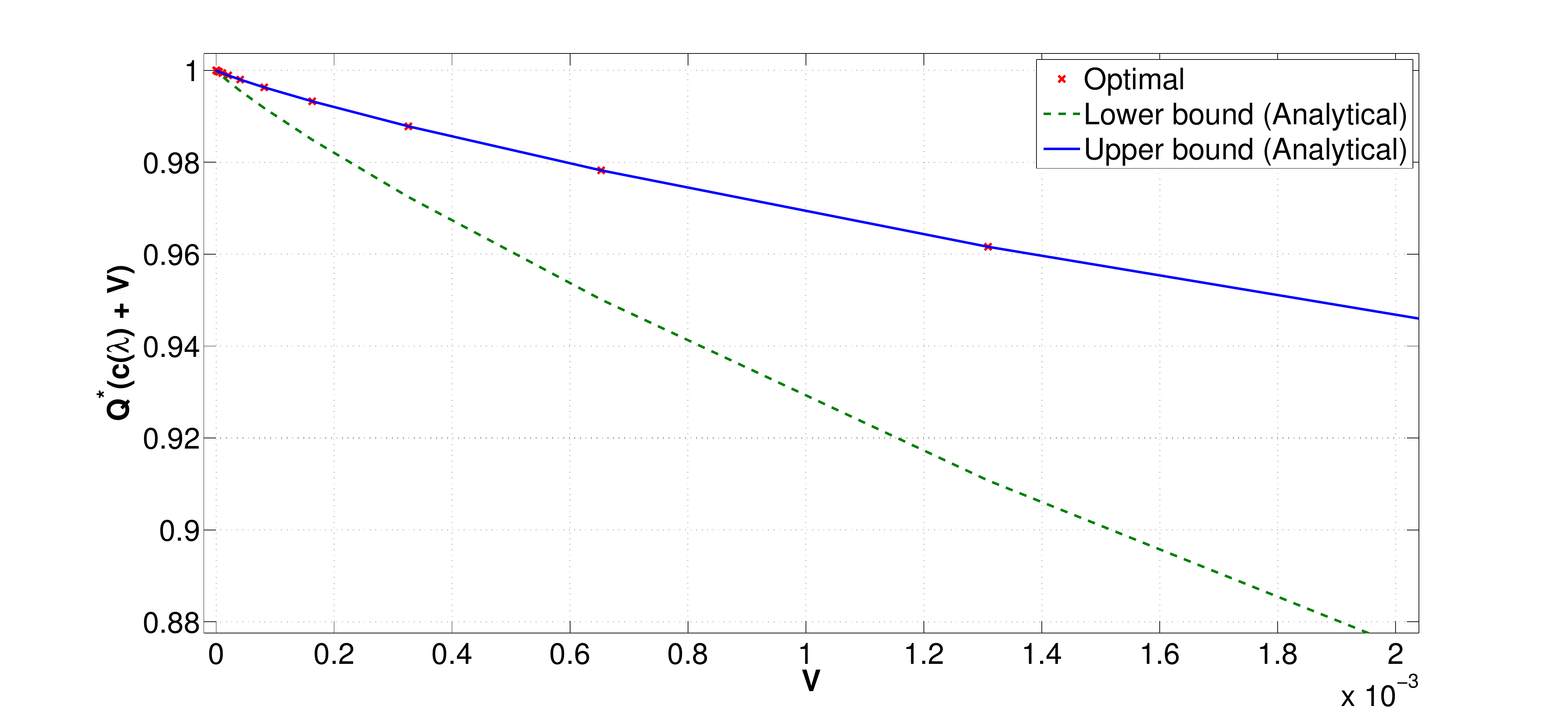}
	\caption{$Q^*(c_{c})$ as a function of $V$, where $c_{c} = c(\lambda) + V$, for $\lambda = 0.25$ and $c(\mu) = \mu^{2}$ for $\mu \in \brac{0, 0.5, 1}$.}
        \label{chap4:fig:111}
\end{figure}
\begin{figure}
	\centering
	\includegraphics[width=140mm,height=70mm]{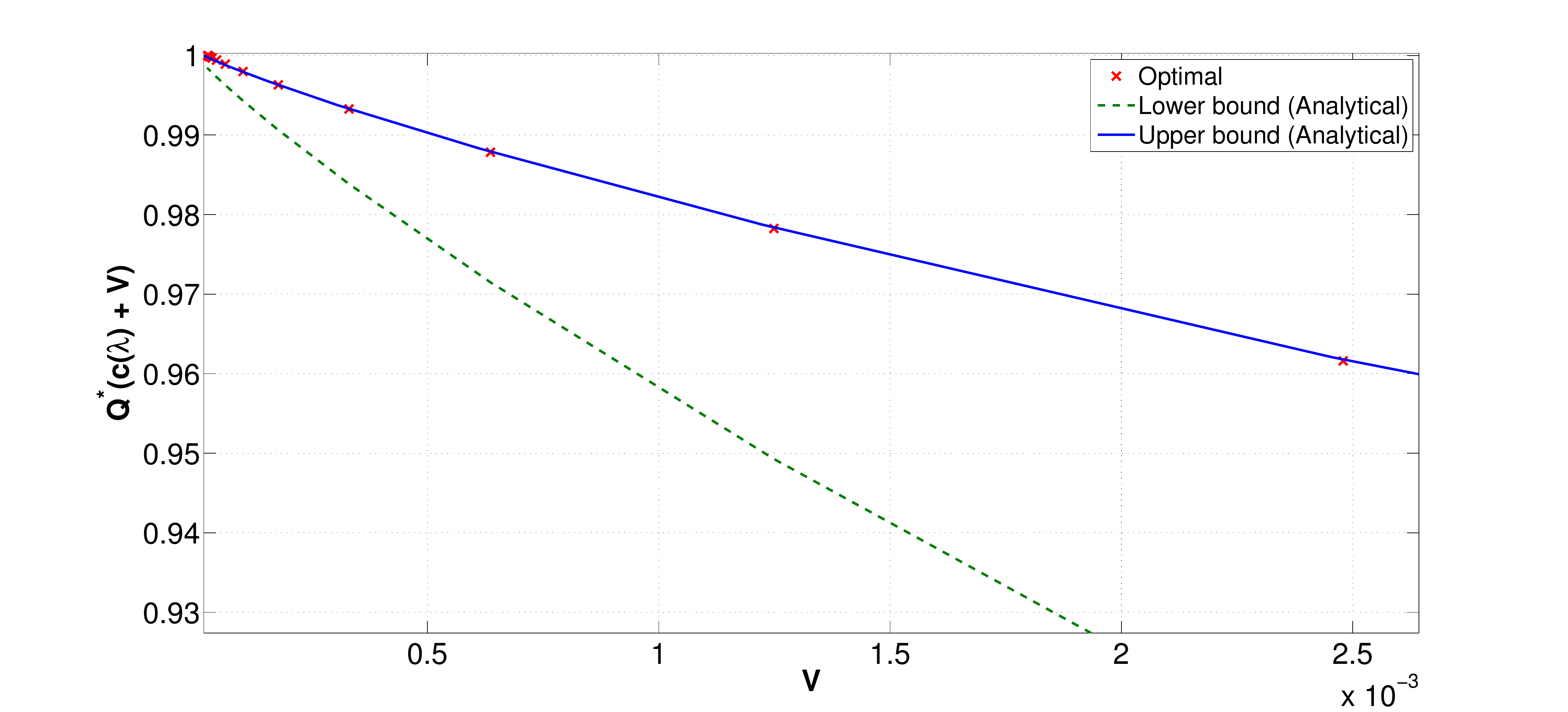}
	\caption{$Q^*(c_{c})$ as a function of $V$, where $c_{c} = c(\lambda) + V$, for $\lambda = 0.25$ and $c(\mu) = \mu^{5}$ for $\mu \in \brac{0, 0.5, 1}$.}
        \label{chap4:fig:121}
\end{figure}

\begin{figure}
	\centering
	\includegraphics[width=140mm,height=70mm]{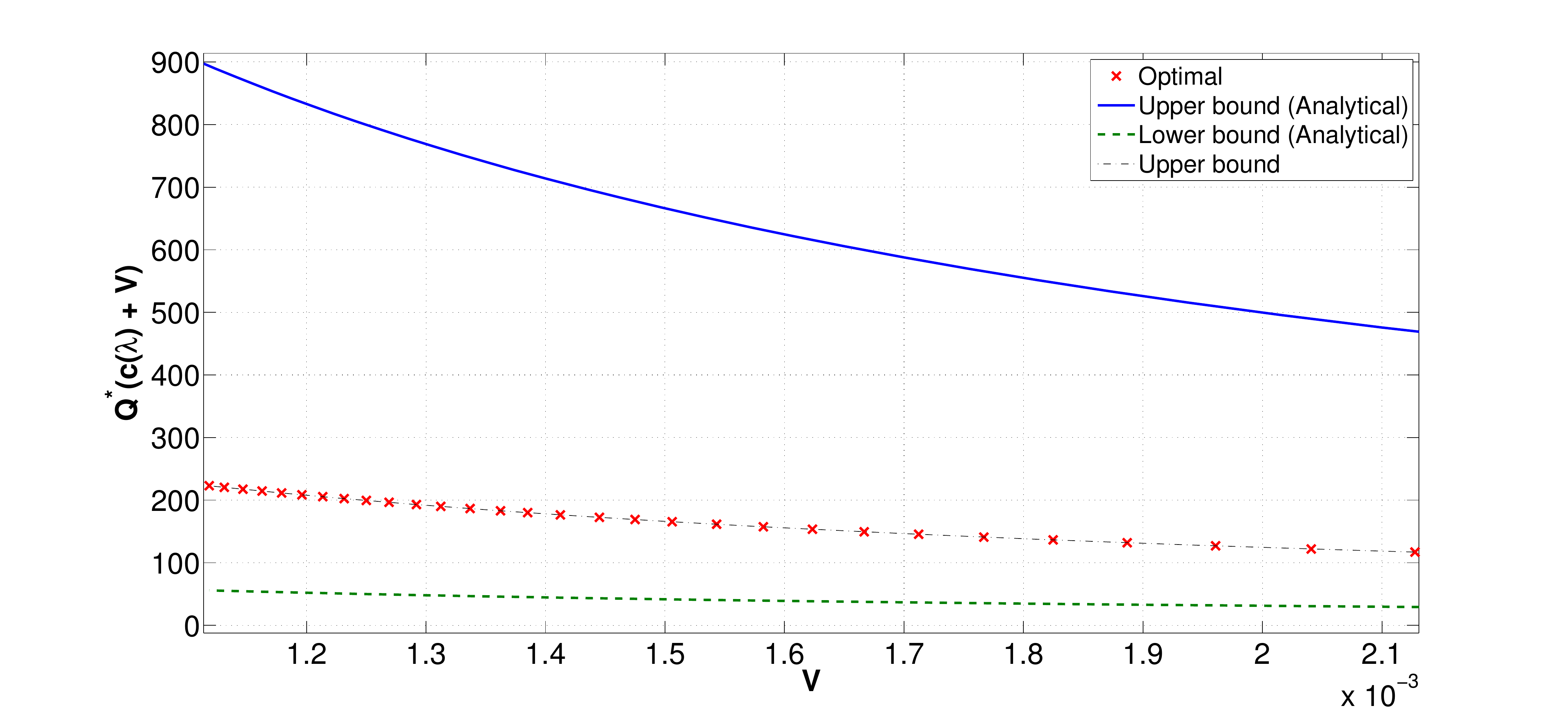}
	\caption{$Q^*(c_{c})$ as a function of $V$, where $c_{c} = c(\lambda) + V$, for $\lambda = 0.50$ and $c(\mu) = \mu^{2}$ for $\mu \in \brac{0, 0.5, 1}$.}
        \label{chap4:fig:112}
\end{figure}
\begin{figure}
	\centering
	\includegraphics[width=140mm,height=70mm]{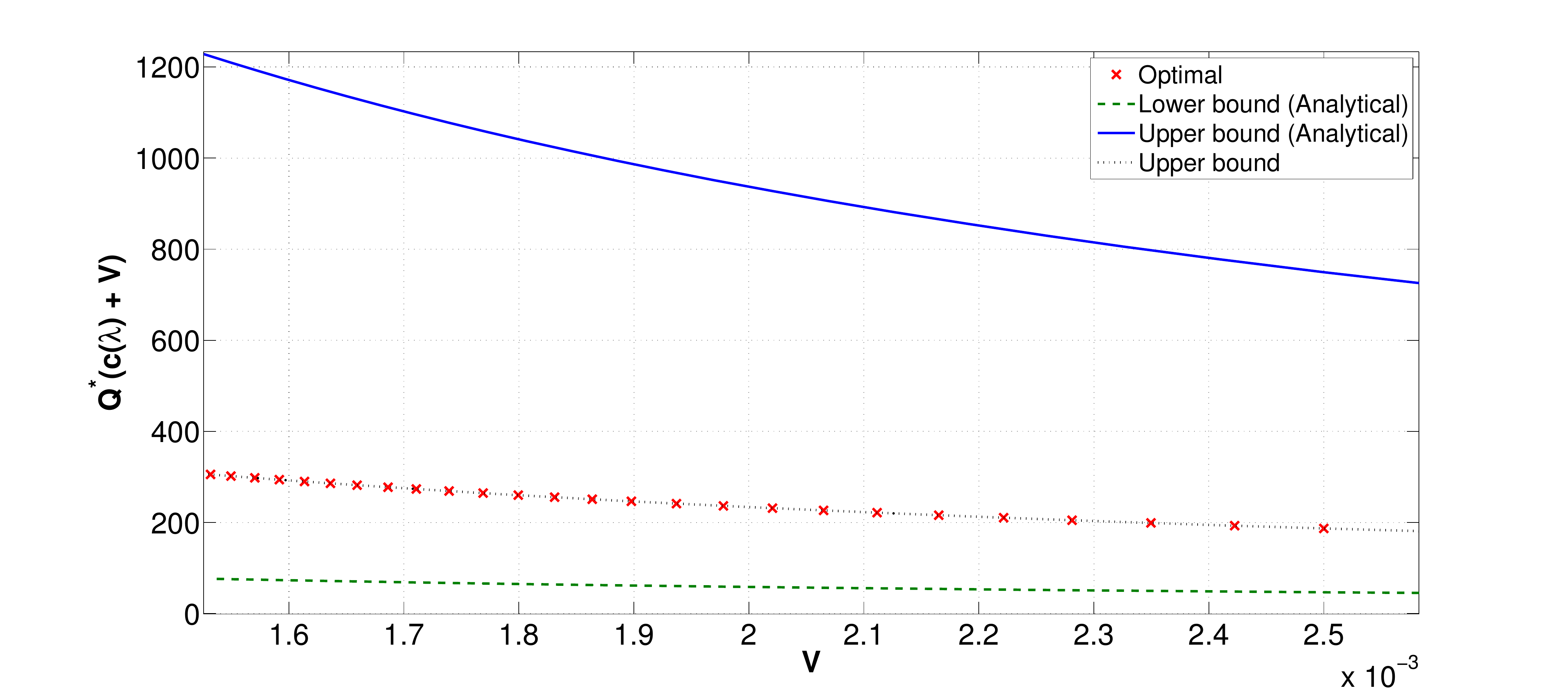}
	\caption{$Q^*(c_{c})$ as a function of $V$, where $c_{c} = c(\lambda) + V$, for $\lambda = 0.50$ and $c(\mu) = \mu^{5}$ for $\mu \in \brac{0, 0.5, 1}$.}
        \label{chap4:fig:122}
\end{figure}

\begin{figure}
	\centering
	\includegraphics[width=140mm,height=70mm]{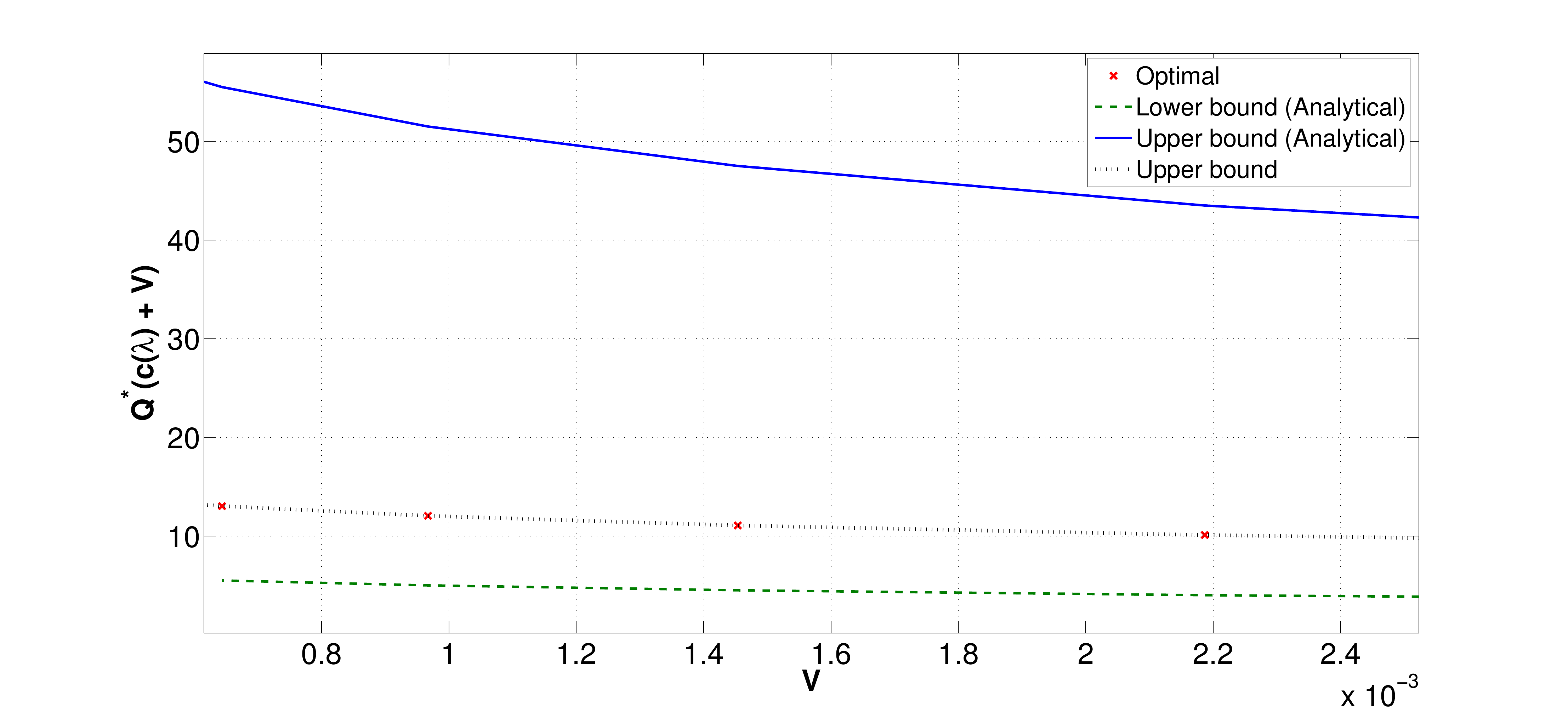}
	\caption{$Q^*(c_{c})$ as a function of $V$, where $c_{c} = c(\lambda) + V$, for $\lambda = 0.75$ and $c(\mu) = \mu^{2}$ for $\mu \in \brac{0, 0.5, 1}$.}
        \label{chap4:fig:113}
\end{figure}
\begin{figure}
	\centering
	\includegraphics[width=140mm,height=70mm]{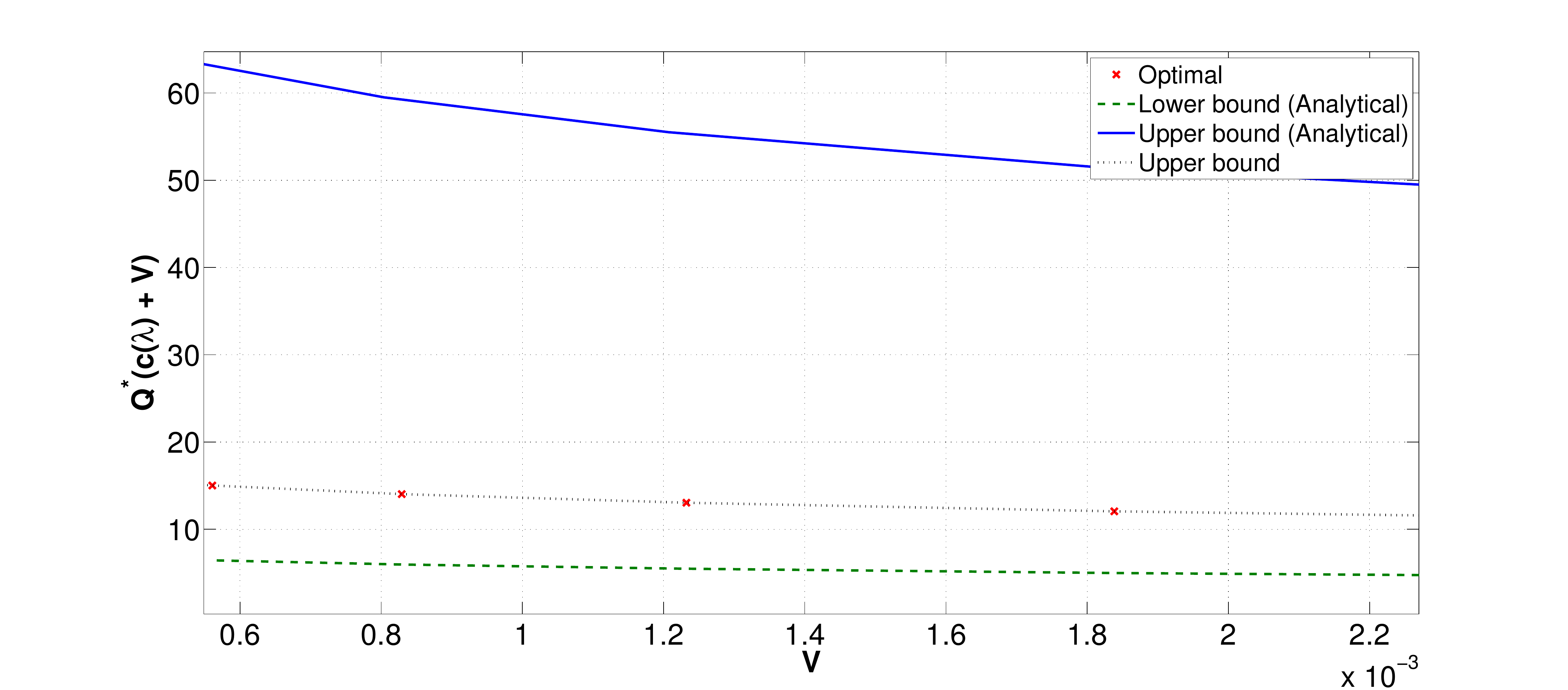}
	\caption{$Q^*(c_{c})$ as a function of $V$, where $c_{c} = c(\lambda) + V$, for $\lambda = 0.75$ and $c(\mu) = \mu^{5}$ for $\mu \in \brac{0, 0.5, 1}$.}
        \label{chap4:fig:123}
\end{figure}

\begin{figure}
	\centering
	\includegraphics[width=140mm,height=70mm]{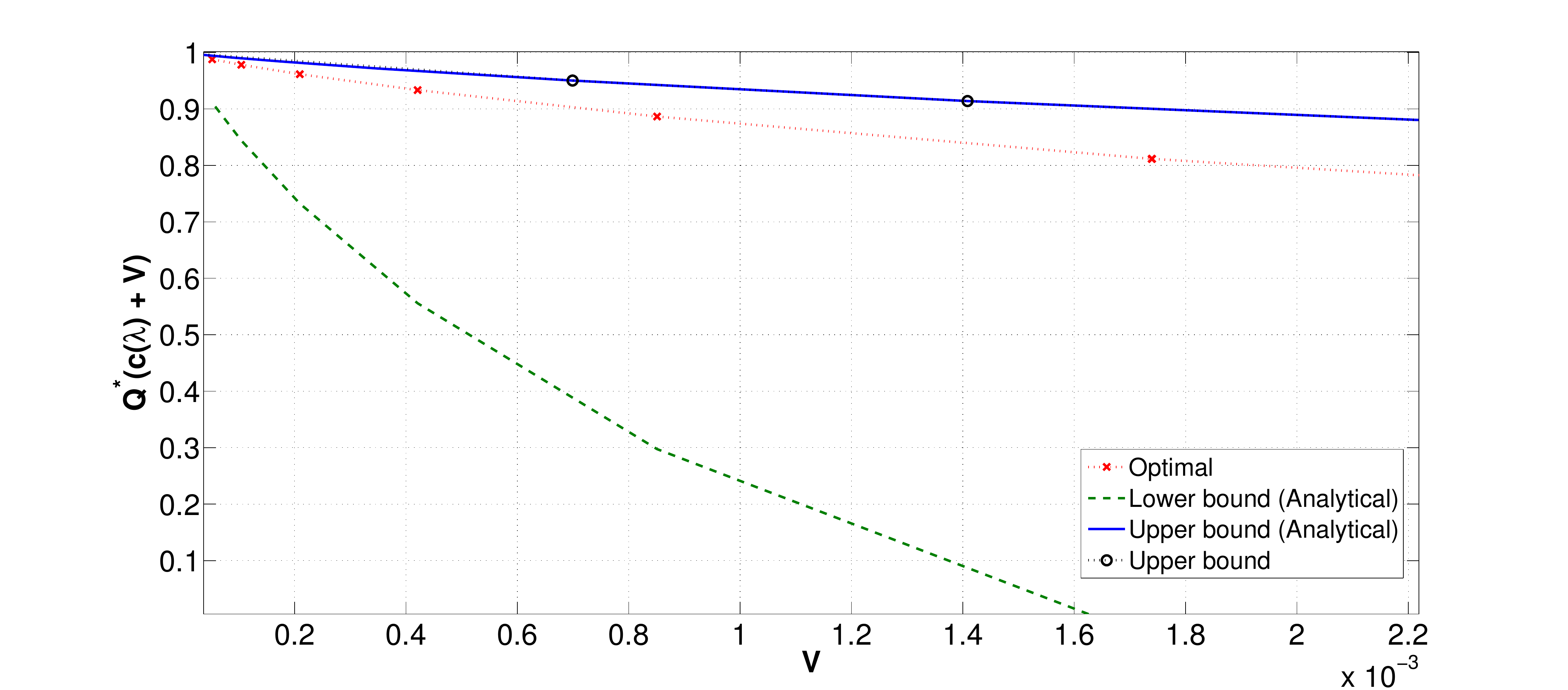}
	\caption{$Q^*(c_{c})$ as a function of $V$, where $c_{c} = c(\lambda) + V$, for $\lambda = 0.10$ and $c(\mu) = \mu^{2}$ for $\mu \in \brac{0, 0.2, 0.4, 0.5, 0.6, 0.8, 1}$.}
        \label{chap4:fig:211}
\end{figure}
\begin{figure}
	\centering
	\includegraphics[width=140mm,height=70mm]{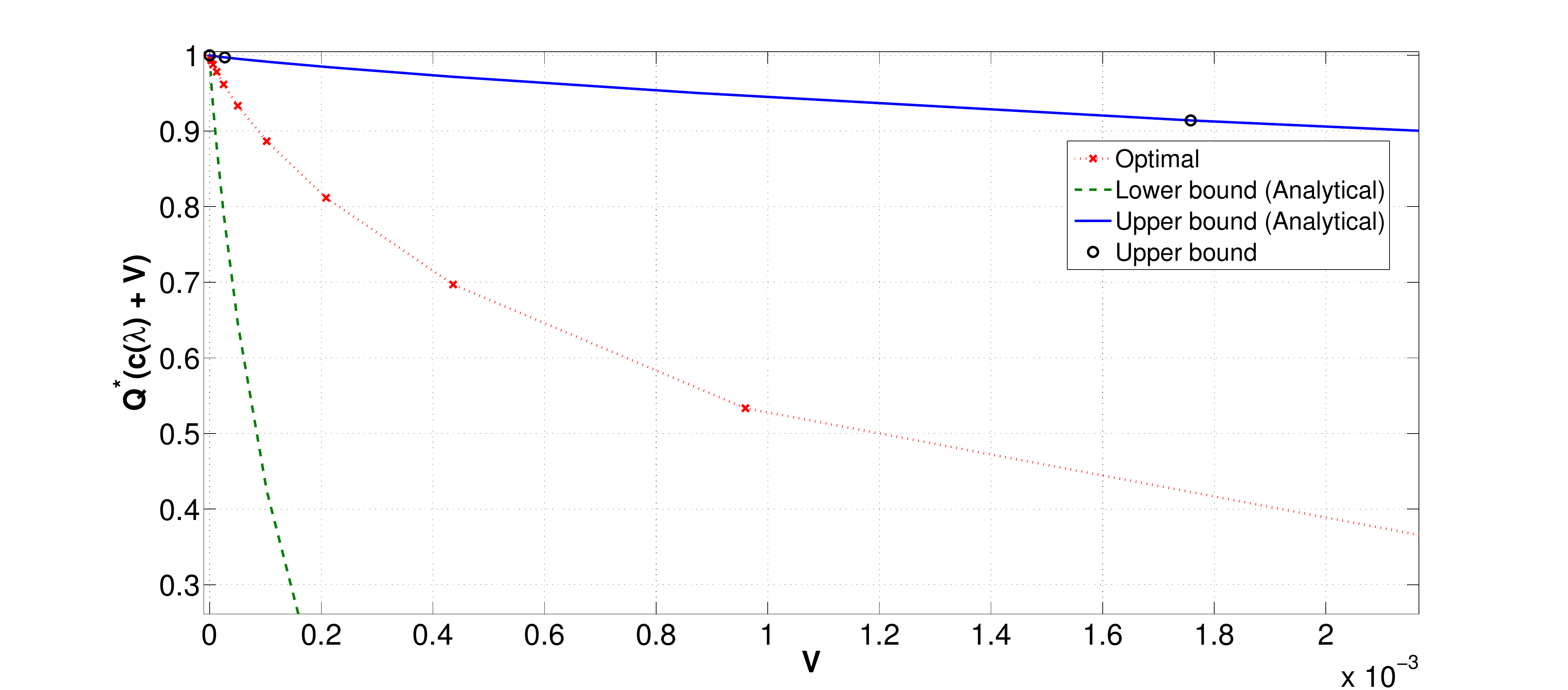}
	\caption{$Q^*(c_{c})$ as a function of $V$, where $c_{c} = c(\lambda) + V$, for $\lambda = 0.10$ and $c(\mu) = \mu^{5}$ for $\mu \in \brac{0, 0.2, 0.4, 0.5, 0.6, 0.8, 1}$.}
        \label{chap4:fig:221}
\end{figure}

\begin{figure}
	\centering
	\includegraphics[width=140mm,height=70mm]{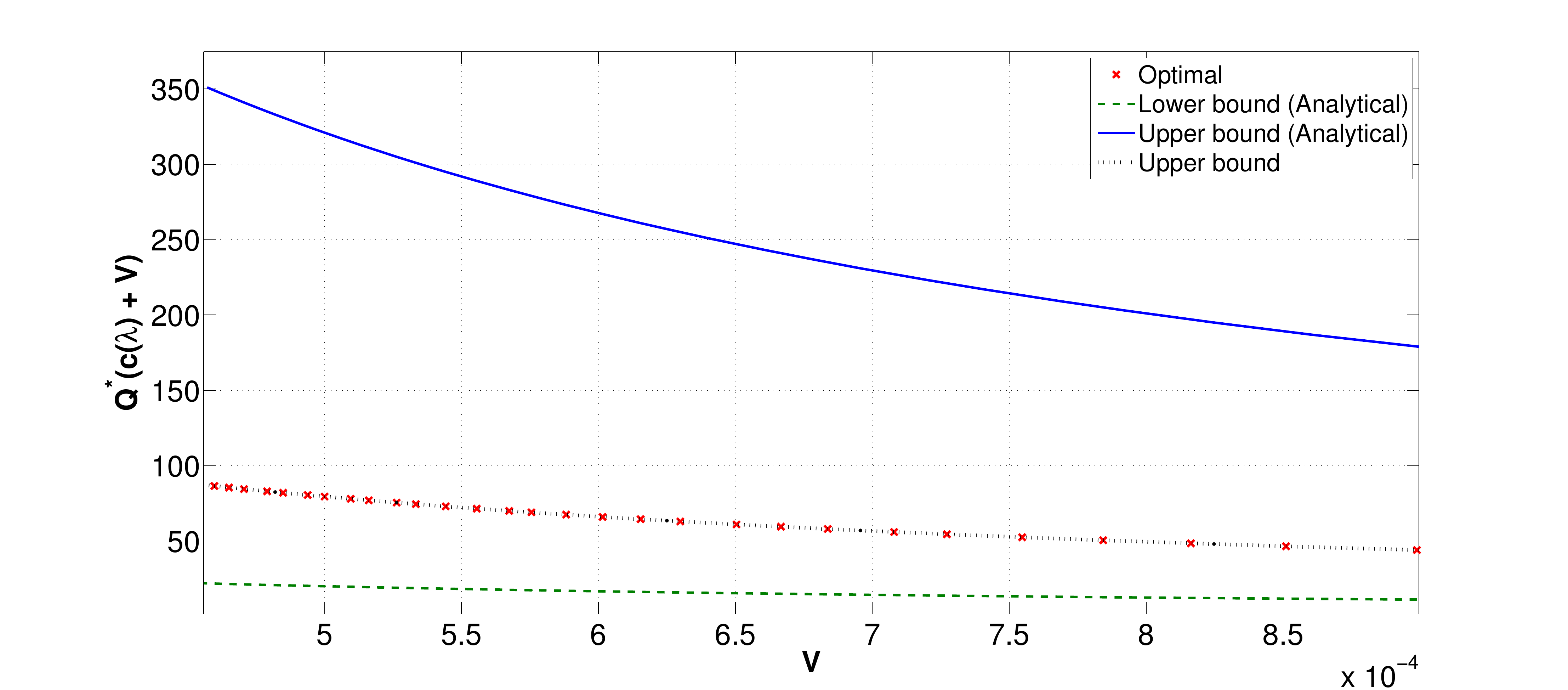}
	\caption{$Q^*(c_{c})$ as a function of $V$, where $c_{c} = c(\lambda) + V$, for $\lambda = 0.20$ and $c(\mu) = \mu^{2}$ for $\mu \in \brac{0, 0.2, 0.4, 0.5, 0.6, 0.8, 1}$.}
        \label{chap4:fig:212}
\end{figure}
\begin{figure}
	\centering
	\includegraphics[width=140mm,height=70mm]{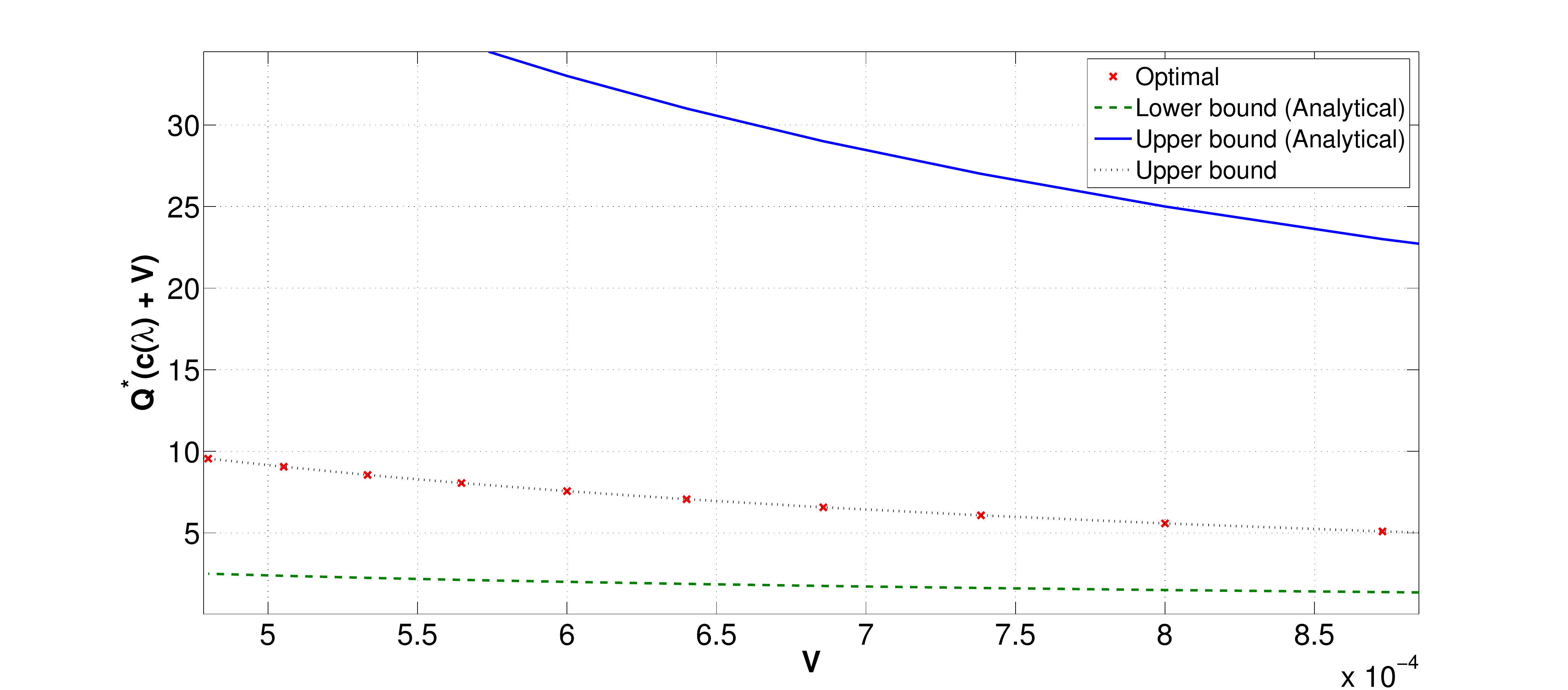}
	\caption{$Q^*(c_{c})$ as a function of $V$, where $c_{c} = c(\lambda) + V$, for $\lambda = 0.20$ and $c(\mu) = \mu^{5}$ for $\mu \in \brac{0, 0.2, 0.4, 0.5, 0.6, 0.8, 1}$.}
        \label{chap4:fig:222}
\end{figure}

\begin{figure}
	\centering
	\includegraphics[width=140mm,height=70mm]{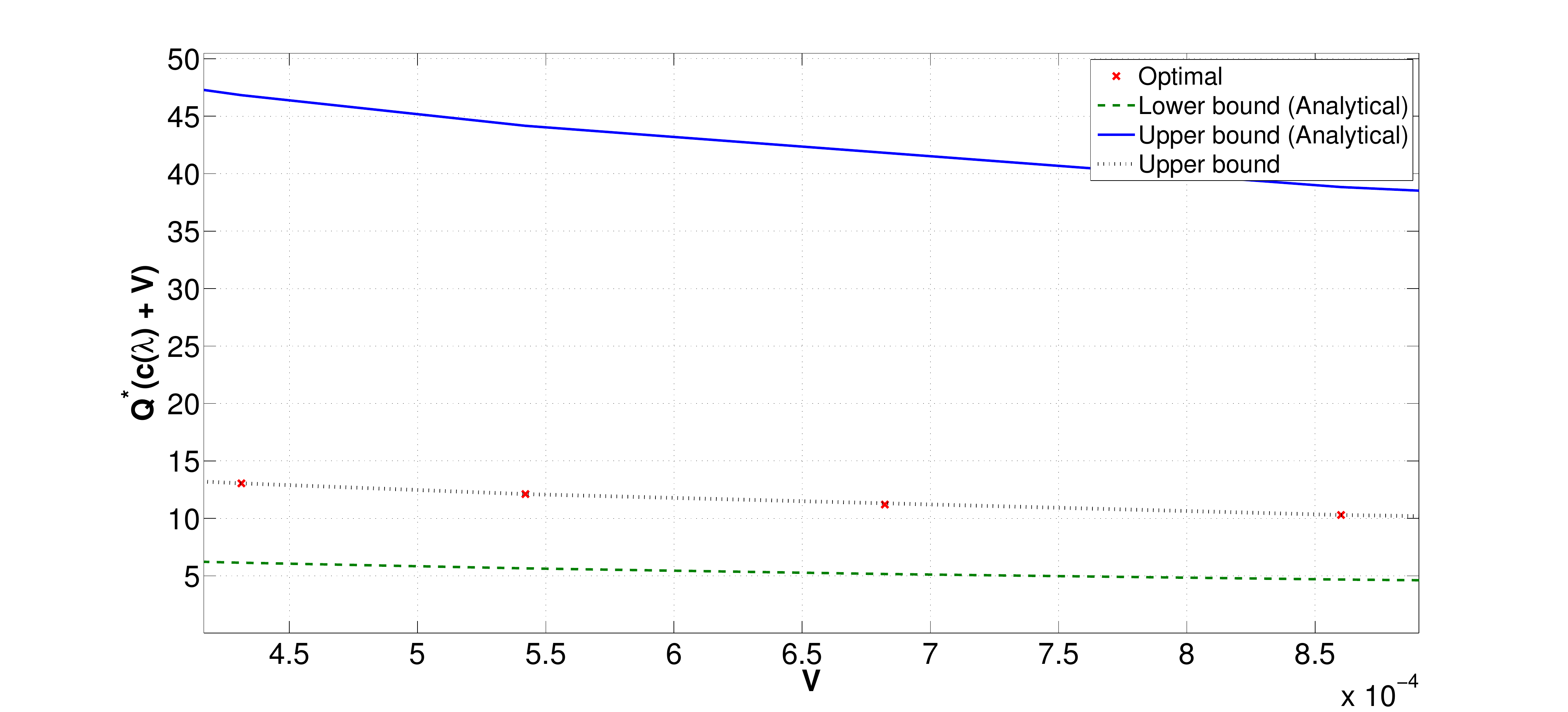}
	\caption{$Q^*(c_{c})$ as a function of $V$, where $c_{c} = c(\lambda) + V$, for $\lambda = 0.25$ and $c(\mu) = \mu^{2}$ for $\mu \in \brac{0, 0.2, 0.4, 0.5, 0.6, 0.8, 1}$.}
        \label{chap4:fig:213}
\end{figure}
\begin{figure}
	\centering
	\includegraphics[width=140mm,height=70mm]{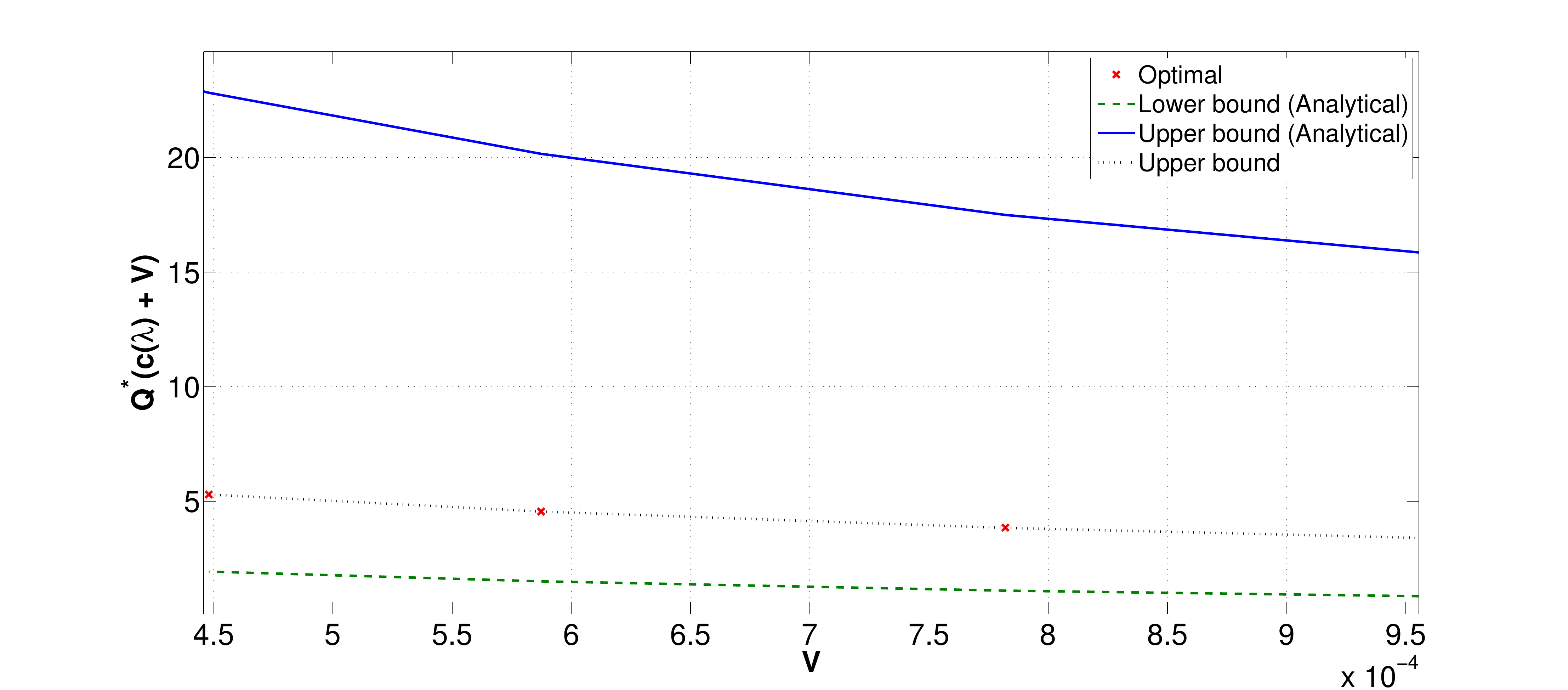}
	\caption{$Q^*(c_{c})$ as a function of $V$, where $c_{c} = c(\lambda) + V$, for $\lambda = 0.25$ and $c(\mu) = \mu^{5}$ for $\mu \in \brac{0, 0.2, 0.4, 0.5, 0.6, 0.8, 1}$.}
        \label{chap4:fig:223}
\end{figure}

\begin{figure}
	\centering
	\includegraphics[width=140mm,height=70mm]{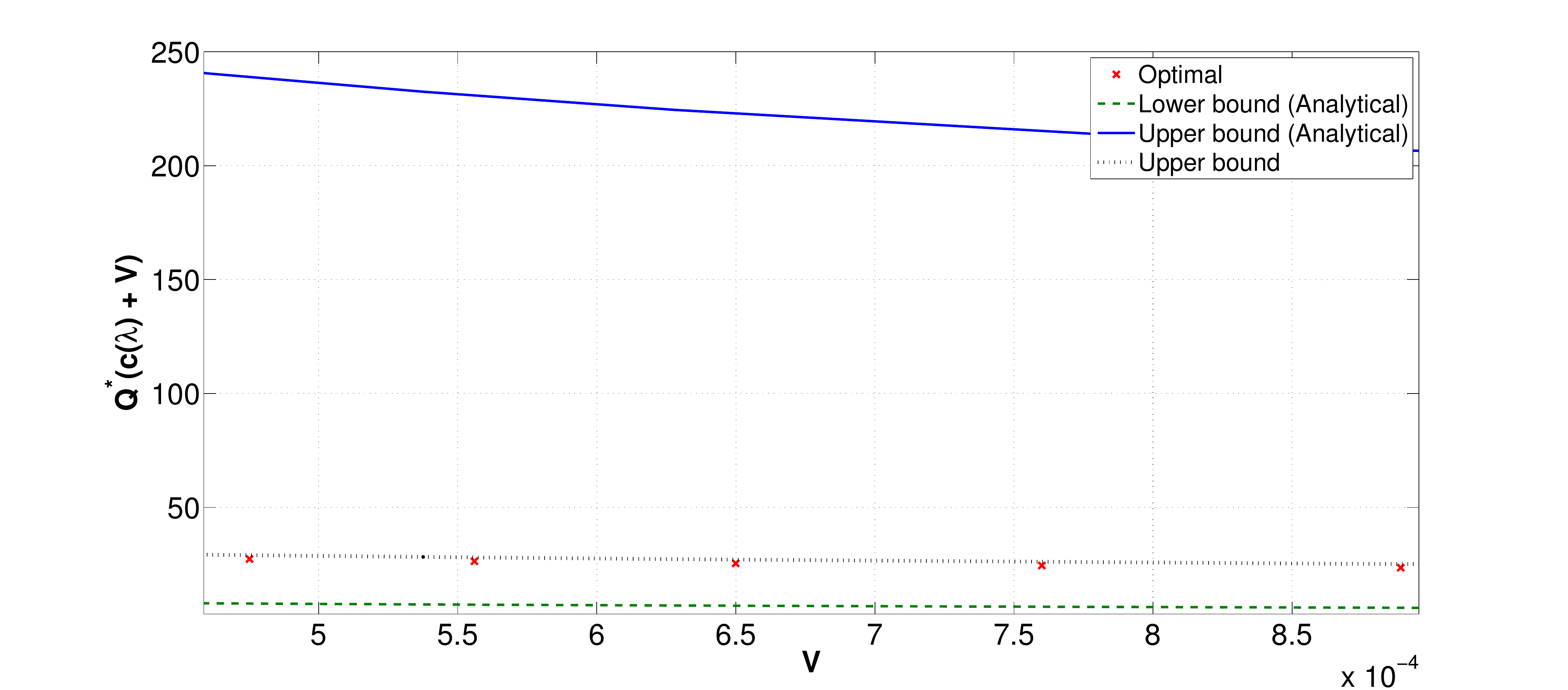}
	\caption{$Q^*(c_{c})$ as a function of $V$, where $c_{c} = c(\lambda) + V$, for $\lambda = 0.70$ and $c(\mu) = \mu^{2}$ for $\mu \in \brac{0, 0.2, 0.4, 0.5, 0.6, 0.8, 1}$.}
        \label{chap4:fig:214}
\end{figure}
\begin{figure}
	\centering
	\includegraphics[width=140mm,height=70mm]{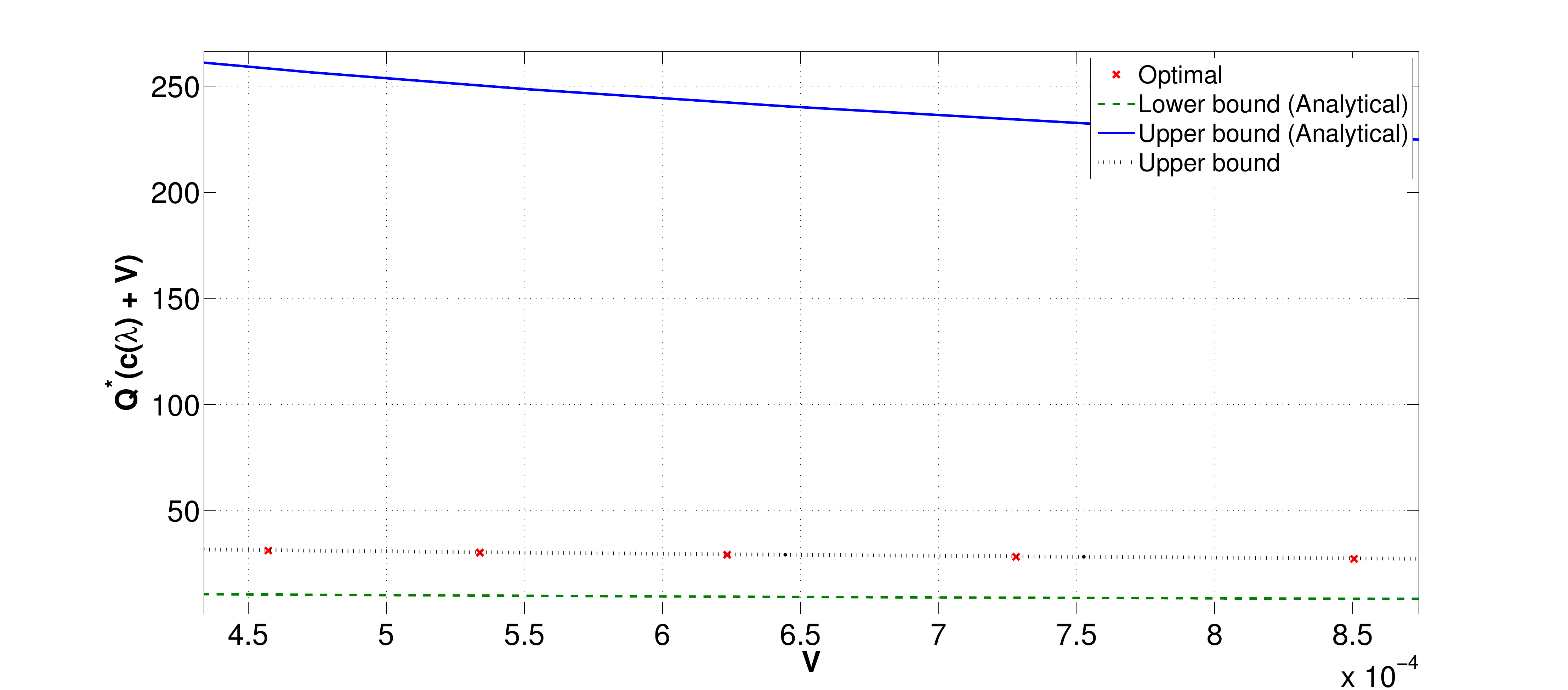}
	\caption{$Q^*(c_{c})$ as a function of $V$, where $c_{c} = c(\lambda) + V$, for $\lambda = 0.70$ and $c(\mu) = \mu^{5}$ for $\mu \in \brac{0, 0.2, 0.4, 0.5, 0.6, 0.8, 1}$.}
        \label{chap4:fig:224}
\end{figure}
We note that in all the cases that we have considered, both the analytical upper and lower bounds are very loose.

\begin{figure}
  \includegraphics[width=160mm,height=60mm]{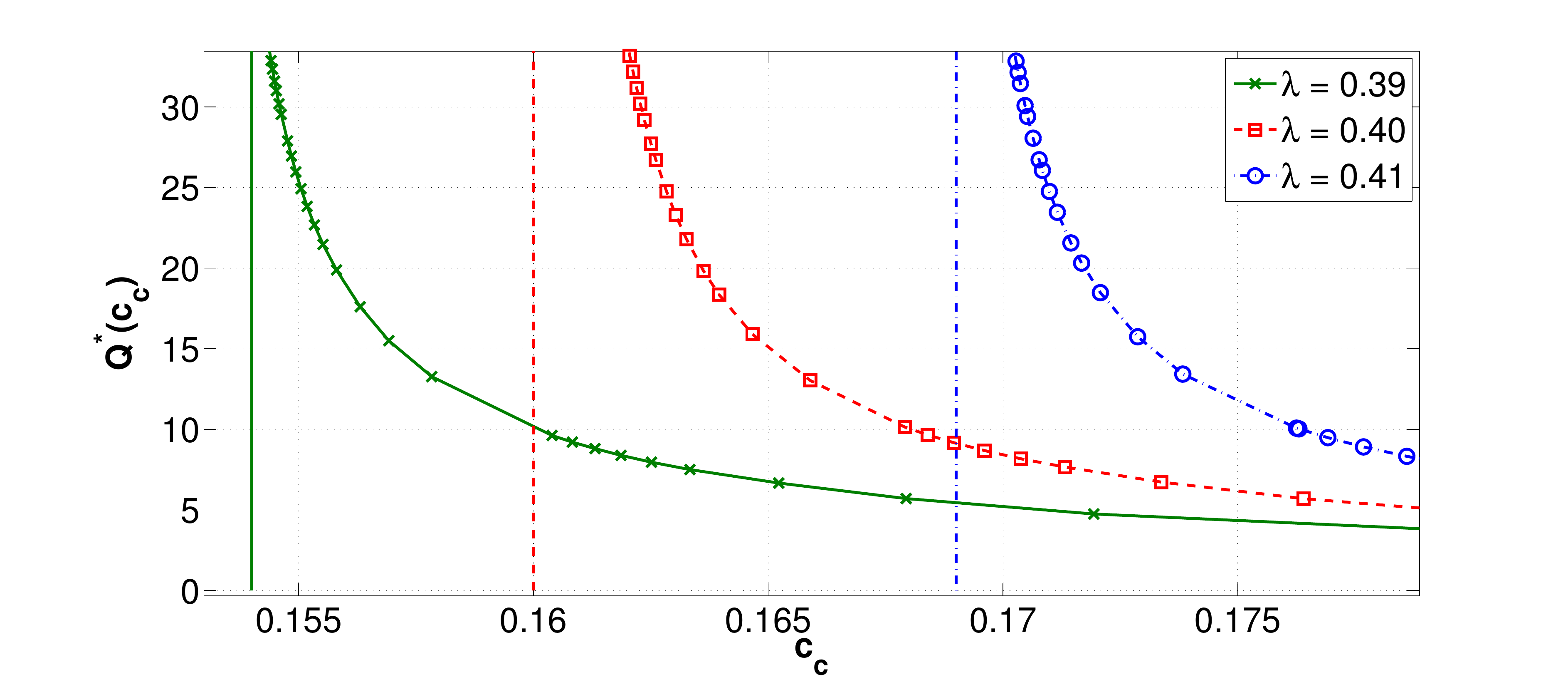}
  \caption{The optimal tradeoff curve for the system with $\mathcal{X}_{\mu} = \brac{0, 0.2, 0.4, 0.5, 0.6, 0.8, 1}$, $c(\mu) = \mu^{2}, \forall \mu \in \mathcal{S}$, and $\lambda = 0.39, 0.40$, and $0.41$. The minimum average service cost rates are $c(0.39) = 0.154, c(0.40) = 0.160$, and $c(0.41) = 0.169$}.
  \label{chap4:fig:comparelambda}
  \vspace{-0.1in}
\end{figure}

The difference in the asymptotic behaviour of $Q^*(c_{c})$ for FINITE-$\mu$CHOICE-2 and FINITE-$\mu$CHOICE-3 is illustrated by the following example.
Consider the following example : we choose the set of service rates $\mathcal{X}_{\mu} = \brac{0, 0.2, 0.4, 0.5, 0.6, 0.8, 1}$, and $c(\mu) = \mu^{2}, \forall \mu \in \mathcal{S}$. In Figure \ref{chap4:fig:comparelambda} we plot the tradeoff curve, numerically obtained from a suitably truncated MDP, for $\lambda = 0.39, 0.40$ and $0.41$.
The minimum average service cost rates corresponding to $\lambda = 0.39$, $0.40$ and $0.41$ are $0.154, 0.160$, and $0.169$.
We note that the difference between the average service cost and $c(\lambda)$ increases when $\lambda$ is changed from $0.39$ to $0.40$ and then decreases when $\lambda$ is increased, since at $\lambda = 0.40$, the average queue length increases at the rate $\frac{1}{V}$.

\begin{figure}
  \includegraphics[width=160mm,height=60mm]{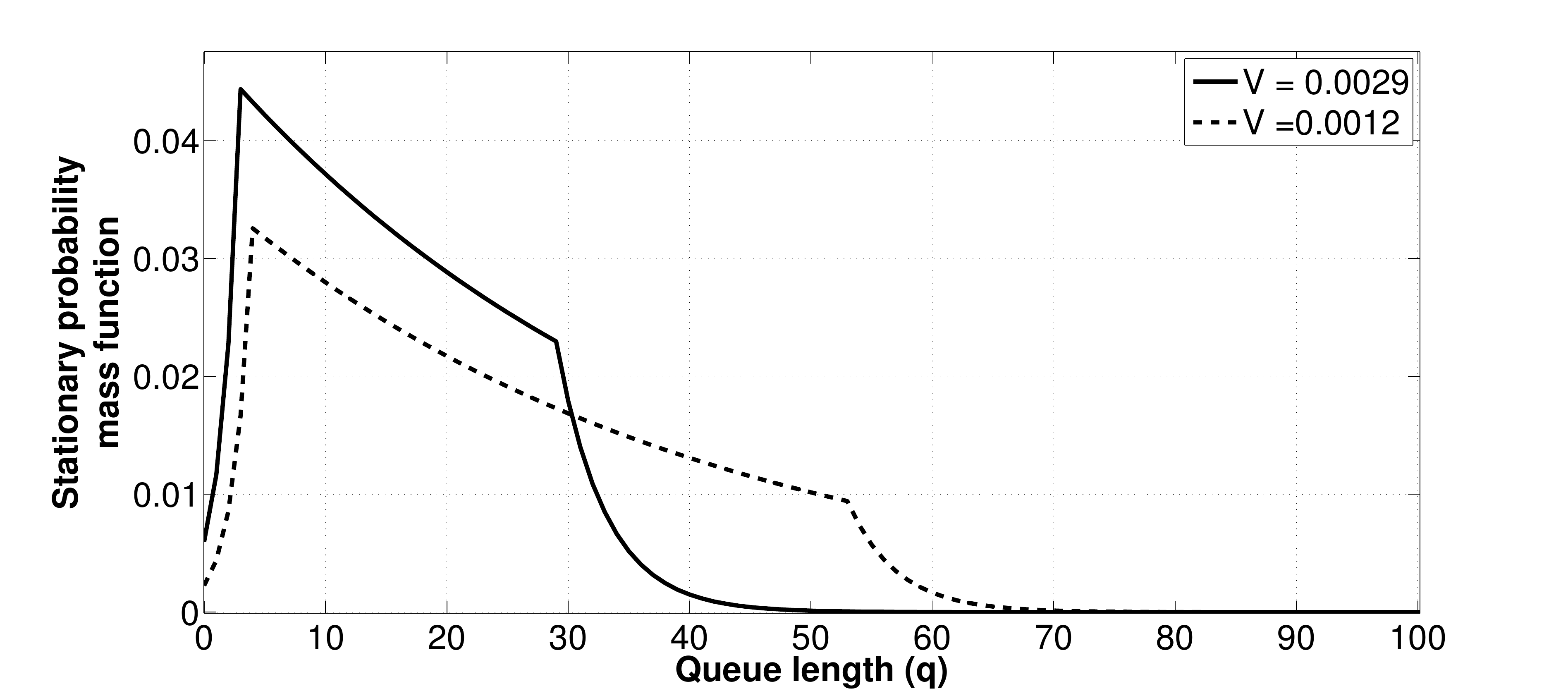}
  \caption{Stationary probability mass functions for optimal policies for the system with $\mathcal{X}_{\mu} = \brac{0, 0.2, 0.4, 0.5, 0.6, 0.8, 1}$, $c(\mu) = \mu^{2}, \forall \mu \in \mathcal{S}$, and $\lambda = 0.39$ (corresponding to FINITE-$\mu$CHOICE-2). The $\Omega\brap{\log\nfrac{1}{V}}$ asymptotic lower bound arises due to the geometrically increasing and decreasing nature of $\pi(q)$ as $V \downarrow 0$.}
  \label{chapter2:fig:pmfcase2}
  \vspace{-0.1in}
\end{figure}

\begin{figure}
  \includegraphics[width=160mm,height=60mm]{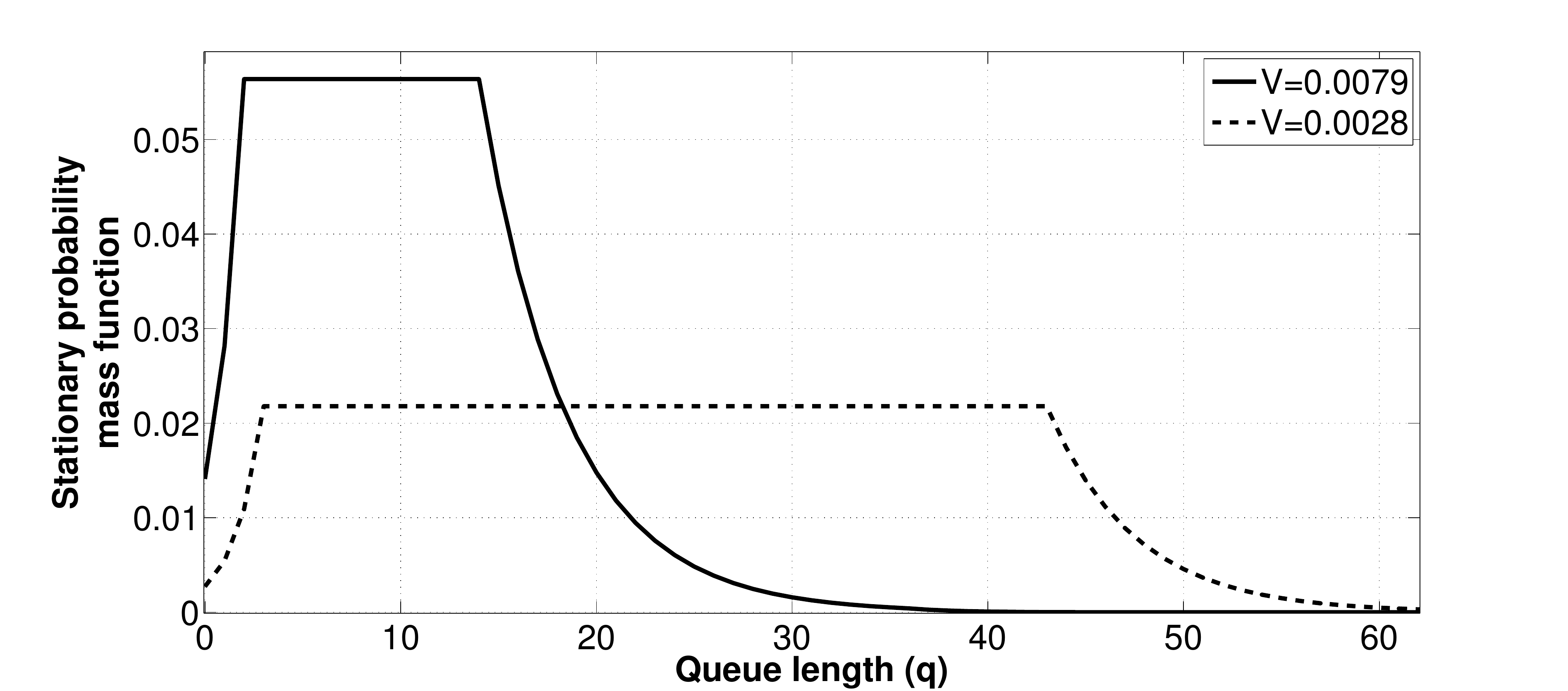}
  \caption{Stationary probability mass functions for optimal policies for the system with $\mathcal{X}_{\mu} = \brac{0, 0.2, 0.4, 0.5, 0.6, 0.8, 1}$, $c(\mu) = \mu^{2}, \forall \mu \in \mathcal{S}$, and $\lambda = 0.40$ (corresponding to FINITE-$\mu$CHOICE-3). The $\Omega\nfrac{1}{V}$ asymptotic lower bound arises due to the constant nature of $\pi(q)$ as $V \downarrow 0$.}
  \label{chapter2:fig:pmfcase3}
  \vspace{-0.2in}
\end{figure}

In Figures \ref{chapter2:fig:pmfcase2} and \ref{chapter2:fig:pmfcase3} we illustrate the stationary probability mass functions $\pi(q)$ for optimal policies for FINITE-$\mu$CHOICE-2 and FINITE-$\mu$CHOICE-3 respectively.

\subsection{An Application}
\label{sec:application}
\begin{figure}
  \hspace{-0.5in}
  \includegraphics[width=180mm,height=60mm]{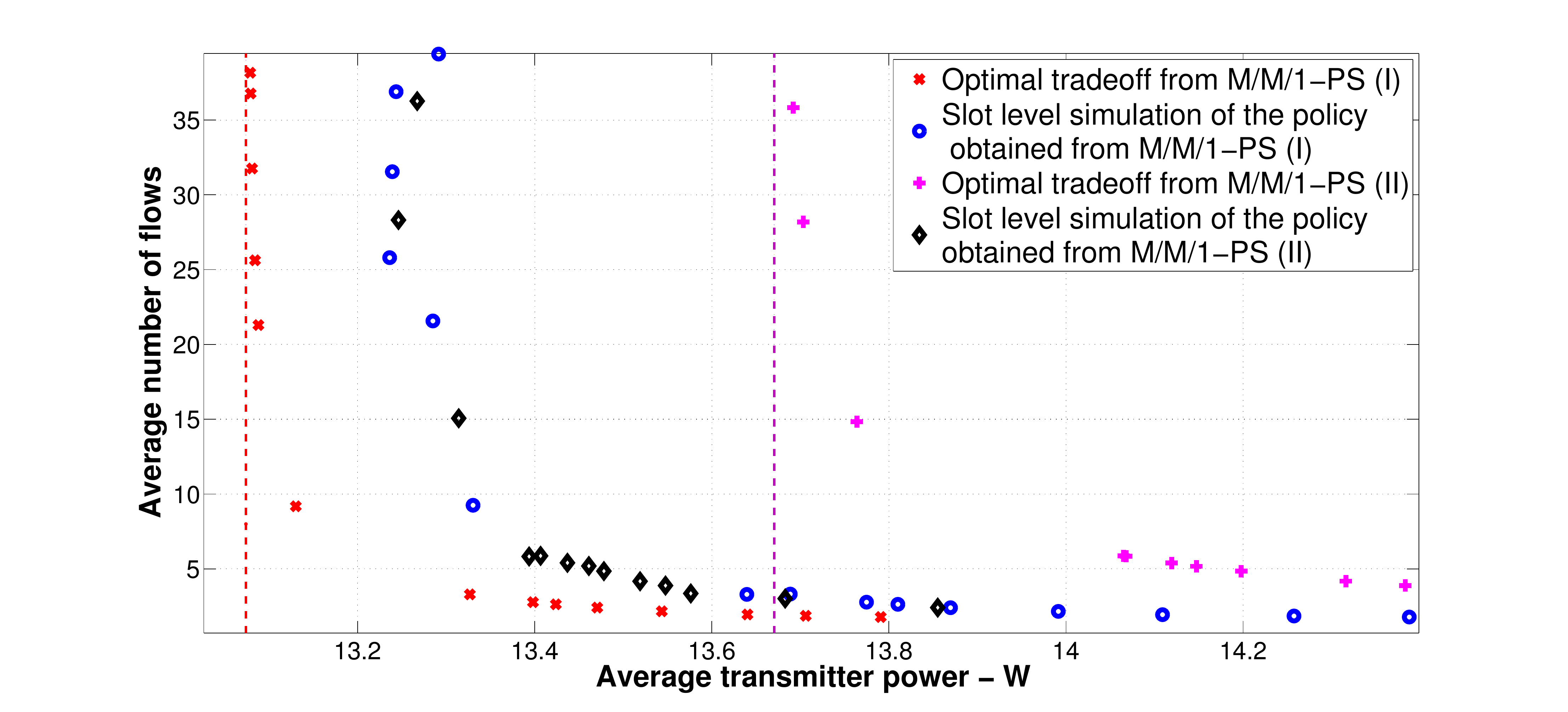}
  \caption{Tradeoff of average number of flows in the downlink scheduler queue with the average transmitter power for the motivating problem considered in Section \ref{chap4:sec:motivatingeg}, for a flow arrival rate $0.2$ flows/sec}
  \label{chap4:fig:tradeoffprac}
\end{figure}

In this section we discuss the application of the above asymptotic results to the example considered in Section \ref{chap4:sec:motivatingeg}.
We obtain a sequence of optimal policies for both the M/M/1-PS (I) and the M/M/1-PS (II) models which trade off the average transmitter power with the average number of flows.
Both of these sequences are obtained by the numerical evaluation of the optimal policy for a suitably truncated MDP, with single stage cost $q + \beta P(m)$ and with state transitions as shown in Figure \ref{chap4:fig:mm1model}, for a sequence of $\beta > 0$.
The optimal tradeoff for both M/M/1-PS (I) and (II) models are shown in Figure \ref{chap4:fig:tradeoffprac}.
We then obtain via simulation, the average power and average number of flows for the discrete time system, for the policies obtained from the M/M/1-PS (I) and M/M/1-PS (II) models, to obtain a possible (sub-optimal) tradeoff curve as shown in Figure \ref{chap4:fig:tradeoffprac}.
We note that the sequence of policies suggested by both (I) and (II) models have similar performance.
For this example, the tradeoff curve obtained from the simulation of the policies suggested by the analysis, demonstrates that with a $1W$ increase in transmitter power, a $36$ fold decrease in latency, i.e., from 3 mins to 5 secs, is possible.
We note that $P^{1}(m) \leq P^{2}(m)$ for $m \in \brac{1.25, 1.50, 1.75, 2}$, i.e., for $\mu \geq \mu_{l}$ given that $\lambda = 0.2$ (we recall that actual service rates are $\frac{m}{6.4}$).
Then for each of the optimal policies for the M/M/1-PS (I) (or (II)) model, for which the average number of flows is large, $P^{1}(m(Q)) \leq P(m(Q)) \leq P^{2}(m(Q))$ with high probability.
Thus, the tradeoff curves obtained from the M/M/1-PS (I) and (II) models are approximate upper and lower bounds to tradeoff curve for the discrete time system, especially when the average number of flows are large.
Then the asymptotic analysis in the above sections, leads to asymptotic upper and lower bounds on the average transmitter power as a function of the average number of flows for the discrete time system.
Let $c^{1}(m)$ and $c^{2}(m)$ be the lower convex envelopes of $P^{1}(m)$ and $P^{2}(m)$ respectively.
Since $\mu_{l} = 1.25/6.4 < \lambda = 0.2 < \mu_{u} = 1.50/6.4$, by considering the sequence of optimal policies from the M/M/1-PS (II) model, the average transmitter power for the discrete time system can bounded below by $c^{1}( 6.4 \lambda ) + \Omega\brap{e^{-\overline{q}}}$ and bounded above by $c^{2}( 6.4 \lambda ) + \mathcal{O}\brap{e^{-\overline{q}}}$, for large $\overline{q}$, where $\overline{q}$ is the average number of flows.

\section{Conclusions}
\label{chap4:sec:conclusion}
The main purpose of this chapter is the illustration of the techniques involved in: (i) the derivation of asymptotic bounds on $Q^*(c_{c})$ and $Q^*_{M}(c_{c})$ for admissible policies and (ii) asymptotic bounds on order-optimal admissible policies in the regime $\Re$, using a simple queueing model.

For FINITE-$\mu$CHOICE, we observe that the constraint on the average service cost leads to a restriction on the stationary probability of service rates which in turn restricts the behaviour of the stationary probability of the queue length in the asymptotic regime $\Re$, where $V = c_{c} - c(\lambda) \downarrow 0$.
For FINITE-$\mu$CHOICE-2 and FINITE-$\mu$CHOICE-3, we note that the stationary probability of service rates less than $\mu_{l}$ or greater than $\mu_{u}$ goes to zero as $V \downarrow 0$.
Since, state $0$ uses a service rate of $0$, the above fact implies that $\pi(0) \downarrow 0$ as $V \downarrow 0$.
Hence, it is intuitive that the average queue length has to increase.
For FINITE-$\mu$CHOICE-3, as $\mu_{l} = \mu_{u} = \lambda$, as $V \downarrow 0$, the stationary probability that any service rate other than $\lambda$ is used approaches zero.
Then we observe that the stationary probability $\pi(q)$ becomes equal and is $\mathcal{O}(V)$ for all $q$ which occur with high probability, as $V \downarrow 0$, since $\pi(q)\lambda = \pi(q + 1)\mu(q), q \geq 0$ and $\mu(q) = \lambda$ for all $q$ occurring with high probability.
The above constant nature of $\pi(q)$ leads to the $\Omega\brap{\frac{1}{V}}$ asymptotic lower bound.
For FINITE-$\mu$CHOICE-2, as $V \downarrow 0$, service rates $\mu_{l} < \lambda$ and $\mu_{u} > \lambda$ could be used.
Therefore, intuitively, one expects that the rate at which $Q^*(c_{c})$ increases is less for \FMC-2 compared to \FMC-3.
Furthermore, since $\pi(q)\lambda = \pi(q + 1)\mu(q), q \geq 0$, we observe that the stationary probability $\pi(q)$ has a geometric growth and decay with growth and decay rates at most $\frac{\lambda}{\mu_{l}}$ and at least $\frac{\lambda}{\mu_{u}}$ respectively, for the set of queue lengths occurring with high probability, as $V \downarrow 0$.
The above geometric growth and decay of $\pi(q)$ leads to the $\Omega\brap{\log\nfrac{1}{V}}$ asymptotic lower bound in Lemma \ref{chap4:prop:p1213lb}.
The above behaviour of the stationary probability in the different cases motivates us to analyse the tradeoff problem for the discrete time queue in Chapter 4, by constructing bounds on the stationary probability of the queue length, which have the same behaviour in the respective cases.

We note that the sequence of policies which achieve the asymptotic order behaviour in Section \ref{chap4:sec:asympbehaviour} for FINITE-$\mu$CHOICE-2 and FINITE-$\mu$CHOICE-3 are similar to buffer partitioning policies.
We note that the buffer partitions that were used for these sequences of policies scaled as $\Omega\brap{\log\nfrac{1}{V}}$ and $\Omega\nfrac{1}{V}$ for FINITE-$\mu$CHOICE-2 and FINITE-$\mu$CHOICE-3 respectively.
This scaling of the buffer partitions was suggested by the asymptotic lower bounds that were derived in Section \ref{chap4:sec:p1_asymp_lb}, and we shall see that similar ideas can be used in the design of buffer partitioning policies for discrete time systems.
Furthermore, we have also derived asymptotic bounds on any sequence of order-optimal policies in Section \ref{chap4:sec:optimalpolicy_aschar}.

We note that FINITE-$\mu$CHOICE-1, where the average queue length increases only to a finite value, even when the average service cost rate is the minimum possible $c(\lambda)$, has been hitherto unidentified in the literature.
To the best of our knowledge, the asymptotic characterizations of the optimal tradeoff curve obtained in this chapter, for all the three cases, were previously not known for the state dependent M/M/1 model.

The development of the asymptotic results in this chapter, partly motivates the definition of admissible policies for the discrete time queueing model in Chapter 4.
The non-idling nature of the optimal policy which has been obtained in Lemma \ref{chap4:prop:optpolicy_nonidling}, motivates us to consider whether the optimal policy for the discrete time queueing model has the same property also.
In Chapter 4, we shall show that in fact it does.
The non-idling property of the optimal policy and its relation with the \emph{place-holder bit} scheduling policies \cite{neely} are discussed in more detail in Chapter 4.
We have also illustrated the utility of the simple state dependent M/M/1 queueing model in the analytical study of scheduling schemes for next generation wireless systems using the example in Section \ref{chap4:sec:motivatingeg}.

\clearpage
\begin{subappendices}
\large{\textbf{Appendix}}
\addcontentsline{toc}{section}{Appendix}
\addtocontents{toc}{\protect\setcounter{tocdepth}{0}}  
\normalsize
\section{Uniformization and a bound on the average queue length}
\label{chap4:app:uniformization}
Let $r_{u} = r_{max} + r_{a,max}$.
Consider a discrete time Markov chain $(Q_{d}[m])$ which is obtained by uniformization of $(Q(t))$ at rate $r_{u}$.
The transition probabilities, $p_{q_{1},q_{2}} = P(Q_{d}[m + 1] = q_{2}|Q_{d}[m] = q_{1}), m \in \mathbb{Z}_+$, of the DTMC are as follows :
\begin{eqnarray*}
  p_{0,0} = 1 - \frac{\lambda(0)}{r_{u}}, \\
  p_{0,1} = \frac{\lambda(0)}{r_{u}}, \\
  p_{q,q + 1} = \frac{\lambda(q)}{r_{u}}, \forall q \geq 1, \\
  p_{q,q - 1} = \frac{\mu(q)}{r_{u}}, \\
  p_{q,q} = 1 - \frac{\lambda(q) + \mu(q)}{r_{u}}.
\end{eqnarray*}
We note that for an admissible policy $\gamma$, the stationary distribution $\pi$ is the same for both the CTMC $Q(t)$ and the DTMC $Q_{d}[m]$.
Thus $\mathbb{E}_{\pi} Q_{d} = \overline{Q}(\gamma)$, $\mathbb{E}_{\pi}u(\lambda(Q_{d})) = \overline{U}(\gamma)$, and $\mathbb{E}_{\pi}c(\mu(Q_{d})) = \overline{C}(\gamma)$.
The following proposition states an upper bound on $\overline{Q}(\gamma)$ for an admissible policy $\gamma$ subject to an assumption about the structure of the policy.

\begin{proposition}
  Assume that the admissible policy $\gamma$ is such that there exists a $q_{\epsilon}$ such that $\mu(q_{\epsilon}) - \lambda(q_{\epsilon}) \geq \epsilon$, for some $\epsilon > 0$. Then 
  \begin{equation}
    \overline{Q}(\gamma) \leq \frac{q_{\epsilon}(\epsilon + r_{a,max})}{\epsilon} + \frac{r_{u}}{2\epsilon}
  \end{equation}
  \label{chap4:app:prop:dotzupperbound}
\end{proposition}

\begin{proof}
  Let $L(q) = q^{2}$.
  We use $L(q)$ as a Lyapunov function to derive the above upper bound.
  The expected Lyapunov drift $\Delta(q)$ = 
  \begin{equation}
    \mathbb{E}\left[ L(Q_{d}[m + 1]) - L(Q_{d}[m]) | Q_{d}[m] = q \right]
  \end{equation}
  We have that
  \begin{eqnarray*}
    \Delta(q) = \frac{-2q}{r_{u}}(\mu(q) - \lambda(q)) + \frac{\mu(q) + \lambda(q)}{r_{u}}, \forall q.
  \end{eqnarray*}
  Note that from the admissibility of $\gamma$, $\forall q \geq q_{\epsilon}$, $\mu(q) - \lambda(q) \geq \epsilon$.
  So for $q \geq q_{\epsilon}$ we have that
  \begin{equation*}
    \Delta(q) \leq \frac{-2q\epsilon}{r_{u}} + \frac{\mu(q) + \lambda(q)}{r_{u}}.
  \end{equation*}
  For $q < q_{\epsilon}$,
  \begin{equation}
    \Delta(q) = \frac{-2q\epsilon}{r_{u}} + \frac{2q\epsilon}{r_{u}} - \frac{2q}{r_{u}}(\mu(q) - \lambda(q)) + \frac{\mu(q) + \lambda(q)}{r_{u}}.
  \end{equation}
  For $q < q_{\epsilon}$, $\mu(q) - \lambda(q) < \epsilon$.
  Therefore
  \begin{eqnarray*}
    \Delta(q) \leq \frac{-2q\epsilon}{r_{u}} + \frac{2q}{r_{u}}(\epsilon - \mu(q) + \lambda(q)) + \frac{\mu(q) + \lambda(q)}{r_{u}}, \\
    \leq \frac{-2q\epsilon}{r_{u}} + \frac{2q_{\epsilon}}{r_{u}}(\epsilon + r_{a,max}) + 1.
  \end{eqnarray*}
  For all $q$, we therefore have that
  \begin{eqnarray*}
    \Delta(q) \leq \frac{-2q\epsilon}{r_{u}} + \frac{2q_{\epsilon}}{r_{u}}(\epsilon + r_{a,max}) + 1.
  \end{eqnarray*}
  Hence from \cite[Theorem A.4.3]{meynctcn} we have that
  \begin{equation*}
    \overline{Q}(\gamma) = \mathbb{E}_{\pi} Q_{d} \leq \frac{q_{\epsilon}(\epsilon + r_{a,max})}{\epsilon} + \frac{r_{u}}{2\epsilon}.
  \end{equation*}
\end{proof}

\end{subappendices}
\addtocontents{toc}{\protect\setcounter{tocdepth}{2}}

\blankpage
\chapter[On the tradeoff of average queue length, average service cost, and average utility for the state dependent M/M/1 queue: Part II]{\textbf{On the tradeoff of average queue length, average service cost, and \\ average utility for the state dependent M/M/1 queue: Part II}}
\section{Introduction}
\label{chap4:sec:intintro}
We continue our analysis of the tradeoff problem for the state dependent M/M/1 model in this chapter.
We note that \FMC\, was primarily motivated by the wireless network problem in Section \ref{chap4:sec:motivatingeg}, for which the set of possible service rates was a finite discrete set.
In this chapter, we consider the INTERVAL-$\mu$CHOICE and INTERVAL-$\lambda\mu$CHOICE problems, which we study with the objective of understanding the tradeoff problem for the discrete time queueing model.
We note that in this chapter, $\mathcal{X}_{\mu}$ and $\mathcal{X}_{\lambda}$ are chosen to be finite intervals.

The method of analysis for \INTMC\, and \INTLMC\, is similar to that in Chapter 2.
We again obtain bounds on the stationary probability distribution of the queue length for admissible policies, leading to an asymptotic characterization of the average queue length as well as order-optimal admissible policies, in the asymptotic regime $\Re$.
We note that the analysis for \INTMC\, can be used to obtain the results for \FMC.
However, in Chapter 4, we will see that some of the steps used in the analysis for \INTMC\, and \INTLMC, which are different from that for \FMC, are essential in the analysis of the discrete time models.
We recall that the consideration of the stationary probability distribution of the queue length in the asymptotic regime $\Re$, for admissible policies (which are monotone), as a method for understanding the asymptotic behaviour of the average queue length, underlies most of the results obtained in this thesis.
We note that in this chapter, since we consider queueing models with arrival rate control as well as other forms of $c(\mu)$ (other than the piecewise linear form in Chapter 2), we are able to obtain new insights in this direction.

The question arises as to how insights about the asymptotic behaviour of the minimum average queue length in the regime $\Re$ for the discrete time model can be obtained from the state dependent M/M/1 model.
It is clear that all the model features for the discrete time model, even with a single environment state, cannot be captured by the state dependent M/M/1 model.
The model features for the state dependent M/M/1 model are the sets $\mathcal{X}_{\mu}$ and $\mathcal{X}_{\lambda}$, and the functions $c(\mu)$ and $u(\lambda)$, with the restriction that $\mu(q)$ and $\lambda(q)$ are deterministic functions of the queue length.
For the discrete time model with a single environment state, we note that in addition to the sets of possible service batch sizes and possible admitted arrival batch sizes, and the cost and utility functions, we also have that the service batch size and the amount of arrivals admitted in a slot are randomized functions of the history of the process, as discussed in Chapter 1.

Suppose we consider only the set of stationary policies for the discrete time model, which choose the service batch size $S$ (or the amount $A$ of arrivals admitted) as a randomized function of the current queue length $q$ only, say with probability distribution $P_{s|q}$ (or distributed as $P_{a|q,r}$ which is a function of the current queue length and number of actual arrivals $r$).
Even then, we have to reduce $P_{s|q}$ (or $P_{a|q,r}$) to a real value, which can then be modelled by $\mu(q)$ (or $\lambda(q)$).
In the following, we take $\Exp_{P_{s|q}}S$ as the quantity which represents $P_{s|q}$.
Thus $\mu(q)$ is assumed to correspond to $\Exp_{P_{s|q}} S$ and therefore $\mathcal{X}_{\mu}$ is the set of all values that $\Exp_{P_{s|q}}S$ can take.
A similar assumption is made for $\lambda(q)$.

We note that for the discrete time model, the service cost incurred in a slot is a random variable, since the batch size $S$ itself is random.
At a queue length $q$, the expected service cost is $\Exp_{P_{s|q}} c(S)$.
Since we have already chosen $\mu(q)$ to correspond to $\Exp_{P_{s|q}} S$, a possibility is to choose $c(\mu)$ to correspond to $c(\Exp_{P_{s|q}} S)$.
A similar assumption is made for $u(\lambda)$.
In retrospect, this turns out to be a good choice of model features for the state dependent M/M/1 model (e.g., we obtain the asymptotic Berry Gallager lower bound for the M/M/1 model which, with the above choice of model features, corresponds to the discrete time model in \cite{berry} with admissible policies).
We can also then surmise that one of the reasons for the asymptotic behaviour of the stationary queue length for admissible policies in the regime $\Re$ for the discrete time model are the behaviours of asymptotic probability distributions of $\Exp_{P_{s|q}}S(Q)$ and $\Exp_{P_{a|q,r}}A(Q,R)$, in the regime $\Re$, since for the M/M/1 model the behaviours of these quantities are significant.

\subsection{Overview}
We present the analysis of INTERVAL-$\mu$CHOICE in Section \ref{chap4:sec:problemp2}.
The motivating discrete time problem for INTERVAL-$\mu$CHOICE is discussed in the same section.
We recall that in defining \INTMC\, we considered admissible policies for which $\lambda(q)$ is a constant $\lambda$ for all $q \in \sZ$.
We obtain asymptotic lower and upper bounds on the tradeoff problem in Sections \ref{chap4:sec:intmc_lb} and \ref{chap4:sec:intmc_ub} respectively.
Asymptotic bounds on order-optimal policies for \INTMC, are presented in Section \ref{chap4:section:aschar_optpolicy_intmc}.
We consider the counterpart \INTLC\, of \INTMC\,, for which $\mu(q)$ is constant over all $q \in \sZ$ and $\lambda(q)$ is the control variable, in Section \ref{chap4:sec:similarintmc}.
The analysis of INTERVAL-$\lambda\mu$CHOICE in presented in Section \ref{chap4:sec:problemp3}, along with the motivating discrete time problem.
We present the main conclusions in Section \ref{chap4:sec:conclusions}, where we discuss the main ideas obtained in the asymptotic analysis of INTERVAL-$\mu$CHOICE, \INTLC, and INTERVAL-$\lambda\mu$CHOICE, and how these ideas can be applied to the discrete time model in Chapters 4, 5, and 6.

\section{Analysis of INTERVAL-$\mu$CHOICE}
\label{chap4:sec:problemp2}
We recall that for INTERVAL-$\mu$CHOICE we restrict to admissible policies $\gamma$ such that $\lambda(q) = \lambda$ and $\mu(q) \in [0,r_{max}]$, $\forall q \in \mathbb{Z}_{+}$.
The tradeoff problem for INTERVAL-$\mu$CHOICE is
\begin{eqnarray*}
  \text{ minimize }_{\gamma \in \Gamma_{a}} & \overline{Q}(\gamma) \nonumber \\
  \text{ such that } & \overline{C}(\gamma) \leq c_{c}, \nonumber
\end{eqnarray*}
whose optimal value is denoted as $Q^*(c_{c})$.
We also note that a tradeoff problem can be defined where we minimize over the set of policies, which includes finite mixtures of pure policies in $\Gamma_{a}$.
The optimal value of this problem is denoted as $Q^*_{M}(c_{c})$.
Then, as for the case of \FMC\, (e.g. as in Corollary \ref{chap4:prop:p11lb_mixed}), asymptotic lower bounds for $Q^*(c_{c})$ can be used to obtain asymptotic lower bounds for $Q^*_{M}(c_{c})$.
In fact, the asymptotic lower bounds in the regime $\Re$ are the same for $Q^*(c_{c})$ and $Q^*_{M}(c_{c})$.
Furthermore, the asymptotic upper bounds that we derive for $Q^*(c_{c})$ are by definition upper bounds for $Q^*_{M}(c_{c})$.
Hence, in the following, we present the results for $Q^*(c_{c})$ only.

The study of INTERVAL-$\mu$CHOICE is classified into:
\begin{description}
\item[INTERVAL-$\mu$CHOICE-1 :] $c(\mu)$ is strictly convex for $\mu \in [0,r_{max}]$, and
\item[INTERVAL-$\mu$CHOICE-2 :] $c(\mu)$ is piecewise linear. 
  That is, (a) there exists a minimal partition of $[0,r_{max}]$ into intervals $\{[a_{i},b_{i}], i \in \{1,\dots,P\}\}$ with $a_{1} = 0$, $b_{P} = r_{max}$, and $b_{i} = a_{i + 1}$, and (b) there are linear functions $f_{i}$ such that $\forall \mu \in [a_{i},b_{i}], f_{i}(\mu) = c(\mu)$.
\end{description}

\begin{remark}
  We first discuss the motivation for INTERVAL-$\mu$CHOICE-1.
  INTERVAL-$\mu$CHOICE-1 corresponds to the tradeoff problem for the following discrete time queueing model.
  Work arrives in a batch, of random size, in every slot, into an infinite buffer queue.
  The state of the queue is the amount of unfinished work.
  We note that for the discrete time model, the amount of unfinished work or the queue state evolves on the set of non-negative real numbers.
  This is approximated by an integer-valued queue evolution process in \INTMC-1.
  The amount of work completed in each slot, or the service batch size, can be chosen as a function (possibly randomized), of the current backlog of unfinished work.
  The choice of the amount of work completed in each slot as a function of the current backlog for the discrete time queue, is modelled by the control of the service rate, $\mu(q)$, in INTERVAL-$\mu$CHOICE-1.
  Hence, the service batch size and the average service batch size (or rate) also takes values in the set of non-negative real numbers.
  As discussed in Section \ref{chap4:sec:intintro} we then assume that $\mu(q)$ takes values in an interval, which for technical reasons is assumed to be finite.
  We assume that there is no admission control in the discrete time model, therefore we assume that the arrival rate is $\lambda$ for every $q$ for INTERVAL-$\mu$CHOICE-1.
  For the discrete time queue, we assume that there is a service cost associated with the amount of work done in each slot.
  This is modelled by the service cost rate function $c(.)$ in INTERVAL-$\mu$CHOICE-1.
  We note that since the amount of work done in a slot can be any real value, the service cost for the discrete time model could be a strictly convex function defined on an interval, which provides the motivation for assuming $c(\mu)$ to be strictly convex for INTERVAL-$\mu$CHOICE-1.
  We note that this discrete time model is similar to the model considered by Berry and Gallager \cite{berry}, but with a single fade state.

  The motivating discrete time queueing model for INTERVAL-$\mu$CHOICE-2 is very similar to the discrete time model discussed above, except that the queue evolution is assumed to be on integers.
  But for stationary randomized policies, the average service batch size could still take any real value in a finite interval, and in light of the discussion in Section \ref{chap4:sec:intintro}, $\mu(q)$ is again assumed to take any value from a finite interval.
  In Chapter 4, we shall see that then $c(\Exp_{P_{s|q}} S)$ is piecewise linear, which is the motivation for the piecewise linear assumption on $c(\mu)$ for INTERVAL-$\mu$CHOICE-2.
\end{remark}
For any admissible policy $\gamma$, from Jensen's inequality, we have that $\overline{C}(\gamma) \geq c(\lambda)$.
We study INTERVAL-$\mu$CHOICE-1 and INTERVAL-$\mu$CHOICE-2 in the asymptotic regime $\Re$ where the service cost constraint $c_{c}$ approaches $c(\lambda)$, since it can be shown that $c(\lambda) = \inf_{\gamma \in \Gamma_{a}} \Cg$.

\begin{figure}[h]
  \centering
  \includegraphics[width=160mm,height=80mm]{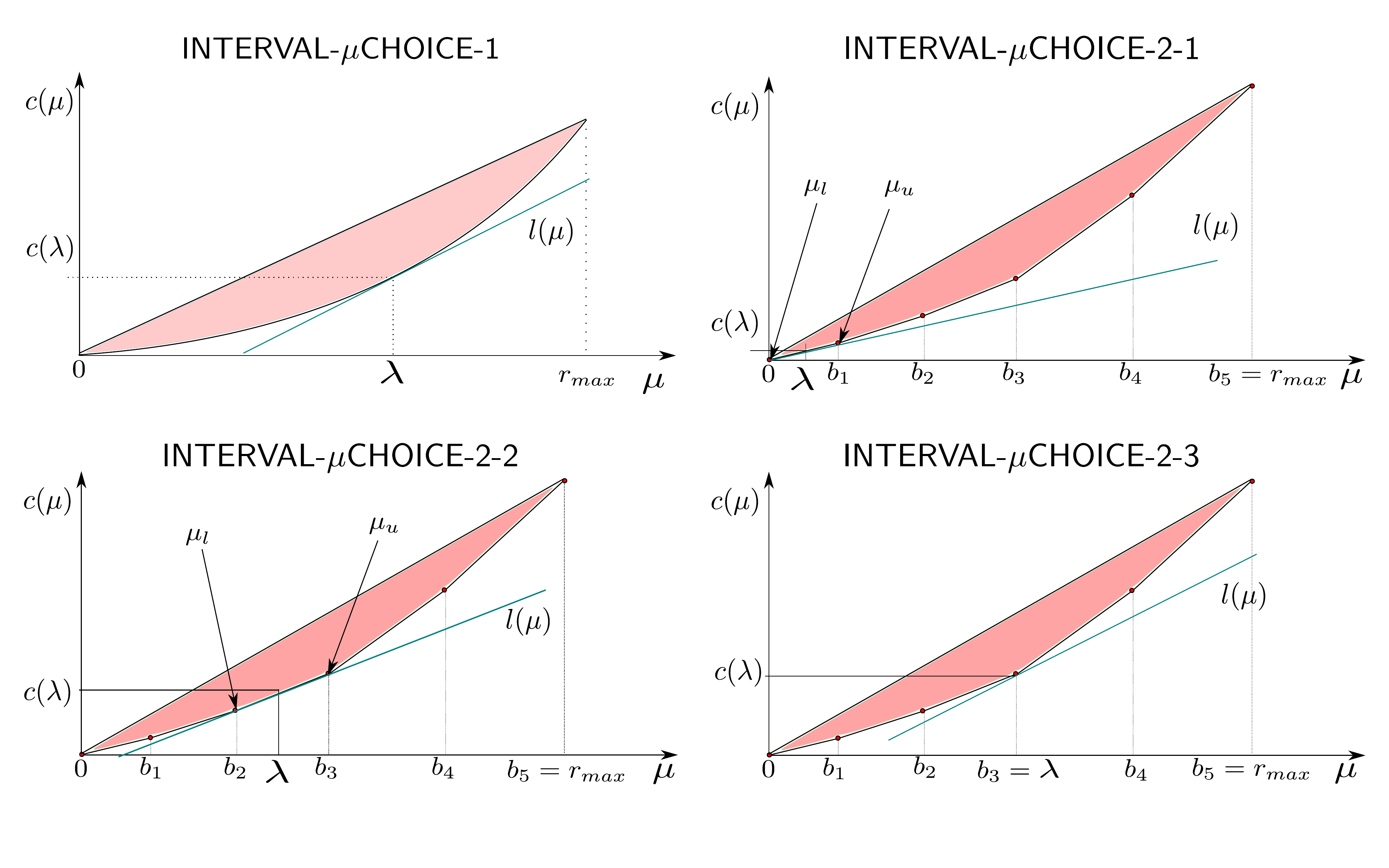}
  \caption{Illustration of the relationship between $\lambda$, $\mu_l$, and $\mu_{u}$ along with the minimum average cost $c(\lambda)$ and the line $l(\mu)$ for the four cases of the INTERVAL-$\mu$CHOICE problem.}
  \label{chap4:fig:p2servicecost_convexhull}
\end{figure}

Similar to \FMC, since the asymptotic behaviour of $Q^*(c_{c})$ for INTERVAL-$\mu$CHOICE-2 depends on the behaviour of $c(\mu)$ in a neighbourhood of $\mu = \lambda$, we consider the following cases for INTERVAL-$\mu$CHOICE-2:
\begin{description}
\item[INTERVAL-$\mu$CHOICE-2-1] : $\lambda \in (0, b_{1} = b_{\lambda})$,
\item[INTERVAL-$\mu$CHOICE-2-2] : $\lambda \in (a_{i} = a_{\lambda},b_{i} = b_{\lambda})$ for some $i \in \{2,\dots,P\}$, and,
\item[INTERVAL-$\mu$CHOICE-2-3] : $\lambda = a_{i} = a_{\lambda}$ for some $i \in \{2,\dots,P\}$.
\end{description}
The motivation for classifying INTERVAL-$\mu$CHOICE-2 into the three cases is the same as that for FINITE-$\mu$CHOICE.
The different cases are illustrated in Figure \ref{chap4:fig:p2servicecost_convexhull}.
We note that for \INTMC-1, the function $c(\mu)$ is strictly convex for every $\mu \in [0, r_{max}]$ and therefore it has no subcases.
We note that the analysis of INTERVAL-$\mu$CHOICE-2 is similar to that of FINITE-$\mu$CHOICE except that now the set of service rates is not a given finite set.
We now present the asymptotic lower bounds for $Q^*(c_{c})$ in the regime $\Re$ for the above cases.

\subsection{Asymptotic lower bounds}
\label{chap4:sec:intmc_lb}
Similar to the definition used for FINITE-$\mu$CHOICE, line $l(\mu)$ in the case of INTERVAL-$\mu$CHOICE is defined as: (a) The tangent to the $c(\mu)$ curve at $\lambda$ for INTERVAL-$\mu$CHOICE-1, (b) the line passing through $(0, 0)$ and $(b_{1},c(b_{1}))$ for INTERVAL-$\mu$CHOICE-2-1, (c) the line passing through $(a_{\lambda},c(a_{\lambda}))$ and $(b_{\lambda},c(b_{\lambda}))$ for INTERVAL-$\mu$CHOICE-2-2, and (d) any line that passes through $(\lambda, c(\lambda))$ with a slope $m$, such that $\frac{dc(\mu)}{d\mu}^{-}\vert_{\mu = \lambda} < m < \frac{dc(\mu)}{d\mu}^{+}\vert_{\mu = \lambda}$ (the left and right derivatives respectively) for INTERVAL-$\mu$CHOICE-2-3.

We note that like in the case of FINITE-$\mu$CHOICE, here we find an asymptotic lower bound on $\overline{Q}(\gamma)$ by (a) obtaining an upper bound on the stationary probability for a certain set of service rates in terms of $\overline{C}(\gamma)$ and $c(\lambda)$, (b) relating the stationary probability of this set of service rates to the stationary probability $\pi(q)$, of a set of queue lengths, and (c) obtaining a lower bound on $\overline{Q}(\gamma)$ in terms of $\pi(q)$.

We first consider INTERVAL-$\mu$CHOICE-1, for which $c(\mu)$ is a strictly convex function of $\mu \in [0, r_{max}]$.
We make the following assumption regarding $c(\mu)$ at $\mu = \lambda$.
\begin{description}
  \item[C2 :] For \INTMC-1, the second derivative of $c(\mu)$ is non-zero at $\mu = \lambda$.
\end{description}
The above assumption has been used in \cite{berry}.
We note that since $c(\mu)$ is strictly convex, the second derivative of $c(\mu)$ is non-zero for all $\mu \in [0, r_{max}]$ except for $\mu$ in a countable set.
We note that even though $\mu(q) \in [0, r_{max}]$, the set $\brac{\mu(q), q \in \mathbb{Z}_{+}}$ is only countable.
Let $\brac{\mu_{0} = 0, \dots, \mu_{k}, \dots}$ denote the set of service rates that is used by a policy $\gamma$.
\begin{lemma}
  For INTERVAL-$\mu$CHOICE-1, for any sequence of non-idling admissible policies $\gamma_{k}$ such that $\overline{C}(\gamma_{k}) - c(\lambda) = V_{k} \downarrow 0$ we have that $\overline{Q}(\gamma_{k}) = \Omega\nfrac{1}{\sqrt{V_{k}}}$.
  \label{chap4:prop:p21lb}
\end{lemma}
\begin{proof}
  Consider a particular policy $\gamma$ in the sequence $\gamma_{k}$ with $V_{k} = V$.
  Let $\mu^* \stackrel{\Delta} = \lambda - \epsilon_{V}$, where $\epsilon_{V} > 0$ is a function of $V$.
  The functional form of $\epsilon_{V}$ will be chosen later.
  We have that
  \begin{eqnarray*}
    V & = & \sum_{k = 0}^{\infty} \left(c(\mu_{k}) - l(\mu_{k})\right)\pi_{\mu}(k) = \sum_{q = 0}^{\infty} \left( c(\mu(q)) - l(\mu(q))\right) \pi(q), \nonumber \\
    & = & \sum_{q = 0}^{\infty} \left( c(\lambda) + \frac{dc(\mu)}{d\mu} \bigg \vert_{\mu = \lambda} (\mu(q) - \lambda) - l(\mu(q)) + G(\mu(q) - \lambda) \right) \pi(q),
  \end{eqnarray*}
  where $G(x)$ is a strictly convex function in $x$ as in \cite[Proposition 4.2]{berry}.
  We note that $c(\lambda) + \frac{dc(\mu)}{d\mu} \vert_{\mu = \lambda} (\mu(q) - \lambda) = l(\mu(q))$.
  Thus we have that $V = \sum_{q = 0}^{\infty} G(\mu(q) - \lambda) \pi(q)$.
  As $G(x)$ is strictly convex in $x$ and $\mu(q) - \lambda$ is bounded, we have from Proposition \ref{chap4:app:prop:quadlowerbound} that $G(\mu(q) - \lambda) \geq a_{1} (\mu(q) - \lambda)^{2}$ for some constant $a_{1} > 0$.
  Thus we have that 
  \begin{equation}
    V \geq a_{1} \sum_{q = 0}^{\infty} (\mu(q) - \lambda)^{2} \pi(q).
    \label{chap4:eq:v_variance}
  \end{equation}
  Define $q_{\mu^*} = \inf\brac{q : \mu(q) \geq \mu^{*}}$.
  From the non-decreasing property of $\mu(q)$ for $\gamma$, we have that  $Pr\brac{\mu(Q) < \mu^*} = Pr\brac{Q < q_{\mu^*}}$.
  Then,
  \begin{eqnarray}
    Pr\brac{\mu(Q) < \mu^*} & = & \sum_{q = 0}^{q_{\mu^*} - 1} \pi(q) \leq \frac{V}{a_{1} \epsilon_{V}^{2}},
    \label{chap4:eq:mu_variance}
  \end{eqnarray}
  where we have used the upper bound \eqref{chap4:eq:v_variance}.
  We choose $\epsilon_{V}$ as $a_{2}\sqrt{V}$, so that $\alpha \stackrel{\Delta} = \frac{2V}{a_{1} \epsilon_{V}^{2}} = \frac{2}{a_{1} a_{2}^{2}}$. 
  We choose $a_{2}$ such that $\alpha < 1$.
  In fact, we note that $\alpha$ can be made arbitrarily close to zero by the choice of $a_{2}$.
  Therefore, $Pr\brac{\mu(Q) \geq \mu^*} \geq 1 - \frac{\alpha}{2}$, which can be made arbitrarily close to one.
  
  We note that for any $q < q_{\mu^*}$, $\pi(q) \leq \frac{\alpha}{2}$.
  Therefore $\pi(q_{\mu^*}) \leq \pi(q_{\mu^*} - 1) \frac{\lambda}{\mu^*} \leq \frac{\lambda\alpha}{2\mu^*}$.
  In order to obtain a lower bound on $\overline{Q}(\gamma)$, we intend to find the largest $\overline{q}$ such that $Pr\brac{Q \leq \overline{q}} \leq \frac{1}{2}$.
  But we note that $Pr\brac{Q < q_{\mu^*}} \leq \frac{V}{a_{1} \epsilon_{V}^{2}} = \frac{\alpha}{2}$.
  Therefore the largest $\overline{q}$ satisfies
  \begin{eqnarray*}
    \sum_{q = 0}^{q_{\mu^*} - 1} \pi(q) + \sum_{q = q_{\mu^*}}^{\overline{q}} \pi(q) \leq \frac{1}{2}
  \end{eqnarray*}
  If $\overline{q}_{1}$ satisfies
  \begin{eqnarray*}
    \sum_{q = q_{\mu^*}}^{\overline{q}_{1}} \pi(q) \leq \frac{1}{2} - \frac{\alpha}{2},
  \end{eqnarray*}
  then $\overline{q}_{1} \leq \overline{q}$, for $\alpha$ sufficiently small.
  As $Q(t)$ is a birth-death process, we have that $\pi(q) \lambda = \pi(q + 1) \mu(q + 1)$.
  Furthermore, if $q \geq q_{\mu^*}$ we have that $\pi(q - 1) \lambda \geq \pi(q) \mu^*$.
  By induction, we obtain that for $q \in \{q_{\mu^*}, \dots\}$
  \begin{eqnarray}
    \pi(q) \leq \pi(q_{\mu^*})\left(\frac{\lambda}{\mu^*} \right)^{q - q_{\mu^*}} \leq \pi(q_{\mu^{*}} - 1) \left(\frac{\lambda}{\mu^*}\right)^{q - q_{\mu^*} + 1}, \\
    \text{and for any } q' \geq q_{\mu^*}, \sum_{q = q_{\mu^*}}^{q'} \pi(q) \leq \pi(q_{\mu^*} - 1) \sum_{m = 1}^{q' - q_{\mu^*} + 1} \left(\frac{\lambda}{\mu^*}\right)^{m}.
    \label{chap4:eq:p2-referfromINTLC_1}
  \end{eqnarray}
  Using the above upper bound on $\sum_{q = q_{\mu^*}}^{q'} \pi(q)$, we obtain a lower bound $\overline{q}_{2}$ to $\overline{q}_{1}$.
  If $\overline{q}_{2}$ is the largest integer such that
  \begin{eqnarray}
    \pi(q_{\mu^*} - 1) \sum_{m = 1}^{\overline{q}_{2} - q_{\mu^*} + 1} \left(\frac{\lambda}{\mu^*}\right)^{m} \leq \frac{1}{2} - \frac{\alpha}{2},
    \label{chap4:eq:p2-3}
  \end{eqnarray}
  then $\sum_{q = 0}^{\overline{q}_{2}} \pi(q) \leq \frac{1}{2}$ and $\overline{q}_{1} \geq \overline{q}_{2}$.

  Now we obtain an upper bound on $\pi(q_{\mu^*} - 1)$, which is tighter than the upper bound $\frac{\alpha}{2}$ derived before.
  From \eqref{chap4:eq:v_variance} we have that
  \begin{eqnarray}
    \frac{V}{a_{1}} & \geq & \sum_{q = 0}^{\infty} (\mu(q) - \lambda)^{2} \pi(q) \geq \sum_{q < q_{\mu^*}} (\mu(q) - \lambda)^{2} \pi(q) \nonumber \\
    & = & \sum_{q < q_{\mu^*}} (\mu(q) - \lambda)^{2} \pi(q) + 0 \sum_{q \geq q_{\mu^*}} \pi(q), \nonumber \\
    & \geq & \left( \sum_{q < q_{\mu^*}} (\mu(q) - \lambda) \pi(q) \right)^{2} \text{(using Jensen's inequality as in \cite{berry})}.
  \end{eqnarray}
  But, as  $\pi(q)\mu(q) = \pi(q - 1)\lambda$, we obtain that
  \begin{eqnarray}
    \sum_{q < q_{\mu^*}} (\mu(q) - \lambda) \pi(q) & = & -\lambda \pi(0) + \sum_{1 \leq q \leq q_{\mu^*} - 1} \brap{\lambda\pi(q - 1) - \lambda\pi(q)} = -\lambda \pi(q_{\mu^{*}} - 1), \nonumber \\
    & & \text{ or } \frac{V}{a_{1}} \geq \lambda^{2} \pi(q_{\mu^*} - 1)^{2}.
    \label{chap4:eq:tighterupperbound}
  \end{eqnarray}
  
  Now we find a lower bound $\overline{q}_{3}$ on $\overline{q}_{2}$ by using the above upper bound on $\pi(q_{\mu^*} - 1)$ in \eqref{chap4:eq:p2-3}.
  Let $\overline{q}_{3}$ be the largest integer such that
  \begin{eqnarray*}
    \frac{1}{\lambda}\sqrt{\frac{V}{a_{1}}} \sum_{m = 1}^{\overline{q}_{3} - q_{\mu^*} + 1} \left(\frac{\lambda}{\mu^*}\right)^{m} \leq \frac{1}{2} - \frac{\alpha}{2}.
  \end{eqnarray*}
  Then $\overline{q}_{3} \leq \overline{q}_{2}$.
  We have that $\overline{q}_{3}$ satisfies
  \begin{eqnarray*}
    \frac{1}{\lambda}\sqrt{\frac{V}{a_{1}}} \sum_{m = 1}^{\overline{q}_{3} - q_{\mu^*} + 1} \left(\frac{\lambda}{\mu^*}\right)^{m} \leq \frac{1}{2} - \frac{\alpha}{2}, \\
    \frac{\left(\frac{\lambda}{\mu^*}\right)^{\overline{q}_{3} - q_{\mu^*} + 1} - 1}{\lambda - \mu^*} \leq \sqrt{\frac{a_{1}}{V}}\left[ \frac{1 - \alpha}{2} \right], \\
    \left(\frac{\lambda}{\mu^*}\right)^{\overline{q}_{3} - q_{\mu^*} + 1} \leq 1 + \left( \lambda - \mu^* \right)\sqrt{\frac{a_{1}}{V}}\left[ \frac{1 - \alpha}{2} \right] \\
    \overline{q}_{3} - q_{\mu^*} + 1 \leq \log_{\frac{\lambda}{\mu^*}} \left[ 1 + \left( \lambda - \mu^*\right)\sqrt{\frac{a_{1}}{V}}\left[ \frac{1 - \alpha}{2} \right]\right].
  \end{eqnarray*}
  Since $q_{\mu^*} > 0$, we note that $\overline{q}_{3}$ is at least
  \begin{eqnarray*}
    \floor{\log_{\frac{\lambda}{\mu^*}} \left[ 1 + \left( \lambda - \mu^* \right)\sqrt{\frac{a_{1}}{V}}\left[ \frac{1 - \alpha}{2} \right]\right] - 1}.
  \end{eqnarray*}
  Therefore,
  \begin{eqnarray*}
  \overline{q}_{3} & \geq & \log_{\frac{1}{1 - \frac{\epsilon_V}{\lambda}}} \left[ 1 + \epsilon_{V}\sqrt{\frac{a_{1}}{V}}\left[ \frac{1 - \alpha}{2} \right]\right] - 2, \\
    & = & \frac{\log \left[ 1 +  \epsilon_{V}\sqrt{\frac{a_{1}}{V}}\left[ \frac{1 - \alpha}{2} \right]\right]}{-\log(1 - \frac{\epsilon_{V}}{\lambda})} - 2.
  \end{eqnarray*}
  Since $\epsilon_{V} = a_{2} \sqrt{V}$, we have
  \begin{eqnarray*}
    \overline{q}_{3} & \geq & \frac{\log \left[ 1 + a_{2}\sqrt{a_{1}}\left[ \frac{1 - \alpha}{2} \right]\right]}{-\log\brap{1 - \frac{a_{2}\sqrt{V}}{\lambda}}} - 2.
  \end{eqnarray*}
  Since $\overline{Q}(\gamma) \geq \frac{\overline{q}}{2} \geq \frac{\overline{q}_{1}}{2} \geq \frac{\overline{q}_{2}}{2} \geq \frac{\overline{q}_{3}}{2}$ we have that
  \begin{eqnarray*}
    \overline{Q}(\gamma) & \geq & \frac{1}{2} \bras{\frac{\log \left[ 1 + a_{2}\sqrt{a_{1}}\left[ \frac{1 - \alpha}{2} \right]\right]}{-\log\brap{1 - \frac{a_{2}\sqrt{V}}{\lambda}}} - 2 }.
  \end{eqnarray*}
  As $V \downarrow 0$, we note that $\log\brap{1 - \frac{a_{2}\sqrt{V}}{\lambda}} = \Theta\brap{\sqrt{V}}$.
  Hence, for the sequence $\gamma_{k}$ with $\overline{C}(\gamma_{k}) - c(\lambda) = V_{k} \downarrow 0$, we have that $\overline{Q}(\gamma_{k}) = \Omega\nfrac{1}{\sqrt{V_k}}$.
\end{proof}

\begin{remark}
  \label{chap4:remark:sqrtvdiscussion}
  We note that as $V \downarrow 0$, $\pi(0) \rightarrow 0$.
  We also note that as $V \downarrow 0$, there exists a set of queue lengths, $\hpq$, occurring with high probability ($1 - V$), such that $\mu(q) \rightarrow \lambda, \forall q \in \hpq$.
  If $\pi(0) \rightarrow 0$, then $|\hpq| \rightarrow \infty$.
  Furthermore, for each $q \in \hpq, \pi(q) = \mathcal{O}(\sqrt{V})$.
  We also note that the stationary probability for each $q \in \hpq$ become equal as $V \downarrow 0$.
  Then the average queue length is $\Omega\nfrac{1}{\sqrt{V}}$.
\end{remark}

\begin{remark}
  \label{chap4:remark:berrygallagerlb}
  We note that the lower bounding technique in \cite{berry} can be used to obtain the $\Omega\nfrac{1}{\sqrt{V}}$ lower bound by considering a uniformized version of $Q(t)$.
  We outline this method in Appendix \ref{chap4:app:berrygallagerlb}.
  Using the stationary probability of the queue length has its advantages, since it gives us additional insights into the form of the optimal policy.
\end{remark}

\begin{lemma}
  For INTERVAL-$\mu$CHOICE-2-1, for any sequence of non-idling admissible policies $\gamma_{k}$ with $\overline{C}(\gamma_{k}) - c(\lambda) = V_{k} \downarrow 0$ we have that $\overline{Q}(\gamma_{k}) = \frac{\lambda}{b_{\lambda} - \lambda} - \mathcal{O}\brap{V^{1 - \delta}\log\nfrac{1}{V}}$, for $0 < \delta < 1$.
  \label{chap4:prop:p221lb}
\end{lemma}
We note that \INTMC-2-1 is very similar to \FMC-1 for which we recall that the asymptotic order was $\mathcal{O}\brap{V\log\nfrac{1}{V}}$.
However, for \INTMC-2-1, we are only able to show that the order is $\mathcal{O}\brap{V^{1 - \delta}\log\nfrac{1}{V}}$, where $\delta$ can be made arbitrarily close to zero.

\begin{proof}
  We note that in this case there exists a policy $\gamma$, for which $\mu(q) = b_{1} = b_{\lambda}, \forall q > 0$, with $\overline{C}(\gamma) = c(\lambda)$ and $\overline{Q}(\gamma) = \frac{\lambda}{b_{\lambda} - \lambda}$.
  For $V = 0$, we note that the above policy is optimal.
  The solution to the tradeoff problem is similar to FINITE-$\mu$CHOICE-1 in which the average queue length increases but only to a finite limit as $V_{k} \downarrow 0$.
  Consider a particular policy $\gamma$ in the sequence $\gamma_{k}$ with $V_{k} = V$.
  We have that
  \begin{eqnarray*}
    V & = & \sum_{k = 0}^{\infty} (c(\mu_{k}) - l(\mu_{k}))\pi_{\mu}(k) = \sum_{\mu_{k} > b_{\lambda}} (c(\mu_{k}) - l(\mu_{k}))\pi_{\mu}(k).
  \end{eqnarray*}
  Let  $\mu^* = b_{\lambda} + \epsilon_{V}$, where $\epsilon_{V}$ is a function of $V$ to be chosen later.
  Then we have that
  \begin{eqnarray}
    V & \geq & \sum_{\mu_{k} \geq \mu^{*}} (c(\mu_{k}) - l(\mu_{k}))\pi_{\mu}(k) \nonumber \\
    & \geq & m_{a} \sum_{\mu_{k} \geq \mu^*} (\mu_{k} - b_{\lambda}) \pi_{\mu}(k) \nonumber \\
    & \geq & m_{a} \epsilon_{V} \sum_{\mu_{k} \geq \mu^*} \pi_{\mu}(k) \nonumber \\
    \frac{V}{m_{a} \epsilon_{V}} & \geq & \sum_{\mu_{k} \geq \mu^*} \pi_{\mu}(k),
    \label{chap4:eq:p221_1}
  \end{eqnarray}
  where $m_{a}$ is the tangent of the angle made by the line passing through $(b_{\lambda}, c(b_{\lambda}))$ and $(b_{2}, c(b_{2}))$ with the line $l(\mu)$.
  We proceed as in the proof of the asymptotic lower bound for problem FINITE-$\mu$CHOICE-1.
  But unlike in FINITE-$\mu$CHOICE-1, we note here that any service rate arbitrarily close to $b_{\lambda}$ might be used by a policy $\gamma$.
  Intuitively, since $\sum_{\mu_{k} \geq \mu^*} \pi_{\mu}(k)$ should approach $0$ as $V \downarrow 0$, we require that the choice of $\epsilon_{V}$ should be such that $\frac{V}{\epsilon_{V}} \downarrow 0$ as $V \downarrow 0$.
  
  Let $q_{\mu^*} = \inf\brac{q : \mu(q) \geq \mu^*}$.
  For $q < q_{\mu^*}$, $\mu(q) < \mu^*$ and therefore $\pi(q) \lambda < \pi(q + 1)\mu^*$.
  Hence, by induction we obtain that
  \begin{eqnarray}
    \pi(q_{\mu^*} - m) < \pi(q_{\mu^*}) \left( \frac{\mu^*}{\lambda}\right)^{m}, \text{ for } m \in \{1,\dots,q_{\mu^*}\}.
    \label{chap4:eq:p221_2}
  \end{eqnarray}
  From \eqref{chap4:eq:p221_1}, we have
  \begin{eqnarray*}
    \sum_{q < q_{\mu^*}} \pi(q) = 1 - \sum_{q \geq q_{\mu^*}} \pi(q) \geq 1 - \frac{V}{m_{a} \epsilon_{V}}.
  \end{eqnarray*}
  Now from \eqref{chap4:eq:p221_2} we have
  \begin{eqnarray*}
    \sum_{q < q_{\mu^*}} \pi(q) \leq \sum_{q = 0}^{q_{\mu^*} - 1} \pi(q_{\mu^*}) \left(\frac{\mu^*}{\lambda}\right)^{q_{\mu^*} - q},\text{ hence we have that,} \\
    1 - \frac{V}{m_{a} \epsilon_{V}} \leq \pi(q_{\mu^*}) \sum_{m = 1}^{q_{\mu^*}} \left(\frac{\mu^*}{\lambda}\right)^{m}, \\
    \frac{1}{\pi(q_{\mu^*})} \left( 1 - \frac{V}{m_{a} \epsilon_{V}} \right) \leq \sum_{m = 1}^{q_{\mu^*}} \left(\frac{\mu^*}{\lambda}\right)^{m} = \frac{\mu^*}{\mu^* - \lambda} \left[ \left( \frac{\mu^*}{\lambda} \right)^{q_{\mu^*}} - 1 \right].
  \end{eqnarray*}
  We note that as $\pi(q_{\mu^*}) \leq \sum_{q \geq q_{\mu^*}} \pi(q) \leq \frac{V}{m_{a} \epsilon_{V}}$, we have that $\frac{1}{\pi(q_{\mu^*})} \geq \frac{m_{a} \epsilon_{V}}{V}$ and therefore
  \begin{eqnarray*}
    \frac{m_{a} \epsilon_{V}}{V} \left( 1 - \frac{V}{m_{a} \epsilon_{V}} \right) \leq  \frac{\mu^*}{\mu^* - \lambda} \left[ \left( \frac{\mu^*}{\lambda} \right)^{q_{\mu^*}} - 1 \right], \nonumber \\
    \log_{\frac{\mu^*}{\lambda}}\left[\frac{\mu^* - \lambda}{\mu^*} \frac{m_{a}\epsilon_{V}}{V}\left( 1 - \frac{V}{m_{a} \epsilon_{V}} \right) + 1\right] \leq q_{\mu^*}.
  \end{eqnarray*}
  By definition, for every $q < q_{\mu^*}$, $\mu(q) < \mu^*$, and for every $q \geq q_{\mu^*}, \mu(q) \leq r_{max}$.
  Let us define 
  \begin{eqnarray*}
    q_{\mu^*,l} = \left\lceil\log_{\frac{\mu^*}{\lambda}}\left[\frac{\mu^* - \lambda}{\mu^*} \frac{m_{a}\epsilon_{V}}{V}\left( 1 - \frac{V}{m_{a} \epsilon_{V}} \right) + 1\right] \right\rceil,
  \end{eqnarray*}
  which is the smallest possible value for $q_{\mu^*}$ for any policy $\gamma$.
  Consider another policy $\gamma'$ defined as follows :
  \begin{eqnarray*}
    \mu(0) & = & 0, \\
    \mu(q) & = & \mu^*, \text{ for } 1 \leq q \leq q_{\mu^*,l}, \\
    \mu(q) & = & r_{max}, \text{ for } q > q_{\mu^*,l}.
  \end{eqnarray*}
  Then $\overline{Q}(\gamma') \leq \overline{Q}(\gamma)$.
  We now obtain a lower bound on $\overline{Q}(\gamma')$ as in FINITE-$\mu$CHOICE-1.
  
  Recall that for $\gamma'$ we have (using the sequence of steps leading to \eqref{chap4:eq:p13ezlb})
  \begin{eqnarray*}
    \overline{Q}(\gamma') & \geq & (1 - a)\left[ \frac{a}{(1 - a)^{2}} \left\{1 - (1-a)(q_{\mu^*,l} + 1) a^{q_{\mu^*,l}} - a.a^{q_{\mu^*,l}}\right\} + a^{q_{\mu^*,l}}\left\{ q_{\mu^*,l} \frac{b}{1 - b} + \frac{b}{(1 - b)^{2}} \right\} \right], \\
    & = & \frac{a}{1 - a} + (1 - a)\left[ a^{q_{\mu^*,l}}\left\{ q_{\mu^*,l} \frac{b}{1 - b} + \frac{b}{(1 - b)^{2}}\right\} - \frac{a}{(1 - a)^{2}} \left\{ (1 - a)(q_{\mu^*,l} + 1)a^{q_{\mu^*,l}} + a.a^{q_{\mu^*,l}} \right\}\right], \\
    & \geq & \frac{a}{1 - a} - \frac{a}{1 - a} a^{q_{\mu^*,l}}\left[1 + (1 - a)q_{\mu^*,l}\right],
  \end{eqnarray*}
  where $a = \frac{\lambda}{\mu^*}$ and $b = \frac{\lambda}{r_{max}}$.
  We note that for $V \downarrow 0$, the term $\frac{a}{1 - a} = \frac{\lambda}{b_{\lambda} - \lambda}\brap{1 - \frac{\epsilon_{V}}{b_{\lambda} - \lambda} + o(\epsilon_{V})}$.
  If $V \downarrow 0$, since we require that $\frac{V}{\epsilon_{V}} \downarrow 0$, $q_{\mu^*,l} \uparrow \infty$ and therefore the second term in the lower bound for $\overline{Q}(\gamma')$ is $a q_{\mu^*,l} a^{q_{\mu^*,l}}$.
  We note that at $V = 0$, since we require that the lower bound is tight, we only consider $\epsilon_{V}$ such that $\epsilon_{V} \downarrow 0$ as $V \downarrow 0$.
  Then it can be shown that $\overline{Q}(\gamma) \geq \overline{Q}(\gamma') = \frac{\lambda}{b_{\lambda} - \lambda} - \mathcal{O}\brap{\epsilon_{V} + \frac{V}{\epsilon_{V}}\log\nfrac{\epsilon_{V}}{V}}$, for any sequence $\epsilon_{V} \downarrow 0$ and $\frac{V}{\epsilon_{V}} \downarrow 0$ as $V \downarrow 0$.
  By choosing $\epsilon_{V} = {V}^{1 - \delta}$, where $0 < \delta < 1$, we obtain that 
  \begin{eqnarray*}
    \overline{Q}(\gamma) = \frac{\lambda}{b_{\lambda} - \lambda} - \mathcal{O}\brap{{V}^{1- \delta}\log\nfrac{1}{V}}.
  \end{eqnarray*}
  For the sequence $\gamma_{k}$, we therefore obtain that $\overline{Q}(\gamma_{k}) = \frac{\lambda}{b_{\lambda} - \lambda} - \mathcal{O}\brap{{V}^{1- \delta}\log\nfrac{1}{V}}$.
\end{proof}

\begin{lemma}
  For INTERVAL-$\mu$CHOICE-2-2, for any sequence of non-idling admissible policies $\gamma_{k}$ with $\overline{C}(\gamma_{k}) - c(\lambda) = V_{k} \downarrow 0$, we have that $\overline{Q}(\gamma_{k}) = \Omega\brap{\log\nfrac{1}{V_{k}}}$.
  \label{chap4:prop:p222lb}
\end{lemma}
\begin{proof}
  Consider a particular policy $\gamma$ in the sequence $\gamma_{k}$ with $V_{k} = V$.
  Let $\mu^{*} \stackrel{\Delta} = a_{\lambda} - \epsilon_{V}$.
  We have that 
  \begin{eqnarray*}
    V & = & \sum_{\mu_{k} < a_{\lambda}} (c(\mu_{k}) - l(\mu_{k})) \pi_{\mu}(k) + \sum_{\mu_{k} > b_{\lambda}} (c(\mu_{k}) - l(\mu_{k})) \pi_{\mu}(k), \\
    & \geq & \sum_{\mu_{k} < a_{\lambda} - \epsilon_{V}} (c(\mu_{k}) - l(\mu_{k})) \pi_{\mu}(k), \\
    & = & m_{a} \sum_{\mu_{k} < a_{\lambda} - \epsilon_{V}} (a_{\lambda} - \mu_{k}) \pi_{\mu}(k), \\
    & \geq & m_{a} \epsilon_{V} Pr\brac{\mu(Q) < \mu^*},
  \end{eqnarray*}
  where $m_{a}$ is the tangent of the angle made by the line passing through $(a_{i} = a_{\lambda}, c(a_{i}))$ and $(a_{i - 1}, c(a_{i - 1}))$ with the line $l(\mu)$.
  Define $q_{\mu^*} = \inf\brac{q : \mu(q) \geq \mu^{*}}$.
  As $\gamma$ is admissible, we have that $Pr\brac{\mu(Q) < \mu^*} = Pr\brac{Q < q_{\mu^*}}$.
  Hence we have
  \begin{eqnarray*}
    Pr\brac{Q < q_{\mu^*}} & \leq & \frac{V}{m_{a} \epsilon_{V}}, \\
    \text{and } \pi(q_{\mu^*} - 1) & \leq & \frac{V}{m_{a} \epsilon_{V}}.
  \end{eqnarray*}
  We now choose $\epsilon_{V} = \epsilon$, a positive constant.
  To find a lower bound on $\overline{Q}(\gamma)$, in the following, we intend to find the largest $\overline{q}$ such that $\sum_{q = 0}^{\overline{q}} \pi(q) \leq \frac{1}{2}$.
  But we note that $Pr\brac{Q < q_{\mu^*}} \leq \frac{V}{m_{a} \epsilon}$ and for any $q < q_{\mu^*}, \pi(q) \leq \frac{V}{m_{a} \epsilon}$.
  Therefore, $\pi(q_{\mu^*}) \leq \pi(q_{\mu^*} - 1) \frac{\lambda}{\mu^*} \leq \frac{\lambda V}{m_{a} \epsilon \mu^*}$.
  Let $\overline{q}_{1}$ be the largest integer such that
  \begin{eqnarray*}
    \sum_{q = q_{\mu^*}}^{\overline{q}} \pi(q) \leq \frac{1}{2} - \frac{V}{m_{a} \epsilon},
  \end{eqnarray*}
  then $\overline{q}_{1} \leq \overline{q}$.
  Proceeding as for problem INTERVAL-$\mu$CHOICE-1, we obtain a lower bound $\overline{q}_{2}$ on $\overline{q}_{1}$ by using an upper bound for $\pi(q)$.
  We note that if $q \geq q_{\mu^*}$ we have that $\pi(q - 1) \lambda \geq \pi(q) \mu^*$.
  By induction, we obtain that for $q \in \{q_{\mu^*}, \dots\}$
  \begin{eqnarray*}
    \pi(q) \leq \pi(q_{\mu^{*}} - 1) \left(\frac{\lambda}{\mu^*}\right)^{q - q_{\mu^*} + 1}, \\
    \text{and for any } q' \geq q_{\mu^*}, \sum_{q = q_{\mu^*}}^{q'} \pi(q) \leq \pi(q_{\mu^*} - 1) \sum_{m = 1}^{q' - q_{\mu^*} + 1} \left(\frac{\lambda}{\mu^*}\right)^{m}.
  \end{eqnarray*}
  Using the above upper bound on $\sum_{q = q_{\mu^*}}^{q'} \pi(q)$, and $\pi(q_{\mu^*} - 1) \leq \frac{V}{m_{a} \epsilon}$, we obtain the following lower bound $\overline{q}_{2}$ to $\overline{q}_{1}$.

  If $\overline{q}_{2}$ is the largest integer such that
  \begin{eqnarray*}
    \frac{V}{m_{a} \epsilon} \sum_{m = 1}^{\overline{q}_{2} - q_{\mu^*} + 1} \left(\frac{\lambda}{\mu^*}\right)^{m} \leq \frac{1}{2} - \frac{V}{m_{a} \epsilon},
  \end{eqnarray*}
  then $\sum_{q = 0}^{\overline{q}_{2}} \pi(q) \leq \frac{1}{2}$ and $\overline{q}_{2} \leq \overline{q}_{1}$.
  Hence, $\overline{q}_{2}$ is the largest integer such that
  \begin{eqnarray*}
    \left( \frac{\lambda}{a_{\lambda} - \epsilon} \right)^{\overline{q}_{2} - q_{\mu^*} + 1} \leq 1 + \left( \frac{\lambda - a_{\lambda} + \epsilon}{{\lambda}} \right) \frac{\epsilon m_{a}}{V} \left( \frac{1}{2} - \frac{V}{m_{a} \epsilon}\right) \\
    \overline{q}_{2} \leq q_{\mu^*} - 1 + \log_{ \frac{\lambda}{a_{\lambda} - \epsilon}}\left[ 1 + \left( \frac{\lambda - a_{\lambda} + \epsilon}{{\lambda}} \right) \frac{\epsilon m_{a}}{V} \left( \frac{1}{2} - \frac{V}{m_{a} \epsilon}\right)\right].
  \end{eqnarray*}
  Since $q_{\mu^*} \geq 0$, $\overline{q}_{2}$ is at least
  \begin{eqnarray*}
    \floor{\log_{ \frac{\lambda}{a_{\lambda} - \epsilon}}\left[ 1 + \left( \frac{\lambda - a_{\lambda} + \epsilon}{{\lambda}} \right) \frac{\epsilon m_{a}}{V} \left( \frac{1}{2} - \frac{V}{m_{a} \epsilon}\right)\right] - 1}.
  \end{eqnarray*}
  Therefore,
  \begin{eqnarray*}
    \overline{Q}(\gamma) \geq \frac{\overline{q}}{2} \geq \frac{\overline{q}_{1}}{2} \geq \frac{\overline{q}_{2}}{2} \geq \frac{1}{2}\bras{\log_{ \frac{\lambda}{a_{\lambda} - \epsilon}}\left[ 1 + \left( \frac{\lambda - a_{\lambda} + \epsilon}{{\lambda}} \right) \frac{\epsilon m_{a}}{V} \left( \frac{1}{2} - \frac{V}{m_{a} \epsilon}\right)\right] - 2}.
  \end{eqnarray*}
  Hence, for any sequence $\gamma_{k}$ with $\overline{C}(\gamma_{k}) - c(\lambda) = V_{k} \downarrow 0$, we obtain that $\overline{Q}(\gamma_{k}) = \Omega\left(\log\left(\frac{1}{V_{k}}\right)\right)$.
\end{proof}

\begin{remark}
  We note that as $V \downarrow 0$, $\pi(0) \downarrow 0$.
  We also note that as $V \downarrow 0$, there exists a set of queue lengths, $\hpq$, occurring with high probability ($1 - V$), such that $\mu(q) \in [a_{\lambda}, b_{\lambda}], \forall q \in \hpq$.
  If $\pi(0) \rightarrow 0$, then $|\hpq| \rightarrow \infty$.
  We note that for $q \not \in \hpq$, $\pi(q) = \mathcal{O}({V})$.
  Then from the birth death structure of $Q(t)$ we obtain that the smallest queue length $q_{min} \in \hpq$ has $\pi(q_{min}) = \mathcal{O}(V)$.
  Since for $q \in \hpq$, $\mu(q) \in [a_{\lambda}, b_{\lambda}]$ and the policies that we consider are monotone, the stationary probability distribution of the queue lengths in $\hpq$ can be observed to be geometrically increasing and then decreasing, which leads to the $\log\nfrac{1}{V}$ growth for the average queue length.
\end{remark}

\begin{remark}
  We note that the lower bounding technique in \cite{neely_utility} can be used to obtain the $\Omega\brap{\log\nfrac{1}{{V}}}$ lower bound by considering a uniformized version of $Q(t)$ as we discussed in Remark \ref{chap4:remark:berrygallagerlb}.
  We again note that obtaining the stationary probability of the queue length, as in the above proof has its advantages, since it gives us additional insights into the form of the optimal policy.
\end{remark}

\begin{lemma}
  For INTERVAL-$\mu$CHOICE-2-3, for any sequence of non-idling admissible policies $\gamma_{k}$ with $\overline{C}(\gamma_{k}) - c(\lambda) = V_{k} \downarrow 0$, we have that $\overline{Q}(\gamma_{k}) = \Omega\nfrac{1}{V_{k}}$.
  \label{chap4:prop:p223lb}
\end{lemma}

\begin{proof}
  Consider a policy $\gamma$ in the sequence $\gamma_{k}$, with $V_{k} = V$.
  Let $\mu^* \stackrel{\Delta} = \lambda - \epsilon_{V} = a_{\lambda} - \epsilon_{V}$.
  Since $c(\mu_{k}) \geq l(\mu_{k})$, we have
  \begin{eqnarray*}
    V & \geq & \sum_{\mu_{k} < \lambda} (c(\mu_k) - l(\mu_k)) \pi_{\mu}(k) \geq \sum_{\mu_{k} < \mu^*} (c(\mu_k) - l(\mu_k)) \pi_{\mu}(k)\\
    & \geq & \sum_{\mu_{k} < \mu^*} m_{a}(\lambda - \mu_{k}) \pi_{\mu}(k) \geq m_{a} \epsilon_{V} Pr\brac{\mu(Q) < \mu^*},
  \end{eqnarray*}
  where $m_{a}$ is the tangent of the angle made by the line passing through $(a_{i} = a_{\lambda}, c(a_{i}))$ and $(a_{i - 1}, c(a_{i - 1}))$ with the line $l(\mu)$.
  Let $\epsilon_{V} \stackrel{\Delta} = a_{2} V$, where $a_{2}$ is chosen so that $\alpha \stackrel{\Delta} = \frac{2V}{m_{a} \epsilon_{V}} < 1$.
  We note that $a_{2}$ can be chosen such that $\alpha$ is arbitrarily close to zero.
  Define $q_{\mu^*} = \inf\brac{q : \mu(q) \geq \mu^{*}}$.
  Since $\gamma$ is admissible, we have that $Pr\brac{Q < q_{\mu^*}} = Pr\brac{\mu(Q) < \mu^*} \leq \frac{V}{m_{a} \epsilon_V}$.
  We have
  \begin{eqnarray*}
    Pr\brac{Q < q_{\mu^*}} & \leq & \frac{\alpha}{2}, \\
    \text{and } \pi(q_{\mu^*} - 1) & \leq & \frac{\alpha}{2}.
  \end{eqnarray*}
  To find a lower bound on $\overline{Q}(\gamma)$, in the following, we intend to find the largest $\overline{q}$, such that $\sum_{q = 0}^{\overline{q}} \pi(q) \leq \frac{1}{2}$.
  But we note that $Pr\brac{Q < q_{\mu^*}} \leq \frac{\alpha}{2}$ and for any $q < q_{\mu^*}, \pi(q) \leq \frac{\alpha}{2}$.
  Therefore, $\pi(q_{\mu^*}) \leq \pi(q_{\mu^*} - 1) \frac{\lambda}{\mu^*} \leq \frac{\lambda \alpha}{2 \mu^*}$.
  If $\overline{q}_{1}$ is the largest integer such that
  \begin{eqnarray*}
    \sum_{q = q_{\mu^*}}^{\overline{q}} \pi(q) \leq \frac{1}{2} - \frac{V}{m_{a} \epsilon_{V}},
  \end{eqnarray*}
  then $\overline{q}_{1} \leq \overline{q}$.
  Proceeding as for problem INTERVAL-$\mu$CHOICE-1, we obtain a lower bound $\overline{q}_{2}$ on $\overline{q}_{1}$ by using an upper bound for $\pi(q)$.
  We note that if $q \geq q_{\mu^*}$ we have that $\pi(q - 1) \lambda \geq \pi(q) \mu^*$.
  By induction, we obtain that for $q \in \{q_{\mu^*}, \dots\}$
  \begin{eqnarray*}
    \pi(q) \leq \pi(q_{\mu^{*}} - 1) \left(\frac{\lambda}{\mu^*}\right)^{q - q_{\mu^*} + 1}, \\
    \text{and for any } q' \geq q_{\mu^*}, \sum_{q = q_{\mu^*}}^{q'} \pi(q) \leq \pi(q_{\mu^*} - 1) \sum_{m = 1}^{q' - q_{\mu^*} + 1} \left(\frac{\lambda}{\mu^*}\right)^{m}.
  \end{eqnarray*}
  Using the above upper bound on $\sum_{q = q_{\mu^*}}^{q'} \pi(q)$ we obtain the following lower bound $\overline{q}_{2}$ to $\overline{q}_{1}$.
  If $\overline{q}_{2}$ is the largest integer such that
  \begin{eqnarray*}
    \pi(q_{\mu^*} - 1) \sum_{m = 1}^{\overline{q}_{2} - q_{\mu^*} + 1} \left( \frac{\lambda}{\mu^*} \right)^{m} \leq \frac{1 - \alpha}{2},
  \end{eqnarray*}
  then $\overline{q}_{2} \leq \overline{q}$.

  We note that 
  \begin{eqnarray*}
    V & \geq & \sum_{\mu_{k} < \mu^*} m_{a}(\lambda - \mu_{k}) \pi_{\mu}(k), \\
    & = & \sum_{q < q_{\mu^*}} m_{a}(\lambda - \mu(q))\pi(q).
  \end{eqnarray*}
  Again, since $\pi(q)\mu(q) = \pi(q - 1)\lambda$, it follows that
  \begin{eqnarray*}
    V \geq m_{a} \lambda \pi(q_{\mu^*} - 1).
  \end{eqnarray*}
  Now if $\overline{q}_{3}$ is the largest integer such that
  \begin{eqnarray*}
    \frac{V}{m_{a} \lambda} \sum_{m = 1}^{\overline{q}_{3} - q_{\mu^*} + 1} \left( \frac{\lambda}{\mu^*} \right)^{m} \leq \frac{1 - \alpha}{2},
  \end{eqnarray*}
  then $\overline{q}_{3} \leq \overline{q}_{2}$.
  We have that $\overline{q}_{3}$ satisfies
  \begin{eqnarray*}
    \overline{q}_{3} \leq q_{\mu^*} - 1 + \log_{\frac{\lambda}{\mu^*}} \left[ 1 + \frac{\epsilon_{V}}{\lambda} \frac{m_{a} \lambda}{V} \frac{1 - \alpha}{2} \right], \\
    \overline{q}_{3} \leq q_{\mu^*} - 1 + \frac{ \log \left[ 1 + \frac{\epsilon_{V}}{\lambda} \frac{m_{a} \lambda}{V} \frac{1 - \alpha}{2} \right]}{-\log\left(1 - \frac{\epsilon_V}{\lambda}\right)}. 
  \end{eqnarray*}
  Since $q_{\mu^*} \geq 0$, and $\epsilon_{V} = a_{2}V$, we have that $\overline{q}_{3}$ is at least
  \begin{eqnarray*}
    \floor{\frac{ \log \left[ 1 + m_{a} a_{2} \frac{1 - \alpha}{2} \right]}{-\log\left(1 - \frac{a_{2} V}{\lambda}\right)} - 1}.
  \end{eqnarray*}
  So that 
  \begin{eqnarray*}
    \overline{Q}(\gamma) \geq \frac{\overline{q}}{2} \geq \frac{\overline{q}_{1}}{2} \geq \frac{\overline{q}_{2}}{2} \geq \frac{\overline{q}_{3}}{2} \geq \frac{1}{2}\bras{\frac{ \log \left[ 1 + m_{a} a_{2} \frac{1 - \alpha}{2} \right]}{-\log\left(1 - \frac{a_{2} V}{\lambda}\right)} - 2}.
  \end{eqnarray*}
  Since $\log\left(1 - \frac{a_{2} V}{\lambda}\right) = \Theta\brap{V}$ as $V \downarrow 0$, we have that for any sequence $\gamma_{k}$ with $\overline{C}(\gamma_{k}) - c(\lambda) = V_{k} \downarrow 0$, $\overline{Q}(\gamma_{k}) = \Omega\nfrac{1}{V_{k}}$.
\end{proof}

\begin{remark}
  We note that as $V \downarrow 0$, $\pi(0) \rightarrow 0$.
  We also note that as $V \downarrow 0$, there exists a set of queue lengths, $\hpq$, occurring with high probability ($1 - V$), such that $\mu(q) \rightarrow \lambda, \forall q \in \hpq$.
  If $\pi(0) \rightarrow 0$, then $|\hpq| \rightarrow \infty$.
  Furthermore, for each $q \in \hpq, \pi(q) = \mathcal{O}({V})$, rather than $\mathcal{O}(\sqrt{V})$ as in Remark \ref{chap4:remark:sqrtvdiscussion}.
  We also note that the stationary probability for each $q \in \hpq$ become equal as $V \downarrow 0$.
  Then the average queue length is $\Omega\nfrac{1}{{V}}$.
  We note that the difference from the behaviour in Remark \ref{chap4:remark:sqrtvdiscussion} arises since $c(\mu)$ is piecewise linear rather than being strictly convex.
\end{remark}

\subsection{Asymptotic behaviour of the tradeoff curve}
\label{chap4:sec:intmc_ub}
In this section, we obtain asymptotic upper bounds for the cases INTERVAL-$\mu$CHOICE-1, INTERVAL-$\mu$CHOICE-2-1, INTERVAL-$\mu$CHOICE-2-2, and INTERVAL-$\mu$CHOICE-2-3.
With the asymptotic lower bounds which were derived in the previous section, these bounds provide an almost complete order characterization of the tradeoff curve $Q^*(c_{c})$.
\begin{lemma}
  For INTERVAL-$\mu$CHOICE-1, there exists a sequence of admissible policies $\gamma_{k}$ with a sequence $V_{k} \downarrow 0$, such that $\overline{Q}(\gamma_{k}) = \mathcal{O}\brap{\frac{1}{\sqrt{V_{k}}}\log\nfrac{1}{V_{k}}}$ and $\overline{C}(\gamma_{k}) - c(\lambda) = V_{k}$.
  \label{chap4:prop:p21ub}
\end{lemma}
\begin{proof}
  We evaluate the average queue length $\overline{Q}(\gamma)$ and average service cost $\overline{C}(\gamma)$ for a policy $\gamma$ defined as follows :
  \begin{eqnarray*}
    \mu(0) & = & 0, \\
    \mu(q) & = & \lambda - \epsilon_{U}, \text{ for } q \in \{1,\dots, q_{1}\}, \\
    \mu(q) & = & \lambda + \epsilon'_{U}, \text{ for } q \in \{q_{1} + 1, \dots, 2q_{1}\}, \\
    \mu(q) & = & \lambda + K, \text{ for } q \in \brac{2q_{1} + 1,\dots}.
  \end{eqnarray*}
  Let $\epsilon_{U} = \sqrt{U}$, $\epsilon'_{U} = \frac{\lambda \epsilon_{U}}{\lambda - \epsilon_{U}}$, and $K$ be some positive constant such that $\lambda + K \leq r_{max}$.
  We also let $q_{1} \stackrel{\Delta} = \left \lfloor \log_{\nfrac{\lambda}{\lambda - \epsilon_{U}}}\brap{1 + \frac{\epsilon_{U}}{U \lambda }} \right \rfloor$.
  The sequence of policies $\gamma_{k}$ is obtained by choosing $U$ from a sequence $U_{k} \downarrow 0$.
  
  For $\gamma$, we have
  \begin{eqnarray*}
    \pi(q) & = 
    \begin{cases}
      \pi(0) \fpow{\lambda}{\lambda - \epsilon_{U}}{q} & \text{ for } q \in \brac{1,\dots,q_{1}}, \\
      \pi(0) \fpow{\lambda}{\lambda - \epsilon_{U}}{q_{1}}\fpow{\lambda}{\lambda + \epsilon'_{U}}{q - q_{1}} & \text{ for } q \in \brac{q_{1} + 1, \dots, 2q_{1}}, \\
      \pi(0) \fpow{\lambda}{\lambda - \epsilon_{U}}{q_{1}}\fpow{\lambda}{\lambda + \epsilon'_{U}}{q_{1}}\fpow{\lambda}{\lambda + K}{q - 2q_{1}} = 
      \pi(0)\fpow{\lambda}{\lambda + K}{q - 2q_{1}} & \text{ for } q \in \brac{2q_{1} + 1,\dots}.
    \end{cases}
  \end{eqnarray*}
  Now, since $\sum_{q = 0}^{\infty} \pi(q) = 1$ we have that
  \begin{eqnarray*}
    \pi(0) \bras{1 + \sum_{q = 1}^{q_{1}}\fpow{\lambda}{\lambda - \epsilon_{U}}{q} + \fpow{\lambda}{\lambda - \epsilon_{U}}{q_{1}}\sum_{q = 1}^{q_{1}}\fpow{\lambda}{\lambda + \epsilon'_{U}}{q} + \sum_{q = 1}^{\infty}\fpow{\lambda}{\lambda + K}{q}} = 1, \\
    \pi(0) \bras{1 + \frac{\lambda}{\epsilon_{U}}\brap{\fpow{\lambda}{\lambda - \epsilon_{U}}{q_{1}} - 1} + \fpow{\lambda}{\lambda - \epsilon_{U}}{q_{1}}\frac{\lambda}{\epsilon'_{U}}\brap{1 - \fpow{\lambda}{\lambda + \epsilon'_{U}}{q_{1}}} + \frac{\lambda}{K}} = 1, \\
    \pi(0) \bras{1 + \frac{\lambda}{\epsilon_{U}}\brap{\fpow{\lambda}{\lambda - \epsilon_{U}}{q_{1}} - 1} + \fpow{\lambda}{\lambda - \epsilon_{U}}{q_{1}}\frac{\lambda}{\epsilon'_{U}} - \frac{\lambda}{\epsilon'_{U}} + \frac{\lambda}{K}} = 1.
  \end{eqnarray*}
  From the above expression, using the lower bound $\log_{\nfrac{\lambda}{\lambda - \epsilon_{U}}}\brap{1 + \frac{\epsilon_{U}}{U\lambda}} - 1$ on $q_{1}$, we have that
  \begin{eqnarray}
    \pi(0) & \leq & \frac{1}{1 + \frac{\lambda}{\epsilon_{U}}\brap{\nfrac{\lambda - \epsilon_{U}}{\lambda}\brap{1 + \frac{\epsilon_{U}}{U \lambda}} - 1} + \frac{\lambda - \epsilon_{U}}{\lambda}\frac{\lambda}{\epsilon'_{U}}\brap{1 + \frac{\epsilon_{U}}{U\lambda}} - \frac{\lambda}{\epsilon'_{U}} + \frac{\lambda}{K}}, \nonumber \\
    & = & \frac{\epsilon'_{U} U \lambda}{(\epsilon'_{U} + \epsilon)(\lambda(1 - U) - \epsilon_{U}) + U\lambda\epsilon'_{U}(1 + \frac{\lambda}{K})}, \nonumber \\
    & = & \mathcal{O}(U) 
    \label{chap4:eq:p21pi0bound}
  \end{eqnarray}

  We now evaluate $\overline{C}(\gamma)$.
  We have that
  \begin{eqnarray*}
    \overline{C}(\gamma) & = & \pi(0).0 + \pi_{\mu}(\lambda - \epsilon_{U}) c(\lambda - \epsilon_{U}) + \pi_{\mu}(\lambda + \epsilon'_{U}) c(\lambda + \epsilon'_{U}) + \pi_{\mu}(\lambda + K) c(\lambda + K), \\
    & = & \pi_{\mu}(\lambda - \epsilon_{U})\brap{c(\lambda) - \epsilon_{U} \frac{dc(\lambda)}{d\mu} + \mathcal{O}(\epsilon_{U}^{2})} + \pi_{\mu}(\lambda + \epsilon'_{U})\brap{c(\lambda) + \epsilon'_{U} \frac{dc(\lambda)}{d\mu} + \mathcal{O}( (\epsilon'_{U})^{2})} \nonumber \\
    & & + \pi_{\mu}(\lambda + K)\brap{c(\lambda) + K\frac{dc(\lambda)}{d\mu} + G(K)},
  \end{eqnarray*}
  where $\frac{dc(\lambda)}{d\mu} \stackrel{\Delta} = \frac{dc(\mu)}{d\mu}\vert_{\mu = \lambda}$, and $G(K) = c(\lambda + K) - c(\lambda) - K\frac{dc(\lambda)}{d\mu}$.
  Combining the $c(\lambda)$ terms, bounding $\pi_{\mu}(\lambda - \epsilon_{U}) + \pi_{\mu}(\lambda + \epsilon'_{U}) + \pi_{\mu}(\lambda + K)$ above by $1$ and using $\epsilon_{U} = \epsilon'_{U} = \mathcal{O}(\sqrt{U})$, we obtain that
  \small
  \begin{eqnarray*}
    \overline{C}(\gamma) & \leq & c(\lambda) + \mathcal{O}(U) + \brap{-\epsilon_{U}\pi_{\mu}(\lambda - \epsilon_{U}) + \epsilon'_{U}\pi_{\mu}(\lambda + \epsilon'_{U}) + K\pi_{\mu}(\lambda + K)}\frac{dc(\lambda)}{d\mu} + G(K)\pi_{\mu}(\lambda + K).
  \end{eqnarray*}
  \normalsize
  In the following, we show that $\overline{Q}(\gamma) < \infty$ and therefore $\gamma$ is admissible.
  Then, since
  \begin{eqnarray*}
    \pi_{\mu}(\lambda - \epsilon_{U})(\lambda -\epsilon_{U}) + \pi_{\mu}(\lambda + \epsilon'_{U})(\lambda + \epsilon'_{U}) + \pi_{\mu}(\lambda + K)(\lambda + K) = \lambda, \text{ therefore } \\
    \pi_{\mu}(\lambda - \epsilon_{U})(-\epsilon_{U}) + \pi_{\mu}(\lambda + \epsilon'_{U})(\epsilon'_{U}) + \pi_{\mu}(\lambda + K)(K) = \pi(0)\lambda.
  \end{eqnarray*}
  Now from \eqref{chap4:eq:p21pi0bound} we have that $\pi(0) = \mathcal{O}(U)$.
  Furthermore, $\pi_{\mu}(\lambda + K) = \pi(0)\frac{\lambda}{K} = \mathcal{O}(U)$.
  Therefore 
  \begin{eqnarray*}
    \overline{C}(\gamma) & \leq & c(\lambda) + \mathcal{O}(U) +\frac{dc(\lambda)}{d\mu} \mathcal{O}(U) + G(K)\mathcal{O}(U).
  \end{eqnarray*}
  For $V \stackrel{\Delta} = \overline{C}(\gamma) - c(\lambda)$, $V = \mathcal{O}(U)$.

  In order to obtain an upper bound on $\overline{Q}(\gamma)$, we use Proposition \ref{chap4:app:prop:dotzupperbound} with ${q}_{\epsilon} = 2q_{1} + 1$ and $\epsilon = K$, to obtain that
  \begin{eqnarray*}
    \overline{Q}(\gamma) & \leq & \frac{(2q_{1} + 1)(K + \lambda)}{K} + \frac{\lambda + r_{max}}{2K}, \\
    \overline{Q}(\gamma) & = & \mathcal{O}\brap{q_{1}} = \mathcal{O}\brap{\frac{1}{\sqrt{U}}\log\nfrac{1}{U}}.
  \end{eqnarray*}
  Hence $\gamma$ is admissible.
  
  Corresponding to the sequence $U_{k}$, we have a sequence $V_{k} = \mathcal{O}(U_{k})$.
  Therefore, we have a sequence of policies $\gamma_{k}$ with $\overline{Q}(\gamma_{k}) = \mathcal{O}\brap{\frac{1}{\sqrt{U_{k}}}\log\nfrac{1}{U_{k}}}$.
  Then $\overline{Q}(\gamma_{k}) = \mathcal{O}\brap{\frac{1}{\sqrt{V_{k}}}\log\nfrac{1}{V_{k}}}$ and $\overline{C}(\gamma) - c(\lambda) = V_{k}$.
\end{proof}
We note that the asymptotic upper bound above for the sequence $\gamma_{k}$ does not match the asymptotic lower bound $\Omega\nfrac{1}{\sqrt{V_{k}}}$, which was derived in Lemma \ref{chap4:prop:p21lb}.

\begin{lemma}
  For INTERVAL-$\mu$CHOICE-2-1, there exists a sequence of admissible policies $\gamma_{k}$, with a sequence of $V_{k} \downarrow 0$, such that $\frac{\lambda}{b_{\lambda} - \lambda} - \overline{Q}(\gamma_{k}) = \Theta\brap{V_{k} \log \nfrac{1}{V_{k}}}$ and $\overline{C}(\gamma_{k}) - c(\lambda) = V_{k}$.
  \label{chap4:prop:p221ub}
\end{lemma}

\begin{proof}
  Consider the policy $\gamma$ defined as follows :
  \begin{eqnarray*}
    \mu(0) & = & 0, \\
    \mu(q) & = & b_{\lambda}, \text{ for } q \in \{1,\dots, q_{b_{\lambda}}\}, \\
    \mu(q) & = & r_{max}, \text{ for } q \in \{q_{b_{\lambda}} + 1, \dots\}.
  \end{eqnarray*}
  The sequence $\gamma_{k}$ is obtained by choosing $q_{b_{\lambda}} = k \in \mathbb{Z}_+$.
  The rest of the proof is similar to the proof of Lemma \ref{chap4:prop:p11ub}.
\end{proof}

\begin{remark}
  We note that the above asymptotic upper bound does not match the asymptotic lower bound in Lemma \ref{chap4:prop:p221lb}.
  Since $\mu(q) \in [0, r_{max}]$, one expects that perhaps another sequence of policies for which $\mu(q) = b_{\lambda} + \epsilon_{V}$, for $q \in \{1,\dots, q_{b_{\lambda}}\}$ and where $\epsilon_{V}$ is a sequence decreasing to zero, achieves a better asymptotic upper bound.
  We have found out that this is not the case, and the asymptotic upper bound is the same as the one above.
\end{remark}

\begin{lemma}
  For INTERVAL-$\mu$CHOICE-2-2, there exists a sequence of admissible policies $\gamma_{k}$, with a sequence $V_{k} \downarrow 0$, such that $\overline{Q}(\gamma_{k}) = \mathcal{O}\brap{\log\nfrac{1}{V_{k}}}$ and $\overline{C}(\gamma_{k}) - c(\lambda) = V_{k}$.
  \label{chap4:prop:p222ub}
\end{lemma}

\begin{proof}
  Consider a policy $\gamma$ defined as follows :
  \begin{eqnarray*}
    \mu(0) & = & 0, \\
    \mu(q) & = & a_{\lambda}, \text{ for } q \in \{1,\dots,q_{1}\}, \\
    \mu(q) & = & b_{\lambda}, \text{ for } q \in \{q_{1} + 1, \dots\},
  \end{eqnarray*}
  where $q_{1} \stackrel{\Delta} = \left\lceil \log_{\left(\frac{\lambda}{a_{\lambda}}\right)} \left(1 + \frac{\lambda - a_{\lambda}}{\lambda} \frac{{1}}{U} \right) \right \rceil$, with $U > 0$.
  The sequence $\gamma_{k}$ is obtained by choosing $U$ from a sequence $U_{k} \downarrow 0$.
  The rest of the proof is similar to the proof of Lemma \ref{chap4:prop:p12ub}.
  Since $V_{k} = \mathcal{O}\brap{U_{k}}$, we obtain $\overline{Q}(\gamma_{k}) = \mathcal{O}\brap{\log\nfrac{1}{V_{k}}}$.
\end{proof}

Using the asymptotic lower bound on $\overline{Q}(\gamma_{k})$ from Lemma \ref{chap4:prop:p222lb}, and the asymptotic upper bound above, and proceeding as in the proof of Proposition \ref{chap4:prop:p11}, we obtain the following result.
\begin{proposition}
  For INTERVAL-$\mu$CHOICE-2-2, we have that the optimal tradeoff curve $Q^*(c_{c,k})$ is $\Theta\left(\log\nfrac{1}{c_{c,k} - c(\lambda)}\right)$, for a sequence $c_{c,k} = \overline{C}(\gamma_{k})$, where $\gamma_{k}$ is the sequence of policies in Lemma \ref{chap4:prop:p222ub}.
  \label{chap4:prop:p222}
\end{proposition}

\begin{lemma}
  For INTERVAL-$\mu$CHOICE-2-3, there exists a sequence of admissible policies $\gamma_{k}$, with a sequence $V_{k} \downarrow 0$, such that $\overline{Q}(\gamma_{k}) = \mathcal{O}{\nfrac{1}{V_{k}}}$ and $\overline{C}(\gamma) - c(\lambda) = V_{k}$.
  \label{chap4:lemma:p223ub}
\end{lemma}

\begin{proof}
  Consider a policy $\gamma$ defined as follows :
  \begin{eqnarray*}
    \mu(0) & = & 0, \\
    \mu(q) & = & \lambda, \text{ for } q \in \brac{1,\dots,q_{1}}, \\
    \mu(q) & = & \lambda + K, \text{ for } q \in \brac{q_{1} + 1, \dots}.
  \end{eqnarray*}
  We note that $\lambda = a_{i}$, for some $i > 1$.
  Here $K$ is a constant such that $\lambda + K \leq a_{i + 1}$.
  We define $q_{1} = \left \lceil \frac{{1}}{U} \right \rceil$, with $U > 0$.
  The sequence $\gamma_{k}$ is obtained by choosing $U$ from a sequence $U_{k} \downarrow 0$.
  The rest of the proof is similar to that of Lemma \ref{chap4:prop:p13ub}.
  Since $V_{k} = \mathcal{O}\brap{U_{k}}$, we obtain $\overline{Q}(\gamma_{k}) = \mathcal{O}\nfrac{1}{V_{k}}$.
\end{proof}

Using the asymptotic lower bound on $\overline{Q}(\gamma_{k})$ from Lemma \ref{chap4:prop:p223lb}, and the asymptotic upper bound above, and proceeding as in the proof of Proposition \ref{chap4:prop:p11}, we obtain the following result
\begin{proposition}
  For INTERVAL-$\mu$CHOICE-2-3, we have that the optimal tradeoff curve $Q^*(c_{c,k})$ is $\Theta\nfrac{1}{c_{c,k} - c(\lambda)}$, for a sequence $c_{c,k} = \overline{C}(\gamma_{k})$, where $\gamma_{k}$ is the sequence of policies in Lemma \ref{chap4:lemma:p223ub}.
  \label{chap4:prop:p223}
\end{proposition}

\subsection{Asymptotic characterization of order-optimal admissible policies}
\label{chap4:section:aschar_optpolicy_intmc}
As in Chapter 2, it is possible to obtain an asymptotic characterization of any sequence of order-optimal admissible policies using the bounds $P_{l}\brac{.}, P_{u}\brac{.}, \pi_{l}(.)$, and $\pi_{u}(.)$ and the inequalities \eqref{chap4:eq:ascharpolicy_obs1} and \eqref{chap4:eq:ascharpolicy_obs2}.
We only consider the cases where the minimum average queue length increases to infinity in the asymptotic regime $\Re$.
The characterization of order-optimal policies for \INTMC-2-2 and \INTMC-2-3 can be obtained using similar methods as \FMC-2 and \FMC-3.
Therefore, in this section we discuss the asymptotic characterization of the optimal policy for \INTMC-1 only.
We first obtain $P_{l}\brac{A}$ and $P_{u}\brac{A}$, where $A \subseteq [0, r_{max}]$ is a set of service rates.

Let $0 \leq \delta \leq 1$ and $a_{2} > 0$.
Let $A \subseteq [0, \lambda - a_{2}V^{\frac{1 - \delta}{2}}] \bigcup [\lambda + a_{2} V^{\frac{1 - \delta}{2}}, r_{max}]$.
From \eqref{chap4:eq:v_variance}, we have that 
\begin{eqnarray}
  Pr\brac{\mu(Q) \in A} \leq P_{u}\brac{A} \Deq \frac{V^{\delta}}{a_{1}a_{2}}, 
  \label{chap4n:eq:aschar1}
\end{eqnarray}
if $\delta > 0$.
Suppose $\delta = 0$, then for $a_{2} > \frac{1}{a_{1}}$, we have that 
\begin{eqnarray}
  Pr\brac{\mu(Q) \in A} \leq P_{u}\brac{A} \Deq \frac{1}{a_{1}a_{2}}.
  \label{chap4n:eq:aschar2}
\end{eqnarray}

We then note that $P_{l}\brac{A^{c}} \Deq 1 - \frac{V^{\delta}}{a_{1}a_{2}}$ if $\delta > 0$ and $P_{l}\brac{A^{c}} \Deq 1 - \frac{1}{a_{1}a_{2}}$ if $\delta = 0$ and $a_{2} > \frac{1}{a_{1}}$.

The upper bound $\pi_{u}(q)$ that we use in the following is obtained as in \eqref{chap4:eq:p2-referfromINTLC_1}.
We now proceed as in Chapter 2 to obtain a lower bound $\pi_{l}(0)$ on $\pi(0)$.
From Lemma \ref{chap4:prop:p21ub}, we have that for any sequence of non-idling order-optimal policies $\gamma_{k}$, $\Qgk = \mathcal{O}\brap{\frac{1}{\sqrt{V_{k}}}\log\nfrac{1}{V_{k}}}$.
We use the above upper bound since we are not able to show that $\Qgk = \mathcal{O}\nfrac{1}{\sqrt{V_{k}}}$.
Therefore, as in Section \ref{chap4:sec:optimalpolicy_aschar}, we have that 
\begin{eqnarray}
  \pi_{l}(0) & = & \Omega\brap{\frac{\sqrt{V}}{\log\nfrac{1}{V}} \brap{1 - \frac{\lambda}{r_{u}}}^{\brap{\frac{1}{\sqrt{V}}\log\nfrac{1}{V}}}},
  \label{chap4n:eq:aschar3} \\
  & = & \Omega\brap{\frac{\sqrt{V}}{\log\nfrac{1}{V}} \frac{1}{\nfrac{1}{1 - \frac{\lambda}{r_{u}}}^{\brap{\frac{1}{\sqrt{V}}\log\nfrac{1}{V}}}}} \nonumber.
\end{eqnarray}
With $k \Deq \frac{1}{\log_{\frac{1}{\brap{1 - \frac{\lambda}{r_{u}}}}} e}$ we then have that
\begin{eqnarray}
  \pi_{l}(0) & = & \Omega\brap{\frac{\sqrt{V}}{\log\nfrac{1}{V}} \frac{1}{\nfrac{1}{1 - \frac{\lambda}{r_{u}}}^{\log_{\frac{1}{\brap{1 - \frac{\lambda}{r_{u}}}}} \bras{\nfrac{1}{V}}^{\brap{\frac{k}{\sqrt{V}}}}}}} = \Omega\nfrac{V^{\brap{\frac{k}{\sqrt{V}} + \frac{1}{2}}}}{\log\nfrac{1}{V}}.
  \label{chap4n:eq:aschar4}
\end{eqnarray}
Furthermore, we have the following lower bound $\pi_{l}(q')$ on $\pi(q')$, where $q' = \max\brac{q : \mu(q) < \lambda}$.

We note that for $\mu < \lambda$, there exists a $m_{1} > 0$ such that $c(\mu) - l(\mu) \leq m_{1}\brap{\lambda - \mu}$.
Also for $\mu \geq \lambda$, there exists a $m_{2} > 0$ such that $c(\mu) - l(\mu) \leq m_{2} \brap{\mu - \lambda}$.
Then we have that 
\[V = \Exp\bras{c(\mu(Q)) - l(\mu(Q))} \leq \sum_{q : \mu(q) < \lambda} m_{1}\brap{\lambda - \mu(q)} \pi(q) + \sum_{q : \mu(q) \geq \lambda} m_{2}\brap{\mu(q) - \lambda}\pi(q).\]
\begin{eqnarray}
  V & \leq & m_{1} \sum_{q \leq q'} \brap{\lambda - \mu(q)} \pi(q) + m_{2} \sum_{q > q'} \brap{\mu(q) - \lambda} \pi(q), \nonumber \\
  & = & m_{1}\bras{ \lambda \pi(0) + \sum_{1 \leq q \leq q'} \brap{\lambda \pi(q) - \lambda \pi(q - 1)}} + m_{2} \sum_{q > q'} \brap{\pi(q - 1) - \pi(q)} \lambda, \nonumber \\
  & \leq & m_{1} \lambda \pi(q') + m_{2} \lambda \pi(q'), \text{ or}, \nonumber \\
  \pi(q') & \geq & \pi_{l}(q') \Deq \frac{V}{\lambda\brap{m_{1} + m_{2}}}.
  \label{chap4n:eq:aschar5}
\end{eqnarray}
We note that $q'$ could be zero, in which case the sum $\sum_{1 \leq q \leq q'} \brap{\lambda \pi(q) - \lambda \pi(q - 1)}$ is defined to be zero.

\begin{lemma}
  For any sequence of non-idling order-optimal admissible policies $\gamma_{k}$, with $\Cgk - c(\lambda) = V_{k} \downarrow 0$, and $Q_{A} = \brac{q : \mu(q) \in A}$ for a $A \subseteq [0, r_{max}]$, we have that
  \begin{eqnarray*}
    |Q_{A}| & = & 
    \begin{cases}
      \mathcal{O}\brap{\frac{1}{V_{k}}\log\nfrac{1}{V_{k}}}, \text{ if } A = \bras{0, \lambda - a_{2} V_{k}^{\nfrac{1 - \delta}{2}}}, \\
      \Omega\nfrac{1}{\sqrt{V_{k}}}, \text{ if } \bras{\lambda - a_{2} V_{k}^{\nfrac{1}{2}}, \lambda + a_{2} V_{k}^{\nfrac{1}{2}}} \subseteq A, \\
      \mathcal{O}\nfrac{1}{V_{k}}, \text{ if } A = \brac{\lambda},
    \end{cases}
  \end{eqnarray*}
  where $0 \leq \delta \leq 1$, and $a_{2} > \frac{1}{a_{1}}$ if $\delta = 0$.
  \label{chap4:lemma:intmcsqrt_policy_aschar}
\end{lemma}

\begin{proof}
  Consider a particular policy $\gamma$ in the above sequence with $\Cgk - c(\lambda) = V$.
  Let $A = \bras{0, \lambda - a_{2} V^{\nfrac{1 - \delta}{2}}}$.
  Then, from \eqref{chap4n:eq:aschar1} we have that $P_{u}\brac{A} = \frac{V^{\delta}}{a_{1}a_{2}}$.
  Consider a $q$ such that $\mu(q) \in A$.
  Then $\pi(q) \geq \pi_{l}(q)$, where
  \begin{eqnarray*}
    \pi_{l}(q) \Deq \pi_{l}(0) \fpow{\lambda}{\lambda - a_{2} V^{\nfrac{1 - \delta}{2}}}{q},
  \end{eqnarray*}
  and $\pi_{l}(0)$ is given in \eqref{chap4n:eq:aschar4}.
  As in \eqref{chap4:eq:ascharpolicy_obs2} if $q_{1} = \max\brac{q : \mu(q) \in A}$, then the smallest integer $q_{1,u}$ such that
  \begin{eqnarray*}
    \sum_{q = 0}^{q_{1,u}} \pi_{l}(q) \geq \frac{V^{\delta}}{a_{1}a_{2}},
  \end{eqnarray*}
  is an upper bound on $q_{1}$.
  We have that 
  \begin{eqnarray*}
    \pi_{l}(0) \brap{\fpow{\lambda}{\lambda - a_{2} V^{\nfrac{1 - \delta}{2}}}{q_{1,u} + 1} - 1} \geq \frac{V^{\nfrac{1 + \delta}{2}}}{a_{1} \brap{\lambda - a_{2} V^{\nfrac{1 - \delta}{2}}}}
  \end{eqnarray*}
  From \eqref{chap4n:eq:aschar4}, we have that for small enough $V$, there exists a $a_{3} > 0$ such that 
  \begin{eqnarray*}
    \pi_{l}(0) & \geq & \nfrac{a_{3}V^{\brap{\frac{k}{\sqrt{V}} + \frac{1}{2}}}}{\log\nfrac{1}{V}}.
  \end{eqnarray*}
  So if $q_{1,u}'$ is the smallest integer such that
  \begin{eqnarray*}
    \brap{\fpow{\lambda}{\lambda - a_{2} V^{\nfrac{1 - \delta}{2}}}{q_{1,u}' + 1} - 1} \geq \frac{V^{\nfrac{1 + \delta}{2}}}{a_{1} \brap{\lambda - a_{2} V^{\nfrac{1 - \delta}{2}}}} \nfrac{\log\nfrac{1}{V}}{a_{3}V^{\brap{\frac{k}{\sqrt{V}} + \frac{1}{2}}}},
  \end{eqnarray*}
  then $q_{1,u}' \geq q_{1,u}$.
  Or we have that $q_{1,u}'$ is the smallest integer such that
  \begin{eqnarray*}
    q_{1,u}' \geq \frac{\log\brap{\frac{V^{\nfrac{1 + \delta}{2}}}{a_{1} \brap{\lambda - a_{2} V^{\nfrac{1 - \delta}{2}}}} \nfrac{\log\nfrac{1}{V}}{a_{3}V^{\brap{\frac{k}{\sqrt{V}} + \frac{1}{2}}}}}}{\log\nfrac{\lambda}{\lambda - a_{2} V^{\nfrac{1 - \delta}{2}}}}
  \end{eqnarray*}
  As $V \downarrow 0$, we have that $q_{1,u}' = \mathcal{O}\brap{\frac{1}{V}\log\nfrac{1}{V}}$.
  If $Q_{A} = \brac{q : \mu(q) \in A}$, then $|Q_{A}| = \mathcal{O}\brap{\frac{1}{V}\log\nfrac{1}{V}}$.

  Now consider $A = \bras{\lambda - a_{2} V^{\nfrac{1}{2}}, \lambda + a_{2} V^{\nfrac{1}{2}}}$, where $a_{2} > \frac{1}{a_{1}}$.
  We have that $P_{l}(A) = 1 - \frac{1}{a_{1}a_{2}}$.
  Let $q_{1} = \max\brac{q : \mu(q) < \lambda - a_{2} V^{\nfrac{1}{2}}}$.
  Then from \eqref{chap4:eq:tighterupperbound} we have that $\pi(q_{1}) \leq \sqrt{\frac{V}{a_{1}\lambda^{2}}}$.
  Then for every $q \in A$, we have that
  \begin{eqnarray}
    \pi(q) \leq \pi_{u}(q) \Deq \sqrt{\frac{V}{a_{1}\lambda^{2}}}\fpow{\lambda}{\lambda - a_{2} V^{\nfrac{1}{2}}}{q - q_{1}}.
  \end{eqnarray}
  Let $Q_{A} = \brac{q : \mu(q) \in A}$.
  We now obtain a lower bound on $|Q_{A}|$.
  From \eqref{chap4:eq:ascharpolicy_obs1}, if $q_{l,a}$ is the largest integer such that
  \begin{eqnarray}
    \sum_{q = 0}^{q_{l,a}} \sqrt{\frac{V}{a_{1}\lambda^{2}}}\fpow{\lambda}{\lambda - a_{2} V^{\nfrac{1}{2}}}{q} \leq 1 - \frac{1}{a_{1}a_{2}},
  \end{eqnarray}
  then $q_{l,a} \leq |Q_{A}|$.
  This is equivalent to finding the largest integer $q_{l,a}$ such that
  \begin{eqnarray*}
    \sqrt{\frac{V}{a_{1}\lambda^{2}}} \brap{\fpow{\lambda}{\lambda - a_{2} V^{\nfrac{1}{2}}}{q_{l,a}} - 1} & \leq & \brap{1 - \frac{1}{a_{1}a_{2}}}\frac{a_{2}V^{\frac{1}{2}}}{\lambda - a_{2} V^{\frac{1}{2}}}, \\
    \fpow{\lambda}{\lambda - a_{2} V^{\nfrac{1}{2}}}{q_{l,a}} - 1 & \leq & \brap{1 - \frac{1}{a_{1}a_{2}}}\frac{a_{1} \sqrt{a_{2} \lambda^{2}}}{\lambda - a_{2} V^{\frac{1}{2}}}.
  \end{eqnarray*}
  Then as in the proof of Lemma \ref{chap4:prop:p21lb} we have that $q_{l,a} = \Omega\nfrac{1}{\sqrt{V}}$.

  Consider $Q_{\lambda} = \brac{q : \mu(q) = \lambda}$.
  We note that $\pi(Q_{\lambda}) \leq 1$.
  We also note that for every $q \in Q_{\lambda}$, $\pi(q) \leq \pi_{l}(q) \Deq \pi_{l}(q')$ defined in \eqref{chap4n:eq:aschar5}.
  As in \eqref{chap4:eq:ascharpolicy_obs2}, if $q_{u,\lambda}$ is the smallest integer such that
  \begin{eqnarray*}
    q_{u,\lambda} \pi_{l}(q') \leq 1,
  \end{eqnarray*}
  then $q_{u,\lambda}$ is an upper bound on $|Q_{\lambda}|$.
  Hence we obtain that $|Q_{\lambda}| = \mathcal{O}\nfrac{1}{V}$.
\end{proof}

\begin{remark}
  From the above proof, the intuition behind choosing the buffer partition to scale as $\Omega\nfrac{1}{\sqrt{V}}$ can be observed.
\end{remark}

\subsection{Tradeoff problems which are similar to INTERVAL-$\mu$CHOICE}
\label{chap4:sec:similarintmc}
In this section, we consider tradeoff problems which are similar to \INTMC, for which an asymptotic characterization can be obtained using the techniques presented above for INTERVAL-$\mu$CHOICE.
We note that for INTERVAL-$\mu$CHOICE, we restricted to admissible policies $\gamma$ for which $\lambda(q) = \lambda, \forall q \in \sZ$, and $\lambda$ was such that $u(\lambda) \geq u_{c}$.
The counterpart \INTLC\, of \INTMC\, is one in which we restrict to admissible policies $\gamma$ for which $\mu(q) = \mu, \forall q \in \brac{1,2,\dots}$ and $\lambda(q) \in [0, r_{a,max}]$.

We note that $\overline{C}(\gamma)$ for such a policy $\gamma$ is $(1 - \pi(0))c(\mu)$, which depends on the policy, unlike \INTMC\, where the choice of $\lambda$ fixed $\overline{U}(\gamma)$ to be $u(\lambda)$.
For \INTLC, we restrict to admissible $\gamma$ such that $\mu(q) = \mu, \forall q > 0$, where $\mu$ is such that $c(\mu) \leq c_{c}$, so that $\overline{C}(\gamma) \leq c_{c}$.
The tradeoff problem INTERVAL-$\lambda$CHOICE is
\begin{eqnarray}
  \text{ minimize }_{\gamma \in \Gamma_{a}} & \overline{Q}(\gamma) \nonumber \\
  \text{ and } & \overline{U}(\gamma) \geq u_{c}.
  \label{chap4:eq:sysmodel:alternate_probstat}
\end{eqnarray}
The optimal value of the above problem is denoted as $Q^*(u_{c})$.
We also note from Lemma \ref{chap4:lemma:feasible_solns}, that the maximum value of $\Ug$ over all admissible $\gamma$ is $u(\mu)$.
We obtain an asymptotic characterization of $Q^*(u_{c})$ in the asymptotic regime $\Re$, where $u_{c} \uparrow u(\mu)$.

We assume that $\mu$ is such that $r_{a,max} > \mu$.
If $r_{a,max} = \mu$, then we note that only the non-admissible policy $\gamma$, with $\lambda(q) = r_{a,max}, \forall q$, can achieve $u(\mu)$.
Intuitively, if $u_{c} \uparrow u(\mu)$, then since $\Exp \lambda(Q) = \Exp \mu(Q) = (1 - \pi(0))\mu \uparrow \mu$, we have that $\pi(0) \downarrow 0$.
Then for any policy which is feasible as $u_{c} \uparrow u(\lambda)$, $\pi(0) \downarrow 0$.
Thus, intuitively for problem \eqref{chap4:eq:sysmodel:alternate_probstat} we do not have a case where $Q^*(u_{c})$ increases only up to a finite value as $u_{c} \uparrow u(\mu)$ (unlike \INTMC-2-1).

As for INTERVAL-$\mu$CHOICE, we consider the following cases for INTERVAL-$\lambda$CHOICE:
\begin{description}
\item[INTERVAL-$\lambda$CHOICE-1:] $u(\lambda)$ is strictly concave for $\lambda \in [0,r_{a,max}]$. 
\item[INTERVAL-$\lambda$CHOICE-2:] $u(\lambda)$ is piecewise linear and concave. 
  That is, (a) there exists a minimal partition of $[0,r_{a,max}]$ into intervals $\{[a_{i},b_{i}], i \in \{1,\dots,P\}\}$ with $a_{1} = 0$, $b_{P} = r_{max}$, and $b_{i} = a_{i + 1}$ and (b) there are linear functions $f_{i}$ such that $\forall \mu \in [a_{i},b_{i}], f_{i}(\lambda) = u(\lambda)$. This is further subdivided into two cases:
  \begin{description}
  \item[1.] $a_{\mu} \Deq a_{i} < \mu < b_{\mu} \Deq b_{i}$, for some $i \geq 1$.
  \item[2.] $\mu = a_{\mu} \Deq a_{i}$, for some $i > 1$.
  \end{description}
\end{description}
\begin{remark}
  We note that for the discrete time queueing model, we only consider the case where the utility function is linear, since for such models we are interested in the average throughput as the performance measure.
  We note that \INTLC-2-1 encompasses the case of linear utility functions.
  For linear utility functions, we can motivate the choice of system parameters for \INTLC-2-1 as done for the case of \INTMC.
  \INTLC-2-1 with linear utility function is a simplified model for a discrete time queueing model, where the service batch size is \emph{fixed} (but if the queue length is less than this fixed batch size, then the service batch size is equal to the queue length) and there is randomized admission control.
  The fixed service batch size is modelled by the fixed $\mu$, while the admission control is modelled by the choice of the arrival rate $\lambda(q)$ as a function of $q$.
  In light of the discussion in Section \ref{chap4:sec:intintro} we assume that $\lambda(q)$ takes values in a finite interval.
\end{remark}

We now present an asymptotic characterization of $Q^*(u_{c})$ in the regime $u_{c} \uparrow u(\mu)$.
We note that for \INTLC-1 as well as for \INTLC-2 it is possible to show (see Lemma \ref{chap4:lemma:intlc_ub}) that there exists a sequence of admissible policies $\gamma_{k}$ such that $\Ugk \uparrow u(\mu)$.
We note that for every $u_{c} < u(\mu)$, $\forall \epsilon > 0$, there exists some feasible $\gamma \in \Gamma_{a}$ such that $\overline{Q}(\gamma) \leq Q^*(u_{c}) + \epsilon$.
Such an admissible policy is called $\epsilon$-optimal in the following.

We first present asymptotic lower bounds on $Q^*(u_{c})$ in the regime $u_{c} \uparrow u(\mu)$.
Asymptotic lower bounds for \INTLC\, are obtained along similar lines as for \INTMC.
For all cases, we first obtain upper bounds on the stationary probability of certain arrival rates (rather than service rates), which go to zero as $u_{c} \uparrow u(\mu)$.
Then as before, these upper bounds on the stationary probability of certain arrival rates lead to constraints on the stationary probability of all queue lengths.
Since the stationary probability of the queue length determines the average queue length, the constraints determine the behaviour of average queue length as $u_{c} \uparrow u(\mu)$.

For the asymptotic analysis of \INTLC, we define a line $l(\lambda)$ whose definition is similar to that of $l(\mu)$ for \INTMC.
For \INTLC-1, $l(\lambda)$ is defined as the tangent to $u(\lambda)$ at $\lambda = \mu$.
For \INTLC-2-1, $l(\lambda)$ is defined as the line through $(a_{\mu}, u(a_{\mu}))$ and $(b_{\mu}, u(b_{\mu}))$, while for \INTLC-2-2, $l(\lambda)$ is any line through $(a_{\mu}, u(a_{\mu}))$ with slope $m$, such that $\frac{du(\lambda)}{d\lambda}^{-}\vert_{\lambda = \mu} < m < \frac{du(\lambda)}{d\lambda}^{+}\vert_{\lambda = \mu}$.
We note that $l(\lambda) \geq u(\lambda)$ and $\Exp l(\lambda(Q)) = l(\Exp\lambda(Q))$.
We also note that the function $l(\lambda) - u(\lambda)$ is a convex function.

We now present a result, which formalizes the intuition that $\pi(0)\downarrow 0$ as $u_{c} \uparrow u(\mu)$.
\begin{lemma}
  For \INTLC, for any sequence of admissible policies such that $u(\mu) - \Ugk = V_{k} \downarrow 0$, we have that $\pi(0) = \mathcal{O}(V_{k})$.
  Therefore, as $u_{c} \uparrow u(\mu)$, $\pi(0) \downarrow 0$, for any sequence of feasible policies for \eqref{chap4:eq:sysmodel:alternate_probstat}.
  \label{chap4:lemma:intlc_pi0ub}
\end{lemma}
\begin{proof}
  Consider a particular policy $\gamma$ in the sequence with $V_{k} = V$.
  We note that $\Ug \leq u(\Exp \lambda(Q))$.
  Since $\gamma$ is admissible, we have that $\Ug \leq u(\Exp \mu(Q))$.
  We then have that $u^{-1}(\Ug) \leq \Exp \mu(Q) = (1 - \pi(0)) \mu$, since $u^{-1}(.)$ exists if $u(\lambda)$ is concave and increasing in $\lambda$.
  Therefore, we have that
  \begin{eqnarray*}
    \pi(0) \leq 1 - \frac{u^{-1}(\Ug)}{\mu} = 1 - \frac{u^{-1}(u(\mu) - V)}{\mu}.
  \end{eqnarray*}
  We note that $u^{-1}(x) \geq l^{-1}(x)$, $x \in \sR$, where $l^{-1}(.)$ is the inverse function of $l(\lambda)$.
  Then we have 
  \begin{eqnarray*}
    \pi(0) \leq 1 - \frac{l^{-1}(u(\mu) - V)}{\mu} = 1 - \frac{l^{-1}(u(\mu)) - mV}{\mu} = \frac{mV}{\mu},
  \end{eqnarray*}
  since $u(\mu) = l(\mu)$ and where $m$ is the slope of $l^{-1}$.
  Therefore, for the sequence $\gamma_{k}$, $\pi(0) = \mathcal{O}(V_{k})$.
  For \INTLC, as $u_{c} \uparrow u(\mu)$, for any sequence $\gamma_{k}$ of feasible policies, $u(\mu) - \Ugk \downarrow 0$ and hence $\pi(0) \downarrow 0$.
\end{proof}

For any policy, the set of arrival rates $\brac{\lambda(q) : q \in \sZ}$ is countable and is denoted as $(\lambda_{0}, \lambda_{1}, \dots)$, with $\lambda_{k} < \lambda_{k + 1}$.
For an admissible policy, let $\pi_{\lambda}(k)$ denote the stationary probability of using an arrival rate $\lambda_{k}$, i.e., $\pi_{\lambda}(k) = Pr\brac{\lambda(Q) = \lambda_{k}} = \sum_{\brac{q : \lambda(q) = \lambda_{k}}} \pi(q)$.

We make the following assumption, which is similar to (C2):
\begin{description}
\item[U2:] For \INTLC-1, the second derivative of $u(\lambda)$ at $\lambda = \mu$ is non-zero.
\end{description}
\begin{lemma}
  For \INTLC-1, for any sequence of non-idling admissible policies $\gamma_{k}$ such that $u(\mu) - \Ugk = V_{k} \downarrow 0$, we have that $\Qgk = \Omega{\nfrac{1}{\sqrt{V_{k}}}}$. Therefore, $Q^*(u_{c}) = \Omega{\nfrac{1}{\sqrt{u(\mu) - u_{c}}}}$.
\end{lemma}
\begin{proof}
  The proof follows that of Lemma \ref{chap4:prop:p21lb}, but with some minor differences.
  We again consider a particular policy in the sequence with $V_{k} = V$.
  Let $\lambda^* \Deq  \mu + \epsilon_{V}$, where $\epsilon_{V} > 0$ is a function of $V$ to be chosen later.
  As in the proof of Lemma \ref{chap4:prop:p21lb}, we have that 
  \begin{eqnarray*}
    V & = & \sum_{q = 0}^{\infty} \pi(q) \bras{\brap{\mu - \lambda(q)}\frac{du(\lambda)}{d\lambda}\vert_{\lambda = \mu} + G(\lambda(q) - \mu)},
  \end{eqnarray*}
  where $G(x)$ is a strictly convex function of $x$.
  As $\lambda(q) - \mu$ is bounded, we again have that there exists a positive $a_{1}$ such that 
  \begin{eqnarray*}
    V & \geq & \sum_{q = 0}^{\infty} \pi(q) \bras{\brap{\mu - \lambda(q)}\frac{du(\lambda)}{d\lambda}\vert_{\lambda = \mu}  + a_{1}(\lambda(q) - \mu)^{2}}.
  \end{eqnarray*}
  Since $\sum_{q = 0}^{\infty} \pi(q) \leq \mu$, we have that
  \begin{eqnarray*}
    V & \geq & \sum_{q = 0}^{\infty} \pi(q) a_{1}(\lambda(q) - \mu)^{2}.
  \end{eqnarray*}
  Let $q_{\lambda^*} \Deq \inf\brac{q : \lambda(q) \leq \lambda^*}$.
  We note that unlike $q_{\mu^*}$ in Lemma \ref{chap4:prop:p21lb}, $q_{\lambda^*}$ could be $0$.
  We proceed as in the proof of Lemma \ref{chap4:prop:p21lb} by choosing $\epsilon_{V} = a_{2} \sqrt{V}$.
  Then $Pr\brac{Q < q_{\lambda^*}} \leq \frac{1}{a_{1}a_{2}}$ if $q_{\lambda^*} > 0$.
  As before, we choose $a_{2}$ such that $Pr\brac{Q < q_{\lambda^*}} \leq \frac{\alpha}{2}$, where $\alpha$ can be made arbitrarily close to zero.
  If $q_{\lambda^*} = 0$, then $Pr\brac{Q < q_{\lambda^*}} = 0 \leq \frac{1}{a_{1}a_{2}}$.
  
  As in the proof of Lemma \ref{chap4:prop:p21lb} we find the largest $\overline{q}$ such that $Pr\brac{Q \leq \overline{q}} \leq \frac{1}{2}$.
  We note that if $q \geq q_{\lambda^*}$, then $\pi(q - 1)\lambda^* \geq \pi(q) \mu$.
  Then by induction we obtain that for any $q \geq q_{\lambda^*}$,
  \begin{eqnarray}
    \sum_{q = q_{\lambda^*}}^{q} \pi(q) \leq \pi(q_{\lambda^*}) \sum_{m = 0}^{q - q_{\lambda^*}} \fpow{\lambda^*}{\mu}{m}.
    \label{chap4:eq:intlc_indeq}
  \end{eqnarray}
  We note that this is similar to \eqref{chap4:eq:p2-referfromINTLC_1}, except that we express the above upper bound in terms of $\pi(q_{\lambda^*})$ rather than $\pi(q_{\mu^*} - 1)$ in \eqref{chap4:eq:p2-referfromINTLC_1}, since $q_{\lambda^*}$ could be zero.

  If $q_{\lambda^*} = 0$, then from Lemma \ref{chap4:lemma:intlc_pi0ub} we have that $\pi(q_{\lambda^*}) = \pi(0) = \mathcal{O}(V)$.
  If $q_{\lambda^*} > 0$, we obtain an upper bound on $\pi(q_{\lambda^*})$, as in the proof of Lemma \ref{chap4:prop:p21lb}.
  We have that 
  \begin{eqnarray*}
    \frac{V}{a_{1}} & \geq & \sum_{q < q_{\lambda^*}} \pi(q) (\lambda(q) - \mu)^{2} \geq \brap{\sum_{q < q_{\lambda^* - 1}}(\lambda(q) - \mu)\pi(q)}^{2}.
  \end{eqnarray*}
  Since for $q > 0$, since $\pi(q)\lambda(q) = \mu \pi(q + 1)$, we proceed as in the proof of Lemma \ref{chap4:prop:p21lb} to obtain that
  \begin{eqnarray*}
    \frac{V}{a_{1}} & \geq & \brap{\mu \pi(q_{\lambda^*}) - \mu \pi(0)}^{2}, \\
    & = & \mu^{2} \pi(q_{\lambda^*})^{2} + \mu^{2} \pi(0)^{2} - 2\mu^{2} \pi(q_{\lambda^*}) \pi(0).
  \end{eqnarray*}
  Since $\pi(0) \geq 0$ and $\pi(0) = \mathcal{O}(V)$ from Lemma \ref{chap4:lemma:intlc_pi0ub}, we have that
  \begin{eqnarray*}
    \frac{V}{a_{1}} + 2\mu^{2} \pi(q_{\lambda^*}) \pi(0) & \geq & \mu^{2} \pi(q_{\lambda^*})^{2}, \\
    \frac{V}{a_{1}} + 2\mu^{2} \mathcal{O}(V) & \geq & \mu^{2} \pi(q_{\lambda^*})^{2}, \\
    \text{or } \pi(q_{\lambda^*}) & = & \mathcal{O}(\sqrt{V}).
  \end{eqnarray*}
  We note that for both $q_{\lambda^*} = 0$ or $q_{\lambda^*} > 0$, we have that $\pi(q_{\lambda^*}) = \mathcal{O}(\sqrt{V})$.

  We now proceed as in the proof of Lemma \ref{chap4:prop:p21lb}, by using \eqref{chap4:eq:intlc_indeq}, to find the largest integer $\overline{q}$ such that 
  \begin{eqnarray*}
    \pi(q_{\lambda^*}) \sum_{m = 0}^{q - q_{\lambda^*}} \fpow{\lambda^*}{\mu}{m} \leq \frac{1}{2} - \frac{\alpha}{2}.
  \end{eqnarray*}
  The rest of the proof is similar to that of Lemma \ref{chap4:prop:p21lb}, and we obtain that $\overline{Q}(\gamma_{k}) = \Omega\nfrac{1}{\sqrt{V_{k}}}$.
  Then given a sequence of $u_{c,k} \uparrow u(\mu)$, we have that there exists a sequence of feasible $\gamma_{k}$ such that $\Qgk \leq Q^*(u_{c,k}) + \epsilon$, for some $\epsilon > 0$.
  Therefore, $Q^*(u_{c,k}) = \Omega\nfrac{1}{\sqrt{u(\mu) - u_{c,k}}}$, since $u_{c,k} \leq \Ugk$.
\end{proof}

\begin{lemma}
  For \INTLC-2-1, if $b_{i} = b_{\mu}$ and $i < P$, then for any sequence of non-idling admissible policies $\gamma_{k}$ such that $u(\mu) - \Ugk = V_{k} \downarrow 0$, we have that $\Qgk = \Omega\brap{\log\nfrac{1}{{V_{k}}}}$. Therefore $Q^*(u_{c}) = \Omega\brap{\log\nfrac{1}{{u(\mu) - u_{c}}}}$.
\end{lemma}
\begin{proof}
  The proof follows that of Lemma \ref{chap4:prop:p222lb}.
  We define $\lambda^* = b_{\mu} + \epsilon$, where $\epsilon > 0$.
  Let $q_{\lambda^*} = \inf\brac{q : \lambda(q) \leq \lambda^*}$.
  We note that $q_{\lambda^*}$ could be $0$, unlike $q_{\mu^*}$ in Lemma \ref{chap4:prop:p222lb}.

  If $q_{\lambda^*} = 0$, then we have that $\pi(q_{\lambda^*}) = \pi(0) = \mathcal{O}(V)$.
  If $q_{\lambda^*} > 0$, then we have that 
  \begin{eqnarray*}
    V & = & \sum_{\lambda_{k} > \lambda^*} \brap{l(\lambda_{k}) - u(\lambda_{k})}\pi_{\lambda}(k), \\
    & \geq & m_{a} \epsilon Pr\brac{\lambda(Q) > \lambda^*},
  \end{eqnarray*}
  where $m_{a}$ is the tangent of the angle made by the line passing through $(b_{i} = b_{\mu}, u(b_{\mu}))$ and $(b_{i + 1}, u(b_{i + 1}))$ with $l(\mu)$.
  Or we have that $Pr\brac{Q < q_{\lambda^{*}}} \leq \frac{V}{m_{a}\epsilon}$ and $\pi(q_{\lambda^*} - 1) \leq \frac{V}{m_{a} \epsilon}$.
  Since $\pi(q_{\lambda^*} - 1)\lambda(q_{\lambda^*} - 1) = \pi(q_{\lambda^*})\mu$ we have that $\pi(q_{\lambda^*}) \leq \pi(q_{\lambda^*} - 1) \frac{r_{a,max}}{\mu}$.
  We note that therefore $\pi(q_{\lambda^*}) = \mathcal{O}(V)$ for both $q_{\lambda^*} = 0$ and $q_{\lambda^*} > 0$.
  
  Now proceeding as in the proof of Lemma \ref{chap4:prop:p222lb} we have that for any $q \geq q_{\lambda^*}$ (we express the bound in terms of $\pi(q_{\lambda^*})$)
  \begin{eqnarray}
    \sum_{q = q_{\lambda^*}}^{q} \pi(q) \leq \pi(q_{\lambda^*}) \sum_{m = 0}^{q - q_{\lambda^*}} \fpow{\lambda^*}{\mu}{m}.
    \label{chap4:eq:intlc_21_1}
  \end{eqnarray}
  We note that independently of whether $q_{\lambda^*}$ is $0$ or not, if we find the largest $\overline{q}$ such that
  \begin{eqnarray*}
    \sum_{q = q_{\lambda^*}}^{\overline{q}} \pi(q) \leq \frac{1}{2} - \frac{V}{a_{1} \epsilon},
  \end{eqnarray*}
  then $\overline{Q}(\gamma) \geq \frac{\overline{q}}{2}$.
  We now proceed as in the proof of Lemma \ref{chap4:prop:p222lb}, using the upper bound in \eqref{chap4:eq:intlc_21_1} and $\pi(q_{\lambda^*}) = \mathcal{O}(V)$ to obtain that $\overline{Q}(\gamma_{k}) = \Omega\brap{\log\nfrac{1}{V_{k}}}$.
  Now given a sequence of $u_{c,k} \uparrow u(\mu)$, we have that there exists a sequence of feasible $\gamma_{k}$ such that $\Qgk \leq Q^*(u_{c,k}) + \epsilon$, for some $\epsilon > 0$.
  Therefore, $Q^*(u_{c,k}) = \Omega\brap{\log\nfrac{1}{u(\mu) - u_{c,k}}}$, since $u_{c,k} \leq \Ugk$.
\end{proof}

\begin{remark}
  We note that the above lemma is used to obtain an asymptotic lower bound on $Q^*(u_{c})$.
  As far as this asymptotic lower bound is concerned, the above lemma can be used even when if $i = P$, where $i$ is such that $b_{i} = b_{\mu}$.
  We consider \INTLC-2-1 for a larger $\mathcal{X}_{\lambda}$ defined as follows.
  We extend $\mathcal{X}_{\lambda}$ to $\overline{\mathcal{X}}_{\lambda} = \mathcal{X}_{\lambda} \cup (b_{P}, b_{P} + \delta]$, for some $\delta > 0$.
  We also extend the definition of $u(.)$ to $\overline{\mathcal{X}}_{\lambda}$, by choosing a piecewise linear function on $(b_{P}, b_{P} + \delta]$ which preserves the strictly increasing concave property of $u(.)$.
  We denote $Q^*(u_{c})$ when $\lambda(q) \in \overline{\mathcal{X}}_{\lambda}$, by $Q^*_{e}(u_{c})$.
  Then we note that $Q^*_{e}(u_{c}) \leq Q^*(u_{c})$.
  The asymptotic lower bound for $Q^*_{e}(u_{c})$ follows from the above lemma, which then also holds for $Q^*(u_{c})$.
  \label{chap4:remark:linearutilityhowto}
\end{remark}

\begin{remark}
  We note that the above asymptotic lower bound holds even in the case where $a_{\mu} = 0$.
  For \INTMC-2, we note that the case with $a_{\lambda} = 0$ corresponds to the case \INTMC-2-1, for which $Q^*(c_{c})$ only increased to a finite value.
\end{remark}

\begin{remark}
  We note that in many cases, for queueing models with a single queue, the utility constraint is on the average throughput.
  Then we have that $u(\lambda)$ is a line segment, with $a_{\mu} = 0$ and $b_{\mu} = r_{a,max}$.
  We note that the asymptotic $\Omega\brap{\log\nfrac{1}{V}}$ lower bound holds for $Q^*(u_{c})$, from the discussion in Remark \ref{chap4:remark:linearutilityhowto}.
\end{remark}
\begin{lemma}
  For \INTLC-2-2, for any sequence of non-idling admissible policies $\gamma_{k}$ such that $u(\mu) - \Ugk = V_{k} \downarrow 0$, we have that $\Qgk = \Omega{\nfrac{1}{{V_{k}}}}$.
\end{lemma}

\begin{proof}
  The proof follows that of Lemma \ref{chap4:prop:p223lb}.
  We choose $\lambda^* = \mu + \epsilon_{V} = a_{\mu} + \epsilon_{V}$.
  Let $q_{\lambda^*} = \inf\brac{q : \lambda(q) \leq \lambda^*}$.
  We note that $q_{\lambda^*}$ could be $0$, unlike $q_{\mu^*}$ in Lemma \ref{chap4:prop:p223lb}.

  If $q_{\lambda^*}$ is $0$, then we note that $\pi(q_{\lambda^*}) = \pi(0) = \mathcal{O}(V)$ from Lemma \ref{chap4:lemma:intlc_pi0ub}.
  If $q_{\lambda^*} > 0$, then we have that 
  \begin{eqnarray*}
    V & = & \sum_{\lambda_{k} > \lambda^*} \brap{l(\lambda_{k}) - u(\lambda_{k})}\pi_{\lambda}(k), \\
    & \geq & m_{a} \epsilon Pr\brac{\lambda(Q) > \lambda^*},
  \end{eqnarray*}
  where $m_{a}$ is the tangent of the angle made by the line passing through $(a_{i} = a_{\mu}, u(a_{mu}))$ and $(a_{i + 1}, u(a_{i + 1}))$ with $l(\mu)$.
  Or we have that $Pr\brac{Q < q_{\lambda^*}} \leq \frac{V}{m_{a} \epsilon_{V}}$ and $\pi(q_{\lambda^*} - 1) \leq \frac{V}{m_{a} \epsilon_{V}}$.
  We also note that since $\pi(q_{\lambda^*} - 1)\lambda(q_{\lambda^*} - 1) = \pi(q_{\lambda^*})\mu$ we have that $\pi(q_{\lambda^*}) \leq \pi(q_{\lambda^*} - 1) \frac{r_{a,max}}{\mu}$.
  Therefore $\pi(q_{\lambda^*}) \leq \frac{V}{m_{a}\epsilon_{V}} \frac{r_{a,max}}{\mu}$.
  We choose $\epsilon_{V} = a_{2} V$, so that $\frac{r_{a,max}}{m_{a} \mu a_{2}} \leq \frac{\alpha}{2}$, where $\alpha << 1$.
  
  Then, as in the proof of Lemma \ref{chap4:prop:p223lb}, if $\overline{q}$ is the largest integer such that 
  \[ \sum_{q = q_{\lambda^*}}^{\overline{q}} \pi(q) \leq \frac{1}{2} - \frac{\alpha}{2},\]
  then $\overline{Q}(\gamma) \geq \frac{\overline{q}}{2}$, independently of whether $q_{\lambda^*} = 0$ or not.

  We note that for any $q \geq q_{\lambda^*}$ we have that
  \begin{eqnarray}
    \sum_{q = q_{\lambda^*}}^{q} \pi(q) \leq \pi(q_{\lambda^*}) \sum_{m = 0}^{q - q_{\lambda^*}} \fpow{\lambda^*}{\mu}{m}.
    \label{chap4:eq:intlc_2231}
  \end{eqnarray}
  
  We also note that if $q_{\lambda^*} > 0$, then we have that
  \begin{eqnarray*}
    V & \geq & \sum_{\lambda_{k} > \lambda^*} m_{a} \brap{\lambda_{k} - \mu}\pi_{\lambda}(k), \\
    & = & m_{a} \sum_{q < q_{\lambda^*}} \brap{\lambda(q) - \mu}\pi(q), \\
    & = & m_{a} \mu\pi(q_{\lambda^*}) - \mu\pi(0).
  \end{eqnarray*}
  Then since $\pi(0) = \mathcal{O}(V)$ we have that $\pi(q_{\lambda^*}) = \mathcal{O}(V)$.
  Thus independently of whether $q_{\lambda^*} = 0$ or not, we have that $\pi(q_{\lambda^*}) = \mathcal{O}(V)$.
  
  Now proceeding as in the proof of Lemma \ref{chap4:prop:p223lb}, using the above upper bound on $\pi(q_{\lambda^*})$ in \eqref{chap4:eq:intlc_2231} we have that $\overline{Q}(\gamma_{k}) = \Omega\nfrac{1}{V_{k}}$.
  Now given a sequence of $u_{c,k} \uparrow u(\mu)$, we have that there exists a sequence of feasible $\gamma_{k}$ such that $\Qgk \leq Q^*(u_{c,k}) + \epsilon$, for some $\epsilon > 0$.
  Therefore, $Q^*(u_{c,k}) = \Omega\nfrac{1}{{u(\mu) - u_{c,k}}}$, since $u_{c,k} \leq \Ugk$.
\end{proof}

We note that as for \INTMC, using policies with similar structure as in Lemmas \ref{chap4:prop:p21ub}, \ref{chap4:prop:p222ub}, and \ref{chap4:lemma:p223ub} it is possible to obtain an asymptotic upper bound for \INTLC-1 and tight asymptotic upper bounds for \INTLC-2-1 and \INTLC-2-2.
Here, we obtain a single asymptotic upper bound for \INTLC-1, \INTLC-2-1, and \INTLC-2-2, which shows that $u(\mu) = \sup_{\gamma \in \Gamma_{a,M}} \Ug$.
\begin{lemma}
  There exists a sequence of admissible policies $\gamma_{k}$ such that $\Qgk = \mathcal{O}\nfrac{1}{V_{k}}$ and $u(\mu) - \Ugk = V_{k}$.
  \label{chap4:lemma:intlc_ub}
\end{lemma}
\begin{proof}
  The proof follows that of Lemma \ref{chap4:lemma:p223ub} (which follows from that of Lemma \ref{chap4:prop:p13ub}).
  Consider a policy $\gamma$ defined as follows:
  \begin{eqnarray*}
    \lambda(q) & = 
    \begin{cases}
      \mu \text{ for } q \in \brac{0, \dots, q_{1}}, \\
      \mu - K \text{ for } q \in \brac{q_{1} + 1, \dots},
    \end{cases}
  \end{eqnarray*}
  where $q_{1} = \frac{1}{U}$ for positive $U$.
  The sequence of policies $\gamma_{k}$ is obtained by choosing $U$ from a sequence $U_{k} \downarrow 0$.
  
  We note that 
  \begin{eqnarray*}
    \overline{U}(\gamma) & = & Pr\brac{Q \leq q_{1}}u(\mu) + Pr\brac{Q > q_{1}} u(\mu - K), \\
    & = & u(\mu) - Pr\brac{Q > q_{1}}(u(\mu) - u(\mu - K)).\\
    \text{Or } u(\mu) - \Ug & = & Pr\brac{Q > q_{1}}\brap{u(\mu) - u(\mu - K)}.
  \end{eqnarray*}
  We also note that $\overline{Q}(\gamma) = \mathcal{O}\brap{q_{1}} = \mathcal{O}\brap{\frac{1}{U}}$.
  Therefore we have that $Pr\brac{Q \leq q_{1}} \mu + Pr\brac{Q > q_{1}} (\mu - K) = (1 - \pi(0))\mu$.
  Or 
  \begin{eqnarray*}
    \brap{1 - Pr\brac{Q > q_{1}}} \mu + Pr\brac{Q > q_{1}} (\mu - K) & = & (1 - \pi(0))\mu, \\
    \mu + Pr\brac{Q > q_{1}}(-K) & = & \mu - \mu \pi(0), \text{ or}\\
    Pr\brac{Q > q_{1}} & = & \frac{\mu \pi(0)}{K}.
  \end{eqnarray*}
  Following the proof of Lemma \ref{chap4:prop:p13ub}, we have that $\pi(0) = \mathcal{O}\brap{U}$.
  Hence, we have that $Pr\brac{Q > q_{1}} = \mathcal{O}\brap{U}$.
  Therefore, for the sequence of policies $\gamma_{k}$, we have that $u(\mu) - \Ugk = \mathcal{O}(U_{k})$. If $V_{k} \Deq u(\mu) - \Ugk$, then we have that $\Qgk = \mathcal{O}\nfrac{1}{V_{k}}$.
\end{proof}

\begin{remark}
  Since the techniques used in the analysis of \INTMC\, and \INTLC\, are similar, we expect that asymptotic bounds on any sequence of order-optimal policies can be obtained for \INTLC as for \INTMC.
  We note that the role of $\mu(q)$ and $\lambda(q)$ are interchanged.
  For example, using a similar sequence of steps as in the proof of Lemma \ref{chap4:lemma:intmcsqrt_policy_aschar}, it is possible to show that for a sequence of non-idling order-optimal admissible policies with $u(\mu) - \Ugk = V_{k} \downarrow 0$, $|Q_{A}| = \Omega\nfrac{1}{\sqrt{V_{k}}}$ for \INTLC-1, where $Q_{A} = \brac{q : \lambda(q) \in [\mu - a_{2} V^{\frac{1}{2}}, \mu + a_{2} V^{\frac{1}{2}}]}$ and $a_{2} > 0$.
\end{remark}

\section{Analysis of INTERVAL-$\lambda\mu$CHOICE}
\label{chap4:sec:problemp3}
\newcommand{\lz}{\lambda(q)}
\newcommand{\mz}{\mu(q)}
We recall that for INTERVAL-$\lambda\mu$CHOICE we restrict to policies $\gamma$ such that $\lambda(q) \in [r_{a,min},r_{a,max}]$ and $\mu(q) \in [0,r_{max}]$, $\forall q \in \mathbb{Z}_{+}$.
The tradeoff problem for INTERVAL-$\lambda\mu$CHOICE is:
\begin{eqnarray}
  \text{ minimize }_{\gamma \in \Gamma_{a}} & \overline{Q}(\gamma) \nonumber \\
  \text{ such that } & \overline{C}(\gamma) \leq c_{c}, \nonumber \\
  \text{ and } & \overline{U}(\gamma) \geq u_{c},
  \label{chap4:eq:intlmc_problem_statement}
\end{eqnarray}
whose optimal value is $Q^*(c_{c}, u_{c})$.
Although it is possible to consider various forms of the function $c(\mu)$ as in the case of FINITE-$\mu$CHOICE and INTERVAL-$\mu$CHOICE, here we obtain a complete analysis for the case $c(\mu)$ being a strictly convex function of $\mu \in [0, r_{max}]$ (with assumption C2) and $u(\lambda)$ being either a strictly concave (with assumption U2) or a piecewise linear function of $\lambda \in [r_{a,min}, r_{a,max}]$.
The reason for this assumption is the motivating discrete time problem, described below.
We then comment on the asymptotic bounds for other forms of $c(\mu)$ in the following discussion.
\begin{remark}
  INTERVAL-$\lambda\mu$CHOICE corresponds to the tradeoff problem for the following discrete time queueing model.
  Work arrives in a batch of random size, in every slot, into an infinite buffer queue.
  The queue length is the amount of unfinished work and evolves on $\sR$.
  The amount of work which is admitted into the queue can be controlled, as a function, possibly randomized, of the current queue length.
  This feature is modelled by the control of the arrival rate, $\lambda(q)$, in INTERVAL-$\lambda\mu$CHOICE.
  The amount of work done by the server in each slot, or the service batch size, can also be chosen as a function, possibly randomized, of the current queue length and is assumed to be a non-negative real number.
  This feature of the discrete time queue is modelled by the control of the service rate, $\mu(q)$, in INTERVAL-$\lambda\mu$CHOICE.
  We note that the drift in the discrete time queueing model is real valued.
  In light of the discussion in Section \ref{chap4:sec:intintro} we assume that $\lambda(q)$ and $\mu(q)$ take any non-negative real values, but in finite intervals.
  In each slot, assume that there is a utility accrued with admitting customers and a service cost incurred in serving them.
  These are modelled by the utility rate and service cost rate functions $u(.)$ and $c(.)$ in INTERVAL-$\lambda\mu$CHOICE.
  We choose $c(\mu)$ to be strictly convex as for INTERVAL-$\mu$CHOICE-1.
  Another motivating factor for considering $c(\cdot)$ to be strictly convex, is the need to explain the logarithmic $\Theta\brap{\log\nfrac{1}{V}}$ behaviour of the average queue length when admission control is allowed, noticed by Neely in \cite{neely_utility}, compared to the $\Omega\brap{\frac{1}{\sqrt{V}}}$ behaviour of the average queue length, with strictly convex $c(\cdot)$, when admission control is not allowed.
  We note that in \cite{neely_utility}, since there was only a constraint on the average throughput, the function $u(.)$ need only be linear, while the following result is presented for the case of strictly convex or piecewise linear $u(.)$.
\end{remark}
\begin{remark}
  In this analysis, we assume that $u_{c} \leq u(r_{a,max})$.
  If $u_{c} > u(r_{a,max})$, then there does not exist any feasible policies for \eqref{chap4:eq:intlmc_problem_statement}
  We note that if $u_{c} = u(r_{a,max})$, then policies which satisfy this utility constraint need to have $\lz = r_{a,max}, \forall q$, in which case the problem is the same as that considered in INTERVAL-$\mu$CHOICE-1 (if $r_{a,max} < r_{max}$).
  We also note that the restriction of analysis to admissible policies implicitly requires that $u^{-1}(u_{c}) < r_{max}$.
\end{remark}
\begin{lemma}
  For INTERVAL-$\lambda\mu$CHOICE, with $r_{a,min} > 0$, the service cost $\overline{C}(\gamma)$ for any admissible policy $\gamma$ is bounded below by $c(u^{-1}(u_{c}))$.
  \label{chap4:prop:p3minservicecost}
\end{lemma}
\begin{proof}
  To find a lower bound on the service cost for any admissible policy $\gamma$, we consider the following equivalent formulation of TRADEOFF \eqref{chap4:eq:sysmodel:probstat} and use a series of relaxations on the constraints.
  The minimum average service cost for a given average queue length constraint $q_{c}$ and a utility constraint $u_{c}$ is given by
  \begin{eqnarray*}
    \min_{\gamma \in \Gamma_{a}} & \mathbb{E}_{\pi} c(\mu(Q)), \\
    \text{such that } & \mathbb{E}_{\pi} u(\lambda(Q)) \geq u_{c}, \text{ and } \mathbb{E}_{\pi} Q \leq q_{c}.\nonumber 
  \end{eqnarray*}
  We note that for every $q_{c} < \infty$, we have that $\mathbb{E}_{\pi} \lambda(Q) = \mathbb{E}_{\pi} \mu(Q)$.
  So the optimal value of the optimization problem above is bounded below by the optimal value of 
  \begin{eqnarray*}
    \min_{\gamma \in \Gamma_{a}} & \mathbb{E}_{\pi} c(\mu(Q)), \\
    \text{such that } & \mathbb{E}_{\pi} u(\lambda(Q)) \geq u_{c}, \text{ and } \mathbb{E}_{\pi} \lambda(Q) = \mathbb{E}_{\pi} \mu(Q) \nonumber. 
  \end{eqnarray*}
  Since $u(\lambda)$ is concave in $\lambda$, we have that for every $\gamma$ such that $\Expp u(\lambda(Q)) \geq u_{c}$, $u(\Expp \lambda(Q)) \geq u_{c}$. Therefore the optimal value of the problem above is bounded below by the optimal value of
  \begin{eqnarray*}
    \min_{\pi} & \mathbb{E}_{\pi} c(\mu(Q)), \\
    \text{such that } & \mathbb{E}_{\pi} \lambda(Q) \geq u^{-1}(u_{c}), \text{ and } \mathbb{E}_{\pi}\lambda(Q) = \mathbb{E}_{\pi}\mu(Q), \nonumber
  \end{eqnarray*}
  where we are considering all possible distributions $\pi$ for $Q$.
  Now since $c(\mu)$ is convex in $\mu$, we obtain that the optimal value of the above problem is $\geq c(u^{-1}(u_{c}))$.
  Therefore $\overline{C}(\gamma) \geq c(u^{-1}(u_{c}))$.
\end{proof}

In the following, as in the case of FINITE-$\mu$CHOICE and INTERVAL-$\mu$CHOICE, we consider INTERVAL-$\lambda\mu$CHOICE in the asymptotic regime $\Re$ where the service cost constraint $c_{c}$ approaches $c(u^{-1}(u_{c}))$, where $u_{c}$ is kept fixed.
\newcommand{\newl}{u^{-1}(u_{c})}
\newcommand{\epv}{\epsilon_{V}}
\newcommand{\zms}{q_{\mu^*}}

\subsection{Asymptotic lower bound}
In this section we find an asymptotic lower bound on $\overline{Q}(\gamma_{k})$ for any sequence of non-idling admissible policies $\gamma_{k}$ for which $\overline{U}(\gamma_{k}) \geq u_{c}$ and $\overline{C}(\gamma_{k}) \downarrow c(u^{-1}(u_{c}))$.
Subsequently, in Lemma \ref{chap4:prop:p3ub} we show that there exists a sequence of non-idling admissible policies $\gamma_{k}$ for which $\overline{C}(\gamma_{k})$ approaches $c(u^{-1}(u_{c}))$ arbitrarily closely.

We note that, as in the case INTERVAL-$\mu$CHOICE, even though the service rate $\mu(q)$ can take any value in $[0,r_{max}]$ and the arrival rate $\lambda(q)$ can take any value in $[r_{a,min},r_{a,max}]$, the sets of service rates and arrival rates used by $\gamma$ are both countable as the queue length takes only integer values.
We first present the asymptotic lower bound for the case when $r_{a,min} > 0$, for which the proof is very similar to the case INTERVAL-$\mu$CHOICE-1, and then consider the case $r_{a,min} = 0$.
\begin{lemma}
  For INTERVAL-$\lambda\mu$CHOICE, for any sequence of non-idling admissible policies $\gamma_{k}$ such that $\overline{C}(\gamma_{k}) - c(u^{-1}(u_{c})) = V_{k} \downarrow 0$ and $\overline{U}(\gamma_{k}) \geq u_{c}$, we have that
  \begin{eqnarray*}
    \overline{Q}(\gamma_{k}) = \Omega\brap{\log\nfrac{1}{V_{k}}}.
  \end{eqnarray*}
  \label{chap4:prop:p3lb}
\end{lemma}
\begin{proof}
  Consider a particular policy $\gamma$ in the sequence $\gamma_{k}$ with $V_{k} = V$.
  Since $\gamma$ is admissible, we have that $\Expp \mu(Q) = \Expp \lambda(Q)$.
  From the concavity of $u(\lambda)$, we have that $\Expp \lambda(Q) \geq \newl$.
  Let $\mu^* = \newl - \epsilon_{V}$, where $\epv$ is a function of $V$ to be chosen later.
  Define $\zms = \inf\brac{q : \mu(q) \geq \mu^*}$.
  We note that $\forall q < \zms$, $\mu(q) < \mu^*$.
  As $\mu(q)$ is non-decreasing, we have that
  \begin{eqnarray*}
    Pr\brac{ Q < \zms} = Pr\brac{\mu(Q) < \mu^*}.
  \end{eqnarray*}
  Let the countable set of service rates be denoted by $\brac{\mu_{0} = 0, \mu_{1}, \dots}$, where $\mu_{i} < \mu_{i + 1}$ and $\mu_{i} \in [0,r_{max}]$.
  Let $l(\mu)$ be the tangent line at $(\newl, c(\newl))$ to the curve $c(\mu)$.
  Then $V = \sum_{q = 0}^{\infty} \bras{c(\mz) - l(\mz)}\pi(q)$.
  From Proposition \ref{chap4:app:prop:quadlowerbound}, we have a positive $a_{1}$ such that
  \begin{eqnarray*}
    V \geq a_{1} \sum_{q = 0}^{\infty} \bras{\mz - \newl}^{2} \pi(q) \geq a_{1} \sum_{q = 0}^{\zms - 1} \bras{\mz - \newl}^{2} \pi(q).
  \end{eqnarray*}
  Hence 
  \begin{eqnarray*}
    Pr\brac{Q \leq \zms - 1} \leq \frac{V}{a_{1}\epv^{2}}, \\
    \text{and } \pi(\zms - 1) \leq \frac{V}{a_{1}\epv^{2}}.
  \end{eqnarray*}
  Now, since $Q(t)$ is a birth death process $\forall q$, we have that $\pi(q) \lz = \pi(q + 1)\mu(q + 1)$.
  For any $q \geq \zms$,
  \begin{eqnarray}
    \pi(q + 1) = \frac{\pi(q)\lz}{\mu(q + 1)} \leq \frac{\pi(q) r_{a,max}}{\mu^*}, \nonumber \\
    \pi(q) \leq \pi(\zms - 1) \fpow{r_{a,max}}{\mu^*}{q - \zms + 1}.
    \label{chap4:eq:p31}
  \end{eqnarray}
  Let $\overline{q}$ be the largest integer such that $\sum_{q = 0}^{\overline{q}} \pi(q) \leq \frac{1}{2}$.
  We find a lower bound on $\overline{q}$ as in the proof of Lemma \ref{chap4:prop:p1213lb}.
  We note that $Pr\brac{Q \geq \zms} \geq 1 - \frac{V}{a_{1}\epv^{2}}$.
  Let $\epv = \epsilon$, where $0 < \epsilon < u^{-1}(u_{c})$.
  For $V$ small, let $\overline{q}_{1}$ be the largest integer such that
  \begin{eqnarray*}
    \sum_{q = \zms}^{\overline{q}_{1}} \pi(q) \leq \frac{1}{2} - \frac{V}{a_{1}\epv^{2}}.
  \end{eqnarray*}
  Then $\overline{q}_{1} \leq \overline{q}$.
  We find a lower bound on $\overline{q}_{1}$ by using the upper bound on $\pi(q)$ from \eqref{chap4:eq:p31}.
  Let $\overline{q}_{2}$ be the largest integer such that
  \begin{eqnarray*}
    \pi(\zms - 1) \sum_{q = 1}^{\overline{q}_{2} - \zms + 1} \fpow{r_{a,max}}{\mu^*}{q} \leq \frac{1}{2} - \frac{V}{a_{1}\epv^{2}}.
  \end{eqnarray*}
  Then $\overline{q}_{2} \leq \overline{q}_{1}$.
  After substituting for $\mu^*$, we have that any $\overline{q}_{2}$ satisfying the above inequality is such that
  \begin{eqnarray*}
  \overline{q}_{2} - \zms + 1 \leq \log_{\nfrac{r_{a,max}}{\newl - \epv}} \brap{1 + \frac{r_{a,max} - \newl + \epv}{r_{a,max}} \frac{1}{\pi(\zms - 1)}\brap{\frac{1}{2} - \frac{V}{a_{1}\epv^{2}}}}.
  \end{eqnarray*}
  Hence we obtain that $\overline{q}_{2}$ is at least
  \begin{eqnarray*}
    \log_{\nfrac{r_{a,max}}{\newl - \epv}} \brap{1 + \frac{r_{a,max} - \newl + \epv}{r_{a,max}} \frac{1}{\pi(\zms - 1)}\brap{\frac{1}{2} - \frac{V}{a_{1}\epv^{2}}}} - 2.
  \end{eqnarray*}
  We note that $\frac{1}{\pi(\zms - 1)} \geq \frac{a_{1}\epsilon^{2}}{V}$ and is the dominant term in the regime where $V \downarrow 0$.
  Since $\overline{q} \geq \overline{q}_{1} \geq \overline{q}_{2}$ and $\overline{Q}(\gamma) \geq \frac{\overline{q}}{2}$, we have that for any sequence of $\gamma_{k}$ with $\overline{C}(\gamma_{k}) - \newl = V_{k} \downarrow 0$, $\overline{Q}(\gamma_{k}) = \Omega\brap{\log\nfrac{1}{V_{k}}}$.
\end{proof}
\begin{remark}
  We note that in this proof, no use was made of the assumption that the sequence of policies satisfies the constraint $\Expp u(\lambda(Q)) \geq u_{c}$.
  The difficulty in problem INTERVAL-$\lambda\mu$CHOICE, is to actually construct such a sequence of policies.

  In the above proof, we note that there exists a set $\hpq$ of queue lengths occurring with high probability, such that $\mu(q) \rightarrow u^{-1}(u_{c})$, for every $q \in \hpq$ as $V \downarrow 0$.
  But for every $q \in \hpq$, it is possible to ensure through arrival rate control that the service rate $\mu(q)$ is not equal to the arrival rate $\lambda(q)$, while for Lemma \ref{chap4:prop:p21lb} $\mu(q) = \lambda$ for all queue lengths with high probability.
  Hence, for admissible policies, for $q \in \hpq$, the stationary probability distribution geometrically grows and then decays which leads to the $\log\nfrac{1}{V}$ behaviour.
\end{remark}

\begin{remark}
\emph{The case when $r_{a,min} = 0$ or $\lambda(q) \in [0,r_{a,max}]$ :}\\
We note that if $r_{a,min} = 0$, then the birth death process may not be irreducible on $\sZ$.
Therefore, in this case, admissible policies are assumed to induce a single positive recurrent class including zero.
In Lemma \ref{chap4:prop:p3lb} we had assumed that $r_{a,min} > 0$.
If $r_{a,min} = 0$, then for an admissible policy $\gamma$, there could exist $q$ such that $\lambda(q) = 0$.
Let $q' = \inf\brac{q : \lambda(q) = 0}$.
Note that $q'$ is in general dependent on the parameters $c_{c}$ and $u_{c}$.
If $q'$ is infinity, then the same approach as in Lemma \ref{chap4:prop:p3lb} holds.
If $q'$ is finite, then as $\gamma$ is admissible, $\forall q > q'$, $\lambda(q) = 0$.
Hence all states $q > q'$ are transient and in steady state we need only consider the CTMC evolving on $\{0,\dots,q'\}$.
Note that $\sum_{q = 0}^{q'} \pi(q) = 1$ and therefore $\overline{q}$ in Lemma \ref{chap4:prop:p3lb} is smaller than $q'$.
Therefore, the same approach as in Lemma \ref{chap4:prop:p3lb} holds even if $q'$ is finite.
\end{remark}

\subsection{Asymptotic behaviour of the tradeoff curve}
In this section, we construct a sequence of admissible policies $\gamma_{k}$ which achieves the minimum average service cost $c(\newl)$ arbitrarily closely with $\overline{Q}(\gamma_{k})$ scaling at the optimal rate as in Lemma \ref{chap4:prop:p3lb}.
However, we are able to obtain an asymptotic upper bound only for the case where $u(\lambda)$ is strictly concave or linear (and not piecewise linear).
\begin{lemma}
  For INTERVAL-$\lambda\mu$CHOICE, with $u(\lambda)$ strictly concave or linear, there exists a sequence of admissible policies $\gamma_{k}$ with a corresponding sequence $V_{k} \downarrow 0$ such that
  \begin{eqnarray*}
    \overline{Q}(\gamma_{k}) & = & \mathcal{O}\brap{\log\nfrac{1}{V_{k}}},\\
    \overline{C}(\gamma_{k}) - c(\newl) & = & V_{k}, \\
    \overline{U}(\gamma_{k}) & \geq & u_{c}.
  \end{eqnarray*}
  \label{chap4:prop:p3ub}
\end{lemma}

The construction of the sequence of admissible policies $\gamma_{k}$ is motivated by the following intuition, that we have obtained from the lower bound in Lemma \ref{chap4:prop:p3lb}.
The sequence of policies should be such that as $V_{k} \downarrow 0$, the service rate used, at a queue length occurring with high probability, should be close to $\newl$.
But the arrival rate $\lambda(q)$ should not exactly equal $\newl$, for all queue lengths $q$ which occur with high probability.
Then it should be possible to have a stationary distribution which is geometrically growing and then decaying, leading to the required $\log\nfrac{1}{V_{k}}$ scaling of $\overline{Q}(\gamma_{k})$.
\begin{proof}
  \renewcommand{\epv}{\epsilon_{U}}
  \newcommand{\lambdao}{\lambda_{1}}
  \newcommand{\lo}{\lambda_{1}}
  \newcommand{\lt}{\lambda_{2}}
  \newcommand{\mo}{\mu_{1}}
  \newcommand{\mt}{\mu_{2}}
  \newcommand{\lambdat}{\lambda_{2}}
  \newcommand{\muo}{\mu_{1}}
  \newcommand{\mut}{\mu_{2}}

  Consider a policy $\gamma$ of the following form :
  \begin{eqnarray*}
    &	\mu(0)  & =  0, \\
    & 	\mu(q)  & = \muo = \newl - \epv, \text{ for } q \in \brac{1,\dots,q_{1}}, \\
    &	\mu(q)  & = \mut = \newl + \epv, \text{ for } q \in \brac{q_{1} + 1,\dots}; \\
    \text{and } & \lambda(q) & = \lo, \text{ for } q \in \brac{0, \dots, q_{1} - 1}, \\
    & \lambda(q) & = \newl, \text{ for } q \in \brac{q_{1}, \dots, q_{1} + K}, \\
    & \lambda(q) & = \lt, \text{ for } q \in \brac{q_{1} + K + 1, \dots}.
  \end{eqnarray*}
  Let $\epv = U$, $\lo > \newl > \lt, \lo > \muo, \lt < \mut$, and $q_{1} = \left \lceil \log_{\nfrac{\lambdao}{\mo}} \brap{1 + \frac{\lambdao - \mo}{\lambdao}\frac{{1}}{U}} \right \rceil$,
  We will specify $K$, $\lambda_{1}$, and $\lambda_{2}$ later.
  Let $\frac{dc(\newl)}{d\mu} \stackrel{\Delta} = \frac{dc(\mu)}{d\mu}\vert_{\mu = \newl}$. 
  We now obtain $\overline{C}(\gamma)$.
  \begin{eqnarray*}
    \overline{C}(\gamma) & = & \pi(0).0 + \pi_{\mu}(\mo) c(\mo) + \pi_{\mu}(\mt) c(\mt), \\
    & = & \pi_{\mu}(\mo) \brap{c(\newl) + (-\epv)\frac{dc(\newl)}{d\mu} + \mathcal{O}(\epv^{2})} + \\ 
    & & \pi_{\mu}(\mt) \brap{c(\newl) + (\epv)\frac{dc(\newl)}{d\mu} + \mathcal{O}(\epv^{2})}, \nonumber \\
    & \leq & c(\newl) + \mathcal{O}(U^{2}) + (-\epv \pi_{\mu}(\mo) + \epv \pi_{\mu}(\mt))\frac{dc(\newl)}{d\mu}, \\
    & \leq & c(\newl) + \epv\frac{dc(\newl)}{d\mu} + \mathcal{O}(U^{2}), \\
    & \leq & c(\newl) + \mathcal{O}(U),
  \end{eqnarray*}
  where $\frac{dc(\newl)}{d\mu} \Deq \frac{dc(\mu)}{d\mu}\vert_{\mu = \newl}$.
  Let $V = \overline{C}(\gamma) - c(\newl)$, then we have that $V = \mathcal{O}(U)$.
  For $\gamma$, the average utility is
  \begin{eqnarray*}
    \overline{U}(\gamma) = u(\lo)\brap{ \sum_{q = 0}^{q_{1} - 1} \pi(q) } + u_{c} \sum_{q = q_{1}}^{q_{1} + K} \pi(q) + u(\lt) \sum_{q = q_{1} + K + 1}^{\infty} \pi(q).
  \end{eqnarray*}
  Let $\lo = \newl + \epsilon$ and $\lt = \newl - \epsilon$, where $\epsilon$ is a small positive constant.
  Then for strictly concave and linear $u(.)$ we have that
  \begin{eqnarray*}
    \overline{U}(\gamma) & \geq & u_{c} + \bras{\epsilon \sum_{q = 0}^{q_{1} - 1} \pi(q) - \epsilon \sum_{q = q_{1} + K + 1}^{\infty} \pi(q)}D(u(\newl)) + O(u(\newl)),
  \end{eqnarray*}
  where $D(u(\newl))$ and $O(u(\newl))$ are defined as follows.
  If $u(.)$ is a strictly concave function, then it is differentiable at $\newl$ and $D(u(\newl)) \Deq \frac{du(\lambda)}{d\lambda}\vert_{\lambda = \newl}$ and $O(u(\newl)) = \mathcal{O}(\epsilon^{2})$, with the above inequality being an equality.
  If $u(.)$ is linear then $D(u(\newl)) \Deq \frac{du(\lambda)}{d\lambda}\vert_{\lambda = \newl}$ and $O(u(\newl)) = 0$, with the above inequality being an equality.

  In the following we show that $\sum_{q = q_{1} + K + 1}^{\infty} \pi(q) \leq \sum_{q = 0}^{q_{1} - 1} \pi(q)$, in which case we have that $\overline{U}(\gamma) \geq u_{c}$ for sufficiently small $\epsilon$ (which is fixed and independent of $V$).
  We have that
  \begin{eqnarray*}
    \pi(q) = \pi(0) \fpow{\lambdao}{\muo}{q}, & \text{ for } q \in \brac{1,\dots,q_{1}}, \\
    \pi(q) = \pi(0) \fpow{\lo}{\mo}{q_{1}} \fpow{\newl}{\mt}{q - q_{1}}, & \text{ for } q \in \brac{q_{1} + 1, \dots, q_{1} + K}, \\
    \pi(q) = \pi(0) \fpow{\lambdao}{\muo}{q_{1}}\fpow{\newl}{\mt}{K}\nfrac{\newl}{\mt}\fpow{\lambdat}{\mut}{q - q_{1} - K - 1}, & \text{ for } q \in \brac{q_{1} + K + 1,\dots}.
  \end{eqnarray*}
  % Since $\sum_{q = 0}^{\infty} \pi(q) = 1$, we have 
  % \begin{eqnarray*}
  %  \pi(0)\bras{1 + \sum_{q = 1}^{q_{1}} \fpow{\lambdao}{\muo}{q} + \fpow{\lambdao}{\muo}{q_{1}}\sum_{q = 1}^{K}\fpow{\newl}{\mut}{q} + \fpow{\lambdao}{\muo}{q_{1}}\fpow{\newl}{\mt}{K}\nfrac{\newl}{\mt}\sum_{q = 0}^{\infty}\fpow{\lt}{\mt}{q}} = 1.
  % \end{eqnarray*}
  Therefore, 
  \begin{eqnarray*}
    \sum_{q = 0}^{q_{1} - 1} \pi(q)  & = & \pi(0) + \pi(0)\sum_{q = 1}^{q_{1} - 1} \fpow{\lo}{\mo}{q}, \\
    & = & \pi(0) + \pi(0)\nfrac{\lo}{\lo - \mo}\brap{\fpow{\lo}{\mo}{q_{1} - 1} - 1}.
  \end{eqnarray*}
  And,
  \begin{eqnarray*}
    \sum_{q_{1} + K + 1}^{\infty}\pi(q) & = & \pi(0)\fpow{\lambdao}{\muo}{q_{1}}\fpow{\newl}{\mt}{K}\nfrac{\newl}{\mt}\sum_{q = 0}^{\infty}\fpow{\lt}{\mt}{q}, \\
    & = & \pi(0)\fpow{\lambdao}{\muo}{q_{1}}\fpow{\newl}{\mt}{K}\nfrac{\newl}{\mt - \lt}.
  \end{eqnarray*}
  If 
  \begin{eqnarray}
    \fpow{\lambdao}{\muo}{q_{1}}\fpow{\newl}{\mt}{K}\nfrac{\newl}{\mt - \lt} \leq \nfrac{\lo}{\lo - \mo}\brap{\fpow{\lo}{\mo}{q_{1} - 1} - 1} + 1
    \label{chap4:eq:p3s}
  \end{eqnarray}
  then $\sum_{q = q_{1} + K + 1}^{\infty} \pi(q) \leq \sum_{q = 0}^{q_{1} - 1} \pi(q)$.
  We note that \eqref{chap4:eq:p3s} can be simplified to the question
  \begin{eqnarray*}
    \frac{\mo}{\lo - \mo} \stackrel{?} \leq \fpow{\lo}{\mo}{q_1}\brap{\frac{\mo}{\lo - \mo} - \fpow{\newl}{\mt}{K}\nfrac{\newl}{\mt - \lt}}, \\
    \frac{\mo}{\lo - \mo} \stackrel{?} \leq \fpow{\lo}{\mo}{q_1}\brap{\frac{\newl - \epv}{\epsilon + \epv} - \fpow{\newl}{\newl + \epv}{K}\nfrac{\newl}{\epsilon + \epv}}, \\
    \frac{\mo}{\lo - \mo} \stackrel{?} \leq \fpow{\lo}{\mo}{q_1}\nfrac{\newl}{\epsilon + \epv}\brap{1 - \frac{\epv}{\newl} - \brap{1 + \frac{\epv}{\newl}}^{-K}}.
  \end{eqnarray*}
  We use the lower bound on $q_{1}$, obtained by removing the ceiling, to arrive at the following question :
  \begin{eqnarray*}
    \frac{\newl - \epv}{\epsilon + \epv} \stackrel{?} \leq \brap{1 + \frac{\epsilon + \epv}{\newl - \epv}\frac{1}{U}} \nfrac{\newl}{\epsilon + \epv}\brap{1 - \frac{\epv}{\newl} - \brap{1 + \frac{\epv}{\newl}}^{-K}}
  \end{eqnarray*}
  For sufficiently small $U$, with $\epv = U$, we have that $\brap{1 + \frac{\epv}{\newl}}^{-K} \leq 1 - \frac{K\epv}{2\newl}$.
  So, instead of the above question we can ask the stronger question
  \begin{eqnarray*}
    \frac{\newl - \epv}{\epsilon + \epv} \stackrel{?} \leq \brap{1 + \frac{\epsilon + \epv}{\newl - \epv}\frac{1}{V}} \nfrac{\newl}{\epsilon + \epv}\brap{\frac{\epv}{\newl}\brap{\frac{K}{2} - 1}}.
  \end{eqnarray*}
  We choose $K > 2\brap{1 + \frac{(\newl)^{2}}{\epsilon}}$.
  Then we can ask the even stronger questions
  \begin{eqnarray*}
    \newl - \epv \stackrel{?} \leq \brap{1 + \frac{\epsilon + \epv}{\newl - \epv}\frac{1}{V}}\brap{\epv\frac{(\newl)^{2}}{\epsilon}}, \\
    \newl - \epv \stackrel{?} \leq \frac{\epsilon + \epv}{\epsilon}\frac{(\newl)^{2}}{\newl - \epv}
  \end{eqnarray*}
  which indeed hold.
  Hence for sufficiently small $V$ and $\epsilon$, $\overline{U}(\gamma) \geq u_{c}$.
  
  We now obtain $\overline{Q}(\gamma)$ using Proposition \ref{chap4:app:prop:dotzupperbound} with $q_{\epsilon} = q_{1} + 1$.
  Hence,
  \begin{eqnarray*}
    \overline{Q}(\gamma) = \frac{(q_{1} + 1)(\epv + \epsilon + r_{a.max})}{\epv + \epsilon} + \frac{r_{max}}{2(\epv + \epsilon)}, \\
    \overline{Q}(\gamma) \leq \frac{(q_{1} + 1)(\epsilon + r_{a,max})}{\epsilon} + \frac{r_{max}}{2\epsilon}.
  \end{eqnarray*}
  As $q_{1} = \mathcal{O}\brap{\log\nfrac{1}{U}}$, we obtain that $\overline{Q}(\gamma) = \mathcal{O}\brap{\log\nfrac{1}{U}}$.
  We note that the policy $\gamma$ is admissible.
  The sequence of policies is obtained by choosing $U_{k} = \frac{1}{k}$.
  We note that then we have a corresponding sequence $V_{k} = \mathcal{O}(U_{k})$.
  Thus, $\overline{Q}(\gamma_{k}) = \mathcal{O}\brap{\log\nfrac{1}{V_{k}}}$, and we have that there exists a sequence of admissible policies $\gamma_{k}$ with a corresponding sequence $V_{k} \downarrow 0$ such that
  \begin{eqnarray*}
    \overline{Q}(\gamma_{k}) & = & \mathcal{O}\brap{\log\nfrac{1}{V_{k}}},\\
    \overline{C}(\gamma_{k}) - c(\newl) & = & V_{k}, \\
    \overline{U}(\gamma_{k}) & \geq & u_{c}.
  \end{eqnarray*}
\end{proof}

\begin{remark}
  We note that the above proof also applies if $u(\lambda)$ is piecewise linear and $(\newl, u_{c})$ lies on a linear segment of the piecewise linear function $u(\lambda)$.
  However, the proof does not apply if $u(\lambda)$ is piecewise linear and $\newl$ is such that the slope of $u(\lambda)$ changes at $(\newl, u_{c})$.
\end{remark}

Using the asymptotic lower bound from Lemma \ref{chap4:prop:p3lb}, the asymptotic upper bound above, and proceeding as in the proof of Proposition \ref{chap4:prop:p11}, we arrive at the following result.

\begin{proposition}
  For INTERVAL-$\lambda\mu$CHOICE, for strictly concave or linear $u(\lambda)$, we have that the optimal tradeoff curve $Q^*(c_{c,k},u_{c}) = \Theta\brap{\log\nfrac{1}{c_{c,k} - c(u^{-1}(u_{c}))}}$, for the sequence $c_{c,k} = \Cgk$, for the sequence of policies $\gamma_{k}$ in Lemma \ref{chap4:prop:p3ub}.
  \label{chap4:prop:p3}
\end{proposition}

\begin{remark}
  For \INTLMC, an admissible policy $\gamma$ can be specified by the sets $Q_{\mu, \lambda} = \brac{ q : \mu(q) = \mu, \lambda(q) = \lambda }$ for all possible $\mu$ and $\lambda$. 
  However, we are only able to obtain bounds on sets of the form $Q_{A} = \brac{q : \mu(q) \in A \subseteq [0, r_{max}]}$ for an asymptotic characterization of a sequence of order-optimal admissible policies.
  These bounds can be derived using similar techniques as in Section \ref{chap4:section:aschar_optpolicy_intmc}.
  For example, if $r_{a,min} > 0$, then it can be shown that if $A \subseteq [0, u^{-1}(u_{c}) - a_{2} V^{\frac{1 - \delta}{2}}] \bigcup [u^{-1}(u_{c}) + a_{2} V^{\frac{1 - \delta}{2}}]$, for $0 < \delta \leq 1$ and $a_{2} > 0$, then $|Q_{A^{c}}| = \Omega\brap{\log\nfrac{1}{V_{k}}}$, for a sequence of non-idling order-optimal admissible policies with $\Cgk - c(u^{-1}(u_{c})) = V_{k} \downarrow 0$.
    
\end{remark}

\subsection{Tradeoff problems which are similar to \INTLMC}
\label{chap4:sec:sim_probs_INTMC}
In this section, we first consider other asymptotic regimes for \eqref{chap4:eq:sysmodel:probstat}, which are similar to \INTLMC.
We note that for \INTLMC, the utility constraint $u_{c}$ was kept fixed while $c_{c,k} \downarrow c(u^{-1}(u_{c}))$.
A similar problem (SP1) is one in which $c_{c}$ is fixed and $u_{c,k} \uparrow u(c^{-1}(c_{c}))$.
Another problem scenario (SP2) is one in which both $c_{c,k}$ and $u_{c,k}$ vary such that (a) $c_{c,k} - c(u^{-1}(u_{c,k})) \downarrow 0$ or (b) $u(c^{-1}(c_{c,k})) - u_{c,k} \downarrow 0$.

We note that SP2(b) encompasses SP1 since the sequence $u_{c,k}$ can be chosen such that $u_{c,k} = u_{c}, \forall k \in \sZ$.
We now show that the asymptotic regime for SP2(b) is equivalent to that for SP2(a), i.e., $c_{c,k} - c\brap{u^{-1}(u_{c,k})} \downarrow 0$.
We note that for any $(u_{c,k})$ and $(c_{c,k})$, for which the problem \eqref{chap4:eq:intlmc_problem_statement} is feasible, and also such that $u(c^{-1}(c_{c,k})) - u_{c,k} \downarrow 0$, we have that $\forall \epsilon > 0$, $\exists K_{\epsilon}$ such that, $\forall k > K_{\epsilon}$, $u(c^{-1}(c_{c,k})) - \epsilon \leq u_{c,k} \leq u(c^{-1}(c_{c,k}))$ (since $u_{c,k} \leq u(c^{-1}(c_{c,k}))$ if the problem \eqref{chap4:eq:intlmc_problem_statement} is feasible).
Then we have that $u^{-1}\brap{u(c^{-1}(c_{c,k})) - \epsilon} \leq u^{-1}(u_{c,k}) \leq c^{-1}(c_{c,k})$.
For every $c_{c,k}$, we define $l_{1,k}(\lambda)$ (as in Section \ref{chap4:sec:similarintmc}) to be (i) the tangent to $u(\lambda)$ at $(c^{-1}(c_{c,k}), u\brap{c^{-1}(c_{c,k})} )$, if $u(\lambda)$ is strictly convex, (ii) the line passing through $(a_{\mu}, u(a_{\mu}))$ and $(b_{\mu}, u(b_{\mu}))$, if $u(\lambda)$ is piecewise linear and $c^{-1}(c_{c,k})$ lies on a linear segment, and (iii) any line through $(a_{\mu}, u(a_{\mu}))$ with slope $m$, such that $\frac{du(\lambda)}{d\lambda}^{-}\vert_{\lambda = \mu} < m < \frac{du(\lambda)}{d\lambda}^{+}\vert_{\lambda = \mu}$, if $u(\lambda)$ is piecewise linear and $c^{-1}(c_{c,k})$ is a corner point of $u(\lambda)$.
We note that $l_{1,k}(c^{-1}(c_{c,k})) = u(c^{-1}(c_{c,k}))$ in all three cases.
Then 
\[u^{-1}\brap{u(c^{-1}(c_{c,k})) - \epsilon} \geq l_{1,k}^{-1}\brap{u(c^{-1}(c_{c,k})) - \epsilon} = l_{1,k}^{-1}\brap{u(c^{-1}(c_{c,k}))} - m_{1,k}\epsilon,\]
where $m_{1,k}$ is the slope of $l_{1,k}^{-1}$.
Since $l_{1,k}^{-1}\brap{u(c^{-1}(c_{c,k}))} = c^{-1}(c_{c,k})$, we have that
\begin{eqnarray*}
  c^{-1}(c_{c,k}) - m_{1,k}\epsilon \leq u^{-1}(u_{c,k}) \leq c^{-1}(c_{c,k}), \\
  c\brap{c^{-1}(c_{c,k}) - m_{1,k}\epsilon} \leq c(u^{-1}(u_{c,k})) \leq c_{c,k}.
\end{eqnarray*}
Let $l_{2,k}(\mu)$ be the tangent to $c(\mu)$ at $(c^{-1}(c_{c,k}), c_{c,k})$.
Then we have that
\begin{eqnarray*}
  l_{2,k}\brap{c^{-1}(c_{c,k}) - m_{1,k}\epsilon} \leq c(u^{-1}(u_{c,k})) \leq c_{c,k}, \\
  c_{c,k} - m_{2,k} m_{1,k} \epsilon \leq c(u^{-1}(u_{c,k})) \leq c_{c,k},
\end{eqnarray*}
where $m_{2,k}$ is the slope of $l_{2,k}$.
We note that $\exists M_{1}, M_{2} \in \sR$ such that $m_{1,k} \leq M_{1}$ and $m_{2,k} \leq M_{2}$ for every $k$, since both $u(\lambda)$ and $c(\mu)$ are defined on bounded domains.
Since the above inequality holds for every $\epsilon > 0$ and $k > K_{\epsilon}$, we have that that $c_{c,k} - c(u^{-1}(u_{c,k})) \downarrow 0$.

We now have the following result, under the stronger assumption that $u(.)$ is $m$-strongly concave, with $m > 0$.
The proof is similar to that of Lemma \ref{chap4:prop:p3lb}.
\begin{lemma}
  For INTERVAL-$\lambda\mu$CHOICE, for any sequence of non-idling admissible policies $\gamma_{k}$ and a sequence $u_{c,k} > 0$ such that $\overline{C}(\gamma_{k}) - c(u^{-1}(u_{c,k})) = V_{k} \downarrow 0$ and $\overline{U}(\gamma_{k}) \geq u_{c,k}$, we have that
  \begin{eqnarray*}
    \overline{Q}(\gamma_{k}) = \Omega\brap{\log\nfrac{1}{V_{k}}}.
  \end{eqnarray*}  
\end{lemma}
\begin{proof}
  The proof follows that of Lemma \ref{chap4:prop:p3lb} closely.
  Hence, we only state the differences here.
  We define $\mu^* = u^{-1}(u_{c,k}) - \epsilon_{V}$ and $q_{\mu^*} = \inf\brac{q : \mu(q) \geq \mu^*}$.
  We note that unlike in the proof of Lemma \ref{chap4:prop:p3lb}, we define a different tangent line $l_{k}(\mu)$ for every $u_{c,k}$.
  Let $l_{k}(\mu)$ be the tangent line to $c(\mu)$ at $(u^{-1}(u_{c,k}), c(u^{-1}(u_{c,k})))$.
  From Proposition \ref{chap4:app:prop:quadlowerbound}, we have a positive $a_{1,k}$ such that
  \begin{eqnarray*}
    V_{k} \geq a_{1,k} \sum_{q = 0}^{\zms - 1} \bras{\mz - u^{-1}(u_{c,k})}^{2} \pi(q).
  \end{eqnarray*}
  We note that unlike the proof of Lemma \ref{chap4:prop:p3lb}, here $a_{1,k}$ depends on the sequence $u_{c,k}$.
  Let $a \Deq \inf_{k} \brac{a_{1,k}}$.
  Since $u(.)$ is $m$-strongly convex, we have that $a \geq m > 0$.
  Then we have that
  \begin{eqnarray*}
    Pr\brac{Q \leq \zms - 1} \leq \frac{V}{a\epv^{2}}, \\
    \text{and } \pi(\zms - 1) \leq \frac{V}{a\epv^{2}}.
  \end{eqnarray*}
  Then, we proceed as in the proof of Lemma \ref{chap4:prop:p3lb} to obtain that $\overline{Q}(\gamma_{k}) = \Omega\brap{\log\nfrac{1}{V_{k}}}$.
\end{proof}
We note that an asymptotic upper bound can be obtained by evaluating $\Qgk, \Cgk$, and $\Ugk$ for a sequence of policies $\gamma_{k}$ as in Lemma \ref{chap4:prop:p3ub}, but with $u_{c}$ now being the sequence $u_{c,k}$.
Then we have the following result
\begin{proposition}
  For INTERVAL-$\lambda\mu$CHOICE, for strongly concave or linear $u(\lambda)$, we have that the optimal tradeoff curve $Q^*(c_{c,k},u_{c,k}) = \Theta\brap{\log\nfrac{1}{c_{c,k} - c(u^{-1}(u_{c,k}))}}$, for the sequence $c_{c,k} = \Cgk$ and $u_{c,k} = \Ugk$, for the sequence of policies $\gamma_{k}$ as above.
\end{proposition}

\begin{remark}
  We note that for \INTLMC\, we have considered the case where $c(\mu)$ is strictly convex and $u(\lambda)$ is either strictly concave or linear (also piecewise linear for the asymptotic lower bound in Lemma \ref{chap4:prop:p3lb}).
  Although we have not presented the analysis for other forms of $c(\mu)$, such as when $c(\mu)$ is piecewise linear, here we outline how the methods presented in Chapter 2 as well as this chapter can be used in obtaining asymptotic lower bounds in these cases, in the asymptotic regime where $c_{c,k} \downarrow c(\newl)$.
  Suppose $c(\mu)$ is piecewise linear and $u_{c}$ is fixed.
  We note that as in \INTMC-2, we can define service rates $\mu_{l}$ and $\mu_{u}$ with respect to $u^{-1}(u_{c})$ rather than $\lambda$.
  Then the asymptotic behaviour of $Q^*(c_{c},u_{c})$ depends upon whether (i) $\mu_{l} < u^{-1}(u_{c}) < \mu_{u}$ and $\mu_{l} = 0$ or (ii) otherwise.
  For case (i), we note that $Q^*(c_{c}, u_{c})$ only increases to a finite value, since we can fix $\lambda(q) = u^{-1}(u_{c})$ and apply the analysis of \INTMC-2-1.
  However, we do not have an asymptotic lower bound in this case.
  For case (ii), we can proceed as in the proof of Lemma \ref{chap4:prop:p3lb}, except that $\mu^* \Deq \mu_{l} - \epsilon$, where $\epsilon > 0$, to obtain that $Q^*(c_{c}, u_{c})$ is $\Omega\brap{\log\nfrac{1}{V}}$.
\end{remark}

\section{Conclusions}
\label{chap4:sec:conclusions}
In this chapter, we have considered the asymptotic characterization of the tradeoff problem for the state dependent M/M/1 model, with model features chosen such that insights can be obtained for the tradeoff problem for discrete time queueing models also.
From the analysis, we see that for INTERVAL-$\mu$CHOICE and INTERVAL-$\lambda\mu$CHOICE the constraint on the average service cost, in the regime $\Re$, leads to constraints on the stationary probability distribution of the service rate $\mu(Q)$ and therefore the stationary probability distribution of the queue length which in turn determines the asymptotic growth of the minimum average queue length as a function of the average service cost.
The exact nature of these constraints and the behaviour of the stationary distribution of the queue length depends on the nature of $c(\mu)$ at $\mu = \lambda$ and the extent of freedom in the choice of $\lambda(q)$ and $\mu(q)$ at a queue length $q$.

We observe that if $c(\mu)$ is strictly convex at $\lambda$, then in the asymptotic regime $\Re$ as $V = c_{c} - c(\lambda) \downarrow 0$, the stationary probability of any queue length $q$ such that $\mu(q) \neq \lambda$ goes to zero.
More precisely, as $V \downarrow 0$, $\mu(q)$ for $q \in \hpq$ (the set of queue lengths with high probability) has to approach $\lambda$, the stationary probabilities for such queue lengths are equal and each is $\mathcal{O}(\sqrt{V})$.
This leads to the $\frac{1}{\sqrt{V}}$ behaviour for strictly convex $c(\mu)$.
We expect that this is the same phenomenon which gives rise to the Berry-Gallager asymptotic lower bound \cite{berry}, but for admissible policies.
In Chapters 4 and 5, we see that this particular behaviour of the stationary probability does carry over to the discrete time model.
Suppose, however that it is possible to control the arrival rates $\lambda(q)$, as in the case of INTERVAL-$\lambda\mu$CHOICE.
Then even though $\mu(q)$ for queue lengths $q \in \hpq$ have to approach a single value $u^{-1}(u_{c})$, the drift for such $q$ need not be zero, since $\lambda(q)$ can be chosen to be different from $u^{-1}(u_{c})$ for such $q$.
In fact it is possible to chose $\lambda(q)$ such that the drift is initially positive and then negative for the set of queue lengths occurring with high probability so that the stationary probability of the queue length has a geometrically increasing and decaying behaviour which leads to the $\log\nfrac{1}{V}$ asymptotic growth for the minimum average queue length.
We expect that this is the reason for the $\mathcal{O}\brap{\log\nfrac{1}{V}}$ behaviour observed by Neely \cite{neely_utility}, but for admissible policies.
In Chapter 5, we shall obtain an asymptotic lower bound for the discrete time model considered in \cite{neely_utility} using the above idea, for admissible policies.

When $c(\mu)$ is piecewise linear in $\mu$, we note that either $(\lambda, c(\lambda))$ can be a corner point as in INTERVAL-$\mu$CHOICE-2-3 or $(\lambda, c(\lambda))$ can lie on a linear portion of $c(\mu)$ as in INTERVAL-$\mu$CHOICE-2-1 or INTERVAL-$\mu$CHOICE-2-2.
For INTERVAL-$\mu$CHOICE-2-3, as in the case of INTERVAL-$\mu$CHOICE-1, we observe that as $V \downarrow 0$, $\mu(q)$ for queue lengths in $\hpq$ has to approach $\lambda$, we again observe that the drift (proportional to $\lambda - \mu(q)$) for such queue lengths approaches zero, and the stationary probabilities for such queue lengths are equal and each is $\mathcal{O}(V)$.
This leads to the $\frac{1}{{V}}$ behaviour for INTERVAL-$\mu$CHOICE-2-3.

For INTERVAL-$\mu$CHOICE-2-1 and INTERVAL-$\mu$CHOICE-2-2, we note that as $V \downarrow 0$ service rates $\mu$ such that $\mu_{l} \leq \mu \leq \mu_{u}$ could be used, so that the drift for the set of queue lengths occurring with high probability is not zero.
For INTERVAL-$\mu$CHOICE-2-2, it is possible to choose $\mu(q)$ so that the drift is initially increasing and then decreasing for the set of queue lengths occurring with high probability.
Then the stationary probability has a geometrically increasing and decaying behaviour which leads to the $\log\nfrac{1}{V}$ behaviour.
We expect that this is the reason behind the $\mathcal{O}\brap{\log\nfrac{1}{V}}$ growth observed for the case of piecewise linear cost functions in \cite{neely_mac}.
For INTERVAL-$\mu$CHOICE-2-1 we note that as $V \downarrow 0$ the service rate $0$ could be used, which for non-idling admissible policies implies that the queue length $0$ has positive stationary probability, even for $V = 0$.
This intuitively implies that the average queue length does not increase to infinity in this case.

Asymptotic bounds on any sequence of non-idling order-optimal admissible policies have been presented in Section \ref{chap4:section:aschar_optpolicy_intmc}.
We have also discussed other variants of the tradeoff problem, such as \INTLC, where only the arrival rate can be controlled.
Even though \INTLC\, is very similar to \INTMC\, we find that there are no cases for \INTLC\, where the minimum average queue length increases to a finite value unlike \INTMC.
We observe that the asymptotic behaviour of the minimum average queue length as the utility constraint is made arbitrarily close to the maximum value of the utility, can be obtained using ideas which are similar to that of \INTMC, which shows that the method of obtaining the asymptotic behaviour of the average queue length through its stationary distribution for monotone policies is sufficiently general and provides a unified method which explains other scenarios also.

We also note that through the analysis of the state dependent M/M/1 model, with proper choice of model features, we get insights as to how to construct asymptotic lower bounds for the discrete time models.
In Chapter 4, we find bounds on the stationary probability distribution of the queue length for the discrete time queueing model in the regime $\Re$, which have the asymptotic behaviour as suggested by the above analysis.
Thus these bounds lead to the \emph{right} asymptotic behaviour of the average queue length in the regime $\Re$ for the discrete time queueing models.

\clearpage
\begin{subappendices}
\large{\textbf{Appendices}}
\addcontentsline{toc}{section}{Appendices}
\addtocontents{toc}{\protect\setcounter{tocdepth}{0}}
\normalsize  
\section{A lower bound for strictly convex functions}
\begin{proposition}
  Let $g(x) : [0,R] \rightarrow \mathbb{R}_{+}$ be a finite strictly convex function, such that: (i) $g(0) = 0$, (ii) $g'(0) \geq 0$, and (iii) $g''(0) > 0$.
  Then there exists a positive constant $a$ such that $g(x) \geq ax^{2}$, $\forall x \in [0,R]$.
  \label{chap4:app:prop:quadlowerbound}
\end{proposition}

\begin{proof}
  We note that at $0$, $g(0) = a.0$.
  If $\exists a > 0$ such that for all $x \in (0,R]$, $g'(x) \geq 2ax$, then $g(x) \geq ax^{2}$.
  So we have to prove that for all $x \in (0,R]$, $g'(x) \geq 2ax$.
  In essence, we have to prove that it is possible to find a positive $a$ such that $\forall x \in (0,R], \frac{g'(x)}{x} \geq a$.
  We note that if $\inf_{x \in (0,R]} \frac{g'(x)}{x} > 0$, then it is possible to find such an $a$.

  Now we prove that $\inf_{x \in (0,R]} \frac{g'(x)}{x} > 0$.
  Since both $g'(x)$ and $x$ are non-negative, $\frac{g'(x)}{x} \geq 0$.
  Suppose we assume that $\inf_{x \in (0,R]} \frac{g'(x)}{x} = 0$.
  Then we have that $\forall \epsilon > 0$, $\exists x$ such that $\frac{g'(x)}{x} \leq \epsilon$.

  Consider a sequence $\epsilon_{n} \downarrow 0$, then there exists a sequence $x_{n}$ such that $\frac{g'(x_{n})}{x_{n}} \leq \epsilon_{n}$.
  Note that as $x_{n} \leq R$, we have that $1) \lim_{n \rightarrow \infty} g'(x_{n}) = 0$ and $2) \lim_{n \rightarrow \infty} \frac{g'(x_{n})}{x_{n}} = 0$.
  The sequence $x_{n}$ may not be convergent.
  So we consider the subsequence $y_{m}$ such that $\lim_{m \rightarrow \infty} y_{m} = \limsup_{n \rightarrow \infty} x_{n} = y$.
  Note that $\lim_{m \rightarrow \infty} g'(y_{m}) = 0$ and $\lim_{m \rightarrow \infty} \frac{g'(y_{m})}{y_{m}} = 0$ as $y_{m}$ is a subsequence of $x_{n}$.
  Now there are two cases : 1) $y > 0$ and 2) $y = 0$.
  Let $y > 0$, then by continuity of $g'(x)$ we have that $g'(y) = 0$.
  But $g'(y) > 0$ for every $y > 0$ and we have a contradiction on the assumption that $\inf_{x \in (0,R]} \frac{g'(x)}{x} = 0$.
  Consider the second case when $y = 0$, then we have that $\lim_{y_{m} \rightarrow 0} \frac{g'(y_{m})}{y_{m}} = 0$.
  But note that $\lim_{x \rightarrow 0} \frac{g'(x)}{x} = g''(0) > 0$.
  Thus we again have a contradiction.
  Hence $\inf_{x \in (0,R]} \frac{g'(x)}{x} > 0$.
  We choose $a = \frac{1}{4} \inf_{x \in (0,R]} \frac{g'(x)}{x}$.
\end{proof}

\section{Application of the Berry-Gallager lower bounding technique \cite{berry} in Remark \ref{chap4:remark:berrygallagerlb} for {\INTMC-1}}
\label{chap4:app:berrygallagerlb}
For \INTMC, since we assume that $\mu(q) \leq r_{max}$, we can obtain a discrete time process $Q_{d}[m], m \in \sZ$, by uniformizing the CTMC $Q(t)$ at rate $r_{u} = \lambda + r_{max}$ as in Appendix \ref{chap4:app:uniformization}.
We now outline how the lower bounding technique in \cite{berry} can be applied to $Q_{d}[m]$ to obtain the $\Omega\nfrac{1}{\sqrt{c_{c} - c(\lambda)}}$ asymptotic lower bound for $Q^*(c_{c})$ as $c_{c} \downarrow c(\lambda)$.
We first show that a slightly modified form of \cite[Lemma 4.1]{berry} holds for the uniformized process $Q_{d}[m]$ under an admissible policy $\gamma$ (we note that we are using admissibility as defined in Chapter 2 and not as in \cite{berry}).
As in \cite[Appendix A]{berry}, we have that $Pr\brac{Q < \ceiling{2\Qg}} > \frac{1}{2}$.
Let $q_{p} = \argmax_{q \in \brac{0, \dots, \ceiling{2\Qg}}} \pi(q)$.
Then we have that $\pi(q_{p}) \geq \frac{1}{2\ceiling{2\Qg}}$.
\newcommand{\wqd}{\widehat{Q}_{d}}
Now we define $\wqd[m] = \max\brap{Q_{d}[m], q_{p}}$.
Then as in \cite[Appendix A]{berry} we can show that
\begin{eqnarray*}
  \pi(q_{p}) \Exp\bras{\wqd[m + 1] - \wqd[m]\vert Q_{d}[m] = q_{p}} + \sum_{q = q_{p} + 1}^{\infty} \pi(q)\brap{\lambda - \mu(q)} \leq 0.
\end{eqnarray*}
We note that $\Exp\bras{\wqd[m + 1] - \wqd[m]\vert Q_{d}[m] = q_{p}} = \frac{\lambda}{r_{u}}$.
Therefore we obtain that
\begin{eqnarray*}
  \pi(q_{p}) \frac{\lambda}{r_{u}} + \sum_{q = q_{p} + 1}^{\infty} \pi(q)\brap{\lambda - \mu(q)} & \leq & 0, \text{ or,} \\
  \sum_{q = q_{p} + 1}^{\infty} \pi(q)\brap{\lambda - \mu(q)} & \leq & -\frac{\lambda}{r_{u} 2 \ceiling{2\Qg}}.
\end{eqnarray*}
Then to obtain the asymptotic lower bound on $Q^*(c_{c})$ we use the above upper bound in step (49) in the proof of \cite[Proposition 4.2]{berry}.

\end{subappendices}
\addtocontents{toc}{\protect\setcounter{tocdepth}{2}}

\blankpage
\newcommand{\Deltat}{\frac{\Delta}{2}}
\chapter[On the tradeoff of average queue length and average service cost for discrete time \\single server queues]{\textbf{On the tradeoff of average queue length and average service cost \\ for discrete time single server queues}}

\section{Introduction}
\label{chap5:sec:introduction}

In this chapter we consider the tradeoff between average queue length and average service cost for the discrete time single server queueing models introduced in Chapter 1.
Such discrete time models are in some cases more appropriate for modelling resource allocation problems in communication networks than the continuous time models considered in the previous chapters.
A context in which such a model and the following analysis may be appropriate is that of wireless networks with fading, where one of the issues, which has been studied by many researchers (see \cite{berry}, \cite{munish}, \cite{bettesh}, and \cite{elif}), is the optimal tradeoff of the average power and average delay, when the service batch size is dynamically chosen as a function of the fade state and the queue length.
The characterization of the tradeoff between average error rate and average delay, for a point-to-point noisy link, when the service batch size is dynamically chosen as a function of the queue length, is another resource allocation problem that motivates the model studied in this chapter.
The tradeoff of average power and average delay is dealt with in more detail in Chapter 5, whereas the tradeoff of average error rate with average delay is considered in Chapter 6.
In this chapter, we consider two simplified discrete time models, with no admission control (i.e. $A[m] = R[m], \forall m$) and a single environment state, to develop the basic techniques for the characterization of such tradeoffs.
The glossary of notation that we use in this chapter is given in Table \ref{chapter4:notationtable}.
We now summarize the methodology that is used for obtaining the asymptotic lower bounds on the tradeoffs.
\begin{table}
\centering
\begin{tabular}{|l|l|}
\hline
Symbol & Description \\
\hline
$m$ & slot index \\
$A[m]$ & random number of arrivals in slot $m$ (after admission control) \\
$A_{max}$ & maximum number of arrivals in any slot \\
$\lambda, \sigma^{2}$ & mean and variance of $A[1]$ \\
$S[m]$ & batch service size in slot $m$ \\
$S_{max}$ & maximum batch service size \\
$\epsilon_{a}$ & probability of $A[m]$ exceeding $S_{max}$ \\
$Q[m]$ & queue length at the start of $(m + 1)^{th}$ slot \\
$\sigma[m]$ & history of queue evolution \\
$\gamma$ & policy - $(S[1], S[2], \cdots)$ \\
$\Gamma$ & set of all policies \\
$\Gamma_{s}$ & set of all stationary policies \\
$\Qg$ & average queue length for a policy $\gamma$ \\
$c(s)$ & service cost for I-model; a function of integer valued batch service size $s$ \\
$c_{R}(s)$ & service cost for R-model; a function of real valued batch service size $s$ \\
$\Cg$ & average service cost for a policy $\gamma$ \\
$c_{c}$ & average service cost constraint \\
$\beta_{c_{c}}$ & non-negative Lagrange multiplier \\
$c_{\beta_{c_{c}}}(q, s)$ & single stage cost; defined as $q + \beta_{c_{c}} c(s)$ \\
$g^*_{\beta}$ & optimal average cost for unconstrained MDP \\
$J_{\beta}(q)$ & relative value function for unconstrained MDP \\
$\gamma^*_{\beta}$ & optimal stationary policy for unconstrained MDP \\
$s^*_{\beta}(q)$ & action at state $q$ for the optimal policy $\gamma^*_{\beta}$ \\
$\mathcal{R}_{\gamma^*_{\beta}}$ & recurrence class for $\gamma^*_{\beta}$ \\
$\Gamma_{a}$ & set of all admissible policies \\
$\overline{s}(q)$ & average service rate at a queue length $q$ \\
$\pi$ & stationary distribution for a policy which is clear from the context \\
$\pi_{\gamma}$ & stationary distribution for policy $\gamma$ \\
$Q^*(c_{c})$ & minimum average queue length over the set $\Gamma_{a}$ under constraint $c_{c}$ \\
$s_{l}$ & largest service batch size $\leq \lambda$ at which slope of $c(.)$ changes \\
$s_{u}$ & smallest service batch size $\geq \lambda$ at which slope of $c(.)$ changes \\
\hline
\end{tabular}
\caption{Notation used in this chapter.}
\label{chapter4:notationtable}
\end{table}

\subsection{Methodology}
\label{chapter4:methodology}
In this section, we summarize a scheme that is used in Chapters 4, 5, and 6 to obtain asymptotic lower bounds on the minimum average queue length in the regime $\Re$ as a function of the average service cost constraint (Chapters 4 and 6) or average power constraint (Chapter 5).
As in Chapters 2 and 3, we obtain a lower bound on the stationary mean queue length using an upper bound on the stationary probability distribution of the queue length under the assumption that the queue length process is a DTMC and ergodic. 
The average service cost (or average power) is then related to the stationary probability distribution.

Let $Q \sim \pi$, where $\pi$ is the stationary probability distribution of the queue length.
From Markov inequality, we have that $\Exp_{\pi} Q \geq \overline{q} Pr\brac{Q \geq \overline{q}}$.
Suppose $Pr_{u}\brac{Q < q}$ is any upper bound on the stationary probability distribution $Pr\brac{Q < q}$.
If $\overline{q}$ is the largest $q$ such that $Pr_{u}\brac{Q < q} \leq \alpha$, then we have that $\Exp_{\pi}Q \geq \overline{q}\brap{1 - \alpha}$.
For convenience, we choose $\alpha = \frac{1}{2}$.

The upper bounds $Pr_{u}\brac{Q < q}$ were obtained for the state dependent M/M/1 models in Chapters 2 and 3 using the detailed balance equations and lower bounds on the service rate $\mu(q)$.
In Chapters 4, 5, and 6 we obtain geometric bounds on $Pr_{u}\brac{Q < q}$, the first (Lemma \ref{chap5:lemma:barq_pi0_relation}) of which has been obtained by assuming certain properties for the transition probability distribution  while the second (Lemma \ref{chap5:lemma:dtmc_stat_prob}) has been obtained by extending the results available in Bertsimas et al. (\cite{gamarnik} and \cite{gamarnik_1}).

We now illustrate this scheme for the case of integer-valued queue evolution through an example.
We will obtain that for the queueing process, the upper bound $Pr_{u}\brac{.}$ has a geometric form, i.e., $Pr_{u}\brac{Q < q} = \pi(0)\rho^{q}$.
For our tradeoff problem, it will turn out that a non-negative function $D(q)$ can be obtained, such that $\Exp_{\pi}D(Q)$ is the difference between the average service cost and $c(\lambda)$ (which can be defined similarly as for \INTMC).
Then if $\Exp_{\pi}D(Q) \leq V$ for a $V > 0$, then we have that $\pi(0)D(0) \leq V$, where $D(0) > 0$.
Therefore, we have a further upper bound on the stationary probability distribution, $Pr_{u}\brac{Q < q} = \frac{V}{D(0)}\rho^{q}$.
Then, the largest $\overline{q}_{1}$ such that $Pr_{u}\brac{Q < q} \leq \frac{1}{2}$ satisfies $\frac{V}{D(0)}\rho^{\overline{q}_{1}} \leq \frac{1}{2}$.
Or we have that $\overline{q}_{1} = \floor{\log_{\rho}\nfrac{D(0)}{2V}}$.
In the asymptotic regime $\Re$, as $V \downarrow 0$, we obtain that $\overline{q}_{1}$ is $\Omega\brap{\log\nfrac{1}{V}}$ and therefore so is the average queue length as a function of $V$.
Variations of this basic method are used throughout Chapters 4, 5, and 6 to obtain the asymptotic lower bounds.

\subsection{System model - Integer valued queue evolution}
\label{chap5:sec:integervalued_setup}
We assume time to be slotted, with slots indexed by $m \in \mathbb{Z}_{+}$.
In each slot $m$, a random number of customers $A[m] \in \mathbb{Z}_{+}$ arrive into the system.
The arrival process $(A[m], m \geq 1)$ is assumed to be IID with $A[1] \leq A_{max}$, batch arrival rate $\mathbb{E}A[1] = \lambda < \infty$, $\text{var}(A[1]) = \sigma^{2} < \infty$.
The customers arrive into an infinite buffer queue.
In slot $m$, a batch of customers of size $S[m] \in \mathbb{Z}_{+}$ is served.
The batch of $S[m]$ customers is removed from the queue at the end of the $m^{th}$ slot just before the new batch of customers which arrive in the $m^{th}$ slot, $A[m]$, is admitted.
We assume that $S[m] \leq S_{max}$, where $S_{max}$ is the maximum batch size that can be served.
We also assume that
\begin{description}
\item[A1 :]{$Pr\brac{A[1] > S_{max}} > \epsilon_{a} > 0$.}
\end{description}
We note that the above assumption is similar to the assumptions made in the definition of admissible policies in \cite{berry} and \cite{neely_mac}.
Furthermore, we note that the above assumption is reasonable, since the maximum number of arrivals usually exceeds the maximum capacity $S_{max}$ of service, but $\lambda < S_{max}$.

The number of customers in the queue at the start of the $(m + 1)^{th}$ slot is denoted by $Q[m]$.
We assume that $Q[0] = q_{0} \in \mathbb{Z}_{+}$ customers.
The queue evolution for $m \geq 0$ is given by:
\begin{equation}
  Q[m + 1] = Q[m] - S[m + 1] + A[m + 1],
  \label{chap5:eq:evolution}
\end{equation}
where $S[m + 1] \leq \min(S_{max}, Q[m])$.
We define a policy $\gamma$ to be the sequence of batch sizes $(S[1], S[2], \dots)$ under which the queue evolves according to \eqref{chap5:eq:evolution}.
The set of all policies is denoted by $\Gamma$.
A policy $\gamma$ is stationary if $S[m + 1] = S(Q[m]), \forall m$, where $S(q)$ is a randomized function of $q$.
The set of all stationary policies is denoted as $\Gamma_{s}$.

We assume that a service cost of $c(s)$ is incurred when serving a batch of size $s$.
For example, this cost could be the expected number of symbols in error for the transmission of a batch of message symbols.
The function $c(s):\brac{0,1,\dots,S_{max}} \rightarrow \mathbb{R}_{+}$ is assumed to satisfy the following properties:
\begin{description}
\item[C1 :]{$c(0) = 0$,}
\item[C2 :]{$c(s)$ is non-decreasing and convex
\footnote{If $S_{max} = 1$, then there is no tradeoff. If $S_{max} \geq 2$, then $c(s + 2) - c(s + 1) \geq c(s + 1) - c(s), \forall s \in \brac{0, \dots, S_{max} - 2}$.}
for $s \in \brac{0, \dots, S_{max} - 2}$.}
\end{description}

We now define the performance measures that we are interested in:
a) The worst case average queue length for a policy $\gamma \in \Gamma$ is
\begin{equation}
  \overline{Q}(\gamma, q_{0}) = \limsup_{M \rightarrow \infty} \frac{1}{M} \mathbb{E}\left[\sum_{m = 0}^{M - 1} Q[m] \middle\vert Q[0] = q_{0} \right],
  \label{chap5:eq:averageq}
\end{equation}
and b) the worst case average service cost for a policy $\gamma \in \Gamma$ is
\begin{equation}
  \overline{C}(\gamma, q_{0}) = \limsup_{M \rightarrow \infty} \frac{1}{M} \mathbb{E}\left[\sum_{m = 1}^{M}c(S[m]) \middle\vert Q[0] = q_0 \right].
  \label{chap5:eq:averagec}
\end{equation}

\subsection{System model - Real valued queue evolution}
\label{chap5:sec:realvaluedproblem}
We state only the differences from the model discussed in Section \ref{chap5:sec:integervalued_setup}.
We assume that the arrival random variable $A[m] \in [0, A_{max}] \subset \mathbb{R}_+$, with mean $\lambda < \infty$ and variance $\sigma^{2} < \infty$.
We also assume that the service batch size $S[m] \in [0, S_{max}] \subset \mathbb{R}_+$.
Similar to assumption A1, we make the assumption:
\begin{description}
\item[RA1 :]{$Pr\brac{A[1] - S_{max} > \delta_{a}} > \epsilon_{a}$,}
\end{description}
where both $\delta_{a}$ and $\epsilon_{a}$ are positive real numbers.
Let the initial queue length be $q_{0} \in \mathbb{R}_+$.
We have that the queue length $Q[m] \in \mathbb{R}_+$ and the evolution of $Q[m]$ is given by \eqref{chap5:eq:evolution} with $S[m] \leq \min(S_{max}, Q[m - 1])$.
We assume that there is a service cost $c_{R}(s)$ associated with the service of a batch of size $s$.
The function $c_{R}(s) : [0,S_{max}] \rightarrow \mathbb{R}_{+}$ is assumed to satisfy the following properties:
\begin{description}
\item[RC1 :]{$c_{R}(0) = 0$,}
\item[RC2 :]{$c_{R}(s)$ is strictly convex and increasing in $s$, for $s \in [0, S_{max}]$.}
\end{description}
In the following, the model with integer valued queue evolution is referred to as the I-model, while the model with real valued queue evolution is referred to as the R-model.
We note that R-model with the strictly convex $c_{R}(s)$ cost function is usually used as an approximation to the I-model, which has $c(s)$ as the cost function.

We note that our I-model is a simplified version of the model studied by Goyal et al. \cite{munish}, wherein there is an additional state variable which is used to model fading, the arrival process is Markov, and $S_{max} = \infty$.
Our R-model is a simplified version of the model studied by Berry and Gallager \cite{berry}, wherein there is an additional fade state variable, and $S_{max} = \infty$.

\subsection{Overview}
\label{chap5:sec:overview}
The tradeoff problem \eqref{introduction:eq:dt_specialTradeoff} for I-model and R-model is formulated as a constrained Markov decision problem in Section \ref{chap5:sec:problemformulation}.
We consider the I-model first.
For I-model, for certain values of the cost constraint, we consider an equivalent unconstrained Markov decision problem (as in \eqref{introduction:eq:dt_specialTradeoff_dual}) in Section \ref{chap5:sec:mdpformulation}.
We also identify several properties that are possessed by any stationary deterministic optimal policy for this problem.
We then define the set of admissible policies, which are policies possessing the above properties (the definition of admissible policies is similar to that in Chapter 2).

From Section \ref{chap5:sec:tradeoffproblem} onwards, we consider the tradeoff problem for the set of admissible policies.
In Section \ref{chap5:sec:asymp_analysis_tprob} we characterize the infimum of the average service cost over all possible admissible policies, which is equal to the minimum average service cost required for mean rate stability.
We identify three cases, which are similar to the three subcases for INTERVAL-$\mu$CHOICE-2, for which the asymptotic behaviour of the minimum average queue length is characterized.
The asymptotic behaviour of the minimum average queue length is obtained as for \INTMC-2, by first obtaining upper bounds on the stationary probability of the queue length.
One of these bounds is a state dependent extension of the geometric bounds on the stationary probability of discrete time Markov chains presented in \cite{gamarnik} and \cite{gamarnik_1}.
In Section \ref{chap5:sec:asympchar} we show that depending on the value of the arrival rate, the minimum average queue length either (i) increases only to a finite value, or (ii) increases as $\log\nfrac{1}{V}$, or (iii) increases as $\frac{1}{V}$, when the average service cost is $V$ more than the infimum of the average service costs for admissible policies.
Asymptotic bounds on order-optimal policies are presented in Section \ref{chap5:sec:aschar}.
We obtain an asymptotic lower bound on the minimum average queue length for ergodic arrival processes in Section \ref{chap5:sec:extension_ergodic}, when the average service cost is $V$ more than the infimum of the average service costs for admissible policies.

For R-model, we present an asymptotic analysis in Section \ref{chap5:sec:realvaluedproblem}.
We obtain that the minimum average queue length is $\Omega\nfrac{1}{\sqrt{V}}$ when the average service cost is $V$ more than the infimum of the average service costs for admissible policies, for strictly convex $c_{R}(s)$.
Then we consider the case where $c_{R}(s)$ is piecewise linear and show that the asymptotic behaviour of the minimum average queue length is similar to that of the I-model.

\section{Problem formulation for I-model and R-model}
\label{chap5:sec:problemformulation}
The tradeoff problem is to obtain $Q^*(c_{c}, q_{0})$, which is the optimal value of the optimization problem
\begin{equation}
  \mini_{\gamma \in \Gamma} \overline{Q}(\gamma, q_{0}) \text{ such that } \overline{C}(\gamma, q_{0}) \leq c_{c},
  \label{chap5:eq:inittradeoffprob}
\end{equation}
where $c_{c} \geq 0$ is the average service cost constraint.
The tradeoff curve $Q^*(c_{c}, q_{0})$ is non-increasing and convex in $c_{c}$ (see \cite{berry_thesis}) for any $q_{0}$.
We note that if $\lambda > S_{max}$ then any feasible policy for \eqref{chap5:eq:inittradeoffprob} is optimal, as the average queue length for any such feasible policy is infinity.
Hence, in the following we assume that $\lambda \leq S_{max}$.

\subsection{A constrained Markov decision process formulation}
\label{chap5:sec:cmdpformulation}
The tradeoff problem \eqref{chap5:eq:inittradeoffprob} can be formulated as a constrained Markov decision problem (CMDP) \cite{altman}.
The state space of the CMDP is the state space of the queue length, which is $\mathbb{Z}_+$ for the I-model and $\mathbb{R}_{+}$ for the R-model.
The action spaces at each state $q$ are the sets $\brac{0,\dots, \min(q, S_{max})}$ for the I-model and $[0, \min(q, S_{max})]$ for the R-model, both of which are compact for every $q$.
The probabilistic evolution of the state of the CMDP from stage $m$ to $m + 1$ is given by \eqref{chap5:eq:evolution}.
Associated with the CMDP there are two single stage costs: (i) the holding cost $q$ at state $q$, and (ii) the service cost $c(s)$ when an action $s$ is taken at state $q$.

We redefine $c(s)$ as the lower convex envelope\footnote{We note that the lower convex envelope can be interpreted as the solution: $c(s) = \mini \Exp c(X)$, such that $\Exp X = s$.} of $c(s), s \in \brac{0, \dots, S_{max}}$.
We note that the redefined cost function $c(s)$ ($c(s):[0, S_{max}] \rightarrow \mathbb{R}_+$) is a piecewise linear convex function.
From \cite{hernandez} and \cite{hernandez_2}, it is possible to show that if $c_{c} > c(\lambda)$ then \eqref{chap5:eq:inittradeoffprob} has an optimal solution and there exists an optimal policy $\gamma \in \Gamma_{s}$.

In the following, we show that for some values of $c_{c}$, there exists a stationary deterministic optimal policy for \eqref{chap5:eq:inittradeoffprob}.
Consider the following MDP:
\begin{eqnarray}
  \mini_{\gamma \in \Gamma} \bras{ \overline{Q}(\gamma, q_{0}) + \beta_{c_{c}}(\overline{C}(\gamma, q_{0}) - c_{c}) }.
  \label{chap5:eq:initial_uncon_mdp}
\end{eqnarray}
We note that the above MDP has a single stage cost of $c_{\beta_{c_{c}}}(q,s) = q + \beta_{c_{c}} c(s)$ in state $q$.
From Ma et al. \cite{ma}, it is known that if there exists a $\beta_{c_{c}} > 0$, such that any stationary deterministic optimal policy for \eqref{chap5:eq:initial_uncon_mdp}, has an average service cost equal to the constraint $c_{c}$, then the same policy is optimal for the constrained problem \eqref{chap5:eq:inittradeoffprob}.
The factor $\beta_{c_{c}}$ can be interpreted as a Lagrange multiplier.
The set of values of $c_{c}$ for which such $\beta_{c_{c}}$ exist is denoted as $\mathcal{O}^{u}$, as in Chapter 2.
We note that the properties of any stationary deterministic optimal policy, which can be obtained from \eqref{chap5:eq:initial_uncon_mdp}, carry over to \eqref{chap5:eq:inittradeoffprob} if $c_{c} \in \mathcal{O}^{u}$.

We note that the development in Altman \cite{altman} which leads to Theorem 12.7, which shows that for every value of the constraint $c_{c}$, there exists a Lagrange multiplier for which there is a stationary deterministic policy which is optimal for both the unconstrained MDP and the CMDP, requires assumption (B1) \cite[Chapter 11]{altman}, which does not hold for our model.

In the next section, for the I-model we study \eqref{chap5:eq:initial_uncon_mdp} in detail.
The properties of any stationary deterministic optimal policy are then used to motivate the definition of a class of admissible policies.
The tradeoff problem \eqref{chap5:eq:inittradeoffprob} is then analysed for the class of admissible policies.
We note that for $c_{c} \in \mathcal{O}^{u}$, there exists at least one optimal admissible policy.

\section{Asymptotic bounds for I-model}
\subsection{An unconstrained MDP formulation}
\label{chap5:sec:mdpformulation}

The unconstrained MDP \eqref{chap5:eq:initial_uncon_mdp}, which is obtained via the above Lagrange multiplier relaxation, is studied in \cite{munish}.
However, we note that the development in \cite{munish} does not lead to an average cost optimality equation (ACOE).
Since the ACOE enables us to obtain some additional properties of the optimal policy, in the next section we use the results from Sennott \cite{sennott_mdp} to show that there exists a stationary deterministic average cost optimal policy, which also satisfies an ACOE, for a single stage cost of $c_{\beta}(q,s)$ for our model.

The state space of the unconstrained MDP \eqref{chap5:eq:initial_uncon_mdp} is the state space $\mathbb{Z}_+$ of the queue length, which is countable.
The action space at each state $q$ is the set $\brac{0,\dots, \min(q, S_{max})}$ which is compact.
The probabilistic evolution of the state of the MDP from stage $m$ to $m + 1$ is given by \eqref{chap5:eq:evolution}.
The single stage cost for the unconstrained MDP is $c_{\beta}(q,s) = q + \beta c(s)$, where $\beta \geq 0$.
We also assume that:
\begin{description}
\item[A2 :]{$Pr\brac{A[1] = a} > 0$, for all $a \in \brac{0,\dots, A_{max}}$,}
\end{description}

\begin{lemma}
If $\lambda < S_{max}$, and if assumptions A1 and A2 hold, then there exists a stationary deterministic optimal policy $\gamma^*_{\beta}$ for the unconstrained MDP, with optimal average cost $g^*_{\beta}$ satisfying the following ACOE:
\begin{equation*}
  g^*_{\beta} + J_{\beta}(q) = \min_{s \in \brac{0,\dots,\min(q,S_{max})}} \bigg \{ c_{\beta}(q, s) + \Exp J_{\beta}(q - s + A[1]) \bigg \}, \forall q \geq 0,
\end{equation*}
with $\gamma^*_{\beta}$ using a batch size $s^*_{\beta}(q)$ at queue length $q$, satisfying 
\begin{equation*}
  s^*_{\beta}(q) = \argmin_{s \in \brac{0,\dots,\min(q,S_{max})}} \bigg \{ c_{\beta}(q, s) + \Exp J_{\beta}(q - s + A[1]) \bigg \}, \forall q \geq 0,
\end{equation*}
where $J_{\beta}(q)$ is the optimal relative value function.
\label{chap5:lemma:statopt}
\end{lemma}
The proof is given in Appendix \ref{chap5:app:statopt}.
From now on, we assume $\lambda < S_{max}$ and that A1 and A2 hold.

The following property of any optimal policy $\gamma^*_{\beta}$ can also be obtained.
\begin{description}
\item[O1 :]{any stationary deterministic optimal policy $\gamma^*_{\beta}$ is such that $s^*_{\beta}(q)$ is non-decreasing in $q$.}
\end{description}
The proof of the above property is similar to that of Theorem 3.2 (iii) of \cite{munish}, and is therefore omitted.
We now state some observations which are obtained from the above lemma and O1.
\begin{description}
\item[O2 :]{The optimal average cost $g^*_{\beta}$ is independent of the initial state $q_{0}$ and is finite.}
\item[O3 :]{For any policy, from assumptions A1 and A2, we note that from state 0 it is possible to reach any other state $q$.
From O1, we obtain that any stationary deterministic optimal policy has a single recurrence class $\mathcal{R}_{\gamma^*_{\beta}}$, of the form $\brac{q_{m},\dots}$, where $q_{m} = \min\brac{q : \exists q' > q, p_{q',q} > 0}$, and $p_{q',q} = Pr\brac{Q[m + 1] = q \vert Q[m] = q'}$ for the optimal policy under consideration.}
\item[O4 :]{We note that $s^*_{\beta}(q_{m}) = 0$ by definition. From A2, we have that $q_{m}$ is an aperiodic state, and therefore the class $\mathcal{R}_{\gamma^*_{\beta}}$ of the Markov chain under $\gamma^*_{\beta}$ is aperiodic.}
\item[O5 :]{From Lemma \ref{chap5:lemma:statopt}, for $\gamma^*_{\beta}$, for $q \in \mathcal{R}_{\gamma^*_{\beta}}$, we have that
\begin{equation*}
  g^*_{\beta} + J_{\beta}(q) = c_{\beta}(q, s^*_{\beta}(q)) + \Exp J_{\beta}(q - s^*_{\beta}(q) + A[1]),
\end{equation*}
which verifies the drift condition (10.13) from \cite{meyn}, with the Lyapunov function $V(q) = J_{\beta}(q)$, for a Markov chain restricted to $\mathcal{R}_{\gamma^*_{\beta}}$.
We also note that $c_{\beta}(q, s_{\beta}^*(q))$ is near-monotone \cite{meyn} in $q$.
Therefore using \cite[Theorem 10.3]{meyn}, we obtain that any stationary deterministic optimal policy is $c_{\beta}$-regular for a Markov chain on $\mathcal{R}_{\gamma^*_{\beta}}$.
Then the Markov chain under $\gamma^*_{\beta}$ is positive recurrent on $\mathcal{R}_{\gamma^*_{\beta}}$ with an associated invariant distribution.
Furthermore the expected total cost of first passage from any state $q \in \mathcal{R}_{\gamma^*_{\beta}}$ to another state $q' \in \mathcal{R}_{\gamma^*_{\beta}}$ is finite \cite[Theorem 10.3]{meyn}. }
\end{description}

Using property O5 of $\gamma^*_{\beta}$ we prove the following lemma, which shows that any optimal policy is in fact non-idling. We note that if $\gamma^*_{\beta}$ is non-idling, then $R_{\beta} = \mathbb{Z}_+$ and the Markov chain under $\gamma^*_{\beta}$ is irreducible.
\begin{lemma}
  Any stationary deterministic optimal policy $\gamma^*_{\beta}$ is non-idling, i.e., $s^*_{\beta}(q) > 0$, for all $q \in \mathcal{R}_{\gamma^*_{\beta}}$ and $q > 0$.
  \label{chap5:prop:nonidling}
\end{lemma}
The proof shows that if $\gamma^*_{\beta}$ is such that there exists a positive $q_{1} \in \mathcal{R}_{\gamma^*_{\beta}}$ with $s^*_{\beta}(q_{1}) = 0$, then it is possible to construct a history dependent non-stationary policy for which the average $c_{\beta}(q,s)$ cost is strictly less, contradicting the optimality of $\gamma^*_{\beta}$. The essential steps in the proof are: a) we consider a particular sample path of the queue evolution, for which it is assumed that at the start of a slot $m$, $Q[m - 1] = q_{1}$ with $S[m] = 0$, b) we obtain a new non-stationary policy which advances the service of one customer\footnote{Since $q_{1}$ is positive at least one customer will be present in the system at slot $m$, who will be served in some slot $> m$.} from one of the succeeding slots to $m$ while keeping the departure times of all other customers unchanged, and c) we show that for this new policy the total $c_{\beta}(q,s)$ cost decreases because i) the delay of the customer whose departure time was advanced has decreased and ii) convexity of the service cost function implies that the service cost at $m$, $c(1)$, is less than or equal to the decrease in service cost at the slot where the customer was being served under policy $\gamma^*_{\beta}$. Extension of the proof to the case of average cost, with the optimal policy being not irreducible, is more technical and is therefore presented in Appendix \ref{chap5:app:nonidling}.

We recall that by solving the unconstrained MDP, we are able to get solutions to problem \eqref{chap5:eq:inittradeoffprob}, with service cost constraint $c_{c}$, only if a Lagrange multiplier $\beta$ exists such that the optimal policy for the unconstrained MDP with single stage cost $c_{\beta}(q,s)$ has an average service cost equal to $c_{c}$, i.e, if $c_{c} \in \mathcal{O}^{u}$.
Thus in general, the properties O1, O2, O3, O4, and O5, as well as Lemma \ref{chap5:prop:nonidling} may not hold for all values of $c_{c}$ in the original problem \eqref{chap5:eq:inittradeoffprob}.

\subsection{The tradeoff problem}
\label{chap5:sec:tradeoffproblem}
As noted in Section \ref{chap5:sec:cmdpformulation}, we consider problem \eqref{chap5:eq:inittradeoffprob} for the set of randomized stationary policies $\Gamma_{s}$, since there exists an optimal stationary policy.
A policy $\gamma \in \Gamma_{s}$, specifies the service batch size $S(q)$ at a queue length $q$.
We note that $S(q)$ is a random variable with support on $\brac{0,\dots,\min(S_{max},q)}$.
We further restrict the study of problem \eqref{chap5:eq:inittradeoffprob} to the set of stationary \emph{admissible} policies, whose definition is motivated by the properties O1, O2, O3, O4, O5, and Lemma \ref{chap5:prop:nonidling}.
We now define the notion of stability for a policy.

\textbf{Stability:} A policy $\gamma \in \Gamma_{s}$ is said to be stable if: a) the Markov chain $Q[m]$ under $\gamma$ is positive recurrent  with stationary distribution $\pi_{\gamma}$ on the recurrence class corresponding to $q_{0}$ and b) $\overline{Q}(\gamma, q_{0}) < \infty$.

\textbf{Admissibility:} A policy $\gamma$ is called admissible if:
\begin{description}
\item[G1 :]{it is stable,}
\item[G2 :]{it induces an aperiodic, irreducible Markov chain $Q[m]$,}
\item[G3 :]{the average service rate at a queue length $q$, $\Exp S(q)$ is non-decreasing in $q$.}
\end{description}
The set of all admissible policies is $\Gamma_{a}$.
We note that the above properties of admissible policies are motivated by the observations about stationary deterministic policies made in Section \ref{chap5:sec:mdpformulation}.
Property G1 is motivated by O2 and O5.
Properties O3, O4 and Lemma \ref{chap5:prop:nonidling} motivates property G2\footnote{We note that the development of asymptotic lower bounds also holds under an assumption weaker than irreducibility. We can assume that under the policy $\gamma$, (i) there is only a single positive recurrent class $R_{\gamma}$ and (ii) the expected cumulative queue length and expected cumulative service cost starting from any state $q_{0}$ until $R_{\gamma}$ is hit is finite. We note under assumption A2 and G3, $R_{\gamma}$ is a contiguous set.}, while G3 is motivated by O1.
We note that the above definition of admissibility differs in the addition of property G3, from the definitions of admissible policies which were used by Berry and Gallager \cite[Section IV]{berry} and Neely \cite[Section III]{neely_mac}.
We also note that assumptions A1 and A2 have been used to motivate G2, through the properties O3 and O4, but as in Berry and Gallager \cite{berry}, we could assume G2.

\begin{remark}
  We now compare our definition of an admissible policy with that of Berry and Gallager \cite{berry}.
  In \cite{berry}, it is required that a sequence of admissible policies form an ergodic Markov chain, i.e., an aperiodic, irreducible and positive recurrent Markov chain. 
  Our admissible policies are also assumed to satisfy the same properties.
  In \cite{berry}, it is required that a sequence of admissible policies $\gamma_{k}$ are such that $\Qgk < \infty$ and $\lim_{k \rightarrow \infty} \Qgk = \infty$.
  We also assume that $\Qgk < \infty$.
  However, we do not assume that $\lim_{k \rightarrow \infty} \Qgk = \infty$.
  In fact, we shall see that for a particular case (Case 1), the optimal sequence of policies $\gamma^*_{k}$ is such that $\lim_{k \rightarrow \infty} \overline{Q}(\gamma^*_{k}) < \infty$.
  The third property in \cite{berry}, that admissible policies are assumed to satisfy is similar to our assumption A1.
  We note that the additional property G3 can be used to obtain additional insights about any stationary deterministic optimal policy.
  Furthermore, for $c_{c} \in \mathcal{O}^{u}$, there exists an admissible optimal policy.
\end{remark}

From G1 and G2, the average queue length as well as the average service cost are independent of the initial queue length since the Markov chain $(Q[m])$ is aperiodic, irreducible and stable \cite{meyn}.
The average queue length and average service cost for $\gamma \in \Gamma_{a}$ are therefore denoted by $\overline{Q}(\gamma)$ and $\overline{C}(\gamma)$.
We note that $\overline{Q}(\gamma) = \Exp_{\pi_{\gamma}} Q$ and $\overline{C}(\gamma) = \Exp_{\pi_{\gamma}} c(S(Q))$, where $Q$ denotes the stationary queue length.
Since, in general, for every $c_{c}$, we do not know if Lemma \ref{chap5:prop:nonidling} holds for the optimal constrained policy, admissible policies are not required to be non-idling.

\paragraph{Objective :}
Our objective is to obtain the optimal tradeoff curve $Q^*(c_{c})$ while restricting our attention to the class of admissible policies $\Gamma_{a}$, where $Q^*(c_{c})$ is the optimal value of the following optimization problem
\begin{equation*}
  \text{TRADEOFF : } \mini_{\gamma \in \Gamma_{a}}\, \overline{Q}(\gamma) \text{ such that } \overline{C}(\gamma) \leq c_{c}.
\end{equation*}

We note that the TRADEOFF problem can be formulated for a larger class of policies, which are obtained by time sharing or mixing of policies $\gamma_{k} \in \Gamma_{a}$.
Let $Q^*_{M}(c_{c})$ denote the optimal tradeoff curve, when we consider time shared policies also.
We note that the tradeoff curve $Q^*_{M}(c_{c})$ which is obtained from time sharing is the lower convex envelope of the points $(c_{c}, Q^*(c_{c}))$.
Since the asymptotic behaviour of $Q^*_{M}(c_{c})$ can be obtained from that of $Q^*(c_{c})$, as in Chapter 2, in the following we analyse $Q^*(c_{c})$ only.

We note that given any $\epsilon > 0$, and for any $c_{c}$ such that TRADEOFF is feasible, by definition there exists an admissible policy $\gamma$ such that $\Qg \leq Q^*(c_{c}) + \epsilon$ and $\Cg \leq c_{c}$.
Such a feasible admissible policy is called $\epsilon$-optimal in the following.

We note that for any $\gamma \in \Gamma_{a}$, we have that $\Exp_{\pi_{\gamma}} S(Q) = \lambda$.
We recall that $c(s)$ was redefined as the piecewise linear lower convex envelope of $c(s), s \in \brac{0, \dots, S_{max}}$.
Then, from Jensen's inequality, we have that for any policy $\gamma \in \Gamma_{a}$, $\overline{C}(\gamma) = \Exp_{\pi_{\gamma}} c(S(Q)) \geq c(\lambda)$.
Therefore, $\inf_{\gamma \in \Gamma_{a}} \overline{C}(\gamma) \geq c(\lambda)$.
We note that TRADEOFF does not have any feasible solutions if $c_{c}$ is less than $c(\lambda)$.

We also note that $c(\lambda)$ is also the minimum average service cost which has to be expended for mean rate stability.
In the following we show that $c(\lambda) = \inf_{\gamma \in \Gamma_{a}} \overline{C}(\gamma)$.
We obtain an asymptotic characterization of $Q^*(c_{c})$ in the asymptotic regime $\Re$ as $c_{c} \downarrow c(\lambda)$ in the next section.

\subsection{Asymptotic analysis of TRADEOFF - Preliminaries}
\label{chap5:sec:asymp_analysis_tprob}
The ideas used in the analysis of TRADEOFF are the same as those used for the analysis of INTERVAL-$\mu$CHOICE-2 in Chapter 3.
However, bounds on the stationary probability of the queue length for the DTMC $(Q[m])$ have to be developed, in order to relate the average service cost to the average queue length.
As in Chapter 3, we identify three different cases based on the nature of the function $c(s)$ at $s = \lambda$.
The cases are defined in terms of the quantities $s_{l}$ and $s_{u}$ defined as follows:
\begin{eqnarray*}
  s_{u} & = &
  \begin{cases}
    \min\brac{s : s \in \brac{\ceiling{\lambda}, \dots, S_{max} - 1}, c(s + 1) - c(s) > c(s) - c(s - 1)} \text{ if this set is non-empty} \\
    S_{max} \text{ otherwise}.
  \end{cases} \\
  s_{l} & = & 
  \begin{cases}
    \max\brac{s : s \in \brac{1, \dots, \floor{\lambda}}, c(s + 1) - c(s) > c(s) - c(s - 1)} \text{ if this set is non-empty} \\
    0 \text{ otherwise}.
  \end{cases}
\end{eqnarray*}
We note that $s_{l}$ and $s_{u}$ are analogous to the service rates $\mu_{l}$ and $\mu_{u}$ defined in chapter 3.
The three cases that we consider are:
\begin{description}
\item[Case 1 :]{$s_{l} = 0, s_{l} < \lambda < s_{u}$}
\item[Case 2 :]{$s_{l} > 0, s_{l} < \lambda < s_{u}$}
\item[Case 3 :]{$s_{l} = \lambda = s_{u}$}
\end{description}

We now show that $c(\lambda)$ can in fact be approached arbitrarily closely.
\begin{lemma}
  There exists a sequence of policies $\gamma_{\epsilon_{k}} \in \Gamma_{a}$ such that $c(\lambda)$ can be approached arbitrarily closely, i.e., $\lim_{k \rightarrow \infty} \overline{C}(\gamma_{\epsilon_{k}}) = c(\lambda)$. Therefore $c(\lambda) = \inf_{\gamma \in \Gamma_{a}} \overline{C}(\gamma)$.
  \label{chap5:lemma:minach_servicecost}
\end{lemma}
The proof is given in Appendix \ref{chap5:app:minach_servicecost}.
We characterize the tradeoff curve $Q^*(c_{c})$ in the asymptotic regime where the cost constraint $c_{c}$ approaches $c(\lambda)$.

\begin{figure}[h]
  \centering
  \includegraphics[width=160mm,height=40mm]{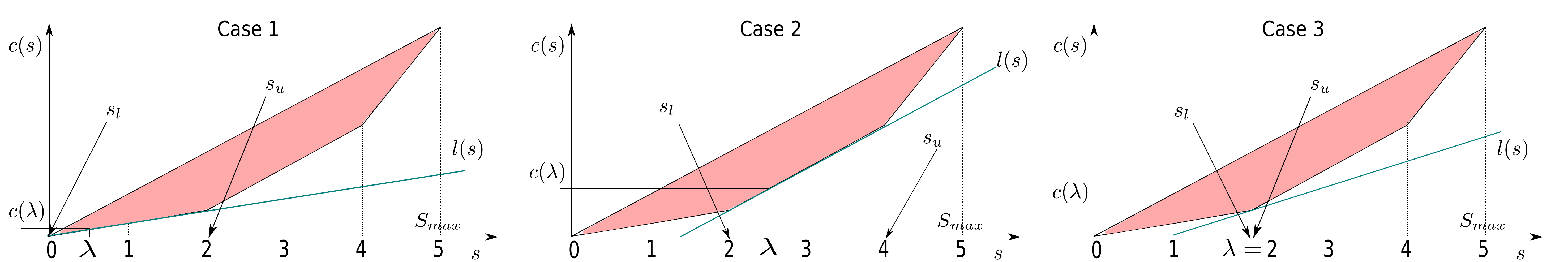}
  \caption{Illustration of the relationship between $\lambda$, $s_{l}$, and $s_{u}$ along with the minimum average cost $c(\lambda)$ and the line $l(s)$ for the three cases}
  \label{chap5:fig:lambdacases}
\end{figure}

Similar to the definition of the line $l(\mu)$ in Chapter 3, here we define the line $l(s) : [0,S_{max}] \rightarrow \mathbb{R}_+$ as follows:
\begin{enumerate}
\item If $s_{l} < \lambda < s_{u}$, then $l(s)$ is the line through $(s_{l},c(s_{l}))$ and $(s_{u},c(s_{u}))$.
\item If $s_{l} = \lambda = s_{u}$, then $l(s)$ is a line through $(\lambda, c(\lambda))$ with slope $m$ chosen such that $c(\lambda) - c(\lambda - 1) < m < c(\lambda + 1) - c(\lambda)$.
\end{enumerate}
The different cases along with the line $l(s)$ are illustrated in Figure \ref{chap5:fig:lambdacases}.
We note that $\mathbb{E}_{\pi_{\gamma}}\bras{c(S) - c(\lambda)} = \mathbb{E}_{\pi_{\gamma}}\bras{c(S) - l(S)}$.

For a particular policy $\gamma$, if there is no source for confusion we use $\pi(q)$ to denote the stationary probability of queue length being $q$.
The stationary probability of using a particular batch size $s \in \brac{0,\dots,S_{max}}$ is denoted by $\pi_{s}(s)$.
We now present two results which are used in the asymptotic characterization of $Q^*(c_{c})$.

\begin{lemma}
  For $\gamma \in \Gamma_{a}$, for some positive $\epsilon < s_{l}$, $q_{s} = \inf\brac{q : \Exp S(q) \geq s_{l} - \epsilon}, \rho_{d} = \nfrac{s_{l} - \epsilon}{S_{max}} Pr\brac{A[1] = 0}$, and $\rho = 1 + \frac{1}{\rho_{d}}$, if $Pr\brac{Q < q_{s}}\brap{1 + \frac{\rho}{\rho_{d}}} < \frac{1}{2}$, we have that 
  \begin{equation*}
    \overline{Q}(\gamma) \geq \frac{1}{2} \left[ \log_{\rho} \left[ \frac{1}{2Pr\brac{Q < q_{s}}} \right] - 1 \right],
  \end{equation*}
  \label{chap5:lemma:barq_pi0_relation}
\end{lemma}
The proof is given in Appendix \ref{chap5:app:barq_pi0_relation}.
The above result is useful in obtaining an asymptotic lower bound to $Q^*(c_{c})$ as $c_{c} \downarrow c(\lambda)$, since in Cases 2 and 3, as $c_{c} \downarrow c(\lambda)$, for any sequence of feasible policies for TRADEOFF, $Pr\brac{Q < q_{s}} \downarrow 0$.

\begin{remark}
  We note that a similar asymptotic lower bound has been derived in \cite[Theorem 2]{neely_utility} (where admission control is allowed) and in \cite[Theorem 2]{botan}, where the assumption G3 has not been used.
  Although, the above result has been derived independently, we note that underlying all the three derivations, there is the idea of bounding the probability of an event by a particular sequence of transitions for a Markov chain, i.e., a sequence of transitions in which the state of the Markov chain becomes successively smaller.
  Furthermore, in our proof, using assumption G3, we obtain geometric bounds on the stationary probability of any queue length, which is not available in \cite{neely_utility} as well as \cite{botan}.
\end{remark}

We note that in Chapter 3, since the queue length process $Q(t)$ was a birth-death process, bounds on the stationary probability of the queue length could be obtained relatively easily.
However, in this chapter, bounds on the stationary probability of discrete time Markov chains (DTMC) are required.
In the following lemma, we present three bounds on the stationary probability of the queue length, one of which has been obtained by Bertsimas et al. \cite{gamarnik} and \cite{gamarnik_1} and the other two are state dependent extensions of the geometric bounds on the stationary probability of discrete time Markov chains presented in \cite{gamarnik} and \cite{gamarnik_1}.
\begin{lemma}
  Let $(Q[m], m \geq 0)$ be as in \eqref{chap5:eq:evolution}, for an admissible policy $\gamma$.
  Let $\epsilon_{a}$ be as defined in assumption A1. Then,
  \begin{description}
  \item[TAIL-PROB \cite{gamarnik} :]{Suppose $\forall q \geq 0$, $\mathbb{E}\bras{Q[m + 1] - Q[m] \vert Q[m] = q} \geq -d$, where $d$ is positive. Then for any finite $q_{1}$ and $k \geq 1$ we have 
      \begin{equation*}
        Pr\brac{Q \geq q_{1} + k} \geq \brap{\frac{\epsilon_{a}}{\epsilon_{a} + d}}^{k} Pr\brac{Q \geq q_{1}}.
      \end{equation*}
    }
  \item[TAIL-PROB-STATE-DEP-1 :]{Suppose there exists a $q_{d}$ such that \[ \forall q \in \brac{0,\dots,q_{d}}, \mathbb{E}\bras{Q[m + 1] - Q[m] \vert Q[m] = q} \geq -d,\] where $d$ is positive. Then for any $q_{1}$, $k \geq 0$ such that $0 \leq q_{1} + k \leq q_{d}$, we have 
      \small
      \begin{eqnarray*}
        Pr\brac{Q \geq q_{1} + k } & \geq & \brap{\frac{\epsilon_{a}}{\epsilon_{a} + d}}^{k} Pr\brac{Q \geq q_{1}} \\
        & & + \brap{1 - \brap{\frac{\epsilon_{a}}{\epsilon_{a} + d}}^{k}}\Bigg[Pr\brac{Q \geq q_{d} + 1} \\
          & & + \frac{1}{d} \sum_{q = q_{d} + 1}^{\infty} \pi(q) \mathbb{E}\bras{{Q}[m + 1] - {Q}[m] | Q[m] = q}\Bigg].
      \end{eqnarray*}
      \normalsize
    }
  \item[TAIL-PROB-STATE-DEP-2 :]{Suppose there exists a $q_{d}$ such that \[ \forall q \in \brac{0,\dots,q_{d}}, \mathbb{E}\bras{Q[m + 1] - Q[m] \vert Q[m] = q} \geq -d_{1},\] \[\text{ and } \forall q \in \brac{q_{d} + 1, \dots}, \mathbb{E}\bras{Q[m + 1] - Q[m] \vert Q[m] = q} \geq -d_{2},\]  where $d_{2} > d_{1} > 0$. Then for any $q_{1}$, $k \geq 0$ such that $0 \leq q_{1} + k \leq q_{d}$, we have 
      \begin{eqnarray*}
        Pr\brac{Q \geq q_{1} + k} & \geq & \brap{\frac{\epsilon_{a}}{\epsilon_{a} + d_{1}}}^{k} Pr\brac{Q \geq q_{1}} - \brap{1 - \brap{\frac{\epsilon_{a}}{\epsilon_{a} + d_{1}}}^{k}}\frac{d_{2} - d_{1}}{d_{1}} Pr\brac{Q \geq q_{d} + 1}.
      \end{eqnarray*}
    }
    \end{description}
  \label{chap5:lemma:dtmc_stat_prob}
\end{lemma}
The proof is presented in Appendix \ref{chap5:app:dtmc_stat_prob}.
We now present the asymptotic characterization of $Q^*(c_{c})$ as $c_{c} \downarrow c(\lambda)$.

\subsection{Asymptotic characterization of $Q^*(c_{c})$ as $c_{c} \downarrow c(\lambda)$}
\label{chap5:sec:asympchar}
We first consider Case 2, the proof of the following asymptotic lower bound has already been briefly discussed in Section \ref{chapter4:methodology}.
We use the geometric bound based lower bound on $\Qg$ from Lemma \ref{chap5:lemma:barq_pi0_relation} to obtain the asymptotic lower bound for Case 2.
\begin{lemma}
  For Case 2, given any sequence of admissible policies $\gamma_{k}$ with $\overline{C}(\gamma_{k}) - c(\lambda) = V_{k} \downarrow 0$, we have that $\overline{Q}(\gamma_{k}) = \Omega\brap{\log\nfrac{1}{V_{k}}}$.
  \label{chap5:lemma:tprob_2}
\end{lemma}
\begin{proof}
  Let us consider a particular policy $\gamma$ in the sequence $\gamma_{k}$ with $V_{k} = V$.
  From the definition of $l(s)$ we have that $\mathbb{E}_{\pi} \ExpS \bras{c(S(Q)) - l(S(Q))} = V$.
  From the convexity of $c(s)$ and the linearity of $l(s)$ we have
  \begin{eqnarray*}
    \sum_{q = 0}^{\infty} \pi(q) \bras{ c(\Exp S(q)) - l(\Exp S(q)) } \leq V.
  \end{eqnarray*}
  Now as $c(s) \geq l(s)$, we have that
  \begin{eqnarray*}
    \sum_{q = 0}^{q_{s} - 1} \pi(q) \brap{ c(\Exp S(q)) - l(\Exp S(q)) } \leq V,
  \end{eqnarray*}
  where $q_{s}$, as in Lemma \ref{chap5:lemma:barq_pi0_relation}, is $\inf\brac{q : \Exp S(q) \geq s_{l} - \epsilon}$ for a positive $\epsilon < s_{l}$.
  We note that $q_{s} \geq 1$.
  We note that for $q < q_{s}$, $c(\Exp S(q)) - l(\Exp S(q)) \geq m \epsilon$, where $m$ is the tangent of the angle made by the line passing through $(s_{l} - 1, c(s_{l} - 1))$ and $(s_{l}, c(s_{l}))$ with $l(s)$.
  Therefore we obtain that 
  \[ Pr\brac{Q < q_{s}} \leq \frac{V}{m\epsilon}. \]
  
  Using the above bound on $Pr\brac{Q < q_{s}}$, we have that for sufficiently small $V$, $Pr\brac{Q < q_{s}} < \frac{1}{2}$.
  Then, from Lemma \ref{chap5:lemma:barq_pi0_relation} we have that
  \begin{equation*}
    \overline{Q}(\gamma) \geq \frac{1}{2} \left[ \log_{\rho} \left[ \frac{m\epsilon}{2V}\right] - 1 \right],
  \end{equation*}
  where the upper bound $\frac{V}{m\epsilon}$ on $Pr\brac{Q < q_{s}}$ is used.
  Therefore for the sequence of policies $\gamma_{k}$ with $V_{k} \downarrow 0$, we have that 
  \begin{equation*}
    \overline{Q}(\gamma_{k}) = \Omega\left(\log\left(\frac{1}{V_{k}}\right)\right).
  \end{equation*}
\end{proof}
\begin{remark}
  For Case 2, as for \INTMC-2-2, we have a set of queue lengths $\mathcal{Q}_{h}$, which occur with high probability, such that $\Exp S(q) \in [s_{l}, s_{u}], \forall q \in \mathcal{Q}_{h}$.
  Let the drift in state $q$ be $\Delta(q) \Deq \Exp \bras{Q[m + 1] - Q[m] | Q[m] = q}$.
  We note that, intuitively the drift $\Delta(q)$ for $q \in \mathcal{Q}_{h}$ is increasing and then decreasing for in $\mathcal{Q}_{h}$.
  Then we expect that the stationary distribution of the queue length is geometrically increasing and then decreasing as for \INTMC-2-2.
  However, unlike \INTMC-2-2, for the discrete time model we are only able to obtain a geometrically increasing upper bound.
  This bound suffices to obtain the $\Omega\left(\log\left(\frac{1}{V_{k}}\right)\right)$ asymptotic lower bound in the above lemma.
\end{remark}

We note that the model that is considered here is a simplified version of Neely's \cite{neely_mac} model - there is only one queue evolving on $\sZ$ and we have the fade state taking only a single value.
The tradeoff optimal control algorithm (TOCA) of \cite{neely_mac} achieves the logarithmic tradeoff stated in the above lemma but for the problem \eqref{chap5:eq:inittradeoffprob}, since TOCA is not admissible.
Therefore, we propose a sequence of admissible policies that achieves the asymptotic logarithmic tradeoff of $Q^*(c_{c})$ in Lemma \ref{chap5:lemma:tprob_2}.
\begin{lemma}
  Let a policy $\gamma$ be defined as follows.
  At a queue length $q$, $\gamma$ serves a batch size $\min(q, \tilde{s}(q))$, where 
  \begin{equation*}
    \tilde{s}(q) = 
    \begin{cases}
      s_{l}, \text{ for } 0 \leq q < q_{v}, \\
      s_{u}, \text{ for } q_{v} \leq q.
    \end{cases}
  \end{equation*}
  where $q_{v} > 0$.
  We obtain a sequence of policies $\gamma_{k}$, by choosing $q_{v} = \log\nfrac{1}{V_{k}}$, where $V_{k} < 1$ is a sequence decreasing to zero.
  Then for Case 2, $\gamma_{k}$ is a sequence of admissible policies, such that $\overline{C}(\gamma_{k}) - c(\lambda) = \mathcal{O}(V_{k})$ and $\Qgk = \mathcal{O}\brap{\log\nfrac{1}{V_{k}}}$.
  \label{chap5:lemma:case2upperbound}
\end{lemma}
The proof of this lemma is given in Appendix \ref{chap5:app:case2upperbound}.
We that the structure of the sequence of policies $\gamma_{k}$ is similar to that in Lemma \ref{chap4:prop:p12ub}.
We note that this proof is motivated by and borrows ideas from the derivation of the asymptotic upper bound for the sequence of TOCA policies in \cite[Corollary 2]{neely_mac}.
This leads to the following asymptotic characterization for Case 2.
\begin{proposition}
  For Case 2, the optimal tradeoff curve $Q^*(c_{c,k}) = \Theta\brap{\log\nfrac{1}{c_{c,k} - c(\lambda)}}$ as $c_{c,k} \downarrow c(\lambda)$, for the sequence $c_{c,k} = \overline{C}(\gamma_{k})$, where $\gamma_{k}$ is the sequence of policies in Lemma \ref{chap5:lemma:case2upperbound}.
  \label{chap5:proposition:case2}
\end{proposition}
\begin{proof}
  For the sequence $c_{c,k} = \overline{C}(\gamma_{k})$, we have that $Q^*(c_{c,k}) \leq \overline{Q}(\gamma_{k}) = \mathcal{O}\brap{\log\nfrac{1}{c_{c,k} - c(\lambda)}}$.
  For $\epsilon > 0$, consider any sequence of feasible $\epsilon$-optimal admissible policies $\gamma'_{k}$ for the sequence $c_{c,k}$.
  We have that $\overline{Q}(\gamma'_{k}) = \Omega\brap{\log\nfrac{1}{c_{c,k} - c(\lambda)}}$ and $\overline{Q}(\gamma'_{k}) \leq Q^*(c_{c,k}) + \epsilon$.
  Therefore, $Q^*(c_{c,k}) = \Omega\brap{\log\nfrac{1}{c_{c,k} - c(\lambda)}}$.
  Hence, for $c_{c,k} = \overline{C}(\gamma_{k})$, we have that $Q^*(c_{c,k}) = \Theta\brap{\log\nfrac{1}{c_{c,k} - c(\lambda)}}$.
\end{proof}

\begin{remark} \textbf{TOCA algorithm :}
\label{chap5:remark:toca}
We note that the set of all available power values (denoted by $\Pi$ in \cite{neely_mac}) can be chosen such that the corresponding rates are $\brac{0,\dots,S_{max}}$.
As required in \cite{neely_mac}, the set is compact.
The TOCA algorithm is parametrized by positive numbers $w, \epsilon, \tilde{q},$ and $\beta$.
The algorithm chooses at each slot $m \geq 1$, the batch size $s_{TOCA}$ such that
\begin{eqnarray*}
  s_{TOCA}[m] & = & \min\brap{\argmin_{s \in \brac{0,\dots,S_{max}}} \bigg \{\beta c(s) - {W}[m] s \bigg\}, Q[m - 1]},
\end{eqnarray*}
where
\small
\begin{eqnarray*}
  W[m] & = & \mathbb{I}\brac{Q[m - 1] \geq \tilde{q}}\bras{w e ^{w(Q[m - 1] - \tilde{q})} + 2 X[m - 1]} + \\
  & & \mathbb{I}\brac{Q[m - 1] < \tilde{q}}\bras{-w e ^{w(\tilde{q} - Q[m - 1])} + 2 X[m - 1]}.
\end{eqnarray*}
\normalsize
We note that $s_{TOCA}[m] = 0$ if $W[m] \leq 0$.
The sequence $X[m], m \geq 0$ is obtained from a \emph{virtual} queue which evolves according to
\begin{equation*}
  X[m + 1] = \max(X[m] - s_{TOCA}[m + 1] + \epsilon \mathbb{I}\brac{Q[m] < \tilde{q}}, 0) + A[m + 1] + \epsilon\mathbb{I}\brac{Q[m] \geq \tilde{q}}.
\end{equation*}
As in \cite{neely_mac}, let $\delta_{max} = \max(A_{max}, S_{max})$.
Let $0 < \epsilon < \min(\lambda - s_{l}, s_{u} - \lambda)$, $w = \frac{\epsilon}{\delta_{max}^{2}} e^{\frac{-\epsilon}{\delta_{max}}}$, and $\tilde{q} = \frac{2}{w}\log\brap{\beta}$.
A sequence of policies $\gamma_{k}$ is generated by choosing a sequence $\beta_{k} = \frac{1}{V_{k}}$, for a sequence $V_{k} \downarrow 0$.
Then from \cite[Corollary 2]{neely_mac}, we have that 
\begin{eqnarray*}
  \overline{Q}(\gamma_{k}) = \mathcal{O}\brap{\log\nfrac{1}{V_{k}}},
  \overline{C}(\gamma_{k}) = c(\lambda) + \mathcal{O}\brap{V_{k}}.
\end{eqnarray*}
Therefore, we obtain that for the sequence of policies $\gamma_{k}$, $\overline{Q}(\gamma_{k}) = \mathcal{O}\brap{\log\nfrac{1}{\overline{C}(\gamma_{k}) - c(\lambda)}}$.
We note that $W[m]$ is a non-decreasing function of $q$, where $Q[m - 1] = q$.
Since $s_{TOCA}[m]$ is a non-decreasing function of $W[m]$, we have that $s_{TOCA}[m]$ is a non-decreasing function of $Q[m - 1]$.
However, we note that $s_{TOCA}[m]$ is stationary only with respect to a state which includes an additional state variable $X[m - 1]$ and hence is not admissible.
Therefore, the above bound is an upper bound to the optimal value of \eqref{chap5:eq:inittradeoffprob} and not TRADEOFF.

If for any subsequence $c_{c,k}$ of $\mathcal{O}^{u}$ such that there exists a constant $0 < m \leq 1$ and a subsequence $c_{TOCA,k}$ of $\overline{C}(\gamma_{k})$ such that $c_{TOCA,k} \leq c_{c,k}$ and $c_{TOCA,k} - c(\lambda) \geq m\brap{c_{c,k} - c(\lambda)}$, then we have that $Q^*(c_{c,k}) \leq Q(\gamma_{k}) = \mathcal{O}\brap{\log\nfrac{1}{c_{TOCA,k} - c(\lambda)}} = \mathcal{O}\brap{\log\nfrac{1}{c_{c,k} - c(\lambda)}}$.
\end{remark}

We now obtain an asymptotic characterization for Case 3.
The method used is the same as that summarized in Section \ref{chapter4:methodology}, except that the geometric bound used is obtained from Lemma \ref{chap5:lemma:dtmc_stat_prob}(TAIL-PROB-STATE-DEP-1).
Furthermore, we will see that in the asymptotic regime $\Re$, the geometric bound reduces to a constant bound, which leads to the specific form of asymptotic lower bound for Case 3.
\begin{lemma}
  For Case 3, given any sequence of admissible policies $\gamma_{k}$ with $\overline{C}(\gamma_{k}) - c(\lambda) = V_{k} \downarrow 0$, we have that $\overline{Q}(\gamma_{k}) = \Omega\nfrac{1}{V_{k}}$.
  \label{chap5:lemma:case3}
\end{lemma}

\begin{proof}
  Let us consider a particular policy $\gamma$ in the sequence $\gamma_{k}$ with $V_{k} = V$.
  We define $q_{d} = \sup\brac{q : \Exp S(q) \leq \lambda + \epsilon_{V}}$, where $\epsilon_{V} > 0$ will be chosen later.
  We note that as $\Exp S(0) = 0$, the above set is non-empty.
  Suppose we assume that $q_{d}$ is finite.

  We note that by the admissibility of $\gamma$, $\forall q \in \brac{0,\dots, q_{d}}$, $\Exp S(q) \leq \lambda + \epsilon_{V}$.
  Hence, using $d = \epsilon_{V}$, we have from Lemma \ref{chap5:lemma:dtmc_stat_prob}(TAIL-PROB-STATE-DEP-1), for a $\bar{q} < q_{d}$ :
  \begin{eqnarray*}
    & & Pr\brac{Q \geq \bar{q} + 1} \geq \brap{\frac{\epsilon_{a}}{\epsilon_{a} + \epsilon_{V}}}^{\bar{q} + 1} \\
    & & + \brap{1 - \brap{\frac{\epsilon_{a}}{\epsilon_{a} + \epsilon_{V}}}^{\bar{q} + 1}}\bras{Pr\brac{Q \geq q_{d} + 1} + \frac{1}{\epsilon_{V}} \sum_{q = q_{d} + 1}^{\infty} \pi(q) \Exp \bras{Q[m + 1] - Q[m] | Q[m] = q}}.
  \end{eqnarray*}
  Or 
  \begin{eqnarray}
    Pr\brac{Q \leq \bar{q}} & \leq & 1 - \brap{\frac{\epsilon_{a}}{\epsilon_{a} + \epsilon_{V}}}^{\bar{q} + 1} - \nonumber \\
    & & \brap{1 - \brap{\frac{\epsilon_{a}}{\epsilon_{a} + \epsilon_{V}}}^{\bar{q} + 1}}\brap{\frac{1}{\epsilon_{V}}\sum_{q =  q_{d} + 1}^{\infty} \mathbb{E}\bras{{Q}[m + 1] - {Q}[m] | Q[m] = q} \pi(q)},
    \label{chap5:eq:case2lb1}
  \end{eqnarray}
  as $Pr\brac{Q \geq q_{d} + 1} \geq 0$.
  For brevity, let $D_{t} \Deq -\brap{\frac{1}{\epsilon_{V}}\sum_{q =  q_{d} + 1}^{\infty} \mathbb{E}\bras{{Q}[m + 1] - {Q}[m] | Q[m] = q} \pi(q)}$.
  We note that $D_{t}$ is positive, as for $q \geq q_{d} + 1$, $\mathbb{E}[Q[m + 1] - Q[m] | Q[m] = q] < -\epsilon_{V}$.
  Consider the expression for $D_{t}$.
  We have that for $q \geq q_{d} + 1$, \[\mathbb{E}\bras{{Q}[m] - {Q}[m + 1] | Q[m] = q} = \Exp S(q) - \lambda\] is positive as $\Exp S(q) > \lambda + \epsilon_{V}$.
  We note that by definition, $c(s)$ is piecewise linear.
  Let $m$ be the tangent of the angle between (i) the line passing through $(\lambda + 1, c(\lambda + 1))$ and  $(\lambda, c(\lambda))$, and (ii) $l(s)$.
  Then $m \sum_{q = q_{d} + 1}^{\infty} \pi(q) \brap{\Exp S(q) - \lambda} \leq \sum_{q = q_{d} + 1}^{\infty} \pi(q) \bras{c(\Exp S(q)) - l(\Exp S(q))}$.
  Furthermore from the convexity of $c(.)$, linearity of $l(.)$, and as $c(s) - l(s) \geq 0$, we have that 
  \[  \sum_{q = q_{d} + 1}^{\infty} \pi(q) \bras{c(\Exp S(q)) - l(\Exp S(q))} \leq \Expp \bras{c(\Exp S(Q)) - l(\Exp S(Q))} \leq \Exp \bras{c(S(Q)) - l(S(Q))} = V.\]
  Therefore \[ D_{t}\ \leq\ \frac{V}{m\epsilon_{V}}. \]
  
  Now, as in the proof of Lemma \ref{chap5:lemma:barq_pi0_relation}, we find a lower bound $\frac{\bar{q}}{2}$ on $\overline{Q}(\gamma)$ by finding the largest $\bar{q}$ such that $Pr\brac{Q \leq \bar{q}} \leq \frac{1}{2}$. 
  A lower bound $\bar{q}_{1}$ to $\bar{q}$ can be obtained by using the upper bound \eqref{chap5:eq:case2lb1} on $Pr\brac{Q \leq \bar{q}}$.
  Let $\bar{q}_{1}$ be the largest integer, if one exists, such that
  \begin{eqnarray*}
    1 - \brap{\frac{\epsilon_{a}}{\epsilon_{a} + \epsilon_{V}}}^{\bar{q}_{1} + 1} - \brap{1 - \brap{\frac{\epsilon_{a}}{\epsilon_{a} + \epsilon_{V}}}^{\bar{q}_{1} + 1}}\brap{\frac{1}{\epsilon_{V}}\sum_{q =  q_{d} + 1}^{\infty} \mathbb{E}\bras{{Q}[m + 1] - {Q}[m] | Q[m] = q} \pi(q)} \leq \frac{1}{2}.
  \end{eqnarray*}
  Then $\bar{q}_{1} \leq \bar{q}$.
  Then we have to find $\bar{q}_{1}$ such that 
  \begin{eqnarray*}
    1 - \brap{\frac{\epsilon_{a}}{\epsilon_{a} + \epsilon_{V}}}^{\bar{q}_{1} + 1} + \brap{1 - \brap{\frac{\epsilon_{a}}{\epsilon_{a} + \epsilon_{V}}}^{\bar{q}_{1} + 1}}D_{t} \leq \frac{1}{2}, \\
    \text{or } \frac{1 + 2D_{t}}{2 + 2D_{t}} \leq \nfrac{\epsilon_{a}}{\epsilon_{a} + \epsilon_{V}}^{\bar{q}_{1} + 1},\\
    \text{or } \brap{1 + \frac{\epsilon_{V}}{\epsilon_{a}}}^{\bar{q}_{1} + 1} \leq \frac{2 + 2D_{t}}{1 + 2D_{t}}
  \end{eqnarray*}
  We note that if $q_{d} = \infty$, then $D_{t} = 0$.
  However, $\bar{q}_{1}$ satisfying the above inequality for finite $q_{d}$ is a lower bound for $\bar{q}_{1}$ for $q_{d} = \infty$.
  Hence, we proceed with finding the above $\bar{q}_{1}$.
  Let $\bar{q}_{2}$ be the largest integer such that 
  \begin{eqnarray}
    \brap{1 + \frac{\epsilon_{V}}{\epsilon_{a}}}^{\bar{q}_{2} + 1} \leq \frac{2}{1 + 2D_{t}}.
    \label{chap5:prop:case31}
  \end{eqnarray}
  Then $\bar{q}_{2} \leq \bar{q}_{1}$.
  From \eqref{chap5:prop:case31} and the upper bound $\frac{V}{m\epsilon_{V}}$ on $D_{t}$, if $\bar{q}_{3}$ is the largest integer such that 
  \begin{eqnarray*}
    & & \brap{1 + \frac{\epsilon_{V}}{\epsilon_{a}}}^{\bar{q}_{3} + 1} \leq \frac{2}{1 + 2\frac{V}{m\epsilon_{V}}}, \\
  \end{eqnarray*}
  then $\bar{q}_{3} \leq \bar{q}_{2}$.
  Or, we have that $\bar{q}_{3}$ is the largest integer such that
  \begin{eqnarray*}
    & & \bar{q}_{3} + 1 \leq \log_{\brap{1 + \frac{\epsilon_{V}}{\epsilon_{a}}}} \brap{ \frac{2}{1 + \frac{2V}{m\epsilon_{V}}}}.
  \end{eqnarray*}
  We note that, as $V \downarrow 0$, if $\frac{V}{\epsilon_{V}} \rightarrow \infty$, then the bound will be negative.
  We choose $\epsilon_{V} = aV$, where $a > \frac{2}{m}$.
  Then we obtain that 
  \begin{eqnarray*}
    & & \bar{q}_{3} \leq \log_{\brap{1 + \frac{a{V}}{\epsilon_{a}}}} \brap{ \frac{2}{1 + \frac{2}{ma}}} - 1,
  \end{eqnarray*}
  where the RHS is positive as $V \downarrow 0$.
  Therefore the maximum $\bar{q}_{3}$ is at least
  \begin{eqnarray*}
    & & \floor{\log_{\brap{1 + \frac{a{V}}{\epsilon_{a}}}} \brap{ \frac{2}{1 + \frac{2}{ma}}} - 1}.
  \end{eqnarray*}
  Since $\overline{Q}(\gamma) \geq \frac{\bar{q}}{2} \geq \frac{\bar{q}_{1}}{2}  \geq \frac{\bar{q}_{2}}{2}  \geq \frac{\bar{q}_{3}}{2}$ and $\log \brap{1 + \frac{a{V}}{\epsilon_{a}}} = \Theta\brap{V}$, we have that for the sequence of policies $\gamma_{k}$, $\overline{Q}(\gamma_{k}) = \Omega\nfrac{1}{V_{k}}$.
\end{proof}

\begin{remark}
  As for \INTMC-2-3, we note that there is a set of queue lengths $\mathcal{Q}_{h}$, which occur with high probability.
  Let the drift in state $q$ be $\Delta(q) \Deq \Exp \bras{Q[m + 1] - Q[m] | Q[m] = q}$.
  Then $\mathcal{Q}_{h}$ is the set of queue lengths such that $\Delta(q) \rightarrow 0$, for $q \in \mathcal{Q}_{h}$ as $c_{c} \downarrow c(\lambda)$.
  Intuitively, we expect that the stationary probabilities of all queue lengths in $\mathcal{Q}_{h}$ are equal.
  With $\epsilon_{V}$ chosen to be $aV$ as in the above proof, we have that
  \begin{eqnarray*}
    Pr\brac{Q \leq \overline{q}} & \leq & \brap{1 - \fpow{\epsilon_{a}}{\epsilon_{a} + aV}{\overline{q} + 1}} \times \text{a constant}, \\
    & \approx & \overline{q} \frac{aV}{\epsilon_{a}} \times \text{ another constant.}
  \end{eqnarray*}
  This suggests that the constant stationary probability for queue lengths in $\mathcal{Q}_{h}$ is $\mathcal{O}(V)$, and therefore we obtain the $\Omega\nfrac{1}{V_{k}}$ asymptotic lower bound in the above lemma.
\end{remark}

\begin{remark}
We note that the sequence of randomized policies $\gamma_{\epsilon_{k}}$ in the proof of Lemma \ref{chap5:lemma:minach_servicecost} is such that $\gamma_{\epsilon_{k}}$ is admissible and $\overline{C}(\gamma_{\epsilon_{k}}) - c(\lambda) = \epsilon_{k}$ and $\overline{Q}(\gamma_{\epsilon_{k}}) = \mathcal{O}\nfrac{1}{\epsilon_{k}}$ as $\epsilon_{k} \downarrow 0$.
\end{remark}
This leads to the following asymptotic characterization of case 3.
The proof is similar to that of Proposition \ref{chap5:proposition:case2}.
\begin{proposition}
  For Case 3, the optimal tradeoff curve $Q^*(c_{c,k}) = \Theta\nfrac{1}{c_{c,k} - c(\lambda)}$ as $c_{c,k} \downarrow c(\lambda)$, for the sequence $c_{c,k} = \overline{C}(\gamma_{\epsilon_{k}})$.
\end{proposition}

\paragraph{\textbf{Case 1:}}
We note that in Case 1, $Q^*(c_{c})$ does not grow to infinity as $c_{c}$ approaches $c(\lambda)$.
In fact, the policy $\gamma_{u}$, which serves $S[m + 1] = \min(Q[m], s_{u})$, has the finite minimum average queue length over all admissible policies which achieve an average service cost of $c(\lambda)$.
First of all, we note that any admissible policy $\gamma$ which has $\bar{C}(\gamma) = c(\lambda)$, will have $Pr\brac{S(q) > s_{u}} = 0$, $\forall q$ and $Pr\brac{S[m] > s_{u}} = 0, \forall m \geq 1$.
For a given realization of the arrival process and the randomization of the batch sizes, let $q^*[m]$ and $q[m]$ be the evolution of the queue process under $\gamma_{u}$ and $\gamma$ respectively.
Then we note that $q^*[m] \leq q[m], \forall m$, and therefore $\gamma_{u}$ has the least average queue length over all policies which have their average service cost equal to $c(\lambda)$.
Furthermore, $Q^*(c_{c}) \leq \overline{Q}(\gamma_{u})$ for $c_{c} \geq c(\lambda)$.

For Case 1, we are only able to obtain a tight lower bound for a restricted case.
We show that if $s_{u} = 1$, then for any sequence of non-idling deterministic $\gamma_{k} \in \Gamma_{a}$, for which $\overline{C}(\gamma_{k}) - c(\lambda) = V_{k}$, we have that $\overline{Q}(\gamma_{k}) = \frac{\sigma^{2}}{2(s_{u} - \lambda)} + \frac{\lambda}{2} + \mathcal{O}\brap{V_{k}\log\nfrac{1}{V_{k}}}$.
We note that if $s_{u} = 1$, then $\overline{Q}(\gamma_{u}) = \frac{\sigma^{2}}{2(s_{u} - \lambda)} + \frac{\lambda}{2}$, from \cite{denteneer}.
Thus as $V \downarrow 0$, we have that the asymptotic lower bound has $\overline{Q}(\gamma_{u})$ as the limit point.
Furthermore, we note that the asymptotic order matches with what that was derived for FINITE-$\mu$CHOICE-1 in Chapter 2.

We present the lower bound on the average queue length in a series of steps.
Consider a particular policy $\gamma$ in the above sequence in the sequence $\gamma_{k}$, with $\overline{C}(\gamma) - c(\lambda) = V$.

Let $q_{u} \stackrel{\Delta} = \sup \brac{ q : s(q) \leq s_{u}}$.
Since we have restricted attention to non-idling deterministic $\gamma$ and $s_{u} = 1$, we have that $s(q) = s_{u} = 1$ for $q \in \brac{1, \dots, q_{u}}$.
In the following lemma, we obtain an upper bound on $Pr\brac{Q > q_{u}}$, which will be used to obtain an upper bound on $q_{u}$.

We note that from Assumption A1, $A_{max} > S_{max}$.
Therefore starting from any queue length $q$, the queue length in the next slot is at most $q - A_{max}$ or $q + A_{max}$.
\begin{lemma}
  For any non-idling deterministic $\gamma \in \Gamma_{a}$, if $q_{u} \geq A_{max}$, then for any $k \geq 0$,
  \begin{eqnarray*}
    Pr\brac{Q > q_{u} + k} \leq \min\brap{\rho_{u}^{\ceiling{\frac{q_{u} + k}{2A_{max}}}}, 1},
  \end{eqnarray*}
  where $\rho_{u} = \frac{\lambda}{s_{u}}$.
  \label{chap5:lemma:qul_pr}
\end{lemma}

The proof is discussed in Appendix \ref{chap5:app:qul_pr}.
The proof is very similar to that of Lemma 1 of Bertsimas \cite{gamarnik} and \cite{gamarnik_1} but with some slight modification.
We note that $c(s) \leq c_{1}(s)$, where 
\begin{eqnarray*}
  c_{1}(s) & = 
  \begin{cases}
    c(s), & \text{ for } s \in [0, s_{u}], \\
    c(s_{u}) + \frac{c(S_{max}) - c(s_{u})}{S_{max} - s_{u}} \brap{s - s_{u}}, & \text{ for } s \in (s_{u}, S_{max}].
  \end{cases}
\end{eqnarray*}
We note that $c_{1}(s) = l(s)$ for $s \in [0, s_{u}]$.
Then,
\begin{eqnarray*}
  V = \overline{C}(\gamma) - c(\lambda) \leq \Expp \bras{c_{1}(s(Q)) - l(s(Q))}, \text{ and},\\
  \Expp \bras{c_{1}(s(Q)) - l(s(Q))} = \sum_{q > q_{u}} \pi(q) \bras{c_{1}(s(q)) - l(s(q))}, \text{ and},\\
  \sum_{q > q_{u}} \pi(q) \bras{c_{1}(s(q)) - l(s(q))} \leq \sum_{q > q_{u}} \pi(q)\bras{m(S_{max} - s_{u})},
\end{eqnarray*}
where $m$ is the tangent of the angle between $c_{1}(s)$ and $l(s)$ at $s_{u}$.
Therefore, we have that
\begin{eqnarray}
  V \leq m(S_{max} - s_{u}) \sum_{q > q_{u}} \pi(q), \text{ or}, \nonumber \\
  Pr\brac{Q > q_{u}} \geq \frac{V}{m(S_{max} - s_{u})}.
  \label{chap5:eq:case11}
\end{eqnarray}
We now proceed to find a upper bound on $q_{u}$ by combining the above lower bound on $Pr\brac{Q > q_{u}}$ with the upper bound derived in Lemma \ref{chap5:lemma:qul_pr}.

We note that if $q_{u} < A_{max}$, then $A_{max}$ is an upper bound on $q_{u}$.
Suppose $q_{u} \geq A_{max}$, then from Lemma \ref{chap5:lemma:qul_pr} and \eqref{chap5:eq:case11}, we have that
\begin{eqnarray*}
  \frac{V}{m(S_{max} - s_{u})} \leq Pr\brac{Q > q_{u}} \leq \rho_{u}^{\ceiling{\frac{q_{u}}{2A_{max}}}} \leq \rho_{u}^{{\frac{q_{u}}{2A_{max}}}}.
\end{eqnarray*}
Since $\rho_{u} < 1$, we have that
\begin{eqnarray}
  \fpow{1}{\rho_{u}}{\frac{q_{u}}{2A_{max}}} \leq \frac{m(S_{max} - s_{u})}{V}, \text{ or}, \nonumber \\
  q_{u} \leq 2A_{max} \log_{\frac{1}{\rho_{u}}} \bras{\frac{m(S_{max} - s_{u})}{V}}.
  \label{chap5:eq:gamma_dash_lb_0}
\end{eqnarray}

\begin{lemma}
  The average queue length for the policy $\gamma$,
  \begin{eqnarray*}
    \overline{Q}(\gamma) \geq \frac{\sigma^{2}}{2(s_{u} - \lambda)} + \frac{\lambda}{2} - \frac{S_{max} - s_{u}}{s_{u} - \lambda}\brap{q_{u} Pr\brac{Q > q_{u}} + \sum_{q = q_{u}}^{\infty} Pr\brac{Q > q}}.
  \end{eqnarray*}
  \label{chap5:lemma:gamma_dash_lb}
\end{lemma}
\vspace{-1in}
\begin{proof}
  We note that the policy $\gamma$ is admissible.
  Squaring both sides of the evolution equation \eqref{chap5:eq:evolution}, taking expectations with respect to the stationary distribution, and simplifying we obtain that
  \begin{eqnarray*}
    2\Exp \bras{Q(S - A)} = \Exp A^{2} + \Exp S^{2} - 2\Exp A \Exp S.
  \end{eqnarray*}
  We note that $\Exp A^{2} - \Exp A \Exp S = \sigma^{2}$.
  Since $s_{u} = 1$, we have that $\Exp S^{2} = \sum_{s = 1}^{S_{max}} \pi_{s}(s) s^{2} \geq s_{u} \sum_{s = 1}^{S_{max}} \pi_{s}(s)s = \lambda s_{u}$.
  \begin{eqnarray}
    2\Exp Q(S - A) & \geq & \sigma^{2} + \lambda(s_{u} - \lambda), \text{ or}, \nonumber \\
    \sum_{q = 0}^{\infty} \pi(q) q (\Exp S(q) - \lambda) & \geq & \frac{\sigma^{2}}{2} + \lambda(s_{u} - \lambda), \nonumber \\
  \end{eqnarray}
  Or, we have that
  \begin{eqnarray}
    \sum_{q = 0}^{\infty} \pi(q) q (s_{u} - \lambda) + \sum_{q = q_{u} + 1}^{\infty} \pi(q) q (S_{max} - s_{u}) & \geq & \frac{\sigma^{2}}{2} + \lambda(s_{u} - \lambda), \nonumber
  \end{eqnarray}
  \begin{eqnarray}
    (s_{u} - \lambda) \sum_{q = 0}^{\infty} \pi(q) q & \geq & \frac{\sigma^{2}}{2} + \lambda(s_{u} - \lambda) - (S_{max} - s_{u}) \sum_{q = q_{u} + 1}^{\infty} \pi(q) q \nonumber \\
    \label{chap5:eq:gamma_dash_lb}
  \end{eqnarray}
  Simplifying the term $\sum_{q = q_{u} + 1}^{\infty} \pi(q)q$, we have
  \begin{eqnarray}
    \sum_{q = q_{u} + 1}^{\infty} \pi(q) q & = & \sum_{q = 1}^{\infty} (q_{u} + q) \pi(q_{u} + q), \nonumber \\
    & = & q_{u} \sum_{q = q_{u} + 1}^{\infty} \pi(q) + \sum_{q = 1}^{\infty} q \pi(q_{u} + q).
    \label{chap5:eq:case12}
  \end{eqnarray}
  Then from \eqref{chap5:eq:gamma_dash_lb} we have that
  \begin{eqnarray*}  
    \overline{Q}(\gamma) & \geq & \frac{\sigma^{2}}{2(s_{u} - \lambda)} + \frac{\lambda}{2} - \frac{S_{max} - s_{u}}{s_{u} - \lambda} \brap{q_{u} Pr\brac{Q > q_{u}} + \sum_{q = q_{u}}^{\infty} Pr\brac{Q > q}}.
  \end{eqnarray*}
\end{proof}

We now use the upper bound on $q_{u}$ from \eqref{chap5:eq:gamma_dash_lb_0} and the upper bound on $Pr\brac{Q \geq q_{u} + k}$ from Lemma \ref{chap5:lemma:qul_pr}, in the above lower bound on $\overline{Q}(\gamma)$ to obtain our final result.

\begin{lemma}
  If $s_{u} = 1$, then for any sequence of non-idling, deterministic $\gamma_{k} \in \Gamma_{a}$ such that $\overline{C}(\gamma_{k}) - c(\lambda) = V_{k} \downarrow 0$, we have that $\overline{Q}(\gamma_{k}) = \frac{\sigma^{2}}{2(s_{u} - \lambda)} + \frac{\lambda}{2} - \mathcal{O}\brap{V_{k} \log\nfrac{1}{V_{k}}}$.
\end{lemma}
\begin{proof}
  We consider a $\gamma$ in the sequence $\gamma_{k}$, with $\overline{C}(\gamma) - c(\lambda) = V$.
  Then, from Lemma \ref{chap5:lemma:gamma_dash_lb} we have that
  \begin{eqnarray*}
    \overline{Q}(\gamma) & \geq & \frac{\sigma^{2}}{2(s_{u} - \lambda)} + \frac{\lambda}{2} - \frac{S_{max} - s_{u}}{s_{u} - \lambda} \brap{q_{u} Pr\brac{Q > q_{u}} + \sum_{q = q_{u}}^{\infty} Pr\brac{Q > q}}.
  \end{eqnarray*}
  Consider $\sum_{q = q_{u}}^{\infty} Pr\brac{Q > q}$.
  We have that
  \begin{eqnarray}
    \sum_{q = q_{u}}^{\infty} Pr\brac{Q > q} = \sum_{m = 0}^{\infty} \sum_{k = 0}^{2A_{max} - 1} Pr\brac{Q > q_{u} + m 2A_{max} + k}.
    \label{chap5:eq:case13}
  \end{eqnarray}
  From the proof of Lemma \ref{chap5:lemma:qul_pr} we have that
  \begin{eqnarray*}
    Pr\brac{Q > q_{u} + m 2A_{max} + k} & \leq & \rho_{u}^{\ceiling{\frac{m 2A_{max} + k}{2 A_{max}}}} Pr\brac{Q > q_{u}}, \\
    & \leq & \rho_{u}^{\ceiling{\frac{m 2A_{max}}{2 A_{max}}}} Pr\brac{Q > q_{u}}, \text{ since } \rho_{u} < 1,\\
    & = & Pr\brac{Q > q_{u}} \rho_{u}^{m}.
  \end{eqnarray*}
  Substituting in \eqref{chap5:eq:case13}, we have that
  \begin{eqnarray}
    \sum_{q = q_{u}}^{\infty} Pr\brac{Q > q} & \leq & 2 A_{max} Pr\brac{Q > q_{u}} \sum_{m = 0}^{\infty} \rho_{u}^{m}, \nonumber \\
    & = & 2 A_{max} Pr\brac{Q > q_{u}} \frac{1}{1 - \rho_{u}}.
    \label{chap5:eq:case14}
  \end{eqnarray}
  
  Now we obtain an upper bound on $Pr\brac{Q > q_{u}}$.
  We note that \[\overline{C}(\gamma) - c(\lambda) = \sum_{q > q_{u}} \pi(q) \bras{ c(s(q)) - l(s(q)) }.\]
  Then 
  \begin{eqnarray*}
    m_{u} \sum_{q > q_{u}} \pi(q) \bras{s(q) - s_{u}} \leq \overline{C}(\gamma) - c(\lambda) = V,
  \end{eqnarray*}
  where $m_{u}$ is the tangent of the angle made by the line through $(c(s_{u} + 1), s_{u} + 1)$ and $(c(s_{u}), s_{u})$ with $l(s)$.
  We note that for $q > q_{u}, s(q) - s_{u} \geq 1$.
  Therefore 
  \begin{eqnarray}
    \sum_{q > q_{u}} \pi(q) = Pr\brac{Q > q_{u}} \leq \frac{V}{m_{u}}.
  \end{eqnarray}
  
  Using the above upper bound in \eqref{chap5:eq:case14}, we obtain that
  \begin{eqnarray*}
    \sum_{q = q_{u}}^{\infty} Pr\brac{Q > q} & \leq & 2 A_{max} \frac{V}{m_{u}(1 - \rho_{u})}.
  \end{eqnarray*}
  From \eqref{chap5:eq:case12}, we also obtain that
  \begin{eqnarray*}
    q_{u} Pr\brac{Q > q_{u}} \leq \frac{2A_{max}V}{m_{u}} \log_{\frac{1}{\rho_{u}}} \bras{\frac{m(S_{max} - s_{u})}{V}}.
  \end{eqnarray*}
  Then we have that
  \begin{eqnarray*}
    \overline{Q}(\gamma) & \geq & \frac{\sigma^{2}}{2(s_{u} - \lambda)} + \frac{\lambda}{2} - \frac{S_{max} - s_{u}}{s_{u} - \lambda} \brap{\frac{2A_{max}V}{m_{u}} \log_{\frac{1}{\rho_{u}}} \bras{\frac{m(S_{max} - s_{u})}{V}} + 2 A_{max} \frac{V}{m_{u}(1 - \rho_{u})}}.
  \end{eqnarray*}
  Thus, for the sequence of policies $\gamma_{k}$, with $V_{k} \downarrow 0$, we obtain that
  \begin{eqnarray*}
    \overline{Q}(\gamma_{k}) = \frac{\sigma^{2}}{2(s_{u} - \lambda)} + \frac{\lambda}{2} - \mathcal{O}\brap{V_{k} \log\nfrac{1}{V_{k}}}.
  \end{eqnarray*}
\end{proof}

\begin{remark}
  We note that the above asymptotic lower bounds can be obtained for admissible policies, even for general holding costs.
  Suppose, the holding cost is $h(q)$ in state $q$, instead of the queue length $q$.
  We assume that $h(q)$ is a strictly increasing function of $q$.
  Then, instead of $\Qg$ for an admissible policy, we are interested in the average holding cost $\overline{H}(\gamma) = \sum_{q = 0}^{\infty} \pi(q) h(q)$.
  We assume that for admissible policies $\overline{H}(\gamma)$ is finite.
  We note that asymptotic lower bounds on $\overline{H}(\gamma)$ can be obtained quite easily, from the above results.
  Consider the random variable $h(Q)$ for a policy $\gamma$.
  Then we obtain a lower bound $\frac{\overline{h}}{2}$ on $\overline{H}(\gamma)$, where $\overline{h}$ is the largest number such that $Pr\brac{h(Q) \leq \overline{h}} \leq \frac{1}{2}$.
  If the inverse function $h^{-1}$ of $h$ exists, then we have that $\overline{h}$ is the largest number such that $Pr\brac{Q \leq h^{-1}(\overline{h})} \leq \frac{1}{2}$.
  We note that we have already obtained lower bounds $\overline{q}_{l}$ to $\overline{q}$, where $\overline{q}$ is the largest integer such that $Pr\brac{Q \leq \overline{q})} \leq \frac{1}{2}$.
  Therefore, we obtain that $\overline{H}(\gamma) \geq \frac{h(\overline{q}_{l})}{2}$.
\end{remark}
\subsection{Asymptotic characterization of admissible policies for TRADEOFF}
\label{chap5:sec:aschar}
We consider a sequence of $c_{c,k} \downarrow c(\lambda)$ for TRADEOFF.
Let $\gamma_{k}$ be any sequence of feasible policies for the sequence $c_{c,k}$.
In this section, we obtain an asymptotic characterization of $\gamma_{k}$.
Our approach is similar to that in Sections \ref{chap4:sec:optimalpolicy_aschar} and \ref{chap4:section:aschar_optpolicy_intmc} for the state dependent M/M/1 model.
However, we are unable to obtain asymptotic upper bounds.
Since, only the asymptotic upper bounds depended on the order-optimality property, the bounds that we derive here hold for any sequence of feasible policies for the sequence $c_{c,k}$.

We first obtain the bounds $P_{l}\brac{A}$ and $P_{u}\brac{A}$ as in Section \ref{chap4:section:aschar_optpolicy_intmc}.
We note that the elements of sets $A$ are average service rates $\Exp S(q)$.
Let $A \subseteq [0, s_{l} - \epsilon_{V}] \bigcup [s_{u} + \epsilon_{V}, S_{max}]$.
Let $Q_{A} = \brac{q : \Exp S(q) \in A_{V}}$.
Proceeding as in the proof of Lemma \ref{chap5:lemma:tprob_2}, we have that $Pr\brac{Q \in Q_{A}} \leq P_{u}(Q_{A}) = \frac{V}{m\epsilon_{V}}$.
Then, we have that
\begin{eqnarray}
  P_{u}(Q_{A}) & = & 
  \begin{cases}
    \frac{V^{1 - \delta}}{m a}, \text{ if } \epsilon_{V} = aV^{\delta}, 0 \leq \delta < 1, \\
    \frac{1}{m a}, \text{ if } \epsilon_{V} = aV.
  \end{cases}
  \label{chap5:eq:aschar1}
\end{eqnarray}

Let $A = [s_{l} - \epsilon_{V}, s_{u} + \epsilon_{V}]$.
Let $Q_{A}$ be defined as before.
Then using \eqref{chap5:eq:aschar1}, we have that $Pr\brac{Q \in Q_{A}} \geq P_{l}(Q_{A})$, where 
\begin{eqnarray}
  P_{l}(Q_{A}) & = & 
  \begin{cases}
    1 - \frac{V^{1 - \delta}}{m a}, \text{ if } \epsilon_{V} = aV^{\delta}, 0 \leq \delta < 1, \\
    1 - \frac{1}{m a}, \text{ if } \epsilon_{V} = aV.
  \end{cases}
  \label{chap5:eq:aschar2}
\end{eqnarray}
We note that the above two bounds hold for cases 2 and 3.

Consider the sets $A_{1} = [0, s_{l} - \epsilon_{V}]$, $A_{2} = (s_{l} - \epsilon_{V}, \tilde{s}]$, $A_{3} = (\tilde{s}, s_{u} + \epsilon_{V}]$, and $A_{4} = [s_{u} + \epsilon_{V}, S_{max}]$.
From \eqref{chap5:eq:aschar1} we have that
\begin{eqnarray*}
  \pi(A_{2}) + \pi(A_{3}) = 1 - \pi(A_{1}) - \pi(A_{4}) & \geq & 1 - \frac{V}{m\epsilon_{V}}, \text{ and} \\
  (s_{l} - \epsilon_{V})\pi(A_{2}) + \tilde{s}\pi(A_{3}) & \leq & \lambda.
\end{eqnarray*}
We have that
\begin{eqnarray*}
  \pi(A_{3}) & \leq & \frac{\lambda - (s_{l} - \epsilon_{V}) \pi(A_{2})}{\tilde{s}}, \text{ and}, \\
  \pi(A_{2}) + \frac{\lambda - (s_{l} - \epsilon_{V}) \pi(A_{2})}{\tilde{s}} & \geq & 1 - \frac{V}{m\epsilon_{V}}.
\end{eqnarray*}
Therefore,
\begin{eqnarray*}
  \pi(A_{2})\bras{\frac{\tilde{s} - (s_{l} - \epsilon_{V})}{\tilde{s}}} \geq \frac{\tilde{s} - \lambda}{\tilde{s}} - \frac{V}{m\epsilon_{V}}, \\
  \pi(A_{2}) \geq \frac{\tilde{s} - \lambda}{\tilde{s} - (s_{l} - \epsilon_{V})} - \frac{\tilde{s} V}{m\epsilon_{V}\brap{\tilde{s} - (s_{l} - \epsilon_{V})}}.
\end{eqnarray*}
The above lower bound can be non-negative only if $\tilde{s} > \lambda$.
In the following, we set $\tilde{s} = \lambda + \epsilon_{V}'$, where $\lambda + \epsilon_{V}'$ is assumed to be less than or equal to $s_{u} + \epsilon_{V}$ and $\lambda + \epsilon_{V}$ for cases 2 and 3 respectively.
Then we have that,
\begin{eqnarray}
  \pi(A_{2}) \geq \frac{\epsilon_{V}'}{\lambda - s_{l} + \epsilon_{V}' + \epsilon_{V}} - \frac{\tilde{s}V}{m\epsilon_{V}\brap{\lambda - s_{l} + \epsilon_{V}' + \epsilon_{V}}}.
  \label{chap5:eq:aschar3}
\end{eqnarray}

We have the following result.
\begin{lemma}
  For any sequence of admissible policies $\gamma_{k}$, with $\overline{C}(\gamma_{k}) - c(\lambda) = V_{k} \downarrow 0$, and $\mathcal{Q}_{A} = \brac{q : \Exp S(q) \in A}$ for a $A \subseteq [0, S_{max}]$, we have that
  \begin{eqnarray*}
    |\mathcal{Q}_{A}| & = & 
    \begin{cases}
      \Omega\brap{\log\nfrac{1}{V_{k}}}, \text{ for Case 2, if } A = [s_{l} - a V^{\delta}, s_{u} + a V^{\delta}], 0 \leq \delta < 1, a > 0, \\
      \Omega\brap{\log\nfrac{1}{V_{k}}}, \text{ for Case 2, if } A = [s_{l} - a V^{\delta}, \lambda + \epsilon], 0 \leq \delta < 1, a > 0, \epsilon > 0, \\
      \Omega\nfrac{1}{V_{k}}, \text{ for Case 3, if } A = [0, \lambda + aV], a > 0. \\
    \end{cases}
  \end{eqnarray*}
\end{lemma}

\begin{proof}
  Consider a particular policy $\gamma$ in the above sequence with $\Cg - c(\lambda) = V$.
  Let us first consider the case where $A = [s_{l} - \epsilon_{V}, s_{u} + \epsilon_{V}]$.
  From \eqref{chap5:eq:aschar2}, we have that $P_{l}(Q_{A}) \geq 1 - \frac{V^{1 - \delta}}{m a}$, by choosing $\epsilon_{V} = a V^{\delta}, 0 \leq \delta < 1$.
  
  We proceed as follows for Case 2.
  Let $q_{s} \Deq \sup\brac{q : \Exp S(q) < s_{l} - \epsilon_{V}}$.
  Then, we have that $Pr\brac{Q \leq q_{s}} \leq \frac{V^{1 - \delta}}{m a}$.
  Suppose $q_{l}$ is the largest integer such that
  \begin{eqnarray*}
    Pr\brac{Q \leq q_{s}} \sum_{q = 1}^{q_{l}} \frac{\rho^{k}}{\rho_{d}} \leq 1 - \frac{V^{1 - \delta}}{m a},
  \end{eqnarray*}
  where $\rho$ and $\rho_{d}$ are as in Lemma \ref{chap5:lemma:barq_pi0_relation}, then $q_{l} \leq |Q_{A_{V}}|$.
  We then obtain that $q_{l} = \Omega\brap{\log\nfrac{1}{V}}$.

  Now we consider $A$ of the form $[s_{l} - \epsilon_{V}, \tilde{s}]$, for $\tilde{s} = \lambda + \epsilon_{V}' \leq s_{u} + \epsilon_{V}$ as in \eqref{chap5:eq:aschar3}.
  For this $A$, since the asymptotic lower bound on $|Q_{A}|$ is $\Omega\nfrac{1}{V}$ and is obtained as in the previous case, we do not present the derivation here.
  For Case 2, we choose $\epsilon_{V} = a V^{\delta}, 0 \leq \delta < 1$, and $\epsilon_{V}' = \epsilon$, where $\epsilon > 0$ is such that $\lambda + \epsilon < s_{u}$.
  Then from \eqref{chap5:eq:aschar3}, we have that
  \begin{eqnarray}
    \pi(A) \geq \frac{\epsilon'}{\lambda - s_{l} + \epsilon' + aV^{\delta}} - \frac{\tilde{s}V^{1 - \delta}}{m\brap{\lambda - s_{l} + \epsilon + aV^{\delta}}}.
  \end{eqnarray}
  Now proceeding as for the case when $A = [s_{l} - \epsilon_{V}, s_{u} + \epsilon_{V}]$ above, we obtain that for $A = [s_{l} - \epsilon_{V}, \lambda + \epsilon]$, $|Q_{A}| = \Omega\brap{\log\nfrac{1}{V}}$.
  
  We now consider Case 3.
  For Case 3, we consider $A$ to be $[0, s_{u} + \epsilon_{V}]$ rather than $[s_{l} - \epsilon_{V}, s_{u} + \epsilon_{V}]$.
  We also choose $\epsilon_{V}$ to be $a V$.
  Then, we have that $P_{l}(Q_{A}) \geq 1 - \frac{1}{m a}$.
  Let $q_{d} \Deq \sup\brac{q : \Exp S(q) \leq s_{u} + \epsilon_{V}}$.
  We note that $Pr\brac{Q \leq q_{d} - 1} = 1 - Pr\brac{Q \geq q_{d}}$.
  From TAIL-PROB-STATE-DEP-2 Lemma \ref{chap5:lemma:dtmc_stat_prob}, we have that
  \begin{eqnarray*}
    Pr\brac{Q \leq q_{d} - 1} \leq \brap{1 - \fpow{\epsilon_{a}}{\epsilon_{a} + \epsilon_{V}}{q_{d}}} - \brap{1 - \fpow{\epsilon_{a}}{\epsilon_{a} + \epsilon_{V}}{q_{d}}}\bras{\frac{1}{\epsilon_{V}}\sum_{q = q_{d} + 1} \bras{\lambda - \Exp S(q)} \pi(q)}.
  \end{eqnarray*}
  Then proceeding as in the proof of Lemma \ref{chap5:lemma:case3}, we have that $q_{d} = \Omega\nfrac{1}{V}$. Therefore, $|Q_{A}| = \Omega\nfrac{1}{V}$.
\end{proof}

\subsection{An asymptotic lower bound for the tradeoff problem \eqref{chap5:eq:inittradeoffprob}}
In this section, using the asymptotic results for TRADEOFF, derived in Section \ref{chap5:sec:asympchar}, we derive lower bounds for the optimal value of \eqref{chap5:eq:inittradeoffprob} for a set of $c_{c}$, via the Lagrange dual of \eqref{chap5:eq:inittradeoffprob}.
We note that the same  approach applies to the tradeoff problems in Chapter 5.
The asymptotic results for \eqref{chap5:eq:inittradeoffprob} are derived only for cases 2 and 3.

For \eqref{chap5:eq:inittradeoffprob}, from Section \ref{chap5:sec:cmdpformulation}, we have that there exists a stationary optimal policy. 
Therefore, the optimal value of \eqref{chap5:eq:inittradeoffprob} is equal to that of
\begin{eqnarray}
  \mini_{\gamma \in \Gamma_{s}} & & \Qg, \nonumber \\
  \text{such that } & & \Cg \leq c_{c}.
  \label{chap5:eq:dual1}
\end{eqnarray}
However, in this chapter, the problem that we have considered is 
\begin{eqnarray}
  \mini_{\gamma \in \Gamma_{a}} & & \Qg, \nonumber \\
  \text{such that } & & \Cg \leq c_{c}.
  \label{chap5:eq:dual2}
\end{eqnarray}
We note that the optimal value of \eqref{chap5:eq:dual1} is lower bounded by the optimal value of its Lagrange dual:
\begin{eqnarray}
  \max_{\beta \geq 0} \bras{\min_{\gamma \in \Gamma_{s}} \bras{\Qg + \beta\brap{\Cg - c_{c}}}}.
  \label{chap5:eq:dual3}
\end{eqnarray}
From Lemma \ref{chap5:lemma:statopt}, we have that for any $\beta \geq 0$, there exists an admissible policy $\gamma^*_{\beta}$ which achieves the minimum for the problem $\min_{\gamma \in \Gamma_{s}} \bras{\Qg + \beta\brap{\Cg - c_{c}}}$.
In the following, we show that an asymptotic lower bound to the solution of \eqref{chap5:eq:dual1} can be obtained using \eqref{chap5:eq:dual3} from the asymptotic behaviour of the optimal solution of \eqref{chap5:eq:dual2}, for certain sequences of $c_{c}$ as $c_{c} \downarrow c(\lambda)$.

We note that for any sequence $\beta \uparrow \infty$, it can be shown that $\overline{C}(\gamma^*_{\beta}) \downarrow c(\lambda)$.
For the following analysis, we also consider the MDP:
\begin{eqnarray}
  \min_{\gamma \in \Gamma_{s}} \bras{\Qg + \beta\brap{\Cg - c(\lambda)}}.
  \label{chap5:eq:dual4}
\end{eqnarray}
We note that for a $\beta \geq 0$, $\gamma^*_{\beta}$ is optimal for both \eqref{chap5:eq:dual3} and \eqref{chap5:eq:dual4}.
We have the following result.

\begin{proposition}
  Suppose $\gamma_{k}$ is any sequence of policies such that $\Cgk - c(\lambda) = V_{k} \downarrow 0$ and
  \begin{eqnarray*}
    \Qgk & = & 
    \begin{cases}
      \mathcal{O}\brap{\log\nfrac{1}{V_{k}}}, \text{ for Case 2,} \\
      \mathcal{O}\nfrac{1}{V_{k}}, \text{ for Case 3.}
    \end{cases}
  \end{eqnarray*}
  Let $c_{c,k}$ be any sequence such that 
  \begin{eqnarray*}
    c_{c,k} & = & 
    \begin{cases}
      \Theta\brap{\Cgk - c(\lambda)}, \text{ for Case 2,} \\
      \Theta\brap{\brap{\Cgk - c(\lambda)}^{2}}, \text{ for Case 3}. \\
    \end{cases}
  \end{eqnarray*}
  Then, for the tradeoff problem \eqref{chap5:eq:dual1}, for the sequence $c_{c,k} \downarrow c(\lambda)$ we have that
  \begin{eqnarray*}
    Q^*(c_{c,k}) & = &
    \begin{cases}
      \Omega\brap{\log\nfrac{1}{c_{c,k} - c(\lambda)}}, \text{ for Case 2,} \\
      \Omega\nfrac{1}{c_{c,k} - c(\lambda)}, \text{ for Case 3}.
    \end{cases}
  \end{eqnarray*}
\end{proposition}

\begin{proof}
We consider Case 2 first.
Let $\gamma_{k}$ be any sequence of policies (admissible or otherwise), which is such that $\overline{C}(\gamma_{k}) - c(\lambda) = V_{k} \downarrow 0$ and $\Qgk = \mathcal{O}\brap{\log\nfrac{1}{V_{k}}}$.
From Lemma \ref{chap5:lemma:case2upperbound}, we note that at least one such sequence exists.
Let $\tilde{\beta}_{k} = \frac{1}{\brap{\overline{C}(\gamma_{k}) - c(\lambda)}}$.
We show that $\overline{C}(\gamma^*_{\tilde{\beta}_{k}}) - c(\lambda) = \mathcal{O}\nfrac{1}{{\tilde{\beta}_{k}}^{\delta}}$, where $0 \leq \delta < 1$.
We proceed by contradiction.
Suppose $\overline{C}(\gamma^*_{\tilde{\beta}_{k}}) - c(\lambda)$ is not $\mathcal{O}\nfrac{1}{{\tilde{\beta}_{k}}^{\delta}}$.
Then $\overline{C}(\gamma^*_{\tilde{\beta}_{k}}) - c(\lambda)$ is $\omega\nfrac{1}{{\tilde{\beta}_{k}}^{\delta}}$.
Therefore, the optimal value of \eqref{chap5:eq:dual4}, $\overline{Q}(\gamma^*_{\tilde{\beta}_{k}}) + \tilde{\beta}_{k} \brap{\overline{C}(\gamma^*_{\tilde{\beta}_{k}}) - c(\lambda)} = \omega\brap{{\tilde{\beta}_{k}}^{1 - \delta}}$.
However, we note that the sequence of policies $\gamma_{k}$ is such that $\Qgk + \tilde{\beta}_{k} (\Cgk - c(\lambda)) = \mathcal{O}\brap{\log\brap{\tilde{\beta}_{k}}}$, which contradicts the optimality of the sequence $\gamma^*_{\tilde{\beta}_{k}}$.
Therefore, $\overline{C}(\gamma^*_{\tilde{\beta}_{k}}) - c(\lambda) = \mathcal{O}\nfrac{1}{{\tilde{\beta}_{k}}^{\delta}}$.

Consider a sequence of $c_{c,k} \downarrow c(\lambda)$ for \eqref{chap5:eq:dual1}.
Suppose $c_{c,k}$ is such that $c_{c,k} - c(\lambda) = \Theta\brap{\Cgk - c(\lambda)}$.
Then \eqref{chap5:eq:dual3} can be bounded below as
\begin{eqnarray*}
  \max_{\beta \geq 0} \min_{\gamma \in \Gamma_{s}} \bras{\Qg + \beta\brap{\Cg - c_{c,k }}} & \geq & \overline{Q}(\gamma^*_{\tilde{\beta}_{k}}) + \tilde{\beta}_{k} \brap{\overline{C}(\gamma^*_{\tilde{\beta}_{k}}) - c(\lambda)} - \tilde{\beta}_{k}\brap{c_{c,k} - c(\lambda)}.
\end{eqnarray*}
We have that $\tilde{\beta}_{k} \brap{\overline{C}(\gamma^*_{\tilde{\beta}_{k}}) - c(\lambda)} \geq 0$.
Since $c_{c,k} - c(\lambda) = \mathcal{O}\brap{\Cgk - c(\lambda)}$, we have that $\tilde{\beta}_{k}\brap{c_{c,k} - c(\lambda)} = \mathcal{O}(1)$.
Furthermore, since $\overline{C}(\gamma^*_{\tilde{\beta}_{k}}) - c(\lambda) = \mathcal{O}\nfrac{1}{{\tilde{\beta}_{k}}^{\delta}}$, we have that $\overline{Q}(\gamma^*_{\tilde{\beta}_{k}}) = \Omega\brap{\log\brap{\tilde{\beta}_{k}}}$.
Since $c_{c,k} - c(\lambda)$ is also $\Omega\brap{\Cgk - c(\lambda)}$ we have that
\begin{eqnarray*}
  \overline{Q}(\gamma^*_{\tilde{\beta}_{k}}) + \tilde{\beta}_{k} \brap{\overline{C}(\gamma^*_{\tilde{\beta}_{k}}) - c(\lambda)} - \tilde{\beta}_{k}\brap{c_{c,k} - c(\lambda)} = \Omega\brap{\log\nfrac{1}{c_{c,k} - c(\lambda)}},
\end{eqnarray*}
which provides an asymptotic lower bound for \eqref{chap5:eq:dual1}.

We now consider Case 3.
Let $\gamma_{k}$ be any sequence of policies, which is such that $\overline{C}(\gamma_{k}) - c(\lambda) = V_{k} \downarrow 0$ and $\Qgk = \mathcal{O}\nfrac{1}{V_{k}}$.
From Lemma \ref{chap5:lemma:minach_servicecost}, we note that at least one such sequence exists.
Let $\tilde{\beta}_{k} = \frac{1}{\brap{\overline{C}(\gamma_{k}) - c(\lambda)}^{2}}$.
We show that $\overline{C}(\gamma^*_{\tilde{\beta}_{k}}) - c(\lambda) = \mathcal{O}{\nfrac{1}{\sqrt{\tilde{\beta}_{k}}}}$.
We proceed by assuming that $\overline{C}(\gamma^*_{\tilde{\beta}_{k}}) - c(\lambda)$ is not $\mathcal{O}{\nfrac{1}{\sqrt{\tilde{\beta}_{k}}}}$.
Then $\overline{C}(\gamma^*_{\tilde{\beta}_{k}}) - c(\lambda)$ is $\omega{\nfrac{1}{\sqrt{\tilde{\beta}_{k}}}}$.
Then we have that the optimal value of \eqref{chap5:eq:dual4}, $\overline{Q}(\gamma^*_{\tilde{\beta}_{k}}) + \tilde{\beta}_{k} \brap{\overline{C}(\gamma^*_{\tilde{\beta}_{k}}) - c(\lambda)} = \omega\brap{\sqrt{\tilde{\beta}_{k}}}$.
We note that the sequence of policies $\gamma_{k}$ is such that $\Qgk + \tilde{\beta}_{k} (\Cgk - c(\lambda)) = \mathcal{O}\brap{\sqrt{\tilde{\beta}_{k}}}$, which contradicts the optimality of the sequence $\gamma^*_{\tilde{\beta}_{k}}$.
Therefore, $\overline{C}(\gamma^*_{\tilde{\beta}_{k}}) - c(\lambda) = \mathcal{O}\nfrac{1}{\sqrt{\tilde{\beta}_{k}}}$.

Then, proceeding as in Case 2, for any sequence of $c_{c,k}$ such that $c_{c,k} - c(\lambda) = \Theta\brap{\brap{\Cgk - c(\lambda)}^{2}}$, we have that for Case 3, the optimal value of \eqref{chap5:eq:dual1} is $\Omega\nfrac{1}{c_{c,k} - c(\lambda)}$.
\end{proof}

We note that since for $c'_{c,k} \in \mathcal{O}^{u}$, there exists an admissible optimal policy for \eqref{chap5:eq:inittradeoffprob}, the asymptotic lower bounds obtained in Lemma \ref{chap5:lemma:tprob_2} and Lemma \ref{chap5:lemma:case3} apply directly.
The above Lagrange dual approach shows that the asymptotic lower bounds also apply to \eqref{chap5:eq:inittradeoffprob} for the sequences $c_{c,k}$ considered above.
Since the sequence of admissible optimal policies for the sequence $c'_{c,k}$ satisfy the properties required for $\gamma_{k}$ stated in the above proposition, we note that the set of $c_{c,k}$ for which the above lower bound holds also contains $\mathcal{O}^{u}$.
However, we are unable to show that for any sequence $c_{c,k} \downarrow c(\lambda)$ the asymptotic lower bounds in the above proposition hold.

\subsection{Asymptotic lower bounds for ergodic $(A[m], m \geq 1)$}
\label{chap5:sec:extension_ergodic}
We note that when the arrival process $(A[m], m \geq 1)$ is an ergodic batch arrival process, the optimal policy for the tradeoff problem \eqref{chap5:eq:inittradeoffprob} may not be stationary.
But in this section, we consider the set of policies $\Gamma_{s}$, which are such that the batch size $S(q)$ used for service in slot $m$ is a function only of the queue length $Q[m - 1] = q$, and is independent of anything else.
We note that $S(Q[m - 1])$ could be a randomized function of $Q[m - 1]$.
The asymptotic lower bound presented here is significant, in that it complements the asymptotic upper bound obtained for Markov batch arrival processes in \cite[Section 4.9]{neely} and \cite{huang_neely}.

We assume that $(A[m])$ is ergodic, so that almost surely
\begin{eqnarray*}
  \lim_{M \rightarrow \infty} \frac{1}{M} \sum_{m = 1}^{M} A[m] = \Exp A[1] = \lambda,
\end{eqnarray*}
and $\lambda < S_{max}$, where $S_{max}$ is the largest batch size which can be served, as defined before.
We also assume that the arrival process $(A[m])$ is such that
\begin{description}
\item[NA1 :]{Let $\sigma[m - 1] = (Q[0] = q_{0}, A[1] = a_{1}, Q[1] = q_{1}, A[2] = a_{2}, \dots, A[m - 1] = a_{m - 1}, Q[m - 1] = q_{m - 1})$. We assume that
\begin{eqnarray*}
\inf_{m \in \sZ} \mathop{\min_{\brac{a_{1}, \dots, a_{m - 1}}}}_{\brac{q_{1}, \dots, q_{m - 1}}} Pr\brac{A[m] = 0 \middle \vert \sigma[m - 1]} = \nu_{a} > 0.
\end{eqnarray*}
}
\end{description}
We restrict to policies $\gamma \in \Gamma_{s}$ for which the following limits exist
\begin{eqnarray}
  \lim_{m \rightarrow \infty} Pr\brac{Q[m] = q \vert Q[0] = q_{0}} & = & \pi(q), \forall q \in \sZ,
  \label{chap5:eq:genA_2}
\end{eqnarray}
with $\sum_{q = 0}^{\infty} \pi(q) = 1$.
We note that for policies $\gamma \in \Gamma_{s}$ for which the above limits exist, also have well defined $\pi_{s}(s), \forall s \in \brac{0,\dots, S_{max}}$, where
\begin{eqnarray}
  \pi_{s}(s)  & = &  \lim_{M \rightarrow \infty} \frac{1}{M} \sum_{m = 1}^{M} Pr\brac{S[m] = s \vert Q[0] = q_{0}}.
  \label{chap5:eq:genA_1}
\end{eqnarray}
For such a policy $\gamma$, the average service cost is $\overline{C}(\gamma) = \sum_{s = 0}^{S_{max}} \pi_{s}(s) c(s)$ and the average queue length is $\overline{Q}(\gamma)$ is defined as $\sum_{q = 0}^{\infty} \pi(q) q$.

To obtain an asymptotic characterization of the tradeoff, we again restrict to a class $\Gamma_{a}$ of admissible policies.
However, we use a weaker definition\footnote{The set of admissible policies in this section contains the set of admissible policies defined in the previous sections. This shows that the development of the $\log\nfrac{1}{V}$ asymptotic lower bound can be obtained under weaker assumptions than what was assumed in the previous sections.} of admissible policies, compared with the definition in Section \ref{chap5:sec:tradeoffproblem}.
A policy $\gamma \in \Gamma_{s}$ is admissible if:
\begin{description}
\item[NG1 :]{the limits in \eqref{chap5:eq:genA_2} exist for $\gamma$,}
\item[NG2 :]{$\gamma$ is \emph{mean rate stable} (see \cite{neely}), i.e., $\sum_{s = 0}^{S_{max}} \pi_{s}(s) s = \lambda$,}
\item[NG3 :]{the average service rate $\Exp S(q)$ is a non-decreasing function of $q$ for $\gamma$.}
\end{description}
Then for any admissible policy $\gamma$, we have that $\overline{C}(\gamma) \geq c(\lambda)$, as before, by applying Jensen's inequality.

Let $U_{V}$ be a random variable with support on $\brac{0, \dots, S_{max}}$ and $\Exp U_{V} = \lambda + V$.
Let $(U_V[m], m \geq 1)$ be an IID sequence with $U_{V}[m] \sim U_{V}$.
Consider a particular policy $\gamma_{V}$ for a $V > 0$, which chooses $S[m] = \min(Q[m - 1], U_{V}[m])$.
Using \cite[Lemma 1]{loynes} we have that the limit $\pi(q) = \lim_{m \rightarrow \infty} Pr\brac{Q[m] = q}$ exists.
We note that then the evolution of the queue can be written as 
\begin{eqnarray*}
  Q[m] = \max\bigg (Q[m - 1] - U_V[m], 0 \bigg ) + A[m], \text{ for } m \geq 1, \text{ with } Q[0] = q_{0}.
\end{eqnarray*}
Let $Q'[1] = \max(q_{0} - U_V[1], 0)$, and
\begin{eqnarray*}
  Q'[m] = \max \bigg (Q'[m - 1] + A[m - 1] - U_V[m], 0 \bigg ), \text{ for } m \geq 2.
\end{eqnarray*}
We note that $Q[m] = Q'[m] + A[m], m \geq 1$.
Since the sequence of random variables $\zeta[m] = A[m] - U_V[m + 1], m \geq 1$ is ergodic and $\Exp \zeta[1] = \lambda - (\lambda + V) < 0$, from \cite[Chapter 1, Theorem 7]{borokov} we have that $\lim_{m \rightarrow \infty} Pr\brac{Q'[m] < \infty} = 1$ and $\lim_{m \rightarrow \infty} Pr\brac{Q'[m] = q} = \pi'(q)$ exists.
Since $A[1] \leq A_{max}$ we have that $\lim_{m \rightarrow \infty} Pr\brac{Q[m] < \infty} = 1$.
Thus $\gamma_{V}$ satisfies properties NG1 and NG2.
We also note that by construction $\gamma_{V}$ satisfies property NG3 and hence $\gamma_{V}$ is admissible.
We note that $\overline{C}(\gamma_{V}) \leq \Exp c(U_{V}[1])$.
Now consider the sequence of admissible policies $\gamma_{V}$ for a sequence $V \downarrow 0$.
Then it is always possible to choose 
\footnote{For small enough $V$, a distribution for $U_{V}$ that gives mass to either (i) $\lambda$ and $\lambda + 1$, if $\lambda$ is an integer or (ii) $\floor{\lambda}$ and $\ceiling{\lambda}$, if $\lambda$ is not an integer, can be chosen such that $\Exp U_{V} = \lambda + V$ and $\Exp c(U_{V}) = c(\lambda) + m V$, where $m$ is the slope of the line joining $(\lambda, c(\lambda))$ and $(\lambda + 1, c(\lambda + 1))$ for (i) or the line joining $(\floor{\lambda}, c(\floor{\lambda}))$ and $(\ceiling{\lambda}, c(\ceiling{\lambda}))$ for (ii).}
the distribution of $U_{V}$ such that $\overline{C}(\gamma_{V}) = c(\lambda) + \mathcal{O}(V)$ as $V \downarrow 0$.
Therefore, as before, $c(\lambda) = \inf_{\gamma \in \Gamma_{a}} \overline{C}(\gamma)$.

We note that $c(\lambda)$ is piecewise linear and three cases arise depending on the value of $\lambda$, as shown in Section \ref{chap5:sec:asymp_analysis_tprob}.
We define $s_{l}$, $s_{u}$, and the line $l(s)$ as in Section \ref{chap5:sec:asymp_analysis_tprob}.
For Case 1, we do not have a tight asymptotic lower bound.
We now obtain an asymptotic lower bound which applies to Cases 2 and 3.
We note that the bound is obtained by generalizing the proofs of Lemmas \ref{chap5:lemma:barq_pi0_relation} and \ref{chap5:lemma:tprob_2}.
We also note that this asymptotic lower bound holds even if $(A[m])$ is just stationary rather than ergodic.
Ergodicity of $(A[m])$ was required for proving that $c(\lambda) = \inf_{\gamma \in \Gamma_{a}} \overline{C}(\gamma)$.

For some positive $\epsilon < s_{l}$, let $q_{s} \stackrel{\Delta} = \inf\brac{q : \Exp S(q) \geq s_{l} - \epsilon}$.
We note that the proof of the following asymptotic lower bound also follows the methodology summarized in Section \ref{chapter4:methodology}, except that the geometric bound on the stationary probability distribution for the ergodic process $Q[m]$ is obtained from the assumptions NA1 and NG3.
\begin{lemma}
  For an ergodic arrival process $(A[m], m \geq 1)$, satisfying NA1, and for any sequence of admissible policies $\gamma_{k}$ (satisfying NG1, NG2, and NG3), with $\overline{C}(\gamma_{k}) - c(\lambda) = V_{k} \downarrow 0$, we have that $\overline{Q}(\gamma_{k}) = \Omega\brap{\log\nfrac{1}{V_{k}}}$, for Cases 2 and 3.
  \label{chap5:lemma:extension_ergodic}
\end{lemma}
\begin{proof}
  For a particular policy $\gamma$ in the above sequence with $V_{k} = V$, as in the proof of Lemma \ref{chap5:lemma:tprob_2}, we have that 
  \begin{eqnarray*}
    \sum_{q} \pi(q) \bigg[c(\Exp S(q)) - l(\Exp S(q))\bigg] \leq V,
  \end{eqnarray*}
  since $c(s)$ is convex and $l(s)$ is linear.
  Therefore, 
  \begin{eqnarray*}
    \sum_{q = 0}^{q_{s} - 1} \pi(q) \leq \frac{V}{m_{1}\epsilon},
  \end{eqnarray*}
  where $m_{1}$ is the tangent of the angle between (i) the line passing through $(s_{l} - 1, c(s_{l} - 1))$ and $(s_{l}, c(s_{l}))$ and (ii) $l(s)$.

  As in the proof of Lemma \ref{chap5:lemma:barq_pi0_relation}, we have that $Pr\brac{S(q) > 0} \geq \frac{s_{l} - \epsilon}{S_{max}}, \forall q \geq q_{s}$, since for $q \geq q_{s}$, $\Exp S(q) \geq s_{l} - \epsilon$.
  Now we relate the stationary probability $\pi(q), q \geq q_{s}$ to $\sum_{q = 0}^{q_{s} - 1} \pi(q)$.
  We have that for a $q \geq q_{s}$ and for every $m \geq 0$
  \begin{eqnarray*}
    Pr\brac{Q[m + 1] < q | Q[0] = q_{0}} & = & Pr\brac{Q[m] - S(Q[m]) + A[m + 1] < q | Q[0] = q_{0}},
  \end{eqnarray*}
  which can be written as
  \begin{eqnarray*}
    & = & \Exp_{S[1],Q[1],\dots,Q[m - 1], S[m]} \bigg [Pr\bigg\{Q[m] - S(Q[m]) + A[m + 1] < q \bigg\vert \\
      & & Q[0] = q_{0}, S[1], Q[1], \dots, Q[m - 1], S[m]\bigg\} \bigg ],
  \end{eqnarray*}
  which is
  \begin{eqnarray*}
    & \geq & \Exp_{S[1],Q[1],\dots,Q[m - 1], S[m]} \bigg [ Pr\brac{Q[m] = q | Q[0] = q_{0}, S[1], Q[1],  \dots, Q[m - 1], S[m]} \times \\
    & & Pr\brac{S(Q[m]) > 0 | Q[0] = q_{0}, S[1], Q[1], \dots, Q[m - 1], S[m], Q[m] = q} \times \\
    & & Pr\brac{A[m + 1] = 0 | Q[0] = q_{0},  S[1], Q[1], \dots, Q[m - 1], S[m], Q[m] = q, \brac{S(Q[m]) > 0}} \bigg ].
  \end{eqnarray*}
  We note that the batch size $S(Q[m])$ is chosen independently of the history of the queue length evolution.
  We also note that the evolution $(Q[0] = q_{0}, S[1], Q[1], \dots, Q[m - 1], S[m], Q[m] = q)$ is equivalent to the evolution $\sigma[m] = (q_{0}, A[1], Q[1], A[2], Q[2], \dots, A[m], Q[m] = q)$.
  Furthermore, $A[m + 1]$ is independent of the batch size $S[m + 1] = S(Q[m])$ given $Q[m]$.
  Therefore, using NA1, the above lower bound can be written as
  \begin{eqnarray*}
    & = & \Exp_{S[1],Q[1],\dots,Q[m - 1], S[m]} \bigg [ Pr\brac{Q[m] = q | Q[0] = q_{0}, Q[1], S[1], \dots, Q[m - 1], S[m]} \times \\
    & & Pr\brac{S(Q[m]) > 0 | Q[m] = q} \times Pr\brac{A[m + 1] = 0 | \sigma[m]} \bigg ].
  \end{eqnarray*}
  Then using the property NA1 and the above lower bound on $Pr\brac{S(q) > 0}$ for $q \geq q_{s}$ we have that $Pr\brac{Q[m + 1] < q \vert Q[0] = q_{0}}$
  \begin{eqnarray*}
    & \geq & \frac{s_{l} - \epsilon}{S_{max}} \nu_{a} \Exp_{S[1],Q[1],\dots,Q[m - 1], S[m]} \bigg [ Pr\brac{Q[m] = q | Q[0] = q_{0}, Q[1], S[1], \dots, Q[m - 1], S[m]} \bigg ], \\
    & = & Pr\brac{Q[m] = q \vert Q[0] = q_{0}} \frac{s_{l} - \epsilon}{S_{max}} \nu_{a}.
  \end{eqnarray*}
  Defining $\rho_{d} = \frac{s_{l} - \epsilon}{S_{max}} \nu_{a}$ and $\rho = 1 + \frac{1}{\rho_{d}}$, and proceeding as in the proof of Lemma \ref{chap5:lemma:barq_pi0_relation}, we have that for any non-negative $k$, $q = q_{s} + k$, and for every $m \geq 0$,
  \begin{eqnarray*}
    Pr\brac{Q[m] = q \vert Q[0] = q_{0}} \leq Pr\brac{Q[m + 1] < q_{s} \vert Q[0] = q_{0}} \frac{\rho^{k}}{\rho_{d}}.
  \end{eqnarray*}
  Therefore,
  \begin{eqnarray*}
    \frac{1}{M} \sum_{m = 0}^{M - 1} Pr\brac{Q[m] = q \vert Q[0] = q_{0} } & \leq & \frac{\rho^{k}}{\rho_{d}} \bigg [\frac{1}{M} \sum_{m = 0}^{M - 1} Pr\brac{Q[m] < q_{s} \vert Q[0] = q_{0} } + \\
    & & \frac{Pr\brac{Q[M] < q_{s} \vert Q[0] = q_{0} } - Pr\brac{Q[0] < q_{s} \vert Q[0] = q_{0} }}{M} \bigg].
  \end{eqnarray*}
  Then as $M \rightarrow \infty$, since $\gamma$ is admissible, we have that
  \begin{eqnarray*}
    \pi(q) \leq \frac{\rho^{k}}{\rho_{d}} \sum_{q = 0}^{q_{s} - 1} \pi(q).
  \end{eqnarray*}
  Proceeding similarly as in the proof of Lemma \ref{chap5:lemma:barq_pi0_relation} (from \eqref{chap5:eq:barpiq1}), we can show that
  \begin{eqnarray*}
    \overline{Q}(\gamma) \geq \frac{1}{2} \bras{\log_{\rho} \bras{\frac{1}{2 \sum_{q = 0}^{q_{s} - 1} \pi(q)}} -1}.
  \end{eqnarray*}
  Since $\sum_{q = 0}^{q_{s} - 1} \pi(q) \leq \frac{V}{m_{1} \epsilon}$, we have that for the sequence of policies $\gamma_{k}$, $\overline{Q}(\gamma_{k}) = \Omega\brap{\log\nfrac{1}{V_{k}}}$.
\end{proof}
We note that the above asymptotic lower bound is not tight for Case 3 for IID $(A[m])$ since we have a $\Omega\nfrac{1}{V_{k}}$ lower bound on $\overline{Q}(\gamma_{k})$.

In \cite[Theorem 4.12]{neely} and \cite{huang_neely}, it has been shown that if $(A[m], m \geq 1)$ is Markov, i.e., 
\[Pr\brac{A[m + 1] = a_{m + 1} | A[m] = a_{m}, \dots, A[1] = a_{1}} = Pr\brac{A[m + 1] = a_{m + 1} | A[m] = a_{m}},\]
 then for a sequence of Quadratic Lyapunov Algorithm (QLA) policies, parametrized by a sequence $V_{k} \downarrow 0$, the average queue length is $\mathcal{O}\nfrac{1}{V_{k}}$ for an average service cost $V_{k}$ more than $c(\lambda)$.

We note that the QLA algorithm chooses a deterministic batch size $s(q)$ for service based on the current queue length only (unlike, say TOCA, for which the batch size is chosen as a function of other auxiliary variables also).
Therefore, the QLA algorithm falls in the restricted class of admissible policies considered in this section.
We note that the batch sizes are chosen deterministically, therefore we have that
\begin{eqnarray*}
  & & Pr\brac{A[m + 1] = a_{m + 1} | A[m] = a_{m}, \dots, A[1] = a_{1}, Q[0] = q_{0}} = \\
  & & Pr\brac{A[m + 1] = a_{m + 1} | Q[m] = q_{m}, A[m] = a_{m}, Q[m - 1] = q_{m - 1}, \dots, A[1] = a_{1}, Q[0] = q_{0}},
\end{eqnarray*}
for the given policy, for $(q_{n}, 0 \leq n \leq m)$, such that $q_{n + 1} = q_{n} - s(q_{n}) + a_{n + 1}$.
Then we note that the above asymptotic lower bound applies, for the QLA algorithm, under the assumption
\begin{eqnarray*}
& & Pr\brac{A[m] = a_{m} \middle \vert A[1] = a_{1}, A[2] = a_{2}, \dots, A[m - 1] = a_{m - 1}, Q[0] = q_{0}} =  \\
& & Pr\brac{A[m] = a_{m} \middle \vert A[1] = a_{1}, A[2] = a_{2}, \dots, A[m - 1] = a_{m - 1}}.
\end{eqnarray*}

\section{Asymptotic bounds for R-model}
\label{chap5:sec:realvalued_setup}
As for the integer valued case, the tradeoff problem is to obtain $Q^*(c_{c}, q_{0})$, which is the optimal value of 
\begin{equation*}
  \mini_{\gamma \in \Gamma} \overline{Q}(\gamma, q_{0}) \text{ such that } \overline{C}(\gamma, q_{0}) \leq c_{c},
\end{equation*}
where $c_{c} \geq 0$ is the average service cost constraint.

We recall that the above problem can again be formulated as a constrained Markov decision problem over the class of policies $\Gamma$.
However, since the R-model is usually used as an approximation for the I-model, we analyse the tradeoff problem only for a restricted class of stationary admissible policies, where the definition of this class of admissible policies is motivated by the definition in Section \ref{chap5:sec:integervalued_setup}.
We assume that $\lambda < S_{max}$ as before.

\textbf{Stability :}
A policy $\gamma \in \Gamma_{s}$ is said to be stable if : a) the Markov chain $Q[m]$ under $\gamma$ is positive Harris recurrent with stationary distribution $\pi_{\gamma}$ on the recurrence class corresponding to $q_{0}$, and b) $\overline{Q}(\gamma, q_{0}) < \infty$.

\textbf{Admissibility :}
In the following we restrict ourselves to the class of admissible policies $\Gamma_{a}$ which is defined below.
A policy $\gamma$ is called admissible if:
\begin{description}
\item[RG1 :]{it is stable,}
\item[RG2 :]{it induces an aperiodic, irreducible Harris Markov chain $Q[m]$, and,}
\item[RG3 :]{the average service rate at a queue length $q$, $\Exp S(q)$ is non-decreasing in $q$.}
\end{description}
We note that the properties RG1 and RG2 are similar to the properties of admissible policies used in Berry and Gallager \cite{berry}.
The additional property RG3 is motivated by the monotonicity property derived in Agarwal et al. \cite{agarwal} as well as the monotonic non-decreasing property of any stationary deterministic optimal policy derived for the integer valued case in Section \ref{chap5:sec:integervalued_setup}.
We note that as in Section \ref{chap5:sec:integervalued_setup}, for $\gamma \in \Gamma_{a}$, the average queue length and average service cost are independent of $q_{0}$ and are therefore denoted by $\overline{Q}(\gamma)$ and $\overline{C}(\gamma)$ respectively.
We also note that $\overline{Q}(\gamma) = \Exp_{\pi_{\gamma}}Q$ and $\overline{C}(\gamma) = \Exp_{\pi_{\gamma}} c_{R}(S(Q))$.

We have that the TRADEOFF problem, for the R-model, is to obtain $Q^*(c_{c})$ which is the optimal value of
\begin{eqnarray*}
  \mini_{\gamma \in \Gamma_{a}} \overline{Q}(\gamma), \text{ such that } \overline{C}(\gamma) \leq c_{c}.
\end{eqnarray*}

From the convexity of $c_{R}(s)$ and Jensen's inequality, we obtain that for any $\gamma \in \Gamma_{a}$, $\overline{C}(\gamma) \geq c_{R}(\lambda)$.
Similar to the proof of Lemma \ref{chap5:lemma:minach_servicecost}, it can be shown that there exists a sequence of policies $\gamma_{\epsilon} \in \Gamma_{a}$ such that $\overline{C}(\gamma_{\epsilon}) \downarrow c_{R}(\lambda)$.
Therefore, we obtain that $\inf_{\gamma \in \Gamma_{a}} \overline{C}(\gamma) = c_{R}(\lambda)$.
In the next section, we obtain an asymptotic characterization of $Q^*(c_{c})$ in the asymptotic regime $\Re$ as $c_{c} \downarrow c_{R}(\lambda)$.

\subsection{Asymptotic lower bound}
We note that this problem can be considered as a special case of the tradeoff problem considered by Berry and Gallager \cite{berry}, with the fade state taking only a single value.
The contribution in this section is a step towards an alternative explanation for the Berry-Gallager lower bound, but with the extra property RG3 for an admissible policy.
We note that asymptotic bounds for any order-optimal policy can be obtained under the additional assumption RG3.
In Chapter 5, we present a lower bound for multiple fade states.
As in \cite{berry} and as for the integer valued service case, we obtain lower bounds on the minimum average queue length for any sequence of admissible policies $\gamma_{k}$ such that $\overline{C}(\gamma_{k}) - c_{R}(\lambda) = V_{k} \downarrow 0$.
The asymptotic lower bound is obtained via a lower bound on the stationary probability for a Markov chain which evolves on $\mathbb{R}_{+}$.
The lower bound is similar to Lemma \ref{chap5:lemma:dtmc_stat_prob}-TAIL-PROB-STATE-DEP-1 but for a Markov chain with state space $\mathbb{R}_{+}$.
\begin{lemma}
  Let $(Q[m])$ be the queue length evolution process for an admissible policy $\gamma$.
  Let $\epsilon_{a}$ and $\delta_{a}$ be as in assumption RA1.
  Suppose there exists a $q_{d}$ such that \[ \forall q \in [0,q_{d}], \mathbb{E}\bras{Q[m + 1] - Q[m] \middle \vert Q[m] = q} \geq -d,\] where $d$ is positive. Then for any $q_{1}$, $k \geq 0$, $\Delta > 0$, $\delta > 0$, $\Delta + \delta < \delta_{a}$, and $0 \leq q_{1} + k\Delta \leq q_{d}$, we have 
  \small
  \begin{eqnarray*}
    Pr\brac{Q \geq q_{1} + k\Delta} \geq \brap{\depn}^{k} Pr\brac{Q \geq q_{1}} + \brap{1 - \brap{\depn}^{k}} \bras{ Pr\brac{Q \geq q_{d}} - \frac{1}{d}\int_{q_{d}}^{\infty} (\mathbb{E}S(q) - \lambda)d\pi(q)}.
  \end{eqnarray*}
  \normalsize
  \label{chap5:lemma:realq_dtmc_stat_prob}
\end{lemma}
The proof is presented in Appendix \ref{chap5:app:realq_dtmc_stat_prob}.
We note that the development of the above lower bound on $Pr\brac{Q \geq q}$ is an extension of the geometric lower bound on stationary probability for countable space DTMCs available in Bertsimas et al. \cite{gamarnik} and \cite{gamarnik_1}, to the case of DTMCs on $\mathbb{R}_+$ with state dependent drift.

Using the above result, we derive the following asymptotic lower bound.
Similar to assumption C2 in Chapter 3 and as in \cite{berry}, we assume that the second derivative of $c_{R}(s)$ is positive at $s = \lambda$.
\begin{proposition}
  For any sequence of policies $\gamma_{k} \in \Gamma_{a}$ with $\overline{C}(\gamma_{k}) - c_{R}(\lambda) = V_{k} \downarrow 0$, $\overline{Q}(\gamma_{k}) = \Omega\nfrac{1}{\sqrt{V_{k}}}$.
  \label{chap5:prop:realvalued_evolution_lb}
\end{proposition}

\begin{proof}
  Consider a particular policy $\gamma$ in the sequence $\gamma_{k}$ with $V_{k} = V$.
  Let $q_{d} = \sup\brac{ q : \Exp S(q) \leq \lambda + \epsilon_{V}}$, where $\epsilon_{V}$ will be chosen later.
  Suppose $q_{d}$ is finite.
  From the admissibility of $\gamma$, we have that $\forall q \in [0, q_{d}], \Exp S(q) \leq \lambda + \epsilon_{V}$.
  Using $d = \epsilon_{V}$ in Lemma \ref{chap5:lemma:realq_dtmc_stat_prob}, we have for a $\bar{q} = k\Delta \leq q_{d}$, $k \geq 0$,
  \begin{eqnarray*}
    Pr\brac{Q \geq \bar{q}} \geq \brap{\dep}^{k} + {1 - \brap{\dep}^{k}} \bras{Pr\brac{Q \geq q_{d}} - \frac{1}{\epsilon_{V}}\int_{q_{d}}^{\infty} (\mathbb{E}S(q) - \lambda)d\pi(q)}.
  \end{eqnarray*}
  Or we have that
  \begin{eqnarray}
    Pr\brac{Q < \bar{q}} \leq \brap{1 - \brap{\dep}^{k}}\bras{1 + \frac{1}{\epsilon_{V}} \int_{q_{d}}^{\infty} (\mathbb{E}S(q) - \lambda)d\pi(q) - Pr\brac{Q \geq q_{d}} }.
    \label{chap5:eq:realvalevol}
  \end{eqnarray}
  We note that for $q \in [q_{d},\infty)$, $\mathbb{E}S(q) - \lambda \geq \epsilon_{V}$. 
  For brevity, we denote $\frac{1}{\epsilon_{V}}\int_{q_{d}}^{\infty} (\mathbb{E}S(q) - \lambda)d\pi(q)$ by $D_{t}$. 
  Now we note that for the policy $\gamma$, $\mathbb{E}_{\pi}c_{R}(S(Q)) - c_{R}(\lambda) = V$.
  Define $l(s)$ as the tangent to the curve $c_{R}(s)$ at $(\lambda, c_{R}(\lambda))$.
  Then we have that $\mathbb{E}_{\pi} \Exp_{S|Q} [c_{R}(S(Q)) - l(S(Q))] = V$.
  Now as $c_{R}(s)$ is convex and $l(s)$ is linear, using Jensen's inequality we have that $\mathbb{E}_{\pi}[c_{R}(\mathbb{E}S(Q)) - l(\mathbb{E}S(Q))] \leq V$.
  As in \cite[step (41)]{berry}, $c_{R}(s) - l(s) = G(s - \lambda)$ where $G(x)$ is a strictly convex function with $G(0) = 0$, $G'(0) = 0$, and $G''(0) > 0$.
  Thus we have that $\mathbb{E}_{\pi} G(\mathbb{E}S(Q) - \lambda) \leq V$.
  Using the sequence of steps (45), (46), (47), and (48) of Berry and Gallager \cite{berry}, we obtain that
  \begin{eqnarray*}
    \bras{\int_{q_{d}}^{\infty} (\mathbb{E}S(q) - \lambda) d\pi(q)}^{2} \leq \frac{V}{a_{1}},
  \end{eqnarray*}
  where $a_{1} > 0$ is such that $G(x) \geq a_{1} x^{2}$ (see Proposition \ref{chap4:app:prop:quadlowerbound}), for $x \in [-\lambda, S_{max} - \lambda]$.
  We note that then $D_{t} \leq \frac{1}{\epsilon_{V}} \sqrt{\frac{V}{a_{1}}}$.
  Choosing $\epsilon_{V} = 4\sqrt{\frac{V}{a_{1}}}$ we obtain that $D_{t} \leq \frac{1}{4}$.

  We note that $\overline{Q}(\gamma) \geq \frac{\bar{q}}{2}$, where $\bar{q} = \sup\brac{q : Pr\brac{Q < q} \leq \frac{1}{2}}$.
  Using the upper bound \eqref{chap5:eq:realvalevol} and non-negativity of $Pr\brac{Q \geq q_{d}}$, if $\bar{q}_{1} = k_{1}\Delta$ where $k_{1}$ is the largest integer such that
  \begin{eqnarray*}
    \brap{1 - \brap{\dep}^{k_{1}}}\bras{1 + D_{t}} \leq \frac{1}{2},
  \end{eqnarray*}
  then $\bar{q}_{1} \leq \bar{q}$.
  Therefore, $k_{1}$ is such that
  \begin{eqnarray*}
    \frac{1 + 2D_{t}}{2 + 2D_{t}} \leq \nfrac{\delta\epsilon_{a}}{\delta\epsilon_{a} + \epsilon_{V}}^{k_{1}},\\
    \brap{1 + \frac{\epsilon_{V}}{\delta\epsilon_{a}}}^{k_{1}} \leq \frac{2 + 2D_{t}}{1 + 2D_{t}}
  \end{eqnarray*}
  Let $k_{2}$ be the largest integer such that 
  \begin{eqnarray}
    \brap{1 + \frac{\epsilon_{V}}{\delta\epsilon_{a}}}^{k_{2}} \leq \frac{2}{1 + 2D_{t}}.
    \label{chap5:eq:real5}
  \end{eqnarray}
  Then $k_{2} \leq k_{1}$.
  We note that even if $q_{d}$ is infinite, $ k_{2} \Delta$ is a lower bound to $\bar{q}$, since $D_{t}$ is $0$ in that case.
  The rest of the proof holds irrespective of whether $q_{d}$ is finite or infinite.

  Then, from \eqref{chap5:eq:real5} and using the upper bound $\frac{1}{4}$ on $D_{t}$, if $k_{3}$ is the largest integer such that 
  \begin{eqnarray*}
    \brap{1 + \frac{\epsilon_{V}}{\delta\epsilon_{a}}}^{k_{3}} \leq \frac{2}{1 + \frac{1}{2}},
  \end{eqnarray*}
  then $k_{3} \leq k_{2}$.
  We obtain that $k_{3}$ is at least 
  \begin{eqnarray*}
    \log_{\brap{1 + \frac{\epsilon_{V}}{\delta\epsilon_{a}}}} \brap{\frac{4}{3}} - 1.
  \end{eqnarray*}
  Since $\overline{Q}(\gamma) \geq \frac{\bar{q}}{2} \geq \frac{\Delta k_{1}}{2} \geq \frac{\Delta k_{2}}{2} \geq \frac{\Delta k_{3}}{2}$, we have that $\overline{Q}(\gamma) \geq \frac{\Delta}{2} \brap{\log_{\brap{1 + \frac{\epsilon_{V}}{\delta\epsilon_{a}}}} \brap{\frac{4}{3}} - 1}$.
  Since $\log\brap{1 + \frac{\epsilon_{V}}{\delta\epsilon_{a}}} = \Theta\brap{\sqrt{V}}$, we have that for the sequence $\gamma_{k}$ as $V_{k} \downarrow 0$, $\overline{Q}(\gamma_{k}) = \Omega\nfrac{1}{\sqrt{V_{k}}}$.
\end{proof}

We note that the problem considered here is a special case of the Berry-Gallager problem (with a single fade state) with admissible policies.
The upper bounds for the average queue length and average service cost for the TOCA policy from Neely \cite{neely_mac}, can be used to obtain asymptotic upper bounds for this problem.
However, we note that these bounds hold only for the problem \eqref{chap5:eq:inittradeoffprob}, since the sequence of TOCA policies is not admissible.
Therefore, as in Lemma \ref{chap5:lemma:case2upperbound}, we present a sequence of admissible policies which achieve the above asymptotic growth rate up to a logarithmic factor.
\begin{lemma}
  Let a policy $\gamma$ be defined as follows.
  At a queue length $q$, $\gamma$ serves a batch size $\min(q, \tilde{s}(q))$, where 
  \begin{equation*}
    \tilde{s}(q) = 
    \begin{cases}
      \lambda - \epsilon_{V}, \text{ for } 0 \leq q \leq q_{v}, \\
      \lambda + \epsilon_{V}, \text{ for } q_{v} < q \leq 2q_{v}, \\
      \lambda + \epsilon, \text{ for } 2q_{v} < q.
    \end{cases}
  \end{equation*}
  where $q_{v} > 0$ and $\lambda + \epsilon \leq S_{max}$.
  We obtain a sequence of policies $\gamma_{k}$ by choosing $\epsilon_{V}$ and $q_{v}$ from the sequence $\epsilon_{V_{k}}$ and $q_{v_{k}}$ defined as follows.
  Let $\omega_{k} = \sqrt{V_{k}}$, where $V_{k} \downarrow 0$.
  Let $\epsilon_{V_{k}} = \omega_{k} A_{max}^{2} e^{\omega A_{max}}$ and $q_{v_{k}} = \frac{1}{\omega_{k}} \log\nfrac{1}{ \epsilon_{V_{k}}^{3}}$.
  Then we have that $\gamma_{k}$ is a sequence of admissible policies, such that $\overline{C}(\gamma_{k}) - c(\lambda) = \mathcal{O}(V_{k})$ and $\Qgk = \mathcal{O}\brap{\frac{1}{\sqrt{V_{k}}}\log\nfrac{1}{V_{k}}}$.
  \label{chap5:lemma:rmodelupperbound}
\end{lemma}
The proof of this lemma is given in Appendix \ref{chap5:app:rmodelupperbound} and is motivated by and borrows ideas from the proof of the asymptotic upper bound for TOCA policies in \cite{neely_mac}.

The TOCA policy is the same as that in Remark \ref{chap5:remark:toca} except that for every $m \geq 1$, we have that the service batch size $s_{TOCA}[m]$ is chosen as
\begin{eqnarray*}
  s_{TOCA}[m] & = & \min\brap{\argmin_{s \in [0,S_{max}]} \bigg \{\beta c(s) - {W}[m] s \bigg\}, Q[m - 1]}.
\end{eqnarray*}
Then, from \cite[Theorem 3 and Corollary 1]{neely_mac}, for the TOCA policy $\gamma$ as above, for $\beta > S_{max}$, $w = \frac{\epsilon}{\delta^{2}_{max}}e^{-\frac{\epsilon}{\delta_{max}}}$, $\epsilon = \frac{1}{\sqrt{\beta}}$, and $\tilde{q} = \frac{6}{w}\log\nfrac{1}{\epsilon}$, we have that
\begin{eqnarray*}
  \overline{Q}(\gamma) = \mathcal{O}\brap{\sqrt{\beta}\log\brap{\beta}}, \\
  \overline{C}(\gamma) = c(\lambda) + \mathcal{O}\nfrac{1}{\beta}.
\end{eqnarray*}
For a sequence of policies $\gamma_{k}$, generated by choosing $\beta_{k} = \frac{1}{V_{k}}$, for a sequence $V_{k} \downarrow 0$, we have that $\overline{Q}(\gamma_{k}) = \mathcal{O}\brap{\frac{1}{\sqrt{V_{k}}}\log\nfrac{1}{V_{k}}}$ and $\overline{C}(\gamma) = c(\lambda) + \mathcal{O}\brap{V_{k}}$.
The above bound is an upper bound for the optimal solution of \eqref{chap5:eq:inittradeoffprob}.
Furthermore, we note that the above upper bound is an upper bound to $Q^*(c_{c,k})$ for a sequence $c_{c,k}$ as in Remark \ref{chap5:remark:toca}.

\subsection{R-model with a piecewise linear cost function}
\label{chap5:eq:rmodel_piecewise_cost}
We note that R-model with a strictly convex $c_{R}(s)$ is usually used as an approximation to I-model.
Usually, the function $c_{R}(s)$ coincides with $c(s)$ for $s \in \brac{0, \dots, S_{max}}$.
But we find that there are differences in the asymptotic behaviour of $Q^*(c_{c})$ for I-model and R-model.
We note that $c_{R}(\lambda) \leq c(\lambda), \forall \lambda \in (0, S_{max})$ and $c_{R}(\lambda) < c(\lambda)$ for $\lambda \not \in \brac{0, \dots, S_{max} - 1}$.
Furthermore, R-model suggests that $Q^*(c_{c})$ increases to infinity for all $\lambda \in (0, S_{max})$ as $c_{c} \downarrow c_{R}(\lambda)$.
However, for Case 1, we see that $Q^*(c(\lambda))$ is finite (note that $c_{R}(\lambda) < c(\lambda)$ in this case).
For Case 3, with $c_{R}(\lambda) = c(\lambda)$ we have that $Q^*(c_{c}) = \Omega\nfrac{1}{\sqrt{c_{c} - c(\lambda)}}$ for the R-model, whereas for the I-model $Q^*(c_{c}) = \Omega\nfrac{1}{{c_{c} - c(\lambda)}}$.
So R-model with a strictly convex $c_{R}(s)$ overestimates the behaviour of $Q^*(c_{c})$ for Case 1 and underestimates $c(\lambda)$ and $Q^*(c_{c})$ for Cases 2 and 3.
In the following, we briefly outline a method to show that a better approximation for I-model, is R-model with a piecewise linear $c_{R}(s)$.
The service cost function $c_{R}(s)$ is chosen as the lower convex envelope of the service cost function $c(s), s \in \brac{0, \dots, S_{max}}$ for the I-model.
With this choice of $c_{R}(s)$, the asymptotic behaviour of I-model and its approximation, R-model, is the same.

We consider Case 2 first.
We define $s_{l}, s_{u}$, and the line $l(s)$ as in Section \ref{chap5:sec:asymp_analysis_tprob}.
Consider any sequence of admissible policies $\gamma_{k}$ with $\Cgk - c_{R}(\lambda) = V_{k} \downarrow 0$.
Then we have that $\Exp c_{R}(\Exp S(Q)) - c_{R}(\lambda) \leq V_{k}$.
For a particular policy $\gamma$ in the sequence, $0 < \epsilon < s_{l}$, and $q_{s} \Deq s_{l} - \epsilon$ we have that $Pr\brac{Q < q_{s}} \leq \frac{V}{m\epsilon}$ as in the proof of Lemma \ref{chap5:lemma:tprob_2}.
We assume that $Pr\brac{A[1] \leq \frac{\Delta}{2}} = \epsilon'_{a} > 0$, for some $0 < \Delta < \lambda$.
We note that for the R-model, the queue evolution is on $\sR$.
We discretize $\sR$ into a countable number of intervals $\brap{[0, \frac{\Delta}{2}), [\frac{\Delta}{2}, \Delta), \dots}$.
Then the proof of Lemma \ref{chap5:lemma:tprob_2} can be modified to show that $\Qgk = \Omega\brap{\log\nfrac{1}{V_{k}}}$.
A complete illustration of this proof technique is given in Lemma \ref{lemma:tradeoffutilitylb}.

For Case 3, we proceed as in the proof of Proposition \ref{chap5:prop:realvalued_evolution_lb} by defining $q_{d}$ to be $\sup\brac{q : \Exp S(q) \leq \lambda + \epsilon_{V}}$, where $\epsilon_{V}$ is a function of $V$ to be chosen in the following.
We recall that $D_{t} \Deq \frac{1}{\epsilon_{V}} \int_{q_{d}}^{\infty} \brap{\Exp S(q) - \lambda} d\pi(q)$.
For the policy $\gamma$ we have that
\begin{eqnarray*}
  \Exp \bras{c_{R}(\Exp S(Q)) - l(\Exp S(Q))} & \leq & V, \\
  \int_{q_{d}}^{\infty} \brap{c_{R}(\Exp S(q)) - l(\Exp S(q))} d\pi(q) & \leq & V, \text{ or,} \\
  \frac{1}{\epsilon_{V}} \int_{q_{d}}^{\infty} \brap{\Exp S(q) - \lambda} d\pi(q) & \leq & \frac{V}{m\epsilon_{V}},
\end{eqnarray*}
where $m$ is the tangent of angle made by the line passing through $(\lambda - 1, c_{R}(\lambda - 1))$ and $(\lambda, c_{R}(\lambda))$ with $l(s)$.
Now we choose $\epsilon_{V} = \frac{4V}{m}$ to obtain that $D_{t} \leq \frac{1}{4}$.
Then we proceed as in the proof of Proposition \ref{chap5:prop:realvalued_evolution_lb} to obtain that $\Qgk = \Omega\nfrac{1}{V_{k}}$.

We note that, by construction R-model has the service cost function $c_{R}(\lambda) = c(\lambda)$.
Furthermore, the asymptotic behaviour for R-model and I-model coincide for Cases 2 and 3.
For Case 1, it can be shown that $\overline{Q}(\gamma_{u})$ is finite for R-model, so that $Q^*(c_{R}(\lambda))$ is also finite.
However, we do not have asymptotic lower bounds for this case.

\section{Conclusions}
In this chapter, we have obtained an asymptotic characterization of the tradeoff curve $Q^*(c_{c})$ in the asymptotic regime $\Re$ for a discrete time queueing model (I-model).
This asymptotic characterization has been obtained using the insights obtained from the analysis of \INTMC\, in Chapter 3.
We also consider a real valued approximation (R-model) to I-model, and compare the asymptotic results which are obtained for R-model with that for I-model. 

For I-model we observe that the cost function $c(s)$ as a function of the average service rate $s$ is piecewise linear.
Then as for INTERVAL-$\mu$CHOICE-2-2, we have three cases.
For Case 2, motivated by INTERVAL-$\mu$CHOICE-2-2, we construct an upper bound to the stationary probability distribution for the queue length which is geometrically increasing, which leads to $\Theta\brap{\log\nfrac{1}{V}}$ asymptotic growth for $Q^*(c_{c})$ as $c_{c} \downarrow c(\lambda)$.
We note that this geometric upper bound on the stationary probability distribution can be obtained in general, even for Cases 1 and 3.
However, for Case 3, motivated by INTERVAL-$\mu$CHOICE-2-3, we expect that there is a set of queue lengths with high probability for which the average drift $\Exp S(q) - \lambda \downarrow 0$ as $V \downarrow 0$.
We then expect that the stationary probability of such queue lengths should be equal and $\mathcal{O}(V)$.
This intuition leads us to a refined bound on the stationary probability of the queue length, obtained by extending the bounds available in Bertsimas et al. \cite{gamarnik_1}, from which we obtained the $\Theta\brap{\frac{1}{V}}$ asymptotic growth for $Q^*(c_{c})$ as $V \downarrow 0$.

We note that Case 1 is similar to INTERVAL-$\mu$CHOICE-2-1, however we are unable to obtain asymptotic lower bounds except for the restricted case where $s_{u} = 1$ and for the set of non-idling admissible policies.
A direct translation of the ideas from INTERVAL-$\mu$CHOICE-2-1 is not possible, since for the discrete time model we cannot obtain a dominating policy $\gamma'$ as for INTERVAL-$\mu$CHOICE-2-1.

We note that R-model is similar to INTERVAL-$\mu$CHOICE-1, and as in Case 3 above, motivated by the observation that for INTERVAL-$\mu$CHOICE-1, the stationary probability of queue lengths occurring with high probability should be equal and $\mathcal{O}\brap{\sqrt{V}}$ as $V \downarrow 0$, we obtain a $\Omega\nfrac{1}{\sqrt{V}}$ asymptotic lower bound on $Q^*(c_{c})$.
We comparing the asymptotic behaviour of $Q^*(c_{c})$ between I-model and R-model in their respective asymptotic regimes in Section \ref{chap5:eq:rmodel_piecewise_cost}.
We observe that R-model with a strictly convex $c_{R}(s)$ overestimates the behaviour of $Q^*(c_{c})$ for Case 1 and underestimates $c(\lambda)$ and $Q^*(c_{c})$ for Cases 2 and 3.
Therefore, we conclude that a more appropriate real valued approximation to I-model should have a cost function $c_{R}(s)$ chosen as the piecewise linear lower convex envelope of $c(s)$.

Since in our approach, we obtain bounds on the stationary probability of the queue length, we are able to obtain asymptotic bounds on any sequence of order-optimal policies in the asymptotic regime $\Re$.
These bounds provide intuition for the design of buffer-partitioning policies and are presented in Section \ref{chap5:sec:aschar}.
We also obtain that the minimum average queue length is $\Omega\brap{\log\nfrac{1}{V}}$ for Cases 2 and 3, when the arrival process $A[m]$ is ergodic.

In \cite[Section 4.8]{neely}, it is observed that the \emph{drift plus penalty} algorithm idles for certain values of the queue length.
However, in Lemma \ref{chap5:prop:nonidling}, we have obtained that any optimal policy $\gamma^*_{\beta}$ should be non-idling.
Hence, the drift plus penalty algorithm has to be modified to be non-idling, for the models considered in this chapter.
So Lemma \ref{chap5:prop:nonidling} can be thought of as providing theoretical motivation for the \emph{place-holder} method in \cite[Section 4.8]{neely}.

In the rest of the thesis, we use the above results to obtain asymptotic characterizations of some resource tradeoff problems arising in point-to-point communication links.

\clearpage
\begin{subappendices}
\large{\textbf{Appendices}}
\addcontentsline{toc}{section}{Appendices}
\addtocontents{toc}{\protect\setcounter{tocdepth}{0}}

\normalsize

% \section{Proof of Lemma \ref{chap5:lemma:optimalsolution}}
% \label{chap5:app:optimalsolution}
% We first verify that the Hypothesis 1 of \cite{hernandez} holds in our case for both I-model and R-model.
% We note that the state spaces and action spaces for both I-model and R-model are Borel spaces.
% \begin{enumerate}
% \item{
% In Lemma \ref{chap5:lemma:minach_servicecost}, we show that there exists a policy $\gamma$ in $\Gamma$ which is feasible for \eqref{chap5:eq:inittradeoffprob} for any $c_{c} > c(\lambda)$. Hence \eqref{chap5:eq:inittradeoffprob} is \emph{consistent} for any $c_{c} > c(\lambda)$.}
% \item{We note that the function $f(q) = q$ is inf-compact, since for any $r$ the set $\brac{q: q \leq r}$ is compact.}
% \item{The function $c(s)$ is non-negative and continuous for both R-model (on $[0, S_{max}]$) and I-model (on $\brac{0, S_{max}}$).}
% \item{The transition probability kernel is weakly continuous for both I-model and R-model.}
% \end{enumerate}
% Hence from \cite[Theorem 1]{hernandez} we have that \eqref{chap5:eq:inittradeoffprob} has an optimal solution in $\Gamma$.

% Again by using Lemma \ref{chap5:lemma:minach_servicecost}, we have that there exists a stationary policy $\gamma$ such that $\overline{Q}(\gamma) < \infty$ for any $c_{c} > c(\lambda)$, furthermore $\gamma$ is a \emph{stable control} in the sense of \cite[Definition 3.8]{hernandez_2}.
% Hence we have verified \cite[Hypothesis 4.1]{hernandez_2} and we have there is a stationary policy $\gamma^*$ with a stationary distribution $\pi^*$ such that $\overline{Q}(\gamma^*) = \Exp_{\pi^*}Q < \infty$.
  
\section{Proof of Lemma \ref{chap5:lemma:statopt}}
\label{chap5:app:statopt}
Our approach is to verify the hypotheses of the single Theorem in Sennott \cite{sennott_mdp} by showing that the Assumptions (1), (2), and (3*) of Sennott \cite{sennott_mdp} are satisfied.
The assumptions in Sennott \cite{sennott_mdp} are as follows (note that the notation is as in \cite{sennott_mdp}):
\begin{enumerate}
\item For every state $i$ and discount factor $\alpha$, the optimal expected total discount cost $V_{\alpha}(i)$ is finite,
\item Let $h_{\alpha}(i) \stackrel{\Delta} = V_{\alpha}(i) - V_{\alpha}(0)$. There exists a non-negative $N$ such that $-N \leq h_{\alpha}(i)$, for all states $i$ and discount factors $\alpha$,
\item There exists non-negative $M_{i}$, such that $h_{\alpha}(i) \leq M_{i}$, for every state $i$ and discount factor $\alpha$. Let the transition probability under action $a$, from state $i$ to $j$ be $P_{i,j}(a)$. Then for all $i$, $\sum_{j} P_{i,j}(a(i)) M_{j} < \infty$ for an action $a(i)$ feasible in state $i$. 
\end{enumerate}
Assumption (3*) assumes that in addition, $\sum_{j} P_{i,j}(a) M_{j} < \infty$, for all $a$ feasible in state $i$.
Let $C(i,a)$ be the single stage cost at state $i$, when action $a$ is taken.
If Assumptions (1), (2) and (3*) hold, the Theorem \cite{sennott_mdp} states that:
\begin{theorem}
  There exists a constant $g$, which is independent of the state $i$, and a function $h(i)$ with $-N \leq h(i) \leq M_{i}$, such that
  \begin{equation*}
    g + h(i) = \min_{a} \brac{C(i,a) + \sum_{j} P_{i,j}(a) h(j) }, i \geq 0.
  \end{equation*}
  A policy $f$ that attains the minimum in the RHS of the above equation is average cost optimal, with optimal average cost $g$.
\end{theorem}
For proving that the Assumptions (1) and (3) are satisfied we use \cite[Proposition 5(i)]{sennott_mdp} which states that:
\begin{proposition}
  Assume that the Markov decision process has a stationary policy $f$ inducing an irreducible, ergodic Markov chain satisfying $\sum_{i} \pi_{i} C(i, f(i)) < \infty$ (\cite[Proposition 4, Condition (i)]{sennott_mdp}), where $\pi_{i}$ is the stationary probability for state $i$ under $f$. Then Assumptions (1) and (3) hold.
\end{proposition}
Consider the stationary deterministic policy $\gamma_{f}$ which uses a batch size $s(q) = \min(q, S_{max})$ when the queue length is $q$.
Then, from assumption A2 we have that from any $q > 0$ the state $0$ can be reached, since $Pr\brac{A[1] = 0} > 0$.
From state 0, any state $q > 0$ can be reached, which follows from assumptions A1 and A2.
Therefore $\gamma_{f}$ is irreducible.
We note that state $0$ is aperiodic, therefore the Markov chain under $\gamma_{f}$ is also aperiodic.

We now verify the drift condition (10.13) in \cite{meyn} by choosing $V(q) = \frac{q^{2}}{2(S_{max} - \lambda)}$ which is non-negative if $\lambda < S_{max}$.
Let $p_{q,q'} \stackrel{\Delta} = Pr\brac{Q[m + 1] = q'\middle\vert Q[m] = q, S[m + 1] = s(q)}$.
% Then we have that under policy $\gamma_{f}$, $\sum_{q'} p_{q,q'} V(q) < \infty$, as $A[1] \leq A_{max}$.
We have that 
\[\sum_{q'} p_{q,q'} (V(q') - V(q)) \leq -q + \frac{3S_{max}^{2} + \lambda^{2} + \sigma^{2}}{2(S_{max} - \lambda)}, \forall q \geq 0.\]
As the function $c(q) = q$ is near-monotone \cite{meyn} and $\frac{3S_{max}^{2} + \lambda^{2} + \sigma^{2}}{2(S_{max} - \lambda)} < \infty$ we have from \cite[Theorem 10.3]{meyn} that the Markov chain under the policy $\gamma_{f}$ is $c$-regular, implying that it is also positive recurrent with invariant distribution $\pi$.
Then under the same policy we have that $\Expp c_{\beta}(q,s) < \infty$ as $c_{\beta}(q,s) \leq q + \beta c(S_{max})$.
This implies that we have verified condition (i) of Proposition 4 of \cite{sennott_mdp} and therefore from \cite[Proposition 5]{sennott_mdp} the first and third assumptions hold.
We note that as $A[1] \leq A_{max}$, once Assumption (3) holds, Assumption (3*) is implied.

To verify Assumption (2) of \cite{sennott_mdp} it is sufficient to show that the optimal discounted cost $V_{\alpha}(q)$ is non-decreasing in $q$ for every discount factor $\alpha \in (0,1)$.
This proposition follows as a special case of \cite[Lemma C.1]{munish} and therefore we claim that $V_{\alpha}(q)$ is non-decreasing in $q$ without proof.

Then from \cite[Theorem]{sennott_mdp} there exists a stationary deterministic optimal policy $\gamma^*_{\beta}$ with optimal average cost $g^*_{\beta}$ satisfying the following ACOE:
\begin{equation*}
  g^*_{\beta} + J_{\beta}(q) = \min_{s \in \brac{0,\dots,\min(q,S_{max})}} \brac{ c_{\beta}(q, s) + \Exp J_{\beta}(q - s + A[1])},
\end{equation*}
with the stationary optimal policy $\gamma^*_{\beta}$ using a batch size $s^*_{\beta}(q)$ at queue length $q$ satisfying 
\begin{equation*}
  s^*_{\beta}(q) = \argmin_{s \in \brac{0,\dots,\min(q,S_{max})}} \brac{ c_{\beta}(q, s) + \Exp J_{\beta}(q - s + A[1])},
\end{equation*}
where $J_{\beta}(q)$ is the optimal relative value function.

\section{Proof of Lemma \ref{chap5:prop:nonidling}}
\label{chap5:app:nonidling}

\begin{proof}
  The proof proceeds by contradiction.
  Let $(Q[m], m \geq 0)$ be the evolution of the queue process under any stationary deterministic optimal policy $\gamma^*_{\beta}$ starting from initial state $q_{0}$.
  We assume that $\gamma^*_{\beta}$ is such that there exists a queue length $q_{1} > 0$, $q_{1} \in \mathcal{R_{\beta}}$, such that $s^*_{\beta}(q_{1}) = 0$.
  Then we present a perturbation to this policy, which leads to a history dependent policy $\tilde{\gamma}$, which has a smaller average queue length as well as average service cost, which contradicts the assumed optimality of $\gamma^*_{\beta}$.
  However, to compare $\overline{Q}(\tilde{\gamma})$ with $\overline{Q}(\gamma^*_{\beta})$ and $\overline{C}(\tilde{\gamma})$ with $\overline{C}(\gamma^*_{\beta})$, we identify a delayed renewal process $(X_{k})$ embedded in $(Q[m])$.
  For $\gamma^*_{\beta}$ and $\tilde{\gamma}$, we associate different reward processes with $(X_{k})$ and use the Renewal Reward theorem to obtain $\overline{Q}(\gamma^*_{\beta})$, $\overline{C}(\gamma^*_{\beta})$, $\overline{Q}(\tilde{\gamma})$, and $\overline{C}(\tilde{\gamma})$.

  Consider the evolution $(Q[m])$.
  The first cycle $X_{1}$ of the renewal process $(X_{k})$ is defined as follows :
  \begin{eqnarray*}
    X_{1} = \min\{m : Q[m] = q_{1} \}.
  \end{eqnarray*}
  At slot $X_{1}$, $Q[m]$ enters the recurrence class $\mathcal{R}_{\gamma^*_{\beta}}$ and does not leave $\mathcal{R}_{\gamma^*_{\beta}}$ again.
  We note that $X_{1}$ depends only on the initial state $q_{0}$ as $(Q[m])$ is Markov.
  If $T_{q}(q_{1})$ is the random time taken to hit $q_{1}$ starting from $q$, then $X_{1} = T_{q_{0}}(q_{1})$.
  Furthermore for $\gamma^*_{\beta}$, $X_{1} < \infty$.

  Now we note that by definition $Q[X_{1}] = q_{1}$ and $S[X_{1} + 1] = 0$.
  Let $T^S_{q_{1}}(1)$ be the smallest positive integer such that $S[X_{1} + T^S_{q_{1}}(1) + 1] > 0$, i.e., there is service of at least one customer in the $(X_{1} + T^S_{q_{1}}(1) + 1)^{th}$ slot.
  As $(Q[m])$ is Markov, the distribution of $T^S_{q_{1}}(1)$ given $Q[X_{1}] = q_{1}$, is independent of $(Q[m], m < X_{1})$.
  Let $\mathcal{S}_{1} \subset R_{\beta}$ be the set of states in which at least one customer is served.
  The distribution of $T^S_{q_{1}}(1)$ is the same as that of the smallest random time $T^S$, to hit $\mathcal{S}_{1}$, starting from $q_{1}$.
  Furthermore, we note that the queue length random variable $Q(1) = Q[X_{1} + T^S_{q_{1}}(1)]$ is distributed as the state $Q^S$ of the Markov chain at the random time $T^S$ when $\mathcal{S}_{1}$ is hit, starting from $q_{1}$.
  The second cycle $X_{2}$ is defined as
  \begin{equation*}
    X_{2} = T^S_{q_{1}}(1) + T_{Q(1)}(q_{1}).
  \end{equation*}
  Similarly, the $k^{th}$ cycle 
  \begin{equation*}
    X_{k} = T^S_{q_{1}}(k - 1) + T_{Q(k - 1)}(q_{1}),
  \end{equation*}
  where $Q(k - 1) = Q[X_{k - 1} + T^S_{q_{1}}(k - 1)]$.
  We note that $T^S_{q_{1}}(k), k \geq 1$ are all IID and have the same distribution as $T^S$.
  The random variables $Q(k), k \geq 1$ are all IID and have the same distribution as $Q^S$.
  Hence, the random variables $X_{k}, k \geq 1$ are independent and $X_{k}, k \geq 2$ are identically distributed.
  Thus $(X_{k})$ constitutes a delayed renewal process.
  From property O5, we have that $\mathbb{E}X_{k} < \infty$.

  We now associate a queue cost and service cost process with the renewal process.
  In each cycle $k$, we define the queue cost as the cumulative expected queue length :
  \begin{eqnarray*}
    C^Q_{k} = \sum_{m = 0}^{X_{k} - 1} \mathbb{E}\left[ Q[T_{k} + m] \middle | Q[T_{k}] = q_{1}\right],
  \end{eqnarray*}
  where $T_{k} = \sum_{j = 1}^{k - 1} X_{j}$.
  Using property O5, we have that $\forall k \geq 2$, $C^Q_{k} < \infty$ with probability $1$, as the optimal policy is $c_{\beta}$-regular \cite{meyn}.
  Furthermore in each cycle $k$, we define the service cost as the cumulative expected service cost :
  \begin{eqnarray*}
    C^C_{k} = \sum_{m = 1}^{X_{k}} \mathbb{E}\left[ c(S[T_{k} + m]) \middle | Q[T_{k}] = q_{1} \right].
  \end{eqnarray*}
  Again using property O5, we have that $\forall k \geq 2$, $C^C_{k} < \infty$ with probability $1$, as the optimal policy is $c_{\beta}$-regular.
  
  We note that the both $C^Q_{k}$ and $C^C_{k}$ are dependent only on $X_{k}$.
  Then from the renewal reward theorem we have that
  \begin{eqnarray*}
    \overline{Q}(\gamma^*_{\beta}) = \frac{\mathbb{E} C^Q_{2}}{\mathbb{E}X_{2}}, \\
    \overline{S}(\gamma^*_{\beta}) = \frac{\mathbb{E} C^C_{2}}{\mathbb{E}X_{2}}.
  \end{eqnarray*}

  In the following, we perturb $\gamma^*_{\beta}$ to obtain the policy $\tilde{\gamma}$.
  The perturbation, as well as the renewal cycle embedded in $Q[m]$ are illustrated in Figure \ref{chap5:fig:nonidlingproof}.

  \begin{figure}
    \centering
    \includegraphics[width=120mm,height=90mm]{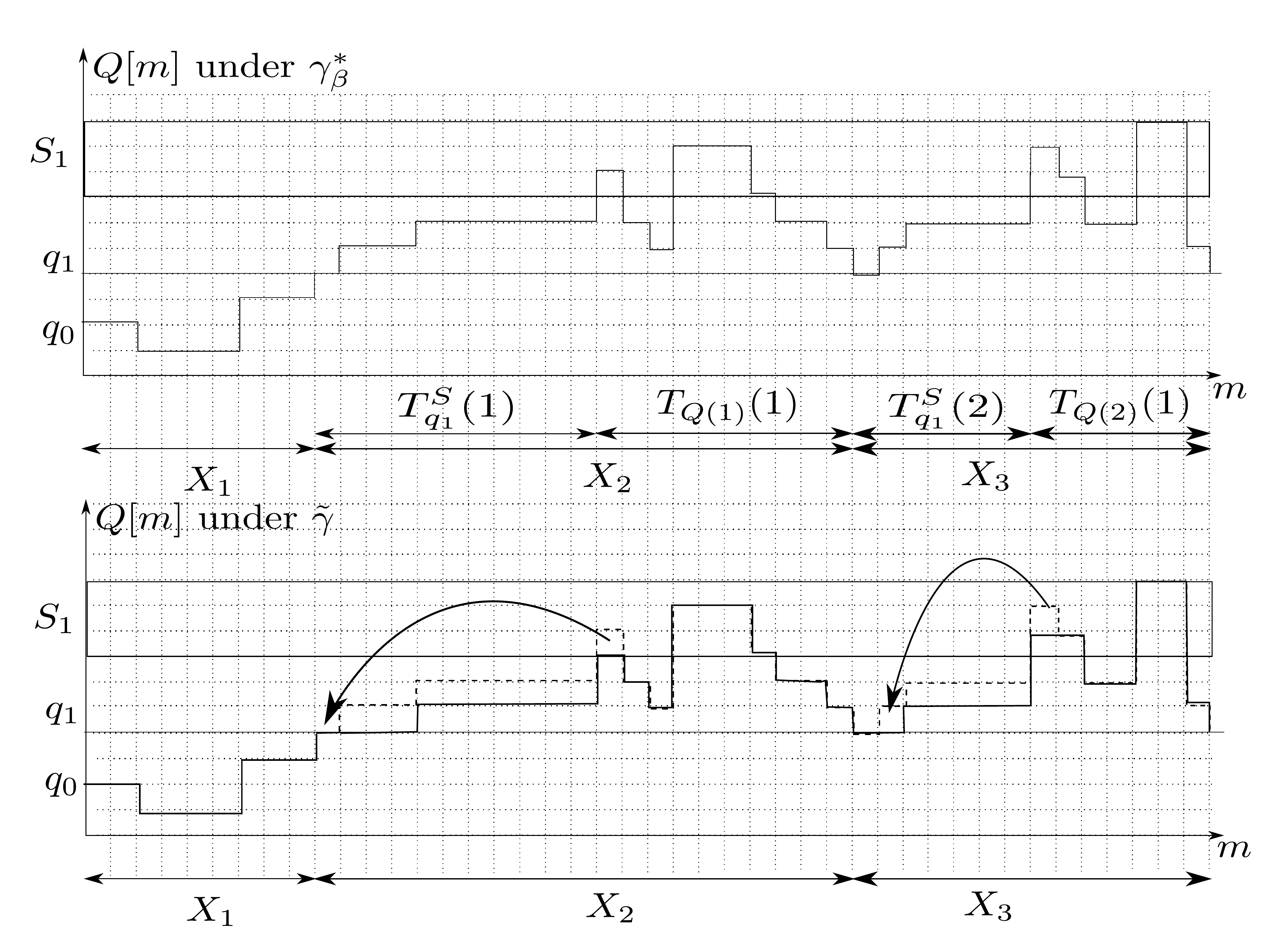}
    \caption{Illustration of the queue evolution under policy $\gamma^*_{\beta}$ and its perturbation $\tilde{\gamma}$. The first three renewal cycles $X_{1}, X_{2},$ and $X_{3}$ are also shown.}
    \label{chap5:fig:nonidlingproof}
  \end{figure}
  
  We note that, given $(S[1], S[2], \dots)$, $\overline{Q}(\gamma)$ and $\overline{C}(\gamma)$ do not depend on the order of service for customers.
  Therefore, we can assume that at least one customer, chosen to be served at $T_{k + 1} + T^S_{q_{1}}(k)$ was present in the queue at $T_{k + 1}$.
  Also, from the definition of $T^S_{q_{1}}(k)$, the system is idle in the slots $\{T_{k + 1}, \dots, T_{k + 1} + T^S_{q_{1}}(k)\}$.
  The perturbed policy $\tilde{\gamma}$ advances the service of one customer served in the $\brap{T_{k} + T^S_{q_{1}}(k) + 1}^{th}$ slot to the $(T_{k} + 1)^{th}$ slot, for every cycle $k$.
  The policy does not change the service batch size used at any other slot.
  To implement $\tilde{\gamma}$, the evolution of the queue under the unperturbed $\gamma^*_{\beta}$ policy is simulated for the same arrival process, in order to ascertain the slots at which the service batch size is to be changed.
  The policy $\tilde{\gamma}$ serves one customer at the $(T_{k+1} + 1)^{th}$ slot, and serves one customer less at the  $\brap{T_{k + 1} + T^S_{q_{1}}(k) + 1}^{th}$ slot.
  To obtain $\overline{Q}(\tilde{\gamma})$ and $\overline{C}(\tilde{\gamma})$, we use the same renewal process $(X_{k})$, but associate different per-cycle queue and service cost costs.
  In cycle $k \geq 2$, the new per-cycle queue cost is the cumulative expected queue cost 
  \begin{eqnarray*}
    \tilde{C^Q_{k}} = q_{1} + \sum_{m = 1}^{X_{k} - 1} \mathbb{E}\left[ Q[T_{k} + m] - 1 | Q[T_{k}] = q_{1}\right].
  \end{eqnarray*}
  As $\tilde{C^Q_{k}} < C^Q_{k}$ using the renewal reward theorem we obtain that $\overline{Q}(\tilde{\gamma}) < \overline{Q}(\gamma^*)$.
  
  For $\tilde{\gamma}$, the cumulative expected service cost for a cycle $k \geq 2$, 
  \begin{eqnarray*}
    \tilde{C^C_{k}} = c(1) + \sum_{m = 2, m \neq T_{k} + T^S_{q_{1}}(k - 1) + 1}^{X_{k}} \mathbb{E}\left[ c(S[T_{k} + m]) | Q[T_{k}] = q_{1} \right] + \mathbb{E}\left[ c(S[T_{k} + T^S_{q_{1}}(k - 1) + 1] | Q[T_{k}] = q_{1} \right].
  \end{eqnarray*}
  The difference $C^C_{k} - \tilde{C^C_{k}}$ is
  \begin{eqnarray*}
    c(0) + c(S[T_{k} + T^S_{q_{1}}(k - 1) + 1]) - c(1) - c(S[T_{k} + T^S_{q_{1}}(k - 1) + 1] - 1).
  \end{eqnarray*}
  From convexity for any $s \geq 1$, $c(s) - c(s - 1) \geq c(1) - c(0)$.
  Therefore, $\tilde{C^C_{k}} \leq C^C_{k}$ and, using the renewal reward theorem for $\tilde{\gamma}$, we obtain that $\overline{C}(\tilde{\gamma}) \leq \overline{C}(\gamma^*_{\beta})$.
  Therefore $\gamma^*_{\beta}$ cannot be optimal.
  Hence, $\forall q \in \mathcal{R}_{\gamma^*_{\beta}}$ such that $q > 0$, $s^*_{\beta}(q) > 0$.
\end{proof}

\section{Proof of Lemma \ref{chap5:lemma:minach_servicecost}}
\label{chap5:app:minach_servicecost}

\begin{proof}
  \emph{Case 1 :}
  We first consider the case when $s_{l} = 0$ and $s_{l} < \lambda < s_{u}$.
  We note that the stationary deterministic policy, $s(q) = \min(q, s_{u})$ achieves the minimum average service cost $c(\lambda)$. 
  Furthermore, the above policy has finite average queue length.

  \emph{Case 2 :}
  We now consider the case when $s_{l} \geq 1$ and $s_{l} < \lambda < s_{u}$.
  Let $\alpha \in (0,1)$ be such that $\alpha s_{l} + (1 - \alpha)s_{u} = \lambda + \epsilon$, where $\epsilon > 0$ and $\lambda + \epsilon < s_{u}$.
  Let $S_{\alpha,m}$ be a sequence of IID random variables with distribution
  \begin{equation*}
    S_{\alpha,m} \sim 
    \begin{cases}
      s_{l} \text{ w.p. } \alpha,\\
      s_{u} \text{ w.p. } 1 - \alpha.
    \end{cases}
  \end{equation*}
  Consider a stationary policy $S(Q[m]) = \min(Q[m], S_{\alpha,m})$.
  We note that as $c(.)$ is monotonically non-decreasing,
  \begin{equation*}
    \mathbb{E}_{\pi}c(S(Q)) \leq \mathbb{E}_{\pi}c(S_{\alpha,1}) \leq c(\lambda + \epsilon) = c(\lambda) + m\epsilon,
  \end{equation*}
  where $m = \frac{c(s_{u}) - c(s_{l})}{s_{u} - s_{l}}$.
  On squaring both sides of the evolution equation \eqref{chap5:eq:evolution} and taking expectations with respect to the stationary distribution, we have that
  \small
  \begin{eqnarray*}
    \mathbb{E}_{\pi} Q[m + 1]^{2} & = & \mathbb{E}_{\pi} \left[ Q[m]^{2} + S(Q[m])^{2} + A[m + 1]^{2} - 2S(Q[m])Q[m] - 2S(Q[m])A[m + 1] + 2Q[m]A[m + 1]\right] \\
    \text{which implies that }& & 2\mathbb{E}_{\pi}\left[ Q[m]S(Q[m]) - Q[m]A[m + 1]\right] \leq \mathbb{E}\bras{S_{\alpha,1}^{2}} + \mathbb{E}\bras{A[1]^{2}}.
  \end{eqnarray*}
  \normalsize
  Let $\Delta_{m} = S_{\alpha,m} - S(Q[m])$, we note that $\Delta_{m} \leq s_{u}$.
  Furthermore $\Delta_{m} = 0$ if $Q[m] > s_{u}$.
  Therefore $Q[m]S(Q[m]) = Q[m](S_{\alpha,m} - \Delta_{m}) \geq Q[m](S_{\alpha,m}) - s_{u}^{2}$.
  Hence we have that 
  \begin{eqnarray*}
    2\mathbb{E}_{\pi}\left[ Q[m]S(Q[m]) - Q[m]A[m + 1]\right] \leq \mathbb{E}S_{\alpha,1}^{2} + \mathbb{E}A[1]^{2}, \\
    2\mathbb{E}_{\pi}\left[ Q[m](S_{\alpha,m} - A[m + 1]) \right] - 2 s_{u}^{2} \leq \mathbb{E}S_{\alpha,1}^{2} + \mathbb{E}A[1]^{2}, \\
    \text{or as } m \rightarrow \infty,  2\epsilon \mathbb{E}_{\pi}Q \leq 2 s_{u}^{2} + \mathbb{E}S_{\alpha,1}^{2} + \mathbb{E}A[1]^{2}, \\ 
    \mathbb{E}_{\pi}Q \leq \frac{1}{2\epsilon}\left[ 2 s_{u}^{2} + \mathbb{E}S_{\alpha,1}^{2} + \mathbb{E}A[1]^{2}\right].
  \end{eqnarray*}
  Since $\Expp Q$ is thus finite, we note that the above policy is admissible.
  By choosing $\epsilon$ from a sequence of $\epsilon_{k} \downarrow 0$ as $k \uparrow \infty$, we obtain a sequence of policies $\gamma_{\epsilon_{k}} \in \Gamma_{a}$ such that $\overline{C}(\gamma_{\epsilon_{k}}) - c(\lambda) \leq m\epsilon_{k}$ and $\overline{Q}(\gamma_{\epsilon_{k}}) = \mathcal{O}\nfrac{1}{\epsilon_{k}}$.
  Therefore $\lim_{k \uparrow \infty} \overline{C}(\gamma_{\epsilon_{k}}) = c(\lambda)$.

  \emph{Case 3 :}
  We next consider the case when $1 \leq s_{l} = \lambda = s_{u} < S_{max}$.
  Let $\alpha \in (0,1)$ be such that $\alpha \lambda + (1 - \alpha) (\lambda + 1) = \lambda + \epsilon$, where $0 < \epsilon < 1$.
  Again we let $S_{\alpha,m}$ be a sequence of IID random variables with distribution
  \begin{equation}
    S_{\alpha,m} \sim 
    \begin{cases}
      \lambda \text{ w.p. } \alpha, \\
      \lambda + 1 \text{ w.p. } 1 - \alpha.
    \end{cases}
  \end{equation}
  Consider a stationary policy $S(Q[m]) = \min(Q[m], S_{\alpha,m})$.
  The rest of the proof for Case 3 is similar to that of Case 2, and we obtain that there exists a sequence of policies $\gamma_{\epsilon_{k}} \in \Gamma_{a}$ such that $\lim_{\epsilon_{k} \downarrow 0} \overline{C}(\gamma_{\epsilon_{k}}) = c(\lambda)$.
  We note that for a sequence of $\epsilon_{k} \downarrow 0$, we obtain a sequence of policies $\gamma_{\epsilon_{k}} \in \Gamma_{a}$ such that $\overline{C}(\gamma_{\epsilon_{k}}) - c(\lambda) \leq m\epsilon_{k}$ and $\overline{Q}(\gamma_{\epsilon_{k}}) = \mathcal{O}\nfrac{1}{\epsilon_{k}}$.
  For both cases 2 and 3, since $\overline{C}(\gamma_{\epsilon_{k}}) - c(\lambda) \leq m\epsilon_{k}$, by redefining $\epsilon_{k}$ we can obtain a sequence of policies $\gamma_{\epsilon_{k}}$ such that $\overline{C}(\gamma_{\epsilon_{k}}) - c(\lambda) = \epsilon_{k}$ and $\overline{Q}(\gamma_{\epsilon_{k}}) = \mathcal{O}\nfrac{1}{\epsilon_{k}}$.
\end{proof}

\section{Proof of Lemma \ref{chap5:lemma:barq_pi0_relation}}
\label{chap5:app:barq_pi0_relation}

\begin{proof}
  For a policy $\gamma \in \Gamma_{a}$, let $q_{s} \stackrel{\Delta} = \inf\brac{ q : \Exp S(q) \geq s_{l} - \epsilon }$ for a fixed positive $\epsilon < s_{l}$.
  Then we note that from assumption G3, for $q \geq q_{s}$, $\Exp S(q) \geq s_{l} - \epsilon$.
  In the following, for $q \geq q_{s}$, we find a lower bound on the probability that at least one customer is served, i.e., a lower bound on $Pr\brac{S(q) > 0}$, given $q \geq q_{s}$.
  We note that for $q \geq q_{s}$.
  \begin{eqnarray}
    \Exp S(q) & \geq & s_{l} - \epsilon, \nonumber \\
    \sum_{s = 0}^{S_{max}} s  Pr\brac{S(q) = s} & \geq & s_{l} - \epsilon, \nonumber \\
    S_{max} Pr\brac{S(q) > 0} & \geq & s_{l} - \epsilon, \text{ or}, \nonumber \\
    Pr\brac{S(q) > 0} & \geq & \frac{s_{l} - \epsilon}{S_{max}}.
    \label{chap5:app:barq0}
  \end{eqnarray}
  
  For the policy $\gamma$, we have that $\pi = \pi \mathbb{P}$ where $\mathbb{P}$ is the transition probability matrix of the Markov chain under policy $\gamma$, i.e. $\mathbb{P}_{q_{1},q_{2}} = Pr\brac{Q[m + 1] = q_{2}|Q[m] = q_{1}}$.
  We note that for a $q \geq q_{s}$, from \eqref{chap5:app:barq0}, we have that there is a positive probability of reaching a state less than $q$, starting from $q$ in one step, i.e.
  \begin{equation*}
    P(Q[m + 1] < q |Q[m] = q) \geq \nfrac{s_{l} - \epsilon}{S_{max}}Pr\brac{A[1] = 0}.
  \end{equation*}
  Let $\rho_{d} \stackrel{\Delta} = \nfrac{s_{l} - \epsilon}{S_{max}}Pr\brac{A[1] = 0}$.
  For $q \geq q_{s}$, from $\pi = \pi \mathbb{P}$, we have
  \begin{eqnarray*}
    \sum_{q' = 0}^{q - 1} \pi(q') & \geq & \pi(q) \sum_{q' = 0}^{q - 1} \mathbb{P}_{q, q'} \geq \pi(q) \rho_{d}.
  \end{eqnarray*}
  For $q = q_{s}$, we have
  \begin{eqnarray*}
    Pr\brac{Q < q_{s}} & \geq & \pi(q_{s}) \rho_{d}.
  \end{eqnarray*}
  For $q = q_{s} + 1$, we have
  \begin{eqnarray}
    Pr\brac{Q < q_{s}} + \pi(q_{s}) & \geq & \pi(q_{s} + 1) \rho_{d}, \nonumber \\
    \text{or } Pr\brac{Q < q_{s}}\brap{1 + \frac{1}{\rho_{d}}} & \geq & \pi(q_{s} + 1) \rho_{d}.
  \end{eqnarray}
  Proceeding similarly, we obtain that for $q = q_{s} + k, k \geq 0$
  \begin{equation*}
    \pi(q) \leq Pr\brac{Q < q_{s}} \frac{\left(1 + \frac{1}{\rho_{d}}\right)^{k}}{\rho_{d}} = Pr\brac{Q < q_{s}} \frac{\rho^{k}}{\rho_{d}},
  \end{equation*}
  where $\rho = 1 + \frac{1}{\rho_{d}} > 1$.

  Recall that in Chapter 3, we obtained a lower bound $\frac{\bar{q}}{2}$ on the average queue length for a policy $\gamma$, where $\bar{q}$ was such that $Pr\brac{Q \leq \bar{q}} \leq \frac{1}{2}$.
  We use the same idea here.
  Let $\bar{q} = \sup\{q : \sum_{q' = 0}^{q} \pi(q') \leq \frac{1}{2}\}$.
  Suppose $Pr\brac{Q < q_{s}}\brap{1 + \frac{\rho}{\rho_{d}}} < \frac{1}{2}$.
  Let $\bar{q}_{1}$ be the largest integer such that
  \begin{equation}
    Pr\brac{Q < q_{s}} + Pr\brac{Q < q_{s}}\sum_{q = q_{s}}^{\bar{q}_{1}} \frac{\rho^{q - q_{s}}}{\rho_{d}} \leq \frac{1}{2}.
    \label{chap5:eq:barpiq1}
  \end{equation}
  Then $\bar{q}_{1} \leq \bar{q}$.
  We note that \eqref{chap5:eq:barpiq1} is equivalent to finding the largest $\bar{q}_{1}$ such that
  \begin{eqnarray*}
    Pr\brac{Q < q_{s}}\left[ 1 + \frac{1}{\rho_{d}} \frac{\rho^{\bar{q}_{1} - q_{s} + 1} - 1}{\rho - 1} \right] \leq \frac{1}{2}, \\
    \text{or } \bar{q}_{1} \leq \log_{\rho} \left[ 1 + \rho_{d}\brap{\rho - 1} \left( \frac{1}{2Pr\brac{Q < q_{s}}} - 1 \right)\right].
  \end{eqnarray*}
  Hence we obtain that the $\bar{q}_{1}$ is at least
  \begin{equation*}
    \log_{\rho} \left[ \frac{1}{2Pr\brac{Q < q_{s}}}\right] - 1.
  \end{equation*}
  Since $\overline{Q}(\gamma) \geq \frac{\bar{q}}{2} \geq \frac{\bar{q}_{1}}{2}$, we have that
  \begin{equation*}
    \overline{Q}(\gamma) \geq \frac{1}{2} \left[ \log_{\rho} \left[ \frac{1}{2Pr\brac{Q < q_{s}}}\right] - 1 \right].
  \end{equation*}
\end{proof}

\section{Proof of Lemma \ref{chap5:lemma:dtmc_stat_prob}}
\label{chap5:app:dtmc_stat_prob}

\emph{TAIL-PROB :}
\begin{proof}
  The proof of TAIL-PROB follows from that of Lemma 2 and Theorem 3 (2) of Bertsimas et al. \cite{gamarnik}, and is presented here for completeness.
  Let us define $\widehat{Q}[m] = \max(q_{1}, Q[m])$ and $\widehat{Q} = \max(q_{1}, Q)$.
  Then we note that as $q_{1}$ is finite, $\Expp \widehat{Q} < \infty$.
  Hence, $\sum_{q = 0}^{\infty} \pi(q) \Exp \bras{ \widehat{Q}[m + 1] - \widehat{Q}[m] \middle |Q[m] = q} = 0$.
  We split this sum into three parts leading to :
  \begin{eqnarray}
    0 & = & \sum_{q = 0}^{q_{1} - 1} \pi(q) \mathbb{E}\bras{\widehat{Q}[m + 1] - \widehat{Q}[m] \middle | Q[m] = q}
    \label{chap5:eq:dsp1} \\
    & & +\ \pi(q_{1}) \mathbb{E}\bras{\widehat{Q}[m + 1] - \widehat{Q}[m] \middle | Q[m] = q_{1}} 
    \label{chap5:eq:dsp2} \\
    & & + \sum_{q = q_{1} + 1}^{\infty} \pi(q) \mathbb{E}\bras{\widehat{Q}[m + 1] - \widehat{Q}[m] \middle | Q[m] = q}.
    \label{chap5:eq:dsp3}
  \end{eqnarray}
  
  Now, as in \cite{gamarnik} we note that for $q \in \brac{0,\cdots, q_{1} - 1}$, $\widehat{Q}[m] = q_{1}$ and $\widehat{Q}[m + 1] \geq q_{1}$.
  Therefore \eqref{chap5:eq:dsp1} $\geq 0$.
  Also we note that for $q \in \brac{q_{1} + 1, \cdots}$, we have that $\widehat{Q}[m] = Q[m]$ and $\widehat{Q}[m + 1] \geq Q[m + 1]$, so that \eqref{chap5:eq:dsp2} is bounded below by
  \begin{eqnarray*}
    & & \sum_{q = q_{1} + 1}^{\infty} \pi(q) \mathbb{E}\bras{{Q}[m + 1] - {Q}[m] \middle | Q[m] = q} \geq -d \sum_{q = q_{1} + 1}^{\infty} \pi(q).
  \end{eqnarray*}
  Now let us consider \eqref{chap5:eq:dsp2}.
  We have that \eqref{chap5:eq:dsp2} $ = $ 
  \begin{eqnarray*}
    \Exp \bras{\widehat{Q}[m + 1] \middle | Q[m] = q_{1}} - q_{1} \geq \Exp \max(A[m + 1] - S(q_{1}), 0) \geq 1.\epsilon_{a},
  \end{eqnarray*}
  where assumption A2 is used.
  Substituting these lower bounds in \eqref{chap5:eq:dsp1} and \eqref{chap5:eq:dsp3} we have that
  \begin{eqnarray*}
    0 & \geq & \pi(q_{1}) \epsilon_{a} - d \sum_{q = q_{1} + 1}^{\infty} \pi(q), \\
    & = & \epsilon_{a} Pr\brac{Q \geq q_{1}} - \epsilon_{a} Pr\brac{Q \geq q_{1} + 1} - d Pr\brac{Q \geq q_{1} + 1}. \\
    \text{Hence } Pr\brac{Q \geq q_{1} + 1} & \geq & \frac{\epsilon_{a}}{\epsilon_{a} + d} Pr\brac{Q \geq q_{1}}.
  \end{eqnarray*}
  By redefining $\widehat{Q}[m] = \max(q_{1} + 1, Q[m])$ we obtain that 
  \begin{equation*}
    Pr\brac{Q \geq q_{1} + 2} \geq \frac{\epsilon_{a}}{\epsilon_{a} + d} Pr\brac{Q \geq q_{1} + 1}.
  \end{equation*}
  Induction leads to the following bound, for $k \geq 1$ :
  \begin{equation*}
    Pr\brac{Q \geq q_{1} + k} \geq \brap{\frac{\epsilon_{a}}{\epsilon_{a} + d}}^{k} Pr\brac{Q \geq q_{1}}.
  \end{equation*}
\end{proof}
\emph{TAIL-PROB-STATE-DEP-1 :}
\begin{proof}
  We note that in this case the lower bound on $Pr\brac{Q \geq q_{1} + k}$ is obtained in terms of expected drift over the tail of the queue length.
  Let us again define $\widehat{Q}[m] = \max(q_{1}, Q[m])$ and $\widehat{Q} = \max(q_{1}, Q)$.
  Then we note that as $q_{1}$ is finite, $\Expp \widehat{Q} < \infty$.
  Hence, we have that $\sum_{q = 0}^{\infty} \pi(q) \Exp \bras{ \widehat{Q}[m + 1] - \widehat{Q}[m] \middle |Q[m] = q} = 0$.
  We again split this sum into three parts leading to :
  \begin{eqnarray}
    0 & = & \sum_{q = 0}^{q_{1} - 1} \pi(q) \mathbb{E}\bras{\widehat{Q}[m + 1] - \widehat{Q}[m] \middle | Q[m] = q}
    \label{chap5:eq:dsp4} \\
    & & +\ \pi(q_{1}) \mathbb{E}\bras{\widehat{Q}[m + 1] - \widehat{Q}[m] \middle | Q[m] = q_{1}} 
    \label{chap5:eq:dsp5} \\
    & & + \sum_{q = q_{1} + 1}^{\infty} \pi(q) \mathbb{E}\bras{\widehat{Q}[m + 1] - \widehat{Q}[m] \middle | Q[m] = q}.
    \label{chap5:eq:dsp6}
  \end{eqnarray}
  As in the case of \eqref{chap5:eq:dsp1} and \eqref{chap5:eq:dsp2} we lower bound \eqref{chap5:eq:dsp4} by zero and \eqref{chap5:eq:dsp5} by $\pi(q_{1}) \epsilon_{a}$.
  Let us consider the case $q_{d} > q_{1}$.
  Then \eqref{chap5:eq:dsp6} can be written as
  \begin{eqnarray*}
    & & \sum_{q = q_{1} + 1}^{q_{d}} \pi(q) \mathbb{E}\bras{\widehat{Q}[m + 1] - \widehat{Q}[m] | Q[m] = q} +  \sum_{q = q_{d} + 1}^{\infty} \pi(q) \mathbb{E}\bras{\widehat{Q}[m + 1] - \widehat{Q}[m] | Q[m] = q}, \\
    & \geq & -d \sum_{q = q_{1} + 1}^{q_{d}} \pi(q) + \sum_{q = q_{d} + 1}^{\infty} \pi(q) \mathbb{E}\bras{Q[m + 1] - Q[m] | Q[m] = q},
  \end{eqnarray*}
  using the definition of $q_{1}$ and $\widehat{Q}[m]$.
  Using the lower bounds above on \eqref{chap5:eq:dsp4}, \eqref{chap5:eq:dsp5}, and \eqref{chap5:eq:dsp6} we obtain that
  \begin{eqnarray}
    0 & \geq & \epsilon_{a} \pi(q_{1}) - d \sum_{q = q_{1} + 1}^{q_{d}} \pi(q) + \sum_{q = q_{d} + 1}^{\infty} \pi(q) \mathbb{E}\bras{{Q}[m + 1] - {Q}[m] | Q[m] = q},  \nonumber \\
    & = & \epsilon_{a} Pr\brac{Q \geq q_{1}} - \epsilon_{a} Pr\brac{Q \geq q_{1} + 1} - d Pr\brac{Q \geq q_{1} + 1} + d Pr\brac{Q \geq q_{d} + 1} \nonumber \\
    & & + \sum_{q = q_{d} + 1}^{\infty} \pi(q) \mathbb{E}\bras{{Q}[m + 1] - {Q}[m] | Q[m] = q}, \text{ or}, \nonumber \\
    Pr\brac{Q \geq q_{1} + 1} & \geq & \frac{\epsilon_{a}}{\epsilon_{a} + d} Pr\brac{Q \geq q_{1}} + \frac{d}{\epsilon_{a} + d} Pr\brac{Q \geq q_{d} + 1} \\
    & & + \frac{1}{\epsilon_{a} + d} \sum_{q = q_{d} + 1}^{\infty} \pi(q) \mathbb{E}\bras{{Q}[m + 1] - {Q}[m] | Q[m] = q},
    \label{chap5:eq:dsp7}
  \end{eqnarray}
  Redefining $\widehat{Q}[m] = \max(q_{1} + 1, Q[m])$, where $q_{1} + 1 < q_{d}$, we obtain that
  \begin{eqnarray*}
  Pr\brac{Q \geq q_{1} + 2} & \geq & \frac{\epsilon_{a}}{\epsilon_{a} + d} Pr\brac{Q \geq q_{1} + 1} + \frac{d}{\epsilon_{a} + d} Pr\brac{Q \geq q_{d} + 1} \\
    & & + \frac{1}{\epsilon_{a} + d} \sum_{q = q_{d} + 1}^{\infty} \pi(q) \mathbb{E}\bras{{Q}[m + 1] - {Q}[m] | Q[m] = q}.
  \end{eqnarray*}
  Induction leads to the following bound for $q_{1} + k \leq q_{d}$, $k \geq 1$,
  \begin{eqnarray*}
    Pr\brac{Q \geq q_{1} + k} & \geq & \brap{\frac{\epsilon_{a}}{\epsilon_{a} + d}}^{k} Pr\brac{Q \geq q_{1}} \\
    & & + \frac{1 - \brap{\frac{\epsilon_{a}}{\epsilon_{a} + d}}^{k}}{\frac{d}{\epsilon_{a} + d}}\Bigg[\frac{d}{\epsilon_{a} + d} Pr\brac{Q \geq q_{d} + 1} + \\
      & & \frac{1}{\epsilon_{a} + d} \sum_{q = q_{d} + 1}^{\infty} \pi(q) \mathbb{E}\bras{{Q}[m + 1] - {Q}[m] | Q[m] = q}\Bigg].
  \end{eqnarray*}
  We note that the bound also holds trivially for $k = 0$.
  Simplifying we obtain that for $q_{1}, k \geq 0$, such that $q_{1} + k \leq q_{d}$,
  \begin{eqnarray*}
    Pr\brac{Q \geq q_{1} + k} & \geq & \brap{\frac{\epsilon_{a}}{\epsilon_{a} + d}}^{k} Pr\brac{Q \geq q_{1}} \\
    & & + \brap{1 - \brap{\frac{\epsilon_{a}}{\epsilon_{a} + d}}^{k}}\Bigg[Pr\brac{Q \geq q_{d} + 1} + \\
      & & \frac{1}{d} \sum_{q = q_{d} + 1}^{\infty} \pi(q) \mathbb{E}\bras{{Q}[m + 1] - {Q}[m] | Q[m] = q}\Bigg].
  \end{eqnarray*}
  
\end{proof}

\emph{TAIL-PROB-STATE-DEP-2 :}
\begin{proof}
  The derivation of this bound is very similar to that of TAIL-PROB-STATE-DEP-1.
  We follow the steps in the proof of TAIL-PROB-STATE-DEP-1 till \eqref{chap5:eq:dsp7} with the drift $d$ replaced by $d_{1}$.
  In this case, for $0 \leq q_{1} < q_{d}$, \eqref{chap5:eq:dsp7} is further simplified to :
  \begin{eqnarray*}
    Pr\brac{Q \geq q_{1} + 1} & \geq & \frac{\epsilon_{a}}{\epsilon_{a} + d_{1}} Pr\brac{Q \geq q_{1}} + \frac{d_{1}}{\epsilon_{a} + d_{1}} Pr\brac{Q \geq q_{d} + 1} - \frac{d_{2}}{\epsilon_{a} + d_{1}} Pr\brac{Q \geq q_{d} + 1},
  \end{eqnarray*}
  since $\forall q \geq q_{d} + 1$, $\Exp \bras{Q[m + 1] - Q[m] \middle | Q[m] = q} \geq -d_{2}$.
  Hence, we have that 
  \begin{eqnarray*}
    Pr\brac{Q \geq q_{1} + 1} & \geq & \frac{\epsilon_{a}}{\epsilon_{a} + d_{1}} Pr\brac{Q \geq q_{1}} - \frac{d_{2} - d_{1}}{\epsilon_{a} + d_{1}} Pr\brac{Q \geq q_{d} + 1}.
  \end{eqnarray*}
  Again by induction as before, we obtain that for $q_{1} + k \leq q_{d}$, $k \geq 1$,
  \begin{eqnarray*}
    Pr\brac{Q \geq q_{1} + k} & \geq & \brap{\frac{\epsilon_{a}}{\epsilon_{a} + d_{1}}}^{k} Pr\brac{Q \geq q_{1}} - \brap{1 - \brap{\frac{\epsilon_{a}}{\epsilon_{a} + d_{1}}}^{k}}\frac{d_{2} - d_{1}}{d_{1}} Pr\brac{Q \geq q_{d} + 1}.
  \end{eqnarray*}
  The above lower bound also holds trivially for $k = 0$.
\end{proof}

\section{An upper bound on $\Qg$ for a policy $\gamma$}
Let $\gamma$ be such that at a queue length $q$, a batch size $\min(q, S(q))$ is served, where $S(q) \leq S_{max}$ is a random function of the current queue length.
Furthermore, let $\gamma$ be such that $\Exp S(q)$ is a monotonically non-decreasing function of $q$ such that there exists a finite queue length $q_{1}$ such that $\Exp S(q_{1}) \geq \lambda + \epsilon$, where $\epsilon > 0$.
Then we have the following upper bound on $\Qg$.
\begin{proposition}
  For a policy $\gamma$ as above, we have that
  \begin{eqnarray*}
    \Qg \leq q_{1} \frac{\lambda + \epsilon}{\epsilon} + \frac{\Exp A[1]^{2} + S_{max}^{2}}{\epsilon}.
  \end{eqnarray*}
  \label{chap5:prop:avgq_driftub}
\end{proposition}
\begin{proof}
  The proof is very similar to that of \ref{chap4:app:prop:dotzupperbound}.
  Let $L(q) = q^{2}$ be a Lyapunov function.
  Then the Lyapunov drift
  \begin{eqnarray*}
    \Delta(q) & = & \Exp\bras{L(Q[m + 1]) - L(Q[m]) | Q[m] = q},\\
    & \leq & -2q(\Exp S(q) - \lambda) + \Exp A[1]^{2} + S_{max}^{2}.
  \end{eqnarray*}
  We note that for $q \geq q_{1}$, since $\Exp S(q) \geq \lambda + \epsilon$, we have that
  \begin{eqnarray*}
    \Delta(q) \leq -2q(\epsilon) + \Exp A[1]^{2} + S_{max}^{2}.
  \end{eqnarray*}
  For $q < q_{1}$, we have that
  \begin{eqnarray*}
    \Delta(q) & \leq & -2q(\epsilon) + 2q\bras{\lambda + \epsilon - \Exp S(q)} + \Exp A[1]^{2} + S_{max}^{2},\\
    & \leq & -2q(\epsilon) + 2q_{1}(\lambda + \epsilon) + \Exp A[1]^{2} + S_{max}^{2}.
  \end{eqnarray*}
  Hence, $\forall q$, we have that
  \begin{eqnarray*}
    \Delta(q) & \leq & -2q(\epsilon) + 2q_{1}(\lambda + \epsilon) + \Exp A[1]^{2} + S_{max}^{2}.
  \end{eqnarray*}
  Now applying \cite[Theorem A.4.3]{meynctcn} we have that
  \begin{eqnarray*}
    \Qg \leq q_{1} \frac{\lambda + \epsilon}{\epsilon} + \frac{\Exp A[1]^{2} + S_{max}^{2}}{\epsilon}.
  \end{eqnarray*}
\end{proof}

\section{Proof of Lemma \ref{chap5:lemma:case2upperbound}}
\label{chap5:app:case2upperbound}
Let $L(q) \Deq e^{\omega(q_{v} - q)}$ be a Lyapunov function.
Since for the policy $\gamma$, the batch size $\tilde{S}(q)$ could be more than $q$, the queue evolution equation under $\gamma$ is written as 
\[ Q[m + 1] = \max(Q[m] - \tilde{S}(Q[m]), 0) + A[m + 1].\]
The expected Lyapunov drift is
\begin{eqnarray*}
  \Delta(q) \Deq \Exp\bras{L(Q[m + 1]) - L(Q[m]) \vert Q[m] = q}.
\end{eqnarray*}
Since $(A[m])$ is assumed to be IID, as in \cite[Lemma 5(a)]{neely_mac} we have that
\begin{eqnarray*}
  \Delta(q) & \leq & e^{\omega(q_{v} - q)}\bras{\Exp e^{\omega\brap{\tilde{S}(q) - A[1]}} - 1}.
\end{eqnarray*}
As in the proof of \cite[Lemma 5(a)]{neely_mac} we have that
\begin{eqnarray*}
  \Exp e^{\omega(\tilde{S}(q) - A)} \leq 1 + \omega(\Exp \tilde{S}(q) - \lambda) + \frac{\omega^{2}A_{max}^{2}}{2} e^{\omega A_{max}}.
\end{eqnarray*}
Hence, we have that 
\begin{eqnarray*}
  \Delta(q) \leq \omega e^{\omega(q_{v} - q)}\bras{(\Exp \tilde{S}(q) - \lambda) + K}.
\end{eqnarray*}
where $K = \frac{\omega A_{max}^{2}}{2} e^{\omega A_{max}}$.

Now by definition, the policy $\gamma$ is such that
\begin{equation*}
  \tilde{S}(q) = 
  \begin{cases}
    s_{l}, \text{ for } 0 \leq q < q_{v}, \\
    s_{u}, \text{ for } q_{v} \leq q.
  \end{cases}
\end{equation*}
Then we have that for $q < q_{v}$
\begin{eqnarray*}
  \Delta(q) & \leq & -\omega e^{\omega(q_{v} - q)}\bras{\lambda - s_{l} - K}.
\end{eqnarray*}
And for $q \geq q_{v}$,
\begin{eqnarray*}
  \Delta(q) & \leq & \omega e^{\omega(q - q_{v})}\bras{(s_{u} - \lambda) + K}, \\
  & = & -\omega e^{\omega(q_{v} - q)}\bras{\lambda - s_{l} - K} + \omega e^{\omega(q_{v} - q)}\bras{s_{u} - s_{l}}, \\
  & \leq & -\omega e^{\omega(q_{v} - q)}\bras{\lambda - s_{l} - K} + \omega \bras{s_{u} - s_{l}}, \\
\end{eqnarray*}
Hence, for all $q$ we have that
\begin{eqnarray*}
  \Delta(q) & \leq & -\omega e^{\omega(q_{v} - q)}\bras{\lambda - s_{l} - K} + \omega \bras{s_{u} - s_{l}}.
\end{eqnarray*}

We choose $\omega$ such that $K < \lambda - s_{l}$.
Proceeding as in the proof of \cite[Theorem 3(c)]{neely_mac}, we have that
\begin{eqnarray*}
  \Exp e^{\omega(q_{v} - Q)} \leq \frac{\bras{s_{u} - s_{l}}}{(\lambda - s_{l}- K)}.
\end{eqnarray*}
Since $\Exp e^{\omega(q_{v} - Q)} \geq \Exp \bras{e^{\omega(q_{v} - Q)} \vert Q < s_{l}} Pr\brac{Q < s_{l}}$, we therefore have that
\begin{eqnarray}
  Pr\brac{Q < s_{l}} \leq e^{-\omega q_{v}} \frac{e^{\omega s_{l}} \bras{s_{u} - s_{l}}}{(\lambda - s_{l}- K)}.
  \label{chap5:eq:qlessslbound}
\end{eqnarray}
Now we note that 
\begin{eqnarray*}
  \overline{C}(\gamma) & = & \sum_{s < s_{l}} \pi_{s}(s) c(s) + \pi_{s}(s_{l}) c(s_{l}) + \pi_{s}(s_{u}) c(s_{u}), \\
  & \leq & Pr\brac{Q < s_{l}} c(s_{l} - 1) + c(\lambda)\brap{1 - \pi_{s}(s_{l}) - \pi_{s}(s_{u})} + m\bras{s_{u} \pi_{s}(s_{u}) + s_{l} \pi_{s}(s_{l}) - \lambda(\pi_{s}(s_{u}) + \pi_{s}(s_{l}))},
\end{eqnarray*}
where $m$ is the slope of $l(s)$.
Then, we have that
\begin{eqnarray*}
  \Cg - c(\lambda) & \leq & \bras{c(s_{l} - 1) + m\lambda - c(\lambda)}Pr\brac{Q < s_{l}}.
\end{eqnarray*}

Now consider the sequence of policies $\gamma_{k}$ for which $q_{v} = \log\nfrac{1}{V_{k}}$ for a sequence $V_{k} < 1$ such that $V_{k} \downarrow 0$.
Then we have that $\Cg - c(\lambda) = \mathcal{O}(V_{k})$.
Furthermore, from Proposition \ref{chap5:prop:avgq_driftub} we have that $\Qg = \mathcal{O}\brap{\log\nfrac{1}{V_{k}}}$.
We note that $\gamma_{k}$ is also a sequence of admissible policies, since $s(q)$ is a non-decreasing function of $q$ and $\overline{Q}(\gamma_{k} < \infty$.

\section{Proof of Lemma \ref{chap5:lemma:qul_pr}}
\label{chap5:app:qul_pr}
\begin{proof}
  Define $\widehat{Q}[m] \stackrel{\Delta} = \max(q_{u}, Q[m])$ and $\widehat{Q} = \max(q_{u}, Q)$ for a $q_{u} \geq 0$.
  Then for any finite $q_{u}$ we have that $\Exp \widehat{Q} < \infty$ and therefore
  \begin{eqnarray*}
    \sum_{q = 0}^{\infty} \pi(q) \Exp \bras{ \widehat{Q}[m + 1] - \widehat{Q}[m] \middle \vert Q[m] = q} = 0.
  \end{eqnarray*}
  We note that this can be written as
  \begin{eqnarray}
    0 & = & \sum_{q \leq q_{u} - A_{max}} \pi(q) \Exp \bras{ \widehat{Q}[m + 1] - \widehat{Q}[m] \middle \vert Q[m] = q} 
    \label{chap5:eq:qul_pr_ub0}\\
    & & + \sum_{q_{u} - A_{max} < q \leq q_{u} + A_{max}} \pi(q) \Exp \bras{ \widehat{Q}[m + 1] - \widehat{Q}[m] \middle \vert Q[m] = q}
    \label{chap5:eq:qul_pr_ub1}\\
    & & + \sum_{q_{u} + A_{max} < q} \pi(q) \Exp \bras{ \widehat{Q}[m + 1] - \widehat{Q}[m] \middle \vert Q[m] = q}.
    \label{chap5:eq:qul_pr_ub2}
  \end{eqnarray}
  Then we note that for \eqref{chap5:eq:qul_pr_ub0} as $q \leq q_{u} - A_{max}$, $\widehat{Q}[m] = q_{u}$ and $Q[m + 1] \leq q_{u}$, so that $\widehat{Q}[m + 1] = q_{u}$.
  Therefore \eqref{chap5:eq:qul_pr_ub0} $= 0$.
  Now consider \eqref{chap5:eq:qul_pr_ub2}.
  We note that for $q > q_{u} + A_{max}$, both $\widehat{Q}[m]$ and $\widehat{Q}[m + 1]$ are equal to $Q[m]$ and $Q[m + 1]$ respectively.
  Therefore we have that 
  \begin{eqnarray*}
    \eqref{chap5:eq:qul_pr_ub2} & = & \sum_{q_{u} + A_{max} < q} \pi(q) \bras{ \lambda - s(q) }, \\
    & \leq & -(s_{u} - \lambda) Pr\brac{Q > q_{u} + A_{max}}.
  \end{eqnarray*}
  We also note that for $q_{u} - A_{max} < q \leq q_{u} + A_{max}$, $\widehat{Q}[m + 1] - \widehat{Q}[m] \leq A[m + 1]$.
  Therefore, 
  \[ \eqref{chap5:eq:qul_pr_ub1} \leq \lambda Pr\brac{q_{u} - A_{max} < Q \leq q_{u} + A_{max}}. \]
  Using these upper bounds in \eqref{chap5:eq:qul_pr_ub0}, \eqref{chap5:eq:qul_pr_ub1}, and \eqref{chap5:eq:qul_pr_ub2} we obtain that
  \begin{eqnarray*}
    0 & \leq & \lambda Pr\brac{q_{u} - A_{max} < Q \leq q_{u} + A_{max}} - (s_{u} - \lambda) Pr\brac{Q > q_{u} + A_{max}},
  \end{eqnarray*}
  which can be written as
  \small
  \begin{eqnarray}
    0 & \leq & \lambda Pr\brac{Q > q_{u} - A_{max}} -s_{u} Pr\brac{Q > q_{u} + A_{max}}, \text{ or}, \nonumber \\
    Pr\brac{Q > q_{u} + A_{max}} & \leq & \frac{\lambda}{s_{u}} Pr\brac{Q > q_{u} + A_{max}}.
    \label{chap5:eq:qul_pr_ub3}
  \end{eqnarray}
  \normalsize
  Let $\rho_{u} \stackrel{\Delta} = \frac{\lambda}{s_{u}}$.
  Using \eqref{chap5:eq:qul_pr_ub3} and inducting we obtain that
  \begin{eqnarray}
    Pr\brac{Q > q_{u}} & \leq & \rho_{u}^{\ceiling{\frac{q_{u}}{2A_{max}}}},
    \label{chap5:eq:qul_pr_ub4}
  \end{eqnarray}
  for any $q_{u} \geq A_{max}$.
  
  By redefining $q_{u}$ to be $q_{u} + k$, $k \geq 0$, we can show that
  \begin{eqnarray*}
    Pr\brac{Q > q_{u,l} + k} \leq \rho_{u}^{\ceiling{\frac{q_{u} + k}{2A_{max}}}}.
  \end{eqnarray*}
\end{proof}

\section{Proof of Lemma \ref{chap5:lemma:realq_dtmc_stat_prob}}
\label{chap5:app:realq_dtmc_stat_prob}

\begin{proof}
  This proof is similar to that of the proof of Lemma \ref{chap5:lemma:dtmc_stat_prob}-TAIL-PROB-STATE-DEP-1.
  We define $\widehat{Q}[m] = \max(q_{1}, Q[m])$ and $\widehat{Q} = \max(q_{1}, Q)$.
  We note that since the policy is admissible and $q_{1}$ is finite, $\Expp \widehat{Q} < \infty$.
  Therefore
  \begin{equation*}
    \int_{0}^{\infty} \Exp \bras{\widehat{Q}[m + 1] - \widehat{Q}[m] \middle \vert Q[m] = q} d\pi(q) = 0.
  \end{equation*}
  We again split the above integral into three parts which leads to 
  \begin{eqnarray}
    0 & = & \int_{0}^{q_{1} - \Delta} \mathbb{E}\bras{\widehat{Q}[m + 1] - \widehat{Q}[m] \middle | Q[m] = q} d\pi(q) 
    \label{chap5:eq:real0} \\
    & & + \int_{q_{1} - \Delta}^{q_{1}} \mathbb{E}\bras{\widehat{Q}[m + 1] - \widehat{Q}[m] \middle | Q[m] = q} d\pi(q) 
    \label{chap5:eq:real1} \\
    & & + \int_{q_{1}}^{\infty} \mathbb{E}\bras{\widehat{Q}[m + 1] - \widehat{Q}[m] \middle | Q[m] = q } d\pi(q).
    \label{chap5:eq:real2}
  \end{eqnarray}
  where $\Delta > 0$ will be chosen in the following.
  We note that for $q \in [0, q_{1} - \Delta)$ we have that $\widehat{Q}[m] = q_{1}$ and $\widehat{Q}[m + 1] \geq q_{1}$, so that \eqref{chap5:eq:real0} $\geq 0$.
  Now as $q_{1} \leq q_{d}$ and $\widehat{Q}[m] \geq Q[m]$ we obtain that \eqref{chap5:eq:real2}
  \begin{eqnarray*}
    \geq - d Pr\{q_{1} \leq Q < q_{d}) + \int_{q_{d}}^{\infty} \mathbb{E}\bras{{Q}[m + 1] - {Q}[m] | Q[m] = q} d\pi(q)
  \end{eqnarray*}
  To obtain a lower bound on \eqref{chap5:eq:real1} we note that for $q \leq q_{1}$, $\widehat{Q}[m] = q_{1}$ and $\widehat{Q}[m + 1] \geq q_{1}$.
  So $\widehat{Q}[m + 1] - \widehat{Q}[m] \geq 0$.
  Then as in \cite[steps (34), (35), and (36)]{berry} we use Markov inequality to lower bound $\mathbb{E}\bras{\widehat{Q}[m + 1] - \widehat{Q}[m] \middle | Q[m] = q} , q \in [q_{1} - \Delta, q_{1})$.
  \begin{eqnarray*}
    \mathbb{E}\bras{\widehat{Q}[{m + 1}] - \widehat{Q}[{m}] \middle | Q[{m}] = q} & \geq & \delta Pr\brac{\widehat{Q}[{m + 1}] - \widehat{Q}[{m}] \geq \delta \middle |Q[{m}] = q}, \\
    & \geq & \delta Pr\brac{{Q}[{m + 1}] - {Q}[{m}] \geq \delta + \Delta \middle | Q[m] = q}, \\
    & = & \delta Pr\brac{A[m + 1] - S[m + 1] \geq \delta + \Delta \middle \vert Q[m] = q}, \\
    & \geq & \delta Pr\brac{A[m + 1] - S_{max} \geq \delta + \Delta \middle \vert Q[m] = q}, \\
    & \geq & \delta \epsilon_{a}.
  \end{eqnarray*}
  We note that $\Delta$ and $\delta$ have to be chosen so that $\Delta + \delta < \delta_{a}$.
  Thus we obtain that \eqref{chap5:eq:real1} $\geq \delta \epsilon_{a} Pr\brac{q_{1} - \Delta \leq Q < q_{1}}$.
  Combining these bounds and using $\mathbb{E}\bras{{Q}[m + 1] - {Q}[m] | Q[m] = q} = \mathbb{E}S(q) - \lambda$, we obtain that 
  \begin{eqnarray*}
    0 \geq \delta \epsilon_{a} Pr\brac{q_{1} - \Delta \leq Q < q_{1}} - d Pr\{q_{1} \leq Q < q_{d}) + \int_{q_{d}}^{\infty} \mathbb{E}\bras{{Q}[m + 1] - {Q}[m] | Q[m] = q} d\pi(q), \\
    Pr\brac{Q \geq q_{1}} \geq \dep Pr\brac{Q \geq q_{1} - \Delta} + \frac{1}{\delta\epsilon_{a} + d}\bras{d Pr\brac{Q \geq q_{d}} - \int_{q_{d}}^{\infty} (\mathbb{E}S(q) - \lambda)d\pi(q)}
  \end{eqnarray*}
  Similarly, if we define $\widehat{Q}[m] = \max(q_{1} + \Delta, Q[m])$ and if $q_{1} + \Delta \leq q_{d}$, we obtain that
  \begin{eqnarray*}
    Pr\brac{Q \geq q_{1} + \Delta} \geq \dep Pr\brac{Q \geq q_{1}} + \frac{1}{\delta\epsilon_{a} + d}\bras{d Pr\brac{Q \geq q_{d}} - \int_{q_{d}}^{\infty} (\mathbb{E}S(q) - \lambda)d\pi(q)}.
  \end{eqnarray*}
  By induction, we obtain that if $k \geq 0$, and $q_{1} + k\Delta \leq q_{d}$, then $Pr\brac{Q \geq q_{1} + k\Delta}$
  \begin{eqnarray*}
    & \geq & \brap{\dep}^{k} Pr\brac{Q \geq q_{1}} + \frac{1 - \brap{\dep}^{k}}{d} \bras{d Pr\brac{Q \geq q_{d}} - \int_{q_{d}}^{\infty} (\mathbb{E}S(q) - \lambda)d\pi(q)}, \\
    & = & \brap{\dep}^{k} Pr\brac{Q \geq q_{1}} + \brap{1 - \brap{\dep}^{k}} \bras{Pr\brac{Q \geq q_{d}} - \frac{1}{d}\int_{q_{d}}^{\infty} (\mathbb{E}S(q) - \lambda)d\pi(q)}
  \end{eqnarray*}
\end{proof}

\section{Proof of Lemma \ref{chap5:lemma:rmodelupperbound}}
\label{chap5:app:rmodelupperbound}

Consider a particular policy $\gamma$ in the sequence of policies $\gamma_{k}$.
We note that the policy $\gamma$ is a stationary deterministic policy.
Then, we have that
\begin{eqnarray*}
  \Cg = \int_{0}^{S_{max}} c_{R}(s) d\pi_{s}(s).
\end{eqnarray*}
Since $c_{R}(s)$ is a non-decreasing function, we have that
\begin{eqnarray*}
  \Cg & \leq & Pr\brac{S < \lambda - \epsilon_{V}} c(\lambda - \epsilon_{V}) + \pi_{s}(\lambda - \epsilon_{V})c(\lambda - \epsilon_{V}) + \pi_{s}(\lambda + \epsilon_{V}) c(\lambda + \epsilon_{V}) + \pi_{s}(\lambda + \epsilon) c(\lambda + \epsilon). \\
  & \leq & Pr\brac{S < \lambda - \epsilon_{V}} c(\lambda) + \pi_{s}(\lambda - \epsilon_{V})\brap{c(\lambda) - \epsilon_{V} \frac{dc(\lambda)}{ds} + \mathcal{O}(\epsilon_{V}^{2})} \\
  & & + \pi_{s}(\lambda + \epsilon_{V})\brap{c(\lambda) + \epsilon_{V} \frac{dc(\lambda)}{ds} + \mathcal{O}(\epsilon_{V}^{2})} + \pi_{s}(\lambda + \epsilon) \brap{c(\lambda) + \epsilon\frac{dc(\lambda)}{ds} + G(\epsilon)},
\end{eqnarray*}
where $G(\epsilon) = c(\lambda + \epsilon) - \brap{c(\lambda) + \epsilon\frac{dc(\lambda)}{ds}}$.

Since there exists a finite queue length at which a service rate greater than $\lambda$ is used, we have that
\begin{eqnarray*}
  \int_{0}^{\lambda - \epsilon_{V}} s d\pi_{s}(s) + \brap{\lambda - \epsilon_{V}} \pi_{s}(\lambda - \epsilon_{V}) + \brap{\lambda + \epsilon_{V}} \pi_{s}(\lambda + \epsilon_{V}) + \brap{\lambda + \epsilon} \pi_{s}(\lambda + \epsilon) = \lambda, \text{ or}, \\
  -\epsilon_{V} \pi_{s}(\lambda - \epsilon_{V}) + \epsilon_{V} \pi_{s}(\lambda + \epsilon_{V}) + \epsilon \pi_{s}(\lambda + \epsilon) = \int_{0}^{\lambda - \epsilon_{V}} s d\pi_{s}(s) \leq \lambda Pr\brac{S < \lambda - \epsilon_{V}}.
\end{eqnarray*}

Hence, we have that
\begin{eqnarray*}
  \Cg \leq c(\lambda) + Pr\brac{S < \lambda - \epsilon_{V}} \lambda \frac{dc(\lambda)}{ds} + G(\epsilon)\pi_{s}(\lambda + \epsilon) + \mathcal{O}\brap{\epsilon_{V}^{2}}.
\end{eqnarray*}
We note that $\pi_{s}(\lambda + \epsilon) = Pr\brac{Q > 2q_{v}}$ and $Pr\brac{S < \lambda - \epsilon_{V}} = Pr\brac{Q < \lambda - \epsilon_{V}}$.
We proceed to find upper bounds on $Pr\brac{Q < \lambda - \epsilon_{V}}$ and $Pr\brac{Q > 2q_{v}}$.

We first obtain an upper bound on $Pr\brac{Q < \lambda - \epsilon_{V}}$ as in the proof of Lemma \ref{chap5:lemma:case2upperbound}.
Using the Lyapunov function $L(q) = e^{\omega(q_{u} - q)}$ we have that the Lyapunov drift is
\begin{eqnarray*}
  \Delta(q) \leq \omega e^{\omega(q_{v} - q)} \brap{\tilde{s}(q) - \lambda + K},
\end{eqnarray*}
where $K = \frac{\omega A_{max}^{2}}{2} e^{\omega A_{max}}$.
Then, for $q \leq q_{v}$, we have that
\begin{eqnarray*}
  \Delta(q) \leq \omega e^{\omega(q_{v} - q)} \brap{-\epsilon_{V} + K}.
\end{eqnarray*}
For $q_{v} < q \leq 2q_{v}$, we have that
\begin{eqnarray*}
  \Delta(q) & \leq & \omega e^{\omega(q_{v} - q)} \brap{\epsilon_{V} + K}, \\
  & = & \omega e^{\omega(q_{v} - q)} \brap{-\epsilon_{V} + K} + 2\omega e^{\omega(q_{v} - q)} \epsilon_{V},\\
  & \leq & \omega e^{\omega(q_{v} - q)} \brap{-\epsilon_{V} + K} + 2\omega \epsilon_{V}.
\end{eqnarray*}
For $2q_{v} < q$, we have that
\begin{eqnarray*}
  \Delta(q) & \leq & \omega e^{\omega(q_{v} - q)} \brap{\epsilon + K}, \\
  & = & \omega e^{\omega(q_{v} - q)} \brap{-\epsilon_{V} + K} + 2\omega e^{\omega(q_{v} - q)} \brap{\epsilon + \epsilon_{V}},\\
  & \leq & \omega e^{\omega(q_{v} - q)} \brap{-\epsilon_{V} + K} + 2\omega \brap{\epsilon + \epsilon_{V}}.
\end{eqnarray*}
Therefore, for every $q$, we have that
\begin{eqnarray*}
  \Delta(q) & \leq & \omega e^{\omega(q_{v} - q)} \brap{-\epsilon_{V} + K} + 2\omega \brap{\epsilon + \epsilon_{V}}.
\end{eqnarray*}
Let $K = \epsilon_{V}/2$.
Or we have that $\epsilon_{V} = {\omega A_{max}^{2}} e^{\omega A_{max}}$.
Then, 
\begin{eqnarray*}
  \Delta(q) & \leq & - \omega e^{\omega(q_{v} - q)} \frac{\epsilon_{V}}{2} + 2\omega \brap{\epsilon + \epsilon_{V}}.
\end{eqnarray*}
Now as in the proof of Lemma \ref{chap5:lemma:case2upperbound}, we have that
\begin{eqnarray*}
  Pr\brac{Q < \lambda - \epsilon_{V}} \leq \frac{2\brap{\epsilon + \epsilon_{V}}}{\epsilon_{V}} e^{-\omega(q_{v} - \lambda + \epsilon_{V})}.
\end{eqnarray*}

Let us choose $q_{v} = \frac{1}{\omega}\log\nfrac{1}{\epsilon_{V}^{3}}$.
Then we have that 
\begin{eqnarray}
  Pr\brac{Q < \lambda - \epsilon_{V}} = \mathcal{O}\brap{\epsilon_{V}^{2}} = \mathcal{O}\brap{\omega^{2}}.
  \label{chap5:eq:realupperbound1}
\end{eqnarray}

To obtain an upper bound on $Pr\brac{Q > 2q_{v}}$, we proceed similarly but with a Lyapunov function $L(q) = e^{\omega\brap{q - q_{v}}}$.
The expected Lyapunov drift is
\begin{eqnarray*}
  \Delta(q) \Deq \Exp\bras{e^{\omega\brap{Q[m + 1] - q_{v}}} - e^{\omega\brap{Q[m] - q_{v}}} | Q[m] = q}.
\end{eqnarray*}
Then we have that
\begin{eqnarray*}
  \Delta(q) = e^{-\omega q_{v}} \Exp \bras{e^{\omega\brap{\brap{q - \tilde{s}(q)}^{+} + A}} - e^{\omega q}}.
\end{eqnarray*}
As in \cite{neely_mac}, we have that
\begin{eqnarray*}
  \Delta(q) \leq e^{-\omega q_{v}} \Exp \bras{e^{\omega\brap{q - \tilde{s}(q) + A}} - e^{\omega q}} + e^{-\omega q_{v} + \omega \delta_{max}},
\end{eqnarray*}
where $\delta_{max} = A_{max} + S_{max}$.
Or we have that
\begin{eqnarray*}
  \Delta(q) & \leq & e^{\omega\brap{q - q_{v}}} \Exp \bras{e^{-\omega\brap{\tilde{s}(q) - A}} - 1} + e^{-\omega q_{v} + \omega \delta_{max}}, \\
  & \leq & -\omega e^{\omega\brap{q - q_{v}}} \Exp \bras{\brap{\tilde{s}(q) - A} - K} + e^{-\omega q_{v} + \omega \delta_{max}}, \\
  & = & -\omega e^{\omega\brap{q - q_{v}}} \bras{\tilde{s}(q) - \lambda - K} + e^{-\omega q_{v} + \omega \delta_{max}},
\end{eqnarray*}
where $K = \frac{\omega A_{max}^{2}}{2} e^{\omega A_{max}}$.

For $q \leq q_{v}$, we have that 
\begin{eqnarray*}
  \Delta(q) & \leq & -\omega e^{\omega\brap{q - q_{v}}} \bras{-\epsilon_{V} - K} + e^{-\omega q_{v} + \omega \delta_{max}}, \\
  & \leq & -\omega e^{\omega\brap{q - q_{v}}} \bras{\epsilon_{V} - K} + 2\omega\epsilon_{V} + e^{-\omega q_{v} + \omega \delta_{max}}.
\end{eqnarray*}
For $q > q_{v}$, we have that
\begin{eqnarray*}
  \Delta(q) & \leq & -\omega e^{\omega\brap{q - q_{v}}} \bras{\epsilon_{V} - K} + e^{-\omega q_{v} + \omega \delta_{max}}.
\end{eqnarray*}
So for all $q$, we have that
\begin{eqnarray*}
  -\omega e^{\omega\brap{q - q_{v}}} \bras{\epsilon_{V} - K} + 2\omega\epsilon_{V} + e^{-\omega q_{v} + \omega \delta_{max}}.
\end{eqnarray*}
Since $K = \epsilon_{V}/2$, we have that
\begin{eqnarray*}
  \omega\frac{\epsilon_{V}}{2} \Exp\bras{e^{\omega\brap{q - q_{v}}}} & \leq & 2\omega\epsilon_{V} + e^{-\omega q_{v} + \omega \delta_{max}}, \\
  \Exp\bras{e^{\omega\brap{q - q_{v}}}} & \leq & 4 + \frac{2e^{-\omega q_{v} + \omega \delta_{max}}}{\omega\epsilon_{V}}.
\end{eqnarray*}
Since $e^{\omega {q_{v}}} Pr\brac{Q > 2q_{v}} \leq \Exp \bras{e^{\omega\brap{q - q_{v}}}}$, we have that
\begin{eqnarray*}
  Pr\brac{Q > 2q_{v}} \leq 4e^{-\omega q_{v}} + e^{-\omega q_{v}} \frac{2e^{-\omega q_{v} + \omega \delta_{max}}}{\omega\epsilon_{V}}.
\end{eqnarray*}

Since $q_{v} = \frac{1}{\omega}\log\fpow{1}{\epsilon_{V}}{3}$.
Then we have that
\begin{eqnarray*}
  Pr\brac{Q > 2q_{v}} \leq 4\epsilon_{V}^{3} + \epsilon_{V}^{4} A_{max}^{2} {2e^{\omega \brap{\delta_{max} + A_{max}}}}.
\end{eqnarray*}
Therefore, from the above upper bound and \eqref{chap5:eq:realupperbound1}, we have that both $Pr\brac{Q < \lambda - \epsilon_{V}}$ and $Pr\brac{Q > 2q_{v}}$ are $\mathcal{O}\brap{\epsilon_{V}^{2}}$.
Since $\mathcal{O}\brap{\epsilon_{V}} = \mathcal{O}\brap{\omega}$, we have that both $Pr\brac{Q < \lambda - \epsilon_{V}}$ and $Pr\brac{Q > 2q_{v}}$ are $\mathcal{O}\brap{\omega^{2}}$.

For the sequence of policies $\gamma_{k}$, we have that $\Cgk - c(\lambda) = \mathcal{O}\brap{\omega_{k}^{2}} = \mathcal{O}\brap{V_{k}}$.
Since $q_{v_{k}} = \mathcal{O}\brap{\frac{1}{\omega_{k}}\log\nfrac{1}{\epsilon_{V}^{3}}}$, we have that $q_{v_{k}} = \mathcal{O}\brap{\frac{1}{\sqrt{V_{k}}}\log\nfrac{1}{V_{k}}}$.
Therefore, using Proposition \ref{chap5:prop:avgq_driftub}, with $q_{1} = 2q_{v}$, we have that $\Qgk = \mathcal{O}\brap{\frac{1}{\sqrt{V_{k}}}\log\nfrac{1}{V_{k}}}$.
We note that the policy $\gamma_{k}$ is admissible.
Therefore, we have a sequence of admissible policies $\gamma_{k}$ such that $\Qgk = \mathcal{O}\brap{\frac{1}{\sqrt{V_{k}}}\log\nfrac{1}{V_{k}}}$. and $\Cgk = \mathcal{O}\brap{V_{k}}$.
\end{subappendices}
\addtocontents{toc}{\protect\setcounter{tocdepth}{2}}

\ifdefined \Deltat \else 
\newcommand{\Deltat}{\frac{\Delta}{2}}
\fi
\chapter[On the tradeoff of average power and average delay for fading wireless links]{\textbf{On the tradeoff of average power and average delay\\ for fading wireless links}}
\section{Introduction}
Minimizing the average power as well as the average delay is a major requirement in current wireless communication networks, which brings the problem of designing good scheduling and power control policies to the forefront.
In this chapter, we consider the characterization of the optimal tradeoff between average power and average queue length for a fading point to point link, with and without admission control, and obtain bounds on the tradeoff of average power and average delay by applying Little's law.
The models that we consider capture some of the important issues underlying the general problem for wireless networks: there is bursty arrival of traffic which can be subjected to admission control, the channel gain varies unpredictably, and the transmitter can dynamically change its transmission rate by varying the transmission power.
The bounds that we derive are obtained using the methods discussed in Chapter 4, and are asymptotic in nature.
However, unlike the models in Chapter 4, where there was only a single environment state and no admission control, here we consider models with multiple environment states as well as with admission control.
We also consider the asymptotic characterization of the tradeoff for models with multiple queues in this chapter.
The glossary of notation that we use in this chapter is given in Table \ref{chapter5:notationtable}.

\begin{table}
\centering
\begin{tabular}{|l|l|}
\hline
Symbol & Description \\
\hline
$m$ & slot index \\
$R[m]$ & random number of arrivals in slot $m$ (before admission control) \\
$A[m]$ & random number of arrivals in slot $m$ (after admission control) \\
$A_{max}$ & maximum number of arrivals in any slot \\
$\lambda, \sigma^{2}$ & mean and variance of $A[1]$ \\
$\mathcal{H}$ & set of fade states \\
$\pi_{H}$ & distribution of fade state \\
$H[m]$ & fade state in slot $m$ \\
$Q[m]$ & queue length at the start of $(m + 1)^{th}$ slot \\
$S[m]$ & batch service size in slot $m$ \\
$S_{max}$ & maximum batch service size \\
$\sigma[m]$ & history of queue evolution \\
$\gamma$ & policy - $(S[1], S[2], \cdots)$ \\
$\Gamma$ & set of all policies \\
$\Gamma_{s}$ & set of all stationary policies \\
$\Qg$ & average queue length for a policy $\gamma$ \\
$P(h, s)$ & power expended as a function of fade state $h$ and batch size $s$ \\
$\Pg$ & average power for a policy $\gamma$ \\
$P_{c}$ & average power constraint \\
$c(\lambda)$ & minimum average power required for queue stability for I-model \\
$c_{R}(\lambda)$ & minimum average power required for queue stability for R-model \\
$\Gamma_{a}$ & set of all admissible policies \\
$\overline{s}(q)$ & average service rate at a queue length $q$ \\
$Q^*(P_{c})$ & minimum average queue length over $\Gamma_{a}$ under constraint $P_{c}$ \\
$\epsilon_{a}$ & probability of $A[m]$ exceeding $S_{max}$ \\
$\Ag$ & average throughput for a policy $\gamma$ \\
$\pi$ & stationary distribution for a policy which is clear from the context \\
$\pi_{\gamma}$ & stationary distribution for policy $\gamma$ \\
\hline
\end{tabular}
\caption{Notation used in this chapter.}
\label{chapter5:notationtable}
\end{table}

\subsection{Methodology}
\label{chapter5:methodology}
As stated in Section \ref{chapter4:methodology}, the scheme for obtaining asymptotic lower bounds in this chapter is very similar to that in Chapter 4.
We briefly summarize the differences.
We again obtain $\overline{q}$ which is the largest $q$ such that $Pr_{u}\brac{Q < q} \leq \frac{1}{2}$.
However, we note that geometric bounds for $Pr_{u}\brac{Q < q}$ are obtained as functions of the average drift of the queue length, where the averaging is done over the slot fade state also.
For the tradeoff problem in this chapter, again a non-negative function $D(q)$ will be obtained, where $\Exp_{\pi}D(Q)$ is the difference between the average power (rather than the average service cost) and $c(\lambda)$.
For illustrating the method, we again consider the case of integer valued queue evolution and obtain a geometric upper bound $Pr_{u}\brac{Q < q}$.
The steps which were followed for obtaining the $\Omega\brap{\log\nfrac{1}{V}}$ asymptotic lower bound for the example in Section \ref{chapter4:methodology} can then be directly applied to obtain an $\Omega\brap{\log\nfrac{1}{V}}$ asymptotic lower bound for the tradeoff of average delay with average power.

\subsection{System model - Integer valued queue length evolution}
\label{chapfading:sysmodel:intval}
We consider a discrete time system with slots indexed by the positive integer $m$.
We assume that there is no admission control, so that $A[m] = R[m], \forall m$.
In each slot $m$, a random number of packets $A[m] \in \mathbb{Z}_{+}$, where each packet is of the same size, arrive into the transmitter queue.
The arrival sequence $(A[m], m \geq 1)$ is assumed to be IID with $A[1] \leq A_{max}$, batch arrival rate $\mathbb{E}A[1] = \lambda < \infty$, $\text{var}(A[1]) = \sigma^{2} < \infty$.
The packets are assumed to arrive into an infinite buffer, in which they wait until they are transmitted over a point to point fading channel.
The fade state is assumed to be constant in a slot.
The fade state takes values in a finite set $\mathcal{H}$, with $\min\brac{\mathcal{H}} > 0$, and the fade state process $(H[m], m \geq 1)$, is assumed to be IID, with $H[1] \sim \pi_{H}$.
The expectation with respect to $\pi_{H}$ is denoted by $\Exp_{\pi_H}$.
The processes $(A[m])$ and $(H[m])$ are assumed to be independent of each other.

\begin{figure}
  \centering
  \includegraphics[width=120mm,height=40mm]{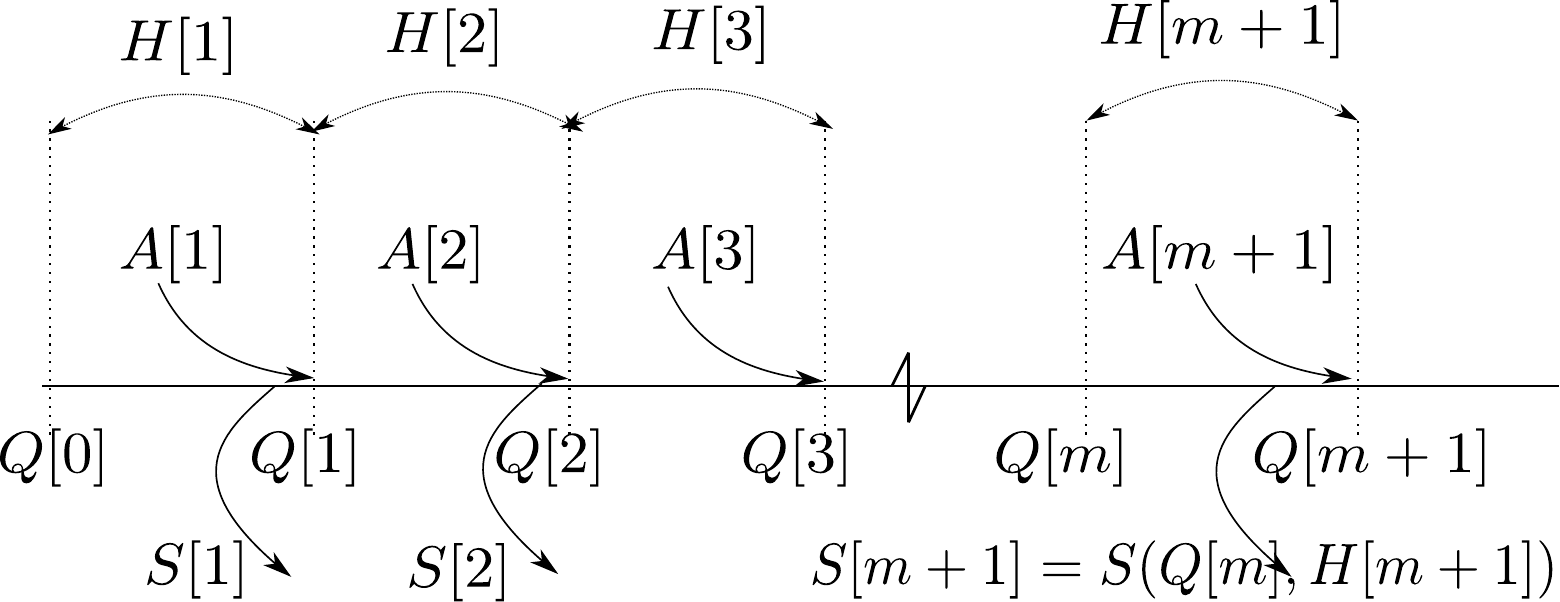}
  \caption{\small{Evolution of the queue length $Q[m]$; the batch size $S[m]$ is chosen as a randomized function of the fade state $H[m]$, the queue length $Q[m - 1]$, and the history of the process $\sigma[m]$.}}
  \label{chap5fading:fig:evolution}
\end{figure}

The number of customers in the queue at the start of the $(m + 1)^{th}$ slot is denoted by $Q[m]$.
The system is assumed to start with $Q[0] = q_{0} \in \mathbb{Z}_{+}$ customers.
At the end of slot $m$, a batch with $S[m] \in \sZ$ packets is removed from the transmitter queue just before the $A[m]$ new packets which arrive in the $m^{th}$ slot are admitted.
We assume that $S[m] \leq \min\brap{Q[m - 1], S_{max}}$, where $S_{max}$ is the maximum batch size that can be served.
The queue evolution sampled at the slots is given by:
\begin{equation}
  Q[m + 1] = Q[m] - S[m + 1] + A[m + 1].
  \label{chap5fading:eq:evolution}
\end{equation}
The evolution of the queue length is illustrated in Figure \ref{chap5fading:fig:evolution}.

At the start of slot $m$, the history of the system is defined as:
\begin{eqnarray*}
  \sigma[m] \Deq (q_{0}, H[1], S[1], Q[1], H[2], S[2], Q[2], \dots, Q[m - 2], H[m - 1]).
\end{eqnarray*}
At the beginning of slot $m$, the scheduler observes  $H[m]$ and chooses a batch service size $S[m] \in \sZ$ as a randomized function of the history $\sigma[m]$, the current queue length $Q[m - 1]$, and the current fade state $H[m]$.

We define a policy $\gamma$ to be the sequence of service batch sizes $(S[1], S[2], \dots)$.
The set of all policies is denoted by $\Gamma$.
If $\gamma$ is such that $S[m + 1] = S(Q[m], H[m + 1])$, then $\gamma$ is a stationary policy.
The set of all stationary policies is denoted as $\Gamma_{s}$.
In this chapter, we restrict attention to $\Gamma_{s}$ in light of the discussion in Section \ref{chap5:sec:cmdpformulation}, which also holds for this model.
We also note that since $H[m]$ is assumed to be IID, the process $(Q[m], m \geq 0)$ is a Markov chain, if $\gamma \in \Gamma_{s}$.

The transmitter expends $P(h, s)$ units of power when transmitting $s$ bits, when the fade state is $h$.
We note that $P(h, s)$ is a function of the fading gain $h^{2}$, when the fade state is $h$.
Motivated by many examples (see \cite{berry} and \cite{elif}) of $P(h,s)$, we assume that $\forall h \in \mathcal{H}$, $P(h,s)$ satisfies the following properties:
\begin{description}
\item[C1 : ]$P(h,0) = 0$, and 
\item[C2 : ]$P(h,s)$ is non-decreasing and convex in $s$, for $s \in \brac{0, \dots, S_{max}}$.
\end{description}
The average power for a policy $\gamma$ is
\begin{equation}
  \overline{P}(\gamma, q_{0}) \stackrel{\Delta} = \limsup_{M \rightarrow \infty} \frac{1}{M} \Exp \bras{\sum_{m = 1}^{M} P(H[m], S[m]) \middle \vert Q[0] = q_{0}}.
  \label{chap5fading:eq:avgpower}
\end{equation}
The average queue length for a policy $\gamma$ is
\begin{equation}
  \overline{Q}(\gamma, q_{0}) \stackrel{\Delta} = \limsup_{M \rightarrow \infty} \frac{1}{M} \Exp \bras{\sum_{m = 0}^{M - 1} Q[m] \middle \vert Q[0] = q_{0}}.
  \label{chap5fading:eq:avgqlength}
\end{equation}
We consider the optimal tradeoff of $\overline{P}(\gamma, q_{0})$ with $\overline{Q}(\gamma, q_{0})$ for this model.
The optimal tradeoff between $\overline{P}(\gamma, q_{0})$ and average delay can be obtained from this using Little's law and is discussed in the following.
We note that I-model considered in Chapter 4 is a special case of this model, with a single fade state.

\subsection{System model - Real valued queue length evolution}
\label{chapfading:sysmodel:realval}
We state only the differences from the model discussed in the previous section.
We assume that for $m \geq 1$, $A[m] \in [0, A_{max}]$, $S[m] \in [0, S_{max}]$, and $q_{0} \in \mathbb{R}_+$.
Hence, the queue length $Q[m] \in \mathbb{R}_+, \forall m \geq 0$.
The function $P(h,s)$ is assumed to satisfy the following properties:
\begin{description}
\item[RC1 : ]{$P(h,0) = 0,$ for every $h \in \mathcal{H}$,}
\item[RC2 : ]{$P(h,s)$ is an increasing, strictly convex function in $s$, for $s \in [0, S_{max}]$, for every $h \in \mathcal{H}$.}
\end{description}
The average power and average queue length are as in \eqref{chap5fading:eq:avgpower} and \eqref{chap5fading:eq:avgqlength} respectively.

The tradeoff of average power and average queue length, for such models without admission control,  has been studied by Berry and Gallager \cite{berry}, Neely \cite{neely_mac}, Goyal et al. \cite{munish}, Bettesh and Shamai \cite{bettesh}, Biyikoglu et al. \cite{elif} as well as many others.
We note that this model is similar to that considered by Berry and Gallager \cite{berry}, except that in \cite{berry}, $S_{max} = \infty$.
We recall that Berry and Gallager obtain that any sequence of policies $\gamma_{k}$, for which $\Pgk$ is at most $V_{k}$ more than the above minimum power required for queue stability, has $\Qgk = \Omega\nfrac{1}{\sqrt{V_{k}}}$, as $V_{k} \downarrow 0$.
We also note this model is similar to R-model considered in Chapter 4, except that in Chapter 4 we considered the case with a single fade state.

In the following, the model in the previous section, where the queue length evolution is assumed to be on the non-negative integers, is called the I-model, while the model described here, where the queue length evolution is assumed to be on the non-negative real numbers, is called the R-model.
We note that as in Chapter 4, R-model with fading and $P(h,s)$ being strictly convex is usually used as an approximation for I-model.
When used as an approximation, $P(h,s)$ for R-model coincides with $P(h,s)$ for I-model for $s \in \brac{0, \dots, S_{max}}, \forall h$.

\subsection{An example }
\label{chap5fading:sec:simpleeg}
Throughout this chapter, to illustrate the results for I-model and R-model, we use the following example.
The number of packets $A[1]$ which arrive in a slot is assumed to be distributed according to a Binomial$(A_{max},p)$ distribution, with arrival rate $r_{a} kb/s$.
The rate of service $r$ in $kb/s$ is assumed to be $200\log_{10}\brap{1 + SNR}$, where $SNR$ is the received signal to noise ratio.
We assume that $SNR = \frac{h^{2} P}{L}$, where $h^{2}$ is the fading gain, $P$ is the transmit power, and $L$ encompasses the loss due to attenuation as well as noise power.

We assume that the slots are of duration $2ms$.
We also assume that each packet has a size of $100$-bits.
Then the arrival rate of packets in a slot is $\lambda = \frac{r_{a}}{50}$.
We assume that if $h^{2} = 1$ and $P = 1W$ then $r = 50$.
Therefore, if $P(h, r)$ is the transmit power as a function of the fade state and the rate, we have that $P(h, r) = \frac{1.28}{h^{2}}\brap{10^{r/200} - 1}$.
We note that in one slot, the number of bits served is $2r$.
We assume that the transmitter, in each slot, can choose its transmission rate in the set $\brac{0, 50, 100} kb/s$.
To fit this example to our model, we express the queue length in units of $100$ bits.
Then in each slot, we have a Binomial arrival process of $100$-bit packets and service of $s$ $100$-bit packets, where $s \in \brac{0, 1, 2}$.
The transmit power as a function of $h$ and $s$ is $P(h,s) = \frac{1.28}{h^{2}}\brap{10^{50s/200} - 1}, s \in \brac{0, 1, 2}$.
We note that the average queue length, as defined, is in units of $100$ bits.

For R-model, we assume the same distribution for $A[1]$.
The set of possible batch sizes is assumed to be $ \in [0, 2]$.
The transmit power as a function of $h$ and $s$ is assumed to be $P(h,s) = \frac{1.28}{h^{2}}\brap{10^{50s/200} - 1}$ but for $s \in [0, 2]$.

The above example uses a similar model for the rate as a function of the $SNR$ as the example in Section \ref{chap4:sec:motivatingeg}.
However, in this chapter, we consider a slot level model, which models the system on a faster time scale compared to the continuous time model in Chapter 2.
Consideration of the faster time scale is necessary since we are interested in the average delay advantage that can be achieved by scheduling the packets in accordance with the channel variations in each slot.

\subsection{Overview}
As in the previous chapter, the objective in this chapter is to obtain an asymptotic characterization of the minimum average queue length as the average power is a small $V$ more than the minimum average power required for stability.
We formulate the tradeoff problem for I-model and R-model in Section \ref{chap5fading:sec:problem} for a set of admissible policies, whose definition is similar to that in Chapter 4.
We also obtain the infimum of the average power over the set of admissible policies, which is also the minimum average power required for stability of the queue, and discuss its properties in the same section.
The asymptotic analysis of the tradeoff problem is then carried out in Section \ref{chap5fading:sec:asymp}.
For I-model, we show that depending on the value of $\lambda$, three cases arise, which are similar to those for the I-model in Chapter 4.
For the first case, through numerical experiments, we show that the minimum average queue length does not increase to infinity for an admissible policy which achieves the infimum of the average power over the set of admissible policies.
For the second and third cases, as in Chapter 4, we show that the minimum average queue length increases as $\Theta\brap{\log\nfrac{1}{V}}$ and $\Theta\nfrac{1}{V}$ as $V \downarrow 0$.
For R-model, we show that the minimum average queue length is $\Omega{\nfrac{1}{\sqrt{V}}}$.
We note that this is the same as the Berry-Gallager lower bound, but the set of admissible policies that we consider is a subset of the set of admissible policies considered in \cite{berry}.
For the example in Section \ref{chap5fading:sec:simpleeg}, we provide some numerical results to illustrate the bounds in Section \ref{chap5fading:sec:numerical}.
As in Chapter 4, we also obtain an asymptotic $\log\nfrac{1}{V}$ lower bound on the minimum average queue length for I-model, when $(A[m])$ and $(H[m])$ are assumed to be ergodic in Section \ref{chap5fading:sec:ergodicextension}.

We then consider queueing models, I-model-U and R-model-U, which are similar to I-model and R-model, but with admission control, in Section \ref{sec:systemmodel_modelus}.
The tradeoff problems for I-model-U and R-model-U, are formulated in Section \ref{sec:problem} and its asymptotic analysis is carried out in Section \ref{sec:lowerbound}.
We show that the minimum average queue length increases as $\Theta\brap{\log\nfrac{1}{V}}$ when the average service cost is $V$ more than the minimum, when $V \downarrow 0$ and with a lower bound constraint on the utility of average throughput.

We consider a single hop network model, which is an extension of R-model, in Section \ref{chap5fading:sec:singlehop}.

\section{{Problem formulation for I-model and R-model}}
\label{chap5fading:sec:problem}
Our objective is to characterize the minimum average queue length for a given constraint $P_{c}$ on the average transmit power.
The following formulation is for both the I-model and the R-model.
The tradeoff problem is 
\begin{equation}
  \mini_{\gamma \in \Gamma} \overline{Q}(\gamma, q_{0}), \text{ such that } \overline{P}(\gamma, q_{0}) \leq P_{c}.
  \label{chap5fading:eq:tradeoffproblem_init}
\end{equation}
We note that as in Section \ref{chap5:sec:cmdpformulation} it is possible to formulate a CMDP for the above problem.
The state space of the CMDP is $\sZ \times \mathcal{H}$, the action space at each $(q,h) \in \sZ \times \mathcal{H}$ is the set of batch sizes, and the evolution of the process is as given in \eqref{chap5fading:eq:evolution}.

Similar to the quantities $c(\lambda)$ and $c_{R}(\lambda)$ defined for I-model and R-model in Chapter 4, it is possible to obtain quantities $c(\lambda)$ and $c_{R}(\lambda)$ for I-model and R-model discussed above.
The quantities $c(\lambda)$ and $c_{R}(\lambda)$ can be interpreted as the minimum average power required for mean rate stability of the queue for I-model and R-model respectively.
Then it is possible to show that if $P_{c} > c(\lambda)$ for I-model (or $P_{c} > c_{R}(\lambda)$ for R-model), then there exists an optimal stationary policy $\gamma^*$ for the above problem with stationary distribution $\pi^*$.
So in the following we consider the above problem for policies in $\Gamma_{s}$.

For I-model, as discussed in Section \ref{chap5:sec:cmdpformulation}, from Ma et al. \cite{ma}, if $P_{c}$ is such that there exists a Lagrange multiplier $\beta_{P_{c}}\geq 0$ and the average cost optimal policy $\gamma_{\beta_{P_{c}}}$ for a MDP with single stage cost $q + \beta_{P_{c}} P(h, s)$ in state $(q,h)$ has $\overline{P}(\gamma_{\beta_{P_{c}}}, q_{0}) = P_{c}$, then $\gamma_{\beta_{P_{c}}}$ is optimal for \eqref{chap5fading:eq:tradeoffproblem_init}.
The set of all such $P_{c}$ is denoted as $\mathcal{O}^{u}$ (similar to $\mathcal{O}^{u}$ in Chapter 4).
It can then be shown that for all $P_{c} \in \mathcal{O}^{u}$, there exist optimal stationary deterministic policies for \eqref{chap5fading:eq:tradeoffproblem_init}.
Furthermore, such optimal stationary deterministic policies are such that the service batch size $s(q,h)$ is monotonically non-decreasing in $q$ for every $h$.
Then, as in Chapter 4, we consider the above tradeoff problem only for policies $\gamma$ in an admissible set $\Gamma_{a}$ for all values of $P_{c}$, where the definition of admissible policies is motivated by the above monotonicity property of any stationary deterministic optimal policy for $P_{c} \in \mathcal{O}^{u}$.
Since R-model is an approximation to the I-model, the definition of admissible policies for R-model is motivated by the monotonicity property of any stationary deterministic optimal policy for $P_{c} \in \mathcal{O}^{u}$ for I-model.

We now state the properties which are satisfied by admissible policies for both I-model and R-model.
A policy $\gamma$ is \emph{stable}, if (i) the Markov chain $(Q[m], m \geq 0)$ under $\gamma$ is irreducible, aperiodic, and positive Harris recurrent with stationary distribution $\pi$, and (ii) $\overline{Q}(\gamma, q_{0}) < \infty$.
A policy $\gamma$ is \emph{admissible} if: 
\begin{description}
\item[G1 :] it is stable, and, 
\item[G2 :] the average service rate in state $q$, $\overline{s}(q) \stackrel{\Delta} = \Exp_{\pi_{H}} \Exp_{S|q,H} S(q, H)$ is non-decreasing in $q$
\footnote{We note for a stationary deterministic policy $s(q, h)$ is non-decreasing in $q$ for every $h$. Therefore, $\Exp_{\pi_{H}} s(q, H)$ is non-decreasing in $q$. Since, we are considering randomized policies, we assume that $\Exp_{\pi_{H}} \Exp_{S|q,H} S(q, H)$ is non-decreasing in $q$.}.
\end{description}

For an admissible policy $\gamma$, we have that the performance measures $\overline{Q}(\gamma, q_{0})$ and $\overline{P}(\gamma, q_{0})$ are independent of the initial queue state $q_{0}$ and exist as limits.
Therefore, in the following, these performance measures are denoted by $\Qg$ and $\Pg$ respectively.

So in the following we consider the problem TRADEOFF:
\begin{equation}
  \mini_{\gamma \in \Gamma_{a}} \overline{Q}(\gamma), \text{ such that } \overline{P}(\gamma) \leq P_{c}.
  \label{chap5fading:eq:tradeoffprob}
\end{equation}
The optimal value of TRADEOFF is denoted as $Q^*(P_{c})$.
We note that whenever $P_{c} \in \mathcal{O}^{u}$, since there exists an optimal admissible policy $Q^*(P_{c})$ is the solution to \eqref{chap5fading:eq:tradeoffproblem_init}.

For an admissible policy $\gamma$, we note that since the arrival rate is constant, from Little's law the average delay for $\gamma$ is $\frac{\overline{Q}(\gamma)}{\lambda}$.
The tradeoff of average delay with average power can be obtained as $\frac{Q^*(P_{c})}{\lambda}$.

As in Chapter 4, it is possible to consider a larger class of policies $\Gamma_{a,M}$, which is obtained by mixing or time sharing of policies in $\Gamma_{a}$.
Let $Q^*_{M}(P_{c})$ denote the optimal value of the above problem, but with the minimization carried out over the set $\Gamma_{a,M}$.
We note that the asymptotic behaviour for $Q^*_{M}(P_{c})$ can be directly obtained from $Q^*(P_{c})$.
Therefore, in the following we consider the asymptotic characterization of $Q^*(P_{c})$ only.

If $P_{c}$ is such that the above problem is feasible, then by definition there exists a feasible admissible policy $\gamma$ such that $\Qg \leq Q^*(P_{c}) + \epsilon$.
Such a policy is called $\epsilon$-optimal in the following.

We note that for any admissible policy $\gamma$, $\overline{Q}(\gamma) = \Exp_{\pi}Q$ and $\overline{P}(\gamma) = \Expp\Exp_{H | Q}  \Exp_{S|Q,H} P(H, S(Q,H))$.
Since $Q$ and $H$ are independent, we also have that $\overline{P}(\gamma) = \Expp\Exp_{\pi_{H}}  \Exp_{S|Q,H} P(H, S(Q,H))$.
For any $\gamma \in \Gamma_{a}$, we note that the average arrival rate $\lambda$ has to be equal to the average service rate, i.e., $\lambda = \Expp \Exp_{H|Q}  \Exp_{S|Q,H} S(Q,H) = \Expp \overline{s}(Q)$.
Therefore, for $\gamma \in \Gamma_{a}$, $\overline{P}(\gamma)$ is lower bounded by the optimal value of
\begin{eqnarray}
  \mini_{\gamma \in \Gamma_{a}} & \Expp \Exp_{{H}|Q}  \Exp_{S|Q,H} P(H, S(Q, H)), \nonumber \\
  \text{such that } & \Expp \Exp_{{H}|Q} \Exp_{S|Q,H} S(Q, H) = \lambda,
  \label{chap5fading:eq:tradeoffprobratec}
\end{eqnarray}
since the only constraint is on the average service rate.
Now we note that $\Expp \Exp_{{H}|Q} \Exp_{S|Q,H} S(Q, H) = \Exp_{\pi_{H}} \Exp_{Q|H} \Exp_{S|Q,H} S(Q, H)$.
Then, we have that
\[\Exp_{Q|H} \Exp_{S|Q,H} S(Q, H) = \int_{q} \int_{s} s. dp_{s,q|H}. d\pi(q) = \int_{s} \int_{q} s .dp_{s,q|H} = \int_{s} s \int_{q} dp_{s,q|H},\] 
we have that $\Exp_{\pi_{H}} \Exp_{Q|H} \Exp_{S|Q,H} S(Q, H) = \Exp_{\pi_{H}} \Exp_{S|H} S$ where the conditional distribution of $S$ given $H$ depends upon the policy.
A similar procedure can be carried out on $\Expp \Exp_{H|Q}  \Exp_{S|Q,H} P(H, S(Q, H))$ which leads to $\Expp \Exp_{H|Q}  \Exp_{S|Q,H} P(H, S(Q, H)) = \Exp_{\pi_{H}} \Exp_{S|H} P(H, S)$.
Then the optimal value of \eqref{chap5fading:eq:tradeoffprobratec} is bounded below by the optimal value of 
\begin{eqnarray}
  \mini & \Exp_{\pi_{H}}  \Exp_{S|H} P(H, S),
  \label{chap5fading:eq:app_bg_0} \\
  \text{such that } & \Exp_{\pi_{H}}  \Exp_{S|H} S = \lambda, \nonumber
\end{eqnarray}
where we minimize over all possible conditional distributions for the batch size $S$ given $H$, irrespective of the policy.
For the I-model, we denote the optimal value of \eqref{chap5fading:eq:app_bg_0} by $c(\lambda)$, while for the R-model we denote the optimal value of \eqref{chap5fading:eq:app_bg_0} by $c_{R}(\lambda)$ (we note that $c(\lambda)$ and $c_{R}(\lambda)$ are the minimum powers required for mean rate stability for the I-model and R-model respectively, see \cite{neely_mac}).
We note that for the R-model, the conditional distribution of the batch size has support on $[0, S_{max}]$, while for the I-model the conditional distribution has support on $\brac{0, \dots, S_{max}}$.
Hence, $c_{R}(\lambda) \leq c(\lambda), \forall \lambda \in [0, S_{max}]$.
We note that feasible solutions exist for the above problem only if $\lambda \leq S_{max}$.

We have that $\forall \gamma \in \Gamma_{a}$, $\overline{P}(\gamma) \geq c(\lambda)$ for the I-model, and $\overline{P}(\gamma) \geq c_{R}(\lambda)$ for the R-model.
From \cite[Theorem 1]{neely_mac}, we have that if $\lambda < S_{max}$, then for every $P_{c} > c(\lambda)$ (or every $P_{c} > c_{R}(\lambda)$ for the R-model), there exists an admissible policy $\gamma_{P_{c}}$, such that $\overline{P}(\gamma_{P_{c}}) \leq P_{c}$ and as $P_{c} \downarrow c(\lambda)$, $\overline{Q}(\gamma_{P_{c}})$ grows without bound.
Since, for an arrival rate of $\lambda$, $c(\lambda)$ (or $c_{R}(\lambda)$ for the R-model) can be approached arbitrarily closely by admissible policies, $c(\lambda)$ (or $c_{R}(\lambda)$ for the R-model) is the infimum of the average power for admissible policies.

\begin{figure}[h]
  \centering
  \includegraphics[width=120mm,height=70mm]{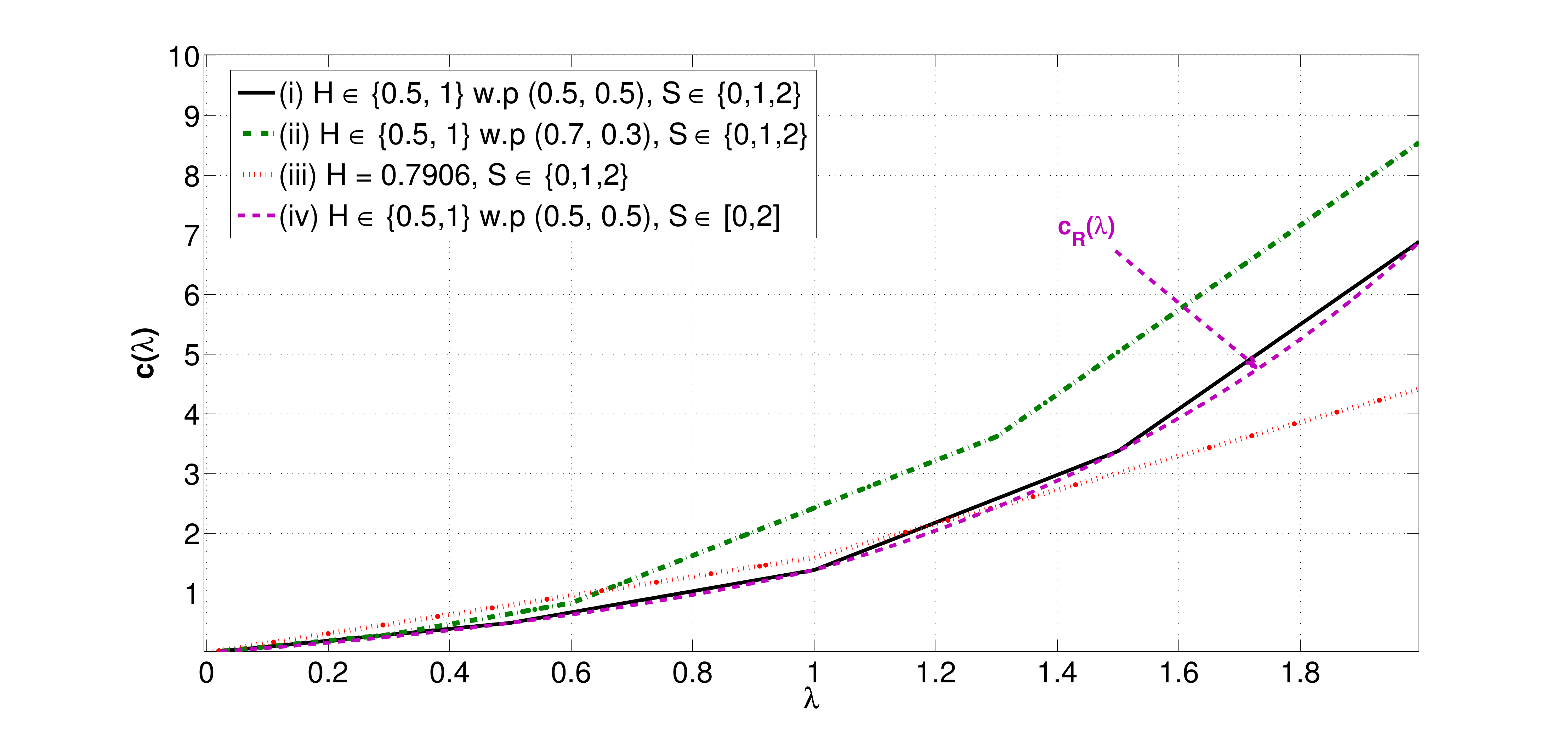}
  \caption{\small{The optimal value $c(\lambda)$ of problem \eqref{chap5fading:eq:app_bg_0} for I-model: i) with $H \in \brac{0.5, 1}$ with $\pi_{H}(0.5) = 0.5$ and $S \in \brac{0,1,2}$, ii) with $H \in \brac{0.5, 1}$ with $\pi_{H}(0.5) = 0.7$ and $S \in \brac{0,1,2}$, and iii) with $S \in \brac{0,1,2}$ and fade state fixed at $0.7906 = \sqrt{0.5\times 0.5^2 + 0.5\times 1^2}$ with the same average fading gain as (i). For R-model: (iv) with $H \in \brac{0.5, 1}$ with $\pi_H(0.5) = 0.5$ and $S \in [0,2]$.}}
  \label{chap5fading:fig:clambdaeg}
\end{figure}

For the I-model, since properties (C1) and (C2) are assumed to hold, from \cite[Section VII]{neely_mac}, we have that $c(\lambda)$ is a piecewise linear (a proof is given in Appendix \ref{chap5fading:app:clambda}), non-decreasing convex function, for $\lambda \in [0, S_{max}]$, with $c(0) = 0$.
Again from \cite{neely_mac}, $c_{R}(\lambda)$ is a non-decreasing, strictly convex function of $\lambda \in [0, S_{max}]$, with $c_{R}(0) = 0$.
For the example discussed in Section \ref{chap5fading:sec:simpleeg}, the function $c(\lambda)$ and $c_{R}(\lambda)$ are illustrated in Figure \ref{chap5fading:fig:clambdaeg}, for different $\mathcal{H}$ and $\pi_{H}$ as well as for the R-model and the I-model.
We observe the following: a) for cases (i) and (iii) in Figure \ref{chap5fading:fig:clambdaeg}, for the same mean fading gain $\ExpH H^{2}$, for $\lambda < 1.2$, having a fading gain larger than the mean with some positive probability leads to a smaller $c(\lambda)$; b) for case (iv) we see that $c_{R}(\lambda)$ for the R-model is strictly convex. The function $c_{R}(\lambda) \leq c(\lambda)$, and coincides with $c(\lambda)$ for $\lambda \in \brac{0.5, 1, 1.5}$; c) for both (i) and (ii) the function $c(\lambda)$ is piecewise linear, but the $\lambda$-s at which the slope changes is different and depends on $\pi_{H}$.
We consider another example of $c(\lambda)$ and $c_{R}(\lambda)$ for the example in Section \ref{chap5fading:sec:simpleeg} with $\mathcal{H} = \brac{0.1, 1}$ and $\pi_{H}(0.1) = 0.6$.
In this case, $c_{R}(\lambda) < c(\lambda)$ for all $\lambda \not \in \brac{0.4, 0.8, 1.4}$.
From these examples, we can conclude for the R-model a smaller minimum average power $c_{R}(\lambda)$ is sufficient for stability, compared with $c(\lambda)$ for the I-model.

\begin{figure}[h]
  \centering
  \includegraphics[width=120mm,height=70mm]{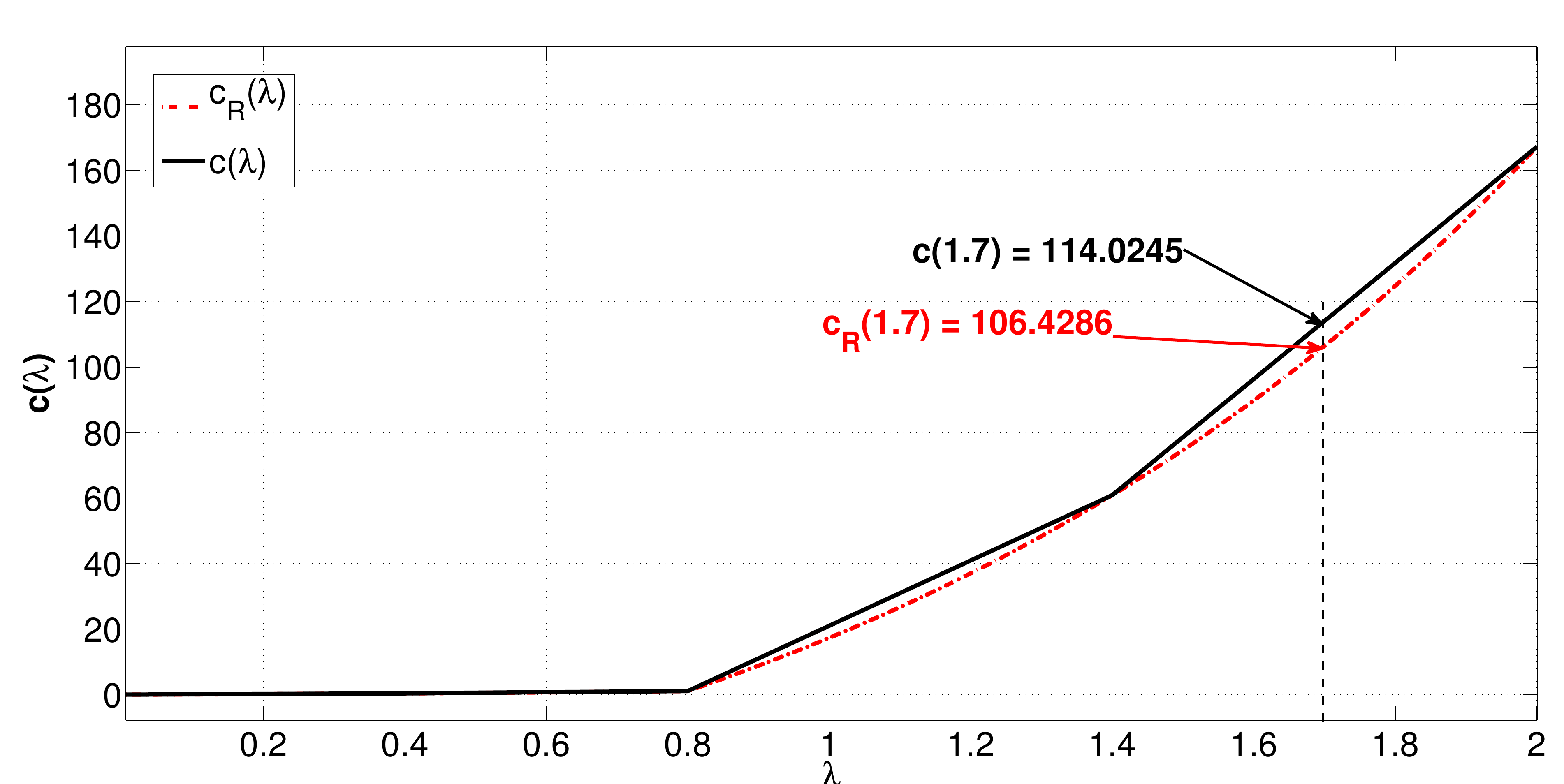}
  \caption{\small{The optimal value of problem \eqref{chap5fading:eq:app_bg_0}: $c(\lambda)$ for I-model and $c_{R}(\lambda)$ for R-model, with $H\in \brac{0.1, 1}$ and $\pi_{H}(0.1) = 0.6$; $c(1.7)$ is $107\%$ of $c_{R}(1.7)$.}}
  \label{chap5fading:fig:clambdaeg_2}
\end{figure}

As shown in Figures \ref{chap5fading:fig:clambdaeg} and \ref{chap5fading:fig:clambdaeg_2}, it is possible that $c_{R}(\lambda_{i}) = c(\lambda_{i})$ for some $\lambda_{i} \in [0, S_{max}]$.
We will see from the asymptotic analysis in the next section, that the asymptotic growth rate of minimum average queue length suggested by the R-model and the I-model for such $\lambda_{i}$ are different.
We note that, in general, it is not known for what $\lambda_{i}$, if any, $c_{R}(\lambda_{i}) = c(\lambda_{i})$.
However, if $|\mathcal{H}| = 1$, it is clear that $c_{R}(s) = c(s), \forall s \in \brac{0, \dots, S_{max}}$.

\section{Asymptotic bounds for I-model}
\label{chap5fading:sec:asymp}
In this section, we obtain an asymptotic characterization of $Q^*(P_{c})$ for I-model and R-model as $P_{c} \downarrow c(\lambda)$ and $P_{c} \downarrow c_{R}(\lambda)$ respectively.
For I-model, as in Chapter 4, we assume 
\begin{description}
\item[A1 :]{$Pr\brac{A[1] > S_{max}} > \epsilon_{a} > 0$.}
\item[A2 :]{$Pr\brac{A[1] = a} > 0$, for all $a \in \brac{0,\dots, A_{max}}$,}
\end{description}
For R-model as in Chapter 4, we assume
\begin{description}
\item[RA1 : ]{$Pr\brac{A[1] - S_{max} > \delta_{a}} > \epsilon_{a}$, for positive $\delta_{a}$ and $\epsilon_{a}$.}
\end{description}

We note that part of this analysis was presented in \cite{vineeth_ncc13_dt}.
We consider the I-model first.
Consider any $\gamma \in \Gamma_{a}$.
To obtain the asymptotic behaviour of $Q^*(P_{c})$ as $P_{c} \downarrow c(\lambda)$, we ascertain the asymptotic behaviour of $\pi(q)$ as $P_{c} \downarrow c(\lambda)$.
We note that $\pi(q)$ determines the average queue length and the average power given the policy. 
As we have seen in Chapter 4, it turns out that the asymptotic behaviour of $\pi(q)$, is determined by the average drift of the queue, $\lambda - \overline{s}(q)$, at a queue length $q$.
So we proceed by relating the average power to the average drift $\lambda - \overline{s}(q)$.
We note that the average power used when the queue length is $q$ is $\Exp_{\pi_{H}}  \Exp_{S|q,H} P(H, S(Q, H))$, which is bounded below by the optimal value of
\begin{eqnarray*}
  \mini & & \Exp_{\pi_{H}}  \Exp_{S|H} P(H, S), \\
  \text{such that } & & \Exp_{\pi_{H}}  \Exp_{S|H} S = \bar{s}(q),
\end{eqnarray*}
where we have considered all possible conditional distributions on the batch size with support on $\brac{0, \dots, S_{max}}$, subject only to the constraint that the average service rate is $\bar{s}(q)$.
The above optimization problem is the same as \eqref{chap5fading:eq:app_bg_0} except that the constraint is now $\bar{s}(q)$ instead of $\lambda$.
Therefore, the average power used when the queue length is $q$ is bounded below by $c(\overline{s}(q))$.
We note that any feasible policy $\gamma$ for TRADEOFF has $\overline{P}(\gamma) \leq P_{c}$.
Then for that $\gamma$, $\Expp c(\overline{s}(Q)) \leq \overline{P}(\gamma) \leq P_{c}$.
We also note that from the convexity of $c(s)$, $\Expp c(\overline{s}(Q)) \geq c(\lambda)$, since $\Expp \overline{s}(Q) = \lambda$.
Now as $P_{c} \downarrow c(\lambda)$, for any sequence of feasible policies for TRADEOFF, $\Exp c(\overline{s}(Q)) \downarrow c(\lambda)$.

The behaviour of ${Q}^*(P_{c})$ as $P_{c} \downarrow c(\lambda)$ is observed to depend on the relationship of $\lambda$ with $c(s), s \in [0, S_{max}]$.
Since $c(s)$ is piecewise linear, we can define a sequence of intervals $[a_{p}, b_{p}]$, $p \in \brac{1,\dots, P}$, with $a_{p + 1} = b_{p}$, $a_{1} = 0$, and $b_{P} = S_{max}$.
The sequence of intervals is such that for $s \in [a_{p}, b_{p}]$, $c(s)$ is linear.
The following three cases arise:
\begin{enumerate}
\item $a_{1} = 0 < \lambda < b_{1}$,
\item $a_{p} < \lambda < b_{p}$, $p > 1$, and,
\item $\lambda = a_{p}, p > 1$.
\end{enumerate}
We note that $c_{R}(a_{p}) = c(a_{p})$ in Figures \ref{chap5fading:fig:clambdaeg} and \ref{chap5fading:fig:clambdaeg_2}.
For Case 1, through numerical examples, we illustrate that $Q^*(P_{c})$ does not grow to infinity as $P_{c} \downarrow c(\lambda)$.
We obtain an asymptotic lower bound for Case 2 which is the asymptotic lower bound to the \emph{super-fast} $\log\nfrac{1}{V}$ upper bound observed for the sequence of policies constructed by Neely in \cite[Corollary 2]{neely_mac}, but for admissible policies.
We also obtain an asymptotic lower bound for Case 3 which is used to illustrate another difference between the asymptotic behaviours of $Q^*(P_{c})$ in the R-model and I-model.

\begin{figure*}
  \centering
  \includegraphics[width=160mm,height=35mm]{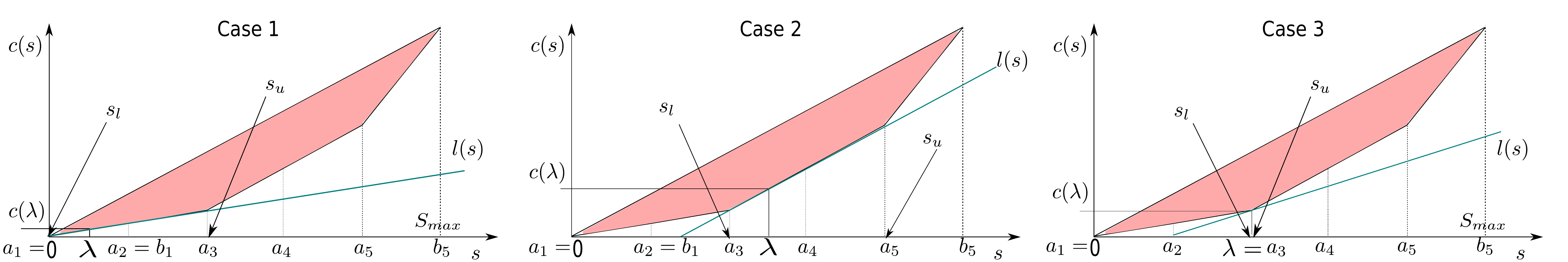}
  \caption{\small{Illustration of the relationship between $\lambda$, $s_{l}$, and $s_{u}$ along with the minimum average cost $c(\lambda)$ and the line $l(s)$ for the three cases}}
  \label{chap5fading:fig:threecases}
\end{figure*}

For Cases 1 and 2, let $s_{l} \stackrel{\Delta} = a_{p}$ and $s_{u} \stackrel{\Delta} = b_{p}$, while for Case 3 let $s_{l} = s_{u} \stackrel{\Delta} = a_{p}$\footnote{We note that unlike in Chapter 4, $s_{l}$ and $s_{u}$ need not be integers.}.
An example is shown in Figure \ref{chap5fading:fig:threecases}.
We define the line $l(s) : [0,S_{max}] \rightarrow \mathbb{R}_+$ as follows:
\begin{enumerate}
\item If $s_{l} < \lambda < s_{u}$, then $l(s)$ is the line through $(s_{l},c(s_{l}))$ and $(s_{u},c(s_{u}))$.
\item If $s_{l} = \lambda = s_{u} = a_{p}$ for some $p > 1$, then $l(s)$ is a line through $(\lambda, c(\lambda))$ with slope $m$ chosen such that $\frac{c(a_{p}) - c(a_{p - 1})}{a_{p} - a_{p - 1}} < m < \frac{c(a_{p + 1}) - c(a_{p})}{a_{p + 1} - a_{p}}$.
\end{enumerate}
We note that $\Exp l(\overline{s}(Q)) = c(\lambda)$.
We now present asymptotic lower bounds for ${Q}^*(P_{c})$, which follow directly from the asymptotic lower bounds in Lemmas \ref{chap5:lemma:tprob_2} and \ref{chap5:lemma:case3}.

\begin{lemma}
  For Case 2, given any sequence of admissible policies $\gamma_{k}$ with $\overline{P}(\gamma_{k}) - c(\lambda) = V_{k} \downarrow 0$, we have that $\overline{Q}(\gamma_{k}) = \Omega\brap{\log\nfrac{1}{V_{k}}}$.
  \label{chap5fading:lemma:case2}
\end{lemma}
\begin{proof}
  We first note that for the sequence $\gamma_{k}$, since $\overline{P}(\gamma_{k}) - c(\lambda) = V_{k}$, we have that $U_{k} \Deq \Expp c(\overline{s}(Q)) - c(\lambda) \downarrow 0$, with $U_{k} \leq V_{k}$.
  The rest of the proof is very similar to that of Lemma \ref{chap5:lemma:tprob_2}.
  With $\overline{s}(q)$ replacing $\Exp S(q)$ and $U_{k}$ replacing $V_{k}$ (with an inequality) in the proof of Lemma \ref{chap5:lemma:tprob_2}, and $m$ being redefined as the tangent of the angle made by the line passing through $(a_{p - 1}, c(a_{p - 1}))$ and $(a_{p}, c(a_{p}))$ with $l(s)$, and proceeding as in the proof of Lemma \ref{chap5:lemma:tprob_2} we have that $\overline{Q}(\gamma_{k}) = \Omega\brap{\log\nfrac{1}{U_{k}}} = \Omega\brap{\log\nfrac{1}{V_{k}}}$, since $U_{k} \leq V_{k}$.
\end{proof}

We note that an asymptotic upper bound for Case 2 can be obtained from a sequence of TOCA policies from \cite{neely_mac}.
However, TOCA policies are not admissible.
Therefore, we now present an asymptotic upper bound for a sequence of policies, which is similar to the sequence of buffer partitioning policies proposed by Berry and Gallager \cite{berry}.
\begin{lemma}
  Let a policy $\gamma$ be defined as follows.
  When the queue length is $q$ and fade state is $h$, $\gamma$ serves a batch size $\min(q, \tilde{S}(q, h))$, where 
  \begin{equation*}
    \tilde{S}(q, h) \sim 
    \begin{cases}
      p_{l}(h), \text{ for } 0 \leq q < q_{v}, \\
      p_{u}(h), \text{ for } q_{v} \leq q.
    \end{cases}
  \end{equation*}
  where $q_{v} > 0$, $p_{l}(h)$ and $p_{u}(h), \forall h \in \mathcal{H}$ are the optimizing distributions for \eqref{chap5fading:eq:app_bg_0} with the constraint being $s_{l}$ and $s_{u}$ respectively.
  We obtain a sequence of policies $\gamma_{k}$, by choosing $q_{v} = \log\nfrac{1}{V_{k}}$, where $V_{k} < 1$ is a sequence decreasing to zero.
  Then for Case 2, $\gamma_{k}$ is a sequence of admissible policies, such that $\overline{P}(\gamma_{k}) - c(\lambda) = \mathcal{O}(V_{k})$ and $\Qgk = \mathcal{O}\brap{\log\nfrac{1}{V_{k}}}$.
  \label{chap5fading:lemma:case2upperbound}
\end{lemma}
The proof of this asymptotic upper bound is given in Appendix \ref{chap5fading:app:case2upperbound}.
We note that the proof is quite similar to that of Lemma \ref{chap5:lemma:case2upperbound}.

We then have the following result.
\begin{proposition}
  For Case 2, we have that $Q^*(P_{c,k}) = \Theta\brap{\log\nfrac{1}{P_{c,k} - c(\lambda)}}$ where $P_{c,k} \downarrow c(\lambda)$ and $P_{c,k} = \Pgk$ for the sequence of policies in Lemma \ref{chap5fading:lemma:case2upperbound}.
\end{proposition}
The proof is very similar to that of Proposition \ref{chap5:proposition:case2} and is therefore omitted.

\begin{remark}
  \label{chap5fading:remark:case2}
  We now consider a sequence of policies generated from a modified form of the tradeoff optimal control algorithm (TOCA) \cite{neely_mac}.
  We note that in \cite{neely_mac}, the control variable is the power allocation, and the service rate is obtained as a function of the power allocation and the current fade state.
  The set of all possible power allocations ($\Pi$ in \cite{neely_mac}) is the same for all fade states.
  However, in our model the control variable is the batch size.
  Therefore, we modify TOCA such that the set of power allocations is a function of the fade state.
  For each fade state, the possible power allocations are such that the batch sizes takes all possible values in $\brac{0,\dots,S_{max}}$.
  
  The modified TOCA algorithm is again parametrized by positive numbers $w, \epsilon, \tilde{q},$ and $\beta$ as in \cite{neely_mac}.
  The algorithm chooses at each slot $m \geq 1$, the batch size $s_{TOCA}$ such that
  \begin{eqnarray*}
    s_{TOCA}[m] & = & \min\brap{\argmin_{s \in \brac{0,\dots,S_{max}}} \bigg \{\beta P(H[m], s) - {W}[m] s \bigg\}, Q[m - 1]},
  \end{eqnarray*}
  where
  \small
  \begin{eqnarray*}
    & & W[m] = \mathbb{I}\brac{Q[m - 1] \geq \tilde{q}}\bras{w e ^{w(Q[m - 1] - \tilde{q})} + 2 X[m - 1]} + \mathbb{I}\brac{Q[m - 1] < \tilde{q}}\bras{-w e ^{w(\tilde{q} - Q[m - 1])} + 2 X[m - 1]}.
  \end{eqnarray*}
  \normalsize
  We note that $s_{TOCA}[m] = 0$ if $W[m] \leq 0$.
  The sequence $X[m], m \geq 0$ is obtained from a \emph{virtual} queue which evolves according to
  \begin{equation*}
    X[m + 1] = \max(X[m] - s_{TOCA}[m + 1] - \epsilon \mathbb{I}\brac{Q[m] < \tilde{q}}, \, 0) + A[m + 1] + \epsilon\mathbb{I}\brac{Q[m] \geq \tilde{q}}.
  \end{equation*}
  As in \cite{neely_mac}, let $\delta_{max} = \max(A_{max}, S_{max})$.
  For our Case 2, let $0 < \epsilon < \min(\lambda - a_{p}, b_{p} - \lambda)$, $w = \frac{\epsilon}{\delta_{max}^{2}} e^{\frac{-\epsilon}{\delta_{max}}}$, and $\tilde{q} = \frac{2}{w}\log\brap{\beta}$.
  A sequence of TOCA policies $\gamma_{k}$ is generated by choosing a sequence $\beta_{k} = \frac{1}{V_{k}}$, for a sequence $V_{k} \downarrow 0$.
  The proof of \cite[Corollary 2]{neely_mac} extends to the above version of the TOCA algorithm, and we have that 
  \begin{eqnarray*}
    \overline{Q}(\gamma_{k}) = \mathcal{O}\brap{\log\nfrac{1}{V_{k}}},
    \overline{P}(\gamma_{k}) = c(\lambda) + \mathcal{O}\brap{V_{k}}.
  \end{eqnarray*}
  Therefore, we obtain that for the sequence of policies $\gamma_{k}$, $\overline{Q}(\gamma_{k}) = \mathcal{O}\brap{\log\nfrac{1}{\overline{P}(\gamma_{k}) - c(\lambda)}}$.
  We note that $W[m]$ is a non-decreasing function of $q$, where $Q[m - 1] = q$.
  Since $s_{TOCA}[m]$ is a non-decreasing function of $W[m]$, we have that $s_{TOCA}[m]$ is a non-decreasing function of $Q[m - 1]$.
  However, $\gamma_{k}$ is not a sequence of admissible policies, since the policies $\gamma_{k}$ depend on the auxiliary variable $X[m - 1]$.
  We note that the above upper bound is an upper bound on the optimal value of \eqref{chap5fading:eq:tradeoffproblem_init}.
  Furthermore, as in Remark \ref{chap5:remark:toca}, for any subsequence $P_{c,k}$ of $\mathcal{O}^{u}$ such that there exists a constant $m \geq 1$ and a subsequence $P_{TOCA,k}$ of $\overline{P}(\gamma_{k})$ such that $P_{TOCA,k} - c(\lambda) \leq m\brap{P_{c,k} - c(\lambda)}$, we have that $Q^*(P_{c,k}) \leq Q(\gamma_{k}) = \mathcal{O}\brap{\log\nfrac{1}{P_{TOCA,k} - c(\lambda)}} = \mathcal{O}\brap{\log\nfrac{1}{P_{c,k} - c(\lambda)}}$.
\end{remark}

\begin{lemma}
  For Case 3, given any sequence of admissible policies $\gamma_{k}$ with $\overline{P}(\gamma_{k}) - c(\lambda) = V_{k} \downarrow 0$, we have that $\overline{Q}(\gamma_{k}) = \Omega\nfrac{1}{V_{k}}$.
  \label{chap5fading:corollary:case3}
\end{lemma}
\begin{proof}
  We again note that for the sequence $\gamma_{k}$, since $\overline{P}(\gamma_{k}) - c(\lambda) = V_{k}$, we have that $U_{k} \Deq \Expp c(\overline{s}(Q)) - c(\lambda) \downarrow 0$, with $U_{k} \leq V_{k}$.
  The rest of the proof is very similar to that of Lemma \ref{chap5:lemma:case3}.
  With $\overline{s}(q)$ replacing $\Exp S(q)$ and $U_{k}$ replacing $V_{k}$ (with an inequality) in the proof of Lemma \ref{chap5:lemma:case3}, and $m$ being redefined as the tangent of the angle made by the line passing through $(a_{p}, c(a_{p}))$ and $(a_{p + 1}, c(a_{p + 1}))$ with $l(s)$, and proceeding as in the proof of Lemma \ref{chap5:lemma:case3} we have that $\overline{Q}(\gamma_{k}) = \Omega\nfrac{1}{U_{k}} = \Omega\nfrac{1}{V_{k}}$, since $V_{k} \leq U_{k}$.
\end{proof}

\begin{remark}
  The derivation of the relationship between the difference of the $\Pg$ and $c(\lambda)$ and the \emph{average} drift defined as $\sum_{q =  q_{d} + 1}^{\infty} \mathbb{E}\bras{{Q}[m + 1] - {Q}[m] | Q[m] = q} \pi(q)$, in the proof of Lemma \ref{chap5:lemma:case3}, is motivated by the approach in \cite{berry}.
  We note that in our proof, $q_{d}$ can be chosen arbitrarily by the choice of $\epsilon_{V}$ and then $\epsilon_{V}$ can be chosen so as to obtain the tightest asymptotic lower bound.
  However, in \cite{berry}, $q_{d}$ cannot be chosen arbitrarily. 
  In fact, $q_{d}$ is chosen as the queue length which has the maximal stationary probability of all queue lengths in the set $\brac{0,\dots,\ceiling{2\Qg}}$ for a policy.
  The freedom in the choice of $q_{d}$ enables us to derive the $\Omega\nfrac{1}{V}$ asymptotic lower bound.
\end{remark}

\begin{remark}
  \label{chap5fading:remark:case3ub}
  As in the analysis of Case 3 in Chapter 4, it is possible to show that a sequence of randomized policies $\brac{\gamma_{k}}$ achieves the asymptotic lower bound derived above for Case 3.
  The sequence of policies is parametrized by a sequence $V_{k} \downarrow 0$.
  For a particular $V_{k}$, the policy chooses 
  \begin{eqnarray*}
    S[m] = \min(S'(H[m]), Q[m - 1]),
  \end{eqnarray*}
  where $S'(H[m])$ is independently chosen for each $m$ and is distributed according to any conditional distribution of batch size given $H[m]$ which is optimal for \eqref{chap5fading:eq:app_bg_0} but with the rate constraint being $\lambda + V_{k}$.
  For a particular $V_{k}$, it can then be shown that $\Pgk = c(\lambda + V_{k})$ and $\Qgk \leq \frac{\sigma^{2} + 2S_{max}^{2} + \ExpH \Exp S(H)^{2} - \lambda^{2}}{2V_{k}}$.
  Then as $k \rightarrow \infty$ and $V_{k} \downarrow 0$, $\overline{Q}(\gamma_{k}) = \mathcal{O}\nfrac{1}{V_{k}}$ and $\overline{P}(\gamma_{k}) = c(\lambda) + \mathcal{O}(V_{k})$.
\end{remark}

We have the following result.
\begin{proposition}
  For Case 3, we have that $Q^*(P_{c,k}) = \Theta\nfrac{1}{P_{c,k} - c(\lambda)}$ where $P_{c,k} \downarrow c(\lambda)$ and $P_{c,k} = \Pgk$ for the sequence of policies in the above remark.
\end{proposition}
The proof is very similar to that of Proposition \ref{chap5:proposition:case2} and is therefore omitted.

\section{Asymptotic bounds for R-model}
We now consider the asymptotic behaviour of $Q^*(P_{c})$ in the asymptotic regime $\Re$ as $P_{c} \downarrow c_{R}(\lambda)$ for the R-model.
Similar to the I-model, it can be shown that the average power used when the queue length is $q$ is bounded below by $c_{R}(\overline{s}(q))$.
We again note that any feasible policy $\gamma$ for TRADEOFF has $\overline{P}(\gamma) \leq P_{c}$, and for that $\gamma$, $\Expp c_{R}(\overline{s}(Q)) \leq \overline{P}(\gamma) \leq P_{c}$.
We also note that from the convexity of $c_{R}(s)$, $\Expp c_{R}(\overline{s}(Q)) \geq c_{R}(\lambda)$, since $\Expp \overline{s}(Q) = \lambda$.
Now as $P_{c} \downarrow c_{R}(\lambda)$, for any sequence of feasible policies for TRADEOFF, $\Exp c_{R}(\overline{s}(Q)) \downarrow c_{R}(\lambda)$.
We note that the following result is similar to the Berry-Gallager lower bound, but is derived with the extra assumption G2.
\begin{lemma}
  For any sequence of admissible policies $\gamma_{k}$ with $\overline{P}(\gamma_{k}) - c_{R}(\lambda) = V_{k} \downarrow 0$, we have that $\overline{Q}(\gamma_{k}) = \Omega\nfrac{1}{\sqrt{V_{k}}}$. Therefore, $Q^*(P_{c}) = \Omega\nfrac{1}{\sqrt{P_{c} - c(\lambda)}}$ as $P_{c} \downarrow c_{R}(\lambda)$.
\label{chap5fading:lemma:realvalued}
\end{lemma}

\begin{proof}
  We note that for the sequence $\gamma_{k}$, since $\overline{P}(\gamma_{k}) - c_{R}(\lambda) = V_{k}$, we have that $U_{k} \Deq \Expp c_{R}(\overline{s}(Q)) - c_{R}(\lambda) \downarrow 0$, with $U_{k} \leq V_{k}$.
  The rest of the proof is similar to that of Proposition \ref{chap5:prop:realvalued_evolution_lb}.
  We follow all the steps in the proof of Proposition \ref{chap5:prop:realvalued_evolution_lb}, with $\bar{s}(q)$ replacing $\Exp S(q)$ and $U_{k}$ replacing $V_{k}$ (with an inequality) in the proof of Proposition \ref{chap5:prop:realvalued_evolution_lb}, to obtain that $\overline{Q}(\gamma_{k}) = \Omega\nfrac{1}{\sqrt{U_{k}}} = \Omega\nfrac{1}{\sqrt{V_{k}}}$.
  Let $\gamma'_{k}$ be a sequence of $\epsilon$-optimal policies for TRADEOFF for the sequence $P_{c,k}$.
  Then we have that $\overline{P}(\gamma'_{k}) \downarrow c_{R}(\lambda)$ and $\overline{Q}(\gamma'_{k}) = \Omega\nfrac{1}{\sqrt{P_{c,k} - c_{R}(\lambda)}}$.
  Since $\gamma'_{k}$ is $\epsilon$-optimal, we have that $Q^*(P_{c,k}) \geq \overline{Q}(\gamma'_{k}) - \epsilon$.
  Therefore, $Q^*(P_{c,k}) = \Omega\nfrac{1}{\sqrt{P_{c,k} - c(\lambda)}}$ as $P_{c,k} \downarrow c_{R}(\lambda)$.
\end{proof}

\begin{remark}
  A sequence of admissible policies $\gamma_{k}$ can be obtained which achieves the above asymptotic lower bound up to a logarithmic factor, as in Lemma \ref{chap5:lemma:rmodelupperbound}.
  The proof of this upper bound is very similar to that of Lemma \ref{chap5fading:lemma:case2upperbound} and is therefore omitted.
  A particular policy $\gamma$ in the sequence is defined as follows.
  The policy $\gamma$ serves a batch size $\min(q, \tilde{S}(q, h)$ when the queue length is $q$ and fade state is $h$, where
  \begin{equation*}
    \tilde{S}(q, h) \sim 
    \begin{cases}
      p_{-}(h), \text{ for } 0 \leq q < q_{v}, \\
      p_{+}(h), \text{ for } q_{v} \leq q < 2q_{v}, \\
      p_{\epsilon}(h), \text{ for } 2q_{v} \leq q, \\
    \end{cases}
  \end{equation*}
  where $q_{v} > 0$, $p_{-}(h)$, $p_{+}(h)$, and $p_{\epsilon}(h), \forall h \in \mathcal{H}$ are the optimizing distributions for \eqref{chap5fading:eq:app_bg_0} with the constraint being $\lambda - \epsilon_{V}$, $\lambda + \epsilon_{V}$, and $\lambda + \epsilon$ respectively.
  We obtain a sequence of policies $\gamma_{k}$, by choosing $\epsilon_{V}$ and $q_{v}$ from sequences $\epsilon_{V_{k}}$ and $q_{v_{k}}$ defined as follows.
  Let $\omega_{k} = \sqrt{V_{k}}$, where $V_{k} \downarrow 0$.
  Let $\epsilon_{V_{k}} = \omega_{k} A^{2}_{max} e^{\omega A_{max}}$ and $q_{v_{k}} = \frac{1}{\omega_{k}}\log\nfrac{1}{\epsilon^{3}_{V_{k}}}$.
\end{remark}

\begin{remark}
  As in Remark \ref{chap5fading:remark:case2}, for a sequence of policies $\gamma_{k}$ generated by the modified form of TOCA, it is possible to show as in \cite{neely_mac}, that for $V_{k} < \frac{1}{S_{max}}$, $\beta_{k} = \frac{1}{V_{k}}$, $\epsilon_{k} = \frac{1}{\sqrt{\beta_{k}}}$, $w_{k} = \frac{\epsilon_{k}}{\delta^{2}_{max}}e^{-\frac{\epsilon_{k}}{\delta_{max}}}$, and $\tilde{q}_{k} = \frac{6}{w_{k}}\log\nfrac{1}{\epsilon_{k}}$,
\begin{eqnarray*}
  \overline{Q}(\gamma) & = & \mathcal{O}\brap{\sqrt{\frac{1}{V_{k}}}\log\brap{\frac{1}{V_{k}}}}, \\
  \overline{P}(\gamma) & = & c(\lambda) + \mathcal{O}\brap{V_{k}},
\end{eqnarray*}
as $V_{k} \downarrow 0$.
We note that the above asymptotic upper bound applies only to problem \eqref{chap5fading:eq:tradeoffproblem_init}, since the sequence of policies considered above is not admissible.
\end{remark}

\section{A numerical example}
\label{chap5fading:sec:numerical}
\begin{figure}[h]
  \includegraphics[width=160mm,height=55mm]{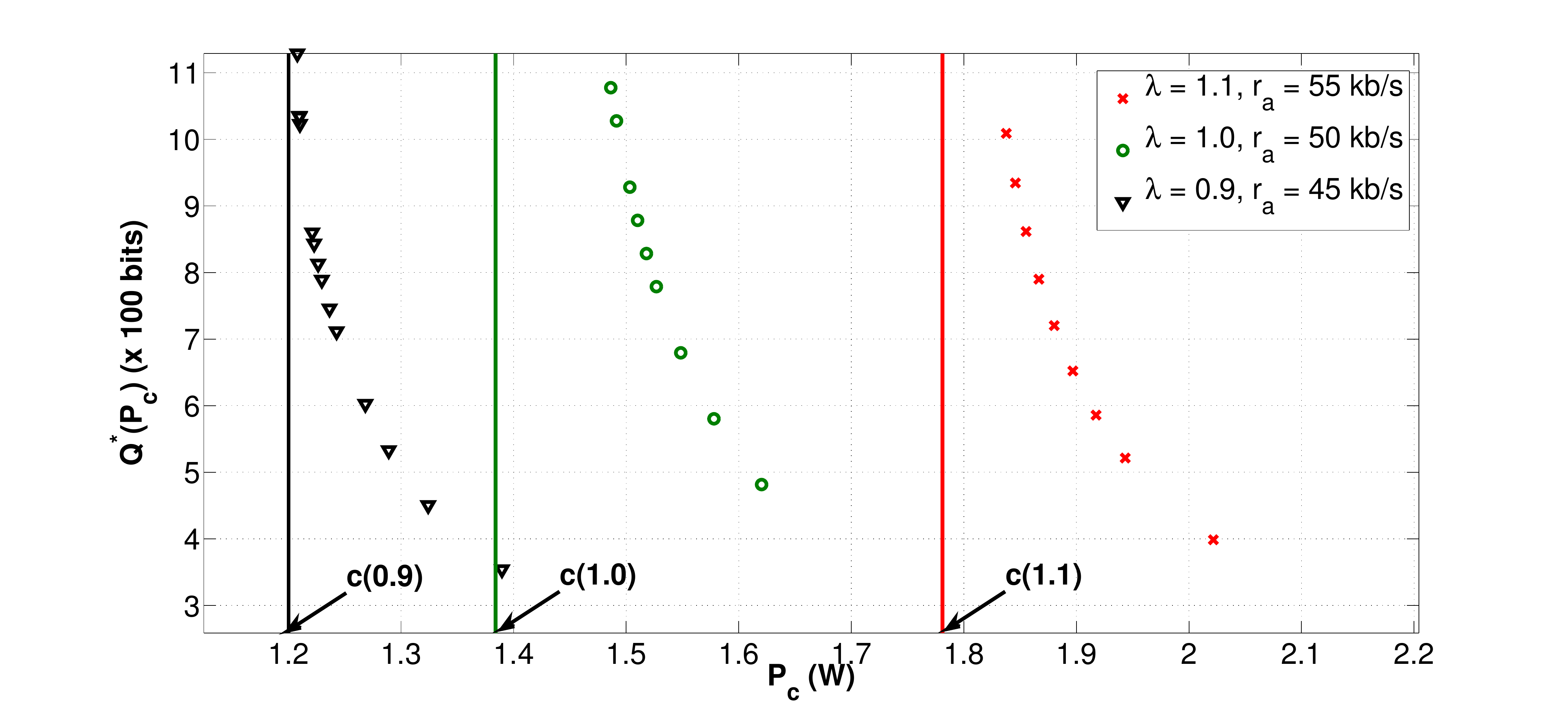}
  \caption{\small{The optimal tradeoff $Q^*(P_{c})$ for the system in Section \ref{chap5fading:sec:simpleeg} with $\mathcal{H} = \brac{0.5, 1}$ and $\pi_{H}(0.5) = 0.5$, for $\lambda \in \brac{0.9, 1.0, 1.1}$ with $c(0.9) = 1.2$, $c(1.0) = 1.384$, and $c(1.1) = 1.781$.}}
  \label{chap5fading:fig:comparelambda}
\end{figure}

\begin{figure}[h]
  \includegraphics[width=160mm,height=55mm]{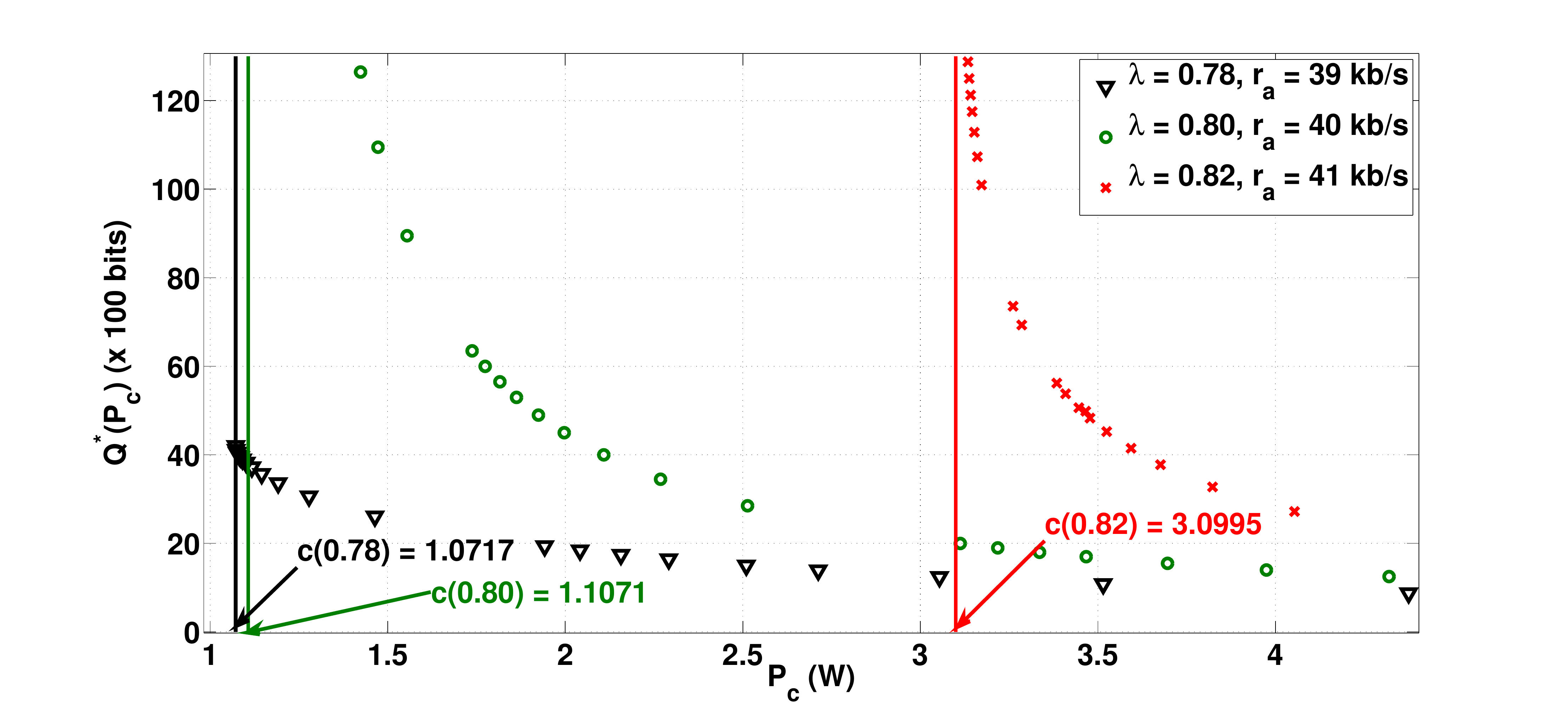}
  \caption{\small{The optimal tradeoff $Q^*(P_{c})$ for the system in Section \ref{chap5fading:sec:simpleeg} with $\mathcal{H} = \brac{0.1, 1}$ and $\pi_{H}(0.1) = 0.6$, for $\lambda \in \brac{0.78, 0.80, 0.82}$ with $c(0.78) = 1.0717$, $c(0.80) = 1.1071$, and $c(0.82) = 3.0995$.}}
  \label{chap5fading:fig:comparelambda_2}
\end{figure}

To illustrate the results obtained in the previous section, we plot the optimal tradeoff curve for the example in Section \ref{chap5fading:sec:simpleeg}.
We note that the results in this section apply only to I-model.
We assume that $A_{max} = 5$.
We consider first a case where the fading gain $\mathcal{H} = \brac{0.5, 1}$ with $\pi_{H}(0.5) = 0.5$.
The tradeoff curves $Q^*(P_{c})$ for $\lambda \in \{0.9, 1.0, 1.1\}$, are shown in Figure \ref{chap5fading:fig:comparelambda}.
We note that these arrival rates correspond to bit arrival rates of $45, 50$, and $55$ kb/s respectively.
Each point in the tradeoff curves is obtained by numerical solution of a suitably truncated MDP with state being the current queue length and the current fade state, and single stage cost $q + \beta P(h,s)$, where $\beta > 0$ is a Lagrange multiplier.
Each tradeoff curve is obtained by varying $\beta$.
From Figure \ref{chap5fading:fig:clambdaeg} we have that $c(0.9) = 1.20$, $c(1.0) = 1.384$, and $c(1.1) = 1.781$.
From the asymptotic characterization of $Q^*(P_{c})$, we have that for $\lambda = 1.0$, $Q^*(P_{c})$ increases as $1/(P_{c} - 1.384)$, while for $\lambda = 0.9$ and $1.1$, $Q^*(P_{c})$ increases as $\log\nfrac{1}{P_{c} - c(\lambda)}$.
In Figure \ref{chap5fading:fig:comparelambda}, we observe that for the same average queue length, the difference between the average service cost and $c(\lambda)$ increases from $\lambda = 0.9$ to $\lambda = 1.0$ and then decreases.
This difference is even more pronounced in Figure \ref{chap5fading:fig:comparelambda_2}, where $\lambda$ is increased from $0.78$ to $0.80$ and then to $0.82$ for the example in Section \ref{chap5fading:sec:simpleeg} with $\mathcal{H} = \brac{0.1, 1}$ and $\pi_{H}(0.1) = 0.6$.

\begin{figure}[h]
  \includegraphics[width=160mm,height=55mm]{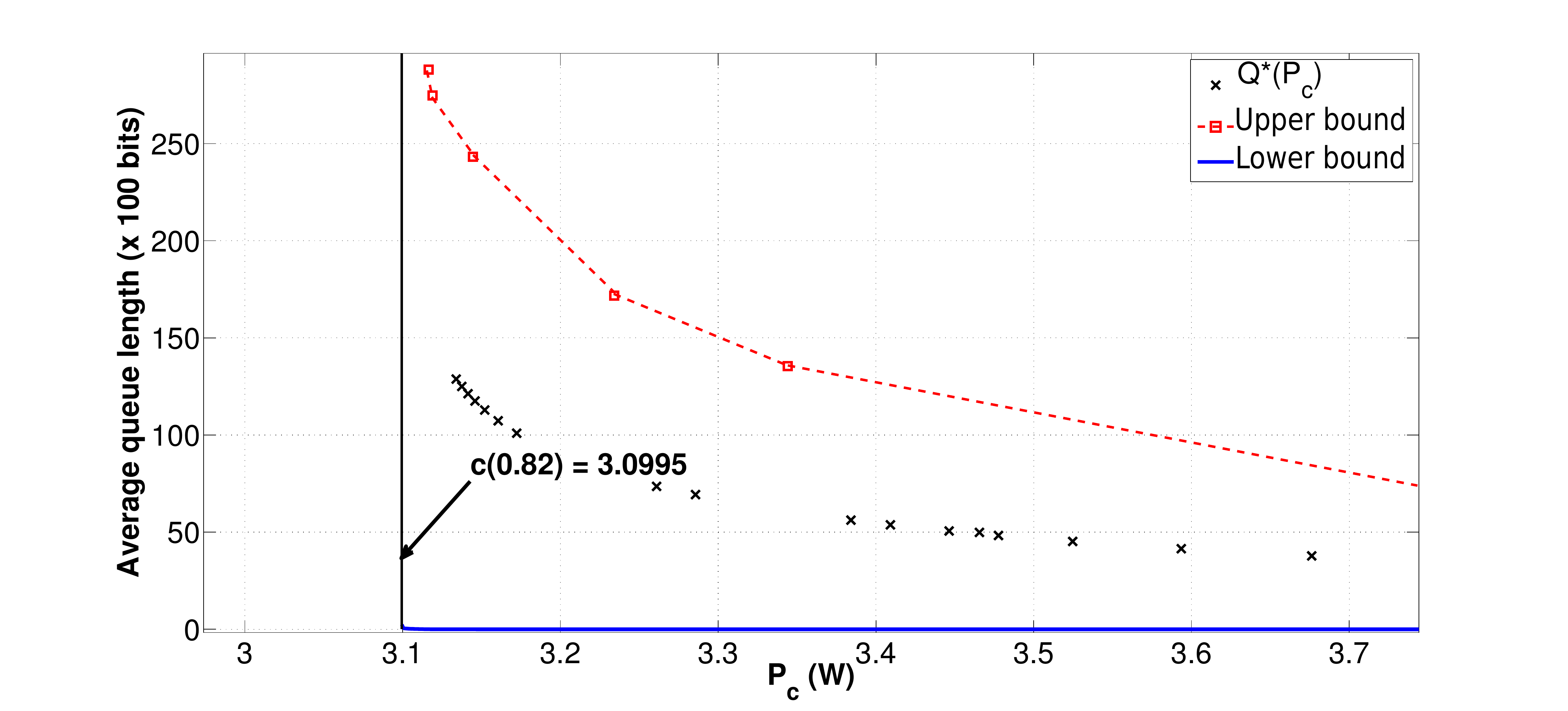}
  \caption{\small{Comparison of the optimal tradeoff $Q^*(P_{c})$ with the lower bound from Lemma \ref{chap5:lemma:tprob_2} and the upper bound from Remark \ref{chap5fading:remark:case2} for the system in Section \ref{chap5fading:sec:simpleeg}, for $\lambda  = 0.82$, for $\mathcal{H} = \brac{0.1, 1}$ and $\pi_{H}(0.1) = 0.6$}}
  \label{chap5fading:fig:case2bounds}
\end{figure}

\begin{figure}[h]
  \includegraphics[width=160mm,height=55mm]{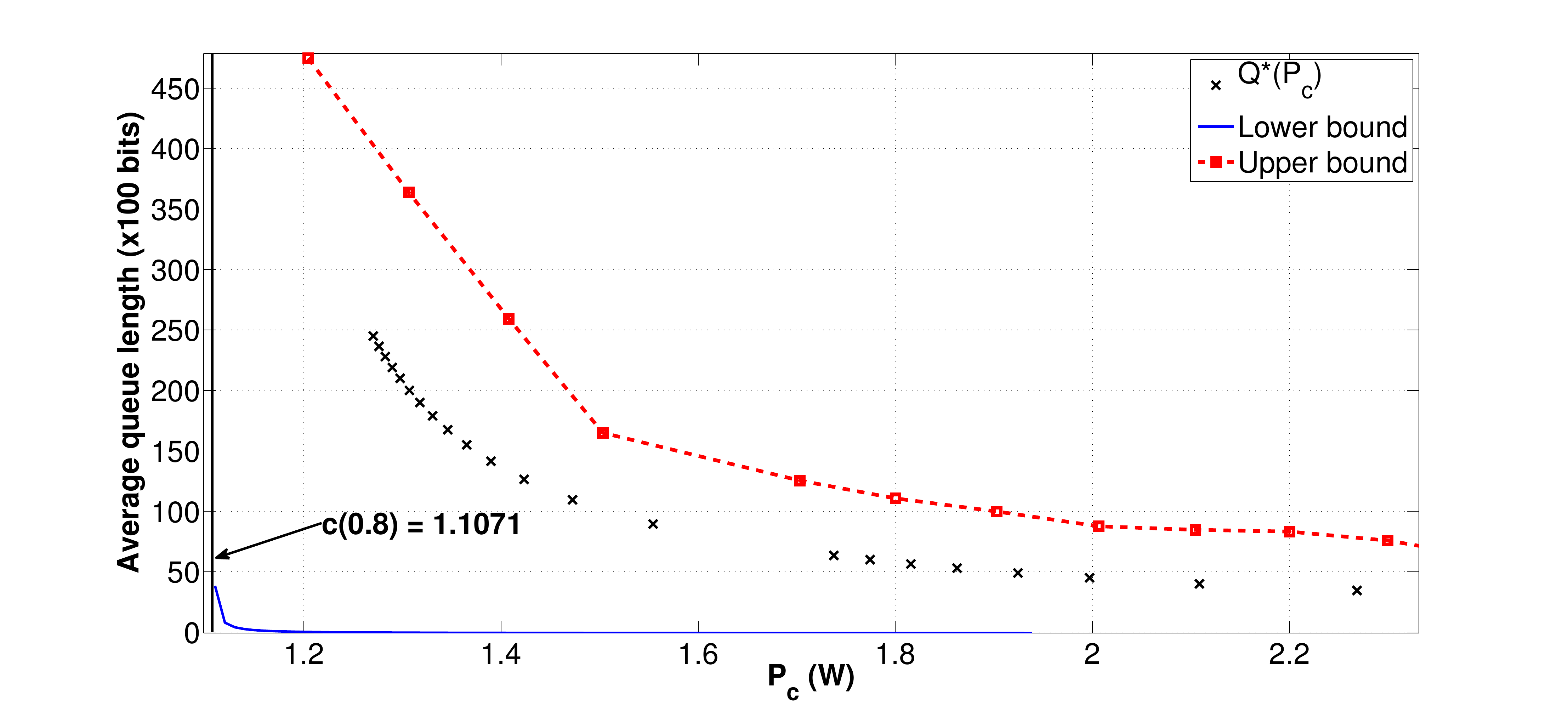}
  \caption{\small{Comparison of the optimal tradeoff $Q^*(P_{c})$ with the lower bound from Lemma \ref{chap5:lemma:case3} and the upper bound from Remark \ref{chap5fading:remark:case3ub} for the system in Section \ref{chap5fading:sec:simpleeg}, for $\lambda  = 0.8$, for $\mathcal{H} = \brac{0.1, 1}$ and $\pi_{H}(0.1) = 0.6$}}
  \label{chap5fading:fig:case3bounds}
\end{figure}

For the example in Section \ref{chap5fading:sec:simpleeg}, with $\mathcal{H} = \brac{0.1, 1}$ and $\pi_{H}(0.1) = 0.6$, for $\lambda = 0.82$, we plot the optimal tradeoff curve $Q^*(P_{c})$, the asymptotic lower bound from Lemma \ref{chap5fading:lemma:case2} (using Lemma \ref{chap5:lemma:tprob_2}), and the upper bound for the sequence of policies from Remark \ref{chap5fading:remark:case2}, in Figure \ref{chap5fading:fig:case2bounds}.
We note that the upper bound is obtained via simulation, for a sequence of policies for which $\tilde{q}$ has been chosen to be $\frac{1}{20}$ of what is suggested in \cite[Theorem 3]{neely_mac}.
This heuristic has been used in \cite{neely_mac}.
The analytical upper bound for the sequence of TOCA policies from \cite[Theorem 3]{neely_mac} is found to be very weak.
We note that the asymptotic bounds, although tight in the order sense, are very weak.
To illustrate the bounds for Case 3, for the system in Section \ref{chap5fading:sec:simpleeg}, with $\mathcal{H} = \brac{0.1, 1}$ and $\pi_{H}(0.1) = 0.6$, for $\lambda = 0.8$, we plot the optimal tradeoff curve $Q^*(P_{c})$, the asymptotic lower bound $\frac{\overline{q}_{3}}{2}$ from Lemma \ref{chap5fading:corollary:case3} (using Lemma \ref{chap5:lemma:case3}), and the upper bound for the sequence of policies from Remark \ref{chap5fading:remark:case3ub}, in Figure \ref{chap5fading:fig:case3bounds}.
Again the upper bound is obtained via simulation.

The numerical results illustrate that the asymptotic bounds which are obtained using the methods in Chapter 4 are weak, although they are tight in the order sense for Cases 2 and 3.

\begin{figure}[h]
  \includegraphics[width=160mm,height=55mm]{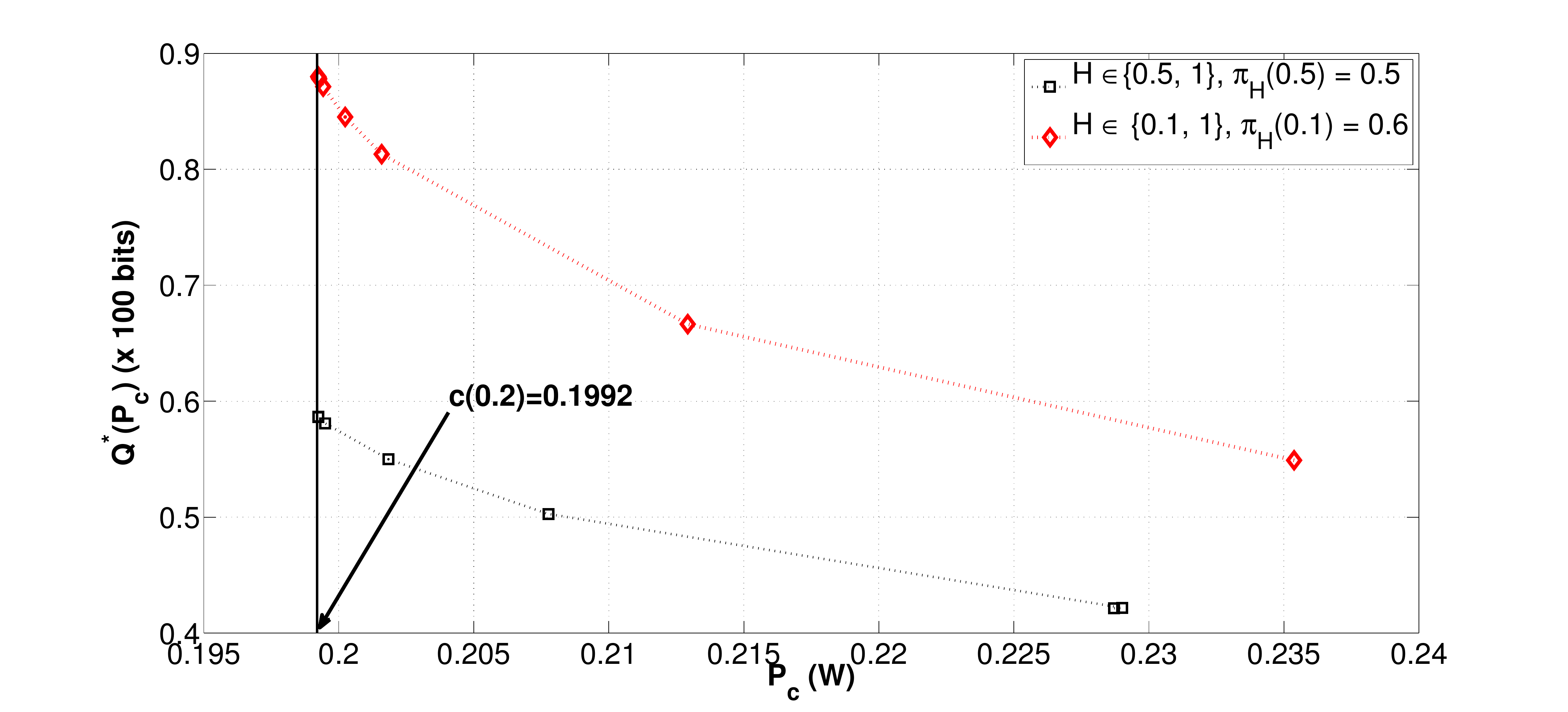}
  \caption{\small{The optimal tradeoff $Q^*(P_{c})$ for the system in Section \ref{chap5fading:sec:simpleeg}, for $\lambda  = 0.2$, for two cases of $\mathcal{H}$; $Q^*(P_{c})$ approaches a finite value in both cases.}}
  \label{chap5fading:fig:case1}
\end{figure}

We consider the case $\lambda = 0.2$ in Figure \ref{chap5fading:fig:case1}, which corresponds to Case 1 for the example in Section \ref{chap5fading:sec:simpleeg}, for both $\mathcal{H} = \brac{0.5,1}$ and $\mathcal{H} = \brac{0.1, 1}$.
We note that $c(0.2)$ is $0.1992$ for both $\mathcal{H} = \brac{0.5,1}$ and $\mathcal{H} = \brac{0.1, 1}$.
% We observe that $Q^*(P_{c})$ approaches a finite value $0.5867$ as $P_{c} \downarrow c(0.2)$, where $c(0.2) = 1.992$.
We observe that $Q^*(P_{c})$ approaches a finite value in both cases.
We note that, for Case 1, if $|\mathcal{H}| = 1$, it is possible to show that there exists a policy $\gamma$ such that $\overline{P}(\gamma) = c(\lambda)$ and $\overline{Q}(\gamma) < \infty$.

\begin{figure}
  \centering
  \includegraphics[width=160mm,height=55mm]{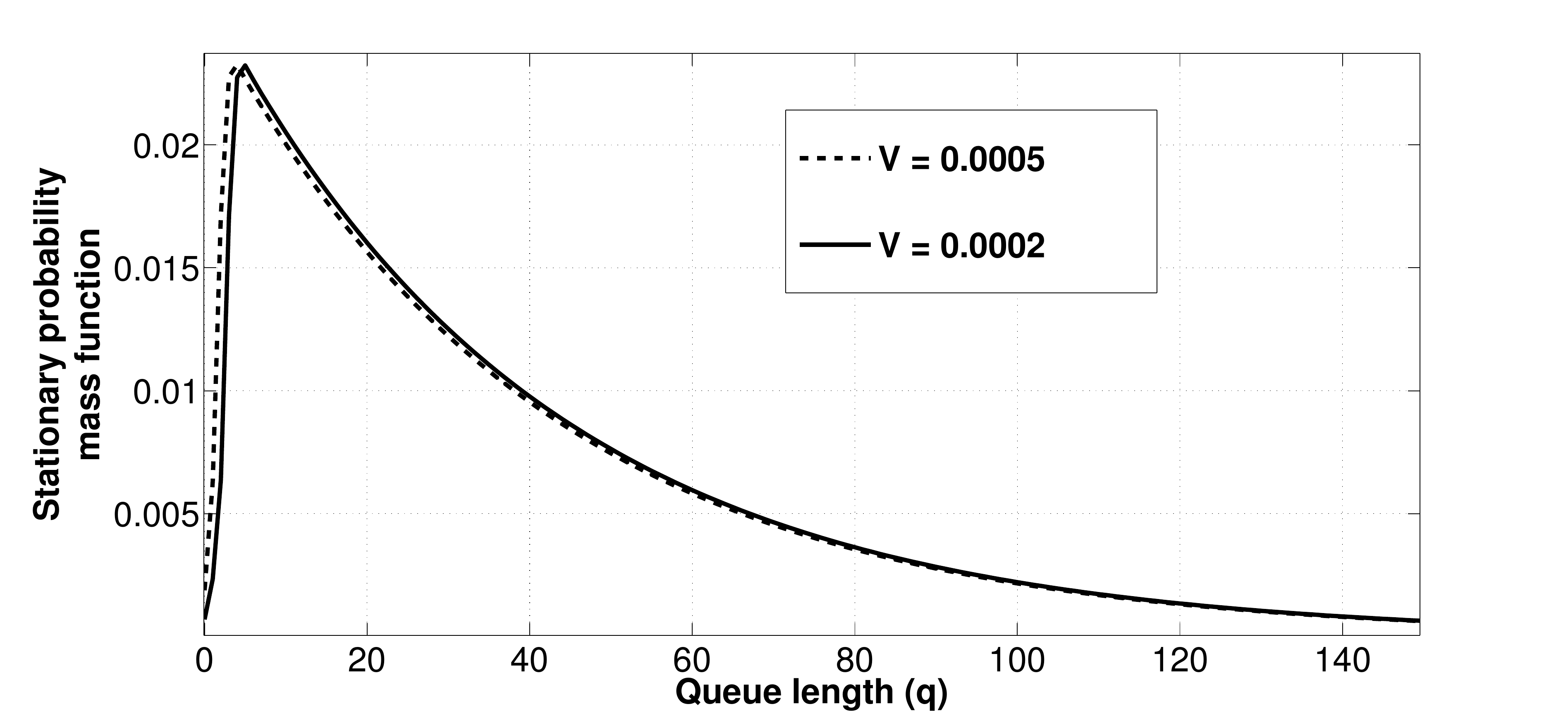}
  \caption{PMF of queue length for optimal policies in the regime $\Re$ for Case 2 for the system in Section \ref{chap5fading:sec:simpleeg}; $\mathcal{H} = \brac{0.1, 1}$ and $\pi_{H}(0.1) = 0.6$; $\lambda = 0.78$ and $c(0.78) = 1.0717$.}
  \label{fig:statprob_case2}
\end{figure}

\begin{figure}
  \centering
  \includegraphics[width=160mm,height=55mm]{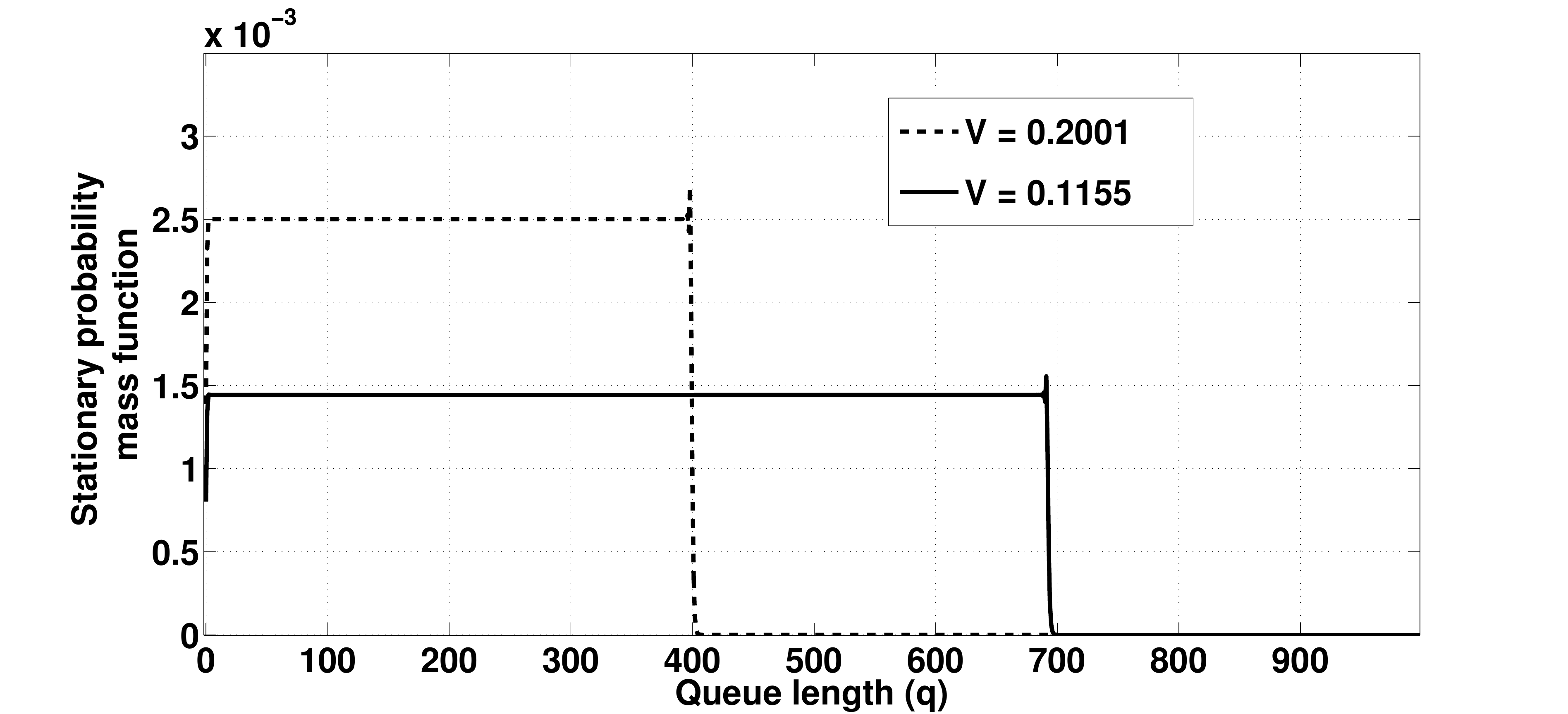}
  \caption{PMF of queue length for optimal policies in the regime $\Re$ for Case 3 for the system in Section \ref{chap5fading:sec:simpleeg}; $\mathcal{H} = \brac{0.1, 1}$ and $\pi_{H}(0.1) = 0.6$; $\lambda = 0.8$ and $c(0.8) = 1.1071$.}
  \label{fig:statprob_case3}
\end{figure}

The exact nature of $\overline{Q}^*(P_{c})$ for the different cases depends on the \emph{shape} of $\pi(q)$ in the regime $\Re$.
We first consider Cases 2 and 3 for I-model.
Intuition for the behaviour of $\overline{Q}^*(P_{c})$ for R-model is similar to that of Case 3 for I-model.
In Figures \ref{fig:statprob_case2} and \ref{fig:statprob_case3} we illustrate the behaviour of the probability mass function (PMF) of the queue length for Cases 2 and 3.
For each case, the PMF has been obtained by solving the global balance equations of the DTMC for the optimal policy for the truncated MDP used in Figure \ref{chap5fading:fig:comparelambda_2}.

We now present our observations about the nature of $\pi(q)$ in the regime $\Re$ using the above example.
We note that the intuition about the shape of $\pi(q)$ in the regime $\Re$, which we discuss below, holds for any sequence of feasible admissible policies, rather than just the sequence of optimal admissible policies.

For I-model, let $\mathcal{Q}_{h} = \brac{q : \mu(q) \in [s_{l} - \epsilon_{V}, s_{u} + \epsilon_{V}]}$, where $\epsilon_{V}$ is $\omega(V)$ as $V \downarrow 0$.
Then it can be shown that $Pr\brac{Q \in \mathcal{Q}_{h}} \uparrow 1$ as $V \downarrow 0$, i.e., as $V \downarrow 0$, the service rates have to be chosen from the set $[s_{l} - \epsilon_{V}, s_{u} + \epsilon_{V}]$ or $[\lambda - \epsilon_{V}, \lambda + \epsilon_{V}]$ for Cases 2 and 3 respectively.

The expected drift of the queue length when the queue length is $q$ is $\Exp\bras{Q[m + 1] - Q[m]|Q[m] = q}$.
We note that for admissible policies the expected drift is a non-increasing function of $q$.

We consider Case 2 first.
We note that since service rates can be chosen from the set $[s_{l} - \epsilon_{V}, s_{u} + \epsilon_{V}]$ as $V \downarrow 0$, intuitively the expected drift in the queue length for small queue lengths in $\mathcal{Q}_{h}$ is positive, while for large queue lengths in $\mathcal{Q}_{h}$ the expected drift is negative.
From the geometric upper bound on $\pi(q)$, intuitively $\pi(q)$ grows geometrically for small queue lengths.
Since the expected drift of the queue length for large queue lengths is negative, from the geometrically decreasing upper bound on $Pr\brac{Q > q - 1}$ and geometrically decreasing lower bound on $Pr\brac{Q > q}$ from \cite[Theorem 3]{gamarnik_1} we have that $\pi(q)$ decreases geometrically to zero for large enough $q$.
We note that this is the geometrically increasing and decreasing behaviour of $\pi(q)$ which is illustrated in Figure \ref{fig:statprob_case2}.
We also note that this behaviour holds for any $V > 0$ and as illustrated in the example in above leads to $\log\nfrac{1}{V}$ behaviour of the average queue length for any sequence of admissible policies.

For Case 3, we note that for $q \in \mathcal{Q}_{h}$, as $V \downarrow 0$, the expected drift approaches $0$.
Suppose we define $q_{d}$ as largest queue length in $\mathcal{Q}_{h}$ and $\epsilon_{V}$ as $aV$ where $a > 0$ is a constant.
Then it can be shown that for any $q \in \mathcal{Q}_{h}$
\[ Pr\brac{Q < q} \leq \brap{1 - \fpow{\epsilon_{a}}{\epsilon_{a} + d}{q}} \times \text{a constant}. \]
We note that the drift $d = \epsilon_{V} = aV$.
Hence, $Pr\brac{Q < q} \approx \mathcal{O}\brap{q \frac{aV}{\epsilon_{a}}}$.
Therefore, using the bounds on the stationary probability distribution, we are able to obtain the intuition that the stationary probability for any $q \in \mathcal{Q}_{h}$ is constant (this is illustrated in Figure \ref{fig:statprob_case3}) and $\mathcal{O}(V)$.
Then we have that the largest $\overline{q}$ such that $Pr\brac{Q < \overline{q}} \leq \frac{1}{2}$ is $\Omega\nfrac{1}{V}$, and therefore so is the asymptotic lower bound on the minimum average queue length.

For R-model, we let $\mathcal{Q}_{h} = \brac{q : \mu(q) \in [\lambda - \epsilon_{V}, \lambda + \epsilon_{V}]}$.
The intuition for the shape of the distribution for R-model is similar to that for Case 3 above.
With $\epsilon_{V} = a\sqrt{V}$ the stationary probability for any $q \in \mathcal{Q}_{h}$ is $\mathcal{O}\brap{\sqrt{V}}$ and thus the asymptotic lower bound on $\overline{q}$ is $\Omega\nfrac{1}{\sqrt{V}}$ and therefore so is the asymptotic lower bound on the minimum average queue length.

\section{Asymptotic lower bounds for ergodic $(A[m])$ and $(H[m])$}
\label{chap5fading:sec:ergodicextension}
In this section, we present an asymptotic lower bound for the optimal value of \eqref{chap5fading:eq:tradeoffproblem_init} when $(A[m])$ and $(H[m])$ are ergodic processes, for I-model.

We restrict to policies in $\Gamma_{s}$, which choose a batch service size $S[m] = S(Q[m - 1], H[m])$ as a function of the current queue length and fade state and independently of anything else.
We note that such policies may not be optimal.
The development for ergodic $(A[m])$ and $(H[m])$ closely follows the development in Section \ref{chap5:sec:extension_ergodic}.

Since $(A[m])$ is assumed to be ergodic, we have that almost surely 
\begin{eqnarray*}
  \lim_{M \rightarrow \infty} \frac{1}{M} \sum_{m = 1}^{M} A[m] = \Exp A[1] = \lambda,
\end{eqnarray*}
and $\lambda < S_{max}$, where $S_{max}$ is the largest batch size which can be served, as defined before.
We also assume that the arrival process $(A[m])$ is such that
\begin{description}
\item[NA1 :]{Let $\sigma[m - 1] = (Q[0] = q_{0}, A[1] = a_{1}, Q[1] = q_{1}, A[2] = a_{2}, \dots, A[m - 1] = a_{m - 1}, Q[m - 1] = q_{m - 1})$. We assume that
\begin{eqnarray*}
\inf_{m \in \sZ} \mathop{\min_{\brac{a_{1}, \dots, a_{m - 1}}}}_{\brac{q_{1}, \dots, q_{m - 1}}} Pr\brac{A[m] = 0 \middle \vert \sigma[m - 1]} = \nu_{a} > 0,
\end{eqnarray*}
}
\item[NA2 :]{$(A[m])$ is independent of $(H[m])$.}
\end{description}

Since $(H[m])$ is assumed to be ergodic, we have that
\begin{eqnarray*}
  Pr\brac{H[m] = h} = \pi_{H}(h), \forall h \in \mathcal{H}, \forall m \geq 1,
\end{eqnarray*}
with $\sum_{h \in \mathcal{H}} \pi_{H}(h) = 1$.

We restrict to policies $\gamma \in \Gamma_{s}$ for which the following limits exist
\begin{eqnarray}
  \lim_{m \rightarrow \infty} Pr\brac{Q[m - 1] = q, H[m] = h} = \pi_{Q,H}(q, h), \forall q \in \sZ \text{ and } h \in \mathcal{H},
  \label{chap5fading:eq:ergodic0}
\end{eqnarray}
with $\sum_{q,h} \pi(q, h) = 1$.
We note that for such a policy the following limits also exist
\begin{eqnarray*}
  \lim_{M \rightarrow \infty} \frac{1}{M} \sum_{m = 0}^{M - 1} Pr\brac{Q[m] = q} & = & \pi(q), \forall q \in \sZ, \\
  \lim_{M \rightarrow \infty} \frac{1}{M} \sum_{m = 1}^{M} Pr\brac{H[m] = h, S[m] = s} & = & \pi_{H, S}(h, s), \forall s \in \brac{0,\dots, S_{max}}, \forall h \in \mathcal{H}, \\
  \lim_{M \rightarrow \infty} \frac{1}{M} \sum_{m = 1}^{M} Pr\brac{S[m] = s} & = & \pi_{s}(s), \forall s \in \brac{0,\dots, S_{max}}.
\end{eqnarray*}
We again restrict to a set of admissible policies $\Gamma_{a} \subseteq \Gamma_{s}$, which are defined similarly as in Section \ref{chap5:sec:extension_ergodic}.
A policy $\gamma \in \Gamma_{s}$ is admissible if :
\begin{description}
\item[NG1 :]{the limits in \eqref{chap5fading:eq:ergodic0} exist}
\item[NG2 :]{$\gamma$ is mean rate stable, i.e., $\sum_{h \in \mathcal{H}} \sum_{s = 0}^{S_{max}} \pi_{H, S}(h, s) s = \lambda$,}
\item[NG3 :]{the average service rate $\overline{s}(q) = \sum_{h} \pi_{H}(h) \Exp S(q, h)$ is a non-decreasing function of $q$ for $\gamma$.}
\end{description}
For any admissible policy $\gamma$,  we have that $\overline{P}(\gamma) = \sum_{h \in \mathcal{H}} \sum_{s \in \brac{0, \dots, S_{max}}} \pi_{H, S} (h, s) P(h, s)$ and $\overline{Q}(\gamma) \Deq \sum_{q} \pi(q) q$.

For any admissible policy $\gamma$, we have that $\overline{P}(\gamma)$ is bounded below by the optimal value of
\begin{eqnarray*}
  \mini_{\gamma \in \Gamma_{a}} & \sum_{h \in \mathcal{H}} \sum_{s = 0}^{S_{max}} \pi_{H,S}(h,s) P(h, s), \nonumber \\
  \text{such that } & \sum_{h \in \mathcal{H}} \sum_{s = 0}^{S_{max}} \pi_{H,S}(h,s) s = \lambda,
\end{eqnarray*}
where $\pi_{H,S}(h,s)$ is determined by the policy $\gamma$.
The optimal value of the above problem is bounded below by the optimal value of
\begin{eqnarray*}
  \mini_{p_{s|h}(s), h \in \mathcal{H}} & \sum_{h \in \mathcal{H}} \pi_{H}(h) \sum_{s = 0}^{S_{max}} p_{s|h}(s) P(h, s), \nonumber \\
  \text{such that } & \sum_{h \in \mathcal{H}} \sum_{s = 0}^{S_{max}} \pi_{H}(h) p_{s|h}(s) s = \lambda,
\end{eqnarray*}
which is $c(\lambda)$.
We note that in the above optimization problem, the minimization is over all possible \emph{conditional distributions} on the batch size.
The conditional distribution $p_{s|h}(s)$ has the interpretation of the fraction of time a batch of size $s$ is used when the fade state is $h$.
  
Let us denote the optimal conditional distribution that achieves $c(\lambda)$ in \eqref{chap5fading:eq:app_bg_0} by $p^*_{s|h}(\lambda), \forall h \in \mathcal{H}$.
Similar to the approach in Section \ref{chap5:sec:extension_ergodic}, we now show that there exists a sequence of admissible policies $\gamma_{V}$ for a sequence $V \downarrow 0$ such that $\lim_{V \downarrow 0} \overline{P}(\gamma_{V}) = c(\lambda)$.
We define random variables $U_{V}(h) \sim p^*_{s|h}(\lambda + V), \forall h \in \mathcal{H}$.
For a particular $V > 0$, the policy $\gamma_{V}$, serves $S[m] = \min(Q[m - 1], U_V[m])$, where for each $m$, $U_V[m]$ is an independent sample of $U_{V}(H[m])$.
Then the queue evolution under $\gamma_{V}$ is
\begin{eqnarray*}
  Q[m + 1] = \max\bigg( Q[m] - U_{V}(H[m + 1]), 0 \bigg ) + A[m + 1], \text{ with } Q[0] = q_{0}, \forall m \geq 0.
\end{eqnarray*}
Applying \cite[Lemma 1]{loynes} we have that the limit $\pi_{Q,H}(q,h) = \lim_{m \rightarrow \infty} Pr\brac{Q[m] = q, H[m + 1] = h}$ exists.
With $Q'[1] = \max\bigg( q_{0} - U_{V}(H[1]), 0 \bigg )$, we can define a sequence $Q'[m]$ which evolves as
\begin{eqnarray*}
  Q'[m + 1] = \max\bigg( Q'[m] + A[m] - U_{V}(H[m + 1]), 0 \bigg ), \forall m \geq 1.
\end{eqnarray*}
We note that $Q[m] = Q'[m] + A[m], \forall m \geq 1$.
Since the sequence $\zeta[m] = A[m] - U_{V}(H[m + 1])$ is ergodic, $\Exp \zeta[1] = -V < 0$, and $A[1] < A_{max}$, we have that $\lim_{m \rightarrow \infty} Pr\brac{Q[m] < \infty} = 1$.

We note that the sequence of policies $\gamma_{V}$ is a sequence of admissible policies, with $\lim_{V \downarrow 0} \overline{P}(\gamma_{V}) = c(\lambda)$.
Therefore $c(\lambda) = \inf_{\gamma \in \Gamma_{a}} \overline{P}(\gamma)$.

Now we obtain the following asymptotic lower bound, the proof of which is similar to that of Lemma \ref{chap5:lemma:extension_ergodic}.
We note that the quantities $s_{l}, s_{u}$ and the line $l(s)$ are defined as in Section \ref{chap5fading:sec:asymp}.
\begin{lemma}
  For an ergodic arrival process $(A[m], m \geq 1)$, satisfying NA1 and NA2, an ergodic fading process $(H[m], m \geq 1)$, and for any sequence of admissible policies $\gamma_{k}$ (satisfying NG1, NG2, and NG3), with $\overline{P}(\gamma_{k}) - c(\lambda) = V_{k} \downarrow 0$, we have that $\overline{Q}(\gamma_{k}) = \Omega\brap{\log\nfrac{1}{V_{k}}}$, for cases 2 and 3.
  \label{chap5fading:lemma:extension_ergodic}
\end{lemma}
\begin{proof}
  For some positive $\epsilon < s_{l}$, let $q_{s} \stackrel{\Delta} = \inf\brac{q : \Exp S(q, H) \geq s_{l} - \epsilon}$.
  Then, as in the proof of Lemma \ref{chap5fading:lemma:case2}, we have that $U_{k} \Deq \Exp_{\pi_{\gamma_{k}}} c(\overline{s}(Q)) - c(\lambda) \leq V_{k}$.
  We consider a particular policy $\gamma$ in the sequence with $\Exp_{\pi} c(\overline{s}(Q)) - c(\lambda) = U$.
  As in the proof of Lemma \ref{chap5:lemma:extension_ergodic} with $U$ in place of $V$, we can then show that $\sum_{q = 0}^{q_{s} - 1} \pi(q) \leq \frac{U}{m_{1} \epsilon}$ and $Pr\brac{S(q, H) > 0} \geq \frac{s_{l} - \epsilon}{S_{max}}, \forall q \geq q_{s}$, since for $q \geq q_{s}$, $\Exp S(q, H) \geq s_{l} - \epsilon$.
  Now we relate the stationary probability $\pi(q), q \geq q_{s}$ to $\sum_{q = 0}^{q_{s} - 1} \pi(q)$.
  We have that for a $q \geq q_{s}$ and for every $m \geq 0$
  \begin{eqnarray*}
    Pr\brac{Q[m + 1] < q | Q[0] = q_{0}} & = & Pr\brac{Q[m] - S(Q[m], H[m + 1]) + A[m + 1] < q | Q[0] = q_{0}}.
  \end{eqnarray*}
  Let $\sigma[m] = (S[1], Q[1], \dots, Q[m - 1], S[m])$.
  
  We have that $Pr\brac{Q[m] - S(Q[m], H[m + 1]) + A[m + 1] < q | Q[0] = q_{0}}$
  \begin{eqnarray*}
    & = & \Exp_{\sigma[m]} \bigg [Pr\brac{Q[m] - S(Q[m], H[m + 1]) + A[m + 1] < q | Q[0] = q_{0}, \sigma[m]} \bigg ], \\
    & \geq & \Exp_{\sigma[m]} \bigg [ Pr\brac{Q[m] = q | Q[0] = q_{0}, \sigma[m]} \times \\
    & & Pr\brac{S(Q[m], H[m + 1]) > 0 | Q[0] = q_{0}, \sigma[m], Q[m] = q} \times \\
    & & Pr\brac{A[m + 1] = 0 | Q[0] = q_{0}, \sigma[m], Q[m] = q, \brac{S(Q[m], H[m + 1]) > 0}} \bigg ].
  \end{eqnarray*}
  We note that since $\sigma[m]$ does not involve the history of the fade process we have that 
  \[Pr\brac{S(q, H[m + 1]) > 0|Q[0] = q_{0}, \sigma[m], Q[m] = q} \geq \frac{s_{l} - \epsilon}{S_{max}}.\]
  Since $(Q[0] = q_{0}, \sigma[m], Q[m] = q, \brac{S(Q[m]) > 0}) = (Q[0] = q_{0}, A[1], Q[1], \dots, A[m], Q[m] = q)$, using property NA1, we obtain the same lower bound as in the proof of Lemma \ref{chap5:lemma:extension_ergodic} on $Pr\brac{Q[m + 1] < q | Q[0] = q_{0}}$.
  Following the rest of the steps in the proof of Lemma \ref{chap5:lemma:extension_ergodic}, we obtain that $\overline{Q}(\gamma_{k}) = \Omega\brap{\log\nfrac{1}{U_{k}}}$.
  Since $U_{k} \leq V_{k}$, we have that $\overline{Q}(\gamma_{k}) = \Omega\brap{\log\nfrac{1}{V_{k}}}$.
\end{proof}

We note that NA2 is not used in the asymptotic lower bound, while it was used in showing that a sequence of admissible policies exist for which $\Pgk \downarrow c(\lambda)$.

The above asymptotic lower bound is weak for Case 3, since we have obtained an $\Omega\nfrac{1}{V}$ asymptotic lower bound when $(A[m])$ is IID.
We note that the above asymptotic lower bound can be applied to the system considered by Huang and Neely \cite{huang_neely}.
We note that a similar asymptotic lower bound can be derived for the R-model also.

\section{Queueing models with admission control}
\label{sec:systemmodel_modelus}
\subsection{System model}
We consider the optimal tradeoff of average queue length and average power for a fading point-to-point link, when the packets arriving to the link can be dropped, subject to a constraint on the utility of the time average throughput of the packets which are transmitted.
We indicate only the differences from the models in Sections \ref{chapfading:sysmodel:intval} and \ref{chapfading:sysmodel:realval}.
In each slot $m$, a random number $R[m]$ of packets arrive into the system.
We assume that $(R[m], m \geq 1)$ are IID, with $R[1] \leq A_{max}$, $\Exp R[1] = \lambda$, and $var(R[1]) = \sigma^{2} < \infty$.
We also denote the expectation with respect to the distribution of $R[1]$ as $\Exp_{R}$.
At the end of slot $m$, $A[m] \leq R[m]$ packets are admitted into the infinite length transmitter queue, while $R[m] - A[m]$ packets are dropped.
The fade state process, $(H[m], m \geq 1)$, is as before.
We also assume that the arrival process $(R[m])$ is independent of $(H[m])$.
At the beginning of slot $m \geq 1$, $Q[m - 1]$ denotes the number of packets in the transmitter queue.
The transmitter starts transmission of $S[m]$ packets at the start of slot $m$.
We assume that just before the end of slot $m$, $A[m] \leq R[m]$ packets are admitted into the transmitter queue.
We note that under our assumptions on $(S[m]), (A[m])$, and $q_{0}$, $Q[m] \in \sR, \forall m \geq 0$.
For this model, a policy $\gamma$ for operation of the transmitter is the sequence of service and arrival batch sizes $(S[1], A[1], S[2], A[2], \dots)$.
The set of all policies is denoted as $\Gamma$.
If $\gamma$ is such that $S[m + 1] = S(Q[m], H[m + 1])$ and $A[m + 1] = A(Q[m], R[m + 1], H[m + 1])$, where $S(q, h)$ and $A(q, r, h)$ are randomized functions, then $\gamma$ is a stationary policy.
The set of stationary policies is denoted by $\Gamma_{s}$.
Since $(R[m], H[m], m \geq 1)$ is assumed to be IID, we have that for a $\gamma \in \Gamma_{s}$, $(Q[m], m \geq 0)$ is a Markov chain evolving on $\sR$.

If we assume that $A[m], R[m], q_{0}, S[m] \in \sZ$, then the queue evolution $Q[m] \in \sZ$ and the model is denoted as I-model-U.
On the other hand, if $A[m], R[m], q_{0}, S[m] \in \sR$, then the queue evolution $Q[m] \in \sR$, and the model is denoted as R-model-U.
Like R-model, R-model-U with a strictly convex $P(h, s)$ function is usually used as an approximation for I-model-U.

We define the average throughput of a policy $\gamma \in \Gamma_{s}$ as
\begin{eqnarray}
  \overline{A}(\gamma, q_{0}) = \liminf_{M \rightarrow \infty} \frac{1}{M} \Exp \bras{\sum_{m = 1}^{M} A[m] \middle \vert Q[0] = q_{0} }.
\end{eqnarray}
Let $U(a) : [0, A_{max}] \rightarrow \sR$ be a strictly concave and increasing function of $a$, with $U(0) = 0$.
The utility of transmitting the packets is $U(\overline{A}(\gamma, q_{0}))$, for a policy $\gamma$.
The average power for a policy $\gamma \in \Gamma_{s}$ is $\overline{P}(\gamma, q_{0})$ and the average queue length is $\overline{Q}(\gamma, q_{0})$, as defined in \eqref{chap5fading:eq:avgpower} and \eqref{chap5fading:eq:avgqlength} respectively.

The model considered by Neely \cite{neely_utility} is the same as R-model-U.
It is shown in \cite{neely_utility} that there exists a sequence of policies $\gamma_{k} \in \Gamma_{s}$ with a corresponding sequence $V_{k} \downarrow 0$, such that $\overline{A}(\gamma_{k}, q_{0}) \geq \rho \lambda$ ($0 < \rho < 1$), $\overline{Q}(\gamma_{k}, q_{0}) = \mathcal{O}\brap{\log\nfrac{1}{V_{k}}}$, and $\overline{P}(\gamma_{k}, q_{0})$ is at most $V_{k}$ more than the minimum average power required for queue stability.
It is also shown in \cite{neely_utility} for $|\mathcal{H}| = 1$, that if $\gamma_{k}$ is any sequence of policies, with $\overline{P}(\gamma_{k}, q_{0})$ at most $V_{k}$ more than the minimum average power required for queue stability and $\overline{A}(\gamma_{k}, q_{0}) \geq \rho \lambda$, then $\overline{Q}(\gamma_{k}, q_{0}) = \Omega\brap{\log\nfrac{1}{V_{k}}}$ as $V_{k} \downarrow 0$.

We consider the optimal tradeoff between $\overline{Q}(\gamma, q_{0})$ and $\overline{P}(\gamma, q_{0})$ subject to the average utility $U(\overline{A}(\gamma, q_{0}))$ being at least a positive $u_{c} < U(A_{max})$, for the class of stationary policies $\Gamma_{s}$, for I-model-U and R-model-U, in this chapter.
The constraint $U(\overline{A}(\gamma, q_{0})) \geq u_{c}$ is equivalent to having the constraint $\overline{A}(\gamma, q_{0}) \geq U^{-1}(u_{c})$, where $U^{-1}$ is the inverse function of $U$.
We note that since the arrival rate is not the same for all $\gamma \in \Gamma_{s}$, minimization of the average queue length does not directly correspond to minimizing the average delay of the packets.
Asymptotic bounds on the average delay can be derived using Little's law and are discussed in this chapter.

\subsection{Problem formulation for I-model-U and R-model-U}
\label{sec:problem}
The general tradeoff problem that we consider is
\begin{eqnarray*}
  \mini_{\gamma \in \Gamma} \overline{Q}(\gamma, q_{0}) \text{ such that } \overline{P}(\gamma, q_{0}) \leq P_{c} \text{ and } \overline{A}(\gamma, q_{0}) \geq \rho \lambda,
  \label{chap5fading:eq:tradeoffutility_init}
\end{eqnarray*}
where $0 < \rho < 1$.
As in Section \ref{chap5:sec:cmdpformulation} we can show that if $P_{c} > c(\rho\lambda)$ for I-model-U or if $P_{c} > c_{R}(\rho\lambda)$ for R-model-U (which are the minimum average powers required for stability while supporting an arrival rate of $\rho\lambda$ rather than $\lambda$) then there exists an optimal stationary policy $\gamma^*$ with stationary probability $\pi^*$.
Therefore, we can restrict ourselves to the set of stationary policies.

As for I-model and R-model, we consider the above tradeoff problem for a set of admissible policies $\Gamma_{a}$.
However, since there is admission control, we relax the irreducibility requirement (for I-model-U and R-model-U) as follows.
For an admissible policy $\gamma$, the Markov chain $(Q[m], m \geq 0)$ has a single positive recurrent class $\mathcal{R}_{\gamma}$ which contains $0$.
Furthermore, the cumulative expected queue cost as well as the cumulative expected power cost starting from any state $q_{0}$ until $\mathcal{R}_{\gamma}$ is hit are finite.

We note that for a $\gamma \in \Gamma_{a}$, $\overline{Q}(\gamma, q_{0}) = \Qg$, $\overline{P}(\gamma, q_{0}) = \Pg$, and $\overline{A}(\gamma, q_{0}) = \Ag$.
A policy $\gamma$ is defined to be admissible if: (i) the Markov process $(Q[m], m \geq 0)$ under $\gamma$ is aperiodic and positive Harris recurrent on a single recurrence class $\mathcal{R}_{\gamma}$ with stationary distribution $\pi$\footnote{We note that $\pi(A) = 0$ for any $A \not \subset \mathcal{R}_{\gamma}$}, (ii) $\Qg < \infty$, and (iii) $\sq$ is non-decreasing in $q$, where $\sq = \ExpH \Exp_{S|q,H}S(q,H)$ is the average service rate at queue length $q$.
We note that for a $\gamma \in \Gamma_{a}$,
\begin{eqnarray*}
  \Qg & = & \Expp Q, \\
  \Pg & = & \Expp \ExpH \Expsqh P(H, S(Q, H)), \\
  \Ag & = & \Expp \ExpH \Exp_{R} \Exp A(Q, R, H).
\end{eqnarray*}
Let the average service rate be $\Sg$, then $\Sg = \Expp \ExpH \Expsqh S(Q, H)$.

The problem TRADEOFF that we consider is 
\begin{eqnarray*}
  \mini_{\gamma \in \Gamma_{a}} \Qg \text{ such that } \Pg \leq P_{c} \text{ and } \Ag \geq \rho \lambda.
\end{eqnarray*}
The optimal value of TRADEOFF is denoted as $Q^*(P_{c}, \rho)$.
Suppose $\gamma$ is feasible for TRADEOFF.
Then
\begin{eqnarray*}
  \Sg = \Ag \geq \rho\lambda.
\end{eqnarray*}
Now we note that $\Pg$ is bounded below by the optimal value of 
\begin{eqnarray}
  \mini_{\gamma \in \Gamma_{a}} & & \Expp \ExpH \Expsqh P(H, S(Q, H)), \nonumber \\
  \text{such that } & & \Expp \ExpH \Expsqh S(Q, H) \geq \rho\lambda.
  \label{eq:mincost_problem1}
\end{eqnarray}
We note that $\Expp \ExpH \Expsqh S(Q,H) = \ExpH \Expp \Expsqh S(Q,H)$.
We have that $\ExpH \Expp \Expsqh S(Q,H) = \ExpH \Exp_{S|H} S$ and $\Expp \ExpH \Expsqh P(H, S(Q,H)) = \ExpH \Exp_{S|H} P(H, S)$, where the conditional distribution of $S$ given $H$ depends on the policy $\gamma$.
Then the optimal value of \eqref{eq:mincost_problem1} is bounded below by the optimal value of
\begin{eqnarray}
  \mini & & \ExpH \Exp_{S|H} P(H, S), \nonumber \\
  \text{such that } & & \ExpH \Exp_{S|H} S \geq \rho\lambda,
  \label{eq:mincost_problem2}
\end{eqnarray}
where we minimize over all possible conditional distributions for $S$ given $h$, irrespective of the policy $\gamma$.
We note that for I-model-U, these distributions have support on $\brac{0, \dots, S_{max}}$, whereas for R-model-U they have support on $[0, S_{max}]$.

We note that \eqref{eq:mincost_problem2} has feasible solutions only if $\rho\lambda \leq S_{max}$.
The optimal value of the above problem is $c(\rho\lambda)$ for I-model-U and $c_{R}(\rho\lambda)$ for R-model-U, since the constraint is satisfied with equality
\footnote{If the distribution which achieves the minimum in \eqref{eq:mincost_problem2} is such that $\ExpH \Exp_{S|H} S > \rho\lambda$, then it is possible to show that there exists another distribution which has a strictly smaller $\ExpH \Exp_{S|H} P(H, S)$.}.
So, we have that for $\gamma \in \Gamma_{a}$, $\Pg \geq c(\rho\lambda)$ for I-model-U and $\Pg \geq c_{R}(\rho\lambda)$ for R-model-U.
Thus, TRADEOFF has feasible solutions only if $P_{c} \geq c(\rho\lambda)$ for I-model-U and $P_{c} \geq c_{R}(\rho\lambda)$ for R-model-U.

We now show that $c(\rho\lambda)$ and $c_{R}(\rho\lambda)$ are both $\inf_{\brac{\gamma : \gamma \in \Gamma_{a}, \Ag \geq \rho\lambda}} \Pg$ for I-model-U and R-model-U respectively.
For I-model-U, we consider a sequence of policies $\gamma_{k}$, where for each $\gamma_{k}$, at each slot $m$, each customer in the batch $R[m]$ is admitted with probability $\rho$ and dropped with probability $1 - \rho$.
Then $\forall \gamma_{k}$ we have that $\Agk \geq \rho\lambda$.
Now as for I-model, $\gamma_{k}$ is such that $\Pgk = c(\rho\lambda) + V_{k}$ and $\Qgk = \mathcal{O}\nfrac{1}{V_{k}}$.
Thus, if $\rho\lambda < S_{max}$, we have that there exists a sequence of admissible policies $\gamma_{k}$, such that $\Pgk = c(\rho\lambda) + V_{k}$, $\Agk \geq \rho\lambda$, and $\Qgk = \mathcal{O}\nfrac{1}{V_{k}}$, for a sequence $V_{k} \downarrow 0$.
For R-model-U, we choose $A[m] = \rho R[m]$, and then as for R-model, serve the customers using a sequence of policies such that $\Pgk = c_{R}(\rho\lambda) + V_{k}$ and $\Qgk = \mathcal{O}\nfrac{1}{V_{k}}$.
Hence, $c(\rho\lambda)$ and $c_{R}(\rho\lambda)$ are $\inf_{\brac{\gamma : \gamma \in \Gamma_{a}, \Ag \geq \rho\lambda}} \Pg$ for I-model-U and R-model-U respectively.

In the following, we obtain an asymptotic characterization of $Q^*(P_{c}, \rho)$ in the asymptotic regimes $\Re$ as $P_{c} \downarrow c(\rho\lambda)$ for I-model-U and $P_{c} \downarrow c_{R}(\rho\lambda)$ for R-model-U, under the assumption that $\rho\lambda < S_{max}$.
We recall that $c(s)$ is a non-decreasing, piecewise linear, and convex function of $s \in [0, S_{max}]$, whereas $c_{R}(s)$ is a non-decreasing strictly convex function of $s \in [0,S_{max}]$, with $c(0)$ and $c_{R}(0)$ both being $0$.

\subsection{Asymptotic bounds}
\label{sec:lowerbound}
We first obtain an asymptotic lower bound for R-model-U.
We then outline the derivation of the asymptotic lower bound for I-model-U, since it can be obtained using very similar techniques as for R-model-U and as in Section \ref{chap5fading:sec:asymp}.

We assume that $R[1]$ satisfies the property:
\begin{description}
\item[RA* :]{$Pr\brac{R[1] \leq \frac{\Delta}{2}} = \epsilon_{a}' > 0$, for some $\Delta$ such that $0 < \Delta < \rho\lambda$.}
\end{description}
Let $\gamma$ be an admissible policy with $\Pg - c_{R}(\rho\lambda) = V$ and $\Ag \geq \rho\lambda$.
For the policy $\gamma$, let $q_{1} \stackrel{\Delta} = \sup\brac{q : \sq \leq \rho\lambda - \epsilon}$, for an $\epsilon$ chosen such that $0 < \epsilon < \rho\lambda - \Delta$.
We note that the average power used when the queue length is $q$ is $\ExpH \Exp_{S|q,H} P(H, S(q, H))$, which is bounded below by the optimal value of
\begin{eqnarray}
  \mini & & \ExpH \Exp_{S|H} P(H, S), \nonumber \\
  \text{such that } & & \ExpH \Exp_{S|H} S \geq \sq,
  \label{eq:mincost_problem3}
\end{eqnarray}
where we minimize over all possible conditional distributions for $S$ given $h$, irrespective of $\gamma$.
We note that problem \eqref{eq:mincost_problem3} is the same as \eqref{eq:mincost_problem2} except that the constraint is $\sq$ instead of $\rho\lambda$.
Therefore $\ExpH \Exp_{S|q,H} P(H, S(q,H)) \geq c_{R}(\sq)$.
Since $\Expp \ExpH \Expsqh P(H, S(Q,H)) = \Pg$, we have that $\Pg \geq \Expp c_{R}(\sQ)$.
Therefore, $\Expp c_{R}(\sQ) - c_{R}(\rho\lambda) \leq \Pg - c_{R}(\rho\lambda) \leq V$.
Since $c_{R}(.)$ is a strictly convex and non-decreasing function, we assume that the second derivative of $c_{R}(s)$ is positive at $s = \rho\lambda$.
As $\Sg = \Expp \sQ \geq \rho\lambda$, we have that
\begin{eqnarray*}
  \Expp c_{R}(\sQ) = \Expp \bras{ c_{R}(\rho\lambda) + \frac{dc_{R}(x)}{dx}\bigg \vert_{\rho\lambda} (\sQ - \rho\lambda) + G(\sQ - \rho\lambda)}
\end{eqnarray*}
where $G(x)$ is a strictly convex function as in \cite[eq (41)]{berry}, with $G(0) = 0$ and $\frac{dG(x)}{dx}\vert_{x = 0} = 0$.
Since $\Expp \sQ \geq \rho\lambda$ we have that
\begin{eqnarray*}
  \Expp c_{R}(\sQ) - c_{R}(\rho\lambda) \geq \Expp G(\sQ - \rho\lambda).
\end{eqnarray*}
Thus, we have that $\Expp G(\sQ - \rho\lambda) \leq V$.
Therefore, for $q_{1}$ as defined before,
\begin{eqnarray*}
  \int_{0}^{q_{1}} G(\sq - \rho\lambda)d\pi(q) \leq V.
\end{eqnarray*}
Since $G(.)$ is strictly convex, for a positive $a_{1}$ we have that
\begin{eqnarray*}
  \bras{\int_{0}^{q_{1}} (\sq - \rho\lambda) d\pi(q)}^{2} \leq \frac{V}{a_{1}},
\end{eqnarray*}
where $a_{1} > 0$.
Since for $q < q_{1}, \sq \leq \rho\lambda - \epsilon$, we have that
\begin{eqnarray}
  Pr\brac{Q < q_{1}} \leq \frac{V}{a_{1} \epsilon^{2}}.
  \label{eq:q1upperb}
\end{eqnarray}

Let $S(q)$ be the random service batch size when in state $q$.
We have that $Pr\brac{S(q) > s} = \int_{h} Pr\brac{S(q, h) > s}d\pi_{H}(h)$.
\begin{lemma}
  For $\gamma$, for $\Delta$ as in RA*, with $q_{1}$ defined as above, we have that $\inf_{q > q_{1}} Pr\brac{S(q) > \Delta} \geq \delta_{s} > 0$, where $\delta_{s} = \frac{\rho\lambda - \epsilon - \Delta}{S_{max} - \Delta}$.
  \label{lemma:deltaservice}
\end{lemma}
\begin{proof}
We note that $q_{1} = \sup\brac{q : \sq \leq \rho\lambda - \epsilon}$, for $0 < \epsilon < \rho\lambda - \Delta$, where $\Delta$ is as in RA*.
Let $P(S(q))$ denote the distribution of $S(q)$.
We note that by definition, $\forall q > q_{1}$, 
\begin{eqnarray}
\sq & > &  \rho\lambda - \epsilon, \text{ or}, \nonumber \\
\int_{0}^{S_{max}} s dP(S(q)) & \geq & \rho\lambda - \epsilon.
\label{chap5:eq:deltaservice0}
\end{eqnarray}
Then, 
\begin{eqnarray*}
\int_{0}^{\Delta} \Delta dP(S(q)) + \int_{\Delta}^{S_{max}} S_{max} dP(S(q)) & \geq & \rho\lambda - \epsilon,\\
\Delta \brap{1 - Pr\brac{S(q)  > \Delta}} + S_{max} Pr\brac{S(q) > \Delta} & \geq & \rho\lambda - \epsilon, \text{ or}, \\
Pr\brac{S(q) > \Delta} & \geq & \frac{\rho\lambda - \epsilon - \Delta}{S_{max} - \Delta}.
\end{eqnarray*}
Thus for any $q > q_{1}$, $Pr\brac{S(q) > \Delta} \geq \delta_{s} > 0$, where $\delta_{s} = \frac{\rho\lambda - \epsilon - \Delta}{S_{max} - \Delta}$.
\end{proof}
From the above result we have that $q_{1} \geq \Delta$.

\begin{lemma}
  For any sequence of admissible policies $\gamma_{k}$ such that $\Agk \geq \rho\lambda$ and $\Pgk - c_{R}(\rho\lambda) = V_{k} \downarrow 0$, we have that $\Qgk = \Omega\brap{\log\nfrac{1}{V_{k}}}$.
  \label{lemma:tradeoffutilitylb}   
\end{lemma}
\begin{proof}
  Let us consider a particular policy $\gamma$ in the sequence $\gamma_{k}$ with $V_{k} = V$.
  As $\gamma$ is admissible we have that
  \[ Pr\brac{0 \leq Q < q_{1}} = \int_{0}^{\infty} P(q, [0,q_{1})) d\pi(q), \]
  where $P(q,\mathcal{Q})$ is the transition kernel of the Markov chain.
  Hence, we have that
  \begin{eqnarray*}
    Pr\brac{0 \leq Q < q_{1}} & \geq & \int_{q_{1}}^{q_{1} + \Deltat} P(q, [0, q_{1})) d\pi(q), \\
    & \geq & \epsilon'_{a} \delta_{s} Pr\brac{q_{1} \leq Q < q_{1} + \Deltat},
  \end{eqnarray*}
  where we have used Lemma \ref{lemma:deltaservice} and the property RA* to lower bound $Pr\brac{Q[m + 1] < q_{1} \vert Q[m] = q}$ by $\epsilon_{a}' \delta_{s}$.
  Let $\rho_{a} = \epsilon_{a}' \delta_{s}$.
  Also, for any $q' > q$, let us denote $Pr\brac{q \leq Q < q'}$ by $\pi[q, q')$.
  
  Then we have obtained that
  \[ \pi[0, q_{1}) \geq \rho_{a} \pi\left[q_{1}, q_{1} + \Deltat\right). \]
  Similarly, we have that
  \[ \pi\left[0, q_{1} + \Deltat\right) \geq \rho_{a} \pi\left[q_{1} + \Deltat, q_{1} + 2\Deltat\right), \]
  which can be written as
  \begin{eqnarray*}
    \pi[0, q_{1}) + \pi\left[q_{1}, q_{1} + \Deltat\right) & \geq & \rho_{a} \pi\left[q_{1} + \Deltat, q_{1} + 2\Deltat\right), \\
    \pi[0, q_{1})\bras{1 + \frac{1}{\rho_{a}}} & \geq & \rho_{a} \pi\left[q_{1} + \Deltat, q_{1} + 2\Deltat\right).
  \end{eqnarray*}
  By induction, for $m \geq 0$, we have that
  \begin{eqnarray*}
    \frac{\pi[0, q_{1})}{\rho_{a}} \brap{1 + \frac{1}{\rho_{a}}}^{m} & \geq & \pi\left[q_{1} + m \Deltat, q_{1} + (m + 1) \Deltat\right).
  \end{eqnarray*}
  Hence, we have that for $m \geq 0$,
  \begin{eqnarray}
    \pi\left[q_{1}, q_{1} + m \Deltat\right) = 
    \begin{cases}
      \sum_{k = 0}^{m - 1} \pi\left[q_{1} + k \Deltat, q_{1} + (k + 1)\Deltat\right), \text{ if } m > 0, \nonumber \\
      0, \text{ if } m = 0, \nonumber \\
    \end{cases} \\
    \leq \frac{\pi[0, q_{1})}{\rho_{a}} \frac{\brap{1 + \frac{1}{\rho_{a}}}^{m} - 1}{1 + \frac{1}{\rho_{a}} - 1} = \pi[0, q_{1}) \bras{\brap{1 + \frac{1}{\rho_{a}}}^{m} - 1}.
  \end{eqnarray}

  Since $Pr\brac{Q < q_{1}} \leq \frac{V}{a_{1} \epsilon^{2}}$ (from \eqref{eq:q1upperb}), if $m$ is the largest integer such that  
  \begin{eqnarray*}
    \pi\left[0, q_{1}\right) + \pi\left[q_{1}, q_{1} + m \Deltat\right) \leq \frac{1}{2},
  \end{eqnarray*}
  then $\overline{Q}(\gamma) \geq \frac{m\Delta}{4}$.
  Suppose $m_{1}$ is the largest integer such that 
  \begin{eqnarray}
    \pi\left[0, q_{1}\right) + \pi[0, q_{1}) \bras{\brap{1 + \frac{1}{\rho_{a}}}^{m_{1}} - 1} = \pi\left[0, q_{1}\right)\brap{1 + \frac{1}{\rho_{a}}}^{m_{1}} \leq \frac{1}{2}.
    \label{eq:lbut0}
  \end{eqnarray}
  Then $m_{1} \leq m$.
  Using \eqref{eq:q1upperb}, if $m_{2}$ is the largest integer such that
  \begin{eqnarray*}
    \brap{1 + \frac{1}{\rho_{a}}}^{m_{2}} & \leq & \frac{a_{1}\epsilon^{2}}{2V}, \text{ or }, \\
    m_{2} & \leq & \log_{\brap{1 + \frac{1}{\rho_{a}}}} \brap{\frac{a_{1}\epsilon^{2}}{2V}},
  \end{eqnarray*}
  then $m_{2} \leq m_{1}$.
  We have that 
  \[ m_{2} = \floor{\log_{\brap{1 + \frac{1}{\rho_{a}}}} \brap{\frac{a_{1}\epsilon^{2}}{2V}}}. \]
  Since $\overline{Q}(\gamma) \geq \frac{m\Delta}{4} \geq \frac{m_{1}\Delta}{4} \geq \frac{m_{2}\Delta}{4}$, we obtain that 
  \[ \overline{Q}(\gamma) \geq \frac{\Delta}{4} \brap{\log_{\brap{1 + \frac{1}{\rho_{a}}}} \brap{\frac{a_{1}\epsilon^{2}}{2V}} - 1}. \]
  So for the sequence of policies $\gamma_{k}$ with $V_{k} \downarrow 0$ we have that $\overline{Q}(\gamma_{k}) = \Omega\brap{\log\nfrac{1}{V_{k}}}$.
\end{proof}

We now outline the derivation of an asymptotic lower bound for I-model-U in the regime $\Re$.
The analysis of I-model-U proceeds in a similar fashion as in Section \ref{chap5fading:sec:asymp}; the piecewise linear function $c(\lambda)$ and the quantities $a_{p}, p \geq 1$ are similarly defined.
The three cases which then arise are : (1) $0 < \rho \lambda < a_{2}$, (2), $a_{p} < \rho \lambda < a_{p + 1}, p > 1$, and (3) $\rho\lambda = a_{p}, p > 1$.
For Cases 2 and 3, proceeding similarly as in the proof of Lemma \ref{chap5fading:lemma:case2}, it is possible to show that, for any sequence of admissible policies $\gamma_{k}$ such that $\Agk \geq \rho\lambda$ and $\Pgk - c(\rho\lambda) = V_{k} \downarrow 0$, we have that $\Qgk = \Omega\brap{\log\nfrac{1}{V_{k}}}$.
We note that we do not have any asymptotic results for Case 1, although numerically it can be shown that $Q^*(P_{c}) < \infty$ even if $P_{c} = c(\rho\lambda)$.

\begin{remark}
  We note that a similar logarithmic asymptotic lower bound can be obtained for a different class of admissible policies.
  The difference is in the definition of the monotonicity property.
  For this new class of admissible policies, the average drift $\Exp S(q, H)$ - $\Exp A(q, R, H)$ is assumed to be monotonically non-increasing in $q$.
\end{remark}

\begin{remark}
  We comment on an asymptotic upper bound for TRADEOFF, which is achieved by the sequence of Dynamic Packet Dropping (DPD) policies in \cite{neely_utility}.
  A DPD policy is parametrized by the quantities $\beta, \epsilon, \omega$, and $\overline{q}$.
  The policy chooses a batch size $s_{DPD}[m]$ in each slot where
  \begin{eqnarray*}
    s_{DPD}[m] = \min\brap{\argmax_{s \in \brac{0,\cdots,S_{max}}} \bras{s \brac{X[m - 1] - \omega e^{\omega(\overline{q} - Q[m - 1])}} - \beta P(H[m], s)}, Q[m - 1]},
  \end{eqnarray*}
  where $(X[m])$ is a \emph{virtual} queue which evolves according to
  \begin{eqnarray*}
    X[m] & = & \max\brap{X[m - 1] - s_{DPD}[m], 0} + (\rho + \epsilon) R[m].
  \end{eqnarray*}
  For the DPD policy, through admission control, the queue length process $(Q[m], m \geq 0)$ evolves as
  \begin{eqnarray*}
    Q[m] & = & \min\bras{\overline{q}, Q[m - 1] - s_{DPD}[m] + R[m]},
  \end{eqnarray*}
  We note that $A[m] = R[m]$ whenever $Q[m] \leq \overline{q}$, otherwise only that fraction of $R[m]$ is admitted so that $Q[m] = \overline{q}$.
  We note that $A(q,r,h)$ for this DPD policy, is a function only of the current queue length $q$ and the current number of arrivals $r$.
  
  From Theorem 1 \cite{neely_utility}, if $0 < \omega$ and $\omega e^{\omega S_{max}} \leq \frac{\lambda(1 - \rho - \epsilon)}{\sigma^{2}}$, $\epsilon = \frac{1 - \rho}{2\beta}$, $B = \frac{S_{max}^{2} + (\rho + \epsilon)^{2} A_{max}^{2}}{2} + 1$, $x \geq \frac{4 S_{max} e^{\omega S_{max}} B}{\lambda^{2} \omega(1 - \rho - \epsilon)(1 - \rho)}$, and $\overline{q} = \frac{\log(x\beta)}{\omega}$, then for the sequence of policies $\gamma$ obtained by a sequence $\beta \uparrow \infty$, we have that $\overline{Q}(\gamma) = \mathcal{O}(\log(\beta))$, $\overline{P}(\gamma) = c_{R}(\rho\lambda) + \mathcal{O}\brap{\frac{1}{\beta}}$, and $\overline{A}(\gamma) \geq \rho\lambda$.

  We note that DPD policies are not stationary, since each policy depends on an auxiliary state $X[m]$.
  However, as noted in \cite[Section III]{neely_utility}, using a sequence of admissible policies which are obtained from the admissible Positive-Drift Algorithm in \cite{neely_utility} by choosing the parameter $Q$ as $\log\nfrac{1}{V}$, where $V \downarrow 0$, it can be shown that the above tradeoff is achievable.
  Therefore, the asymptotic lower bound derived in Lemma \ref{lemma:tradeoffutilitylb} is tight.
\end{remark}

\subsection{Discussion}
\label{sec:discussion}
\paragraph{Minimization of average delay:}
When average delay is the performance measure under consideration, then the problem that we are interested in is
\begin{eqnarray*}
  \mini_{\gamma \in \Gamma_{a}} \frac{\Qg}{\Ag} \text{ such that } \Pg \leq P_{c} \text{ and } \Ag \geq \rho \lambda.
\end{eqnarray*}
Let the optimal value of the above problem be $D^*(P_{c}, \rho)$.

We note that, since $c(s)$ or $c_{R}(s)$ is a convex and non-decreasing function in $s \in [0, S_{max}]$, for any admissible policy $\gamma$, if $\gamma$ is feasible for the above problem, we have that
\begin{eqnarray*}
  \Expp c(\sQ) \leq \Pg \leq P_{c}, \\
  c(\Expp \sQ) \leq P_{c}, \text{ or}, \\
  \Sg = \Expp \sQ \leq c^{-1}(P_{c}),
\end{eqnarray*}
where $c^{-1}$ is the inverse function of $c$ for I-model-U.
Consider any sequence $P_{c,k} \downarrow c(\rho\lambda)$ as $k \uparrow \infty$.
Since $\Ag = \Sg$, the objective function in the above optimization problem can be bounded above by $\frac{\Qg}{\rho\lambda}$ and bounded below by $\frac{\Qg}{c^{-1}(P_{c,1})}$.
A similar bound can be obtained for R-model-U.
Then, it follows that the asymptotic behaviour of $D^*(P_{c,k}, \rho)$ is the same as that of $Q^*(P_{c,k},\rho)$ as $P_{c,k} \downarrow c(\rho\lambda)$ for I-model-U and $P_{c,k} \downarrow c_{R}(\rho\lambda)$ for R-model-U.

\paragraph{Relation to the asymptotic order optimal tradeoff in \cite{neely_superfast}:}
Neely \cite{neely_superfast} considers a system, with both admission control and service rate control, in which the arrival rate $\lambda$ is larger than the maximum service rate $S_{max}$.
The objective is to obtain a sequence of policies $\gamma_{k}$ which achieve an order optimal minimum average queue length $\Qgk$ as the average utility $U(\Sgk)$ approaches the maximum utility value $u_{max} = U(S_{max})$.
We note that there is no cost associated with the service of packets in \cite{neely_superfast}.
It is shown that for any sequence of policies $\gamma_{k}$ such that $u_{max} - U(\Sgk) = V_{k} \downarrow 0$, $\Qgk = \Omega\brap{\log\nfrac{1}{V_{k}}}$.
A sequence of policies $\gamma_{k}$ such that $\Qgk = \mathcal{O}\brap{\log\nfrac{1}{V_{k}}}$ and $U(\Sgk) = u_{max} - V_{k}$ is also obtained.
We note that as the utility function is assumed to be strictly concave and increasing, the throughput value that maximizes the utility is $S_{max}$ itself.
For any sequence $\gamma_{k}$, if $\Ugk \uparrow u_{max} = U(S_{max})$ it can be shown that the probability of using a service rate less than $S_{max}$ decreases to zero.
That is, with $q_{1} \stackrel{\Delta} = \sup\brac{q : \sq \leq S_{max} - \epsilon}$, $Pr\brac{Q < q_{1}} \downarrow 0$.
Therefore, the proof of Lemma \ref{lemma:tradeoffutilitylb} can be applied to obtain an alternate proof for the asymptotic logarithmic lower bound on the average queue length obtained in \cite{neely_superfast}, but for admissible policies.

\paragraph{Asymptotic bounds for a model with just admission control:}
We consider a queueing model in which there is no service batch size control and no service cost.
The queue evolution is assumed to be as follows:
\begin{eqnarray*}
  Q[m + 1] = \max(Q[m] - s_{b}, 0) + A[m + 1],
\end{eqnarray*}
where $A[m + 1] \leq R[m + 1]$ and $s_{b} \leq S_{max}$ is a fixed batch size.
The queue evolution is assumed to be on $\sZ$.
We assume that $\Exp R[1] = \lambda > s_{b}$.
We consider the case with a single fade state.
We consider the tradeoff of average queue length and average throughput in the asymptotic regime where the average throughput approaches its maximum value $s_{b}$.

As in the previous sections, we consider this tradeoff problem for a class of admissible policies $\Gamma_{a}$.
An admissible policy for this problem is a stationary policy which is stable (as defined in Section \ref{sec:problem}).
However, we note that there is no service batch size control, and it is assumed that the average admitted rate at a queue length $q$, $\overline{a}(q) = \Exp A(q, R)$ is a non-increasing function of $q$.
The tradeoff problem that we consider is:
\begin{eqnarray}
  \mini_{\gamma \in \Gamma_{a}} \Qg \text{ such that } \Ag \geq t_{c}.
  \label{chap5fading:eq:tradeoffnoservicecontrol}
\end{eqnarray}

We note for any admissible policy $\gamma$, $\Ag \leq s_{b}$.
Consider a sequence of policies $\gamma_{k}$ defined as follows.
A policy $\gamma_{k}$, in slot $m$, admits a packet from the batch of size $R[m]$ with probability $\rho_{k}$ or rejects the packet with probability $1 - \rho_{k}$.
Let the sequence $\rho_{k} \Deq \frac{s_{b} - V_{k}}{\lambda}$, for a sequence $V_{k} \downarrow 0$.
Since $\Agk = \rho_{k}\lambda < s_{b}$ we have that $\gamma_{k}$ is a sequence of admissible policies with $\Agk \uparrow s_{b}$.
Therefore, we have that $s_{b} = \sup_{\gamma \in \Gamma_{a}} \Ag$.
We consider the tradeoff problem \eqref{chap5fading:eq:tradeoffnoservicecontrol} in the asymptotic regime $\Re$ where $t_{c} \uparrow s_{b}$.

We have the following asymptotic lower bound.
\begin{lemma}
  For any sequence of admissible policies $\gamma_{k}$, with $s_{b} - \Agk = V_{k} \downarrow 0$, we have that $\Qgk = \Omega\brap{\log\nfrac{1}{V_{k}}}$.
\end{lemma}
\begin{proof}
  Consider a policy $\gamma$ in the sequence $\gamma_{k}$ with $s_{b} - \Ag = V$.
  Since $\Sg = \Ag$, we have that
  \begin{eqnarray*}
    (s_{b} - 1)Pr\brac{s(Q) \neq s_{b}} + s_{b} Pr\brac{s(Q) = s_{b}} \geq s_{b} - V, \text{ or,}\\
    Pr\brac{s(Q) \neq s_{b}} = Pr\brac{s(Q) < s_{b}} \leq V.
  \end{eqnarray*}
  We note that $s(q) = \min(q, s_{b})$.
  
  We now proceed as in the proof of Lemma \ref{chap5:lemma:tprob_2}, but with the following changes.
  We define $q_{s}$ to be $s_{b}$.
  Then, we have that $Pr\brac{Q < q_{s}} = Pr\brac{s(Q) < s_{b}} \leq V$.
  
  Now we obtain a geometric upper bound on $\pi(q)$ for $q \geq q_{s}$ as in the proof of Lemma \ref{chap5:lemma:barq_pi0_relation}.
  We note that for $q \geq q_{s}$, we have that
  \begin{eqnarray*}
    Pr\brac{Q[m + 1] < q| Q[m] = q} \geq Pr\brac{A[1] = 0},
  \end{eqnarray*}
  since $s(q) = s_{b}$.
  Therefore, with $\rho_{d}$ redefined to be just $Pr\brac{A[1] = 0}$ in the proof of Lemma \ref{chap5:lemma:barq_pi0_relation}, we obtain that $\pi(q) \leq Pr\brac{Q < q_{s}} \frac{\rho^{k}}{\rho_{d}}$ for $q = q_{s} + k$ and $k \geq 0$.
  Now proceeding as in the proof of Lemmas \ref{chap5:lemma:barq_pi0_relation} and \ref{chap5:lemma:tprob_2}, we have that for the sequence $\gamma_{k}$, $\Qgk = \Omega\brap{\log\nfrac{1}{V_{k}}}$.
\end{proof}
We note that this model is the discrete time equivalent of \INTLC.
The problem considered here corresponds to \INTLC-2-1.

\section{Single hop networks}
\label{chap5fading:sec:singlehop}
In this section, we illustrate how asymptotic lower bounds for the tradeoff of average power and total average queue length can be derived for a system with $N$ source destination pairs communicating over single hop links.
For example, this could be a $N$ user multiple access or $N$ user broadcast channel.
The model and the associated tradeoff problem that we consider is motivated by Neely \cite{neely_mac}, who considered the problem of optimally trading off average power with average delay for a wireless downlink system, with no admission control.

\subsection{System model}
We first consider a model with real-valued queue evolution.
We note that the model for the single hop network, with no admission control, is a straightforward extension of R-model in Section \ref{chapfading:sysmodel:realval}.
We assume that there is an IID arrival process $(A_{n}[m], m \geq 1)$ to the queue for the $n^{th}$ link, $n \in \brac{1, \dots, N}$.
We assume that $A_{n}[m] \leq A_{max}, \forall n$.
The arrival processes to different links are assumed to be independent of each other.
The $n^{th}$ link is subjected to an IID fading process $(H_{n}[m], m \geq 1)$.
The fading processes are assumed to be independent across links.
The queue length at the start of the $m^{th}$ slot for the $n^{th}$ link is denoted as $Q_{n}[m - 1]$.
In the following, we use the notation $\bS{X}$ to denote the vector $(X_{1}, X_{2}, \cdots, X_{N})$.
The arrival rate vector is $\bS{\lambda} = (\Exp A_{1}[1], \Exp A_{2}[1], \cdots, \Exp A_{N}[1])$.
The fade state is assumed to take values in $\bS{\mathcal{H}}$, with $\min\brac{\min_{n \in \brac{1,\dots, N}}\brac{h^{2}_{n}}, \bS{h} \in \bS{\mathcal{H}}} > 0$.
The distribution of the fade state is denoted as $\pi_{\bS{H}}$ and the expectation with respect to this distribution as $\Exp_{\pi_{\bS{H}}}$.
We assume that $|\bS{\mathcal{H}}| < \infty$.

In each slot $m$, a service batch size vector $\boldsymbol{S}[m]$ is chosen as a randomized vector function $\boldsymbol{S}(\boldsymbol{Q}[m - 1], \boldsymbol{H}[m])$ of the current queue length vector $\bS{Q}[m - 1]$ and the current fade state vector $\bS{H}[m]$.
We assume that $S_{n}[m] \leq S_{max}, \forall n, m$.
A policy $\gamma$ is the choice of the randomized function $\bS{S}(\bS{q}, \bS{h}), \forall \bS{q}, \bS{h}$.
The evolution of the $N$ queues under $\gamma$ is given by:
\begin{eqnarray}
  \bS{Q}[m + 1] = \bS{Q}[m] - \bS{S}[m + 1] + \bS{A}[m + 1],
  \label{chap5fading:eq:vectorevol}
\end{eqnarray}
with $\bS{Q}[0] = \bS{q}_{0}$.
We note that the process $(\bS{Q}[m])$ is a Markov process evolving on $\sR^{N}$.
The use of the service vector $\boldsymbol{s}$ incurs a power $P(\bS{h}, \bS{s})$ when the fade state is $\bS{h}$.
Similar to the properties (C1) and (C2), we assume that for every $\bS{h}$, $P(\bS{h}, \bS{0}) = 0$ and $P(\bS{h}, \bS{s})$ is a strictly convex function of $\bS{s}, \forall \bS{s} \in [0, S_{max}]^{N}$.
These assumptions can be motivated by the properties of $P(\bS{h}, \bS{s})$ obtained in \cite[Chapter 7]{berry_thesis}.
The average power $\overline{P}(\gamma, \bS{q}_{0})$ and total average queue length $\overline{Q}(\gamma, \bS{q}_{0})$ are defined as:
\begin{equation*}
  \overline{P}(\gamma, \bS{q}_{0}) \stackrel{\Delta} = \lim_{M \rightarrow \infty} \frac{1}{M} \Exp \bras{\sum_{m = 1}^{M} P(\bS{H}[m], \bS{S}[m]) \middle \vert \bS{Q}[0] = \bS{q}_{0}},
\end{equation*}
\begin{equation*}
  \overline{Q}(\gamma, \bS{q}_{0}) \stackrel{\Delta} = \lim_{M \rightarrow \infty} \frac{1}{M} \Exp \bras{\sum_{m = 0}^{M - 1} \sum_{n = 1}^{N} Q_{n}[m] \middle \vert \bS{Q}[0] = \bS{q}_{0}}.
\end{equation*}

\subsection{Problem formulation}
We study the tradeoff of $\overline{P}(\gamma, \bS{q}_{0})$ with $\overline{Q}(\gamma, \bS{q}_{0})$ for a restricted set $\Gamma_{a}$ of admissible policies.
A policy $\gamma$ is admissible if (i) it is stable with stationary distribution $\pi(\bS{q})$ (where stability is defined similarly as in Section \ref{chap5fading:sec:problem}), and (ii) instead of property G2, $\gamma$ is such that 
\begin{description}
\item[MG2 :]{the average service rate for queue $n$ as a function of its queue length, $\overline{s}_{n}(q)$ is non-decreasing in $q$.}
\end{description}
For example, for $N = 2$ and for the first queue, for an admissible policy $\gamma$ we require that $\overline{s}_{1}(q_{1}) = \Exp_{Q_{2}|q_{1}}\Exp_{\pi_{\bS{H}}}\Exp_{S_{1}|q_{1},Q_{2}, \bS{H}} S_{1}$ is non-decreasing in $q_{1}$.
We note that for a $\gamma \in \Gamma_{a}$, $\overline{P}(\gamma, \bS{q}_{0}) = \Pg$ and $\overline{Q}(\gamma, \bS{q}_{0}) = \Qg$.
As for the single link case, we study TRADEOFF for the single hop network, defined as:
\begin{equation*}
  \mini_{\gamma \in \Gamma_{a}} \overline{Q}(\gamma), \text{ such that } \overline{P}(\gamma) \leq P_{c}.
\end{equation*}
We define the function $c_{R}(\bS{\lambda})$ as the optimal value of
\begin{eqnarray}
  \mini & \Exp_{\pi_{\bS{H}}}  \Exp_{\bS{S}|\bS{H}} P(\bS{H}, \bS{S}),
  \label{chap5fading:eq:singlehop1} \\
  \text{such that } & \Exp_{\pi_{\bS{H}}}  \Exp_{\bS{S}|\bS{H}} \bS{S} = \bS{\lambda}. \nonumber
\end{eqnarray}
We note that $c_{R}(\bS{\lambda})$ is similar to $c_{R}(\lambda)$ for the single link case.
From \cite{neely_mac}, we have that $c_{R}(\bS{\lambda}) = \inf_{\gamma \in \Gamma_{a}} \Pg$, and $c_{R}(\bS{\lambda})$ is a strictly convex function of $\bS{\lambda}$, for $\bS{\lambda} \in [0, S_{max})^{N}$, with $c_{R}(\bS{0}) = 0$.
We now derive an asymptotic lower bound on the average queue length $\Qgk$ for any sequence of admissible policies $\gamma_{k}$ for which $\Pgk - c_{R}(\bS{\lambda}) \downarrow 0$.
We note that the multiuser Berry-Gallager asymptotic lower bound \cite[Theorem 2]{neely_mac}, is rederived in the following, with the extra assumption MG2.
The method of derivation illustrates how the asymptotic lower bounding technique can be extended to network scenarios.
For ease of exposition, the asymptotic lower bound is derived for the case $N = 2$, but can be extended to any finite $N$.

\subsection{Asymptotic lower bound}
We assume that $A_{n}[1]$ satisfies 
\begin{description}
\item[MA1 :]{$Pr\brac{A_{n}[1] - S_{max} > \delta_{n,a}} > \epsilon_{n,a}$, for some positive $\delta_{n,a}$ and $\epsilon_{n,a}$, $\forall n$.}
\end{description}
We first obtain a lower bound on the marginal stationary probability $Pr\brac{Q_{1} \geq q} = \int_{q_{2} = 0}^{\infty} \int_{q_{1} = q}^{\infty} d\pi(q_{1}, q_{2})$ of the two-dimensional Markov process, which is similar to Lemma \ref{chap5:lemma:realq_dtmc_stat_prob}.
The lower bound on the marginal stationary probability is obtained for the first queue, but can be obtained for the second queue also by interchanging the indices of the two queues.
\begin{lemma}
  Let $(\bS{Q}[m], m \geq 0)$ be as in \eqref{chap5fading:eq:vectorevol}, evolving on $\sR^{2}$, with stationary probability $\pi$, for a $\gamma \in \Gamma_{a}$.
  Suppose there exists a $q_{1,d}$ such that 
  \[ \forall q_{1} \in [0,q_{1,d}], \Exp_{Q_{2}, \bS{H}}\bigg[\Exp\bras{ Q_{1}[m + 1] - Q_{1}[m] \bigg | Q_{2}[m], \bS{H}[m + 1], Q_{1}[m] = q_{1}}\bigg] \geq -d, \]
  where $d$ is positive. Then for any $\bar{q}_{1}$, $k \geq 0$, $\Delta > 0$, $\delta > 0$, $\Delta + \delta < \delta_{1, a}$, and $0 \leq \bar{q}_{1} + k\Delta \leq q_{1,d}$,
  \small
  \begin{eqnarray*}
    Pr\brac{Q_{1} \geq \bar{q}_{1} + k\Delta} & \geq & \brap{\frac{\delta\epsilon_{1,a}}{\delta\epsilon_{1,a} + d}}^{k} Pr\brac{Q_{1} \geq \bar{q}_{1}} \\
    & & + \bras{1 - \brap{\frac{\delta\epsilon_{1,a}}{\delta\epsilon_{1,a} + d}}^{k}}\Bigg [ Pr\brac{Q_{1} \geq q_{1,d}} + \frac{1}{d} \int_{q_{1,d}}^{\infty} (\lambda_{1} - \bar{s}_{1}(q_{1})) d\pi(q_{1}) \Bigg ],
  \end{eqnarray*}
  \normalsize
  where $\pi(q_{1})$ is the marginal probability distribution of $q_{1}$.
  \label{chap5:lemma:multdim_statprob_lb}
\end{lemma}
The proof is very similar to that of Lemma \ref{chap5:lemma:realq_dtmc_stat_prob} and is presented in Appendix \ref{chap5:app:multdim_statprob_lb}. Utilizing the above lower bound, we obtain the following result which is the extension of Lemma \ref{chap5fading:lemma:realvalued} to single hop networks.
As for the single link case, we first express the average power $\Pg$ in terms of the function $c_{R}(\bS{s})$.
We note that the average power used when the queue length vector is $\bS{q}$ is bounded below by the optimal value of
\begin{eqnarray*}
  \mini & \Exp_{\pi_{\bS{H}}}  \Exp_{\bS{S}|\bS{H}} P(\bS{H}, \bS{S}),\\
  \text{such that } & \Exp_{\pi_{\bS{H}}}  \Exp_{\bS{S}|\bS{H}} \bS{S} = (\overline{s}_{1}(q_{1}), \overline{s}_{2}(q_{2})). \nonumber
\end{eqnarray*}
By definition, the optimal value of the above problem is $c_{R}(\overline{s}_{1}(q_{1}), \overline{s}_{2}(q_{2}))$.
Then, we note that $\Expp c_{R}(\overline{s}_{1}(Q_{1}), \overline{s}_{2}(Q_{2})) \leq \Pg$.
Furthermore, since $c_{R}(.)$ is convex, $\Expp c_{R}(\overline{s}_{1}(Q_{1}), \overline{s}_{2}(Q_{2})) \geq c_{R}(\bS{\lambda})$.

\begin{proposition}
  For any sequence of admissible policies $\gamma_{k}$ with $\Pgk - c_{R}(\lambda_{1}, \lambda_{2}) = V_{k} \downarrow 0$, we have that $\overline{Q}(\gamma_{k}) = \Omega\nfrac{1}{\sqrt{V_{k}}}$.
  \label{chap5fading:prop:multiqbg}
\end{proposition}

\begin{proof}
  We note that for the sequence $\gamma_{k}$, since $\overline{P}(\gamma_{k}) - c_{R}(\bS{\lambda}) = V_{k}$, we have that $\Expp c_{R}(\overline{s}_{1}(Q_{1}), \overline{s}_{2}(Q_{2})) - c_{R}(\bS{\lambda}) = U_{k} \downarrow 0$, with $U_{k} \leq V_{k}$.
  Consider a particular policy $\gamma$ in the sequence $\gamma_{k}$ with $U_{k} = U$.
  Let $q_{1,d} = \sup \brac{ q_{1} : \bar{s}_{1}(q_{1}) \leq \lambda_{1} + \epsilon_{U}}$, where $\epsilon_{U}$ will be chosen later.
  From the admissibility of $\gamma$ we have that $\forall q \leq q_{1,d}$, $\bar{s}_{1}(q_{1}) \leq \lambda_{1} + \epsilon_{U}$.
  Assume that $q_{d} < \infty$.
  Using $d = \epsilon_{U}$ in Lemma \ref{chap5:lemma:multdim_statprob_lb}, we have for a $0 < \Delta + \delta < \delta_{a}$, $k \geq 1$, and $\bar{q}_{1} = k\Delta \leq q_{1,d}$
  \small
  \begin{eqnarray*}
    Pr\brac{Q_{1} \geq \bar{q}_{1}} \geq \brap{\frac{\delta\epsilon_{1,a}}{\delta\epsilon_{1,a} + \epsilon_{U}}}^{k} + \bras{1 - \brap{\frac{\delta\epsilon_{1,a}}{\delta\epsilon_{1,a} + \epsilon_{U}}}^{k}}\bras{Pr\brac{Q_{1} \geq q_{1,d}} + \frac{1}{\epsilon_{U}} \int_{q_{1,d}}^{\infty} (\lambda_{1} - \bar{s}_{1}(q_{1})) d\pi(q_{1})}.
  \end{eqnarray*}
  \normalsize
  Or we have that
  \begin{eqnarray*}
    Pr\brac{Q_{1} < \bar{q}_{1}} \leq \brap{1 - \brap{\frac{\delta\epsilon_{1,a}}{\delta\epsilon_{1,a} + \epsilon_{U}}}^{k}}\bras{1 - \frac{1}{\epsilon_{U}}\int_{q_{1,d}}^{\infty} (\lambda_{1} - \bar{s}_{1}(q_{1})) d\pi(q_{1})},
  \end{eqnarray*}
  where we have used the non-negativity of $Pr\brac{Q_{1} \geq q_{1,d}}$.
  Let $D_{t} \Deq \frac{1}{\epsilon_{U}}\int_{q_{1,d}}^{\infty} (\bar{s}_{1}(q_{1}) - \lambda)d\pi(q)$.
  For $\gamma$ we have that $\Expp c_{R}(\overline{s}_{1}(Q_{1}), \overline{s}_{2}(Q_{2})) - c_{R}(\bS{\lambda}) = U$.
  Let $l(s_{1},s_{2})$ be the tangent plane to $c_{R}(s_{1},s_{2})$ at $(\lambda_{1}, \lambda_{2})$.
  Then $U = \Expp \bras{c_{R}(\overline{s}_{1}(Q_{1}),\overline{s}_{2}(Q_{2})) - l(\overline{s}_{1}(Q_{1}),\overline{s}_{2}(Q_{2}))}$, since $\Expp l(\overline{s}_{1}(Q_{1}),\overline{s}_{2}(Q_{2})) = c_{R}(\bS{\lambda})$.
  Let $G(x_{1},x_{2}) \Deq c_{R}(x_{1},x_{2}) - l(x_{1},x_{2})$.
  We note that $G(x_{1}, x_{2})$ is strictly convex in $x_{1}$ and $x_{2}$.
  Also $G(0,0) = 0$ and $\frac{\partial G}{\partial x_{1}}(0,0) = \frac{\partial G}{\partial x_{2}}(0,0) = 0$.
  Proceeding as in steps (42)-(46) in \cite[Appendix A]{neely_mac} we obtain similarly as in the proof of Lemma \ref{chap5:prop:realvalued_evolution_lb} that
  \begin{equation}
    D_{t} \leq \frac{1}{\epsilon_{U}}\sqrt{\frac{U}{a_{1}}},
  \end{equation}
  for a positive $a_{1}$.
  We choose $\epsilon_{U} = 4\sqrt{\frac{U}{a_{1}}}$. 
  Let $k_{1}$ be the largest integer such that
  \begin{eqnarray*}
    \brap{1 - \brap{\frac{\delta\epsilon_{1,a}}{\delta\epsilon_{1,a} + \epsilon_{U}}}^{k_{1}}}\bras{1 - \frac{1}{\epsilon_{U}}\int_{q_{1,d}}^{\infty} (\lambda_{1} - \bar{s}_{1}(q_{1})) d\pi(q_{1})} \leq \frac{1}{2}.
  \end{eqnarray*}
  Then $Pr\brac{Q_{1} < k_{1}\Delta} \leq \frac{1}{2}$ and $\overline{Q}(\gamma) \geq \Expp Q_{1} \geq \frac{k_{1}\Delta}{2}$.
  The same approach also holds if $q_{d} = \infty$.
  The rest of the proof is similar to that of Proposition \ref{chap5:prop:realvalued_evolution_lb} and we obtain that $k_{1} = \Omega\nfrac{1}{\sqrt{U}}$, for small $U$.
  Therefore, $\overline{Q}(\gamma_{k}) =  \Omega\nfrac{1}{\sqrt{U_{k}}} = \Omega\nfrac{1}{\sqrt{V_{k}}}$, since $U_{k} \leq V_{k}$.
\end{proof}

We note that the proof of the above lemma illustrates how the lower bounding technique can be applied to a multiqueue case by considering each queue on its own, even though the service vector $\bS{S}(\bS{Q}, \bS{H})$ is chosen as a function of the queue length vector $\bS{Q}$.
The proof depends on the upper bound on the marginal stationary probability of a particular queue, which can be obtained from the average drift for that queue, conditioned on its own queue length rather than on the queue length vector.

\subsection{Discussion}

\paragraph{Model with integer valued queue evolution:}

\begin{figure}[h]
  \includegraphics[width=160mm,height=65mm]{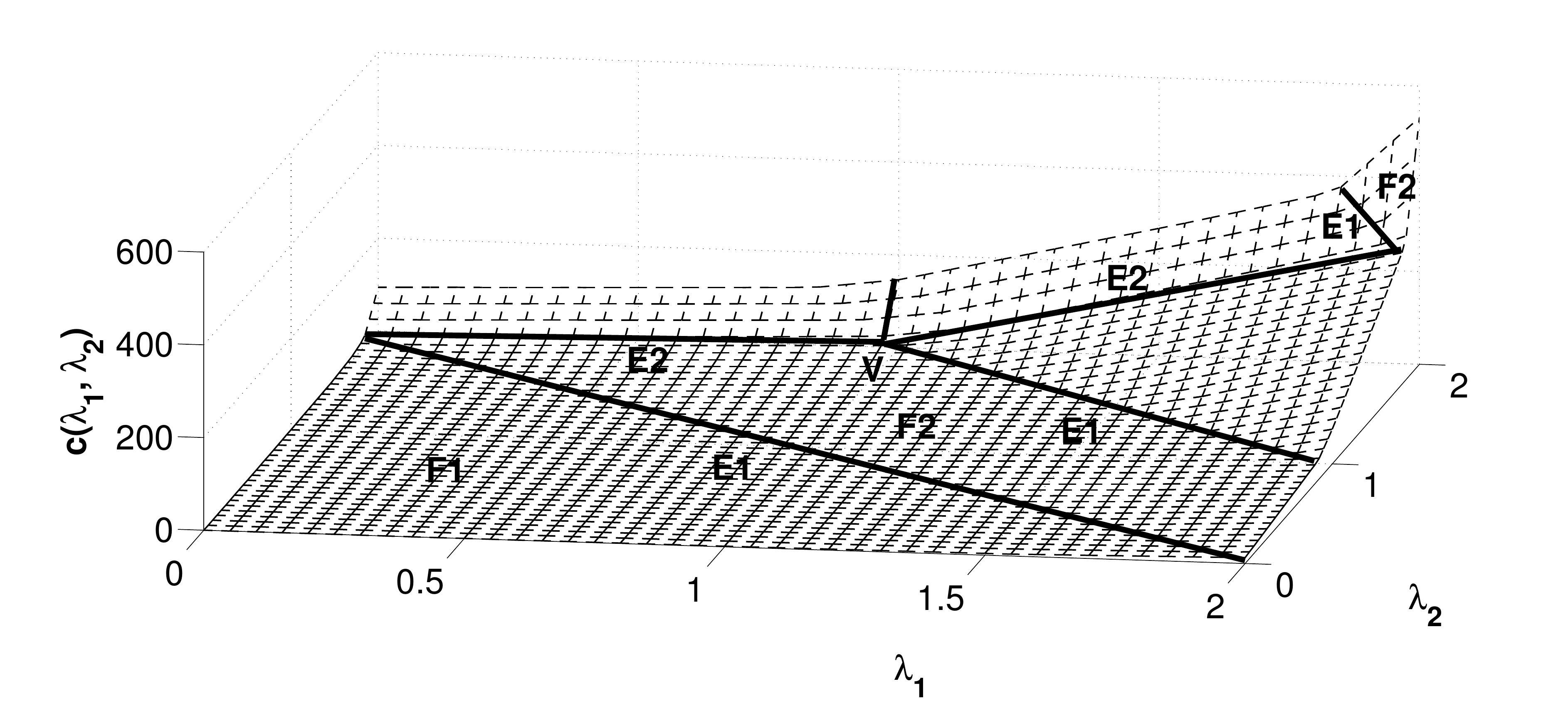}
  \caption{Illustration of $c(\lambda_{1}, \lambda_{2})$ as a function of the arrival rates $\lambda_{1}$ and $\lambda_{2}$.}
  \label{chap5fading:fig:clambdanetwork}
\end{figure}
We note that a single hop queueing network model with $N$ users can be set up similarly as above, where $(\bS{A}[m], m \geq 1)$, $(\bS{S}[m], m \geq 1)$, and $(\bS{Q}[m], m \geq 0)$ are assumed to evolve on $\sZ^{N}$.
Similar to the function $c(\lambda)$ defined for the single link case, a function $c(\bS{\lambda})$ (which is the counterpart of $c_{R}(\bS{\lambda})$) can be obtained as the optimal solution of the problem \eqref{chap5fading:eq:singlehop1}, but with the conditional distributions of service batch sizes having support on $\brac{0, 1, \cdots, S_{max}}$.
We note that $c(\bS{\lambda})$ has a polyhedral structure \cite[Section VII]{neely_mac} (we recall that $c(\lambda)$ was piecewise linear).
Then several cases may arise, e.g., for $N = 2$ we have the following cases, which are illustrated in Figure \ref{chap5fading:fig:clambdanetwork} : (1) $\bS{\lambda}$ lies on the interior of a face of $c(\lambda_{1}, \lambda_{2})$ of type F1, which includes $\bS{0}$ (which corresponds to case 1 for the single link case), (2) $\bS{\lambda}$ lies on the interior of a face of $c(\lambda_{1}, \lambda_{2})$ of type F2, not including $\bS{0}$ (which corresponds to Case 2 for the single link case), (3) $\bS{\lambda}$ lies on a vertex V (not $\bS{0}$) (which corresponds to Case 3 for the single link case), and (4) $\bS{\lambda}$ lies on an edge, which is (i) of type E2, having a projection on the $(\lambda_{1}, \lambda_{2})$ plane which is perpendicular to the $\lambda_{2}$ (or $\lambda_{1}$) axis or (ii) of type E1, having a projection on the $(\lambda_{1}, \lambda_{2})$ plane which is not perpendicular to either the $\lambda_{1}$ or $\lambda_{2}$ axes.

We again consider the TRADEOFF problem for this model in the asymptote of small $V$, where $V$ is the difference between $P_{c}$ and $c(\bS{\lambda})$.
As illustrated in the proof of Proposition \ref{chap5fading:prop:multiqbg}, the asymptotic lower bound for the total average queue length can be obtained by separate lower bounds on the average queue lengths of the individual queues.
Therefore, we expect that the asymptotic lower bounds for Cases 2 and 3 can be obtained from straightforward extensions of Lemmas \ref{chap5fading:lemma:case2} and \ref{chap5fading:corollary:case3} respectively.
As for the single-link case, we do not have any analytical results for Case 1.

For case 4(i) we expect that the total average queue length will grow as $\Omega\nfrac{1}{V}$ since the probability of the average service rate $\overline{s}_{2}(Q)$ being not equal to the arrival rate $\lambda_{2}$ should go to zero as $V \downarrow 0$.
We note that in case 4(i) the average queue length of the first queue is expected to grow as $\Omega\brap{\log\nfrac{1}{V}}$ since the probability of the average service rate $\overline{s}_{1}(Q)$ being greater than the arrival rate $\lambda_{1}$ does not go to zero in the asymptote of small $V$.
We note that for the case of integer valued queue evolution, different queues may have different growth rates in the asymptote of small $V$.

For case 4(ii) we expect that the total average queue length will grow as $\Omega\brap{\log\nfrac{1}{V}}$, since the probability of the average service rate $\overline{s}_{1}(Q)$ and $\overline{s}_{2}(Q)$ being greater than $\lambda_{1}$ and $\lambda_{2}$ respectively is positive as the average power constraint approaches $c(\lambda_{1}, \lambda_{2})$.

\paragraph{Model with admission control:}
We now comment on how the asymptotic lower bound can be derived for a $N$ user model as above, but with admission control.
We consider a model, which is a straightforward extension of R-model-U to the $N$ user case.

We assume that the arrival rate vector into the system is $\bS{R}[m]$ in slot $m$, with $\Exp \bS{R}[1] = \bS{\lambda}$.
For stationary policies, the number of packets admitted into the queue is a function $\bS{A}(\bS{Q}[m - 1], \bS{R}[m], \bS{H}[m])$.
For an admissible policy, defined as above, the average throughput is $\bS{\overline{A}}(\gamma) = \Expp \ExpH \Exp_{R} \Exp \bS{A}(\bS{Q}, \bS{R}, \bS{H})$.
The problem that we are interested in is:
\begin{equation*}
  \mini_{\gamma \in \Gamma_{a}} \overline{Q}(\gamma), \text{ such that } \overline{P}(\gamma) \leq P_{c} \text{ and } \bS{\overline{A}}(\gamma) \geq \rho \bS{\lambda},
\end{equation*}
for a $\rho < 1$.
Similar to the single link case, it can be shown that $\inf_{\gamma \in \Gamma_{a}} \overline{P}(\gamma) = c_{R}(\rho\bS{\lambda})$.
Then under the assumption that $\forall n \in \brac{1,\dots,N}$, $Pr\brac{R_{n}[1] \leq \frac{\Delta_{n}}{2}} = \epsilon'_{a,n} > 0$ for some $\Delta_{n}$ such that $0 < \Delta_{n} < \rho\lambda_{n}$ we can proceed as in the proof of Lemma \ref{lemma:tradeoffutilitylb} to prove that $\Expp Q_{n} = \Omega\brap{\log\nfrac{1}{V_{k}}}, \forall n$, for any sequence of admissible policies $\gamma_{k}$ with $\Pgk - c_{R}(\rho\bS{\lambda}) = V_{k} \downarrow 0$.
Therefore, the total average queue length $\Qgk$ is also $\Omega\brap{\log\nfrac{1}{V_{k}}}$.

Similar results can be obtained for the case where the queue evolution is assumed to be integer valued, whenever $(\bS{\lambda}, c(\bS{\lambda}))$ does not lie on a face which contains $(\bS{0}, \bS{0})$.

\section{{Conclusions}}
We recall that R-model and R-model-U, both with strictly convex $P(h, s)$ functions, are usually used as approximations for I-model and I-model-U respectively.
However, as in Chapter 4, we find that the asymptotic behaviours of $Q^*(P_{c})$ in the asymptotic regimes $\Re$ are different for the approximate models and the original models.

We note that the R-model suggests that a strictly smaller minimum average power $c_{R}(\lambda)$ is sufficient for stability, compared with $c(\lambda)$ for the I-model, for all $\lambda \neq a_{p}, p > 1$.
From Lemma \ref{chap5fading:corollary:case3}, we observe that the asymptotic behaviour of the minimum average queue length is quite different for the I-model and R-model, for $\lambda = a_{p}, p > 1$, for which $c_{R}(\lambda) = c(\lambda)$.
For such a $\lambda$, for the R-model, from Lemma \ref{chap5fading:lemma:realvalued}, we have that the minimum average queue length is $\Omega\nfrac{1}{\sqrt{V}}$, if (a) the average power is $V$ more than the minimum average power $c_{R}(\lambda)$ and (b) $P(h,s)$ is a strictly convex function of $s \in [0, S_{max}]$, for every $h \in \mathcal{H}$.
In contrast, for the I-model, the asymptotic lower bound in Lemma \ref{chap5fading:corollary:case3} shows that the minimum average queue length is $\Omega\nfrac{1}{V}$.
We note that the minimum average queue length for the R-model is always a lower bound to the minimum average queue length obtained from the I-model, for a given constraint on the average power, but the rate of increase of the minimum average queue length as the average power constraint is reduced is strictly smaller.

For Case 1, we find that the $Q^*(P_{c})$ for R-model (as well as R-model-U) increases to infinity in the regime $\Re$, while for I-model we have numerically illustrated that $Q^*(c(\lambda)) < \infty$ (therefore, also for I-model-U).
However, for R-model (as well as R-model-U), if we use a piecewise linear $P(h,s)$ function (as in Section \ref{chap5:eq:rmodel_piecewise_cost}), which is the lower convex envelope of the service cost function defined for the I-model (as well as I-model-U) on $\brac{0, \dots, S_{max}}$, then the asymptotic behaviour of both the R-model (R-model-U) and I-model (I-model-U) matches.
Therefore, a more appropriate approximation for I-model (or I-model-U), is a R-model (or R-model-U) with the above piecewise linear $P(h, s)$ function.

We note that Lemma \ref{chap5fading:lemma:case2} provides the asymptotic lower bound to the $\mathcal{O}\brap{\log\nfrac{1}{V}}$ upper bound observed by Neely in \cite[Corollary 2]{neely_mac} for the I-model.
This asymptotic lower bound was earlier shown only for a specific example (\cite[Section VII-A]{neely_mac}).
We note that if $(A[m])$ and $(H[m])$ are ergodic sequences, independent of each other, then for Cases 2 and 3 for the I-model, we have obtained a $\Omega\brap{\log\nfrac{1}{V}}$ asymptotic lower bound on the minimum average queue length in Lemma \ref{chap5fading:lemma:extension_ergodic}.
As stated, for Case 1, if $\vert \mathcal{H}\vert > 1$, we do not have tight asymptotic upper or lower bounds on $Q^*(P_{c})$, as $P_{c} \downarrow c(\lambda)$.
Lemma \ref{lemma:tradeoffutilitylb} provides an asymptotic lower bound for R-model-U even if $|\mathcal{H}| > 1$, for admissible policies, while earlier an asymptotic lower bound was obtained only for the case $|\mathcal{H}| = 1$.
We also illustrate how the asymptotic lower bound can be obtained for a N user single hop network and identify a case in which average queue lengths for different queues can have different asymptotic behaviours for integer valued queue evolution.

\clearpage
\begin{subappendices}
\large{\textbf{Appendices}}
\addcontentsline{toc}{section}{Appendices}
\addtocontents{toc}{\protect\setcounter{tocdepth}{0}}
\normalsize
\vspace{-0.4in}
\section{The function $c(\lambda)$ in Section \ref{chap5fading:sec:problem}}
\label{chap5fading:app:clambda}
In this section, we show that the function $c(\lambda)$ defined as the optimal value of
\begin{eqnarray*}
  \mini & & \ExpH \Exp_{S|H} P(H, S), \\
  \text{such that } & & \ExpH \Exp_{S|H} S = \lambda,
\end{eqnarray*}
is piecewise linear if $|\mathcal{H}|$ is finite.
Let $p_{s|h}$ be the conditional distribution of the batch size $s$ given the fade state $h$.
Then the above problem can be written as
\begin{eqnarray*}
  \mini & & \sum_{h} \pi_{H}(h) \sum_{s = 0}^{S_{max}} p_{s|h} P(h, s), \\
  \text{such that } & & \sum_{h} \pi_{H}(h) \sum_{s = 0}^{S_{max}} p_{s|h} s = \lambda, \\
  & & \sum_{s = 0}^{S_{max}} p_{s|h} = 1, \forall h, \text{ and } p_{s|h} \geq 0, \forall s, h,
\end{eqnarray*}
which is a linear program in the variables $p_{s|h}$.

For ease of exposition, in the following we consider the case where $|\mathcal{H}| = 2$, but the approach holds for any finite $|\mathcal{H}|$.
Let $\bS{s} = (0, 1, \dots, S_{max})$, $P(h_{i}, \bS{s}) = (P(h_{i}, 0), \dots, P(h_{i}, S_{max}))$, $p_{\bS{s}|h} = (p_{0|h}, \dots, p_{S_{max}|h})$, and let $\bS{1}$ be a row vector of all ones and $\bS{0}$ a row vector of all zeros, both of size $S_{max}$.
Then the above linear program can be written as
\begin{eqnarray*}
  \mini & & [\pi_{H}(h_{1}) P(h_{1}, \bS{s}), \pi_{H}(h_{2}) P(h_{2}, \bS{s})]
  \bras{\begin{array}{c}
      p_{\bS{s}|h_{1}}^{t} \\
      p_{\bS{s}|h_{2}}^{t}
    \end{array}
  } \\
  \text{such that } & & \bras{\begin{array}{cc}
      \pi_{H}(h_{1})\bS{s} & \pi_{H}(h_{2})\bS{s} \\
      \bS{1} & \bS{0} \\
      \bS{0} & \bS{1}
    \end{array}
  }\bras{\begin{array}{c}
      p_{\bS{s}|h_{1}}^{t} \\
      p_{\bS{s}|h_{2}}^{t}
    \end{array}
  } = \bras{\begin{array}{c}
      \lambda \\
      1 \\
      1
    \end{array}
  },\\
  \text{and } & & p_{s|h} \geq 0, \forall s, h.
\end{eqnarray*}
We note that the dual of this problem is 
\begin{eqnarray*}
  \maxi & & \bras{y_{1}, y_{2}, y_{3}}\bras{\begin{array}{c}
      \lambda \\
      1 \\
      1
    \end{array}
  } \\
  \text{such that} & & \bras{y_{1}, y_{2}, y_{3}}\bras{\begin{array}{cc}
      \pi_{H}(h_{1})\bS{s} & \pi_{H}(h_{2})\bS{s} \\
      \bS{1} & \bS{0} \\
      \bS{0} & \bS{1}
    \end{array}
  } \leq [\pi_{H}(h_{1}) P(h_{1}, \bS{s}), \pi_{H}(h_{2}) P(h_{2}, \bS{s})],
\end{eqnarray*}
where $y_{1}, y_{2}, y_{3} \in \mathbb{R}$.
Let $y_{1}^*(\lambda), y_{2}^*(\lambda)$, and $y_{3}^*(\lambda)$ be any optimizers for the dual for $\lambda \in [0, S_{max}]$.
Since $c(\lambda)$ is convex in $\lambda$ (\cite{neely_mac}), it is differentiable at all $\lambda \in [0, S_{max}]$ except for $\lambda \in \mathcal{D}$, where $\mathcal{D}$ is at most countable \cite{william}.
Now consider $c(\lambda)$ for a $\lambda \not \in \mathcal{D}$.
Then we have that $\frac{dc(\lambda)}{d\lambda}$ is well defined.
Furthermore, since strong duality holds for the linear programs above, we also have that $\frac{dc(\lambda)}{d\lambda} = -y_{1}^*(\lambda)$ \cite[Section 5.6]{boyd}.
Therefore, for $\lambda \not \in \mathcal{D}$, $y_{1}^*(\lambda)$ is unique.

We note that the constraint set in the dual problem does not depend on $\lambda$ and has finite number of vertices.
Therefore, there are only finitely many ways in which $y_{1}^*(\lambda)$ can be unique.
Hence, $\frac{dc(\lambda)}{d\lambda}$ can take only finitely many values.
Since $c(\lambda)$ is also non-decreasing in $\lambda$, we have that $c(\lambda)$ is piecewise linear.
We note that the above approach generalizes to any finite $\mathcal{H}$.

\section{Proof of Lemma \ref{chap5fading:lemma:case2upperbound}}
\label{chap5fading:app:case2upperbound}
Let $L(q) \Deq e^{\omega(q_{v} - q)}$ be a Lyapunov function.
Since for the policy $\gamma$, the batch size $\tilde{S}(q)$ could be more than $q$, the queue evolution equation under $\gamma$ is written as 
\[ Q[m + 1] = \max(Q[m] - \tilde{S}(Q[m]), 0) + A[m + 1].\]
The expected Lyapunov drift is
\begin{eqnarray*}
  \Delta(q) \Deq \Exp\bras{L(Q[m + 1]) - L(Q[m]) \vert Q[m] = q}.
\end{eqnarray*}
We note that the randomness in $\tilde{S}(q)$ arises from both the randomness in the fade state as well as the randomization of the batch size.
The expectation of $\tilde{S}(q)$ is therefore with respect to this distribution.
Proceeding as in the proof of Lemma \ref{chap5:app:case2upperbound} we have that
\begin{eqnarray*}
  \Delta(q) \leq \omega e^{\omega(q_{v} - q)}\bras{(\Exp \tilde{S}(q) - \lambda) + K}.
\end{eqnarray*}
where $K = \frac{\omega A_{max}^{2}}{2} e^{\omega A_{max}}$.

Now by definition, the policy $\gamma$ is such that
\begin{equation*}
  \Exp \tilde{S}(q) = 
  \begin{cases}
    s_{l}, \text{ for } 0 \leq q < q_{v}, \\
    s_{u}, \text{ for } q_{v} \leq q.
  \end{cases}
\end{equation*}
Then we have that for $q < q_{v}$
\begin{eqnarray*}
  \Delta(q) & \leq & -\omega e^{\omega(q_{v} - q)}\bras{\lambda - s_{l} - K}.
\end{eqnarray*}
And for $q \geq q_{v}$,
\begin{eqnarray*}
  \Delta(q) & \leq & \omega e^{\omega(q - q_{v})}\bras{(s_{u} - \lambda) + K}, \\
  & = & -\omega e^{\omega(q_{v} - q)}\bras{\lambda - s_{l} - K} + \omega e^{\omega(q_{v} - q)}\bras{s_{u} - s_{l}}, \\
  & \leq & -\omega e^{\omega(q_{v} - q)}\bras{\lambda - s_{l} - K} + \omega \bras{s_{u} - s_{l}}, \\
\end{eqnarray*}
Hence, for all $q$ we have that
\begin{eqnarray*}
  \Delta(q) & \leq & -\omega e^{\omega(q_{v} - q)}\bras{\lambda - s_{l} - K} + \omega \bras{s_{u} - s_{l}}.
\end{eqnarray*}

We choose $\omega$ such that $K < \lambda - s_{l}$.
Proceeding as in the proof of \cite[Theorem 3(c)]{neely_mac}, we have that
\begin{eqnarray*}
  \Exp e^{\omega(q_{v} - Q)} \leq \frac{\bras{s_{u} - s_{l}}}{(\lambda - s_{l}- K)}.
\end{eqnarray*}
Since $\Exp e^{\omega(q_{v} - Q)} \geq \Exp \bras{e^{\omega(q_{v} - Q)} \vert Q < S_{max}} Pr\brac{Q < S_{max}}$, we therefore have that
\begin{eqnarray}
  Pr\brac{Q < S_{max}} \leq e^{-\omega q_{v}} \frac{e^{\omega S_{max}} \bras{s_{u} - s_{l}}}{(\lambda - s_{l}- K)}.
\end{eqnarray}

Now we note that
\begin{eqnarray*}
  \overline{P}(\gamma) & = & Pr\brac{Q < S_{max}} \Exp \bras{P(H, S) \vert Q < S_{max}} + \\
  & & Pr\brac{S_{max} \leq Q < q_{v}} \Exp \bras{P(H, S) \vert S_{max} < Q < q_{v}} + Pr\brac{q_{v} \leq Q} \Exp \bras{P(H, S) \vert q_{v} \leq Q}.
\end{eqnarray*}
We note that for $\gamma$, for $S_{max} \leq q < q_{v}$, $\Exp P(H, S) = c(s_{l})$ and for $q \geq q_{v}$, $\Exp P(H, S) = c(s_{u})$.
Furthermore for $q < S_{max}$, $\Exp P(H, S) \leq c(s_{l})$.
Hence, we have that
\begin{eqnarray*}
  \overline{P}(\gamma) & \leq & Pr\brac{Q < S_{max}} c(s_{l}) + Pr\brac{S_{max} \leq Q < q_{v}} c(s_{l}) + Pr\brac{q_{v} \leq Q} c(s_{u}).
\end{eqnarray*}
Proceeding as in the proof of Lemma \ref{chap5:lemma:case2upperbound} we have that
\begin{eqnarray*}
  \overline{P}(\gamma) - c(\lambda) & \leq & \bras{c(s_{l}) + m\lambda - c(\lambda)} Pr\brac{Q < S_{max}},
\end{eqnarray*}
where $m$ is the slope of $l(s)$.

Now consider the sequence of policies $\gamma_{k}$ for which $q_{v} = \log\nfrac{1}{V_{k}}$ for a sequence $V_{k} < 1$ such that $V_{k} \downarrow 0$.
Then we have that $\Pgk - c(\lambda) = \mathcal{O}(V_{k})$.
Furthermore, from Proposition \ref{chap5:prop:avgq_driftub} we have that $\Qg = \mathcal{O}\brap{\log\nfrac{1}{V_{k}}}$.
We note that $\gamma_{k}$ is also a sequence of admissible policies, since $s(q)$ is a non-decreasing function of $q$ and $\overline{Q}(\gamma_{k}) < \infty$.

\section{Proof of Lemma \ref{chap5:lemma:multdim_statprob_lb}}
\label{chap5:app:multdim_statprob_lb}

\begin{proof}
  Define $\widehat{Q}_{1}[m] = \max(\bar{q}_{1}, Q_{1}[m])$ and $\widehat{Q}_{1} = \max(\bar{q}_{1}, Q)$.
  As the policy is admissible and $\bar{q}_{1}$ is finite we have that $\Expp \widehat{Q} < \infty$.
  Therefore
  \[ \int_{q_{1},q_{2}} \mathbb{E}\bras{\widehat{Q}_{1}[m + 1] - \widehat{Q}_{1}[m] \middle \vert Q_{1}[m] = q_{1}, Q_{2}[m] = q_{2}}d\pi(q_{1},q_{2}) = 0. \]
  We split the integral over $q_{1}$ into three terms which leads to :
  \begin{eqnarray}
    0 & = & \int_{0}^{\bar{q}_{1} - \Delta} \int_{0}^{\infty} \mathbb{E}\bras{\widehat{Q}_{1}[m + 1] - \widehat{Q}_{1}[m] \middle \vert Q_{1}[m] = q_{1}, Q_{2}[m] = q_{2}} d\pi(q_{2}|q_{1}) d\pi(q_{1})
    \label{chap5:eq:multiq1} \\
    & & + \int_{\bar{q}_{1} - \Delta}^{\bar{q}_{1}} \int_{0}^{\infty} \mathbb{E}\bras{\widehat{Q}_{1}[m + 1] - \widehat{Q}_{1}[m] \middle \vert Q_{1}[m] = q_{1}, Q_{2}[m] = q_{2}} d\pi(q_{2}|q_{1})d\pi(q_{1}) 
    \label{chap5:eq:multiq2} \\
    & & + \int_{\bar{q}_{1}}^{\infty} \int_{0}^{\infty} \mathbb{E}\bras{\widehat{Q}_{1}[m + 1] - \widehat{Q}_{1}[m] \middle \vert Q_{1}[m] = q_{1}, Q_{2}[m] = q_{2}} d\pi(q_{2}|q_{1})d\pi(q_{1}) 
    \label{chap5:eq:multiq3}
  \end{eqnarray}
  We note that for $q_{1} \leq \bar{q}_{1} - \Delta$, \eqref{chap5:eq:multiq1} $\geq 0$.
  Consider \eqref{chap5:eq:multiq3}, for which $\widehat{Q}_{1}[m] = Q_{1}[m]$ and $\widehat{Q}_{1}[m + 1] \geq Q_{1}[m + 1]$.
  Therefore
  \begin{equation*}
    \eqref{chap5:eq:multiq3} \geq \int_{\bar{q}_{1}}^{\infty} (\lambda_{1} - \bar{s}_{1}(q_{1})) d\pi(q_{1}).
  \end{equation*}
  Using the assumption MG3, the above integral can be further bounded below by
  \begin{equation*}
    -d Pr\brac{\bar{q}_{1} \leq Q_{1} < q_{1,d}} + \int_{q_{1,d}}^{\infty} (\lambda_{1} - \bar{s}_{1}(q_{1})) d\pi(q_{1}).
  \end{equation*}
  Consider \eqref{chap5:eq:multiq2}, we have that $\widehat{Q}_{1}[m + 1] - \widehat{Q}_{1}[m] \geq 0$ for $q_{1} \in [\bar{q}_{1} - \Delta, \bar{q}_{1}]$.
  Hence as in the proof of Lemma \ref{chap5:lemma:realq_dtmc_stat_prob} we use Markov inequality to lower bound \eqref{chap5:eq:multiq2}.
  \small
  \begin{eqnarray*}
    \mathbb{E}\bras{\widehat{Q}_{1}[m + 1] - \widehat{Q}_{1}[m] \vert Q_{1}[m] = q_{1}, Q_{2}[m] = q_{2}} & \geq & \delta \Pr\brac{\widehat{Q}_{1}[m + 1] - \widehat{Q}_{1}[m] \geq \delta \middle \vert Q_{1}[m] = q_{1}, Q_{2}[m] = q_{2}}, \\ 
    & \geq & \delta Pr\brac{{Q}_{1}[m + 1] - {Q}_{1}[m] \geq \delta + \Delta \middle \vert Q_{1}[m] = q_{1}, Q_{2}[m] = q_{2}}, \\
    & \geq & \delta \epsilon_{1, a},
  \end{eqnarray*}
  \normalsize
  where $\Delta > 0$ and $\delta > 0$ are chosen such that $\Delta + \delta < \delta_{1,a}$.
  Thus we obtain that
  \begin{equation*}
    \eqref{chap5:eq:multiq2} \geq \delta\epsilon_{1,a} Pr\brac{\bar{q}_{1} - \Delta \leq Q_{1} < \bar{q}_{1}}.
  \end{equation*}
  Combining the obtained lower bounds on \eqref{chap5:eq:multiq1}, \eqref{chap5:eq:multiq2}, and \eqref{chap5:eq:multiq3} we obtain that
  \begin{eqnarray*}
    0 & \geq & \delta\epsilon_{1,a} Pr\brac{\bar{q}_{1} - \Delta \leq Q_{1} < \bar{q}_{1}} -d Pr\brac{\bar{q}_{1} \leq Q_{1} < q_{1,d}} + \int_{q_{1,d}}^{\infty} (\lambda_{1} - \bar{s}_{1}(q_{1})) d\pi(q_{1}), \\
    & = & \delta\epsilon_{1,a} Pr\brac{Q_{1} \geq \bar{q}_{1} - \Delta} - (d + \delta \epsilon_{a}) Pr\brac{Q_{1} \geq \bar{q}_{1}} + d Pr\brac{Q_{1} \geq q_{1,d}} + \int_{q_{1,d}}^{\infty} (\lambda_{1} - \bar{s}_{1}(q_{1})) d\pi(q_{1}).
  \end{eqnarray*}
  Hence
  \begin{eqnarray*}
    Pr\brac{Q_{1} \geq \bar{q}_{1}} & \geq & \frac{\delta\epsilon_{1,a}}{\delta\epsilon_{1,a} + d} Pr\brac{Q_{1} \geq \bar{q}_{1} - \Delta} + \\
    & & \frac{1}{\delta\epsilon_{1,a} + d} \bras{d Pr\brac{Q_{1} \geq q_{1,d}} + \int_{q_{1,d}}^{\infty} (\lambda_{1} - \bar{s}_{1}(q_{1})) d\pi(q_{1})}.
  \end{eqnarray*}
  By induction, as in the proof of Lemma \ref{chap5:lemma:realq_dtmc_stat_prob}, we obtain that if $k \geq 0$ and $\bar{q}_{1} + k\Delta \leq q_{1,d}$, then
  \begin{eqnarray*}
    Pr\brac{Q_{1} \geq \bar{q}_{1} + k\Delta} & \geq & \brap{\frac{\delta\epsilon_{1,a}}{\delta\epsilon_{1,a} + d}}^{k} Pr\brac{Q_{1} \geq \bar{q}_{1}} \\
    & & + \bras{1 - \brap{\frac{\delta\epsilon_{1,a}}{\delta\epsilon_{1,a} + d}}^{k}}\bras{Pr\brac{Q_{1} \geq q_{1,d}} + \frac{1}{d} \int_{q_{1,d}}^{\infty} (\lambda_{1} - \bar{s}_{1}(q_{1})) d\pi(q_{1})}.
  \end{eqnarray*}
\end{proof}

\end{subappendices}

\addtocontents{toc}{\protect\setcounter{tocdepth}{2}}

\blankpage
\ifdefined \Deltaa \else 
\newcommand{\Deltaa}{\frac{\Delta_{a}}{2}}
\fi
\chapter[On the tradeoff of average error rate and average delay for point to point links]{\textbf{On the tradeoff of average error rate and average delay \\for point to point links}}

\section{Introduction}
\label{chap2:sec:introduction}

In this chapter, we consider the transmission of a bursty information source over a noisy point to point link.
Our objective is to transmit the randomly arriving source message symbols such that the message symbols are decoded \emph{reliably} and with minimum \emph{delay}.
Reliability and delay performance are measured by the average symbol error rate and the average symbol delay respectively.
Understanding the fundamental tradeoff between reliability and delay is significant due to the increasing use of cross layer scheduling, for allocation of highly constrained wireless resources, in modern high rate communication networks.
We assume that the transmitter and receiver use a block code (such as a LDPC code) to reliably communicate the message symbols, as in many practical scenarios.

It is known \cite[Chapter 24]{elgamal_notes} that if the average information arrival rate into the system is less than the capacity of the channel, then arbitrarily low probability of error can be achieved using block coding with long codeword lengths and \emph{finite} average delay.
In this chapter, we first characterize how the minimum possible average delay grows when the average error rate is made arbitrarily small.
However, there are cases where arbitrarily long codewords cannot be used.
Then there is a positive infimum for achievable error rate.
Then for this case, we characterize how the minimum possible average delay grows as the average error rate is made arbitrarily close to the above positive infimum of achievable error rates, using the techniques discussed in Chapter 4.
We note that the service cost function in this chapter, which is the expected number of message symbols in error when a batch of message symbols is transmitted, turns out to be non-convex unlike the convex service cost functions in the previous chapters.
The notation that we use in this chapter is summarized in Tables \ref{chapter6:notationtable1} and \ref{chapter6:notationtable2}.

\begin{table}
\centering
\begin{tabular}{|l|l|}
\hline
Symbol & Description \\
\hline
$n$ & slot index \\
$A_{s}[n]$ & random number of message symbol arrivals in slot $n$ \\
$A_{max}$ & maximum number of message symbol arrivals in a slot \\
$\lambda, \sigma^{2}$ & mean and variance of $A_{s}[n]$ \\
$\mathcal{X}$ & input alphabet of a point to point channel \\
$\mathcal{Y}$ & output alphabet of a point to point channel \\
$P_{Y|X}$ & channel transition probability function \\
$\mQ$ & distribution on the channel input symbol $\in \mathcal{X}$ \\
$s$ & a particular batch size \\
$\tau$ & a particular transmission duration \\
$T_{a,i}$ & arrival time of $i^{th}$ message \\
$T_{d,i}$ & departure time of $i^{th}$ message \\
$E_{i}$ & event that the $i^{th}$ symbol is in error \\
$m$ & decision epoch index \\
$S[m]$ & service batch size for the $m^{th}$ transmission \\
$S_{max}$ & maximum service batch size \\
$\mathcal{T}[m]$ &  duration of the $m^{th}$ transmission \\
$\gamma[m]$ & $ = (S[m], \mathcal{T}[m])$ \\
$Q[m]$ & queue length at decision epoch $m$ \\
$Q_{s}[n]$ & queue length at start of slot $n$ \\
$A[m]$ & random number of message symbol arrivals between $(m - 1)^{th}$ and $m^{th}$ decision epochs \\
$\Gamma$ & set of all policies \\
$\Gamma_{s}$ & set of stationary policies \\
$P_{e}(\gamma)$ & average error prob. for policy $\gamma$ \\
$D(\gamma)$ & average delay for policy $\gamma$ \\
$P_{e,c}$ & constraint on the probability of error of message symbols \\
$D_{c}$ & constraint on the average delay of message symbols \\
$\mathcal{M}$ & source message alphabet \\
$|\mathcal{M}|$ & cardinality of $\mathcal{M}$ \\
$N_{c}$ & fixed codeword length \\
$h(q,\tau)$ & holding cost; is $q\tau + \frac{\tau(\tau - 1)\lambda}{2}$ \\
$\Gamma_{s,f}$ & set of all policies with transmission time being a fixed parameter \\
$\Gamma_{s, N_{c}}$ & set of all policies with transmission time being $N_{c}$ \\
\hline
\end{tabular}
\caption{Notation used in this chapter (part I).}
\label{chapter6:notationtable1}
\end{table}

\begin{table}
\centering
\begin{tabular}{|l|l|}
\hline
Symbol & Description \\
\hline
$\Qg$ & average queue length for a policy $\gamma$ \\
$c_{s}(s, \tau)$ & expected number of messages in error for a random block code \\
$E_{0}(\rho, \mQ)$ & Gallager's random coding error exponent \\
$P_{e}^*(D_{c})$ & minimum average error probability over $\Gamma_{s}$ under delay constraint $D_{c}$ \\
$P_{e,f}^*(D_{c})$ & minimum average error probability over $\Gamma_{s, f}$ under delay constraint $D_{c}$ \\
$P_{e,N_{c}}^*(D_{c})$ & minimum average error probability over $\Gamma_{s, N_{c}}$ under delay constraint $D_{c}$ \\
$P_{N_{c}}(\lambda)$ & minimum achievable average error probability for the system to be stable \\
$R_{c}$ & cutoff rate of a point to point channel \\
$C$ & capacity of a point to point channel \\
$m_{s}$ & a particular message symbol, $m_{s} \in \mathcal{M}$ \\
$\tau_{0}$ & duration of time the transmitter idles \\
$\tau(s)$ & transmission time as a function of s \\
\hline
\end{tabular}
\caption{Notation used in this chapter (part II).}
\label{chapter6:notationtable2}
\end{table}

\subsection{System model}
\label{chap2:sec:general_model}
The system is assumed to evolve in discrete time units of slots.
The slots are indexed by $n \in \{1,2,\cdots\}$.
Each slot corresponds to one channel use.
We note that these slots may be thought of as subslots in the slots for the discrete time models in Chapter 4 (this is why the slots are indexed by $n$ rather than $m$ in this chapter).
The channel is assumed to be a memoryless channel, with input alphabet $\mathcal{X}$ and output alphabet $\mathcal{Y}$.
The transition probability function of the channel is denoted by $P_{Y|X}$.

We assume that there is no admission control throughout this chapter.
The source generates a random number $A_{s}[n] \leq A_{max}$ of message symbols in each slot $n$.
We assume that $A_{s}[n]$ is IID with $\mathbb{E}A_{s}[1] = \lambda$ and $var(A_{s}[1]) = \sigma^{2}$, both of which are finite.
Each message symbol is generated independently and uniformly from a message alphabet $\mathcal{M}$ of finite cardinality $|\mathcal{M}|$. 
The message symbols are assumed to enter the queue just before the slot boundary and reside in the transmitter buffer until they are encoded and transmitted. 
Transmission of a message symbol is assumed to require at least one slot.
The transmitter buffer size is assumed to be infinite.

The transmitter is assumed to use random block coding \cite[Chapter 5]{gallager_text}.
Each symbol of the transmitted channel codeword is picked IID from a distribution $\mQ(.)$ on $\mathcal{X}$.
At a decision epoch, which occurs at, say the start of slot $n$, if the transmitter decides to transmit, it uses a random block codebook, with codewords generated as above.
The codebook is characterized by two parameters, the number of message symbols ($s$) which are encoded and transmitted by the codeword and the length ($\tau$) of the codeword.
Starting from slot $n$, $s$ message symbols are removed from the transmitter queue after $\tau$ slots.
We assume that there is no path delay in the transmission of the codeword symbols.
The receiver is assumed to decode the $s$ message symbols jointly by using maximum likelihood decoding of the codeword.
If the transmitter decides not to transmit, then the receiver is made aware of the idle state through the control channel and the transmitter idles for $\tau_{0}$ slots.
A decision epoch occurs after every transmission period or idle period.
We assume that at the end of a slot $n$, events occur in the following order: a) if the service of a batch of symbols ends then the batch is removed from the queue, b) new arrivals in slot $n$ are admitted into the queue, and c) the queue length state at the beginning of the next slot is obtained.

For the $i^{th}$ message symbol, let $T_{a,i}$ denote the slot in which the symbol arrives into the transmitter queue.
Let $T_{d,i}$ denote the slot in which the $i^{th}$ symbol departs from the receiver.
We note that $T_{d,i}$ is the slot in which transmission of the batch containing the $i^{th}$ symbol finishes.
The delay of the $i^{th}$ symbol in the queue is $T_{d,i} - T_{a,i}$.
Let $E_{i}$ denote the event that the $i^{th}$ symbol is in error.

\begin{figure}[h]
  \centering
  \includegraphics[width=140mm,height=80mm]{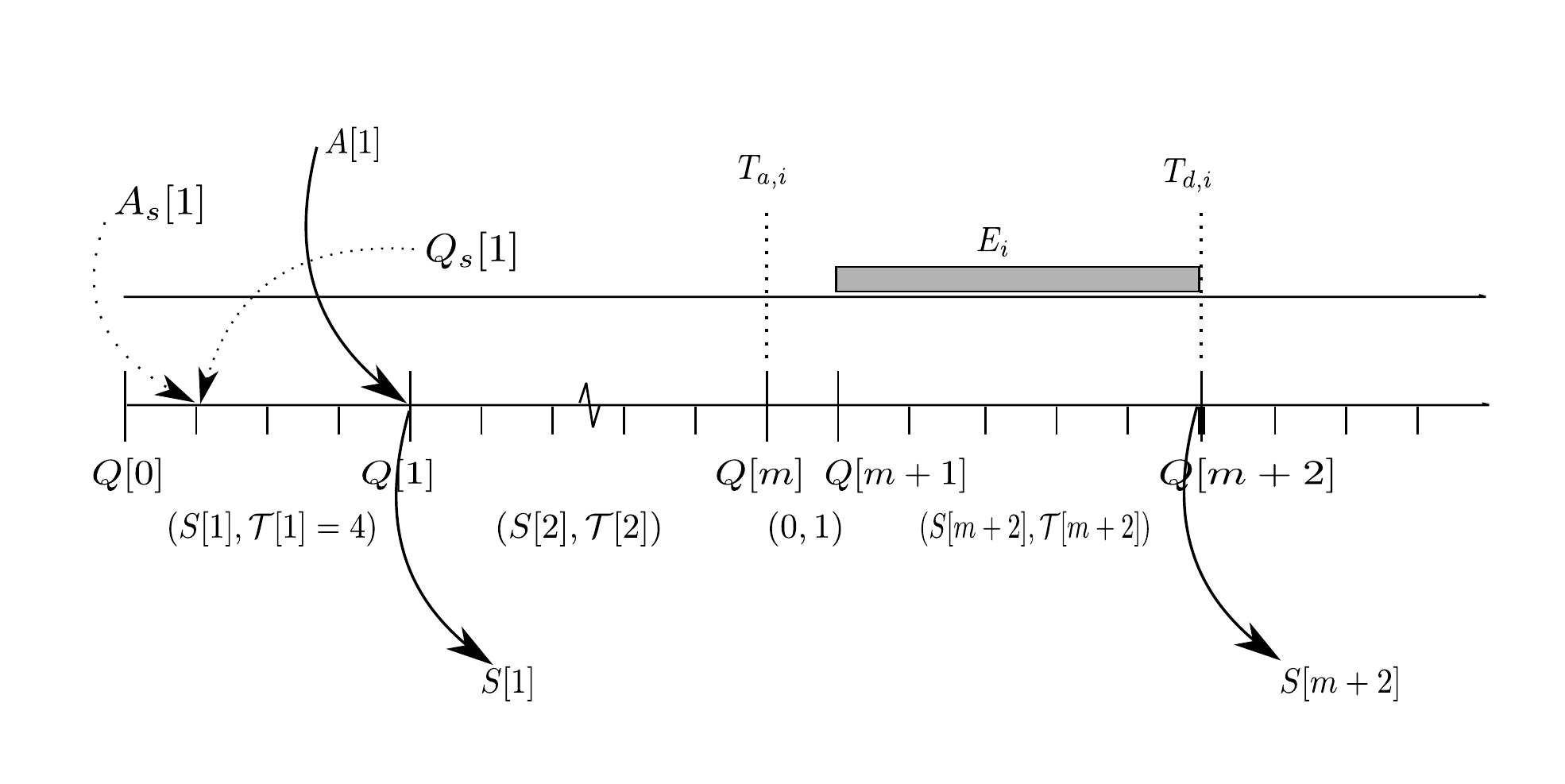}
  \caption{Evolution of the system for a block coding scheme}
  \label{chap2:fig:system_evolution}
\end{figure}

A policy $\gamma$ for operation of the transmitter consists of a sequence $((S[m],\Tau[m]), m \geq 1)$, where $m \in \{1,2,3,\cdots\}$.
If $S[m] > 0$, then $S[m]$ is the number of message symbols which start transmission at the $(m-1)^{th}$ decision epoch using a codeword of length $\mathcal{T}(m)$.
If $S[m] = 0$, then the system idles for $\Tau[m] = \tau_{0}$ slots.
Let $\gamma[m]=(S[m],\Tau[m])$. 
The evolution of the system is illustrated in Figure \ref{chap2:fig:system_evolution}.
For every $m$, $\gamma[m]$ can be a randomized function of (i) the history $(\gamma[1],\gamma[2],\cdots, \gamma[m - 1])$, (ii) the initial number of message symbols in the transmitter queue $q_{0}$, and (iii) the arrival process up to the start of the $(m-1)^{th}$ decision epoch.
The length of the transmitter queue at a decision epoch $m$ is denoted by $Q[m]$.
The queue length at the start of a slot $n$ is denoted by $Q_{s}[n - 1]$. 
The evolution of the system sampled at the decision epochs, for a policy $\gamma$, is given by the following equation:
\begin{equation}
  Q[m + 1] = Q[m] - S[m + 1] + A[m + 1],
  \label{chap2:eq:evolution_equation}
\end{equation}
where $S[m + 1] \leq Q[m]$ and $A[m]$ is the random number of arrivals which have occurred in the period between the $m^{th}$ and $(m + 1)^{th}$ decision epochs.
We note that $A[m] \sim \star_{\mathcal{T}[m]}A_{s}[1]$, the $\mathcal{T}[m]$ convolution of $A_{s}[1]$.
Let $\Gamma$ denote the set of all policies.
The class of all stationary policies $\Gamma_{s}$ is such that for any $\gamma \in \Gamma_{s}$, $S[m] = S(Q[m - 1])$ and $\mathcal{T}[m] = \Tau(Q[m - 1])$, where $S$ and $\Tau$ are functions (possibly randomized) of the queue length $q$.
We note that if $\gamma \in \Gamma_{s}$, then $Q[m]$ is a Markov chain embedded in the random process $Q_{s}[n]$.

In the following, we consider two separate models, set up as follows.
\begin{description}
\item[R-model-A :]{$\forall m \geq 1, A[m] \in [0,A_{max}]$, $S[m] \in \mathbb{R}_+$, and $q_{0} \in \mathbb{R}_+$,}
\item[R-model-B :]{Same as the R-model-A, except that $S[m] \in [0, S_{max}]$, and $\Tau[m] = N_{c}$, $m \geq 1$.}
\end{description}
We note that for R-model-A and R-model-B, the queue length evolution $(Q[m], m \geq 0)$ is on the non-negative real numbers.
Furthermore, for R-model-B, the maximum batch size is bounded by $S_{max}$ and the decision epochs occur every $N_{c}$ slots.
Hence, for R-model-B, all codewords are of length $N_{c}$.

For R-model-A and R-model-B, for a $\gamma \in \Gamma$, we define the average delay as
\begin{equation}
  D(\gamma) = \limsup_{I \rightarrow \infty} \frac{\sum_{i = 1}^{I} (T_{d,i} - T_{a,i})}{I}.
  \label{chap2:eq:d_definition}
\end{equation}
We also define the average error rate as
\begin{equation}
  P_{e}(\gamma) = \limsup_{I \rightarrow \infty} \frac{\sum_{i = 1}^{I} \mathbb{I}_{E_{i}}}{I},
  \label{chap2:eq:pe_definition}
\end{equation}
where $\mathbb{I}_{E_{i}}$ is the indicator function for the event $E_{i}$.

Our problem is then to
\begin{center}
  \begin{tabular}{ccc}
    $\mini_{\gamma \in \Gamma} \limits P_{e}(\gamma)$ & or  & $\mini_{\gamma \in \Gamma} \limits D(\gamma)$ \\
    such that $D(\gamma) \leq D_{c}$ &  & such that $P_{e}(\gamma) \leq P_{e,c}$.
  \end{tabular}
\end{center}
for every $D_{c} \geq 1$ or $0 \leq P_{e,c} \leq 1$.
For an average error rate constraint $P_{e,c} > \frac{|\mathcal{M}| - 1}{|\mathcal{M}|}$, we have the following optimal solution: transmit all the arrivals in a slot in the succeeding slot, using a channel input symbol picked independently of the message symbols, and pick the message symbol estimates uniformly at the receiver (maximum likelihood decoding).
In this chapter, we analyse the above problem, for the class of stationary policies $\Gamma_{s}$, in the asymptotic regime where (i) arbitrarily large reliability is required, that is, as $P_{e,c} \downarrow 0$ for R-model-A,  and (ii) $P_{e,c}$ approaches the minimum probability of error for $N_{c}$, when the codeword length $\tau$ is fixed to be $N_{c}$ for R-model-B.

\subsection{Related work}
\label{chap2:sec:related_work}
We note that for obtaining the average error rate of a policy, we would need to know how the error events $E_{i}$ are related to the scheduling policy $\gamma$ which chooses the parameters $(S[m], \Tau[m])$ for the random block codebook at every $m \geq 1$.
It is intuitive that, for a stationary policy $\gamma$, apart from the noise introduced by the channel, $E_{i}$ is a function only of the size $s$ and the length $\tau$ of the codebook, from which the codeword used to transmit the $i^{th}$ symbol is selected.
Gallager \cite[Chapter 5]{gallager_text} has provided an upper bound on the average error probability of a random block code as a function of $s$ and $\tau$.
The same reference also provides a lower bound on the average error probability of any block code as a function of $s$ and $\tau$.
The above upper and lower bounds show that the average error probability of the random block code decays exponentially with the block length for codeword rates less than the capacity of the channel.
The upper bound which is obtained as an ensemble average over random block codes (Gallager's random coding upper bound) and the lower bound (sphere packing bound) are found to coincide in the exponent for rates greater than the cutoff rate for the channel.
So for large values of the block length and for codeword rates greater than the cutoff rate, the Gallager random coding upper bound can be used as a reasonable approximation for the average error probability of the block codeword.
Recently Polyanskiy \cite{polyanskiy} has obtained upper and lower bounds as well as an analytically tractable approximation for the average error probability of block codewords as a function of the rate and block length of the codebook.
The approximation has been observed to be tight even for small values of the block length.
We provide a detailed review of the results from Gallager \cite{gallager_text} in Section \ref{chap2:sec:service_cost} wherein the Gallager random coding upper bound is used to approximate the average number of symbols in error, as a function of $s$ and $\tau$, when random block codewords are used, as in our model.

Now we provide a survey of prior work on characterizing the tradeoff between reliability and delay for point to point channels.
Javidi and Swamy \cite{tarablock} consider a point to point link with a random stationary arrival process of bits at rate $\lambda$ into an infinite transmitter buffer. 
Every $N_{c}$ slots, a batch of $s = rN_{c}$ bits (with zero padding, if required) is encoded and transmitted over the noisy channel using a block code of fixed rate $r$ and length $N_{c}$.
The transmission of a bit is said to fail if : 1) the bit is part of a codeword that is decoded in error or 2) the bit is decoded past a given delay deadline $D$. 
The authors study the probability of a bit transmission failure as a function of $r, N_{c}$, and $D$. 
The analysis is asymptotic in nature - in the regime of large delay deadline $D$. 
It is assumed that $b > 2$ and the codeword length $N_{c} = \frac{D}{b}$ as $D \uparrow \infty$. 
The constant $b$ can be interpreted as a parameter controlling the division of the overall delay budget $D$ between queueing delay and transmission delay. 
A large deviations result is used to show that the delay deadline violation probability of a bit decays exponentially with the deadline $D$, as $D \uparrow \infty$.
Using Gallager's random coding upper bound, it is then shown that bit transmission failure probability also decays exponentially with the exponential rate being the minimum of $(1/b)$ times the Gallager exponent, and the delay exponent. 
The results obtained are in a similar vein as ours, as the exponential decay of the error rate is shown with respect to the delay deadline.
However, we consider average delay as the metric and as we shall see, we also obtain the best decay rate over the class of stationary policies, using codewords of fixed length. 
We also show that scaling the codeword length linearly with average delay is optimal, with the scaling factor being $2/3$.

Musy and Telatar \cite{musy} consider a continuous time queueing model with a Poisson message symbol arrival process, for a point-to-point link with random block coding.
The block coding scheme is assumed to be such that the codeword length can be varied as a function of the number of message symbols which are jointly encoded in a codeword.
The codeword length is chosen such that the block error probability for every transmission is at most a fixed constant.
The authors obtain upper and lower bounds on the minimum average delay for a fixed upper bound on the constant block error probability.
A joint scheduling-coding scheme for a point to point link, with ARQ, has been analysed in \cite{tara_2} by Swamy and Javidi.
The dependence of average error rate on delay has been considered for non block coding schemes with Poisson arrivals in \cite{yoon} by Yoon, and for a periodic source in \cite{negi} by Negi and Goel.

\subsection{Overview}
\label{chap2:sec:overview}
We formulate the tradeoff problem in Section \ref{chap2:sec:system_model}.
For the class of {stationary} policies, we express the average error rate and average delay, in terms of quantities which are analytically more tractable, in the same section.
In Section \ref{chap2:sec:service_cost} we present a discussion on error exponents for discrete memoryless channels.
In this chapter, the Gallager random coding upper bound is used to obtain an upper bound for the expected number of message symbols in error in every transmission, which is then used to approximate the average error rate of stationary policies.
We then consider the asymptotic behaviour of the minimum average error rate subject to a constraint on the average delay, for R-model-A in Section \ref{chap2:sec:rmodela}.
We show that the minimum average error rate decays exponentially to zero, as the constraint on the average delay increases to infinity.
The exponential decay rate is shown to be two-thirds of the Gallager random coding exponent.
We then comment on the exponential decay rate for a queueing model in which the queue length evolution is assumed to be on the set of non-negative integers.
We also consider the asymptotic behaviour of the average error rate for a class of policies with codeword length dependent on the batch size, in the same section.

The asymptotic behaviour of the minimum average delay subject to a constraint on the average error rate is then characterized for R-model-B in Section \ref{chap2:sec:rmodelb}.
This particular problem is similar to the TRADEOFF problem, analysed in Chapter 4, except that the service cost function is non-convex.
For R-model-B, as all codewords are of finite length $N_{c}$, it is intuitive that the infimum of achievable average error rates, for any finite average delay, is bounded away from zero.
The non-convex nature of the service cost function leads to the result that depending on the value of the arrival rate $\lambda$, the minimum average delay is either $\Omega\nfrac{1}{\sqrt{V}}$ or $\Theta\brap{\log\nfrac{1}{V}}$, when the average error rate constraint is $V$ more than the above positive infimum of achievable average error rates.
We then consider the tradeoff problem for a similar model, where the queue length evolution is on integers, for which the results from Chapter 4 directly applies.

\section{Problem formulation}
\label{chap2:sec:system_model}
In the following we restrict ourselves to a special subset of $\Gamma_{s}$, which is the set of \emph{admissible} policies.
A policy $\gamma \in \Gamma_{s}$ is admissible if:
\begin{enumerate}
\item the embedded Markov chain (EMC) $Q[m]$ is positive Harris recurrent, with stationary distribution $\pi_{\gamma}$, and,
\item the embedded average $\Exp_{\pi_{\gamma}} \ExpT h(Q, \Tau(Q)) < \infty$, where $h(q, \tau) \stackrel{\Delta} = q \tau + \frac{\tau(\tau - 1)\lambda}{2}$.
\end{enumerate}
We note that the function $h(q, \tau)$ can be interpreted as the cumulative expected queue length in a deterministic transmission/idle period of duration $\tau$, when $q$ message symbols are present in the queue at the beginning of that period.
Every admissible policy is stable, according to the definition in Section \ref{chap5:sec:realvaluedproblem} of Chapter 4.
To avoid unnecessary notation, and since in the following we consider only admissible policies
\footnote{We note that the requirement of monotonicity as in the previous chapters is imposed only for R-model-B}, we redefine $\Gamma_{s}$ as the set of all admissible stationary policies.
We recall that $\Tau[m]$ is the duration of the transmission/idle period beginning at the $(m-1)^{th}$ decision epoch.
The set of all $\gamma \in \Gamma_{s}$, for which all transmission and idle periods are of a fixed duration $N_{c}$, i.e., $\mathcal{T}[m] = \tau_{0} = N_{c}$, is denoted as $\Gamma_{s,N_{c}}$.
Let $\Gamma_{s,f} \stackrel{\Delta} = \bigcup_{N_{c}} \Gamma_{s,N_{c}}$.
We note that when operating with a policy $\gamma \in \Gamma_{s,f}$, a transmission duration $N_{c}(\gamma)$ is chosen at the first decision epoch, which is kept fixed at all the other decision epochs.
We have that $\Gamma \supset \Gamma_{s} \supset \Gamma_{s,f} \supset \Gamma_{s,N_{c}}$.
We now express $D(\gamma)$ and $P_{e}(\gamma)$ as averages of functions of the queue length and block coding parameters at the decision epochs when $\gamma \in \Gamma_{s}$.

The average queue length $\overline{Q}(\gamma)$ for a policy $\gamma \in \Gamma_{s}$ is defined as
\footnote{We note that this is a sample path definition, which is different from the definition in Chapter 4. However, for $\gamma \in \Gamma_{s}$, the two definitions are equivalent.}
\begin{equation}
  \overline{Q}(\gamma) = \lim_{N \rightarrow \infty} \frac{\sum_{n = 0}^{N - 1} Q_{s}[n]}{N}.
\end{equation}
For $\gamma \in \Gamma_{s}$, we note that $(Q[m], \mathcal{T}[m + 1])$ is a Markov renewal process, with the renewal instants corresponding to the decision epochs.
In order to express the time average $\overline{Q}(\gamma)$ in terms of a function of $Q[m]$, we use the Markov renewal reward theorem \cite[Theorem D.16]{anurag}. 
We associate a reward $R[m]$ with the $m^{th}$ renewal cycle.
The reward $R[m]$ is the cumulative queue length over the renewal cycle, with the queue length at the beginning of the cycle being $Q[m - 1]$ and cycle length being $\mathcal{T}[m]$.
Since $Q[m]$ is Markov and $\gamma \in \Gamma_{s}$ we have that, conditioned on $Q[m - 1]$ and $Q[m' - 1]$, $R[m]$ is independent of $R[m']$ for any $m' \neq m$.
Let $T[m]$ be the number of slots up to the start of the $m^{th}$ renewal cycle $(T[1] = 0)$.
We have that the expected reward $\Exp \bras{R[m] | Q[m - 1]}$, in cycle $m$, is
\begin{equation*}
  r[m] = \mathbb{E} \left.\left[ Q[m - 1]\mathcal{T}[m] + \sum_{n = 1}^{\Tau[m] - 1} \sum_{n' = 1}^{n} A_{s}[T[m] + n'] \right\vert Q[m - 1] \right].
\end{equation*}
Since $A_{s}[n]$ are assumed to be IID we have that
\begin{equation*}
  r[m] = \mathbb{E} \left.\left[ Q[m - 1]\mathcal{T}[m] + \frac{\Tau[m](\Tau[m] - 1)\lambda}{2} \right\vert Q[m - 1]\right].
\end{equation*}
We define the holding cost at a decision epoch, where the queue length is $q$ and a deterministic transmission/idle time $\tau$ is chosen, as a function $h(q, \tau)$ defined as
\begin{equation*}
  h(q,\tau) = q\tau + \frac{\tau(\tau - 1)\lambda}{2}.
\end{equation*}
We note that $r[m] = \Exp h(q, \Tau(q))$, if $Q[m - 1] = q$, and the expectation is over the randomized choice of $\tau$.
We note that for any $\gamma \in \Gamma_{s}$, since $\mathbb{E}_{\pi_{\gamma}} \Exp h(Q,\Tau(Q)) < \infty$, we also have that $\Exp_{\pi_{\gamma}} \Exp \Tau(Q) < \infty$.
Then, we have from \cite[Theorem D.16]{anurag} that
\begin{equation*}
  \lim_{N \rightarrow \infty} \frac{\sum_{n = 0}^{N - 1} Q_{s}[n]}{N} \stackrel{a.s} = \frac{\mathbb{E}_{\pi_{\gamma}}\ExpT h(Q,\Tau(Q))}{\mathbb{E}_{\pi_{\gamma}} \ExpT \Tau(Q)}.
\end{equation*}
Hence, for $\gamma \in \Gamma_{s}$, using Little's law \cite{whitt}, we have 
\begin{equation}
  D(\gamma) = \frac{\overline{Q}(\gamma)}{\lambda} = \frac{\Exp_{\pi_{\gamma}} \ExpT h(Q, \Tau(Q))}{\lambda \Exp_{\pi_{\gamma}} \ExpT \Tau(Q)}.
  \label{chap2:eq:delay_queuelength}
\end{equation}

To express $P_{e}(\gamma)$ as a time average we use the generalized $H = \lambda G$ form of Little's law \cite{whitt}.
Define $F_{i}[n] = 0$ for $n \in \brac{0,\dots, T_{d,i} - 1}$ and $F_{i}[n] = \mathbb{I}_{E_{i}}$ for $n \in \brac{T_{d,i}, \dots}$.
Let $G_{I}$ denote the fraction of the first $I$ message symbols that are in error, and $H_{N}$ denote the time-rate of message symbols errors over the first $N$ slots, i.e.,
\begin{eqnarray*}
  G_{I} & = & \frac{1}{I} \sum_{i = 1}^{I} \sum_{n = 1}^{\infty} \brap{F_{i}[n] - F_{i}[n - 1]}, \\
  H_{N} & = & \frac{1}{N} \sum_{n = 0}^{N - 1}\sum_{\brac{i: T_{d,i} = n} } \mathbb{I}_{E_{i}}.
\end{eqnarray*}
We note that $\lim_{I \rightarrow \infty} G_{I} = P_{e}(\gamma)$.
If $H = \lim_{N \rightarrow \infty} H_{N}$, then $H = \lambda P_{e}(\gamma)$ from \cite[Section 6]{whitt}.
For the batch transmission scheme, the set $\{ i : T_{d,i} = n\}$ is empty, except for those $n$ such that $n = T[m] = \sum_{m' = 1}^{m - 1} \Tau[m']$.
We also have that $|\{ i : T_{d,i} = T[m]\}| = S[m - 1]$.
We note that since the channel is memoryless and the decoding of a message batch of size $S[m - 1]$ is done based on the channel outputs received only in the transmission time $\Tau[m - 1]$, given $S[m - 1]$ and $\Tau[m - 1]$, the events $\brac{E_{i} : T_{d,i} = T[m]}$ are independent of any other error event of a message symbol transmitted in a period other than $m - 1$.
As in the previous case we associate a reward $R[m]$ with the $m^{th}$ renewal cycle, which is the total number of message symbols in error during the $m^{th}$ transmission period, starting with $Q[m - 1]$ message symbols in the queue.
The expected reward $\Exp \bras{R[m] \vert Q[m - 1]}$ in the $m^{th}$ renewal cycle is
\begin{eqnarray*}
  r[m] & = & \mathbb{E} \left.\left[ \sum_{\{i:T_{d,i} = T[m + 1]\}} I_{E_{i}}\right| Q[m - 1] \right], \\
  & = & \Exp_{S[m], \Tau[m]|Q[m - 1]} \bras{\sum_{\{i : T_{d,i} = T[m + 1]\}} Pr\brac{E_{i} | S[m], \Tau[m]} \middle\vert Q[m - 1]}
\end{eqnarray*}
where $Pr\brac{E_{i} | S[m] = s, \Tau[m] = \tau}$ is the probability of error of the $i^{th}$ symbol transmitted in the $m^{th}$ period using a block code with size $|\mathcal{M}|^{s}$ and of codeword length $\tau$.
We define the error cost at a decision epoch to be $c_{s}(s,\tau) = \sum_{i \in \mathcal{C}(s,\tau)} Pr\brac{E_{i} | S = s, \Tau = \tau}$, where at that decision epoch, $\mathcal{C}(s,\tau)$ is the set of $s$ symbols which have been jointly encoded into a codeword of length $\tau$.
We note that for $\gamma \in \Gamma_{s}$, $\Exp_{\pi_{\gamma}} \ExpT \Tau(Q) < \infty$.
Applying the Markov renewal reward theorem \cite[Theorem D.16]{anurag} we obtain that
\begin{eqnarray*}
  H \stackrel{a.s} = \frac{\Exp_{\pi_{\gamma}} \ExpST c_{s}(S(Q), \Tau(Q))}{\Exp_{\pi_{\gamma}} \ExpT \Tau(Q)}.
\end{eqnarray*}
And therefore, 
\begin{equation}
  P_{e}(\gamma) = \frac{1}{\lambda}\frac{\Exp_{\pi_{\gamma}} \ExpST c_{s}(S(Q), \Tau(Q))}{\Exp_{\pi_{\gamma}} \ExpT \Tau(Q)}.
  \label{chap2:eq:prob_servicecost}
\end{equation}
In the next section we discuss some approximations for $c_{s}(s,\tau)$ which are used in further analysis of the tradeoff problem.

\subsection{The error cost $c_{s}(s,\tau)$}
\label{chap2:sec:service_cost}
We note that $c_{s}(s,\tau) = \sum_{i \in \mathcal{C}(S,\Tau)} \Exp \bras{\mathbb{I}_{E_{i} | S = s, \Tau = \tau}}$, is the expected number of message symbols in error when a random block code is used to transmit $s$ message symbols in $\tau$ slots.
In the following, the error cost $c_{s}(s,\tau)$ is approximated using the Gallager random coding upper bound, since we are interested in an asymptotic characterization of the tradeoff curve as the error rate constraint $P_{e,c} \downarrow 0$.
It is intuitive that policies $\gamma \in \Gamma_{s,f}$, for which $P_{e}(\gamma) \leq P_{e,c}$, have block lengths growing to infinity as $P_{e,c} \downarrow 0$.
Therefore in the regime of $P_{e,c} \downarrow 0$, approximating $c_{s}(s,\tau)$ using the Gallager random coding upper bound is not unreasonable \cite{gallager_rcode} 
\footnote{In \cite{gallager_rcode} Gallager has shown that the random coding upper bound is tight for the probability of block error for an IID ensemble of codes, by showing that the lower bound on the ensemble error probability is at most $\mathcal{O}\brap{\frac{1}{\tau}\ln\nfrac{1}{\sqrt{\tau}}}$ away from the random coding upper bound for large block length $\tau$, for code rates less than the capacity of the channel $P_{Y|X}$.}. 
Let $P(s,\tau) \Deq Pr\brac{\exists i \in \{1,\cdots,s\} \text{ s.t } m_{s,i} \neq \hat{m}_{s,i}|S = s, \Tau = \tau}$, where $m_{s,i}$ and $\hat{m}_{s,i}$ are the transmitted and decoded symbols respectively. 

For a random block code of length $\tau$ transmitting $s$ symbols, we have from \cite[Theorem 5.6.2]{gallager_text} that the codeword error probability $P(s,\tau)$ is
\begin{equation}
    \leq \min\brap{1,e^{-\tau(E_{0}(\rho, \mQ) - \rho \frac{s}{\tau} \ln |\mathcal{M}|)}} = \min\brap{1, e^{-\tau E_{r}(\rho, s, \tau, \mQ)}},
\end{equation}
where $E_{0}(\rho, \mQ) = -\ln \brap{ \int_{y \in \mathcal{Y}} \brap{ \int_{x \in \mathcal{X}} d\mQ(x)\brap{dP_{Y|X} (y|x)}^{\frac{1}{1 + \rho}}}^{1 + \rho}}$, $\rho \in [0,1]$ and $\mQ$ is the channel input distribution.
The best upper bound is obtained by minimizing the above bound over the parameter $\rho \in [0,1]$ and the distribution $\mQ$.
We note that this can be done by maximising the exponent $E_{r}(\rho, s, \tau, \mQ)$ with respect to $\rho$ and $\mQ$.
Depending on whether we do not choose/choose to optimise over $\rho$ and/or $\mQ$ there are four different exponents : (a) $E_{r}(\rho,s,\tau,\mQ) = (E_{0}(\rho, \mQ) - \rho \frac{s}{\tau} \ln |\mathcal{M}|)$, (b) $E_{r}(\rho,s,\tau) = \max_{Q} (E_{0}(\rho, \mQ) - \rho \frac{s}{\tau} \ln|\mathcal{M}|)$, (c) $E_{r}(s,\tau,\mQ) = \max_{\rho \in [0,1]} (E_{0}(\rho, \mQ) - \rho \frac{s}{\tau} \ln|\mathcal{M}|)$, and (d) $E_{r}(s,\tau) = \max_{\rho \in [0,1]} \max_{\mQ} (E_{0}(\rho,\mQ) - \rho\frac{s}{\tau} \ln|\mathcal{M}|)$.

If a symbol is in error then the decoded codeword is also in error, therefore $Pr\brac{E_{i} | S = s, \Tau = \tau} \leq P(s,\tau)$.
Using the union bound on $P(s,\tau)$ and the above inequality we also obtain that
\begin{equation}
    P(s,\tau) \leq \sum_{i = 1}^{s} Pr\brac{E_{i}|S = s, \Tau = \tau} \leq sP(s,\tau).
    \label{chap2:eq:servicecost_from_unionbd}
\end{equation}
In the following, we approximate $c_{s}(s, \tau)$ by assuming that all symbols are decoded incorrectly if the codeword is in error.
Different upper bounds on the codeword error probability (from the different random coding exponents) can be used to obtain approximations for $c_{s}(s, \tau)$.
The following approximations are possible : $c_{s}(s, \tau) = $ 
(a)~$s \min(1, e^{-\tau E_{0}(\rho, \mQ) + \rho s \ln \cM})$ where both $\rho$ and $\mQ$ are fixed, 
(b)~$s \min(1, e^{\min_{\mQ} (-\tau E_{0}(\rho, \mQ) + \rho s \ln \cM)})$, where $\rho$ is fixed, 
(c)~$s e^{\min_{\rho \in [0,1]} (-\tau E_{0}(\rho, \mQ) + \rho s \ln \cM)}$, where $\mQ$ is fixed, and 
(d)~$s e^{\min_{\rho \in [0,1], \mQ} (-\tau E_{0}(\rho, \mQ) + \rho s \ln \cM)}$.
We note that approximation (a) is found to be analytically tractable as $\rho$ and $\mQ$ are fixed and do not depend on $s$ and $\tau$. 
In (b) the optimal $\mQ$ is a function of the fixed $\rho$ and the following analysis applies, as it would be similar to that of (a) with $\mQ$ being fixed at the optimal $\mQ$ for the fixed $\rho$. 
We note that the minimizing $\rho$ and $\mQ$ are not known explicitly as a function of $s$ and $\tau$ even for simple channel models \cite[Chap 5]{gallager_text}.
However, for R-model-A, we shall see that the exponential decay, of the minimum average error rate with the constraint on the average delay, can be shown to be governed by the best exponent (d), even with the following assumption.
For R-model-B, we comment on how the asymptotic results can be extended to other approximations of $c_{s}(s, \tau)$.
We make the following assumption :
\begin{description} 
\item[C1 :]{The error cost or the expected number of symbols in error for a random block code of codeword length $\tau$ transmitting $s$ symbols is $c_{s}(s, \tau) = s\min\left(1,e^{(- \tau E_{0}(\rho,\mQ)+ \rho s\ln \cM)}\right)$, where $\rho$ and $\mQ$ are fixed and are such that $\frac{E_{0}(\rho, \mQ)}{\rho} > \lambda \ln \cM$.}
\end{description}
The tradeoff problem that we consider is defined separately for R-model-A and R-model-B in the following sections.

\section{Asymptotic analysis for R-model-A}
\label{chap2:sec:rmodela}
\label{chap2:sec:peasymp}

\subsection{Problem Statement}
\label{chap2:sec:problem_statement}
We consider the tradeoff of average error rate with average delay for $\gamma \in \Gamma_{s}$.
The following analysis has been presented in \cite{vineeth_isit} and \cite{vineeth_drdo_3}.
The TRADEOFF problem is:
\begin{eqnarray}
  & \mini_{\gamma \in \mathcal{S}_{b}} & \frac{\Exp_{\pig} \ExpST c_{s}(S(Q), \Tau(Q))}{\lambda\Exp_{\pig}\ExpT \Tau(Q)} \nonumber \\
  & \text{such that } & \frac{\Exp_{\pig}\ExpT h(Q, \Tau(Q))}{\lambda\Exp_{\pig}\ExpT \Tau(Q)} \leq D_{c},
  \label{chap2:eq:tradeoffproblem}
\end{eqnarray}
where we have used \eqref{chap2:eq:delay_queuelength} and \eqref{chap2:eq:prob_servicecost}.
If $\mathcal{S}_{b} = \Gamma_{s}$, then let $P_{e}^{*}(D_{c})$ denote the optimal value of TRADEOFF.
If $\mathcal{S}_{b} = \Gamma_{s,f}$, then the optimal value of TRADEOFF is denoted by $P_{e,f}^{*}(D_{c})$, and if $\mathcal{S}_{b} = \Gamma_{s,N_{c}}$, then the optimal value of TRADEOFF is denoted by $P_{e,N_{c}}^{*}(D_{c})$.
We can also consider the equivalent problem EQT-TRADEOFF:
\begin{eqnarray}
  & \mini_{\gamma \in \mathcal{S}_{b}} & \frac{\Exp_{\pig}\ExpT h(Q, \Tau(Q))}{\lambda\Exp_{\pig}\ExpT \Tau(Q)} \nonumber \\
  & \text{such that } & \frac{\Exp_{\pig} \ExpST c_{s}(S(Q), \Tau(Q))}{\lambda\Exp_{\pig}\ExpT \Tau(Q)}  \leq P_{e,c}. 
  \label{chap2:eq:eqtradeoffproblem}
\end{eqnarray}
If $\mathcal{S}_{b} = \Gamma_{s}$, then let $D^{*}(P_{e,c})$ denote the optimal value of TRADEOFF.
If $\mathcal{S}_{b} = \Gamma_{s,f}$, then the optimal value of TRADEOFF is denoted by $D_{f}^{*}(P_{e,c})$, and if $\mathcal{S}_{b} = \Gamma_{s,N_{c}}$, then the optimal value of TRADEOFF is denoted by $D_{N_{c}}^{*}(P_{e,c})$.
In the next section we discuss the asymptotic behaviour of $P_{e,f}^{*}(D_{c})$ and ${P}_{e}^{*}(D_{c})$ as $D_{c} \rightarrow \infty$.
We note that as in Section \ref{chap5:sec:cmdpformulation}, we can show that there exists an optimal policy for the above problem.

\subsection{Asymptotic analysis}
\label{chap2:sec:minerrcost}
In the following lemma, we obtain $\inf_{D_{c}} P_{e,N_{c}}^{*}(D_{c})$.
\begin{lemma}
  We have that $\inf_{D_{c}} P_{e,N_{c}}^*(D_{c}) = P_{N_{c}}(\lambda)$, where 
  \begin{eqnarray*}
    P_{N_{c}}(\lambda) = 
    \begin{cases}
      e^{-N_{c}(E_{0}(\rho, \mQ) - \rho \lambda \ln \cM)}, \text{ if } 0 \leq \lambda N_{c} \leq s', \\
      \frac{1}{\lambda N_{c}} \left( (\lambda N_{c} - s') + s'e^{-N_{c}E_{0}(\rho,\mQ) + \rho s' \ln \cM} \right),  \text{ otherwise},
      \label{chap2:eq:asymplb}
    \end{cases}
  \end{eqnarray*}
  where $s'$ is the unique solution of the equation $\brap{1 + \rho s' \ln \cM} e^{\rho s' \ln \cM} = e^{N_{c} E_{0}(\rho, \mQ)}$.
  \label{chap2:prop:asymplb}
\end{lemma}
\begin{proof}
We first show that $P_{N_{c}}(\lambda) \leq P_{e,N_{c}}^*(D_{c})$, $\forall D_{c}$.
We note that as we are considering policies $ \gamma \in \Gamma_{s,N_{c}}$, the expected transmission time is always $N_{c}$ and therefore \eqref{chap2:eq:delay_queuelength} and \eqref{chap2:eq:prob_servicecost} simplify to $\frac{\Exp_{\pig} h(Q, N_{c})}{\lambda N_{c}}$ and $\frac{\Exp_{\pig} \ExpS c_{s}(S,N_{c})}{\lambda N_{c}}$ respectively.
We consider the following optimization problem:
\begin{eqnarray}
  \mini_{\pi} & & \frac{1}{\lambda N_{c}}\Expp \left[S\min\left(1, e^{-N_{c}E_{0}(\rho, \mQ)+\rho S\ln \cM}\right)\right],
  \label{chap2:eq:asymp_minprob} \\
  \text{such that } & & \Expp S \geq \lambda N_{c} + \epsilon, \, \epsilon \geq 0, \nonumber
\end{eqnarray}
where $\pi$ is any distribution for $S$ with support on $\mathbb{R}_+$, irrespective of the policy $\gamma$.
With $\epsilon = 0$ in \eqref{chap2:eq:asymp_minprob} leading to the constraint $\mathbb{E}_{\pi}S \geq \lambda N_{c}$, the optimal value of the minimization problem \eqref{chap2:eq:asymp_minprob} is a lower bound to ${P}_{e,N_{c}}^{*}(D_{c})$ for any finite $D_{c}$ as the minimization is over all possible distributions of $S$ and we impose only the constraint that the average service rate $\frac{\Expp S}{N_{c}}$ has to be greater than or equal to the arrival rate $\lambda$.
From Appendix \ref{chap2:app:optimization_problem} we obtain that
\begin{eqnarray}
P_{e,N_{c}}^*(D_{c}) \geq
\begin{cases}
e^{-N_{c}(E_{0}(\rho, \mQ) - \rho \lambda \ln \cM)} \text{ if } 0 \leq \lambda N_{c} \leq s', \\
\frac{1}{\lambda N_{c}} \left( (\lambda N_{c} - s') + s'e^{-N_{c}E_{0}(\rho, \mQ) + \rho s' \ln \cM} \right)  \text{ otherwise},
\label{chap2:eq:asymplb1}
\end{cases}
\end{eqnarray}
where $s'$ is the unique solution of the equation $\brap{1 + \rho s' \ln \cM} e^{\rho s' \ln \cM} = e^{N_{c} E_{0}(\rho, \mQ)}$, for every finite $D_{c}$.
Let ${P}_{N_{c}}(\lambda)$ denote the RHS of \eqref{chap2:eq:asymplb1}.
We note that $P_{N_{c}}(\lambda)$ is equal to the value of the lower convex envelope of $\frac{c_{s}(s, N_{c})}{\lambda N_{c}}$ as a function of $s$, at $s = \lambda N_{c}$.

Now we show that for any $\epsilon > 0$, it is possible to construct a policy $\gamma \in \Gamma_{s,N_{c}}$ such that $P_{e}(\gamma) \leq {P}_{N_{c}}(\lambda) + \epsilon$.
Then we have that $P_{e, N_{c}}^*(D(\gamma)) \leq {P}_{N_{c}}(\lambda) + \epsilon$ and therefore $P_{N_{c}}(\lambda) = \inf_{D_{c}} P_{e,N_{c}}^*(D_{c})$.
In the first step of the construction we consider the optimization problem \eqref{chap2:eq:asymp_minprob} but with an $\epsilon > 0$.
The optimal value of the problem \eqref{chap2:eq:asymp_minprob} as a function of $\epsilon$ is denoted by $s(\epsilon)$.
A feasible distribution for $S$ that achieves $s(\epsilon) + \delta$ is called $(\delta,\epsilon)$-optimal for a $\delta > 0$.
In Appendix \ref{chap2:app:optimization_problem} it is shown that there exists a distribution with finite support that is $(\delta,\epsilon)$-optimal for any $\epsilon > 0$ and $\delta > 0$.

Now we proceed to the second step of the construction.
For an $\epsilon > 0$ we use any $(\delta,\epsilon)$-optimal distribution to construct a policy as follows.
Let $U[m]$ be a sequence of IID random variables distributed as the $(\delta,\epsilon)$-optimal distribution.
The policy $\gamma$ uses batch sizes $S[m] = \min(Q[m - 1], U[m])$.
We note that $\gamma$ is stationary.
Since $\epsilon > 0$, $\Exp U > \lambda N_{c}$.
Therefore, the EMC $Q[m]$ under $\gamma$ is positive recurrent.
It can also be shown that the average delay is finite as $U[m]$ has finite support.
Therefore, $\gamma$ is admissible, i.e., $\gamma \in \Gamma_{s}$.
Evaluating the average error cost for $\gamma$ we have that $P_{e}(\gamma) \leq$
\small
\begin{eqnarray*} 
\frac{\Exp c_{s}(U, N_{c})}{\lambda N_{c}} =
\begin{cases}
e^{\epsilon \rho \ln \cM} e^{-N_{c}E_{0}(\rho, \mQ) + (\rho \ln \cM)\lambda N_{c}} + \epsilon e^{-N_{c}E_{0}(\rho, \mQ) + (\rho \ln \cM)(\lambda N_{c} + \epsilon )}, \text{ if }\lambda N_{c} \in [0, s' - \epsilon],\\
\frac{1}{\lambda N_{c}}\left(\lambda N_{c}  -s' + s' e^{-N_{c}E_{0}(\rho, \mQ) + \rho s'\ln \cM} \right) + \frac{\epsilon + \delta}{\lambda N_{c}},\text{ otherwise}.
\end{cases}
\end{eqnarray*}
\normalsize
Let $\delta = \epsilon$.
Therefore, $P_{e}(\gamma) \leq P_{N_{c}}(\lambda) + \mathcal{O}(\epsilon)$.
Hence, by an appropriate choice of $\epsilon$ in \eqref{chap2:eq:asymplb}, we can show that $\forall \epsilon > 0$, there exists $D_{c} = D(\gamma)$ such that $P_{e,N_{c}}^*(D_{c}) \leq P_{e}(\gamma) \leq P_{N_{c}}(\lambda) + \epsilon$.
Hence, $P_{N_{c}}(\lambda) = \inf_{D_{c}} P_{e,N_{c}}^*(D_{c})$.
\end{proof}

\begin{remark}
Since any policy $\gamma \in \Gamma_{s,N_{c}}$ is feasible for TRADEOFF for any $D_{c} \geq D(\gamma)$, we also have that $P_{N_{c}}(\lambda) = \inf_{\gamma \in \Gamma_{s, N_{c}}} P_{e}(\gamma)$.
We note that $P_{e,N_{c}}^*(D_{c})$ is a non-increasing function of $D_{c}$.
Therefore, ${P}_{N_{c}}(\lambda) = \lim_{D_{c} \rightarrow \infty} P_{e,N_{c}}^{*}(D_{c})$.
We note that $P_{N_{c}}(\lambda)$ is similar to $c(\lambda)$ in Chapter 4.
\end{remark}
We now present a lemma which formalizes our intuition that $P_{N_{c}}(\lambda)$ approaches zero as the codeword length $N_{c}$ approaches infinity.
\begin{lemma}
  If $\lambda \ln \cM < \frac{E_{0}(\rho, \mQ)}{\rho}$, then $P_{N_{c}}(\lambda) = e^{-N_{c}(E_{0}(\rho, \mQ) - \rho \lambda \ln \cM)}$ for large enough $N_{c}$, and therefore, $\lim_{N_{c} \rightarrow \infty} P_{N_{c}}(\lambda) = 0$.
  \label{chap2:prop:asymplb_Nclimit}
\end{lemma}

\begin{proof}
  From Lemma \ref{chap2:prop:asymplb} we have that
  \begin{eqnarray*}
    P_{N_{c}}(\lambda) = 
    \begin{cases}
      e^{-N_{c}(E_{0}(\rho, \mQ) - \rho \lambda \ln \cM)} \text{ if } 0 \leq \lambda N_{c} \leq s', \\
      \frac{1}{\lambda N_{c}} \left( (\lambda N_{c} - s') + s'e^{-N_{c}E_{0}(\rho, \mQ) + \rho s' \ln \cM} \right)  \text{ otherwise},
    \end{cases}
  \end{eqnarray*}
  where $s'$ solves the equation $\brap{1 + \rho s' \ln \cM} e^{\rho s' \ln \cM} = e^{N_{c} E_{0}(\rho, \mQ)}$.
  We note that if we show that, $\lambda \ln \cM < \frac{E_{0}(\rho, \mQ)}{\rho}$ implies that $\lambda N_{c} \leq s'$ for large enough $N_{c}$, then we have that $P_{N_{c}}(\lambda) = e^{-N_{c}(E_{0}(\rho, \mQ) - \rho \lambda \ln \cM)}$ for large enough $N_{c}$, which leads to both results of the lemma.
  We have that
  \begin{eqnarray}
    e^{-N_{c} E_{0}(\rho, \mQ)} \bras{e^{\rho s' \ln \cM} \brac{1 + \rho s' \ln \cM}} & = & 1 \text{ or }, \nonumber \\
    \rho s' \ln \cM + \ln \brap{1 + s' \rho \ln \cM} & = & N_{c} E_{0}(\rho, \mQ).
    \label{chap2:eq:asymplb_Nclimit_proof2}
  \end{eqnarray}
  We note that the LHS of \eqref{chap2:eq:asymplb_Nclimit_proof2} is an increasing continuous function of $s'$ with LHS $= 0$ at $s' = 0$.
  Thus for any $N_{c}$, there is a unique $s'$ which solves the equation \eqref{chap2:eq:asymplb_Nclimit_proof2}.
  Also as $N_{c}$ increases this unique solution $s'(N_{c})$, as a function of $N_{c}$, also increases, and as $N_{c} \uparrow \infty$, $s'(N_{c}) \uparrow \infty$.
  We multiply both sides of \eqref{chap2:eq:asymplb_Nclimit_proof2} by $\lambda$, divide by $s'(N_{c})$ and $E_{0}(\rho, \mQ)$ to yield
  \begin{eqnarray}
    \lambda \ln \cM \frac{\rho}{E_{0}(\rho, \mQ)} + \frac{\lambda}{E_{0}(\rho, \mQ)} \frac{\ln \brap{1 + s'(N_{c}) \rho \ln \cM}}{s'(N_{c})} & = & \frac{\lambda N_{c}}{s'(N_{c})}
  \end{eqnarray}
  Since we have assumed that $\lambda \ln \cM < \frac{E_{0}(\rho, \mQ)}{\rho}$, $\delta \Deq 1 - \lambda \ln \cM \frac{\rho}{E_{0}(\rho, \mQ)} > 0$.
  For sufficiently large $N_{c}$, $s'(N_{c})$ can be made sufficiently large, so that $\frac{\ln \brap{1 + s'(N_{c}) \rho \ln \cM}}{s'(N_{c})} < \frac{\delta}{2} \frac{E_{0}(\rho, \mQ)}{\lambda}$ which implies that $\frac{\lambda N_{c}}{s'(N_{c})} \leq 1 - \frac{\delta}{2}$.
  Hence for large $N_{c}$, $\frac{\lambda N_{c}}{s'(N_{c})} < 1$ and therefore $P_{N_{c}}(\lambda) = e^{-N_{c}(E_{0}(\rho, \mQ) - \rho \lambda \ln \cM)}$.
  Therefore, $P_{N_{c}}(\lambda) \downarrow 0$ as $N_{c} \uparrow \infty$ if $\lambda \ln \cM < \frac{E_{0}(\rho, \mQ)}{\rho}$.
\end{proof}

We now consider the asymptotic behaviour of $P_{e,f}^*(D_{c})$, as $D_{c} \uparrow \infty$.
We first show that if $\lambda \ln \cM < \frac{E_{0}(\rho, \mQ)}{\rho}$, then $P_{e,f}^*(D_{c}) \downarrow 0$ as $D_{c} \uparrow \infty$.
This result is obtained by showing that for a sequence of policies $\mu_{k} \in \Gamma_{s,f}$, for which $D(\mu_{k}) \uparrow \infty$, $P_{e}(\mu_{k}) \downarrow 0$ .
Then $P_{e,f}^*(D(\mu_{k})) \downarrow 0$. 
Furthermore, since $D_{\mu_{k}} \uparrow \infty$, $P_{e,f}^*(D_{c}) \downarrow 0$ for any sequence $D_{c} \uparrow \infty$.
The sequence of policies $\mu_{k} \in \Gamma_{s,f}$ is parametrized by the codeword length $N_{c,k}$, where $N_{c,k}$ is an unbounded increasing sequence.
A fixed rate of transmission $r$ is chosen such that $\lambda \ln \cM < r \ln \cM < \frac{E_{0}(\rho, \mQ)}{\rho}$.
Let $R_{k} = r N_{c,k}$.
The policy $\mu_{k}$ chooses $S[m] = \min(Q[m - 1], R_{k})$.
We note that since $r > \lambda$, it can be shown that $\mu_{k} \in \Gamma_{s,f}, \forall k$.
\begin{proposition}
  If $\lambda \ln \cM < r \ln \cM  < \frac{E_{0}(\rho, \mQ)}{\rho}$, for the sequence of policies $\mu_{k}$ defined above, we have that $\lim_{k \rightarrow \infty} \frac{-\ln P_{e}(\mu_{k})}{D(\mu_{k})} = \frac{r - \lambda}{r - \lambda + r\lambda}\brap{E_{0}(\rho, \mQ) - \rho r \ln \cM}$, with $D(\mu_{k}) \uparrow \infty$. Therefore $P_{e}(\mu_{k}) \downarrow 0$.
  \label{chap2:prop:mukub}
\end{proposition}
\begin{proof}
  For $\mu_{k}$, we note that $P_{e}(\mu_{k}) \leq \frac{\Exp_{\pi_{\mu_{k}}} \bras{S e^{-N_{c,k} E_{0}(\rho, \mQ) + \rho S \ln \cM}}}{\lambda N_{c,k}}$.
  As $S \leq r N_{c,k}$ we have that 
  \begin{eqnarray}
    P_{e}(\mu_{k}) & \leq & \frac{r}{\lambda} \Exp_{\pi_{\mu_{k}}} \bras{e^{-N_{c,k} E_{0}(\rho, \mQ) + \rho S \ln \cM}},
    \label{chap2:eq:mukub0} \\
    & \leq & \frac{r}{\lambda} \brap{e^{E_{0}(\rho, \mQ) - \rho r \ln \cM}}^{-N_{c,k}}.
    \label{chap2:eq:mukub1}
  \end{eqnarray}
  From Appendix \ref{chap2:app:ubformu} we have that
  \begin{eqnarray*}
    \Exp_{\mu_{k}} Q \leq \frac{\sigma^{2}}{2(r - \lambda)} + \frac{\lambda N_{c,k}}{2} + \frac{r \lambda N_{c,k}}{r - \lambda}.
  \end{eqnarray*}
  We note that in our case, there is an extra time average holding cost of $\frac{\lambda (N_{c,k} - 1)}{2}$ due to the customers waiting during a transmission period of $N_{c,k}$ slots.
  Therefore, 
  \begin{eqnarray}
    D(\mu_{k}) \leq \frac{\sigma^{2}}{2\lambda(r - \lambda)} - \frac{1}{2} + N_{c,k}\brap{1 + \frac{r \lambda}{r - \lambda}}, \nonumber \\
    \frac{r - \lambda}{r - \lambda + r\lambda} \brap{D(\mu_{k}) - \frac{\sigma^{2}}{2\lambda(r - \lambda)} + \frac{1}{2}} \leq N_{c,k}.
    \label{chap2:eq:mukub}
  \end{eqnarray}
  Since $D(\mu_{k})$ is at least $N_{c,k}$ we have that $D(\mu_{k}) \uparrow \infty$ as $N_{c,k} \uparrow \infty$.

  Substituting the lower bound on $N_{c,k}$ from \eqref{chap2:eq:mukub} in \eqref{chap2:eq:mukub1}, we have that 
  \begin{equation*}
    P_{e}(\mu_{k}) \leq \frac{r}{\lambda} \brap{e^{E_{0}(\rho, \mQ) - \rho r \ln \cM}}^{-\frac{r - \lambda}{r - \lambda + r\lambda} \brap{D(\mu_{k}) - \frac{\sigma^{2}}{2\lambda(r - \lambda)} + \frac{1}{2}}}.
  \end{equation*}
  Hence $\lim_{k \rightarrow \infty} \frac{-\ln P_{e}(\mu_{k})}{D(\mu_{k})} = \frac{r - \lambda}{r - \lambda + r\lambda}\brap{E_{0}(\rho, \mQ) - \rho r \ln \cM}$, and $P_{e}(\mu_{k}) \downarrow 0$.

\end{proof}

We now show that any sequence of policies $\gamma_{k} \in \Gamma_{s,f}$, for which $P_{e}(\gamma_{k}) \rightarrow 0$, has $N_{c}(\gamma_{k}) \rightarrow \infty$.
\begin{lemma}
  For any sequence of policies $\gamma_{k} \in \Gamma_{s,f}$, $N_{c}(\gamma_{k}) \rightarrow \infty$ if $P_{e}(\gamma_{k}) \rightarrow 0$.
  \label{chap2:prop:necessity_largeNc}
\end{lemma}

\begin{proof}
  Assume that there exists a sequence of policies $\gamma_{k} \in \Gamma_{s,f}$ for which $P_{e}(\gamma_{k}) \rightarrow 0$, but $N_{c}(\gamma_{k}) \not \rightarrow \infty$.
  Therefore $\limsup_{k \rightarrow \infty} N_{c}(\gamma_{k}) = N_{c,l} < \infty$.
  From Lemma \ref{chap2:prop:asymplb} we have that $P_{e}(\gamma_{k}) \geq P_{N_{c,k}}(\lambda), \forall k$.
  If $N_{c,l} < \infty$ we have that $\lim_{k \rightarrow \infty} P_{e}(\gamma_{k}) \geq P_{N_{c,l}}(\lambda) > 0$, which contradicts the assumption that $P_{e}(\gamma_{k}) \downarrow 0$.
  Therefore $N_{c}(\gamma_{k})$ should necessarily grow to infinity for the sequence of policies $\gamma_{k}$.
\end{proof}

From the above result and since $P_{e,f}^*(D_{c}) \downarrow 0$ as $D_{c} \uparrow \infty$, we know that to obtain an asymptotic characterization of $P_{e,f}^*(D_{c})$ as $D_{c} \uparrow \infty$, we need only consider sequences of policies $\gamma_{k} \in \Gamma_{s,f}$ with $N_{c}(\gamma_{k}) \uparrow \infty$.
Now for any sequence $\gamma_{k} \in \Gamma_{s,f}$, with $N_{c}(\gamma_{k}) \rightarrow \infty$, we present an upper bound on the decay rate $\lim_{D_{c} \rightarrow \infty} - \frac{\ln({P}_{e, f}^*(D_{c}))}{D_{c}}$.

\begin{proposition} 
  If $\lambda \ln \cM < \frac{E_{0}(\rho, \mQ)}{\rho}$, the exponential decay rate of ${P}_{e,f}^{*}(D_{c})$ has the following upper bound:
  \begin{equation*} 
    \lim_{D_{c} \rightarrow \infty} \frac{-\ln({P}_{e, f}^*(D_{c}))}{D_{c}} \leq \frac{2}{3}\brap{E_{0}(\rho, \mQ) - \rho \lambda \ln \cM},
  \end{equation*}
  for a fixed $\rho$ and $\mQ$.
  The best upper bound on the decay rate is obtained by fixing $\rho$ and $\mQ$ to be $\rho^*$ and $\mQ^*$ respectively, where \[(\rho^*, \mQ^*) = \argmax_{\rho \in [0,1], \mQ } \frac{2}{3}\brap{E_{0}(\rho, \mQ) - \rho \lambda \ln \cM}.\]
  \label{chap2:prop:decayrate}
\end{proposition}

\begin{proof}
  From Proposition \ref{chap2:prop:mukub}, we have that $P_{e,f}^*(D_{c}) \downarrow 0$ as $D_{c} \uparrow \infty$.
  From Lemma \ref{chap2:prop:necessity_largeNc}, we need only consider any sequence of $\gamma_{k} \in \Gamma_{s,f}$ such that $N_{c,k} = N_{c}(\gamma_{k}) \rightarrow \infty$.
  We have from Lemma \ref{chap2:prop:asymplb} that $P_{e}(\gamma_{k}) \geq P_{N_{c,k}}(\lambda)$.
  Then from Lemma \ref{chap2:prop:asymplb_Nclimit} for large enough $k$, as $\lambda \ln \cM < \frac{E_{0}(\rho, \mQ)}{\rho}$ we have that ${P}_{e}(\gamma_{k}) \geq  e^{-N_{c,k}(E_{0}(\rho) - \rho \lambda \ln \cM)}$.
  From \eqref{chap2:eq:delay_queuelength}, for a fixed codeword length $N_{c,k}$ we have that $D(\gamma_{k}) = \frac{\Expp Q}{\lambda} + \frac{N_{c,k} - 1}{2}$.
  For any $\gamma_{k}$, $\Expp Q \geq \Expp \ExpS S = \lambda N_{c,k}$ as $Q \geq S$. 
  Therefore $D(\gamma_{k}) \geq \frac{3 N_{c,k}}{2} - \frac{1}{2}$ or $\frac{2}{3}(D(\gamma_{k}) + \frac{1}{2}) \geq N_{c,k}$.
  Hence ${P}_{e}(\gamma_{k}) \geq e^{-\frac{2}{3}({D}(\gamma_k) + \frac{1}{2})(E_{0}(\rho, \mQ) - \rho \lambda \ln \cM)}$ or $\lim_{k \rightarrow \infty} \frac{-\ln(P_{e}(\gamma_{k}))}{{D}(\gamma_{k})} \leq \frac{2}{3}(E_{0}(\rho, \mQ) - \rho \lambda \ln \cM)$.
  We note that the above upper bound on the exponential decay rate holds for any sequence of policies $\gamma_{k}$.
  For \eqref{chap2:eq:tradeoffproblem}, we have by definition that for $D_{c} \geq 1$ and $\epsilon > 0$, there exists a $\gamma$ such that $D(\gamma) \leq D_{c}$ and $P_{e}(\gamma) \leq P_{e,f}^*(D_{c}) + \epsilon$.
  For any sequence $\gamma_{k}$, choose $\epsilon$ from the sequence $\epsilon_{k} = e^{-e^{N_{c,k}}}$.
  Then, we have that $\lim_{D_{c} \rightarrow \infty} \frac{-\ln(P_{e,f}^*(D_{c}))}{{D_{c}}} \leq \frac{2}{3}(E_{0}(\rho, \mQ) - \rho \lambda \ln \cM)$.
  The best upper bound on decay rate is then obtained by choosing the fixed $\rho$ and $\mQ$ to be $\rho^*$ and $\mQ^*$ respectively.
\end{proof}

Now we consider the decay rate, $\lim_{D_{c} \uparrow \infty} -\frac{\ln P_{e}^*(D_{c})}{D_{c}}$.
Unlike the class of policies considered above, where the transmission duration was always fixed to be a parameter $N_{c}$, for $\gamma \in \Gamma_{s}$ we have that at each decision epoch $m$, the transmitter can choose both the batch size $S[m]$ and the transmission duration $\Tau[m]$.
We obtain a lower bound to the exponential decay rate of $P_{e}^*(D_{c})$ as $D_{c} \uparrow \infty$, by obtaining the exponential decay rate $\lim_{k \rightarrow \infty} -\frac{\ln P_{e}(e_{k})}{D(e_{k})}$, for  a sequence of exhaustive service (EXH) policies $e_{k}$, with $D(e_{k}) \uparrow \infty$.
Consider an EXH policy $e$.
The policy $e$ chooses $S[m] = Q[m - 1]$ and $\Tau[m]$ such that the codeword error probability is at most $P_{e,b}$.
Intuition suggests that the codeword length $\Tau[m]$ has to be at least a minimum value, say $\tau(S[m])$, which is a function of the batch size $S[m]$, to guarantee that the codeword error probability $P(S[m], \Tau[m])$ is at most $P_{e,b}$ for every $m$.
The message alphabet size is $\cM$, therefore the alphabet size of the batch of $s$ message symbols is $\cM^{s}$.
The rate in nats is $R = \frac{s \ln \cM}{\tau(s)}$. 
Consider a randomly generated codebook, in which each codeword symbol is chosen independently according to the distribution $\mQ$ on the input alphabet $\mathcal{X}$.
The receiver is assumed to do maximum likelihood decoding of the joint message. 
From \cite[Theorem 5.6.2]{gallager_text}, we have that the average probability of codeword error is bounded above as:
\begin{eqnarray*}
  P(s,\tau) & \leq & e^{-\tau \left\{ -\rho R + E_{0}(\rho,\mQ) \right\}},\\
  \text{or } \ln {P(s,\tau)} & \leq & \rho s \ln \cM - \tau E_{0}(\rho,\mQ),
\end{eqnarray*}
where $\rho \in [0,1]$, $E_{0}(\rho,\mQ) = \displaystyle -\ln \int_{y \in \mathcal{Y}} \left[\int_{x \in \mathcal{X}}\mQ(x)P_{Y\vert X}(y\vert x)^{\nfrac{1}{1 + \rho}}dx\right]^{1 + \rho}dy$. 
To guarantee the average error rate requirement, we constrain $P(s,\tau)$ to be $\leq P_{e,r}$.
If 
\begin{eqnarray*}
  e^{-\tau \left\{ -\rho R + E_{0}(\rho,\mQ) \right\}} \leq P_{e,r},
\end{eqnarray*}
then $P(s,\tau) \leq P_{e,r}$.
For every $s$, if $\tau$ is chosen as a function $\tau(s)$ to satisfy the above inequality, we have that
\begin{eqnarray*}
  \frac{- \ln P_{e,r}}{E_{0}(\rho,\mQ)} + \frac{\rho s \ln \cM}{E_{0}(\rho,\mQ)} & \leq & \tau(s).
\end{eqnarray*}
Thus $\tau(s)$ has to be chosen as the smallest integer greater than or equal to $as + b$, where $ a = \frac{\rho \ln M}{E_{0}(\rho,\mQ)}$ and $b = \frac{- \ln P_{e,r}}{E_{0}(\rho,\mQ)}$.
Therefore $\tau(s) = \ceiling{as + b}, s > 0$, so that $P(s, \tau) \leq P_{e,r}$.

We assume that $e$ chooses $\Tau[m] = \tau(S[m])$ if $S[m] > 0$ and $\Tau[m] = 1$ if $S[m] = 0$.
Let the stationary distribution corresponding to policy $e$ be $\pi$.
Then for $e$, $P_{e}(e)$ is less than or equal to $\frac{\Expp S P_{e,b}}{\lambda \Expp \Tau}$.
Since $\Expp S = \lambda \Expp \Tau$ we have that $P_{e}(e) \leq P_{e,b}$.
For a given $P_{e,b}$ let the average delay for ${e}$ be denoted by $D_{EXH}(P_{e,b})$.
We have the following upper bound $D_{u}(P_{e,b})$ on the average delay $D_{EXH}(P_{e,b})$.
\begin{proposition}
  $D_{EXH}(P_{e,b}) \leq D_{u}(P_{e,b})$, where
  \[D_{u}(P_{e,b}) = \nfrac{b + 1}{b} \brap{\frac{a\sigma^{2}}{2(1 - a\lambda)} + \frac{3(b + 1)\lambda}{2(1 - a\lambda)} + \frac{a\sigma^{2}}{2(1 - a^{2}\lambda^{2})} - \frac{\lambda}{2}}.\]
  \label{chap2:prop:exhub}
\end{proposition}
The proof is given in Appendix \ref{chap2:app:exhub}.
We now compare the above upper bound with the upper bound obtained by Musy and Telatar \cite{musy}.
We note that the model considered by Musy and Telatar in \cite{musy} is a continuous time model with Poisson arrivals.
In \cite{musy}, if a batch size $s > 0$ is used, then the service time is $D + ks$.
For simplicity, assume that $D$ and $k$ are integers.
Let $\delta$ be such that $b\delta = D$ and $a\delta = k$, where $a$ and $b$ are integers.
Then if we assume that each slot in our model is of $\delta$ duration, then the time taken for service of $s$ customers in our model and the model in \cite{musy} is the same.
Let the Poisson arrival rate of customers in Musy's model be $\lambda_{m}$.
Now assume that the arrival process in our model is Bernoulli with an arrival probability (rate) of $\lambda = \lambda_{m}\delta$ in each slot.
We note that if $a$ and $b$ are integers, then $\tau(s) = as + b$ for every $s > 0$.
Then the upper bound in Proposition \ref{chap2:prop:exhub} simplifies to
\begin{equation}
  \frac{a\sigma^{2}}{2(1 - a\lambda)} + \frac{3b\lambda}{2(1 - a\lambda)} + \frac{a\sigma^{2}}{2(1 - a^{2}\lambda^{2})} - \frac{\lambda}{2}.
\end{equation}
We note that $\sigma^{2} = \lambda - \lambda^{2}$ for a Bernoulli arrival process.
Substituting $\lambda = \lambda_{m} \delta$, $a = \frac{k}{\delta}$ and $b = \frac{D}{\delta}$ and taking the limit as $\delta \downarrow 0$ (along a sequence such that $a$ and $b$ are integers) we obtain that the average queue length in the limit is 
\begin{equation}
  \frac{3D\lambda_{m}}{2(1 - \lambda_{m}k)} + \frac{\lambda_{m}k(2 + \lambda_{m}k)}{2(1 - \lambda_{m}k)}.
\end{equation}
We note that this is the same as that obtained by Musy in Section 2.3.2 of his thesis \cite{musythesis}.

The above upper bound leads to the following characterization of the exponential decay rate of the average error rate.

\begin{proposition}
  The exponential decay rate : $\lim_{D_{c} \rightarrow \infty} \frac{- \ln P_{e}^*(D_{c})}{D_{c}} \geq \frac{2}{3}(E_{0}(\rho^*, \mQ^*) - \rho^* \lambda \ln \cM)$.
  \label{chap2:prop:exhasymub}
\end{proposition}

\begin{proof}
%  We note that, corresponding to policy $e$, we have an upper bound $D_{u}(P_{e,c})$ on the minimum average delay $D^*(P_{e,c})$, since policy $P_{e}(e) \leq P_{e,c}$ and is feasible for EQT-TRADEOFF.
  Let $P_{e,u}(D_{c}) \Deq \inf \brac{ p \in [0,1]: D_{u}(p) \leq D_{c}}$.
  We note that if $D_{u}(p) \leq D_{c}$, then the $e$ policy corresponding to $p$ has $D(e) \leq D_{c}$ and therefore $P_{e}^*(D_{c}) \leq P_{e}(e) \leq p$.
  Therefore, $P_{e}^*(D_{c}) \leq P_{e,u}(D_{c})$.
  We have that $P_{e,u}(D_{c})$ is such that  
  \begin{eqnarray*}
    \nfrac{b_u + 1}{b_{u}\lambda} \brap{\frac{3(b_{u} + 1)\lambda}{2(1 - a\lambda)} + \frac{a\sigma^{2}}{2(1 - a\lambda)} + \frac{a\sigma^{2}}{2(1 - a^{2}\lambda^{2})} - \frac{\lambda}{2}} & = & D_{c},
  \end{eqnarray*}
  where $b_{u} = \frac{- \ln P_{e,u}(D_{c})}{E_{0}(\rho, \mQ)}$.
  We obtain the following quadratic equation in $b_{u}$:
  \begin{eqnarray*}
    (b_{u} + 1)\brap{ (b_{u} + 1) c_{1}  + c_{2}} & = & b_{u}\lambda D_{c},
  \end{eqnarray*}
  where $c_{1} = \frac{3\lambda}{2(1 - a\lambda)}$ and $c_{2} = \frac{a\sigma^{2}(2 + a\lambda)}{2(1 - a^{2}\lambda^{2})} - \frac{\lambda}{2}$.
  Or
  \begin{eqnarray*}
    \frac{- \ln P_{e,u}(D_{c})}{E_{0}(\rho, \mQ)} & = & \frac{1}{2c_{1}}\bras{\lambda D_{c} - 2c_{1} - c_{2} + \sqrt{(\lambda D_{c} - 2c_{1} - c_{2})^{2} - 4c_{1}(c_{1} + c_{2})}},
  \end{eqnarray*}
  since we want the largest exponent.
  Let $D_{c,k}$ be a sequence such that $D_{c,k} \uparrow \infty$.
  Then, we have a sequence $p_{k} = \inf \brac{ p : D_{u}(p) \leq D_{c,k}}$.
  As above, we obtain
  \begin{eqnarray*}
    \frac{- \ln p_{k}}{E_{0}(\rho, \mQ)} & = & \frac{1}{2c_{1}}\bras{\lambda D_{c,k} - 2c_{1} - c_{2} + \lambda D_{c,k} \sqrt{{\brap{1 - \frac{2c_{1} + c_{2}}{D_{c,k}}}^{2} - \frac{4c_{1}(c_{1} + c_{2})}{D_{c,k}^{2}}}}}.
  \end{eqnarray*}
  Therefore, we have that $\lim_{D_{c,k} \rightarrow \infty} \frac{- \ln p_{k}}{D_{c,k}} = \frac{\lambda E_{0}(\rho, \mQ)}{c_{1}} = \frac{2}{3}(E_{0}(\rho, \mQ) - \rho \lambda \ln \cM)$, after substituting for $a$ in $c_{1}$.
  Hence, we have that $\lim_{D_{c} \rightarrow \infty} \frac{- \ln P_{e}^*(D_{c})}{D_{c}} \geq \frac{2}{3}(E_{0}(\rho^*, \mQ^*) - \rho^* \lambda \ln \cM)$, by fixing $\rho$ and $\mQ$ to be $\rho^*$ and $\mQ^*$ respectively.
\end{proof}
We note that we do not have any upper bound on the exponential decay rate $\lim_{D_{c} \rightarrow \infty} \frac{-\ln P_{e}^*(D_{c})}{D_{c}}$, over the class of policies $\Gamma_{s}$.

\subsection{{Integer valued queue evolution}}
\label{chap2:sec:rmodela_intvalue}
We note that for R-model-A, the queue length process evolves on $\mathbb{R}_+$.
Suppose we consider the case, where the arrival process $A[m] \in \brac{0,1,\dots,A_{max} \in \sZ}$, $q_{0} \in \mathbb{Z}_+$, and $S[m] \in \mathbb{Z}_+$.
Then the queue length process $(Q[m], m \geq 0)$ would evolve on $\mathbb{Z}_+$.
Let us denote this model as I-model-A.
Intuitively, $P_{e}^*(D_{c}), P_{e,f}^*(D_{c})$, or $P_{e,N_{c}}^*(D_{c})$ (or $D^*(P_{e,c})$, $D_{f}^*(P_{e,c})$, or $D_{N_{c}}^*(P_{e,c}))$ for  R-model-A (with $Pr\brac{A[m] = a} > 0$ only for $a \in \brac{0, 1, \dots, A_{max}}$) would be a lower bound to the same performance measures for I-model-A, as the set of feasible policies for R-model-A would always be a superset of the set of feasible policies for I-model-A for the optimization problem \eqref{chap2:eq:tradeoffproblem} (or \eqref{chap2:eq:eqtradeoffproblem}), from which $P_{e}^*(D_{c}), P_{e,f}^*(D_{c})$, or $P_{e,N_{c}}^*(D_{c})$ (or $D^*(P_{e,c})$, $D_{f}^*(P_{e,c})$, or $D_{N_{c}}^*(P_{e,c}))$ is obtained as the optimal value.

For I-model-A, we note that $P_{e,f}^*(D_{c}) \downarrow 0$ as $D_{c} \uparrow \infty$, since the sequence of policies $\mu_{k}$ can be restricted to use only non-negative integer valued $R_{k} = r N_{c,k}$ by an appropriate choice of the sequence $N_{c,k}$ and a rational number $r > \lambda$.
Furthermore, in this case, the following tighter bound on the exponential decay rate can be obtained for the sequence $\mu_{k}$.
\begin{proposition}
  If $\lambda \ln \cM < r \ln \cM  < \frac{E_{0}(\rho, \mQ)}{\rho}$, for the sequence of policies $\mu_{k}$ defined above, we have that $\lim_{k \rightarrow \infty} \frac{-\ln P_{e}(\mu_{k})}{D(\mu_{k})} = \frac{2}{3}\brap{E_{0}(\rho, \mQ) - \rho r \ln \cM}$, with $D(\mu_{k}) \uparrow \infty$. Therefore $P_{e}(\mu_{k}) \downarrow 0$.
  \label{chap2:prop:mukubint}
\end{proposition}
\begin{proof}
  For $\mu_{k}$, we note that $P_{e}(\mu_{k}) \leq \frac{\Exp_{\pi_{\mu_{k}}} \bras{S e^{-N_{c,k} E_{0}(\rho, \mQ) + \rho S \ln \cM}}}{\lambda N_{c}}$.
  As $S \leq r N_{c,k}$ we have that 
  \begin{eqnarray}
    P_{e}(\mu_{k}) & \leq & \frac{r}{\lambda} \Exp_{\pi_{\mu_{k}}} \bras{e^{-N_{c,k} E_{0}(\rho, \mQ) + \rho S \ln \cM}},
    \label{chap2:eq:mukub0int} \\
    & \leq & \frac{r}{\lambda} \brap{e^{E_{0}(\rho, \mQ) - \rho r \ln \cM}}^{-N_{c,k}}.
    \label{chap2:eq:mukub1int}
  \end{eqnarray}
  From Denteneer et al. \cite[equation (9) and the upper bound in (12)]{denteneer} we have that
  \begin{equation*}
    \Exp_{\pi_{\mu_{k}}} Q \leq \frac{\sigma^{2}}{2(r - \lambda)} + \frac{\lambda N_{c,k}}{2} + \frac{1}{2} \min(\lambda N_{c,k}, r N_{c,k} - 1).
  \end{equation*}
  For $k$ large enough, as $N_{c,k} \rightarrow \infty$ and $\lambda < r$, we have that
  \begin{equation*}
    \Exp_{\pi_{\mu_{k}}} Q \leq \frac{\sigma^{2}}{2(r - \lambda)} + \lambda N_{c,k}.
  \end{equation*}
  We note that in our case, there is an extra time average holding cost of $\frac{\lambda (N_{c,k} - 1)}{2}$ due to the customers waiting during a transmission period of $N_{c,k}$ slots.
  Therefore an upper bound on the time average queue length is 
  \begin{equation*}
    \frac{\sigma^{2}}{2(r - \lambda)} + \frac{3\lambda N_{c,k}}{2} - \frac{\lambda}{2}.
  \end{equation*}
  Therefore
  \begin{eqnarray}
    D(\mu_{k}) \leq \frac{\sigma^{2}}{2\lambda(r - \lambda)} + \frac{3 N_{c,k}}{2} - \frac{1}{2}, \nonumber \\
    \frac{2}{3} \brap{D(\mu_{k}) - \frac{\sigma^{2}}{2\lambda(r - \lambda)} + \frac{1}{2}} \leq N_{c,k}.
    \label{chap2:eq:mukubint}
  \end{eqnarray}
  Since $D(\mu_{k})$ is at least $N_{c,k}$ we have that $D(\mu_{k}) \uparrow \infty$ as $N_{c,k} \uparrow \infty$.
  Substituting the lower bound on $N_{c,k}$ from \eqref{chap2:eq:mukubint} in \eqref{chap2:eq:mukub1int}, we have that 
  \begin{equation*}
    P_{e}(\mu_{k}) \leq \frac{r}{\lambda} \brap{e^{E_{0}(\rho, \mQ) - \rho r \ln \cM}}^{-\frac{2}{3} \brap{D(\mu_{k}) - \frac{\sigma^{2}}{2\lambda(r - \lambda)} + \frac{1}{2}}}.
  \end{equation*}
  Hence $\lim_{k \rightarrow \infty} \frac{-\ln P_{e}(\mu_{k})}{D(\mu_{k})} = \frac{2}{3}\brap{E_{0}(\rho, \mQ) - \rho r \ln \cM}$, and $P_{e}(\mu_{k}) \downarrow 0$.
\end{proof}

\begin{remark}
We note that Proposition \ref{chap2:prop:mukubint} provides a lower bound on the exponential decay rate of $\lim_{k \rightarrow \infty} \frac{-\ln(P_{e,f}^*(D_{c}))}{{D_{c}}}$, since for any $D_{c,k} \uparrow \infty$, we have a subsequence of $\mu_{k}$, such that $D(\mu_{k}) \leq D_{c,k}$ and therefore $P_{e,f}^*(D_{c,k}) \leq P_{e}(\mu_{k})$.
We note that the capacity $(C)$ of the discrete memoryless channel is given by $\frac{d E_{0}(\rho,\mQ^*)}{d\rho}\vert_{\rho = 0}$. 
Let us consider the case when $C - \delta \leq \lambda \ln \cM < C$, where $\delta$ is a small positive constant.
In the following, we show that the lower bound to the exponential decay rate of $P_{e,f}^*(D_{c})$ achieved by the sequence of policies $\mu_{k}$ approximately matches with the upper bound on the exponential decay rate of $P_{e,f}^*(D_{c})$, $\frac{2}{3}\brap{E_{0}(\rho^*, \mQ^*) - \rho \lambda \ln \cM}$, obtained in Proposition \ref{chap2:prop:decayrate}.
We note that if $C - \delta \leq \lambda \ln \cM < C$, for small positive $\delta$, then any $\rho$ satisfying $\lambda \ln \cM < \frac{E_{0}(\rho, \mQ^*)}{\rho}$ is approximately zero.
Furthermore, $\rho^*$ is such that $\lambda \ln \cM < \frac{E_{0}(\rho^*, \mQ^*)}{\rho^*}$ and is therefore approximately zero.
Then, from \eqref{chap2:eq:mukub0int}, since $\Exp_{\pi_{\mu_{k}}} e^{\rho^* S \ln \cM} \approx e^{\rho^* \lambda N_{c,k} \ln \cM}$, we have that $P_{e}(\mu_{k}) \lessapprox \frac{r}{\lambda} \brap{e^{E_{0}(\rho^*, \mQ^*) - \rho^* \lambda \ln \cM}}^{-N_{c,k}}$, which yields an approximate lower bound on the exponential decay rate $\lim_{k \rightarrow \infty} \frac{-\ln P_{e}(\mu_{k})}{D(\mu_{k})} \approx \frac{2}{3}\brap{E_{0}(\rho^*, \mQ^*) - \rho^* \lambda \ln \cM}$, which matches with the upper bound in Lemma \ref{chap2:prop:decayrate}.
\end{remark}

Recall that $P_{N_{c}}(\lambda) = \inf_{D_{c}} P_{e,N_{c}}^*(D_{c})$ for both R-model-A and I-model-A, where we have restricted to the set of policies $\Gamma_{s,N_{c}}$ for both models.
We have that Proposition \ref{chap2:prop:decayrate} also holds for I-model-A.
The proof of Proposition \ref{chap2:prop:decayrate} holds for I-model-A as: (a) $P_{e,f}^*(D_{c}) \downarrow 0$ as $D_{c} \uparrow \infty$, from Proposition \ref{chap2:prop:mukubint}, and (b) $P_{e}(\gamma_{k}) \geq P_{N_{c,k}}(\lambda)$, which holds since $P_{N_{c}}(\lambda)$ for R-model-A is a lower bound to $P_{N_{c}}(\lambda)$ defined for I-model-A.

We also note that the upper bound $D_{u}(P_{e,c})$ holds for I-model-A under a EXH policy which serves only integer number of message symbols.
Hence Proposition \ref{chap2:prop:exhub} and therefore Proposition \ref{chap2:prop:exhasymub} also holds for I-model-A.

\begin{remark}
  In the next section, we present an asymptotic analysis for R-model-B.
  We recall that for R-model-B, the codeword length $N_{c}$ is fixed.
  The asymptotic analysis of R-model-B is significant, since the tradeoff problem for R-model-B is a subproblem for R-model-A with the restriction to policies in $\Gamma_{s,f}$.
  In fact, we study how $D^*_{N_{c}}(P_{e,c})$ behaves as $P_{e,c} \downarrow P_{N_{c}}(\lambda)$.
\end{remark}

\section{Asymptotic analysis for R-model-B}
\label{chap2:sec:rmodelb}
\label{chap2:sec:otherprob_errorprob}

\subsection{Problem Statement}
\label{chap2:sec:problem_statement_other}
We state the tradeoff problem for R-model-B so that it is similar to the definition of the problem TRADEOFF in Chapter 4.
The TRADEOFF problem for R-model-B is
\begin{eqnarray*}
  & \mini_{\gamma \in \Gamma_{s}} & \frac{\Exp_{\pig}\Exp h(Q, N_{c})}{\lambda N_{c}}  \\
  & \text{such that } & \frac{\Exp_{\pig} \ExpS c_{s}(S(Q), N_{c})}{\lambda N_{c}}  \leq P_{e,c},
\end{eqnarray*}
where $h(q,N_{c}) = q N_{c} + \frac{\lambda N_{c}(N_{c} - 1)}{2}$ as defined before, and $P_{e,c}$ is a constraint on the average error rate.
As in Section \ref{chap5:sec:cmdpformulation}, we can show that there exists a set $\mathcal{O}^{u}$ of $P_{e,c}$ such that there exists a stationary deterministic optimal policy for $P_{e,c} \in \mathcal{O}^{u}$.
This stationary deterministic policy is optimal for an unconstrained MDP with single stage cost $h(q, N_{c}) + \beta_{P_{e,c}} c_{s}(s, N_{c})$, where $\beta_{P_{e,c}} \geq 0$ is a Lagrange multiplier.

However, we consider the above problem only for a subset $\Gamma_{a} \subset \Gamma_{s}$, which is the set of monotone admissible policies.
We redefine the set of admissible policies for R-model-B as follows.
A policy $\gamma \in \Gamma_{a}$ if:
\begin{description}
\item[RG1 :]{$\gamma \in \Gamma_{s}$,}
\item[RG2 :]{it induces an aperiodic, irreducible Harris Markov chain $Q[m]$,}
\item[RG3 :]{the average service rate at a queue length $q$, $\Exp S(q)$ is non-decreasing in $q$.}
\end{description}
We note the above properties are similar to those defined in Section \ref{chap5:sec:realvalued_setup} of Chapter 4.

We note that the function $c_{s}(s, N_{c})$ is not convex.
Consider a $P_{e,c} \in \mathcal{O}^{u}$ and the corresponding unconstrained MDP with Lagrange multiplier $\beta_{P_{e,c}}$.
Existing proofs of the monotonicity property of the optimal policy for the unconstrained MDP, such as those in Goyal et al. \cite{munish}, require that the function $c_{s}(s, N_{c})$ be convex.
We are not able to prove that the batch size is a monotonically non-decreasing function of the queue length for the optimal policy even if $P_{e,c} \in \mathcal{O}^{u}$, unlike in Chapter 4.
However, we observe that the optimal policy prescribes a batch size which is monotonically increasing in the queue length in numerical solutions of the MDP.
This is the only motivation for assuming RG3.

The TRADEOFF problem for R-model-B is to obtain $D^*_{N_{c}}(P_{e,c})$ which is the optimal value of 
\begin{eqnarray*}
  & \mini_{\gamma \in \Gamma_{a}} & \frac{\Exp_{\pig}\Exp h(Q, N_{c})}{\lambda N_{c}}  \\
  & \text{such that } & \frac{\Exp_{\pig} \ExpS c_{s}(S(Q), N_{c})}{\lambda N_{c}}  \leq P_{e,c}.
\end{eqnarray*}
We note that for every $P_{e,c}$ such that the above problem is feasible, for every $\epsilon > 0$, by definition there is a feasible admissible policy $\gamma$ such that $D(\gamma) \leq D^*_{N_{c}}(P_{e,c}) + \epsilon$.
We call such an admissible policy $\epsilon$-optimal for $P_{e,c}$.

\begin{figure}
  \centering
  \includegraphics[width=80mm,height=50mm]{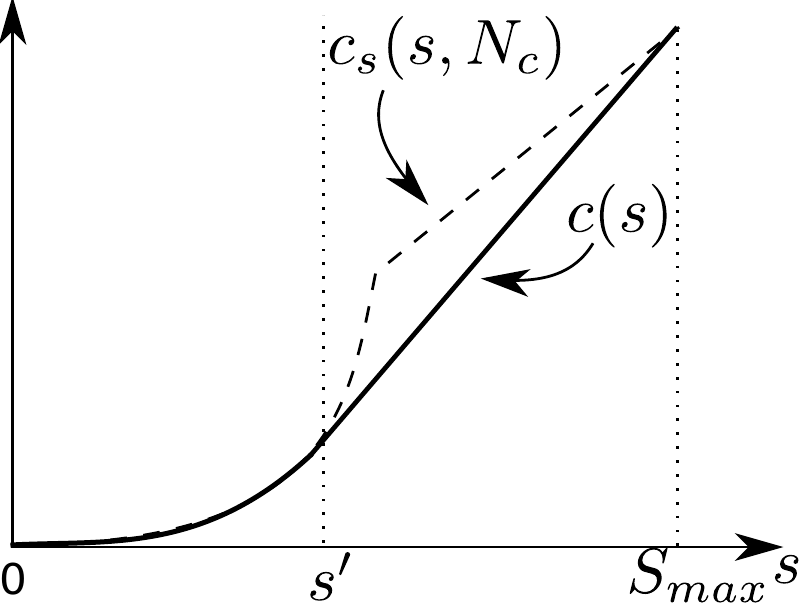}
  \caption{Illustration of the error cost function $c_{s}(s, N_{c})$ and the lower convex envelope $c(s)$. The lower convex envelope coincides with $c_{s}(s,N_{c})$ for all $s \in [0,s']$.}
  \label{chap2:fig:csrmodelb}
\end{figure}

For R-model-B, suppose that $S_{max} > \frac{N_{c} E_{0}(\rho, \mQ)}{\rho \ln \cM}$.
As in Chapter 4, we define $c(s): [0,S_{max}] \rightarrow \mathbb{R}_{+}$ as the lower convex envelope of $\brac{(s,c_{s}(s, N_{c})), s \in [0,S_{max}]}$.
Then there exists a $s' < S_{max}$, which satisfies the following equation:
\begin{eqnarray*}
        S_{max} e^{N_{c}(E_{0}(\rho,\mQ))} = e^{\rho s'\ln \cM} \bras{S_{max} + \rho \ln \cM s' (S_{max} - s')}.
\end{eqnarray*}
We note that the tangent drawn from $(S_{max}, c_{s}(S_{max}, N_{c}) = S_{max})$ touches the $c_{s}(s,N_{c})$ curve at $(s', c_{s}(s',N_{c}))$.
We also note that the definition of $s'$ is similar to that for R-model-A, except that for R-model-B, the slope of $c_{s}(s, N_{c})$ at $s = s'$ is not one.
Furthermore $s'$ for R-model-B is always greater than or equal to $s'$ for R-model-A.
Figure \ref{chap2:fig:csrmodelb} shows an example.
We note that the lower convex envelope $c(s) = c_{s}(s, N_{c})$, for $s \in [0,s']$ and for $s \in (s', S_{max})$, $c(s)$ is the tangent line segment drawn from $(s', c_{s}(s',N_{c}))$ to $(S_{max}, S_{max})$.

For R-model-B, the infimum of all achievable average probabilities of error, $\inf_{\gamma \in \Gamma_{a}} P_{e}(\gamma)$, is again denoted as $P_{N_{c}}(\lambda)$.
Similar to the proof of Lemma \ref{chap2:prop:asymplb}, we can show that $P_{N_{c}}(\lambda) = \frac{c(\lambda)}{\lambda N_{c}}$.
In the following, we obtain an asymptotic lower bound to $D^*_{N_{c}}(P_{e,c})$ as $P_{e,c} \downarrow P_{N_{c}}(\lambda)$.

\subsection{Asymptotic analysis}
We assume that the following properties hold for $A[1]$.
\begin{description}
\item[RA1 :]{$Pr\brac{A[1] - S_{max} > \delta_{a}} > \epsilon_{a}$,}
\item[RA2 :]{$Pr\brac{A[1] \leq \frac{\Delta_{a}}{2}} = \epsilon'_{a} > 0$, for some $0 < \Delta_{a} < s'$.}
\end{description}
We note that RA1 is the same as that defined in Section \ref{chap5:sec:realvalued_setup} of Chapter 4.
We also note that the property RA2, is similar to the property A2 assumed in Section \ref{chap5:sec:integervalued_setup} of Chapter 4, in that there is a positive probability of the number of arrivals being in an interval including zero.

We note that the service cost function $c(s)$, as defined above, satisfies the following properties, which are similar to the properties RC1 and RC2 defined in Section \ref{chap5:sec:realvalued_setup} of Chapter 4.
\begin{description}
\item[RC1 :]$c(0) = 0$, and,
\item[RC2 :]$c(s)$ is strictly convex for $s \in [0, s')$ and linear for $s \in [s',S_{max}]$.
\end{description}

The asymptotic behaviour of $D^*_{N_{c}}(P_{e,c})$ as $P_{e,c} \downarrow P_{N_{c}}(\lambda)$ is different depending on whether $\lambda \leq \frac{s'}{N_{c}}$ or $\lambda > \frac{s'}{N_{c}}$.
We characterize the asymptotic behaviour in both of these cases in the following.

\begin{lemma}
  For $\lambda \leq \frac{s'}{N_{c}}$, and for any sequence of admissible policies $\gamma_{k}$ with $P_{e}(\gamma_{k}) - P_{N_{c}}(\lambda) = V_{k} \downarrow 0$ we have that 
  \begin{eqnarray*}
    \overline{Q}(\gamma_{k}) = \Omega\nfrac{1}{\sqrt{V_{k}}}.
  \end{eqnarray*}
  \label{chap2:lemma:rmodelbcase1}
\end{lemma}

\begin{proof}
  For an admissible policy $\gamma$, define $\overline{C}(\gamma) = \Exp_{\pi_{\gamma}} c(S(Q))$.
  We note that as $c_{s}(s,N_{c}) \geq c(s)$, $P_{e}(\gamma_{k}) = \frac{\Exp_{\pig} c_{s}(S(Q), N_{c})}{\lambda N_{c}} \geq \frac{\overline{C}(\gamma_{k})}{\lambda N_{c}}$.
  Then we have that there exists a sequence $U_{k}$ such that $U_{k} = \overline{C}(\gamma_{k}) - c(\lambda) \leq \lambda N_{c}\brap{P_{e}(\gamma_{k}) - P_{N_{c}}(\lambda)} = \lambda N_{c} V_{k} \downarrow 0$.
  Let us consider a particular policy $\gamma$ in the sequence $\gamma_{k}$ with $U_{k} = U$ and $\pi_{\gamma_{k}} = \pi$.
  To obtain a lower bound on $\overline{Q}(\gamma)$ we proceed, as in the proof of Proposition \ref{chap5:prop:realvalued_evolution_lb}, with $q_{d} = \sup\brac{q : \Exp S(q) \leq \lambda N_{c} + \epsilon_{U}}$ where $\epsilon_{U}$ will be chosen in the following.
  We follow all the steps in the proof of Proposition \ref{chap5:prop:realvalued_evolution_lb}, with $q_{d}$ as above and $\epsilon_{U}$ in place of $\epsilon_{V}$, up to the step where $\overline{Q}(\gamma)$ is bounded below by $\frac{k_{2}\Delta}{2}$, where the $k_{2}$ is the largest integer satisfying the inequality \eqref{chap5:eq:real5}.
  That is, we have that $\overline{Q}(\gamma) \geq \frac{k_{2}\Delta}{2}$, where $k_{2}$ is the largest integer, such that
  \begin{eqnarray*}
    \brap{1 + \frac{\epsilon_{U}}{\delta \epsilon_{a}}}^{k_{2}} \leq \frac{2}{1 + 2D_{t}},
  \end{eqnarray*}
  where $D_{t} = \frac{1}{\epsilon_{U}} \int_{q_{d}}^{\infty} (\Exp S(q) - \lambda N_{c})d\pi(q)$.
  We note that in Proposition \ref{chap5:prop:realvalued_evolution_lb}, a lower bound on $k_{2}$ was obtained via the upper bound $\frac{1}{4}$ on $D_{t}$.
  We note that, unlike in Proposition \ref{chap5:prop:realvalued_evolution_lb}, the function $c(s)$ is not strictly convex for $s \in [0,S_{max}]$.
  So we derive an upper bound on $D_{t}$ in a slightly different way.
  Define $q_{\lambda} = \sup\brac{q : \Exp S(q) \leq \lambda N_{c}}$.
  Let $l(s)$ be the tangent line to the curve $c(s)$ at $(\lambda, c(\lambda))$.
  We have that $\Expp \brac{c(S(Q)) - l(S(Q))} = U$.
  That is, 
  \[ \int_{0}^{\infty} \Exp \brac{c(S(q)) - l(S(q))} d\pi(q) = U. \]
  As $c(s) \geq l(s)$ we have that
  \[ \int_{0}^{q_{\lambda}} \Exp \brac{c(S(q)) - l(S(q))} d\pi(q) \leq U. \]
  As $c(s)$ is convex and $l(s)$ is linear we have that
  \[ \int_{0}^{q_{\lambda}} \brac{c(\Exp S(q)) - l(\Exp S(q))} d\pi(q) \leq U. \]
  We note that for $\lambda \leq \frac{s'}{N_{c}}$, there exists an $a_{1} > 0$ such that $c(s) - l(s) \geq a_{1} (s - \lambda N_{c})^{2}$ for $s \leq \lambda N_{c}$.
  Furthermore as for $q \leq q_{\lambda}, \Exp S(q) \leq \lambda N_{c}$, we have that
  \[ \int_{0}^{q_{\lambda}} \brap{\Exp S(q) - \lambda N_{c}}^{2} d\pi(q) \leq \frac{U}{a_{1}}, \]
  which can be written as
  \[ \int_{0}^{q_{\lambda}} \brap{\Exp S(q) - \lambda N_{c}}^{2} d\pi(q) + \int_{q_{\lambda}}^{\infty} 0 d\pi(q) \leq \frac{U}{a_{1}}. \]
  By Jensen's inequality we then have that
  \[ \brap{ \int_{0}^{q_{\lambda}} \brap{\Exp S(q) - \lambda N_{c}} d\pi(q) + \int_{q_{\lambda}}^{\infty} 0 d\pi(q)  }^{2} \leq \frac{U}{a_{1}}. \]
  Hence we obtain that
  \begin{equation}
    \int_{0}^{q_{\lambda}} \brap{\Exp S(q) - \lambda N_{c}} d\pi(q) \geq -\sqrt{\frac{U}{a_{1}}}.
    \label{chap5:eq:pedrealval0}
  \end{equation}
  Now we note that for an admissible policy $\gamma$,
  \begin{eqnarray*}
    & & \int_{0}^{\infty} \brap{\Exp S(q) - \lambda N_{c}} d\pi(q) = 0, \text{ or,}\\
    & & \int_{0}^{q_{\lambda}} \brap{\Exp S(q) - \lambda N_{c}} d\pi(q) + \int_{q_{\lambda}}^{q_{d}} \brap{\Exp S(q) - \lambda N_{c}} d\pi(q) + \int_{q_{d}}^{\infty} \brap{\Exp S(q) - \lambda N_{c}} d\pi(q) = 0.
  \end{eqnarray*}
  For $q > q_{\lambda}$, $\Exp S(q) > \lambda N_c$, so that $\int_{q_{\lambda}}^{q_{d}} \brap{\Exp S(q) - \lambda N_c} d\pi(q) > 0$, which implies that
  \begin{eqnarray*}
    \int_{0}^{q_{\lambda}} \brap{\Exp S(q) - \lambda N_c} d\pi(q) + \int_{q_{d}}^{\infty} \brap{\Exp S(q) - \lambda N_c} d\pi(q) & \leq & 0, \\
    \int_{q_{d}}^{\infty} \brap{\Exp S(q) - \lambda N_c} d\pi(q) \leq - \int_{0}^{q_{\lambda}} \brap{\Exp S(q) - \lambda N_c} d\pi(q) & \leq & \sqrt{\frac{U}{a_{1}}},
  \end{eqnarray*}
  from \eqref{chap5:eq:pedrealval0}.
  Therefore we obtain that $D_{t} \leq \frac{1}{\epsilon_{U}} \sqrt{\frac{U}{a_{1}}}$.
  We choose $\epsilon_{U} = 4 \sqrt{\frac{U}{a_{1}}}$, and proceed as in the proof of Proposition \ref{chap5:prop:realvalued_evolution_lb} to obtain that $\overline{Q}(\gamma) \geq \frac{\Delta}{2} \brap{\log_{\brap{1 + \frac{\epsilon_{U}}{\delta \epsilon_{a}}}} \brap{\frac{4}{3}} - 1}$.
  Therefore, for the sequence of policies $\gamma_{k}$, we have that $\overline{Q}(\gamma_{k}) = \Omega\nfrac{1}{\sqrt{U_{k}}} = \Omega\nfrac{1}{\sqrt{V_{k}}}$ as $U_{k} \leq \lambda N_{c} V_{k}$.
\end{proof}

\begin{remark}
We note that if $S_{max}$ is such that $S_{max} \leq \frac{N_{c} E_{0}(\rho, \mQ)}{\rho \ln \cM}$, then for all $\lambda < \frac{S_{max}}{N_{c}}$, the above asymptotic lower bound would hold.
\end{remark}

\begin{proposition}
  For $\lambda \leq \frac{s'}{N_{c}}$, as $P_{e,c} \downarrow P_{N_{c}}(\lambda)$, we have that 
  \begin{eqnarray*}
    D^*(P_{e,c}) = \Omega\nfrac{1}{\sqrt{P_{e,c} - P_{N_{c}}(\lambda)}}.
  \end{eqnarray*}  
  \label{chap2:prop:dpec_case1}
\end{proposition}
\begin{proof}
  Consider the sequence of policies $\gamma^*_{k}$ which are $\epsilon$-optimal for TRADEOFF for a sequence $p_{k}$ of $P_{e,c}$ such that $p_{k} \downarrow P_{N_{c}}(\lambda)$ for some $\epsilon > 0$.
  Since $P_{e}(\gamma^*_{k}) \leq p_{k}$, we have that from Lemma \ref{chap2:lemma:rmodelbcase1} that $\overline{Q}(\gamma^*_{k}) = \Omega\nfrac{1}{\sqrt{p_{k} - P_{N_{c}}(\lambda)}}$.
  Since $D^*(p_{k}) \geq \frac{1}{\lambda N_{c}} \bras{ \overline{Q}(\gamma^*_{k}) + \frac{\lambda}{2} N_{c}(N_{c} - 1)} - \epsilon$, we obtain that $D^*(p_{k}) = \Omega\nfrac{1}{\sqrt{p_{k} - P_{N_{c}}(\lambda)}}$.
\end{proof}

\begin{remark}
We note that an asymptotic upper bound, which is tight upto a logarithmic factor, can be obtained for the above case, using a sequence of policies as in Lemma \ref{chap5:lemma:rmodelupperbound}.
We note that for this sequence of policies, it is possible to choose batch sizes $s$ such that $c_{s}(s, N_{c}) = c(s)$.
\end{remark}

In the following, we obtain an asymptotic lower bound to $D^*_{N_{c}}(P_{e,c})$ as $P_{e,c} \downarrow P_{N_{c}}(\lambda)$ for $\lambda > \frac{s'}{N_{c}}$.
This lower bound is obtained by extending Lemma \ref{chap5:lemma:tprob_2} to the case when the state space of the Markov chain is the set of non-negative real numbers.
As in Lemma \ref{chap5:lemma:barq_pi0_relation} we first obtain a lower bound on the queue length as a function of the stationary probability of the queue length being in a certain set.
Then we relate the stationary probability of the queue length being in the above set to the average error rate.

Let $q_{s'} \stackrel{\Delta} = \inf\brac{q : \Exp S(q) \geq s' - \epsilon }$, where $\epsilon$ is a small positive constant, chosen such that $\Delta_{a} < s' - \epsilon$, where $\Delta_{a}$ is as in RA2.
We note that for an admissible policy $\gamma$, $\Expp \Exp S(Q) = \lambda N_{c}$, and therefore there would exist finite $q$, for which $\Exp S(q) \geq \lambda N_{c}$.
Thus, $q_{s'}$ is finite.
The following lemma shows that for a queue length $q \geq q_{s'}$, there is a positive minimum probability of serving at least a certain number of customers.

\begin{lemma}
  For an admissible policy $\gamma$, with $q_{s'}$ defined as above, for $\Delta_{a}$ as in RA2, we have that $\inf_{q \geq q_{s'}} Pr\brac{S(q) > \Delta_{a}} \geq \delta_{s} > 0$, where $\delta_{s} = \frac{s' - \epsilon - \Delta_{a}}{S_{max} - \Delta_{a}}$.
  \label{chap5:lemma:Deltaservice}
\end{lemma}
The proof of this lemma is similar to that of Lemma \ref{lemma:deltaservice} and is presented in Appendix \ref{chap5:app:Deltaservice}.

\begin{lemma}
  For $\lambda > \frac{s'}{N_{c}}$, and for any sequence of admissible policies $\gamma_{k}$ with $P_{e}(\gamma_{k}) - P_{N_{c}}(\lambda) = V_{k} \downarrow 0$, we have that 
  \begin{eqnarray*}
    \overline{Q}(\gamma_{k}) = \Omega\brap{\log\nfrac{1}{V_{k}}}.
  \end{eqnarray*}
\end{lemma}

\begin{proof}
  For an admissible policy $\gamma$, define $\overline{C}(\gamma) = \Exp_{\pi_{\gamma}} c(S(Q))$.
  We note that as $c_{s}(s,N_{c}) \geq c(s)$, we have that there exists a sequence $U_{k}$ such that $U_{k} = \overline{C}(\gamma_{k}) - c(\lambda) \leq \lambda N_{c}\brap{P_{e}(\gamma_{k}) - P_{N_{c}}(\lambda)} = \lambda N_{c} V_{k} \downarrow 0$.
  We consider a particular policy $\gamma$ in the sequence $\gamma_{k}$ with $U_{k} = U$ and $\pi_{\gamma_{k}} = \pi$.
  As in the proof of Lemma \ref{lemma:tradeoffutilitylb}, since $\gamma$ is admissible, we have that if $m_{1}$ is the largest integer such that 
  \begin{eqnarray}
    \pi\left[0, q_{s'}\right) + \pi[0, q_{s'}) \bras{\brap{1 + \frac{1}{\rho}}^{m_{1}} - 1} = \pi\left[0, q_{s'}\right)\brap{1 + \frac{1}{\rho}}^{m_{1}} \leq \frac{1}{2},
    \label{chap5:eq:pedtradeoff0}
  \end{eqnarray}
  then $\overline{Q}(\gamma) \geq \frac{m_{1} \Delta}{4}$.
  Let us define the line $l(s)$ as the line passing through $(s', c(s'))$ and $(S_{max}, c(S_{max}))$.
  For the policy $\gamma$ we have that 
  \[ \int_{0}^{\infty} \Exp \brac{c(S(q)) - l(S(q))} d\pi(q) = U. \]
  As $c(s)$ is convex and $l(s)$ is linear we have that
  \[ \int_{0}^{\infty} \brac{c(\Exp S(q)) - l(\Exp S(q))} d\pi(q) \leq U.\]
  Also as $c(s) \geq l(s)$,
  \[ \int_{0}^{q_{s'}} \brac{c(\Exp S(q)) - l(\Exp S(q))} d\pi(q) \leq U.\]
  We note that for $q < q_{s'}$ we have that $\Exp S(q) < s' - \epsilon$ and there exists $a_{1} > 0$ such that $c(\Exp S(q)) - l(\Exp S(q)) \geq a_{1} (\Exp S(q) - s')^{2}$.
  Hence we obtain that
  \[ \int_{0}^{q_{s'}} \brap{\Exp S(q) - s'}^{2} \leq \frac{U}{a_{1}},\]
  and as $q < q_{s'}$, $\brap{\Exp S(q) - s'}^{2} > \epsilon^{2}$.
  Therefore we obtain that
  \[ \int_{0}^{q_{s'}} \epsilon^{2} d\pi(q) \leq \frac{U}{a_{1}}.\]
  And hence $\pi[0, q_{s'}) \leq \frac{U}{a_{1} \epsilon^{2}}$.
  
  From \eqref{chap5:eq:pedtradeoff0}, if $m_{2}$ is the largest integer such that
  \begin{eqnarray*}
    \brap{1 + \frac{1}{\rho}}^{m_{2}} & \leq & \frac{a_{1}\epsilon^{2}}{2U}, \text{ or }, \\
    m_{2} & \leq & \log_{\brap{1 + \frac{1}{\rho}}} \brap{\frac{a_{1}\epsilon^{2}}{2U}},
  \end{eqnarray*}
  then $m_{2} \leq m_{1}$.
  We note that $m_{2}$ is at least
  \[ \floor{\log_{\brap{1 + \frac{1}{\epsilon}}} \brap{\frac{a_{1}\epsilon^{2}}{2U}}}. \]
  Since $\overline{Q}(\gamma) \geq \frac{m\Delta_{a}}{4} \geq \frac{m_{1}\Delta_{a}}{4} \geq \frac{m_{2}\Delta_{a}}{4}$, we obtain that 
  \[ \overline{Q}(\gamma) \geq \frac{\Delta_{a}}{4} \brap{\log_{\brap{1 + \frac{1}{\epsilon}}} \brap{\frac{a_{1}\epsilon^{2}}{2U}} - 1}. \]
  So for the sequence of policies $\gamma_{k}$ with $U_{k} \downarrow 0$ we have that $\overline{Q}(\gamma_{k}) = \Omega\brap{\log\nfrac{1}{U_{k}}} = \Omega\brap{\log\nfrac{1}{V_{k}}}$.
\end{proof}

\begin{proposition}
  For $\lambda > \frac{s'}{N_{c}}$, as $P_{e,c} \downarrow P_{N_{c}}(\lambda)$, we have that 
  \begin{eqnarray*}
    D^*(P_{e,c}) = \Omega\brap{\log\nfrac{1}{{P_{e,c} - P_{N_{c}}(\lambda)}}}.
  \end{eqnarray*}  
  \label{chap2:prop:dpec_case2}
\end{proposition}
The proof of the above result is similar to that of Proposition \ref{chap2:prop:dpec_case1}.

\begin{remark}
\label{chap2:remark:asympub_rmodelb_2}
An asymptotic upper bound to $D_{N_{c}}^*(P_{e,c})$, as $P_{e,c} \downarrow P_{N_{c}}(\lambda)$ can be obtained from Lemma \ref{chap5:lemma:case2upperbound}.
We note that in this case, $s_{l} = s'$ and $s_{u} = S_{max}$.
Then, as in Lemma \ref{chap5:lemma:case2upperbound} we can show that there exists a sequence of admissible policies $\gamma_{k}$, for which $D^*(P_{e,c}) = \mathcal{O}\brap{\log\nfrac{1}{V_{k}}}$ and $P_{e}(\gamma_{k}) - P_{N_{c}}(\lambda) = V_{k} \downarrow 0$.
We note that $P_{e}(\gamma_{k})$ can be computed as in Lemma \ref{chap5:lemma:case2upperbound} since $c_{s}(s', N_{c}) = c(s')$ and $c_{s}(S_{max}, N_{c}) = c(S_{max})$.
\end{remark}

\begin{remark}
We can set up a queueing model I-model-B, which is similar to R-model-B, except that $q_{0} \in \mathbb{Z}_+$, $A[m] \in \brac{0,1,\dots,A_{max} \in \sZ}$, and $S[m] \in \brac{0, 1, \dots,S_{max}}, m \geq 1$.
Therefore, for I-model-B the queue length evolution is on $\mathbb{Z}_+$.
We note that queueing model I-model-B is analogous to I-model-A.
Asymptotic lower bounds to $D_{N_{c}}^*(P_{e,c})$ can be obtained as in Section \ref{chap5:sec:asymp_analysis_tprob} of Chapter 4, with $c(s)$ defined as the piecewise linear lower convex envelope of $\brac{(s, c_{s}(s, N_{c})), s \in \brac{0, \dots, S_{max}}}$.
Asymptotic upper bounds to $D_{N_{c}}^*(P_{e,c})$ can be obtained as in Remark \ref{chap2:remark:asympub_rmodelb_2}.
\end{remark}

\begin{remark}
\label{chap2:remark:otherapproximations}
For R-model-B, we note that asymptotic lower bounds can be derived for other approximations for $c_{s}(s, N_{c})$, if the lower convex envelopes for such approximations have the same form as $c(s)$ above.
We note that the asymptotic nature of the bounds only depended on: (a) $c(s)$ being strictly convex in $[0, s']$, and (b) $c(s)$ being linear in $(s', S_{max}]$.
In Appendix \ref{chap2:app:examples} we consider some examples for the approximation $\tilde{c}_{s}(s, N_{c})$ instead of $c_{s}(s, N_{c})$ for the error cost, where 
\begin{eqnarray*}                    
\tilde{c}_{s}(s , N_{c}) = \min_{\rho \in [0,1]} s e^{-N_{c} E_{0}(\rho, Q) + \rho s \ln \cM},
\end{eqnarray*}
and illustrate that the lower convex envelope $\tilde{c}(s)$ of $\tilde{c}_{s}(s, N_{c})$ as a function of $s$ has a similar form as $c(s)$.

We note that instead of using Gallager's random coding upper bound we could also use Polyanskiy's normal approximation for the codeword error probability to derive an approximation for $c_{s}(s,N_{c})$.
From \cite{polyanskiy}, if $0 \leq s \leq \frac{C N_{c}}{\log_{2} \cM}$, then we have that the codeword error probability $P_{e,b}$ satisfies the following approximation:
\begin{eqnarray*}
N_{c} \approx \fpow{\mathbb{Q}^{-1}(P_{e,b})}{1 - \frac{s \log_{2} \cM}{N_{c} C}}{2} \frac{V}{C^{2}},
\end{eqnarray*}
where $\mathbb{Q}$ is the Gaussian Q function, $C$ is the channel capacity (in bits/channel use), and $V$ is the channel dispersion.
Then, we have that 
\begin{eqnarray*}
P_{e,b} \approx \mathbb{Q}\brap{\sqrt{\frac{N_{c}C^{2}}{V}}\brap{1 - \frac{s \log_{2} \cM}{N_{c}C}}},
\end{eqnarray*}
and the approximation
\begin{eqnarray*}
\tilde{c}_{s}(s,N_{c}) = s\mathbb{Q}\brap{\sqrt{\frac{N_{c}C^{2}}{V}}\brap{1 - \frac{s \log_{2} \cM}{N_{c}C}}}.
\end{eqnarray*}
We note that if $s > \frac{C N_{c}}{\log_{2} \cM}$, then we have the approximation $\tilde{c}_{s}(s,N_{c}) = s$.
\end{remark}
It can be shown that if $\frac{S_{max} \log_{2} \cM}{N_{c}} \leq C$, then $\tilde{c}_{s}(s, N_{c})$ is strictly convex in $s$.
Then, as in Proposition \ref{chap2:prop:dpec_case1}, we have that for any $\lambda < \frac{C}{N_{c}}$, as $P_{e,c} \downarrow P_{N_{c}}(\lambda)$, we have that $D^*(P_{e,c}) = \Omega\nfrac{1}{\sqrt{P_{e,c} - P_{N_{c}}(\lambda)}}$.

\section{Conclusions}
\label{chap2:sec:conclusions}
We have shown that for R-model-A (from Proposition \ref{chap2:prop:decayrate}) as well as I-model-A (the discussion in Section \ref{chap2:sec:rmodela_intvalue}), the exponential decay of average error rate with average delay is at most two-thirds of the Gallager random coding exponent, when we restrict attention to the set $\Gamma_{s,f}$ of policies.
We also note that for I-model-A, for information arrival rate $\lambda \ln \cM$ approaching the capacity of the channel, a sequence of policies $\in \Gamma_{s,f}$, that uses a fixed service rate, approximately achieves the best exponential decay rate of two-thirds of the Gallager random coding exponent.
We note that the $\frac{2}{3}$ factor arises because of the fundamental limitation of block codes; message symbols arriving in a transmission period of duration $N_{c}$ have an average waiting period of $\frac{N_{c}}{2}$ until the succeeding transmission period and have to wait for at least an additional $N_{c}$ slots before leaving the queue.
Therefore, the average delay is at least $\frac{3}{2} N_{c}$.
So naturally the question arises whether \emph{streaming codes} have a better exponential decay rate.
In \cite{vineeth_streaming}, we show that a sequence of fixed rate \emph{randomly time varying} streaming codes, with increasing \emph{constraint lengths} achieve an exponential decay rate of average error rate with average delay which is equal to the Gallager random coding exponent.

We then considered the exponential decay rate for a sequence of exhaustive service policies, which ensure a constant block error probability per transmission by varying the codeword length.
We obtain that for the above sequence of policies, the exponential decay rate is at least two-thirds of the Gallager random coding exponent.
The performance of such policies has been studied in detail in \cite{vineeth_wiopt}, \cite{vineeth}, \cite{vineeth_drdo_1}, \cite{vineeth_drdo_2}, and \cite{vineeth_3}.
We note that the above exponential decay rate provides a lower bound to the exponential decay rate achievable by a sequence of policies in $\Gamma_{s}$.
However, we do not have an upper bound on the exponential decay rate over any sequence of policies in $\Gamma_{s}$.

The analysis of R-model-B illustrates the application of the lower bounding technique in Chapter 4 to cases where the service cost function is not convex.
We observe that the asymptotic lower bound depends on the nature of the lower convex envelope of the service cost function at $\lambda$.
Since the lower convex envelope $c(s)$ is convex, the analysis and observations from Chapter 4 apply.
We note that the form of $c(s)$ considered in this chapter is not strictly convex or piecewise linear, but piecewise convex.
The asymptotic upper bounds show that a sequence of admissible policies using only the service rates $s$ for which the service cost $c_{s}(s, N_{c}) = c(s)$ is order optimal (only for the $\log\nfrac{1}{V}$ case).
We note that this is reminiscent of Crabill's exclusion principle\footnote{which states that for a M/M/1 queue with controllable service rates, the stationary optimal policy that minimizes the time average of the single stage cost $Q(t) + \beta c_{s}(\mu(Q(t)))$ ($\beta > 0$), uses only service rates which are such that $c_{s}(s) = c(s)$, where $c(.)$ is the lower convex envelope of the service cost function $c_{s}(.)$.}  for the control of M/M/1 queues with non-convex service costs \cite{george}, but optimality in our case is only in the asymptotic order sense.
From this analysis, we can conclude that an asymptotic characterization of the minimum average queue length in the asymptotic regime $\Re$ for admissible policies, can be obtained from the techniques in Chapter 4 by considering the lower convex envelope of the service cost function, even if the service cost function is not convex.

\clearpage
\begin{subappendices}
\large{\textbf{Appendices}}
\addcontentsline{toc}{section}{Appendices}
\addtocontents{toc}{\protect\setcounter{tocdepth}{0}}

\normalsize

\section{Optimization problem \eqref{chap2:eq:asymp_minprob}}
\label{chap2:app:optimization_problem}

\begin{figure}[h]
\centering
\subfigure[Illustration of the convex hull (the region between and including the curves AB and AC) and the optimal values $p_{e}(\epsilon_{1})$ and $p_{e}(\epsilon_{2})$.]{\includegraphics[width=60mm,height=40mm]{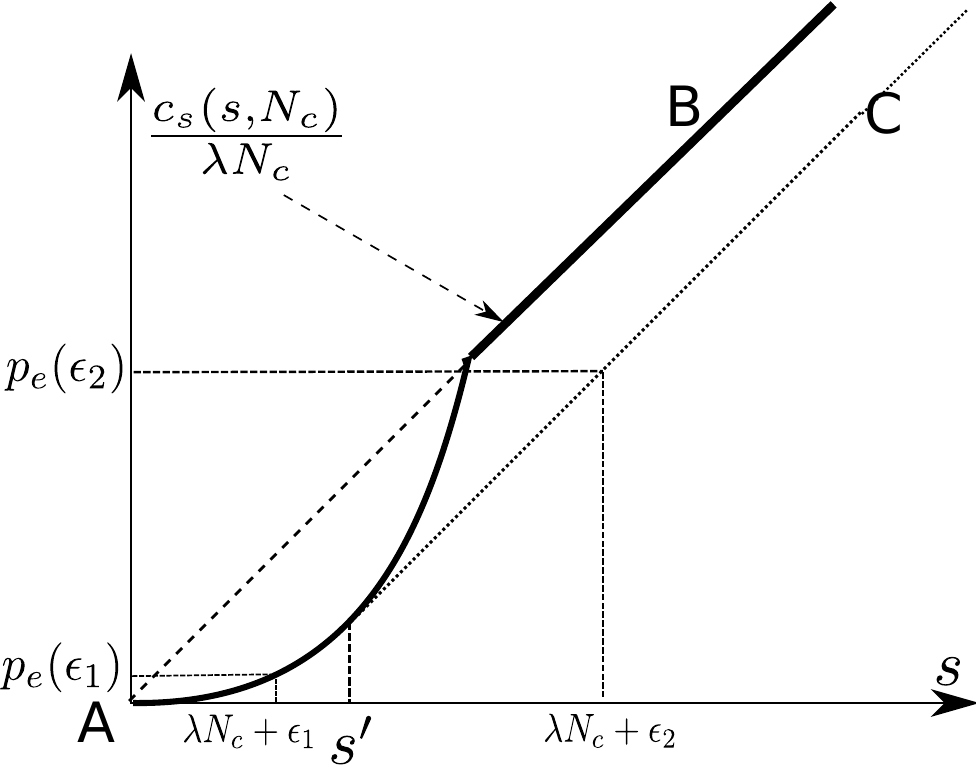}
\label{chap2:fig:opt_prob_a}}
\subfigure[Illustration of the $(\delta,\epsilon)$-optimal distribution : (i) for $\epsilon_{1}$, the distribution has $Pr\brac{S = \lambda N_{c} + \epsilon_{1}} = 1$, and (ii) for $\epsilon_{2}$, the distribution puts mass at $s'$ and $s_{1}$.]{\includegraphics[width=60mm,height=40mm]{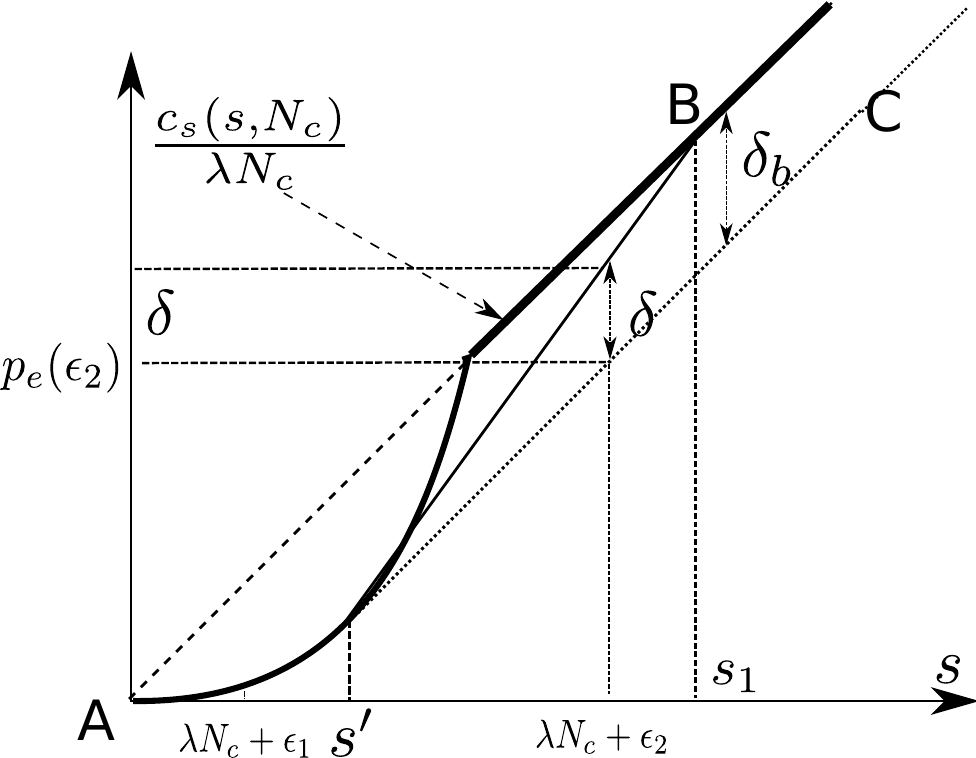}\label{chap1:fig:optprobII}
\label{chap2:fig:opt_prob_b}}
\end{figure}

Consider the optimization problem \eqref{chap2:eq:asymp_minprob} :
\begin{eqnarray}
\mini_{\pi} & & \frac{1}{\lambda N_{c}}\mathbb{E}_{\pi}\left[S\min\left(1, e^{-N_{c}E_{0}(\rho, \mQ)+\rho S\ln \cM}\right)\right]
\label{chap2:eq:app_asymp_minprob_1} \\
\text{such that } & & \mathbb{E}_{\pi}S \geq \lambda N_{c} + \epsilon \nonumber,
\end{eqnarray}
where $\pi$ is any distribution for $S$ and $\epsilon \geq 0$.
We denote the optimal value of the above problem by $p_{e}(\epsilon)$.
For any distribution $\pi$, we have that the point
\[ \bigg( \Expp S, \frac{1}{\lambda N_{c}}\mathbb{E}_{\pi}\left[S\min\left(1, e^{-N_{c}E_{0}(\rho, \mQ)+\rho S\ln \cM}\right)\right] \bigg), \]
lies in the convex hull of the set of points
\[ \bigg (s, \frac{1}{\lambda N_{c}}\left[s\min\left(1, e^{-N_{c}E_{0}(\rho, \mQ)+\rho s\ln \cM}\right)\right] \bigg), s \in \mathbb{R}_+. \]
With the constraint $\Expp S \geq \lambda N_{c} + \epsilon$, it is clear that the there would exist some distribution $\pi'$ such that the point $(\Exp_{\pi'}S, \frac{1}{\lambda N_{c}}\mathbb{E}_{\pi'}\left[S\min\left(1, e^{-N_{c}E_{0}(\rho, \mQ)+\rho S\ln \cM}\right)\right])$, with $\Exp_{\pi'} S = \lambda N_{c} + \epsilon$, lies on the lower convex envelope of $\frac{c_{s}(s, N_{c})}{\lambda N_{c}}$ (the curve AC, as shown in Figure \ref{chap2:fig:opt_prob_a}).
Therefore, $\pi'$ is optimal, and we have that
\begin{eqnarray}
p_{e}(\epsilon) =
\begin{cases}
\frac{\lambda N_{c} + \epsilon}{\lambda N_{c}} e^{-N_{c}E_{0}(\rho, \mQ) + \rho (\lambda N_{c} + \epsilon) \ln \cM} \text{ if } 0 \leq (\lambda N_{c} + \epsilon) \leq s', \\
\frac{1}{\lambda N_{c}} \left( ((\lambda N_{c} + \epsilon) - s') + s'e^{-N_{c}E_{0}(\rho, \mQ) + \rho s' \ln \cM} \right)  \text{ otherwise},
\end{cases}
\end{eqnarray}
where $s'$ is such that $\frac{d \brac{s e^{-N_{c}E_{0}(\rho, \mQ) + \rho s \ln \cM}}}{ds} \vert_{s = s'} = 1$.

Now we obtain distributions which are $(\delta,\epsilon)$-optimal, where $\delta > 0$.
A distribution $d$ is said to be $(\delta,\epsilon)$-optimal for \eqref{chap2:eq:app_asymp_minprob_1} if
\begin{eqnarray*}
  \Exp_{d} S & = & \lambda N_{c} + \epsilon, \text{ and}, \\
  \frac{1}{\lambda N_{c}}\Exp_{d}\left[S\min\left(1, e^{-N_{c}E_{0}(\rho, \mQ)+\rho S\ln \cM}\right)\right] & = & p_{e}(\epsilon) + \delta.
\end{eqnarray*}
We note that if : (i) $\lambda N_{c} + \epsilon \leq s'$ then the distribution that gives probability $1$ to $\lambda N_{c} + \epsilon$ is $(\delta, \epsilon)$-optimal $\forall \delta \geq 0$, (ii) $\lambda N_{c} + \epsilon  > s'$ and $\delta \geq \delta_{b} = s' -  s'e^{-N_{c}E_{0}(\rho, \mQ) + \rho s' \ln \cM}$ then a distribution that gives probability one to $\lambda N_{c} + \epsilon$ is $\delta$-optimal, and (iii) $\lambda N_{c} + \epsilon > s'$ and $\delta < \delta_{b}$, then a distribution that gives mass to two points $s'$ and $s_{1}$, characterized in the following lemma, is $(\delta, \epsilon)$-optimal.
\begin{lemma}
If $\lambda N_{c} + \epsilon > s'$ and $\delta < \delta_{b}$, then the distribution $d$ defined as follows is $(\delta, \epsilon)$-optimal.
Let $s_{1} = \frac{s'}{\delta}\bras{(\lambda N_{c} + \epsilon - s') + \delta} - \frac{c_{s}(s', N_{c})(\lambda N_{c} + \epsilon - s')}{\delta}$.
\begin{eqnarray*}
Pr\brac{S = s'} = \frac{s_{1} - (\lambda N_{c} + \epsilon) }{s_{1} - s'}, \\
Pr\brac{S = s_{1}} = \frac{\lambda N_{c} + \epsilon - s'}{s_{1} - s'}.
\end{eqnarray*}
\end{lemma}
We provide an outline of the proof.
We note that if the distribution needs to achieve only $p_{e}(\epsilon) + \delta$, then we can choose the points $p_{1} = (s_{1}, c_{s}(s_{1}, N_{c}))$ (as shown in Figure \ref{chap2:fig:opt_prob_b}) and $p_{2} = (s', c_{s}(s', N_{c}))$. 
The point $(\lambda N_{c} + \epsilon, p_{e}(\epsilon) + \delta)$ lies on the line joining these two points and is therefore a convex combination of the points $p_{1}$ and $p_{2}$.
Solving for $s_{1}$ and the convex combination leads to the proof of the above lemma.

We note that the $(\delta,\epsilon)$-optimal distribution has finite support for any $\delta > 0$ and $\epsilon > 0$.

\section{Proof of Proposition \ref{chap2:prop:exhub}}
  \label{chap2:app:exhub}
  \begin{proof}
    For the policy $e$, under stationary conditions, we have that $Q = s(Q)$ and $\tau(s) = \ceiling{as + b}, \forall s$.
    Suppose $\Exp_{\pi_{e}} Q < \infty$ for $e$, then we have that $\Exp_{\pi_{e}} Q = \lambda \Exp_{\pi_{e}} \tau(Q)$.
    For brevity, in this proof we use $\Exp\bras{.}$ to denote $\Exp_{\pi_{e}}\bras{.}$.
    Since $as + b \leq \tau(s) \leq as + b + 1$, we obtain that $\Exp Q \geq \frac{b\lambda}{1 - a\lambda}$ and $\Exp Q \leq \frac{(b + 1)\lambda}{1 - a\lambda}$.
    Also $\frac{b}{1 - a\lambda} \leq \Exp \tau(Q) \leq \frac{b + 1}{1 - a\lambda}$.
    We also have that
    \begin{eqnarray*}
      \Exp Q^{2} & = & \sigma^{2} \Exp \tau(Q) + \lambda^{2} \Exp \tau(Q)^{2}, \\
      & \leq & \sigma^{2} (a \Exp Q + b + 1) + \lambda^{2}\brap{a^{2} \Exp Q^{2} + (b + 1)^{2} + 2a(b + 1)\Exp Q} \text{ or, } \\
      \Exp Q^{2} & \leq & \frac{(b + 1)\sigma^{2} + (b + 1)^{2}\lambda^{2} + \frac{(b + 1)\lambda}{1 - a\lambda}(a\sigma^{2} + 2a(b + 1)\lambda^{2})}{(1 - a^{2}\lambda^{2})}.
    \end{eqnarray*}
    For the policy $e$,  $\Exp Q \leq \frac{\Exp Q(aQ + b + 1) + \Exp \frac{(aQ + b)(aQ + b + 1)\lambda}{2}}{\Exp \tau(Q)}$. 
    Substituting the upper bounds for $\Exp Q^{2}$, and $\Exp Q$, and the lower bound for $\Exp \tau(Q)$ in the above expression leads to the upper bound on the average queue length for the EXH policy.
  \end{proof}

\section{Proof of Lemma \ref{chap5:lemma:Deltaservice}}
\label{chap5:app:Deltaservice}
\begin{proof}
We note that $q_{s'} = \inf\brac{q : \Exp S(q) \geq s' - \epsilon}$, for $\epsilon$ such that $0 < \epsilon < s' - \Delta$, for $\Delta$ in RA2.
We note that by definition, $\forall q \geq q_{s'}$, 
\begin{eqnarray}
\Exp S(q) & \geq &  s' - \epsilon, \nonumber \\
\int_{0}^{S_{max}} S(q). dP(S(q)) & \geq & s' - \epsilon.
\label{chap5:eq:deltaservice0}
\end{eqnarray}
Then, 
\begin{eqnarray*}
\int_{0}^{\Delta} \Delta dP(S(q)) + \int_{\Delta}^{S_{max}} S_{max} dPS(q) & \geq & s' - \epsilon,\\
\Delta \brap{1 - Pr\brac{S(q)  > \Delta}} + S_{max} Pr\brac{S(q) > \Delta} & \geq & s' - \epsilon, \text{ or}, \\
Pr\brac{S(q) > \Delta} & \geq & \frac{s' - \epsilon - \Delta}{S_{max} - \Delta}.
\end{eqnarray*}
Thus for any $q \geq q_{s'}$, $Pr\brac{S(q) > \Delta} \geq \delta_{s} > 0$, where $\delta_{s} = \frac{s' - \epsilon - \Delta}{S_{max} - \Delta}$.
\end{proof}

\section{Upper bound for average queue length for the policy $\mu$ in Proposition \ref{chap2:prop:mukub}}
\label{chap2:app:ubformu}
For the policy $\mu$ with parameters $N_{c}$ and $r$, we have that a batch of size $s(q) = \min(q, r N_{c})$ is served when the queue length is $q$.
We note that if the arrival distribution is such that $A_{max} < r$, so that $A[1] < r N_{c}$, then the average queue length $\Exp_{\mu} Q = \lambda N_{c}$.

The following upper bound, holds for any arrival distribution with $\Exp A^{2} = \Exp \brap{A[1]^{2}} < \infty$.
From \eqref{chap2:eq:evolution_equation} we have that
\begin{eqnarray*}
  \Exp\brac{Q[m + 1]^{2} - Q[m]^{2} \middle \vert Q[m] = q} & = & -2q(s(q) - \lambda N_{c}) + \Exp A^{2} + s(q)^{2} - 2s(q)\lambda N_{c}, \\
  & = & -2q N_{c}(r - \lambda) + 2q(rN_{c} - s(q)) + \Exp A^{2} + s(q)^{2} - 2s(q)\lambda N_{c}.
\end{eqnarray*}
Now, by taking expectations with respect to the stationary distribution $\pi$ of $\mu$, we have that
\begin{eqnarray}
  \Expp Q & = & \frac{1}{2 N_{c}(r - \lambda)} \bras{ 2 \Expp \bras{Q(r N_{c} - s(Q))} + \Exp A^{2} + \Expp s(Q)^{2} - 2 \lambda^{2} N_{c}^{2}}, \nonumber \\
  & = & \frac{1}{2 N_{c} (r - \lambda)} \bras{\sigma^{2} N_{c} + 2 \Expp \bras{Q(r N_{c} - s(Q))} + \Expp s(Q)^{2} - \lambda^{2} N_{c}^{2}}.
  \label{chap2:eq:ubformu0}
\end{eqnarray}
We now simplify and provide an upper bound for $2 \Expp \bras{Q(r N_{c} - s(Q))} + \Exp_{\mu} s(Q)^{2} - \lambda^{2} N_{c}^{2}$.

We denote the stationary probability $Pr\brac{s(Q) = r N_{c}} = Pr\brac{Q > rN_{c}}$ by $p_{r}$.
Then we have that $2 \Expp \bras{Q(r N_{c} - s(Q))} + \Expp s(Q)^{2} - \lambda^{2} N_{c}^{2}$
\begin{eqnarray}
  & = & 2\int_{q < rN_{c}} q(rN_{c} - q)d\pi(q) + \int_{q < rN_{c}} q^{2} d\pi(q) + p_{r} r^{2} N_{c}^{2} - \lambda^{2} N_{c}^{2}, \nonumber \\
  & = & 2\int_{q < rN_{c}} q r N_{c} d\pi(q) - \int_{q < rN_{c}} q^{2} d\pi(q) + p_{r} r^{2} N_{c}^{2} - \lambda^{2} N_{c}^{2}, \nonumber \\
  & = & 2rN_{c}\bras{\lambda N_{c} - p_{r} r N_{c}} - \int_{q < rN_{c}} q^{2} d\pi(q) + p_{r}r^{2} N_{c}^{2} - \lambda^{2} N_{c}^{2}, \nonumber \\
  & = & 2r\lambda N_{c}^{2} - \int_{q < rN_{c}} q^{2} d\pi(q) - p_{r} r^{2} N_{c}^{2} - \lambda^{2} N_{c}^{2}, \nonumber \\
  & = & \lambda N_{c}^{2} (r - \lambda) + r\lambda N_{c}^{2} - p_{r} r^{2} N_{c}^{2} - \int_{q < rN_{c}} q^{2} d\pi(q).
  \label{chap2:eq:ubformu2}
\end{eqnarray}
Therefore, 
\begin{eqnarray}
  2 \Expp \bras{Q(r N_{c} - s(Q))} + \Expp s(Q)^{2} - \lambda^{2} N_{c}^{2}  & \leq & \lambda N_{c}^{2} (r - \lambda) + r\lambda N_{c}^{2}.
  \label{chap2:eq:ubformu}
\end{eqnarray}
Substituting in \eqref{chap2:eq:ubformu0}, we obtain that
\begin{eqnarray*}
  \Expp Q \leq \frac{\sigma^{2}}{2(r - \lambda)} + \frac{\lambda N_{c}}{2} + \frac{r \lambda N_{c}}{2(r - \lambda)}.
\end{eqnarray*}

We now present another upper bound on $2 \Expp \bras{Q(r N_{c} - s(Q))} + \Expp s(Q)^{2} - \lambda^{2} N_{c}^{2}$, obtained by lower bounding $p_{r}$ in \eqref{chap2:eq:ubformu2}.
The lower bound on $p_{r}$ is obtained under the assumption that $Pr\brac{A[1] \geq r N_{c}} > 0$.
Let $p_{l} = 1 - p_{r}$.
We note that $p_{l}$ is the fraction of time a batch size less than $r N_{c}$ is used.
We note that the evolution of $(Q[m])$ can be divided into cycles of random duration.
Each cycle comprises of two periods, where each period is also of random duration.
The first period starts in slot $m$, if $Q[m - 1] \geq r N_{c}$ and $Q[m] < r N_{c}$.
The duration of the first period is distributed according to a Geometric distribution with mean $\frac{1}{Pr\brac{A[1] \geq rN_{c}}}$.
Following the first period, we have the second period which starts in a slot $m$ such that $Q[m - 1] < rN_{c}$ and $Q[m] \geq r N_{c}$.
We note that throughout the second period the queue length is greater than or equal to $r N_{c}$ and a batch service of $r N_{c}$ occurs in each slot.

Let $Q_{p}$ be the random queue length at the start of the second period.
We note that $Q_{p} \sim A[1]$ conditioned on $\brac{A[1] \geq r N_{c}}$.
Let $Q_{p} = q_{p}$.
Since $(Q[m])$ is Markov, the duration of the second period $T_{p}(q_{p})$ is then
\begin{eqnarray*}
  T_{p}(q_{p}) = \min\brac{t : q_{p} + \sum_{m = 1}^{t} \bras{A[m] - r N_{c}} < r N_{c}}.
\end{eqnarray*}
Then, applying Wald's lemma for $q_{p}$, and taking expectations over the distribution of $Q_{p}$, we have that
\begin{eqnarray*}
  \Exp T_{p}(Q_{p}) \geq \frac{\Exp_{\bras{A[1]|A[1] \geq rN_{c}}}\bras{ Q_{p} - r N_{c}}}{r N_{c} - \lambda N_{c}}.
\end{eqnarray*}
We note that 
\begin{eqnarray*}
  p_{l} & = & \frac{\frac{1}{Pr\brac{A[1] \geq rN_{c}}}}{\frac{1}{Pr\brac{A[1] \geq rN_{c}}} + \Exp T_{p}(Q_{p})}, \\
  & \leq & \frac{\frac{1}{Pr\brac{A[1] \geq rN_{c}}}}{\frac{1}{Pr\brac{A[1] \geq rN_{c}}} + \frac{\Exp_{\bras{A[1]|A[1] \geq rN_{c}}} \bras{Q_{p} - r N_{c}}}{r N_{c} - \lambda N_{c} }}.
\end{eqnarray*}
Since $\Exp_{\bras{A[1]|A[1] \geq rN_{c}}} Q_{p} = \frac{\int_{a \geq rN_{c}} a dP_{A[1]}(a)}{Pr\brac{A[1] \geq rN_{c}}}$, we have that
\begin{eqnarray*}
  p_{l} & \leq & \frac{1}{1 + \frac{\bras{\int_{a \geq rN_{c}} (a - r N_{c})dP_{A[1]}(a)}}{r N_{c} - \lambda N_{c}}}, \\
  & = & \frac{N_{c} ( r - \lambda)}{N_{c}(r - \lambda) + \bras{\int_{a \geq rN_{c}} (a - r N_{c})dP_{A[1]}(a)}}.
\end{eqnarray*}
Therefore,
\begin{eqnarray*}
  p_{r} & \geq & \frac{\bras{\int_{a \geq rN_{c}} (a - r N_{c})dP_{A[1]}(a)}}{N_{c}(r - \lambda) + \bras{\int_{a \geq rN_{c}} (a - r N_{c})dP_{A[1]}(a)}}.
\end{eqnarray*}

Substituting the above lower bound in the RHS of \eqref{chap2:eq:ubformu2} and using $\int_{q < rN_{c}} q^{2} d\pi(q) \geq 0$, we have that $2\lambda N_{c}^{2} (r - \lambda) + \lambda N_{c}^{2} \brap{\lambda - \frac{p_{r} r^{2}}{\lambda}}$
\begin{eqnarray*}
  \leq 2\lambda N_{c}^{2} (r - \lambda) + N_{c}^{2} \brap{\lambda^{2} - \frac{\bras{\int_{a \geq rN_{c}} (a - r N_{c})dP_{A[1]}(a)}}{N_{c}(r - \lambda) + \bras{\int_{a \geq rN_{c}} (a - r N_{c})dP_{A[1]}(a)}} r^{2}}.
\end{eqnarray*}
The above expression can then be simplified to
\begin{eqnarray}
  2\lambda N_{c}^{2} (r - \lambda) + (r - \lambda) N_{c}^{2} \, \frac{\brap{\lambda^{2}N_{c} - \bras{\int_{a \geq rN_{c}} (a - r N_{c})dP_{A[1]}(a)}(r + \lambda)}}{N_{c}(r - \lambda) + \bras{\int_{a \geq rN_{c}} (a - r N_{c})dP_{A[1]}(a)}}.
  \label{chap2:eq:ubformu1}
\end{eqnarray}

Therefore, $2 \Expp \bras{Q(r N_{c} - s(Q))} + \Expp s(Q)^{2} - \lambda^{2} N_{c}^{2}$
\small
\begin{eqnarray}
  \leq  2\lambda N_{c}^{2} (r - \lambda) + (r - \lambda) N_{c}^{2} \, \frac{\brap{\lambda^{2}N_{c} - \bras{\int_{a \geq rN_{c}} (a - r N_{c})dP_{A[1]}(a)}(r + \lambda)}}{N_{c}(r - \lambda) + \bras{\int_{a \geq rN_{c}} (a - r N_{c})dP_{A[1]}(a)}}.
\end{eqnarray}
\normalsize
Substituting in \eqref{chap2:eq:ubformu0}, we obtain that
\begin{eqnarray*}
  \Expp Q \leq \frac{\sigma^{2}}{2(r - \lambda)} + \lambda N_{c} + \frac{N_{c}}{2} \, \frac{\brap{\lambda^{2}N_{c} - \bras{\int_{a \geq rN_{c}} (a - r N_{c})dP_{A[1]}(a)}(r + \lambda)}}{N_{c}(r - \lambda) + \bras{\int_{a \geq rN_{c}} (a - r N_{c})dP_{A[1]}(a)}}.
\end{eqnarray*}

\section{Examples for $\tilde{c}_{s}(s, N_{c})$ in Remark \ref{chap2:remark:otherapproximations}}
\label{chap2:app:examples}
\begin{figure}
  \centering
  \includegraphics[width=120mm,height=60mm]{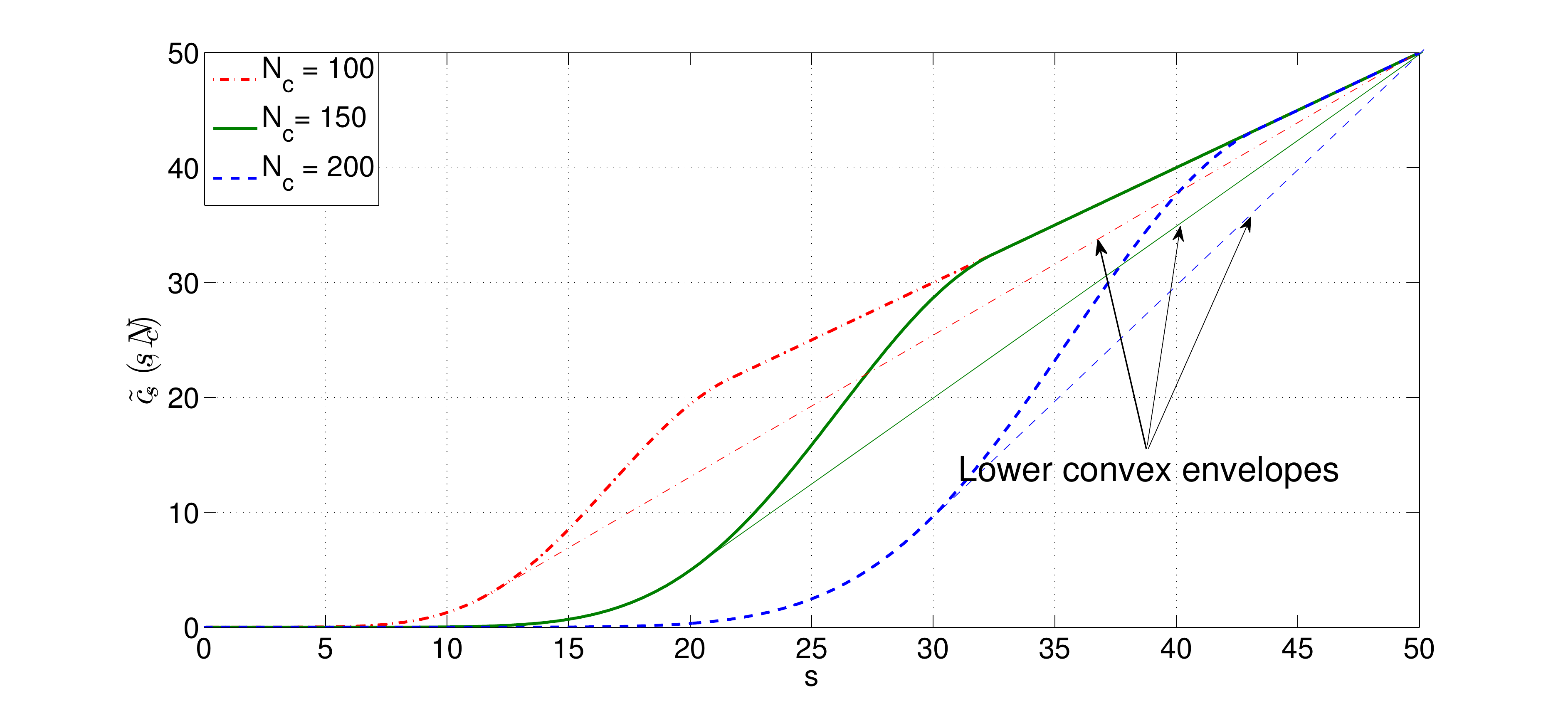}
  \caption{Illustration of $\tilde{c}_{s}(s,N_{c})$ and its lower convex envelope, for $N_{c} \in \brac{100, 150, 200}$, for the parameters in Example 1}
  \label{chap2:fig:p5fig1}
\end{figure}

In this section we provide representative numerical examples which illustrate that the lower convex envelope $\tilde{c}(s)$ of $\tilde{c}_{s}(s, N_{c})$ has a similar form as $c(s)$.
The transition probability matrices and the input distribution $Q$ have been generated randomly for these examples.
\paragraph{Example 1 : }
For this example we take $\cM = 2$, and a DMC with $\vert \mathcal{X} \vert = 5$ and $\vert \mathcal{Y} \vert = 10$.
The transition probability matrix $P_{Y|X}$ is :
\[ \bras{
  \begin{array}{cccccccccc}
    0.0459 &  0.2101 &  0.1339 &  0.1138 &  0.1094 &  0.1212 &  0.0348 &  0.1859 & 0.0309 &  0.0140 \\
    0.0108 &  0.2152 &  0.0646 &  0.0593 &  0.0210 &  0.0341 &  0.2328 &  0.1601 & 0.0612 &  0.1409 \\
    0.1251 &  0.1038 &  0.0403 &  0.1645 &  0.1146 &  0.1375 &  0.0872 &  0.0727 & 0.0657 &  0.0886 \\
    0.0728 &  0.1252 &  0.1291 &  0.1038 &  0.0062 &  0.1191 &  0.1417 &  0.1202 & 0.1211 &  0.0607 \\ 
    0.1000 &  0.1678 &  0.0060 &  0.1042 &  0.0149 &  0.1505 &  0.1352 &  0.0174 & 0.1673 &  0.1368 \\
\end{array}} \]
The input distribution $Q$ is $(0.2018, 0.2551, 0.2515, 0.0943, 0.1972)$.
Then the numerically obtained $\tilde{c}_{s}(s,N_{c})$ for three values of $N_{c}$ is shown in Figure \ref{chap2:fig:p5fig1}.

\begin{figure}
  \centering
  \includegraphics[width=120mm,height=60mm]{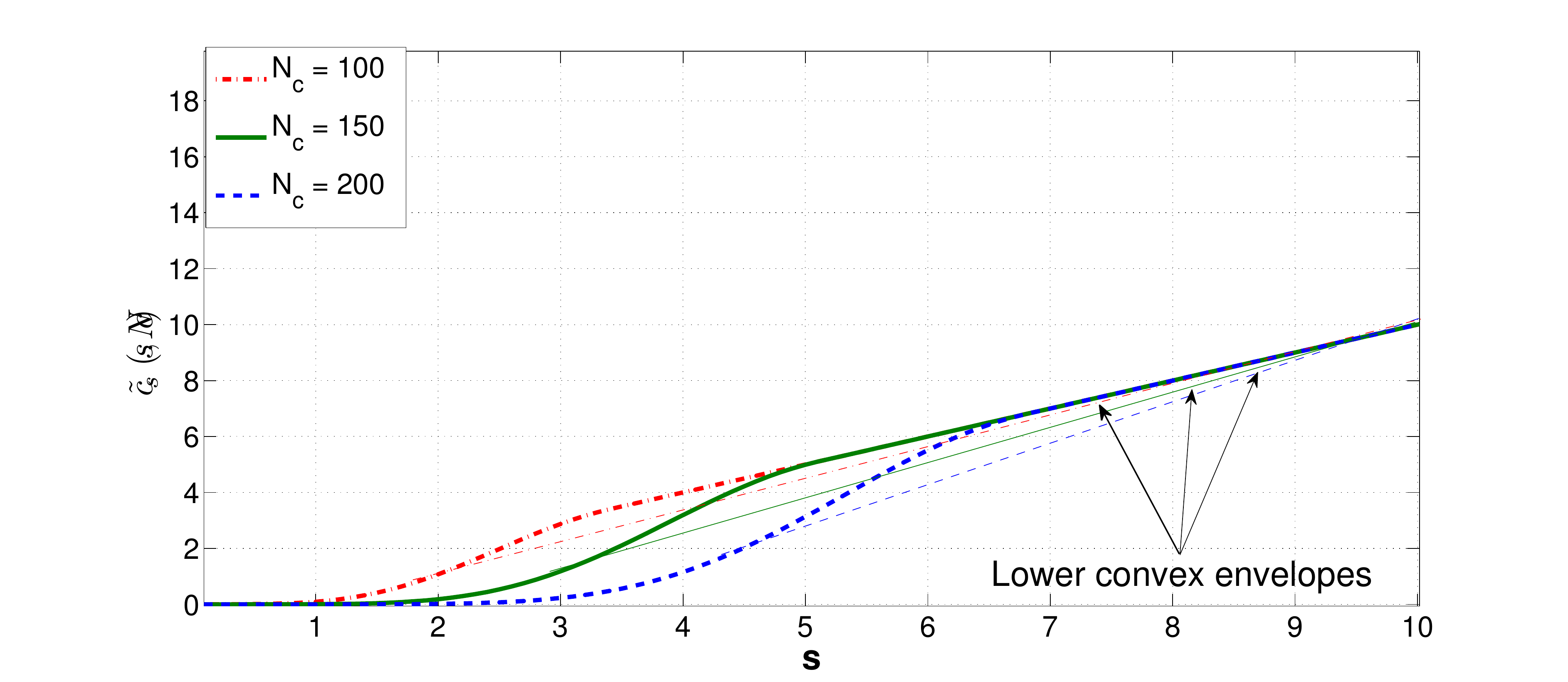}
  \caption{Illustration of $\tilde{c}_{s}(s,N_{c})$ and its lower convex envelope, for $N_{c} \in \brac{100, 150, 200}$, for the parameters in Example 2}
  \label{chap2:fig:p5fig2}
\end{figure}

\paragraph{Example 2 : }
For this example we take $\cM = 20$, and a DMC with $|\mathcal{X}| = 5$ and $|\mathcal{Y}| = 5$.
The transition probability matrix $P_{Y|X}$ is :
\[ \bras{
  \begin{array}{ccccc}
    0.0649 &  0.2915 &  0.1642 &  0.1986 &  0.2808 \\
    0.1557 &  0.2722 &  0.0509 &  0.1063 &  0.4148 \\ 
    0.1883 &  0.1538 &  0.2409 &  0.1187 &  0.2984 \\ 
    0.4160 &  0.0809 &  0.0959 &  0.2615 &  0.1457 \\ 
    0.0860 &  0.2355 &  0.2115 &  0.1459 &  0.3212 \\
\end{array}} \]
The input distribution $Q$ is $(0.0739,0.0141,0.2423,0.3808,0.2889)$.
Then the numerically obtained $\tilde{c}_{s}(s,N_{c})$ for three values of $N_{c}$ is shown in Figure \ref{chap2:fig:p5fig2}.
We note that in all these cases the form of the lower convex envelope of $\tilde{c}_{s}(s, N_{c})$ is the same as that illustrated in Figure \ref{chap2:fig:csrmodelb}.

\end{subappendices}

\addtocontents{toc}{\protect\setcounter{tocdepth}{2}}

\blankpage
\chapter{\textbf{Conclusions and Scope for Future Work}}
We first summarize the main results which are obtained in this thesis.
Motivated by the already available results on the monotonicity property of any stationary deterministic optimal policy for the unconstrained tradeoff problem, in this thesis we consider the constrained tradeoff problem for the set of admissible policies, which are defined to be monotone.

In Chapters 2 and 3, for the state dependent M/M/1 model, using geometric bounds on the stationary probability distribution of the queue length for admissible policies, we obtain the asymptotic behaviour of the solution to the constrained tradeoff problem for admissible policies in the regime $\Re$, as well as asymptotic bounds on any sequence of order optimal admissible policies.
We identify a case in which the average queue length grows only to a finite value in the regime $\Re$.
For the cases for which the average queue length grows without bound, we show that the asymptotic behaviour of the average queue length is either $\Theta\brap{\log\nfrac{1}{V}}$, $\Theta\nfrac{1}{V}$, or $\Theta\nfrac{1}{\sqrt{V}}$.
The asymptotic behaviour of the average queue length is determined by: (i) the nature of the service cost function, i.e., whether it is piecewise linear, a corner point, or strictly convex, at the value of $\lambda$ or $u^{-1}(u_{c})$, and (ii) the extent of freedom that we have in the control of the arrival rates $(\lambda(q))$ and service rates $(\mu(q))$.

Guided by the analysis in Chapters 2 and 3, we obtain asymptotic bounds for the solution of the constrained tradeoff problem for a discrete time model with a fixed environment state in Chapter 4.
We again consider admissible policies, for which geometric bounds on the stationary probability distribution for the queue length are obtained.
The forms of these geometric bounds are motivated by the geometric bounds which were derived for state dependent M/M/1 model in Chapters 2 and 3.
Using these geometric bounds, asymptotic lower bounds are derived for the solution to the constrained tradeoff problem in the regime $\Re$.
Asymptotic upper bounds are also derived, using which a complete asymptotic characterization of the tradeoff in the regime $\Re$ is obtained in two cases.
Asymptotic bounds on any sequence of order-optimal admissible policies are also derived.

In Chapter 5, we first consider the tradeoff of average power and average delay for a noisy point-to-point link with fast fading.
Asymptotic bounds for the solution to the constrained tradeoff problem, for admissible policies, in the regime $\Re$ are then derived by extending the results of Chapter 4.
We also compare the asymptotic lower bounds which are obtained from a real valued approximate queueing model to those from the original integer valued queueing model.
We find that the real valued approximate queueing model with a strictly convex cost function underestimates the average service cost and average delay for the original model.
We show that a more appropriate approximate real valued queueing model is one in which the service cost function is the piecewise linear lower convex envelope of the service cost function for the original model.
We also obtain asymptotic lower bounds for the constrained tradeoff problem for: (i) a model with admission control, and (ii) a single hop network model.
Asymptotic lower bounds are also obtained for the case when the arrival process and fading process are ergodic.

In Chapter 6, we consider the tradeoff of average error rate and average delay for a noisy point-to-point link.
We obtain the exponential decay rate of average error rate with respect to average delay, in the regime of large average delay, for fixed length random block coding schemes, with control on the codeword length parameter $N_{c}$.
Using results from Chapter 4, we then obtain an asymptotic characterization of the tradeoff of average error rate and average delay, for the set of admissible policies, for a fixed $N_{c}$.
One of the main contributions in Chapter 6 is the analysis of the constrained tradeoff problem for a non-convex service cost function, unlike the convex service cost functions in the earlier chapters.
We show that the asymptotic behaviour of the tradeoff in the regime $\Re$ is determined by the lower convex envelope for any non-convex service cost function.

\section{Scope for future work}
We note that all the asymptotic bounds obtained in this thesis are order bounds.
From numerical computation, we have observed that the constants which are involved in the upper and lower bounds are weak.
Tight non-asymptotic bounds for the constrained tradeoff problem are still not available.

We now discuss some specific problems that arise from the analysis in the previous chapters, for which some initial results have been obtained.
\subsection{Order optimality of policies obtained from a fluid model}
\label{chap7:sec:fluid}
A simpler deterministic model for the evolution of the queue length, for the state dependent M/M/1 model in Chapter 2, is the trajectory $q(t)$ obtained from the fluid model:
\begin{equation}
  \frac{dq(t)}{dt} = \lambda - \mu(t), q(0) = q_{0},
  \label{eq:fluidevol}
\end{equation}
where $\mu(t)$ is the service rate at time $t$.
The fluid model can be interpreted as a \emph{limiting} form of the evolution of the integer valued queue length, when both time as well as queue length are scaled.
Such fluid models can also be obtained for the discrete time model in Chapter 4, as well as for general networks, e.g. \cite[Chapter 10]{meynctcn}.
In \cite[Chapter 10]{meynctcn} and \cite{chen}, it is shown that the policy $\mu(t)$ which minimizes the total \emph{cost} $J(q_{0})$ for the deterministic fluid model, can be used to obtain \emph{good} policies for the unconstrained MDP for the original stochastic state dependent M/M/1 model.
Our objective is to investigate the order optimality of policies which are obtained from fluid models in the asymptotic regime $\Re$.

The state transition diagram for the unconstrained MDP is the same as that in Figure \ref{chap4:fig:mm1model}, with the action at each state $q$ being the service rate $\mu(q) \in \{0, \mu_{1}, \cdots, \mu_{K}\}$.
We also note that the optimal policy for the MDP with single stage cost $q + \beta c(\mu)$ is the same as that for the MDP with single stage cost $q + \beta (c(\mu) - c(\lambda))$.
We construct a heuristic policy for the model, when $c(\mu) = \mu^{2}$, for $\mu \in \mathcal{S} = \brac{0, \mu_{1}, \dots, \mu_{K}}$, based upon the analysis in Chen et al. \cite{chen}.
As in Chen et al. \cite{chen}, we consider the fluid model \eqref{eq:fluidevol}, with a modified single stage cost $q + \beta\bras{c(\mu) - c(\lambda)}^+$ and the service rate at time $t$, $\mu(t) \in \mathbb{R}_+$.
Then \[J(q_{0}) = \inf_{\mu(t)} \int_{0}^{\infty} \brac{q(t) + \beta \bras{c(\mu(t)) - c(\lambda)}^+ }dt.\]
If $c(\mu) = \mu^{2}$, then from \cite{chen} we have that $J(q) = \frac{2}{3} q^{\frac{3}{2}}  \brap{\frac{1}{\beta}}^{\frac{1}{2}}$.
Furthermore the optimal service rate function $\mu^*(t)$ for which 
\[J(q_{0}) = \int_{0}^{\infty} \brac{q(t) + \beta \bras{c(\mu^*(t)) - c(\lambda)}^+}dt,\]
is given by the state dependent policy $\mu^*(q) = (q/\beta)^{\frac{1}{2}} + c(\lambda)$.
Motivated by this development, we define the fluid policy $\gamma_{F, \beta}$ for our M/M/1 model as choosing the service rate 
\[\mu_{F, \beta}(q) = \bras{ (q/\beta)^{\frac{1}{2}} }_{\mathcal{S}},\]
where $\bras{x}_{\mathcal{S}}$ means that we pick a service rate in $\mathcal{S}$ which is closest to the argument $x$.
Interestingly, the policy $\gamma_{F,\beta}$ does not depend on $\lambda$ but varies only as a function of $\beta$.
We note that $\gamma_{F,\beta}$ is admissible for any $\lambda < \mu_{K}$, since there exists a queue length $q$ such that $\mu_{F,\beta}(q) = \mu_{K}$ and from the definition, $\mu_{F,\beta}(q)$ is non-decreasing in $q$.
We illustrate the order optimality of $\gamma_{F,\beta}$ only for case 2.
We first obtain an upper bound on $\overline{Q}(\gamma_{F,\beta})$ using Proposition \ref{chap4:app:prop:dotzupperbound}.
It can be shown that for $\epsilon > 0$, if there exists a ${q}_{\epsilon}$ such that $\mu_{F,\beta}(q_{\epsilon}) - \lambda \geq \epsilon$, then $\overline{Q}(\gamma_{F,\beta}) \leq \frac{{q}_{\epsilon}(\epsilon + \lambda)}{\epsilon} + \frac{\lambda + \mu_{K}}{2\epsilon}$.

We note that if $\lambda < \mu_{K}$, for case 2, we can pick $\epsilon = \mu_{u} - \lambda$.
Then $q_{\epsilon} \leq \beta \brap{\frac{\mu_{k_{u} - 1} + \mu_{u}}{2}}^{2} + 1$.
Thus we obtain that $\overline{Q}(\gamma_{F,\beta}) = \mathcal{O}(\beta)$.

We have that $\overline{C}(\gamma_{F,\beta})$
\small
\begin{eqnarray}
  & = &  \sum_{k < k_{l}} \pi_{\mu}(k) c(\mu_{k}) + \sum_{k > k_{u}} \pi_{\mu}(k) c(\mu_{k}) + \sum_{k = k_{l}}^{k_{u}} \pi_{\mu}(k) c(\mu_{k}), \nonumber \\
  & \leq & c(\mu_{k_{l} - 1}) Pr\brac{\mu_{F,\beta}(Q) < \mu_{l}} + c(\mu_{K})Pr\brac{\mu_{F,\beta}(Q) > \mu_{u}} \nonumber \\
  & & + c(\lambda) + \sum_{k = k_{l}}^{k_{u}} \pi_{\mu}(k) m(\lambda - \mu_{k}),
  \label{eq:fluidub1}
\end{eqnarray}
\normalsize
where $m$ is the slope of the line joining $(\mu_{l}, c(\mu_{l}))$ and $(\mu_{u}, c(\mu_{u}))$.
We note that since $\gamma_{F,\beta}$ is admissible, $\sum_{k = 0}^{K} \pi_{\mu}(k) \mu_{k} = \lambda$.
Hence, we have that $\sum_{k < k_{l}} \pi_{\mu}(k)\brap{\mu_{k} - \lambda} + \sum_{k > k_{u}} \pi_{\mu}(k)\brap{\mu_{k} - \lambda} = \sum_{k = k_{l}}^{k_{u}} \pi_{\mu}(k)\brap{\lambda - \mu_{k}}$.
Since for $k < k_{l}, \mu_{k} < \lambda$, we have that $\sum_{k = k_{l}}^{k_{u}} \pi_{\mu}(k)\brap{\lambda - \mu_{k}} \leq \brap{\mu_{K} - \lambda} Pr\brac{\mu_{F,\beta}(Q) > \mu_{u}}$.
We note that $Pr\brac{\mu_{F,\beta}(Q) > \mu_{u}} = \sum_{q > q_{k_{u}}} \pi(q)$ and $Pr\brac{\mu_{F,\beta}(Q) < \mu_{l}} = \sum_{q \leq q_{k_{l} - 1}} \pi(q)$.
From the birth-death structure of $Q(t)$ under $\gamma_{F,\beta}$, we can show that
\begin{eqnarray*}
  \sum_{q > q_{k_{u}}} \pi(q) \leq \pi(q_{k_{u}} + 1)\nfrac{\mu_{k_{u} + 1}}{\mu_{k_{u} + 1} - \lambda},
\end{eqnarray*}
\vspace{-0.2in}
\begin{eqnarray*}
  \sum_{q \leq q_{k_{l} - 1}} \pi(q) \leq \pi(q_{k_{l} - 1}) \nfrac{1 - \fpow{\mu_{k_{l} - 1}}{\lambda}{q_{k_{l} - 1} + 1}}{1 - \frac{\mu_{k_{l} - 1}}{\lambda}}.
\end{eqnarray*}
We note that $\pi(q_{k_{u}} + 1) = \pi(q_{k_{u} - 1})\fpow{\lambda}{\mu_{u}}{q_{k_{u}} - q_{k_{u} - 1} + 1}$ and $\pi(q_{k_{l} - 1}) = \pi(q_{k_{l}})\fpow{\mu_{l}}{\lambda}{q_{k_{l}} - q_{k_{l} - 1}}$.
From the definition of $\mu_{F,\beta}$ we have that $q_{k_{u}} - q_{k_{u} - 1}$ and $q_{k_{l}} - q_{k_{l} - 1}$ are both $\Theta(\beta)$, as $\beta \uparrow \infty$.
Since $\pi(q_{k_{u} - 1})$ and $\pi(q_{k_{l} - 1})$ are both bounded above by one, we have that $Pr\brac{\mu_{F,\beta}(Q) > \mu_{u}}$ as well as $Pr\brac{\mu_{F,\beta}(Q) < \mu_{l}}$ are both $\rho^{\Theta(\beta)}$, where $0 < \rho < 1$.
Then we have that $\overline{C}(\gamma_{\beta}) - c(\lambda) = \rho^{\Theta(\beta)}$ and $\overline{Q}(\gamma_{\beta}) = \mathcal{O}(\beta)$.
In summary, we have shown a new optimality property of heuristic policies, obtained from a fluid model analysis.
This raises the question whether such fluid policies are order optimal even for general network scenarios.

\subsection{Extensions to systems with service time control }
We note that the service rate control variable has been the service batch size for all the discrete time queueing models considered in this thesis.
However, there are scenarios where both batch size and batch service time can be controlled, e.g., for a noisy point to point link which uses block coding, both the number of encoded message symbols (the batch size) and the codeword length can be dynamically controlled (the service time for the batch) to tradeoff the average error rate with the average delay of the message symbols.
Such models also arise as a special case, when \emph{single decision policies with observed initial information} are used, for the general class of renewal models considered by Neely \cite{neely_renewal}.
We note that for such models the service cost is then modelled as a function of both batch size and batch service time.

We now comment on how asymptotic lower bounds can be derived for such models, using the methods in Chapter 4.
We note that for stationary policies, that decide on the batch size and batch service time as a function of the current queue length, the queue length evolution sampled at the decision epochs evolves as a semi-Markov process.
By uniformization \cite{tijms} the average cost and average queue length for a semi-Markov process can be obtained via an equivalent Markov process.
The stagewise drift of this equivalent Markov process depends on both the batch size as well as the batch service time.
We expect that the analysis of this equivalent Markov process is similar to the analysis carried out in Chapter 4, but with the above drift function, under the restriction to admissible policies.

We consider a simplified model in \cite{vineeth_servicetimecontrol}, wherein asymptotic lower bounds are derived for a continuous time queueing model with Poisson arrivals and service time control with service batch size fixed to be $1$.
For this model, as the service batch size is fixed to be $1$, whenever there is service, the service cost is a function only of the batch service time, which is real valued.
For the above continuous time model, where the queue evolution is on the integers, if the service cost per unit batch service time is a strictly convex function of the batch service time, then we obtain that the minimum average queue length grows as $\Omega\nfrac{1}{\sqrt{V}}$ when the average service cost constraint is $V$ more than the minimum average service cost required for stability.
Obtaining asymptotic bounds for queueing models where both batch size and batch service time can be controlled, is a problem which has scope for future work.

\blankpage
\bibliographystyle{plain}
\bibliography{thesis}
\end{document}